\documentclass[11pt,fleqn,a4paper,polish]{report}
\usepackage{color}
\usepackage{amssymb}
\usepackage{amsmath}

\usepackage{esint}
\makeatletter

\usepackage{amsfonts}\@ifundefined{definecolor}
 {\@ifundefined{definecolor}
 {\@ifundefined{definecolor}
 {\@ifundefined{definecolor}
 {\usepackage{color}}{}
}{}
}{}
}{}
\usepackage{indentfirst}\usepackage{array}\usepackage{pstricks}\usepackage{xcolor}\usepackage{axodraw}\usepackage{subfigure}

\usepackage{latexsym}

\makeatother

\usepackage{babel}
\usepackage{pifont}
\usepackage[OT4]{polski}
\usepackage[cp1250]{inputenc}

\usepackage
[dvips,
colorlinks=true,
linkcolor=webgreen,
filecolor=webbrown,
citecolor=webgreen,
bookmarksopen=false,
]{hyperref}
\definecolor{webgreen}{rgb}{0,.5,0}
\definecolor{webbrown}{rgb}{.6,0,0}
\usepackage[T1]{fontenc}

\usepackage{array}
\usepackage{graphicx}
\usepackage{antpolt}
\usepackage{srcltx}
\usepackage{makeidx}
\usepackage{amsmath}

\usepackage{pstricks}
\usepackage{pst-node}
\usepackage[cp1250]{inputenc}
\usepackage{amssymb}
\date{}
\title{Metoda największej wiarygodności i informacja Fisher'a w fizyce \\i ekonofizyce \\
\vspace{2mm}
\normalsize{(Maximum likelihood method and Fisher's information in physics and econophysics 
)}\\}
\author{ Jacek Syska\footnote{\large jacek.syska@us.edu.pl} \\
\\ 
Institute of Physics, University of Silesia,  Uniwersytecka 4, 40-007
Katowice, Poland  \\ \\ \\ \\ \\ \\ \\ \\ \\ 
\\ 
Skrypt dla studentów ekonofizyki\footnote{\large Współfinansowanie projektu ``Uniwersytet partnerem gospodarki opartej na wiedzy'' w ramach Europejskiego Funduszu Społecznego. Kapitał Ludzki Narodowa Strategia Spójności.
} \\
\\
\\
\\
}

\makeatletter

\begin{document}
\thispagestyle{empty}
\maketitle
\tableofcontents
\newpage
\setcounter{equation}{0}

\addcontentsline{toc}{chapter}{Abstract}
\chapter*{Abstract}

Three steps in the development of the maximum likelihood (ML) method are presented. At first,  the application of the ML method and Fisher information notion in the model selection analysis is described (Chapter 1).  The fundamentals of differential geometry in the construction of the statistical space are introduced, illustrated also by examples of the estimation of the exponential models. \\
At second, the notions of the relative entropy and the information channel capacity are introduced (Chapter 2). The observed and expected structural information principle (IP) and the variational IP of the modified extremal physical information (EPI) method of Frieden and Soffer are presented and discussed (Chapter 3). The derivation of the structural IP based on the analyticity of the logarithm of the likelihood function and on the metricity of the statistical space of the system is given. \\
At third, the use of the EPI method is developed (Chapters 4-5).  The information channel capacity is used for the field theory models classification. Next, the modified Frieden and Soffer EPI method, which is a nonparametric estimation that enables the statistical selection of the equation of motions of various field theory models (Chapter 4) or the distribution generating equations of statistical physics models (Chapter 5) is discussed.
The connection between entanglement of the momentum degrees of freedom and the mass of a particle is analyzed. The connection between the Rao-Cram{\'e}r inequality, the causality property of the  processes in the Minkowski space-time and the nonexistence of tachions is shown. The generalization of the Aoki-Yoshikawa sectoral productivity econophysical model is also presented (Chapter 5). Finally, the Frieden EPI method of the  analysis of the EPR-Bhom experiment is presented. It differs from the Frieden approach by the use of the information geometry methods. This work is destined mainly for students in physics and econophysics.  (At present only Polish version is available). 

\newpage

\addcontentsline{toc}{chapter}{Wstęp}
\chapter*{Wstęp}


Tematem skryptu jest metoda największej wiarygodności (MNW) oraz  informacja Fishera (IF) w fizyce i statystyce. Problem dotyczy bardzo aktualnego sposobu  konstrukcji modeli fizycznych, który wywodzi się ze statystycznego opisu zjawisk, którego formalizm pozwala na opis całego spektrum różnych teorii pola, klasycznych i kwantowych. Kluczowe w tym podejściu pojęcie (oczekiwanej) IF wprowadził Fisher na gruncie własnych rozważań związanych z oszacowywaniem parametrów modeli, podlegających badaniu statystycznemu w ramach ogólnej metody ekstremalnej wartości funkcji wiarygodności $L$. 
IF opisuje lokalne własności funkcji wiarygodności $L \equiv P(y|\Theta)$, która formalnie jest łącznym prawdopodobieństwem (lub łączną gęstością prawdopodobieństwa) danych $y$,  lecz jest rozumiana jako funkcja zbioru parametrów $\Theta$, który z kolei tworzy współrzędne w przestrzeni statystycznej. Analiza statystyczna modeli fizycznych idzie jak dotąd dwoma nurtami. 

Pierwszy z nich, geometryczny, próbuje opisać metodę statystyczną wprowadzoną w latach 20 poprzedniego wieku przez Fishera  \cite{Fisher}, twórcę podstaw techniki statystycznej otrzymywania dobrych estymatorów MNW, w jak się okazało naturalnym dla niej środowisku geometrii różniczkowej. Rozwijając MNW, już w 1945 roku C.R. Rao \cite{Rao} zauważył, że macierz informacji Fishera określa metrykę Riemanna i badał strukturę modeli statystycznych z punktu wiedzenia geometrii Riemanowskiej. Z kolei B. Efron \cite{Efron} badając jednoparametrowe modele i analizując ich asymptotyczne własności dla procedur estymacyjnych, wprowadził i odkrył użyteczność pojęcia statystycznej krzywizny. A.P. Dawid \cite{Dawid} wprowadził pojęcie koneksji na przestrzeni wszystkich dodatnio określonych rozkładów prawdopodobieństwa, pokazując, że ze względu na tą koneksję statystyczna krzywizna jest krzywizną zewnętrzną.  
Jednak problemem Dawida był nieskończony wymiar przestrzeni rozkładów. W roku 1980 S. Amari \cite{Amari} opublikował systematyczne ujęcie teorii Dawida dla modeli skończenie wymiarowych i podał spójne określenie $\alpha$-koneksji (wprowadzonej wcześniej poza kontekstem statystycznej estymacji przez  N.N. Chentsova). W 1982 S. Amari wraz z H. Nagaoka \cite{Amari Nagaoka book} wykazali dualność płaskich przestrzeni modeli eksponencjalnych z e-koneksją i modeli mieszanych z m-koneksją. 
\\
Procedury estymacyjne statystycznego opisu mechaniki kwantowej (falowej) poszły dwoma drogami. Pierwsza z nich związana jest z naturalnym dla mechaniki kwantowej formalizmem macierzy gęstości, druga z konstrukcją zasad informacyjnych (entropijnych). 
W przypadku formalizmu macierzy gęstości, ich zbiór $S = \bigcup_{r=1}^{k} S_{r}$  ($S_{r} \cup S_{i} =\emptyset $, $ i \neq r $) dla przypadku skończenie wymiarowych przestrzeni Hilberta ${\cal H}$, tworzy zbiór wypukły. Dla stanów czystych, podzbiór $S_{1}$ tego zbioru tworzą punkty ekstremalne, a przestrzeń stanów czystych związana z nim może być utożsamiona z zespoloną przestrzenią rzutową $CP^{k-1}$, ($k=dim$ ${\cal H}$). Na przestrzeni tej można wprowadzić (z dokładnością do stałej) metrykę Riemannowską nazywaną metryką Fubiniego-Study, która jest kwantową wersją metryki Rao-Fishera. Statystyczną estymacją w modelach dla stanów czystych zajmowali się między innymi Fujiwara, Nagaoka  i Matsumoto \cite{Fujiwara Nagaoka Matsumoto}. Natomiast w przypadku podzbioru $S_{k}$ zbioru $S$ dualna struktura z metryką możne być traktowana jako kwantowy analog metryki Rao-Fishera z  $\pm \alpha$-koneksją.  

Drugim nurtem, który wyłonił się w ostatnich kilkunastu latach i którym szedł rozwój zastosowań MNW oraz pojęcia obserwowanej i oczekiwanej IF w fizyce jest formalizm ekstremalnej fizycznej informacji (EFI) opracowany przez Friedena i jego współpracowników, w szczególności Soffera  \cite{Frieden}. Konstrukcję modeli fizycznych z wykorzystaniem informacji Fishera zapoczątkował Frieden, podając metodę wyprowadzenia z informacji Fishera członu kinetycznego modeli fizycznych. Następnie zapostulował wprowadzenie dwóch zasad informacyjnych służących do ustalenia związku pomiędzy pojemnością  kanału informacyjnego $I$ oraz informacją strukturalną $Q$, tzn. poprzez zapostulowaną nową strukturalną zasadę informacyjną skonstruował on człony strukturalne rozważanych przez siebie modeli. 
W odróżnieniu od Friedena stosujemy jednak inne  \cite{Dziekuje informacja_1}, bardziej fizyczne a mniej informacyjne, podejście do konstrukcji podstawowych zasad informacyjnych, posługując się pojęciem całkowitej fizycznej informacji $K = I + Q$, a nie wprowadzonym przez Friedena pojęciem zmiany fizycznej informacji. Różnica ta,  chociaż nie powoduje zasadniczo rachunkowych zmian w sposobie wyprowadzenia równań ruchu bądź równań generujących rozkład dla rozważanych do tej pory problemów, jednak zmieniając pojęcie informacji fizycznej oraz jej rozkładu na kinetyczne i strukturalne stopnie swobody, idzie w linii prowadzonych ostatnio badań nad konstrukcją zasady ekwipartycji dla entropii. To inne niż Friedenowskie podejście do pojęcia fizycznej informacji powoduje również zmiany w pojmowaniu istoty przekazu informacji 
w procesie po\-miaru przy jej przekazie od strukturalnych do kinetycznych stopni swobody. Pomimo różnic samą metodę będziemy dalej nazywać podejściem Friedenowskim.  
Gdyby pominąć chwilowo proces pomiaru, w metodzie Friedena próbkowanie przestrzeni jest wykonywane przez układ, który poprzez właściwe dla niego pole (i związane z nim amplitudy) o randze $N$ będącej wielkością próby, próbkuje jego kinetycznymi (Fisherowskimi) stopniami swobody dostępną mu przestrzeń konfiguracyjną. Następnie, ponieważ IF jest 
infinitezymalnym typem entropii Kulbacka-Leiblera, to zauważając, że entropia Kulbacka-Leiblera jest wykorzystywana w statystyce do przeprowadzania testów wyboru modeli, pojawia się przypuszczenie, że IF może poprzez narzucenie na nią odpowiednich dodatkowych ograniczeń, zapostulowanych w postaci wspomnianych dwóch zasad informacyjnych, wariacyjnej (skalarnej) oraz  strukturalnej (wewnętrznej), doprowadzić do wyprowadzenia równań ruchu bądź równań stanu układów fizycznych, najlepszych z punktu widzenia owych informacyjnych zasad. Na tym zasadza się Friedenowska idea estymacji fizycznych modeli.  

Zasady informacyjne mają uniwersalną postać, jednak ich konkretne realizacje zależą od fizyki rozważanego zagadnienia. Pierwsza z zasad informacyjnych, strukturalna, opisuje wewnętrzne charakterystyki układu związane np. z jego spinem. Druga,  wariacyjna, prowadzi do otrzymania właściwej relacji dyspersyjnej dla układu.  Ciekawą sprawą jest, że wiele rachunków można wykonać dla przypadku, dla którego całkowita fizyczna informacja układu (oraz jej gęstość) dzieli się na dwie równe (lub z czynnikiem $1/2$) części, tzn. pojemność kanału informacyjnego oraz informację strukturalną, mając swoją całkowitą wartość równą zero. Frieden podał informacyjne wyprowadzenie równania Kleina-Gordona dla ogólnego modelu pola z rangą $N$, z szczególnym uwzględnieniem przypadku pola skalarnego z $N$=2. Dla pola spinorowego z $N$=8 otrzymał równanie Diraca a dla  $N$=4 równania Maxwella. Procedura jest na tyle ogólna, że umożliwia opis pól Rarity-Schwingera, ogólnej teorii względności oraz wprowadzenie transformacji cechowania \cite{Frieden}. 
W oparciu o wprowadzone zasady informacyjne Frieden podał również informacyjne wyprowadzenie zasady  nieoznaczoności Heisenberga oparte ze statystycznego punktu widzenia o twierdzenie Rao-Cram{\'e}ra dla informacji Fishera oraz jej relację z pojemnością informacyjną układu zapisaną w reprezentacji pędowej, czyli po dokonaniu transformacji Fouriera. Transformacja Fouriera pełni zresztą w całym formalizmie Friedenowskim rolę wyjątkową, będąc jednym z typów samosplątania wewnątrz przestrzeni konfiguracyjnej układu, o czym wspomnimy nieco poniżej. 
Frieden podał również wyprowadzenie klasycznej fizyki statystycznej, tzn. jej podstawowych rozkładów, Boltzmanna dla energii oraz Maxwella-Boltzmanna dla pędu jak również pewnych rozkładów, które zinterpretował jako odpowiadające przypadkom nierównowagowym. Kolejną sprawą było wyprowadzenie górnego ograniczenia na tempo zmiany entropii układu dla przykładów klasycznego strumienia cząstek, gęstości rozkładu ładunku, czteropotencjału elektrodynamicznego oraz strumienia cząstek o spinie $1/2$ \cite{Frieden}. Podał również opis teorii pomiaru z szumem wykorzystując wariacyjny formalizm EFI pozwalający na opis redukcji funkcji falowej w trakcie pomiaru urządzeniem dającym swój własny szum. Mianowicie po dokonaniu ekstremalizacji sumy informacji fizycznej $K$ niemierzonego układu oraz funkcjonału opisującego własności układu pomiarowego (a będącego splotem funkcji log-wiarygodności dla funkcji przyrządu splecionej nieliniowo z rozkładem układu), otrzymał równanie ruchu, które (po przejściu do nierelatywistycznej granicy Schrödingera) daje równanie typu Feynmana-Mensky'ego z nieliniowym członem opisującym kolaps funkcji falowej w pomiarze. Ciekawe jest to, że w tym przypadku w pełni ujawnia się  traktowanie czasu na równi ze zmiennymi przestrzennymi, czyli jako zmiennej losowej z rozkładem prawdopodobieństwa. Przedstawiona w skrypcie postać zasad informacyjnych \cite{Dziekuje informacja_2} daje formalnie te same równania ewolucji funkcji falowej układu oplecionej funkcją pomiarową przyrządu, jednak otrzymana interpretacja jest zdecydowanie bardziej spójna niż Friedenowska, pozwalając na jednoznaczne rozróżnienie układu poza pomiarem od układu w pomiarze.  
\\
Przedstawione w skrypcie, fundamentalna postać {\it obserwowanej} strukturalnej zasady informacyjnej oraz jej postać {\it oczekiwana}, 
%
\cite{Dziekuje informacja_1,Dziekuje informacja_2},  
wykorzystywane za Friedenem dla każdego omawianego problemu, zostały ostatnio wyprowadzone dla wartości tzw. współczynnika efektywności $\kappa=1$ \cite{Dziekuje informacja_2}. Zasada strukturalna 
sugeruje splątanie przestrzeni danych obserwowanych z nieobserwowaną przestrzenią konfiguracyjną układu \cite{Dziekuje informacja_2}. 
Zatem informacja strukturalna $Q$ \cite{Dziekuje informacja_2} 
reprezentuje również informację o splątaniu widocznym w korelacji danych w przestrzeni pomiarowej, a EFI może być wykorzystywana jako mechanizm w estymacji stanów splątanych. Np. w przypadku problemu EPR-Bohma, splatanie zachodzi pomiędzy rzutem spinu obserwowanej cząstki i nieobserwowaną konfiguracją łączną układu, a w przypadku relatywistycznych równań ruchu otrzymujemy splątanie kinetycznych i strukturalnych (masa) stopni swobody, czego wyrazem jest związek stanu obserwowanej cząstki w czasoprzestrzeni z jej własnym stanem w przestrzeni energetyczno-pędowej. Ten drugi przypadek jest przykładem wspomnianego samosplątania opisanego  transformatą Fouriera. Ponieważ $Q$ związane jest tu z masą cząstki, zatem w podejściu informacyjnym można wyciągnąć również wniosek, że samosplątanie powinno pomóc w odczytaniu struktury wewnętrznej cząstek. W końcu pojęcie informacji Fishera i jej reinterpretacja przez Friedena jako członu kinetycznego teorii, pozwoliła na przeprowadzenie informacyjnego dowodu \cite{Dziekuje informacja_1} o niewyprowadzalności mechaniki kwantowej (falowej) oraz każdej teorii pola, dla której ranga pola $N$ jest skończona, z mechaniki klasycznej. \\

{\it Temat skryptu} dotyczy więc fundamentalnego zagadnienia związanego z określeniem statystycznej procedury estymacji modeli fizycznych. Jego realizacja wymaga znajomości problemów zwią\-za\-nych z stosowaniem statystycznej MNW oraz fizycznej EFI dla konstrukcji modeli fizycznych, jak również podstaw metod geometrii różniczkowej. 
\\  
Na koniec uwaga słownikowa i podsumowanie treści metody EFI. Pojęcie ``likelihood function'' zostało wprowadzony przez Fishera jako mające związek z prawdopodobieństwem. Również słownikowo powinno być ono przetłumaczone jako ``funkcja możliwości''. Zastosowano jednak tłumaczenie ``funkcja wiarygodności''. Jako posumowanie istoty przedstawionej metody, powiedzmy, że jest ona wyrazem {\it  zastosowania informacji Fishera  w  teorii pola w ujęciu Friedena, którego inspiracja pochodzi z obszaru optyki.} 

\vspace{20mm}

\hspace{5mm} Temat skryptu związany jest z dociekaniami, które dane mi było prowadzić wspólnie ze Sławomirem Manią, Dorotą Mroziakiewicz, Janem Sładkowskim, Robertem Szafronem i Sebastianem Zającem, którym za te dociekania i rozmowy dziękuję. \\
\\
Dziękuję mojej żonie Grażynie za uważne przeczytanie tekstu skryptu.

\chapter[Metoda największej wiarygodności]{Metoda największej wiarygodności}


\label{MNW}

Z powodu możliwości zastosowania {\it metody największej wiarygodności} (MNW) do rozwiązania wielu, bardzo różnych
problemów estymacyjnych, stała się ona obecnie zarówno metodą
podstawową jak również punktem wyjścia dla różnych metod analizy statystycznej. 
Jej wszechstronność związana jest, po pierwsze z możliwością przeprowadzenia analizy 
statystycznej dla małej próbki, opisu zjawisk nieliniowych oraz zastosowania zmiennych
losowych posiadających zasadniczo dowolny {\it rozkład prawdopodobieństwa} \cite{Nowak},  
oraz po drugie, szczególnymi własnościami otrzymywanych przez nią estymatorów, 
które okazują się być zgodne, asymptotycznie nieobciążone, efektywne oraz dostateczne \cite{Nowak}. 
MNW zasadza się na intuicyjnie jasnym postulacie przyjęcia za prawdziwe
takich wartości parametrów rozkładu prawdopodobieństwa 
zmiennej losowej, które maksymalizują funkcję wiarygodności realizacji konkretnej próbki. 

\section[Podstawowe pojęcia MNW]{Podstawowe pojęcia MNW} 

Rozważmy zmienną losową $Y$ \cite{Nowak}, która przyjmuje wartości ${\bf y}$ zgodnie z rozkładem praw\-dopodo\-bieństwa $p\left({\bf y}|\theta \right)$, gdzie $\theta = (\vartheta_{1},\vartheta_{2},...,\vartheta_{k})^{T}\equiv(\vartheta_{s})_{s=1}^{k}$, jest zbiorem $k$ parametrów tego rozkładu ($T$ oznacza transpozycję). 
Zbiór wszystkich możliwych wartości ${\bf y}$ zmiennej $Y$ oznaczmy przez ${\cal Y}$. \\
Gdy $k>1$ wtedy $\theta$ nazywamy parametrem {\it wektorowym}. W szczególnym przypadku $k=1$ mamy $\theta=\vartheta$. Mówimy wtedy, że parametr $\theta$ jest parametrem {\it skalarnym}. \\
\\
{\bf Pojęcie próby i próbki}: 
Rozważmy {\it zbiór danych} 
$\,{{\bf y}_{1},{\bf y}_{2},...,{\bf y}_{N}}$
otrzymanych w $N$ obserwacjach zmiennej losowej $Y$. \\
Każda z danych ${\bf y}_{n}$, $n=1,2,...,N$, jest generowana z rozkładu $p_{n}({\bf y}_{n}|\theta_{n})$ zmiennej losowej $Y$ w populacji, którą charakteryzuje wartość parametru  wektorowego  $\theta_{n} =  (\vartheta_{1},\vartheta_{2},...,\vartheta_{k})_{n}^{T}\equiv((\vartheta_{s})_{s=1}^{k})_{n}$, $n=1,2,...,N$. Stąd zmienną $Y$ w $n$-tej populacji oznaczymy $Y_{n}$. Zbiór zmiennych losowych $\widetilde{Y} = (Y_{1},Y_{2},...,Y_{N}) \equiv( Y_{n})_{n=1}^{N}$ nazywamy $N$-wymiarową {\it próbą}.  \\
Konkretną realizację 
$y=({{\bf y}_{1},{\bf y}_{2},...,{\bf y}_{N}})\equiv ({\bf y}_{n})_{n=1}^{N}$ próby $\widetilde{Y}$ nazywamy {\it próbką}. 
Zbiór wszystkich możliwych realizacji $y$ próby $\widetilde{Y}$ tworzy  przestrzeń próby (układu) oznaczaną jako ${\cal B}$. \\
{\bf Określenie}: Ze względu na to, że $n$ jest indeksem konkretnego punktu pomiarowego próby, rozkład $p_{n}({{\bf y}_{n}|\theta_{n}})$ będziemy nazywali rozkładem {\it punktowym} (czego nie należy mylić z np. rozkładem dyskretnym).

\newpage

{\bf Określenie funkcji wiarygodności}: $\;\;\;$ Centralnym pojęciem MNW jest {\it funkcja wiarygodności} $L\left(y;\Theta\right)$ (pojawienia się)  próbki $y = ({\bf y}_{n})_{n=1}^{N}$, nazywana  też {\it wiarygodnością próbki}. Jest ona funkcją parametru $\Theta$. \\
Przez wzgląd na zapis stosowany w 
fizyce, będziemy stosowali oznaczenie $P(y\,|\Theta)\equiv L\left(y;\Theta\right)$,
które podkreśla, że formalnie {\it funkcja wiarygodności  jest  łącznym rozkładem  
prawdopodobieństwa}\footnote{{\bf Miara produktowa}: 
Niech będzie danych $N$ przestrzeni probabilistycznych $\left\{\Omega_{1}, {\cal F}_{1}, P_{1}\right\}$, ... , $\left\{\Omega_{N}, {\cal F}_{N}, P_{N}\right\}$, gdzie  $\Omega_{n}$, ${\cal F}_{n}$, $P_{n}$ są, dla każdego $n=1,2,...,N$, odpowiednio $n$-tą przestrzenią zdarzeń, $\sigma$ - ciałem na $\Omega_{n}$ oraz miarą probabilistyczną. Wprowadźmy na produkcie $\Omega = \Omega_{1}\times ... \times \Omega_{N}$ tzw. $\sigma$ - ciało produktowe ${\cal F}\equiv {\cal F}_{1} \otimes ... \otimes{\cal F}_{N}$ będące najmniejszym $\sigma$ - ciałem zawierającym zbiory postaci $A_{1}\times ... \times A_{N}$ gdzie $A_{1} \in {\cal F}_{1}$, ... , $A_{N} \in {\cal F}_{N}$. \\
Na produkcie $\Omega$ można zdefiniować miarę produktową $P$ taką, że: 
\begin{eqnarray}
\label{Miara produktowa} 
P(A_{1} \times A_{2} \times... \times A_{N}) = P_{1}(A_{1}) \, P_{2}(A_{2}) \, ... \, P_{N}(A_{N}) \; .
\end{eqnarray}
{\bf Zmienne niezależne}: Zmienne losowe $Y_{1}, ... , Y_{N}$ o wartościach w $\mathbb{R}$ określone odpowiednio na $\left\{\Omega_{1}, {\cal F}_{1}, P_{1}\right\}$, ... , $\left\{\Omega_{N}, {\cal F}_{N}, P_{N}\right\}$ nazywamy {\it niezależnymi}, gdy dla każdego ciągu zbiorów borelowskich $B_{1}, ... , B_{N}$ zachodzi równość:
\begin{eqnarray}
\label{Miara produktowa zmienne losowe} 
P(Y_{1} \in B_{1},  \, Y_{2} \in B_{2},  ...  \, Y_{N} \in B_{N}) = P_{1}(Y_{1} \in B_{1}) \, P_{2}(Y_{2} \in B_{2}) \, ... \, P_{N}(Y_{N} \in B_{N}) \; .
\end{eqnarray}
Ponieważ po prawej stronie (\ref{Miara produktowa zmienne losowe}) stoją zdarzenia losowe $Y_{n}^{-1}(B_{n})$, gdzie $B_{n}$ należy do $\sigma$ - ciała $\mathbb{B(R)}$, zatem stwierdzenie (\ref{Miara produktowa zmienne losowe}) 
oznacza, że {\it zmienne losowe są niezależne wtedy i tylko wtedy, gdy $\sigma$ - ciała ${\cal F}_{n}$ generowane przez zmienne losowe $Y_{n}$, $n=1,2,...,N$,  są niezależne} \cite{Jakubowski-Sztencel}. 
} 
pojawienia się realizacji $y \equiv ({\bf y}_{n})_{n=1}^{N}$  próby $\widetilde{Y} \equiv( Y_{n})_{n=1}^{N}$, to znaczy:
\begin{eqnarray}
\label{funkcja wiarygodności proby - def}
P(\Theta) \equiv P\left({y|\Theta}\right) = \prod\limits_{n=1}^{N} {p_{n}\left({{\bf y}_{n}|\theta_{n}}\right)} \; .
\end{eqnarray}
Zwrócenie uwagi w (\ref{funkcja wiarygodności proby - def}) na występowanie $y$ w argumencie funkcji wiarygodności oznacza, że może  być ona rozumiana jako statystyka $P\left({\widetilde{Y}|\Theta}\right)$. 
Z kolei skrócone oznaczenie $P(\Theta)$  podkreśla, że centralną sprawą w MNW jest fakt, że funkcja wiarygodności jest funkcją  nieznanych parametrów: 
\begin{eqnarray} 
\label{parametr Theta}
\Theta = (\theta_{1},\theta_{2},...,\theta_{N})^{T} \equiv (\theta_{n})_{n=1}^{N} \;\;\;\; {\rm przy\; czym} \;\;\; \theta_{n} = (\vartheta_{1n},\vartheta_{2n},...,\vartheta_{kn})^{T} \equiv ((\vartheta_{s})_{s=1}^{k})_{n} \; ,
\end{eqnarray}
gdzie $\theta_{n}$ jest wektorowym parametrem populacji określonej przez indeks próby $n$. 
W toku analizy chcemy oszacować wektorowy parametr $\Theta$. \\
Zbiór  wartości parametrów $\Theta=(\theta_{n})_{n=1}^{N}$
tworzy współrzędne rozkładu prawdopodobieństwa rozumianego jako punkt
w $d=k \times N$ - wymiarowej (podprzestrzeni) przestrzeni statystycznej ${\cal S}$ \cite{Amari Nagaoka book}.
Temat ten rozwiniemy w Rozdziale~\ref{alfa koneksja}. \\
\\
{\bf Uwaga o postaci rozkładów punktowych}: W skrypcie zakładamy, że  ``{\it punktowe}'' rozkłady  $p_{n}\left({{\bf y}_{n}|\theta_{n}}\right)$ dla poszczególnych pomiarów $n$ w $N$ elementowej próbie są 
{\it niezależne}\footnote{W przypadku 
analizy jednej zmiennej losowej $Y$, rozkłady te obok niezależności spełniają dodatkowo warunek: 
\begin{eqnarray}
\label{rozklady pn} 
p_{n} \left({{\bf y}_{n}|\theta_{n}}\right) = p \left({{\bf y}|\theta} \right) \; ,
\end{eqnarray}
co oznacza, że próba jest {\it prosta}.
}. \\
W ogólności w treści skryptu, rozkłady punktowe  $p_{n} \left({\bf y}_{n}|\theta_{n} \right)$ zmiennych $Y_{n}$  chociaż są {\it tego samego typu}, jednak  nie spełniają warunku (\ref{rozklady pn})  charakterystycznego dla próby prostej. Taka ogólna sytuacja ma np. miejsce w analizie regresji (Rozdział~\ref{regresja klasyczna}).   \\
\\
\\
{\bf Pojęcie estymatora parametru}: 
Załóżmy, że dane $y = ({\bf y}_{n})_{n=1}^{N}$ są generowane losowo z punktowych rozkładów prawdopodobieństwa $p_{n}({\bf y}_{n}|\theta_{n})$, $n=1,2,...,N$, które chociaż nie są znane, to jednak założono o nich, że dla każdego $n$ należą do określonej, tej samej klasy modeli. Zatem funkcja wiarygodności 
(\ref{funkcja wiarygodności proby - def}) należy do określonej, $d = k \times N$ - wymiarowej,  przestrzeni statystycznej ${\cal S}$. \\
Celem analizy jest oszacowanie nieznanych parametrów $\Theta$, (\ref{parametr Theta}), poprzez funkcję:   
\begin{eqnarray} 
\label{estymator parametrow Theta}
\!\!\!\!\!\! \hat{\Theta} \equiv \hat{\Theta}(\widetilde{Y})  =(\hat{\theta}_{1},\hat{\theta}_{2},...,\hat{\theta}_{N})^{T}\equiv (\hat{\theta}_{n})_{n=1}^{N} \;  \;\;\; {\rm gdzie} \;\;\; \hat{\theta}_{n} = (\hat{\vartheta}_{1n},\hat{\vartheta}_{2n},...,\hat{\vartheta}_{kn})^{T} \equiv ((\hat{\vartheta}_{s})_{s=1}^{k})_{n} \; , \;
\end{eqnarray}
mającą $d = k \times N$ składowych. \\
Każda z funkcji $\hat{\vartheta}_{kn} \equiv \hat{\vartheta}_{kn}(\widetilde{Y})$  jako funkcja próby jest {\it statystyką}, którą przez wzgląd na to, że służy do oszacowywania wartości parametru $\vartheta_{kn}$ nazywamy estymatorem tego parametru. {\it Estymator parametru nie może zależeć od parametru, który oszacowuje}\footnote{Natomiast 
rozkład estymatora oszacowywanego parametru, zależy od tego parametru. 
}. \\
\\
Podsumowując, odwzorowanie:
\begin{eqnarray} 
\label{estymator jako odwzorowanie}
\hat{\Theta}: {\cal B} \rightarrow \mathbf{R}^{d} \; ,
\end{eqnarray}
gdzie  ${\cal B}$ jest przestrzenią próby, jest estymatorem parametru (wektorowego) $\Theta$.  \\
\\
{\bf Równania wiarygodności}: Będąc funkcją $\Theta=(\theta_{n})_{n=1}^{N}$, funkcja wiarygodności
służy do konstrukcji estymatorów $\hat{\Theta}=(\hat{\theta}_{1},\hat{\theta}_{2},...,\hat{\theta}_{N})^{T}\equiv(\hat{\theta}_{n})_{n=1}^{N}$
parametrów $\Theta \equiv (\theta_{n})_{n=1}^{N}$. Procedura polega na wyborze takich $(\hat{\theta}_{n})_{n=1}^{N}$ , dla których funkcja wiarygodności przyjmuje maksymalną wartość, skąd   statystyki te nazywamy estymatorami MNW.  \\
Zatem warunek konieczny  otrzymania estymatorów $\hat{\Theta}$ MNW sprowadza 
się do znalezienia rozwiązania układu $d=k \times N$ tzw. {\it równań wiarygodności}  \cite{Fisher}:
\begin{eqnarray}
\label{rown wiaryg}
S\left(\Theta\right)_{\left|\Theta = \hat{\Theta} \right.} \equiv\frac{\partial}{\partial\Theta}\ln P(y\,|\Theta)_{\left|\Theta = \hat{\Theta} \right.} = 0 \; ,
\end{eqnarray}
gdzie zagadnienie maksymalizacji funkcji wiarygodności $P(y\,|\Theta)$ sprowadzono do (na ogół)  analitycznie 
równoważnego mu problemu maksymalizacji jej logarytmu 
$\ln P(y\,|\Theta)$. \\
 \\
{\bf Określenie funkcji wynikowej}: Funkcję $S\left(\Theta\right)$ będącą gradientem
logarytmu funkcji wiarygodności: 
\begin{eqnarray}
\label{funkcja wynikowa}
S\left(\Theta\right)\equiv\frac{\partial}{\partial\Theta}\ln P(y\,|\Theta) = \left(%
\begin{array}{c}
  \frac{\partial \ln P(y|\Theta)}{\partial \theta_{1}} \\
  \vdots \\
  \frac{\partial \ln P(y|\Theta)}{\partial \theta_{N}} \\
\end{array}%
\right) \; \;\;\;\; {\rm gdzie} \;\;\;\;  \frac{\partial \ln P(y|\Theta)}{\partial \theta_{n}} = \left(%
\begin{array}{c}
  \frac{\partial \ln P(y|\Theta)}{\partial \vartheta_{1n}} \\
  \vdots \\
  \frac{\partial \ln P(y|\Theta)}{\partial \vartheta_{kn}} \\
\end{array}%
\right) \; , 
\end{eqnarray}
nazywamy {\it funkcją wynikową}. \\
\\
Po otrzymaniu (wektora) estymatorów $\hat{\Theta}$, {\it zmaksymalizowaną} 
wartość funkcji wiarygodności definiujemy jako numeryczną wartość
funkcji wiarygodności powstałą przez podstawienie do $P(y \,|\Theta)$
wartości oszacowanej $\hat{\Theta}$ w miejsce parametru $\Theta$.
\\
 \\
{\bf Przykład}: Rozważmy problem estymacji skalarnego parametru,
tzn. $\Theta = \theta$ (tzn. $k=1$ oraz $N=1$), dla zmiennej losowej $Y$ opisanej rozkładem dwumianowym (Bernoulliego):
\begin{eqnarray}
\label{Bernoulliego rozklad}
P \left( y|\theta \right) = \left( \begin{array}{l}
m   \\
y 
\end{array} \right) \theta^{y} \left(1 - \theta \right)^{m-y} \; .
\end{eqnarray}
Estymacji parametru $\theta$ dokonamy na podstawie {\it pojedynczej}
obserwacji (długość próby $N=1$) zmiennej $Y$, której iloraz $Y/m$ nazywamy {\it częstością}. Parametr $m$ charakteryzuje rozkład zmiennej Bernoulliego $Y$ (i nie ma związku z długością $N$ próby). \\ 
Zatem ponieważ $y \equiv \left( {\bf y}_{1} \right)$,
więc $P\left(y|\theta \right)$ jest funkcją wiarygodności dla $N=1$
wymiarowej próby. Jej logarytm wynosi:  
\begin{eqnarray}
\label{ln wiaryg dla Bernoulliego}
\ln P\left(y|\theta\right)=\ln\left(\begin{array}{l}
m\\
y \end{array}\right) + y \ln \theta + 
{\left(m-y \right)\ln} \left({1-\theta}\right) \; .
\end{eqnarray}
\\
W rozważanym przypadku otrzymujemy jedno równanie wiarygodności (\ref{rown wiaryg}): 
\begin{eqnarray}
\label{r.wiaryg dla Bernoulliego}
S(\theta)=\frac{1}{\theta}{y}-\frac{1}{{1-\theta}}{\left({m-y}\right)\left|\begin{array}{l}
_{\theta=\hat{\theta}}\end{array}\right.} = 0 
\end{eqnarray}
a jego rozwiązanie  daje estymator MNW parametru $\theta$ rozkładu dwumianowego, równy: 
\begin{eqnarray}
\hat{\theta}=\frac{y}{m}\label{estymator theta dla Bernoulliego}\end{eqnarray}
Ilustracją powyższej procedury znajdowania
wartości estymatora parametru $\theta$ jest Rysunek~1.1 (gdzie przyjęto $m=5$), przy czym na skutek pomiaru zaobserwowano wartość $Y$ równą $y=1$.
\begin{figure}[top]
\label{rysunek dla rozk dwumianowego}
\includegraphics[width=65mm]{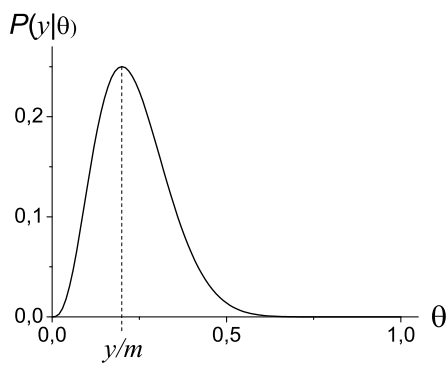} 
\caption{Graficzna ilustracja metody największej wiarygodności dla $P\left(y|\theta\right)$
określonego wzorem (\ref{Bernoulliego rozklad}) dla rozkładu dwumianowego. Przyjęto wartość parametru $m=5$. Na skutek pomiaru zaobserwowano wartość $Y$ równą $y=1$. 
Maksimum $P\left(y|\theta\right)$ przypada
na wartość $\theta$ równą punktowemu oszacowaniu $\hat{\theta}=y/m=1/5$
tego parametru. Maksymalizowana wartość funkcji wiarygodności wynosi $P\left(y|\hat{\theta}\right)$.}
\label{PBernoulli} 
\end{figure}

\newpage

\section[Wnioskowanie w MNW]{Wnioskowanie w MNW}

\label{Wnioskowanie w MNW}

Z powyższych rozważań wynika, że konstrukcja punktowego oszacowania parametru w MNW oparta jest o postulat 
maksymalizacji funkcji wiarygodności przedstawiony powyżej. Jest on
wstępem do statystycznej procedury wnioskowania. Kolejnym krokiem
jest konstrukcja przedziału wiarygodności. Jest on odpowiednikiem
przedziału ufności, otrzymywanego w częstotliwościowym podejściu statystyki
klasycznej do procedury estymacyjnej. Do jego konstrukcji niezbędna
jest znajomość rozkładu prawdopodobieństwa estymatora parametru, co
(dzięki {}``porządnym'' granicznym własnościom stosowanych estymatorów)
jest możliwe niejednokrotnie jedynie asymptotycznie, tzn. dla wielkości
próby dążącej do nieskończoności. Znajomość rozkładu estymatora jest
też niezbędna we wnioskowaniu statystycznym odnoszącym się do weryfikacji hipotez. 
\\
W sytuacji, gdy nie dysponujemy wystarczającą ilością danych,
potrzebnych do przeprowadzenia skutecznego częstotliwościowego wnioskowania,
Fisher \cite{Pawitan} zaproponował do określenia niepewności dotyczącej
parametru $\Theta$ wykorzystanie maksymalizowanej wartość funkcji
wiarygodności. \\
 \\
{\bf Przedział wiarygodności}  jest zdefiniowany
jako zbiór wartości parametru $\Theta$, dla których funkcja 
wiarygodności osiąga (umownie) wystarczająco wysoką wartość, tzn.:
\begin{eqnarray}
\label{przedzial wiarygodnosci}
\left\{ {\Theta, \; \frac{{P\left({y|\Theta}\right)}}{{P\left({y|\hat{\Theta}}\right)}}> c} \right\} \; ,
\end{eqnarray}
dla pewnego {\it parametru obcięcia} $c$, nazywanego {\it poziomem wiarygodności}. \\
  \\
{\bf Iloraz wiarygodności}: 
\begin{eqnarray}
\label{iloraz wiarygodnosci}
P(y|\Theta)/P(y|\hat{\Theta})
\end{eqnarray} 
reprezentuje pewien typ unormowanej wiarygodności i jako taki jest
wielkością skalarną. Jednak z powodu niejasnego znaczenia określonej
wartości parametru obcięcia $c$ pojęcie to wydaje się być na pierwszy
rzut oka za słabe, aby dostarczyć taką precyzję wypowiedzi jaką daje
analiza częstotliwościowa. \\
Istotnie, wartość $c$ nie odnosi się do żadnej wielkości obserwowanej,
tzn. na przykład $1\%$-we ($c=0,01$) obcięcie nie ma ścisłego probabilistycznego znaczenia. Inaczej ma się sprawa dla częstotliwościowych przedziałów
ufności. W tym przypadku wartość współczynnika $\alpha=0,01$ oznacza,
że gdybyśmy rozważyli realizację przedziału ufności na poziomie ufności
$1-\alpha=0,99$, to przy pobraniu nieskończonej (w praktyce wystarczająco
dużej) liczby próbek, $99\%$ wszystkich wyznaczonych przedziałów
ufności pokryłoby prawdziwą (teoretyczną) wartość parametru $\Theta$ w populacji generalnej (składającej się z $N$ podpopulacji).
Pomimo tej słabości MNW zobaczymy, że rozbudowanie
analizy stosunku wiarygodności okazuje się być istotne we  wnioskowaniu statystycznym analizy doboru modeli i to aż po konstrukcję równań teorii pola.

\subsection{Wiarygodnościowy przedział ufności} 

\label{Wiarygodnosciowy przedzial ufnosci}

{\bf Przykład rozkładu normalnego z jednym estymowanym parametrem}: 
Istnieje przypadek pozwalający na prostą {\it interpretację przedziału wiarygodnościowego jako przedziału ufności}. Dotyczy on zmiennej $Y$ posiadającej rozkład Gaussa oraz sytuacji gdy (dla próby prostej)  interesuje nas  estymacja skalarnego parametru $\theta$ będącego wartością oczekiwaną $E(Y)$ zmiennej $Y$. Przypadek ten omówimy  poniżej. W ogólności, przedział wiarygodności posiadający określony poziom ufności jest nazywany przedziałem  ufności.  \\
Częstotliwościowe wnioskowanie o nieznanym parametrze $\theta$ wymaga
określenia rozkładu jego estymatora, co jest zazwyczaj możliwe  jedynie granicznie \cite{Amari Nagaoka book}. Podobnie w MNW, o ile to możliwe, korzystamy przy dużych próbkach z twierdzeń granicznych  dotyczących rozkładu ilorazu wiarygodności \cite{Amari Nagaoka book,Pawitan}. 
W przypadku rozkładu normalnego i parametru skalarnego okazuje się, że możliwa jest konstrukcja skończenie wymiarowa. \\
\
Niech więc zmienna $Y$ ma rozkład normalny 
$N\left({\theta,\sigma^{2}}\right)$:
\begin{eqnarray}
\label{rozklad norm theta sigma2}
p\left({\bf y}|\theta, \sigma^{2}\right) =  \frac{1}{\sqrt{2 \pi \, \sigma^2}} \; \exp \left( - \, {\frac{({\bf y} - \theta)^{2}}{2 \, \sigma^2}} \right) \;\, .
\end{eqnarray}
Rozważmy próbkę $y \equiv ({\bf y}_{1},\ldots, {\bf y}_{N})$, która 
jest realizacją próby prostej $\widetilde{Y}$ i załóżmy, że {\it wariancja $\sigma^{2}$ jest znana}. 
Logarytm funkcji wiarygodności dla $N\left({\theta,\sigma^{2}}\right)$
ma postać: 
\begin{eqnarray}
\label{log wiaryg rozklad norm jeden par}
\ln P\left(y|\theta\right) = - \frac{N}{2}\ln(2 \pi \sigma^{2}) - \frac{1}{{2\sigma^{2}}} \sum\limits_{n=1}^{N}{\left({{\bf y}_{n} - \theta}\right)^{2}} \;\, ,
\end{eqnarray}
gdzie ze względu na próbę prostą, w argumencie funkcji wiarygodności wpisano w miejsce $\Theta~\equiv~(\theta)_{n=1}^{N}$ parametr
$\theta$, jedyny który podlega estymacji. \\
\\
Korzystając z  funkcji wiarygodności (\ref{log wiaryg rozklad norm jeden par}) otrzymujemy oszacowanie MNW parametru $\theta$  
równe\footnote{
{\bf Postać estymatora parametru skalarnego $\theta$ rozkładu $N\left({\theta,\sigma^{2}}\right)$}: Korzystając z równania wiarygodności (\ref{rown wiaryg})  dla  przypadku skalarnego parametru $\theta$, otrzymujemy:
\begin{eqnarray}
\label{rown wiaryg skal}
S\left(\theta\right)_{|_{\theta = \hat{\theta}}}  \equiv\frac{\partial}{\partial\theta} \ln P(y\,|\theta)_{|_{\theta = \hat{\theta}}} = 0 \; , 
\end{eqnarray}
skąd dla log-funkcji wiarygodności (\ref{log wiaryg rozklad norm jeden par}), otrzymujemy:
\begin{eqnarray}
\label{srednia arytmet z MNW}
\hat{\theta} = \bar{{\bf y}} = \frac{1}{{N}}\sum\limits_{n=1}^{N} {\bf y}_{n} \; .
\end{eqnarray}
Zatem estymatorem parametru $\theta$ jest średnia arytmetyczna: 
\begin{eqnarray}
\label{srednia arytmet estymator z MNW}
\hat{\theta} = \overline{Y} = \frac{1}{{N}}\sum\limits_{n=1}^{N} Y_{n} \; .
\end{eqnarray}
Estymator i jego realizowaną  wartość będziemy oznaczali tak samo, tzn. $\hat{\theta}$ dla przypadku skalarnego i $\hat{\Theta}$ dla  wektorowego.
} 
$\hat{\theta} = \bar{{\bf y}} = \frac{1}{{N}}\sum\limits_{n=1}^{N} {\bf y}_{n}$, co pozwala na zapisanie równości $\sum_{n=1}^{N}{\left({{\bf y}_{n} - \theta}  \right)^{2}}$ $ =  \sum_{n=1}^{N}{\left({({\bf y}_{n} - \hat{\theta}) + (\hat{\theta} - \theta)} \right)^{2}}$ $= \sum_{n=1}^{N}({\bf y}_{n} - \hat{\theta})^{2} + \sum_{n=1}^{N} (\hat{\theta} - \theta)^{2}$.   
W końcu, nieskomplikowane  przekształcenia prowadzą do następującej postaci  logarytmu ilorazu wiarygodności (\ref{iloraz wiarygodnosci}):  
\begin{eqnarray}
\label{iloraz wiaryg dla normalnego}
\ln \frac{{P\left( {y|\theta}\right)}}{{P( {y|\hat{\theta}} )}} =  - \frac{N}{{2\sigma^{2}}}\left({\hat{\theta} - \theta}\right)^{2}  \; .
\end{eqnarray}
\\
{\bf Statystyka  Wilka}: Widać, że po prawej stronie (\ref{iloraz wiaryg dla normalnego})  otrzymaliśmy wyrażenie kwadratowe. 
Ponieważ $\bar{Y}$ jest nieobciążonym estymatorem parametru $\theta$, co
oznacza, że wartość oczekiwana ${E(\bar{Y})=\theta}$, zatem (dla rozkładu $Y \sim N\left( \theta, \sigma^{2} \right)$) średnia arytmetyczna $\bar{Y}$ ma rozkład normalny  $N\left({\theta,\frac{\sigma^{2}}{N}}\right)$. \\
Z normalności rozkładu $\bar{Y}$ wynika, że tzw. {\it statystyka ilorazu wiarygodności Wilka}: 
\begin{eqnarray}
\label{1}
W \equiv 2 \ln \frac{{P\left({\widetilde{Y}|\hat{\theta}}\right)}}{{P\left({\widetilde{Y}|\theta}\right)}}\sim\chi_{1}^{2} \; ,
\end{eqnarray}
ma rozkład $\chi^{2}$, w tym przypadku  z jednym stopniem swobody \cite{Pawitan}. \\
 \\
{\bf Wyskalowanie statystyki Wilka w przypadku normalnym}: Wykorzystując (\ref{1}) możemy wykonać wyskalowanie wiarygodności oparte o możliwość powiązania przedziału wiarygodności z jego częstotliwościowym odpowiednikiem. \\
\\
Mianowicie z (\ref{1}) otrzymujemy, że dla ustalonego (chociaż nieznanego) parametru $\theta$  prawdopodobieństwo, że iloraz
wiarygodności znajduje się w wyznaczonym dla parametru obcięcia $c,$
wiarygodnościowym przedziale ufności, wynosi: \\
\begin{eqnarray}
\label{rownosc prawdop dla zdarzenia z c}
P\left({\frac{{P\left({\widetilde{Y}|\theta}\right)}}{{P\left({\widetilde{Y}|\hat{\theta}}\right)}}>c}\right)=P\left({2\ln\frac{{P\left({\widetilde{Y}|\hat{\theta}}\right)}}{{P\left({\widetilde{Y}|\theta}\right)}}<-2\ln c}\right)=P\left({\chi_{1}^{2}<-2\ln c}\right) \; .
\end{eqnarray}\\
Zatem jeśli dla jakiegoś $0<\alpha<1$ wybierzemy parametr obcięcia: 
\begin{eqnarray}
\label{2}
c = e^{-\frac{1}{2}\chi_{1,\left({1-\alpha}\right)}^{2}} \; ,
\end{eqnarray}
gdzie ${\chi_{1,\left({1-\alpha}\right)}^{2}}$ jest kwantylem rzędu
$100(1-\alpha)\%$ rozkładu $\chi$-kwadrat, to spełnienie przez $\theta$ związku: 
\begin{eqnarray}
\label{1 minus alfa}
P\left( {\frac{{P\left({\widetilde{Y}|\theta}\right)}}{
{P\left({\widetilde{Y}|\hat{\theta}}\right)}}>c} \right) = P \left({\chi_{1}^{2}<\chi_{1,\left({1-\alpha}\right)}^{2}}\right) = 1-\alpha \;
\end{eqnarray}
oznacza, że przyjęcie wartości $c$ zgodnej z (\ref{2}) daje zbiór możliwych wartości parametru $\theta$:
\begin{eqnarray}
\label{przedzial wiarygod theta dla rozkl norm}
\left\{ {\theta,\frac{{P\left({\widetilde{Y}|\theta}\right)}}{{P\left({\widetilde{Y}|\hat{\theta}}\right)}}>c}\right\} \; , 
\end{eqnarray}
nazywany 100$(1-\alpha)\%$-owym  {\it wiarygodnościowym 
przedziałem ufności}. Jest on odpowiednikiem wyznaczonego na  poziomie ufności $(1-\alpha)$ częstotliowściowego przedziału ufności dla $\theta$. 
Dla analizowanego przypadku rozkładu normalnego z estymacją skalarnego parametru $\theta$ oczekiwanego poziomu zjawiska, otrzymujemy po skorzystaniu z wzoru (\ref{2}) wartość parametru obcięcia  równego $c=0.15$ lub $c=0.04$  dla odpowiednio $95\%$-owego ($1-\alpha=0.95$) bądź $99\%$-owego ($1-\alpha=0.99$) przedziału ufności. Tak więc w przypadku, {\it gdy przedział wiarygodności da się wyskalować rozkładem prawdopodobieństwa, parametr obcięcia $c$ posiada własność wielkości obserwowanej interpretowanej częstotliwościowo poprzez związek z poziomem ufności}.   \\
Zwróćmy uwagę, że chociaż konstrukcje częstotliwościowego i  wiarygodnościowego przedziału ufności są różne, to {\it ich losowość wynika} w obu przypadkach {\it z rozkładu prawdopodobieństwa estymatora}  $\hat{\theta}$.  \\
  \\
{\bf Ćwiczenie}: W oparciu o powyższe rozważania wyznaczyć, korzystając z (\ref{iloraz wiaryg dla normalnego}) ogólną postać przedziału wiarygodności dla skalarnego parametru $\theta$ 
rozkładu normalnego.

\subsection{Rozkłady regularne}

\label{Rozklady regularne}

Dla zmiennych o innym rozkładzie niż rozkład normalny, statystyka Wilka $W$ ma w ogólności inny rozkład niż $\chi^{2}$ \cite{Pawitan}. Jeśli więc zmienne nie mają dokładnie rozkładu normalnego 
lub dysponujemy za małą próbką by móc odwoływać się do (wynikających z twierdzeń granicznych)  
rozkładów granicznych dla estymatorów parametrów, wtedy związek (\ref{1}) (więc i (\ref{2}))  daje jedynie przybliżone wyskalowanie przedziału wiarygodności rozkładem $\chi^{2}$. 
\\
Jednakże w przypadkach wystarczająco {\it regularnych rozkładów},  zdefiniowanych jako takie, w których możemy zastosować przybliżenie kwadratowe: 
\begin{eqnarray}
\label{log wiaryg dla regularnego}
\ln\frac{{P\left({y|\theta}\right)}}{{P\left({y|\hat{\theta}}\right)}} \approx - \frac{1}{2} \texttt{i\!F} \left(\hat{\theta}\right)\left(\hat{\theta}-\theta\right)^{2} \; , 
\end{eqnarray}
powyższe rozumowanie oparte o wyskalowanie wiarygodności rozkładem $\chi^{2}_{1}$ jest w przybliżeniu słuszne. Wielkość
$\texttt{i\!F}\left(\hat{\theta}\right)$, która pojawiła się powyżej jest {\it obserwowaną} informacją Fishera, a powyższa formuła stanowi poważne narzędzie w analizie doboru modeli. Można powiedzieć, że cały skrypt koncentruje 
się na analizie zastosowania (wartości oczekiwanej) tego wyrażenia
i jego uogólnień. Do sprawy tej wrócimy dalej. \\
  \\
{\bf Przykład}: Rozważmy przypadek parametru skalarnego $\theta$
w jednym eksperymencie ($N=1$) ze zmien\-ną $Y$ posiadającą rozkład Bernoulliego z $m=15$. W wyniku pomiaru zaobserwowaliśmy wartość $Y={\bf y}=3$. Prosta 
analiza 
pozwala wyznaczyć wiarygodnościowy przedział ufności dla parametru
$\theta$. Ponieważ przestrzeń $V_{\theta}$ parametru $\theta$ wynosi
$V_{\theta}=(0,1)$, zatem łatwo pokazać, że dla $c=0,01$, $c=0,1$
oraz $c=0,5$ miałby on realizację odpowiednio $(0,019;0,583)$, $(0,046;0,465)$
oraz $(0,098;0,337)$. 
Widać, że wraz ze wzrostem wartości $c$, przedział wiarygodności zacieśnia
się wokół wartości oszacowania punktowego $\hat{\theta}=y/m=1/5$
parametru $\theta$ i nic dziwnego, bo wzrost wartości $c$ oznacza
akceptowanie jako możliwych do przyjęcia tylko takich {\it modelowych wartości parametru} $\theta$, które gwarantują wystarczająco wysoką
wiarygodność próbki.
\\
Powyższy przykład pozwala nabyć pewnej intuicji co do sensu stosowania ilorazu funkcji wiarygodności. Mianowicie po otrzymaniu w pomiarze określonej wartości $y/m$ oszacowującej parametr $\theta$, jesteśmy 
skłonni preferować model z taką wartością parametru $\theta$, która
daje większą wartość (logarytmu) ilorazu wiarygodności $P(y|\theta)/P(y|\hat{\theta})$. 
Zgodnie z podejściem statystyki klasycznej {\it nie
oznacza to jednak}, że uważamy, że parametr $\theta$ ma jakiś rozkład. 
Jedynie wobec niewiedzy co do modelowej (populacyjnej) wartość parametru
$\theta$ preferujemy ten model, który daje większą wartość ilorazu
wiarygodności w próbce.

\subsection{Weryfikacja hipotez z wykorzystaniem ilorazu wiarygodności}

\label{weryfikacja hipotez z ilorazem wiaryg}

Powyżej wykorzystaliśmy funkcję wiarygodności do {\it estymacji wartości parametru} $\Theta$. Funkcję wiarygodności można również wykorzystać w drugim typie  wnioskowania statystycznego, tzn. w {\it weryfikacji  hipotez statystycznych}. \\
\\
Rozważmy prostą hipotezę zerową  $H_{0}: \Theta = \Theta_{0}$ wobec złożonej hipotezy alternatywnej $H_{1}: \Theta \neq \Theta_{0}$. W celu przeprowadzenia {\it testu statystycznego} wprowadźmy unormowaną funkcję wiarygodności: 
\begin{eqnarray}
\label{unorm fun wiaryg}
\frac{{P\left({y|\Theta_{0}}\right)}}{{P\left({y|\hat{\Theta}}\right)}} \;\, ,
\end{eqnarray}
skonstruowaną przy założeniu prawdziwości hipotezy zerowej. 
Hipotezę zerowa $H_{0}$ odrzucamy na rzecz hipotezy alternatywnej, jeśli jej wiarygodność $P\left({y|\Theta_{0}}\right)$ jest ``za mała''.  Sugerowałoby to, że złożona hipoteza alternatywna $H_{1}$  zawiera pewną hipotezę prostą, która jest lepiej poparta przez dane otrzymane w próbce, niż hipoteza zerowa. \\
Jak o tym wspomnieliśmy powyżej, np. $5\%$-owe obcięcie $c$ w zagadnieniu estymacyjnym, samo w sobie nie mówi nic o frakcji liczby przedziałów wiarygodności pokrywających nieznaną wartość szacowanego parametru. Potrzebne jest wyskalowanie ilorazu wiarygodności. Również dla weryfikacji hipotez skalowanie  wiarygodności jest istotne. Stwierdziliśmy, że takie skalowanie jest możliwe wtedy gdy mamy do czynienia z jednoparametrowym przypadkiem rozkładu Gaussa, a przynajmniej z przypadkiem wystarczająco regularnym. \\
 \\
{\bf Empiryczny poziom istotności}: W przypadku jednoparametrowego, regularnego problemu z  ($\Theta \equiv (\theta)_{n=1}^{N})$ jak w Przykładzie z Rozdziału~\ref{Wiarygodnosciowy przedzial ufnosci}, skalowanie poprzez wykorzystanie statystki Wilka służy otrzymaniu empirycznego poziomu istotności $p$. Ze związku (\ref{1}) otrzymujemy wtedy  przybliżony (a dokładny dla rozkładu normalnego) {\it empiryczny poziom istotności}: 
\begin{eqnarray}
\label{poziom istotnosci wiaryg}
\!\!\!\!\!\!\!\!\!\!\!
p &\approx&   P\left(\frac{{P\left({\widetilde{Y}|\hat{\theta}}\right)}}{{P\left({\widetilde{Y}|\theta_{0}}\right)}} \geq  \frac{{P\left({y|\hat{\theta}_{obs}}\right)}}{{P\left({y|\theta_{0}}\right)}} \right) = P\left({2\ln\frac{{P\left({\widetilde{Y}|\hat{\theta}}\right)}}{{P\left({\widetilde{Y}|\theta_{0}}\right)}} \geq -2\ln c_{obs}}\right) \nonumber  \\ 
&=& P\left({\chi_{1}^{2} \geq -2\ln c_{obs}}\right) \; , \;\;\;\; {\rm gdzie} \;\;\;\;\; c_{obs} \equiv \frac{{P\left({y|\theta_{0}}\right)}}{{P\left({y|\hat{\theta}_{obs}}\right)}} 
 \; ,
\end{eqnarray}
przy czym $\hat{\theta}_{obs}$ jest wartością estymatora MNW $\hat{\theta}$ wyznaczoną w obserwowanej (obs) próbce $y$. 
Powyższe określenie empirycznego poziomu istotności $p$ oznacza, że w przypadku wystarczająco regularnego problemu \cite{Pawitan}, istnieje typowy związek pomiędzy prawdopodobieństwem 
(\ref{1 minus alfa}),  a empirycznym poziomem istotności $p$,  
podobny do związku jaki istnieje pomiędzy poziomem ufności $1-\alpha$, a poziomem istotności $\alpha$ w analizie częstotliwościowej. I tak, np. w przypadku jednoparametrowego rozkładu normalnego możemy  wykorzystać wartość empirycznego 
poziomu istotności $p$ do stwierdzenia, że gdy $p \leq \alpha$ to hipotezę $H_{0}$ odrzucamy na rzecz hipotezy $H_{1}$,  a w przypadku $p > \alpha$ nie mamy podstawy do odrzucenia $H_{0}$. \\
 \\
{\bf Problem błędu pierwszego i drugiego rodzaju}: Jednakże podobne skalowanie ilorazu wiarygodności okazuje się być znacznie trudniejsze już chociażby tylko w przypadku dwuparametrowego rozkładu normalnego, gdy obok $\theta$ estymujemy $\sigma^{2}$  \cite{Pawitan}. Wtedy określenie 
co oznacza sformułowanie ,,zbyt mała'' wartość $c$ jest dość dowolne i zależy od rozważanego
problemu lub wcześniejszej wiedzy wynikającej z innych źródeł niż
prowadzone statystyczne wnioskowanie. Wybór dużego parametru obcięcia $c$ spowoduje, że 
istnieje większe prawdopodobieństwo popełnienia {\it błędu pierwszego rodzaju} polegającego na odrzuceniu hipotezy zerowej w przypadku, gdy jest ona prawdziwa. Wybór małego $c$  spowoduje zwiększenie prawdopodobieństwa popełnienia {\it błędu drugiego rodzaju}, tzn. przyjęcia hipotezy zerowej w sytuacji, gdy  jest ona błędna. 

\newpage

\section[MNW w analizie regresji]{MNW w analizie regresji}

\label{regresja klasyczna}

Analiza zawarta w całym Rozdziale~\ref{regresja klasyczna} oparta jest na przedstawieniu metody MNW w analizie regresji klasycznej podanym w 
\cite{Kleinbaum} i 
\cite{Mroz}. \\
{\bf W metodzie regresji klasycznej}, estymatory parametrów strukturalnych modelu regresji są otrzymane arytmetyczną metodą najmniejszych  kwadratów (MNK). 
Zmienne objaśniające $X_{n} = x_{n}\,$, $n=1,...,N$,  
nie mają wtedy charakteru stochastycznego, co oznacza, że eksperyment jest ze względu na nie kontrolowany. \\
 \\
{\bf MNK polega na} minimalizacji sumy kwadratów odchyleń 
obserwowanych wartości zmiennej objaśnianej (tzw. odpowiedzi) od ich wartości teoretycznych spełniających równanie regresji.
MNK ma znaczenie probabilistyczne tylko w przypadku  analizy standardowej, gdy zmienna objaśniana $Y$ ma rozkład normalny.  Jej estymatory pokrywają się wtedy z estymatorami MNW. Pokażemy, że tak się sprawy mają. \\
\\
Załóżmy, że zmienne $\,Y_{1},Y_{2},...,Y_{N}\,$ odpowiadające kolejnym wartościom zmiennej objaśniającej, $x_{1},$ $x_{2},...,x_{N}$, są względem siebie niezależne i mają rozkład normalny
ze średnią   $\mu_{n}=E\left({Y\left|{x_{n}}\right.}\right) = E\left(Y_{n}\right)$ zależną od wariantu zmiennej objaśniającej $x_{n}$, oraz 
taką samą wariancję $\sigma^{2}(Y_{n})=\sigma^{2}(Y)$. \\
Funkcja wiarygodności próbki  $\left({y_{1},y_{2},...,y_{N}}\right)$ dla normalnego klasycznego
modelu regresji z parametrem $\Theta = \mu \equiv (\mu_{n})_{n=1}^{N}$, ma postać:
\begin{eqnarray} 
\label{wiaryg dla regr klas} 
P(\mu) \equiv P\left({y}|\mu \right) &=& \prod\limits_{n=1}^{N} {f\left({{\bf y}_{n}|\mu_{n}} \right)} = \prod\limits_{n=1}^{N}{\frac{1}{{\sqrt{2\pi\sigma^{2}}}} \, \exp\left\{ {-\frac{1}{{2\sigma^{2}}} \left({{\bf y}_{n} - \mu_{n}}\right)^{2}} \right\} } \nonumber \\
&=& \frac{1}{{\left({2\pi\sigma^{2}}\right)^{N/2}}}\exp\left\{ { - \frac{1}{{2\sigma^{2}}} \sum\limits_{n=1}^{N}{\left({{\bf y}_{n}-\mu_{n}}\right)^{2}}}\right\} \; ,
\end{eqnarray}
gdzie $f\left({{\bf y}_{n}|\mu_{n}} \right)\,$, $n=1,2...,N$,  są punktowymi  rozkładami gestości prawdopodobieństwa Gaussa. 
Widać, że maksymalizacja $P(\mu)$ ze względu na $(\mu_{n})_{n=1}^{N}$ pociąga za sobą minimalizację
sumy kwadratów reszt\footnote{SSE w literaturze angielskiej.} ($SKR$): 
\begin{eqnarray}
\label{SKR}
SKR = \sum\limits_{n=1}^{N}{\left({{\bf y}_{n}-\mu_{n}}\right)^{2}} \; , 
\end{eqnarray}
gdzie $\mu_{n}=E\left({Y\left|{x_{n}}\right.}\right)$ 
jest {\it postulowanym 
modelem regresji}. Zatem w standardowej, klasycznej 
analizie regresji, estymatory MNW pokrywają się
z estymatorami MNK. Widać, że procedura minimalizacji dla $SKR$ prowadzi do liniowej w $Y_{n}$ postaci estymatorów $\hat{\mu}_{n}$ parametrów $\mu_{n}$. \\
 \\
{\bf Problem z nieliniowym układem równań wiarygodności}: Jednak rozwiązanie układu równań wiarygodności (\ref{rown wiaryg}) jest zazwyczaj nietrywialne. Jest tak, gdy otrzymany w wyniku ekstremizacji układ algebraicznych równań wiarygodności dla  estymatorów jest nieliniowy, co w konsekwencji oznacza, że możemy nie otrzymać ich w zwartej analitycznej postaci. Przykładem
może być analiza regresji Poissona, w której do rozwiązania równań
wiarygodności wykorzystujemy metody iteracyjne. W takich sytuacjach wykorzystujemy na ogół  jakiś program komputerowy do analizy  statystycznej, np. zawarty w pakiecie SAS. Po podaniu postaci funkcji wiarygodności, program komputerowy dokonuje jej maksymalizacji rozwiązując układ (\ref{rown wiaryg}) np. metodą Newton-Raphson'a  \cite{Pawitan,Mroz},  wyznaczając numerycznie wartości  estymatorów parametrów modelu. \\  
 \\
{\bf Testy statystyczne}: Logarytm ilorazu wiarygodności jest również wykorzystywany w analizie regresji do przeprowadzania testów statystycznych przy  weryfikacji hipotez o 
nie występowaniu braku dopasowania modelu mniej złożonego, tzw. ``niższego'', o mniejszej liczbie parametrów, w stosunku do bardziej złożonego modelu ``wyższego'', posiadającego większą liczbę  parametrów. Statystyka wykorzystywana do tego typu testów 
ma postać \cite{Kleinbaum,Pawitan,Mroziakiewicz}:
\begin{eqnarray}
\label{prawie dewiancja}
-2\ln\frac{{P\left({\widetilde{Y}|\hat{\Theta}_{1}}\right)}}{{P\left({\widetilde{Y}|\hat{\Theta}_{2}}\right)}}
\end{eqnarray}
gdzie ${P\left({\widetilde{Y}|\hat{\Theta}_{1}}\right)}$ jest maksymalizowaną
wartością funkcji wiarygodności dla modelu mniej złożonego, a ${P\left({\widetilde{Y}|\hat{\Theta}_{2}}\right)}$ dla modelu bardziej
złożonego. Przy prawdziwości hipotezy zerowej $H_{0}$
o braku konieczności rozszerzania modelu niższego do wyższego, statystyka (\ref{prawie dewiancja}) ma asymptotycznie  rozkład $\chi^{2}$ z liczbą stopni swobody równą różnicy liczby parametrów modelu wyższego i niższego. \\
 \\
{\bf Analogia współczynnika determinacji}: Maksymalizowana wartość funkcji wiarygodności zachowuje się podobnie jak {\it współczynnik determinacji} $R^{2}$ \cite{Kleinbaum,Mroz}, tzn. rośnie wraz ze wzrostem liczby parametrów w modelu, zatem wielkość pod logarytmem należy do przedziału $\left(0,1\right)$ i statystyka (\ref{prawie dewiancja}) przyjmuje wartości z przedziału $\left(0,+\infty\right)$. Stąd (asymptotycznie) zbiór krytyczny dla $H_{0}$ jest prawostronny. 
Im lepiej więc model wyższy dopasowuje się do danych empirycznych w stosunku 
do modelu niższego, tym większa jest wartość statystyki ilorazu
wiarygodności (\ref{prawie dewiancja}) i większa szansa, że wpadnie ona w przedział
odrzuceń hipotezy zerowej $H_{0}$, który leży w prawym ogonie wspomnianego rozkładu $\chi^{2}$ \cite{Kleinbaum,Mroz}. 

\newpage

\subsection{Dewiancja jako miara dobroci dopasowania.  Rozkład  Poissona.}

\label{Dewiancja jako miara dobroci dopasowania}

Rozważmy zmienną losową $Y$ posiadającą rozkład Poissona.  Rozkład ten jest wykorzystywany do modelowania zjawisk związanych z rzadko zachodzącymi zdarzeniami, jak na przykład z liczbą rozpadających się niestabilnych jąder w czasie {\it t}.
Ma on postać: 
\begin{equation} 
\label{rozklad Poissona} 
p \left(Y={\bf y}|\mu \right)=\frac{\mu ^{{\bf y}} e^{-\mu } }{{\bf y}\, !} \; , \;\;\; {\rm oraz} \;\;\;   {\bf y} = 0,1,...,\infty  \; ,
\end{equation} 
gdzie $\mu $ jest parametrem rozkładu. 
Zmienna losowa podlegająca rozkładowi Poissona może przyjąć tylko nieujemną wartość całkowitą. 
Rozkład ten można wyprowadzić z rozkładu dwumianowego, bądź wykorzystując rozkłady Erlanga i wykładniczy \cite{Nowak}. \\
\\
{\bf Związek wariancji z wartością oczekiwaną rozkład Poissona}: Rozkład Poissona posiada pewną interesującą właściwość statystyczną, mianowicie jego wartość oczekiwana, wariancja i trzeci moment centralny są równe parametrowi rozkładu $\mu$:
\begin{eqnarray}
\label{E sigma trzeci Poissona}
E(Y) = \sigma ^{2} (Y) = \mu _{3} =\mu \; .
\end{eqnarray}
Aby pokazać dwie pierwsze równości w (\ref{E sigma trzeci Poissona}) skorzystajmy bezpośrednio z definicji odpowiednich momentów, otrzymując:
\begin{eqnarray}
\label{E Y dla rozkl Poissona}
E\left(Y\right)& =& \sum _{{\bf y}=0}^{\infty }{\bf y}\cdot p\left(Y={\bf y}|\mu \right) =\sum _{{\bf y}=0}^{\infty }{\bf y}\cdot \frac{\mu ^{{\bf y}} e^{-\mu } }{{\bf y}!}  = e^{-\mu } \sum _{{\bf y}=1}^{\infty }\frac{\mu ^{{\bf y}} }{\left({\bf y}-1\right)!} \nonumber \\ 
& = & e^{-\mu } \mu \sum _{{\bf y}=1}^{\infty }\frac{\mu ^{{\bf y}-1} }{\left({\bf y}-1\right)!} =e^{-\mu } \mu \sum _{l=0}^{\infty }\frac{\mu ^{l} }{l!} = e^{-\mu } \mu \, e^{\mu } =  \mu  \; ,  
\end{eqnarray}
oraz, korzystając z (\ref{E Y dla rozkl Poissona}):
\begin{eqnarray}
\label{sigma2 dla rozkl Poissona}
& &\sigma^{2} \left(Y\right) = E\left(Y^{2} \right)-\left[E\left(Y\right)\right]^{2} = E\left(Y^{2} \right) - \mu^{2} =\sum _{{\bf y}=0}^{\infty }{\bf y}^{2} \cdot p\left(Y={\bf y}|\mu \right) - \mu^{2} 
\nonumber \\ 
& & = \sum _{{\bf y}=0}^{\infty }{\bf y}^{2} \cdot \frac{\mu ^{{\bf y}} e^{-\mu } }{{\bf y}!} - \mu^{2} = e^{-\mu } \sum _{{\bf y}=1}^{\infty }{\bf y}\frac{\mu ^{{\bf y}} }{\left({\bf y}-1\right)!} - \mu^{2} = \,  e^{-\mu } \mu \sum _{l=0}^{\infty }\left(l+1\right)\frac{\mu ^{l} }{l!}  - \mu^{2}  
\nonumber \\ 
& & = e^{-\mu } \mu \, \left[\sum _{l=0}^{\infty }l\frac{\mu ^{l} }{l!} + \, e^{\mu } \right] - \mu^{2} = e^{-\mu } \mu \, \left[e^{\mu } \mu +e^{\mu } \right] - \mu^{2} = (\mu ^{2} +\mu ) - \mu^{2} = \mu  \; . 
\end{eqnarray}
\\
{\bf Uwaga}: {\it Zatem otrzymaliśmy ważną własność rozkładu Poissona}, która mówi, że stosunek dyspersji $\sigma$ do wartości oczekiwanej $E(Y)$  maleje pierwiastkowo wraz ze wzrostem poziomu zmiennej $Y$ opisanej tym rozkładem: 
\begin{eqnarray}
\label{sigma do E Poissona}
\frac{\sigma}{E(Y)} =  \frac{1}{\sqrt{\mu}} \;\; .
\end{eqnarray}
Fakt ten oznacza z założenia {\it inne zachowanie się odchylenia standardowego} w modelu regresji Poissona niż w klasycznym modelu regresji normalnej (w którym zakładamy jednorodność wariancji zmiennej objaśnianej w różnych wariantach zmiennej objaśniającej). \\
  \\
{\bf Ćwiczenie}: Pokazać (\ref{E sigma trzeci Poissona}) dla trzeciego momentu. \\ 
  \\ 
{\bf Przyczyna nielosowej zmiany wartości zmiennej objaśnianej}:  Rozważmy model regresji dla zmiennej objaśnianej $Y$  posiadającej rozkład Poissona. Zmienne $Y_{n}$,  $n=1, 2,...,N$ posiadają więc również rozkład Poissona i zakładamy, że są {\it parami wzajemnie niezależne}. 
Niech $X$ jest zmienną objaśniającą (tzw. czynnikiem) kontrolowanego eksperymentu, w którym $X$ nie jest zmienną  losową,  ale {\it jej zmiana}, jest rozważana jako możliwa przyczyna warunkująca  {\it nielosową zmianę wartości zmiennej} $Y$. \\
Gdy czynników $X_{1} ,X_{2} ,...X_{k}$ jest więcej, wtedy dla każdego punktu $n$ próby  podane są wszystkie ich wartości: 
\begin{equation} 
\label{wartości czynnikow x} 
x_{1n} ,x_{2n} ,...x_{kn} \; , \;\;\; {\rm gdzie}\;\;\; n=1,2,...,N\; ,
\end{equation} 
gdzie pierwszy indeks  w $x_{in}$, $i=1,2,...,k$, numeruje zmienną objaśniającą.  \\
\\
{\bf Brak możliwości eksperymentalnej separacji podstawowego kanału $n$}: Niech $x_{n} = (x_{1n} ,x_{2n} ,...,$ $x_{kn})$ oznacza zbiór wartości jednego wariantu zmiennych $\left(X_{1} ,X_{2} ,...,X_{k} \right)$, tzn. dla jednej konkretnej podgrupy $n$. Zwróćmy uwagę, że {\it indeks próby} $n$ numeruje podgrupę, co oznacza, że w pomiarze wartości $Y_{n}$ nie ma możliwości eksperymentalnego sięgnięcia ``w głąb'' indeksu $n$ - tego kanału, tzn. do rozróżnienia wpływów na wartość ${\bf y}_{n}$ płynących z różnych ``pod-kanałów'' $i$, gdzie $i=1,2,...,k$. \\
 \\
{\bf Model podstawowy}: Zakładając brak zależności zmiennej $Y$ od czynników $X_{1} ,X_{2} ,...X_{k}$, rozważa się tzw. {\it model podstawowy}. Dla rozkładu (\ref{rozklad Poissona}) i próby $\widetilde{Y} \equiv (Y_{n})_{n=1}^{N}$,  funkcja wiarygodności przy parametrze $\Theta = \mu \equiv (\mu_{n})_{n=1}^{N}$, ma postać:  
\begin{equation} 
\label{f wiaryg Poissona dla modelu podstawowego} 
P\left(\widetilde{Y}|\mu \right) = \prod_{n=1}^{N} \frac{\mu_{n}^{Y_{n} } e^{-\mu_{n} } }{Y_{n} !}  = \frac{\left(\prod_{n=1}^{N}\mu_{n}^{Y_{n} }  \right)\exp \left(-\sum_{n=1}^{N} \mu_{n}  \right)}{\prod_{n=1}^{N} Y_{n} ! }  \; ,
\end{equation} 
jest więc wyrażona jako funkcja wektorowego  parametru $\mu  \equiv (\mu_{n})_{n=1}^{N}$, gdzie każdy z parametrów $\mu_{n} = E(Y_{n})$ jest parametrem skalarnym. Rozważmy układ równań MNW:
\begin{equation} 
\label{uklad MNW rozklad Poissona podstawowy} 
\frac{\partial }{\partial \mu _{n} } 
\left[\ln P \left(\widetilde{Y}|\mu \right) \right] = 0 \; , \;\;\; n=1,2,...,N \; .
\end{equation} 
Dla funkcji wiarygodności (\ref{f wiaryg Poissona dla modelu podstawowego}) otrzymujemy:
\begin{equation} \label{log f wiaryg Poissona dla modelu podstawowego} 
\ln P \left(\widetilde{Y}|\mu\right)=\sum_{n=1}^{N} Y_{n} \ln \mu_{n}  - \sum_{n=1}^{N} \mu_{n}  - \sum_{n=1}^{N} \ln Y_{n} !  \;\, .
\end{equation} 
Zatem rozwiązanie układu (\ref{uklad MNW rozklad Poissona podstawowy}) daje:
\begin{eqnarray}
\label{estymatory modelu podst dla Poissona}
\mu_{n} = \hat{\mu}_{n} = Y_{n} \; , \;\;\; n=1,2,...,N \; ,
\end{eqnarray} 
jako estymatory modelu postawowego. 
%
%
%
%
Zatem funkcja wiarygodności (\ref{f wiaryg Poissona dla modelu podstawowego}) modelu podstawowego przyjmuje w punkcie $\mu$ zadanym przez estymatory (\ref{estymatory modelu podst dla Poissona}) wartość maksymalną: 
\begin{equation} \label{wiaryg zmaksym rozklad Poissona model podstawowy} 
P \left(\widetilde{Y}|\hat{\mu}\right)=\frac{\left(\prod_{n=1}^{N} Y_{n}^{Y_{n}} \right)\exp \left(-\sum _{n=1}^{N} Y_{n}  \right)}{\prod _{n=1}^{N} Y_{n} ! }  \; ,
\end{equation} 
gdzie zastosowano oznaczenie $\hat{\mu }=\left(\hat{\mu }_{1} ,\hat{\mu }_{2} ,...\hat{\mu }_{N} \right)$.

\vspace{5mm}

\subsection{Analiza regresji Poissona.} 

\label{Analiza regresji Poissona}

Niech zmienna zależna $Y$ reprezentuje liczbę zliczeń badanego zjawiska (np. przypadków awarii określonego zakupionego sprzętu), 
otrzymaną dla każdej z $N$ podgrup (np. klienckich). Każda z tych podgrup wyznaczona jest przez komplet wartości zmiennych objaśniających $X \equiv \left(X_{1} ,X_{2} ,...,X_{k} \right) = x \equiv \left(x_{1} ,x_{2} ,...,x_{k} \right)$ (np. wiek, poziom  wykształcenia, cel nabycia sprzętu). Zmienna $Y_{n} $ określa liczbę zliczeń zjawiska 
w $n$-tej podgrupie,  $n=1,2,...,N$. W konkretnej próbce $(Y_{n})_{n=1}^{N} = ({\bf y}_{n})_{n=1}^{N}$. \\
 \\
{\bf Określenie modelu regresji Poissona}: Rozważmy następujący model regresji Poissona: 
\begin{equation} 
\label{regresja Poisson} 
\mu_{n} \equiv E\left(Y_{n} \right) = \ell_{n} \, r\left(x_{n}, \beta \right) \; , \;\;\; n=1,2,...,N \; ,
\end{equation} 
opisujący zmianę wartości oczekiwanej liczby zdarzeń $Y_{n}$ (dla rozkładu Poissona) wraz ze zmianą {\it wariantu} $x_{n} = \left(x_{1n} ,x_{2n} ,...,x_{kn} \right)$.  \\
\\
Funkcja regresji po prawej stronie (\ref{regresja Poisson}) ma dwa czynniki. Funkcyjny czynnik funkcji regresji, $r\left(x_{n}, \beta \right)$, opisuje {\it tempo 
zdarzeń} określanych mianem porażek (np. awarii) w~$n$-tej podgrupie  (tzn. jest częstością tego zjawiska), skąd $r\left(x_{n}, \beta \right)>0$, gdzie  $\beta \equiv \left(\beta _{0} ,\beta _{1} ,...,\beta _{k} \right)$ jest zbiorem nieznanych parametrów  tego modelu regresji.  Natomiast czynnik $\ell_{n}$ jest  współczynnikiem  określającym  {\it dla każdej $\,n$-tej podgrupy}  (np. klientów) {\it skumulowany czas prowadzenia badań kontrolnych dla wszystkich jednostek tej podgrupy}. \\
Ponieważ funkcja 
regresji\footnote{Czynnik $r\left(x_{n}, \beta \right)$ nazywany dalej funkcją regresji, chociaż właściwie nazwa ta odnosi się do całej $E\left(Y_{n} \right)$.
} 
$r\left(x_{n}, \beta \right)$ przedstawia typową liczbę porażek na jednostkę czasu,  zatem nazywamy ją {\it ryzykiem}. \\
  \\
{\bf Uwaga o postaci funkcji regresji}: Funkcję $r\left(x_{n}, \beta \right)$ można zamodelować na różne sposoby \cite{Pawitan}. Wprowadźmy oznaczenie:
\begin{eqnarray}
\label{oznaczenie dla lambda regresji}
\lambda_{n}^{*} \equiv \beta _{0} + \sum _{j=1}^{k}\beta _{j} \, x_{jn} \; . 
\end{eqnarray}
Funkcja regresji $r\left(x_{n}, \beta \right)$ ma różną postać w zależności od typu danych. Może mieć ona postać charakterystyczną dla regresji liniowej (wielokrotnej), $r\left(x_{n}, \beta \right) = \lambda_{n}^{*}$, którą stosujemy  szczególnie wtedy gdy zmienna $Y$ ma {\it rozkład normalny}. 
Postać $r\left(x_{n} \beta \right) = 1/\lambda_{n}^{*}$ jest stosowana w analizie z danymi pochodzącymi z {\it rozkładu eksponentialnego}, natomiast  $r\left(x_{n}, \beta \right) = 1/(1+ \exp(-\lambda_{n}^{*}))$ w modelowaniu regresji logistycznej dla opisu  zmiennej {\it dychotomicznej} \cite{Kleinbaum,Pawitan}. \\
\\
{\bf Postać funkcji regresji użyteczna w regresji Poissona} jest następująca:
\begin{eqnarray}
\label{funkcja reg Poissona}
r\left(x_{n},  \beta \right) = \exp(\lambda_{n}^{*}) \; , \;\;\; \lambda_{n}^{*} =  \beta _{0} +\sum _{j=1}^{k}\beta _{j} x_{jn}  \; .
\end{eqnarray}
Ogólniej mówiąc analiza regresji odnosi się do modelowania wartości oczekiwanej zmiennej zależnej (objaśnianej) jako funkcji pewnych czynników. Postać funkcji wiarygodności stosowanej do estymacji współczynników regresji $\beta$ odpowiada założeniom dotyczącym rozkładu zmiennej zależnej. Tzn. zastosowanie konkretnej funkcji regresji $r(x_{n}, \beta )$, np. jak w (\ref{funkcja reg Poissona}), wymaga określenia postaci funkcji częstości $r(x_{n}, \beta )$, zgodnie z jej postacią dobraną do charakteru losowej zmiennej $Y$ przy której generowane są dane w badanym zjawisku. Na ogół przy konstrukcji $r(x_{n}, \beta )$ pomocna jest uprzednia wiedza dotyczącą relacji między rozważanymi zmiennymi. \\
\\
{\bf Funkcja wiarygodności dla analizy regresji Poissona}: Ponieważ $Y_{n} $ ma rozkład Poissona (\ref{rozklad Poissona}) ze średnią $\mu _{n}$,  $p\left(Y_{n}|\mu_{n} \right) = \frac{\mu _{n}^{Y_{n} } }{Y_{n} !} \, e^{-\mu _{n} }$, $n=1,\, 2,...,N$, zatem dane $Y_{n} =0, 1,...,\infty $ dla określonego  $n=1, 2,...,N$ są generowane z rozkładów warunkowych: 
\begin{equation} 
\label{prawd Poissona model regresji z li} 
p\left(Y_{n}| \beta \right)=\frac{\left[\ell_{n} \, r\left(x_{n}, \beta \right)\right]^{\, Y_{n} } }{Y_{n} !} e^{-\ell_{n}\, r\left(x_{n}, \beta \right)}  \; ,
\end{equation} 
wokół funkcji regresji, (\ref{regresja Poisson}), $\mu_{n} = \ell_{n} \, r(x_{n}, \beta )$, dla  $n=1, 2,...,N$. 
Funkcja wiarygodności dla analizy regresji Poissona ma więc postać:
\begin{eqnarray} 
\label{funkcja wiaryg regresja Poisson} 
P \left(\widetilde{Y}|\beta \right) &=& \prod_{n=1}^{N} p\left(Y_{n}|\beta \right) = \prod _{n=1}^{N}\frac{\left(\ell_{n} \, r\left(x_{n}, \beta \right)\right)^{Y_{n} } e^{-\ell _{n} \, r\left(x_{n}, \beta \right)} }{Y_{n} !} \nonumber \\
&=& \frac{ \prod _{n=1}^{N}\left(\ell _{n} \, r\left(x_{n}, \beta \right) \right)^{Y_{n} }  \, \exp \left[ -\sum _{n=1}^{N} \ell_{n} \, r\left(x_{n}, \beta \right) \right] }{\prod_{n=1}^{N} Y_{n} ! }  \; .
\end{eqnarray} 
\\
Aby w praktyce posłużyć się funkcją regresji $r(x_{n}, \beta )$ będącą określoną funkcją zmiennej $\lambda_{n}^{*} =  \beta_{0} + \sum_{j=1}^{k} \beta_{j} x_{jn}$, parametry $\beta_{0}, \beta_{1} ,...,\beta_{k}$ muszą być oszacowane.  Estymatory  MNW,  $\hat{\beta}_{0} ,\hat{\beta}_{1} ,..., \hat{\beta}_{k}$, tych parametrów  otrzymuje się rozwiązując $k+1$ równań wiarygodności:
\begin{eqnarray} 
\label{ukl row MNW dla beta fun regersji} 
\frac{\partial}{\partial \beta_{j} } \ln P \left(\widetilde{Y}| \beta \right) = 0 \; ,   \;\;\; j=0 , 1, 2,...,k \; .
\end{eqnarray} 
W przypadku regresji Poissona $ P\left(\widetilde{Y}| \beta \right) $ jest określona zgodnie z  (\ref{funkcja wiaryg regresja Poisson}). \\
\\
{\bf Algorytmy IRLS}:  Zauważmy, że dla rozkładu Poissona zachodzi zgodnie z (\ref{E sigma trzeci Poissona}) oraz (\ref{regresja Poisson}), $\sigma^{2} \left(Y_{n} \right) = E\left(Y_{n} \right) = \ell_{n} \, r\left(x_{n}, \beta \right)$, {\it co oznacza, że wariancja $\sigma^{2} \left(Y_{n} \right)$ zmiennej objaśnianej nie jest stała lecz zmienia się jako funkcja $\ell_{n} $ oraz $x_{n} $, wchodząc w analizę z różnymi wagami wraz ze zmianą 
$n$}. Na fakt ten zwróciliśmy już uwagę przy okazji związku (\ref{sigma do E Poissona}). 
Ponieważ układ równań wiarygodności (\ref{ukl row MNW dla beta fun regersji}) jest na ogół rozwiązywany iteracyjnymi metodami numerycznymi \cite{Kleinbaum}, 
a wariancja $\sigma^{2} \left(Y_{n} \right)$ jest również funkcją $\beta$, zatem na każdym kroku procesu iteracyjnego {\it wagi} te zmieniają się jako funkcja zmieniających się składowych estymatora $\hat{\beta}$. Algorytmy takiej analizy określa się ogólnym mianem  {\it algorytmów 
najmniejszych kwadratów\footnote{Należy jednak pamiętać, że zwrotu ``najmniejszych kwadratów'' nie należy tu brać dosłownie, gdyż metoda najmniejszych kwadratów ma sens jedynie wtedy, gdy rozkład zmiennej $Y$ jest 
normalny (por. Rozdział~\ref{regresja klasyczna}).} iteracyjnie ważonych} (IRLS).   \\
%
\\
{\bf Uwaga o programach}: Różne programy do analiz statystycznych, w tym SAS wykorzystujący procedurę PROC GENMOD, mogą być użyte do znajdowania estymatorów $\hat{\beta }$ MNW dla funkcji wiarygodności \eqref{funkcja wiaryg regresja Poisson}.  Również {\it obserwowana macierz kowariancji estymatorów}\footnote{Obserwowana macierz (wariancji-) kowariancji $\hat{V}(\hat{\beta })$ estymatorów  $\hat{\beta }$ MNW  jest zdefiniowana jako odwrotność macierzy obserwowanej informacji Fishera (\ref{I obserwowana}) \cite{Pawitan}:
\begin{eqnarray}
\label{macierz kowariancji estymatorów}
\hat{V}(\hat{\beta }) :=    \texttt{i\!F}^{-1}(\hat{\beta})  \; .
\end{eqnarray}
} 
oraz miary dobroci dopasowania modelu, takie jak omówiona dalej  dewiancja,
mogą być otrzymane przy użyciu powyżej wspomnianych programów. 

\subsubsection{Test statystyczny dla doboru modelu w regresji Poissona}

\label{Testy statystyczne doboru modelu}

{\bf Uwaga o większej wiarygodności modelu podstawowego}: Maksymalna wartość funkcji wiarygodności $P \left(y|\mu \right)$ wyznaczona w oparciu o \eqref{wiaryg zmaksym rozklad Poissona model podstawowy} będzie, dla każdego zbioru danych i dla liczby parametrów $k+1< N $, większa niż otrzymana przez maksymalizację funkcji wiarygodności \eqref{funkcja wiaryg regresja Poisson}. Jest tak, ponieważ w wyrażeniu \eqref{wiaryg zmaksym rozklad Poissona model podstawowy} na funkcję wiarygodności modelu podstawowego {\it nie narzuca się żadnych ograniczeń na postać} $\mu _{n} $, natomiast \eqref{funkcja wiaryg regresja Poisson} wymaga aby $\mu_{n} =\ell _{n} r\left(x_{n}, \beta \right)$. \\
 \\
{\bf Hipoteza zerowa o nie występowaniu braku dopasowania w modelu niższym}: Zgodnie z powyższym zdaniem, analizę doboru modelu regresji można  rozpocząć od postawienia hipotezy zerowej wobec alternatywnej. W hipotezie zerowej wyróżnimy proponowany model regresji. Wybór modelu badanego oznacza wybór funkcji wiarygodności \eqref{funkcja wiaryg regresja Poisson} z nim związanej. 
%
Stawiamy więc hipotezę zerową: 
\begin{equation} 
\label{Ho dla regresji Poissona} 
{\rm H}_{0} :\mu _{n} =\ell _{n} r\left(x_{n}, \beta \right),  n=1, 2,...,N,                   
\end{equation} 
która odpowiada wyborowi modelu z funkcją wiarygodności \eqref{funkcja wiaryg regresja Poisson}, 
wobec hipotezy alternatywnej:
\begin{equation} 
\label{H1 dla regresji Poissona} 
{\rm H}_{A}: \;\; \mu _{n} \; {\rm nie \; ma \; ograniczonej\; postaci}, \; n=1, 2,...,N \; ,
\end{equation} 
która odpowiada wyborowi modelu podstawowego zawierającego tyle parametrów $\mu_{n}$  ile jest punktów pomiarowych, tzn. \textit{N}, z funkcją wiarygodności \eqref{wiaryg zmaksym rozklad Poissona model podstawowy}.    \\
  \\
Niech więc $P \left(\widetilde{Y}|\hat{\beta }\right)$ jest maksymalną wartością funkcji wiarygodności określoną jak w \eqref{funkcja wiaryg regresja Poisson}. Oznacza to, że w miejsce parametrów $\beta  = \left(\beta _{0} ,\beta _{1} ,...,\beta _{k} \right)$ podstawiono  ich estymatory $\hat{\beta } = \left( \hat{\beta}_{0}, \hat{\beta}_{1},..., \hat{\beta}_{k} \right)$ wyznaczone przez MNW, jako te które maksymalizują funkcję wiarygodności \eqref{funkcja wiaryg regresja Poisson}.  Podobnie rozumiemy funkcję wiarygodności $P\left(\widetilde{Y}|\hat{\mu }\right)$ modelu podstawowego.\\
 \\
Ponieważ celem każdej analizy jest otrzymanie możliwie najprostszego opisu danych, model $\mu _{n} =\ell _{n} r\left(x_{n}, \beta \right)$ zawierający $k+1$ parametrów $\beta $, będzie uznany za dobry, jeśli maksymalna wartość funkcji wiarygodności wyznaczona dla niego, będzie prawie tak duża, jak funkcji wiarygodności dla nie niosącego żadnej informacji modelu podstawowego z liczbą parametrów $\mu_{n}$  równą licznie punktów pomiarowych $N$. Sformułowanie  „prawie tak duża'' oznacza, że wartość funkcji wiarygodności $P(y|\hat{\beta })$ nie może być istotnie statystycznie mniejsza od $P\left(y|\hat{\mu }\right)$. Zasadniczo powinno to oznaczać, że musimy podać miary pozwalające na określenie statystycznej istotności przy posługiwaniu się intuicyjnym parametrem obcięcia $c$ (Rozdział~\ref{Wnioskowanie w MNW}). Okazuje się, że dla dużej próby, miary typu (\ref{stat ilorazu wiaryg modelu reg}), podane poniżej, uzyskują cechy pozwalające na budownie wiarygodnościowych  obszarów krytycznych nabywających charakteru standardowego (częstotliwościowego). \\ 
%
%
\\
{\bf Określenie dewiancji}: Wprowadźmy  {\it statystykę typu ilorazu wiarygodności}:
\begin{equation} 
\label{stat ilorazu wiaryg modelu reg} 
D\left(\hat{\beta }\right) = 
-2\ln \left[\frac{P(\widetilde{Y}|\hat{\beta })}{P\left(\widetilde{Y}|\hat{\mu }\right)} \right] 
\end{equation} 
nazywaną \textit{dewiancją} (deviance) dla modelu regresji, w tym przypadku dla modelu Poissona z określoną postacią $\mu _{n} =\ell _{n} r\left(x_{n} , \beta \right)$. Służy ona do badania dobroci dopasowania modelu z zadaną postacią 
$\mu_{n} =\ell _{n} r\left(x_{n}, \beta \right)$ w stosunku do modelu podstawowego, bez narzuconej postaci na $\mu _{n} $, tzn. do stwierdzenia, czy $P(y|\hat{\beta })$ jest istotnie {\it mniejsza} od $P\left(y|\hat{\mu }\right)$, co sugerowałoby istotny  statystycznie  brak dopasowania badanego modelu $\mu _{n} =\ell _{n} r\left(x_{n}, \beta \right)$, do danych empirycznych. 
Jak pokażemy poniżej dewiancja może być rozumiana jako \textit{miara zmienności reszt} (tzn. odchylenia wartości obserwowanych w próbie od wartości szacowanych przez model)\textit{ wokół linii regresji}.  \\
\\  
Przy prawdziwości hipotezy $H_{0} :\mu _{n} =\ell _{n} r\left(x_{n}, \beta \right)$,  rozkład dewiancji  $D\left(\hat{\beta }\right)$ dla regresji Poissona, można asymptotycznie  
przybliżyć rozkładem chi-kwadrat (por. dyskusja w \cite{Pawitan,Kleinbaum}) z $N-k-1$ stopniami swobody. \\
\\
Zatem  statystyczny test  dobroci dopasowania, 
tzn. niewystępowania braku dopasowania badanego modelu $H_{0}: \mu_{n} =\ell _{n} r\left(x_{n} , \beta \right)$, w stosunku do modelu podstawowego, przebiega  w regresji Poissona  następująco:  Porównujemy otrzymaną w próbie wartość statystyki $D(\hat{ \beta })$ z wartością krytyczną leżącą w prawym ogonie  rozkładu chi-kwadrat (o $N-k-1$ stopniach swobody). Przyjęcie przez  $D\left(\hat{\beta }\right)$ wartości równej lub większej od krytycznej skutkuje odrzuceniem hipotezy zerowej. \\
\\
{\bf Wyznaczenie liczby stopni swobody dewiancji}: 
Podana liczba stopni swobody dewiancji $D\left(\hat{\beta }\right)$ wynika z następującego rozumowania. 
Zapiszmy (\ref{stat ilorazu wiaryg modelu reg}) w postaci: 
\begin{equation} 
\label{stat ilorazu wiaryg modelu reg 2} 
D\left(\hat{\beta }\right) + 2\ln P\left(\widetilde{Y}|\hat{\beta }\right) = 2\ln P\left(\widetilde{Y}|\hat{\mu }\right) \; ,
\end{equation} 
co po skorzystaniu z (\ref{funkcja wiaryg regresja Poisson}) dla $\beta=\hat{\beta}$ ma postać:
\begin{equation} 
\label{stat ilorazu wiaryg modelu reg 3} 
D\left(\hat{\beta }\right) - 2\sum _{n=1}^{N} \ell _{n} r\left(x_{n}, \hat{ \beta }\right) = 2\ln P \left(\widetilde{Y}|\hat{\mu }\right) + 2\ln \left(\prod_{n=1}^{N} Y_{n}! \right) - 2\ln \left(\prod _{n=1}^{N} \left(\ell _{n} r \left(x_{n} ,\hat{\beta }\right)\right)^{Y_{n}}  \right) \; .
\end{equation} 
Można zauważyć, że prawa strona tego równania ma $N$-stopni swobody. Istotnie, ze względu na (\ref{estymatory modelu podst dla Poissona})\footnote{W przyjętym przedstawieniu danych jak dla diagramu punktowego,  $N$ jest 
ogólnie liczbą punktów pomiarowych (równą liczbie wariantów czy  komórek). Tylko dla modelu podstawowego jest $N$ również liczbą 
parametrów.},  
$\hat{\mu} \equiv (\hat{\mu}_{n}) = (Y_{n}) \,$, $n=1,2,...,N$, liczba niezależnych zmiennych po prawej strony powyższego równania, których  wartości trzeba określić z eksperymentu, wynosi $N$.  Natomiast drugi składnik po lewej stronie ma liczbę stopni swobody równą $k+1$, co jest liczbą estymatorów parametrów strukturalnych  $\hat{ \beta }$ modelu regresji, których wartości trzeba określić z eksperymentu. Ponieważ liczba stopni swobody po prawej i lewej stronie równania musi być taka sama, zatem liczba stopni swobody dewiancji $D(\hat{\beta })$ wynosi  $N-k-1$. \\
  \\
{\bf Testy ilorazu wiarygodności}: Dewiancje dla hierarchicznych klas modeli mogą służyć do budowy testów stosunku wiarygodności. Zwróćmy szczególnie uwagę na funkcję wiarygodności \eqref{funkcja wiaryg regresja Poisson} zawierającą zbiór parametrów $\beta =\left(\beta_{0} ,\beta_{1} ,.....,\beta_{k} \right)$ z \textit{dewiancją} $D\left(\hat{\beta }\right)$ daną wyrażeniem \eqref{stat ilorazu wiaryg modelu reg}. Przypuśćmy, że chcemy zweryfikować hipotezę o tym, że $k-r$ (gdzie $0<r<k$) ostatnich parametrów będących składowymi wektora $\beta $ jest równych {\it zeru}. \\
\\
{\it Hipoteza zerowa}, o nieistotności rozszerzenia modelu niższego  do wyższego, ma wtedy postać:
\begin{equation} 
\label{hip zerowa w testach hierarchicznych} 
{\rm H}_{0}: \beta _{r+1} =\beta _{r+2} =...=\beta _{k} = 0 \; ,     
\end{equation} 
{\it Hipoteza alternatywna} $H_{A}$ mówi, że przynajmniej jeden z parametrów strukturalnych $\beta_{r+1}, \beta_{r+2},$ $...,\beta_{k}\,$ jest różny od {\it zera}. \\
  \\
Funkcja wiarygodności przy prawdziwości hipotezy zerowej $H_{0}$,   \eqref{hip zerowa w testach hierarchicznych}, ma postać taką jak w \eqref{funkcja wiaryg regresja Poisson}, tyle, że zastąpiono w niej  parametr $\beta$ parametrem $\beta_{(r)}$:
\begin{equation} 
\label{parametr modelu nizszego} 
\beta_{(r)} \equiv \left(\beta _{0} ,\beta _{1} ,...,\beta _{r}; \, 0,0,...,0 \right) \;\;{\rm gdzie \;\, liczba \;\, zer \;\, wynosi} \;\, k-r \; .
\end{equation} 
Oznaczmy funkcje wiarygodności tego modelu jako $P(\widetilde{Y}|\beta_{(r)})$,  a $\hat{ \beta}_{(r)} $ niech będzie estymatorem MNW wektorowego parametru $\beta_{(r)}$, wyznaczonym przez rozwiązanie odpowiadającego mu układu równań wiarygodności (oczywiście dla niezerowych parametrów  $\beta_{0} ,\beta_{1} ,...,\beta_{r}$).  Estymator $\hat{ \beta}_{(r)} = (\hat{\beta}_{0}, \hat{\beta}_{1} ,...,$ $\hat{\beta }_{r};$ $0,0,...,0) $ maksymalizuje funkcję wiarygodności  $P\left(\widetilde{Y}|\beta_{(r)} \right)$.  \\
  \\
{\it Test ilorazu wiarygodności} dla weryfikacji hipotezy $H_{0} $ przeprowadzamy posługując się \textit{statystyką ilorazu wiarygodności}:
\begin{equation} 
\label{statystyka ilorazu wiaryg} 
- 2 \ln \left[\frac{P\left(\widetilde{Y}|\hat{\beta}_{(r)} \right)}{P\left(\widetilde{Y}|\hat{\beta }\right)} \right] \; ,                  
\end{equation} 
która przy prawdziwości hipotezy zerowej ma asymptotycznie rozkład chi-kwadrat z $k-r$ stopniami swobody, 
co widać, gdy zapiszemy (\ref{statystyka ilorazu wiaryg}) jako  różnicę dewiancji: 
\begin{equation} 
\label{roznica dewiancji} 
-2\ln \left[\frac{P\left(\widetilde{Y}|\hat{\beta}_{(r)} \right)}{P\left(\widetilde{Y}|\hat{\beta }\right)} \right] = - 2\ln \left[\frac{P\left(\widetilde{Y}|\hat{\beta}_{(r)} \right)}{P\left(\widetilde{Y}|\hat{\mu }\right)} \right] + 2\ln \left[\frac{P\left(\widetilde{Y}|\hat{\beta}\right)}{P\left(\widetilde{Y}|\hat{\mu }\right)} \right]  = D\left(\hat{\beta}_{(r)} \right) - D\left(\hat{\beta }\right) \; ,   
\end{equation} 
oraz skorzystamy z podobnej analizy jak dla (\ref{stat ilorazu wiaryg modelu reg 3}). \\
 \\
Zatem, przy prawdziwości hipotezy zerowej \eqref{hip zerowa w testach hierarchicznych}, którą można zapisać jako ${\rm H}_{0}: \beta_{r+1} =\beta_{r+2} =...=\beta_{k} =0$, różnica $D(\hat{\beta}_{(r)}) - D(\hat{\beta })$ ma dla dużej próby w przybliżeniu rozkład chi-kwadrat z $k-r$ stopniami swobody. \\ 
\\
{\bf Wniosek}: Jeśli używamy regresji Poissona do analizowania danych empirycznych, modele tworzące hierarchiczne  klasy mogą być porównywane miedzy sobą poprzez wyznaczenie statystyki ilorazu wiarygodności (\ref{statystyka ilorazu wiaryg}), lub co na jedno wychodzi, poprzez wyznaczenie różnicy (\ref{roznica dewiancji})  między parami dewiancji dla tych modeli. Należy przy tym pamiętać o wniosku jaki już znamy z analizy dewiancji, że {\it im model gorzej dopasowuje się do danych empirycznych tym jego dewiancja jest większa.}

\subsubsection{Podobieństwo dewiancji do SKR analizy częstotliwościowej}

\label{Podobienstwo dewiancji do SKR}

Warunkowe wartości oczekiwane $\mu_{n} \equiv E(Y_{n}) = \ell _{n} \, r(x_{n}, \beta)$,  $n=1,2,...,N$, (\ref{regresja Poisson}), są w analizie regresji przyjmowane jako teoretyczne przewidywania modelu regresji dla wartości zmiennej objaśnianej $Y_{n}$, zwanej odpowiedzią (układu). W próbie odpowiadają im oszacowania, oznaczone jako $\hat{Y}_{n}$,  które  w $n$-tej komórce są następujące: 
\begin{eqnarray} 
\label{przewidywania modelu regresji Poissona} 
\hat{Y}_{n} = \ell _{n} \, r(x_{n} ,\hat{\beta }) \; , \;\;\; n=1,2,...,N \; ,
\end{eqnarray} 
zgodnie z  wyestymowaną postacią modelu regresji. 
Wykorzystując (\ref{przewidywania modelu regresji Poissona}) możemy zapisać dewiancję modelu (\ref{stat ilorazu wiaryg modelu reg}) nastepująco: 
\begin{eqnarray} 
\label{dewiancja poprzez przewidywania rachunek}
D\left(\hat{ \beta }\right) &=& - 2 \ln \left[\frac{P\left(\widetilde{Y}|\hat{\beta}\right)}{P\left(\widetilde{Y}|\hat{\mu }\right)} \right] = -2 \ln \left[\frac{\prod _{n=1}^{N} \hat{Y}_{n}^{Y_{n} } \exp \left(-\sum _{n=1}^{N} \hat{Y}_{n}  \right) }{\prod_{n=1}^{N}Y_{n}^{Y_{n} } \exp \left(-\sum _{n=1}^{N}Y_{n}  \right) } \right] \nonumber \\  
&=& -2\left[\sum _{n=1}^{N}Y_{n} \ln \hat{Y}_{n}  - \sum_{n=1}^{N}\hat{Y}_{n}  - \sum_{n=1}^{N}Y_{n} \ln Y_{n}  + \sum_{n=1}^{N} Y_{n}  \right]
\end{eqnarray}
tzn: 
\begin{eqnarray} 
\label{dewiancja poprzez przewidywania} 
D\left(\hat{\beta }\right) = 2\sum _{n=1}^{N} \left[Y_{n} \ln \left(\frac{Y_{n} }{\hat{Y}_{n} } \right) - \left(Y_{n} -\hat{Y}_{n} \right)\right] \;  .  
\end{eqnarray} 
{\bf Podobieństwo $D$ do $SKR$}: Powyższa postać dewiancji oznacza, że $D(\hat{ \beta })$ zachowuje się w poniższym sensie jak suma kwadratów reszt $SKR = \sum_{n=1}^{N} (Y_{n} -\hat{Y}_{n})^{2}$ w standardowej wielorakiej regresji liniowej.  Otóż, gdy dopasowywany model dokładnie przewiduje obserwowane wartości, tzn. $\hat{Y}_{n} =Y_{n} ,\; n=1,2,..,N$ wtedy, jak $SKR$ w analizie standardowej, tak $D(\hat{ \beta })$ w analizie wiarygodnościowej jest  równe zeru \cite{Kleinbaum,Czerwik}. 
Z drugiej strony wartość $D(\hat{ \beta })$ jest tym większa im większa jest różnica między wartościami obserwowanymi $Y_{n}$ i wartościami przewidywanymi  $\hat{Y}_{n} $ przez oszacowany model. \\
\\
{\bf Asymptotyczna postać $D$}: 
W analizowanym modelu $Y_{n}$, $n=1,2,...,N$ są niezależnymi zmiennymi Poissona (np. zmiennymi częstości), natomiast wartości $\hat{Y}_{n} $  są ich przewidywaniami. Nietrudno przekonać się, że gdy wartości przewidywane mają rozsądną 
wartość\footnote{Zauważmy, 
że  statystyka (\ref{dewinacja jak chi}) może mieć myląco dużą wartość gdy wielkości $\hat{Y}_{n} $ są bardzo małe.
}, 
np. $\hat{Y}_{n} >3$ oraz $(Y_{n} -\hat{Y}_{n}) << Y_{n}\,$, $n=1,2,..,N\,$ tak, że $(Y_{n} -\hat{Y}_{n})/Y_{n} << 1$, wtedy wyrażenie w nawiasie kwadratowym w (\ref{dewiancja poprzez przewidywania}) można przybliżyć przez $(Y_{n} -\hat{Y}_{n})^{2}/(2 \,Y_{n})$, a statystykę (\ref{dewiancja poprzez przewidywania}) można przybliżyć statystyką o postaci:
\begin{equation} \label{dewinacja jak chi} 
\chi ^{2} =\sum _{n=1}^{N}\frac{\left(Y_{n} -\hat{Y}_{n} \right)^{2} }{\hat{Y}_{n} }   \;  , 
\end{equation} 
która (dla dużej próby) ma rozkład chi-kwadrat z $N-k-1$ stopniami swobody \cite{Kleinbaum}.


\newpage

\section[Zasada niezmienniczości ilorazu funkcji wiarygodności]{Zasada niezmienniczości ilorazu funkcji wiarygodności}

Z powyższych rozważań wynika, że funkcja wiarygodności reprezentuje niepewność dla ustalonego parametru. Nie jest ona jednak  gęstością  rozkładu prawdopodobieństwa  dla tego parametru. Pojęcie takie byłoby całkowicie obce statystyce klasycznej (nie włączając procesów stochastycznych). Inaczej ma się sprawa w tzw. statystyce Bayesowskiej. 
Aby zrozumieć różnicę pomiędzy podejściem klasycznym i Bayesowskim \cite{Marek_statyst_Bayes} rozważmy transformację parametru. \\
{\bf Przykład transformacji parametru}: Rozważmy eksperyment, w którym dokonujemy jednokrotnego pomiaru zmiennej o  rozkładzie dwumianowym (\ref{Bernoulliego rozklad}). Funkcja wiarygodności ma więc postać $P(\theta)=\left(\!\! \begin{array}{l}
m\\
x\end{array} \!\!\! \right) \theta^{x} (1-\theta)^{m-x}$. Niech  parametr $m=12$ a w pomiarze otrzymano $x=9$. Testujemy model, dla którego $\theta = \theta_{1} = 3/4$ wobec modelu z $\theta = \theta_{2} = 3/10$. Stosunek wiarygodności wynosi:
\begin{eqnarray}
\label{stosunek L skalowanie} 
\frac{P(\theta_{1} = 3/4)}{P(\theta_{2} = 3/10)} = \frac{\left(\begin{array}{l}
m\\
x\end{array}\right) \theta_{1}^{9} \, (1-\theta_{1})^{3}}{\left(\begin{array}{l}
m\\
x\end{array}\right) \theta_{2}^{9} \, (1-\theta_{2})^{3}} \; = \; 173.774
\end{eqnarray}
Dokonajmy hiperbolicznego wzajemnie jednoznacznego przekształcenia parametru:
\begin{eqnarray}
\label{transf parametru}
\psi = 1/\theta  \; .
\end{eqnarray}
Funkcja wiarygodności po transformacji parametru ma postać $\tilde{P}(\psi)\!=\!\left( \!\! \begin{array}{l}
m\\
x
\end{array} \!\!\! \right) (1/\psi)^{x} (1-1/\psi)^{m-x}$. Wartości parametru $\psi$ odpowiadające wartościom $\theta_{1}$ i  $\theta_{2}$ wynoszą odpowiednio $\psi_{1}=4/3$ oraz $\psi_{2}=10/3$. 
Łatwo sprawdzić, że transformacja (\ref{transf parametru}) nie zmienia stosunku wiarygodności, tzn.:
\begin{eqnarray}
\label{stosunek L po transformacji skalowanie} 
\frac{\tilde{P}(\psi_{1} = 4/3)}{\tilde{P}(\psi_{2} = 10/3)} = \frac{P(\theta_{1} = 3/4)}{P(\theta_{1} = 3/10)} = 173.774 \; .
\end{eqnarray}
\\
{\bf Niezmienniczość stosunku wiarygodności}: Zatem widać, że stosunek wiarygodności jest niezmienniczy ze względu na wzajemnie jednoznaczą transformację parametru. Gdyby transformacja parametru była np. transformacją ``logit'' $\psi = \ln(\theta/(1 - \theta))$ lub paraboliczną $\psi = \theta^{2}$, to sytuacja także nie uległaby zmianie. Również w ogólnym przypadku transformacji parametru własność {\it niezmienniczości stosunku wiarygodności} pozostaje słuszna. Oznacza to, że informacja zawarta w próbce jest niezmiennicza ze względu na wybór parametryzacji, tzn. powinniśmy być w takiej samej sytuacji niewiedzy niezależnie od tego jak zamodelujemy zjawisko, o ile różnica w modelowaniu sprowadza się jedynie do transformacji parametru. W omawianym przykładzie powinniśmy  równie dobrze móc  stosować parametr $\theta$, jak $1/\theta$, $\theta^{2}$, czy $\ln(\theta/(1 - \theta))$. \\
 \\
{\bf Uwaga o transformacji parametru w statystyce Bayesowskiej}:  Natomiast sytuacja ma się zupełnie inaczej w przypadku Bayesowskiego podejścia do funkcji wiarygodności \cite{Marek_statyst_Bayes}, w którym funkcja wiarygodności uwzględnia (Bayesowski) rozkład prawdopodobieństwa  $f(\theta|x)$ parametru~$\theta$. Oznacza to, że 
Jakobian transformacji $\theta \rightarrow \psi$ parametru, modyfikując rozkład parametru, zmienia również funkcję wiarygodności. Zmiana ta zależy od wartości parametru, różnie zmieniając licznik i mianownik w (\ref{stosunek L skalowanie}), co niszczy  {\it intuicyjną} własność niezmienniczości ilorazu wiarygodności ze względu na  transformację parametru \cite{Pawitan}. 

\chapter[Entropia względna i informacja Fishera]{Entropia względna i informacja Fishera}

\label{Entropia wzgledna i IF}

W pozostałych częściach skryptu nie będziemy zajmowali sie modelami regresyjnymi. Oszacowywany wektorowy parametr  $\Theta$, od którego zależy  funcja wiarygodności $P(\Theta)$ ma postać 
$\Theta = (\theta_{1},\theta_{2},...,\theta_{N})^{T} \equiv (\theta_{n})_{n=1}^{N}$ jak w (\ref{parametr Theta}), gdzie $\theta_{n} = (\vartheta_{1n},\vartheta_{2n},...,\vartheta_{kn})^{T} \equiv ((\vartheta_{s})_{s=1}^{k})_{n}$, skąd liczba wszystkich parametrów wynosi $d=k \times N$. {\it Aby nie komplikować zapisu będziemy stosowali oznaczenie} $\Theta \equiv  (\theta_{1},\theta_{2},...,\theta_{d})^{T} \equiv (\theta_{i})_{i=1}^{d}$, gdzie indeks $i=1,2,...,d$ zastapił parę indeksów $sn$. \\
 \\
Niech $\hat{\Theta}=\left({\hat{\theta}_{1}, \hat{\theta}_{2},..., \hat{\theta}_{d}}\right)$ jest estymatorem  MNW  wektorowego parametru $\Theta=\left(\theta_{1}, \theta_{2},..., \theta_{d}\right)$,  otrzymanym po rozwiązaniu układu równań wiarygodności (\ref{rown wiaryg}). Rozwiązanie to, jako maksymalizujące funkcję wiarygodności, musi spełniać warunek ujemnej określoności formy 
kwadratowej\footnote{W pełnym 
zapisie indeksów, ze względu na 
to, że jeden punktowy parametr $\theta_{n} = (\vartheta_{1n},\vartheta_{2n},...,\vartheta_{kn})^{T} \equiv ((\vartheta_{s})_{s=1}^{k})_{n}$ może być parametrem wektorowym, co ma miejsce gdy $k>1$, zapis (\ref{forma kw dla P}) oznacza:
\begin{eqnarray}
\label{forma kw dla P wszystkie parametry}
\sum_{n, \,n'=1}^{N} \; \sum_{s, s'=1}^{k} \frac{\partial^{2} \ln P}{\partial \vartheta_{sn} \partial \vartheta_{s'n'}} {\left|_{\Theta = \hat{\Theta}} \right.} \, \Delta\vartheta_{sn} \, \Delta\vartheta_{s'n'} \; .
\end{eqnarray}
Wkrótce i tak ograniczymy się do sytuacji gdy $k=1$, tzn. $\theta_{n} = \vartheta_{1n}=\vartheta_{n}$. 
}:
\begin{eqnarray}
\label{forma kw dla P}
\sum_{i, \,j=1}^{d} \frac{\partial^{2} \ln P}{\partial \theta_{i} \partial \theta_{j}} {\left|_{\Theta = \hat{\Theta}} \right.} \Delta\theta_{i}\Delta\theta_{j} \; ,
\end{eqnarray}
gdzie przyrosty $\Delta\theta_{i}$, $\Delta\theta_{j}$ nie zerują
się jednocześnie. 
W przypadku skalarnym (tzn. jednego parametru $\Theta=\theta=\vartheta$) warunek ten oznacza  ujemność drugiej pochodnej  
logarytmu funkcji wiarygodności w punkcie $\theta=\hat{\theta}$. Większa wartość $-\frac{{\partial^{2}}}{{\partial{\theta}^{2}}} \ln P\left(y|\theta\right)\left|_{\theta=\hat{\theta}} \right.$ oznacza węższe maksimum  $\ln P$ w punkcie $\theta=\hat{\theta}$, tzn. większą krzywiznę funkcji wiarygodności, a co za tym idzie mniejszą niepewność określenia parametru $\theta$.

\section[Obserwowana i oczekiwana informacja Fishera]{Obserwowana i oczekiwana informacja Fishera}

\label{iF oraz I_definicje}

Ponieważ $\Theta$ może być w ogólności parametrem wektorowym, więc jako uogólnienie przypadku skalarnego zdefiniujmy  $d \times d \,$-wymiarową macierz\footnote{
Będziemy powszechnie stosowali skrócony zapis typu:
$\,\Theta^{2} \equiv \Theta \Theta^{T} = (\theta_{i} \theta_{j})_{d \times d}$ oraz $\frac{\partial^{2}}{\partial \Theta^{2}} \equiv \frac{\partial^{2}}{\partial  \Theta \,\partial \Theta^{T}} = (\frac{\partial^{2}}{\partial \theta_{i} \partial \theta_{j}})_{d \times d}$.
}: 
\begin{eqnarray}
\label{I jako krzywizna dla P}
\texttt{i\!F}(\Theta)  \equiv -\frac{{\partial^{2}}}{{\partial{\Theta}} \partial{\Theta}^{T}} \ln P\left(y|\Theta\right) \equiv - \left( \frac{\partial^{2}\ln P}{\partial\theta_{i}\partial\theta_{j}} \right)_{d \times d} \; .
\end{eqnarray}
%
{\bf Określenie obserwowanej informacji Fishera}: Wartość statystyki $\texttt{i\!F}(\Theta)$ zdefiniowanej jak w (\ref{I jako krzywizna dla P}) w punkcie $\Theta=\hat{\Theta}$ (oznaczoną po prostu jako $\texttt{i\!F}$):
\begin{eqnarray}
\label{I obserwowana}
\texttt{i\!F} = \texttt{i\!F}\left(\hat{\Theta}\right) = \left. \texttt{i\!F}\left(\Theta \right)\right|_{\Theta=\hat{\Theta}} \; ,
\end{eqnarray} 
nazywamy  {\it obserwowaną informacją 
Fishera}. 
W teorii wiarygodności odgrywa ona kluczową rolę. Jako statystyka, czyli funkcja próby $\widetilde{Y}$,  jej realizacja w  próbce $y \equiv ({\bf y}_{1}, {\bf y}_{2}, ..., {\bf y}_{N})$ jest macierzą liczbową. Z faktu wyznaczenia $\texttt{i\!F}$ w punkcie estymatora MNW wynika, że jest ona dodatnio określona, natomiast z (\ref{I jako krzywizna dla P}) widać również, że $\texttt{i\!F}$ jest macierzą symetryczną.   \\
\\
We współczesnych wykładach   $\texttt{i\!F}$ jest zazwyczaj zapisane jako: 
\begin{eqnarray}
\label{observed IF Amari}
\widetilde{\texttt{i\!F}} = \left(\frac{\partial \ln P(\Theta)}{\partial\theta_{i'}}\frac{\partial \ln P(\Theta)}{\partial\theta_{i}}\right) \; . 
\end{eqnarray}
Obie definicje, tzn. (\ref{observed IF Amari})
oraz (\ref{I jako krzywizna dla P}), prowadzą na poziomie oczekiwanym do tych samych konkluzji, o ile 
$\int \! dy$ $P(\Theta)$ $\frac{\partial \ln P(\Theta)}{\partial\theta_{i}} = 0\,$, $\,i=1,2,..,d$ (por. (\ref{IF 2 poch na kwadrat pierwszej})).
Zasadniczą zaletą zdefiniowania  $\texttt{i\!F}$ poprzez  (\ref{observed IF Amari}) jest to, że bardzo naturalne  staje się wtedy  wprowadzenie tzw. $\alpha$-koneksji na przestrzeni statystycznej  ${\cal S}$  \cite{Amari Nagaoka book}.  Pojęcie  $\alpha$-koneksji omówimy w Rozdziale~\ref{alfa koneksja}. \\
  \\
{\bf Przykład estymacji wartości oczekiwanej w rozkładzie normalnym}: Jako przykład ilustrujący związek  wielkości obserwowanej informacji Fishera (IF) z niepewnością oszacowania parametru, rozważny realizację próby prostej $y$ dla zmiennej $Y$ posiadającej rozkład  $N\left({\theta,\sigma^{2}}\right)$. \\
\\
Załóżmy, że {\it wariancja $\sigma^{2}$ jest znana}, a {\it estymowanym parametrem
jest jedynie wartość oczekiwana} $\theta=E\left( Y \right)$. Logarytm funkcji wiarygodności ma postać: 
\begin{eqnarray}
\label{log wiaryg dla norm}
\ln P\left(y|\theta\right) = -\frac{N}{2} \ln (2 \pi \sigma^{2})-\frac{1}{{2\sigma^{2}}} \sum\limits_{n=1}^{N}{\left({{\bf y}_{n}-\theta}\right)^{2}}
\end{eqnarray}
skąd funkcja wynikowa (\ref{funkcja wynikowa}) jest równa: 
\begin{eqnarray}
\label{f wynikowa_1 wym N_1 par z definicji}
S\left(\theta\right) = \frac{\partial}{{\partial\theta}}\ln P\left({y|\theta}\right) = \frac{1}{{\sigma^{2}}}\sum\limits_{n=1}^{N}{\left({{\bf y}_{n}-\theta}\right)} \; .
\end{eqnarray}
Rozwiązując jedno równanie wiarygodności, $S\left(\theta\right)_{|\theta=\hat{\theta}} =0$,  otrzymujemy postać estymatora parametru $\theta$ (por. (\ref{srednia arytmet z MNW})): 
\begin{eqnarray}
\label{rozw r wiaryg dla mu w N}
\hat{\theta} = \bar{{\bf y}} = \frac{1}{N}\sum_{n=1}^{N}{\bf y}_{n} \; ,
\end{eqnarray}
skąd:
\begin{eqnarray}
\label{f wynikowa_1 wym N_1 par}
S\left(\theta\right) = \frac{N}{{\sigma^{2}}} \left(\hat{\theta} - \theta \right) \; .
\end{eqnarray}
Natomiast z (\ref{I obserwowana}) oraz (\ref{I jako krzywizna dla P}) otrzymujemy, że obserwowana IF jest równa: 
\begin{eqnarray}
\label{I obserw dla N_parametr mu}
\texttt{i\!F}\left({\hat{\theta}}\right) =  - \frac{\partial^{2}\, \ln P\left({y|\theta}\right)}{{\partial \theta^{2}}}\left. \right|_{\theta=\hat{\theta}} = - \frac{\partial\, S\left(\theta \right)}{{\partial \theta}}\left. \right|_{\theta=\hat{\theta}} = \frac{N}{{\sigma^{2}}} \; .
\end{eqnarray}
Z (\ref{rozw r wiaryg dla mu w N}) oraz z klasycznej analizy\footnote{Ponieważ dla próby prostej wszystkie $Y_{n}$, $n=1,2,...,N$, maja taki sam rozkład jak $Y$ oraz  dla $n \neq n'$ zachodzi ${\rm cov}(Y_{n},Y_{n'}) = 0$, zatem: $\, 
\sigma^{2}({\bar{Y}}) = \sigma^{2}(\frac{1}{N}\sum_{n=1}^{N} Y_{n})  $ $=  \frac{1}{N^{2}}\sigma^{2}(\sum_{n=1}^{N} Y_{n}) =  \frac{1}{N^{2}}\sum_{n=1}^{N} \sigma^{2}(Y_{n})  = \frac{1}{N^{2}} N \,\sigma^{2}(Y)  = \frac{\sigma^{2}}{N}\, $. 
} \cite{Nowak} wiemy, że wariancja: 
\begin{eqnarray}
\label{wariancja dla sredniej y}
{\sigma^{2}}( \hat{\theta} ) = {\sigma^{2}}({\bar{{\bf y}}})=\sigma^{2}/N \; ,
\end{eqnarray}
zatem: 
\begin{eqnarray}
\label{rn}
\texttt{i\!F}({\hat{\theta}}) = \frac{1}{{\sigma^{2}}({\hat{\theta}})} \; .
\end{eqnarray}
{\bf Wniosek}: Otrzymaliśmy więc ważny związek mówiący, że większa obserwowana IF parametru $\theta$ oznacza mniejszą wariancję jego estymatora $\hat{\theta}$. \\
\\
Równanie (\ref{rn}) można zapisać w postaci: 
\begin{eqnarray}
\label{RC dla 1 N z 1 par}
{\sigma^{2}}({\hat{\theta}}) \texttt{i\!F}({\hat{\theta}}) = 1 \; ,
\end{eqnarray}
co w przypadku estymacji wartości oczekiwanej rozkładu normalnego,  jest sygnałem 
osiągnięcia {\it dolnego ograniczenia nierówności Rao-Cram{\'e}ra} \cite{Amari Nagaoka book}. Temat ten będziemy rozwijać dalej. \\
\\
{\bf Określenie oczekiwanej IF}: Zdefiniujmy  oczekiwaną informację Fishera następująco:
\begin{eqnarray}
\label{infoczekiwana}
I_F \left(\Theta\right) \equiv E_{\Theta} \left(\texttt{i\!F}(\Theta)\right) = \int_{\cal B} dy P(y|\Theta) \, \texttt{i\!F}(\Theta) \; ,
\end{eqnarray}
gdzie  ${\cal B}$ jest przestrzenią próby (układu). 
Oznaczenie $\Theta$ w indeksie wartości oczekiwanej mówi, że $\Theta$ jest prawdziwą wartością parametru, przy której generowane są dane $y \equiv ({\bf y}_{1}, {\bf y}_{2}, ..., {\bf y}_{N})$, natomiast element różniczkowy $dy$ oznacza:
\begin{eqnarray}
\label{element rozniczkowy dy}
dy \equiv d^{N}{\bf y} = d{\bf y}_{1} d{\bf y}_{2} ... d{\bf y}_{N}  \; .
\end{eqnarray}
{\bf Oczekiwana v.s. obserwowana IF}: Istnieją znaczące różnice pomiędzy oczekiwaną a obserwowaną informacją 
Fishera \cite{Pawitan,Mania}. Oczekiwana informacja Fishera $I_F$ ma sens jako funkcja dopuszczalnych  wartości $\Theta$, należących do  przestrzeni $V_{\Theta}$ wartości $\Theta$. Natomiast jak to wynika z MNW, obserwowana informacja Fishera, $\texttt{i\!F}(\Theta)$, ma
zasadniczo sens tylko w pobliżu $\hat{\Theta}$. Jako związana
z obserwowaną wartością wiarygodności, $\texttt{i\!F}$ odnosi
się do pojedynczego zestawu danych i zmienia się od próbki do próbki. Oznacza to, że należy o niej myśleć jako o pojedynczej realizacji statystyki $\texttt{i\!F}(\hat{\Theta})$ w próbce, a nie jako o funk\-cji parametru $\Theta$. Natomiast
oczekiwana informacja $I_F$ jest średnią wartością dla wszystkich możliwych  zestawów danych w całej przestrzeni próby ${\cal B}$,  generowanych przy prawdziwej wartości parametru. Zatem  $I_F(\Theta)$ jest nie tyle użytecznym wskaźnikiem informacji dla konkretnego zbioru danych, ile  funkcją $\Theta$ mówiącą  jak trudno jest estymować $\Theta$, co oznacza, że parametr z większą $I_F$ wymaga mniejszej próbki do osiągnięcia wymaganej precyzji jego oszacowania.\\
 \\
Kontynuując rozpoczęty powyżej Przykład rozkładu normalnego z estymacją skalarnego parametru $\theta$, otrzymujemy po skorzystaniu z  (\ref{I obserw dla N_parametr mu}) oraz unormowaniu funkcji wiarygodności:
\begin{eqnarray}
\label{unormowanie dla f wiaryg P}
\int dy\, P(y|\theta)  = 1  \; ,
\end{eqnarray} 
wartość oczekiwaną IF dla parametru $\theta$ rozkładu $N(\theta, \sigma^2$):
\begin{eqnarray}
\label{I oczekiwana dla N_parametr mu}
I_F(\theta) = \int  dy \, P(y|\theta) \,  \texttt{i\!F}\left({\theta}\right)  = \int  dy\, P(y|\theta)  \frac{N}{{\sigma^{2}}}  = \frac{N}{{\sigma^{2}}} \; .
\end{eqnarray} 
Ze względu na (\ref{wariancja dla sredniej y}) wynik ten oznacza, że zachodzi:
\begin{eqnarray}
\label{RC dla 1 N z 1 par oczekiwana IF}
{\sigma^{2}}({\hat{\theta}}) I_F(\theta) = 1 \; ,
\end{eqnarray}
co mówi, że w  
estymacji wartości oczekiwanej  rozkładu normalnego osiągamy  {\it dolne ograniczenie nierówności Rao-Cram{\'e}ra}  \cite{Amari Nagaoka book}.  Nierówność ta  daje najważniejsze ograniczenie statystyki informacyjnej na jakość estymacji w pojedynczym kanale informacyjnym. Sprawie nierówności~Rao-Cramera~poświęcimy  część rozważań skryptu. \\
\\
{\bf Przykład estymacji obu parametrów rozkładu normalnego}: Rozważmy dwuparametrowy ($d=2$) rozkład normalny. Niech ${\bf y}_{1},...,{\bf y}_{N}$ jest realizacją 
próby prostej dla zmiennej $Y$ o rozkładzie normalnym $N\left({\mu,\sigma}\right)$. Ponieważ $\theta_{i}^{n}=\theta_{i}$ dla $n=1,2,...,N$ zatem wektor parametrów przyjmujemy jako $\Theta = ( (\theta_{i}^{\,n})_{i=1}^{2} )_{n=1}^{N} \equiv (\theta_{i})_{i=1}^{2} = \left( \mu, \sigma \right)^{T}$. Funkcja wiarygodności próby ma wtedy postać:
\begin{eqnarray}
\label{fun wiaryg r normalnego 2 par}
\ln P \left({y|\Theta} \right) = -N \ln (\sqrt{2 \pi} \; \sigma) - \frac{1}{{2\, \sigma^{2}}} \sum_{n=1}^{N}{\left({{\bf y}_{n}-\mu} \right)^{2}} \; .
\end{eqnarray}
Funkcja wynikowa z nią związana jest równa: 
\begin{eqnarray}
\label{fun wyn r normalnego 2 par}
S \left(\Theta \right) = \left( \begin{array}{l}
\frac{\partial}{{\partial \mu}} \ln  P \left({y|\Theta} \right)\\
\frac{\partial}{{\partial \sigma}} \ln P \left({y|\Theta} \right) \end{array} \right) = \left( \begin{array}{l}
\frac{N}{{\sigma^{2}}} \left({\bar{\bf y}-\mu} \right) \\
-\frac{N}{{\sigma}} + \frac{{\sum_{n=1}^{N} {\left({{\bf y}_{n}-\mu}\right)^{2}}}}{{\sigma^{3}}}
\end{array}
\right) \; , 
\end{eqnarray}
gdzie $\bar{\bf y} = \frac{1}{N} \sum_{n=1}^{N} {\bf y}_{n}$ jest średnią arytmetyczną w próbie. 
Zatem postacie estymatorów  MNW, tzn. parametru wartości oczekiwanej $\mu$ oraz odchylenia standardowego $\sigma$ zmiennej $Y$,  otrzymujemy rozwiązując układ równań wiarygodności:
\begin{eqnarray}
\label{fun wyn r normalnego 2 par = 0}
S \left(\Theta \right)|_{\Theta={\hat{\Theta}}} =  \left( \begin{array}{l}
\frac{N}{{\sigma^{2}}} \left({\bar{\bf y}-\mu} \right) \\
-\frac{N}{{\sigma}} + \frac{{\sum_{n=1}^{N} {\left({{\bf y}_{n}-\mu}\right)^{2}}}}{{\sigma^{3}}} 
\end{array}
\right)|_{\Theta={\hat{\Theta}}} = 
\left( \begin{array}{l}
0 \\
0
\end{array}
\right) \; , 
\end{eqnarray}
gdzie $\hat{\Theta}=\left( \hat{\mu}, \hat{\sigma} \right)^{T}$ i  rozwiązanie to ma postać:
\begin{eqnarray}
\label{rozw ukl row MNW dla 2-wym rozkl norm}
{\hat{\mu}} = \bar{\bf y} \;\;\;\;\; {\rm oraz} \;\;\;\;\; \hat{\sigma} =  \sqrt{\frac{1}{N}  \sum_{n=1}^{N} \left({{\bf y}_{n} - \bar{\bf y}}\right)^{2}} \; .
\end{eqnarray}
Obserwowana informacja Fishera (\ref{I jako krzywizna dla P}) w punkcie ${\hat{\Theta}}$ wynosi więc:
\begin{eqnarray}
\label{obserw iF r normalnego 2 par}
\texttt{i\!F}({\hat{\Theta}}) = \left(
\begin{array}{cc}
{\frac{N}{{\sigma^{2}}}} & {\frac{2 N}{{\sigma^{3}}} \left({\bar{\bf y}-\mu} \right)} \\
{\frac{2 N}{{\sigma^{3}}} \left({\bar{\bf y}-\mu} \right)} & \;\; -\frac{N}{{\sigma^{2}}} + \frac{3}{\sigma^{4}} \sum_{n=1}^{N} \left({{\bf y}_{n}-\mu} \right)^{2}
\end{array}
\right)|_{\Theta={\hat{\Theta}}} = \left({\begin{array}{cc}
{\frac{N}{{\hat{\sigma}^{2}}}} & 0\\
0 & {\frac{2 N}{{\hat{\sigma}^{2}}}} \end{array}}\right) \; ,
\end{eqnarray}\\
W końcu oczekiwana informacja Fishera jest równa:
\begin{eqnarray}
\label{oczekiw iF r normalnego 2 par}
I_{F} \left(\Theta \right) = E_{\Theta} \texttt{i\!F}\left(\Theta \right) = \int_{\cal B} dy \, P(y|\Theta) \, \texttt{i\!F}\left(\Theta \right)  = \left({\begin{array}{cc}
{\frac{N}{{\sigma^{2}}}} & 0\\
0 & {\frac{2 N}{{\sigma^{2}}}} \end{array}}\right)  \; .
\end{eqnarray}
\\
{\bf Uwaga o estymatorze $\hat{\sigma^{2}}$ MNW}: Widząc powyższy rozkład jako $N\left({\mu,\sigma^{2}}\right)$ i w konsekwencji przyjmując wektor parametrów jako $(\mu, \sigma^{2})^{T}$, oraz przeprowadzając analogiczny rachunek jak powyżej, można pokazać, że estymator wartości oczekiwanej ma postać $\hat{\mu} = \bar{\bf y}$, czyli tak jak w (\ref{rozw ukl row MNW dla 2-wym rozkl norm}), natomiast estymator wariancji  wynosi $\hat{\sigma^{2}}=\frac{1}{N}  \sum_{n=1}^{N} \left({{\bf y}_{n} - \bar{\bf y}}\right)^{2}$. W jego mianowniku  występuje czynnik $N$ a nie $N-1$ jak to ma miejsce w przypadku nieobciążonego estymatora wariancji. Estymator $\hat{\sigma^{2}}$ nie jest też efektywny (tzn. nie posiada najmniejszej z możliwych wariancji). Jednak własności te posiada on asymptotycznie ($N \rightarrow\infty$), co jest charakterystyczne dla wszystkich estymatorów MNW \cite{Nowak}. \\
Sprawdzenie, że przy  odpowiednio dobranych tzw. oczekiwanych parametrach rozkładu normalnego, estymacja obu parametrów jest efektywna dla skończonego $N$, pozostawiamy jako ćwiczenie na koniec tego rozdziału, po tym jak zapoznamy się z pojęciem dualnych układów wpółrzędnych.

\subsection{Wartość oczekiwana i wariancja funkcji wynikowej}

\label{E i var funkcji wynikowej}
  
Pokażmy, że wartość oczekiwana funkcji wynikowej $S(\Theta) \equiv S(\widetilde{Y}|\Theta)$, tzn. gradientu logarytmu naturalnego funkcji wiarygodności, jest równa zeru: 
\begin{eqnarray}
\label{znikanie ES}
E_{\Theta}S\left(\Theta\right) = 0 \; .
\end{eqnarray}
Istotnie, gdy skorzystamy z interpretacji funkcji wiarygodności jako łącznego rozkładu prawdopodobieństwa i jej unormowania do jedności,  wtedy:
\begin{eqnarray}
\label{dow ES=0}
E_{\Theta}S\left(\Theta\right) &=& \int_{\cal B}{dy P\left(y|\Theta\right) S\left(\Theta\right)} = \int_{\cal B} dy P\left({y|\Theta} \right) \left(\frac{\partial}{{\partial\Theta}} \ln P\left({y|\Theta}\right)\right)   \nonumber \\
&=& \int_{\cal B} dy P\left({y|\Theta} \right) \frac{{\frac{\partial}{{\partial\Theta}}P\left(\Theta\right)}}{{P\left(\Theta\right)}} = \int_{\cal B}{ dy  \frac{\partial}{{\partial\Theta}} P\left({y|\Theta} \right)} = \frac{\partial}{{\partial\Theta}} \int_{\cal B}{d{y}\, P\left({y|\Theta}\right)}=0 \; ,
\end{eqnarray}
gdzie zakres całkowania obejmuje całą przestrzeń próby ${\cal B}$.
W przypadku gdy pierwotna zmienna $Y$ jest dyskretna, powyższy dowód przebiega podobnie.  \\
 \\
{\bf Założenie o regularności rozkładu}: W (\ref{dow ES=0}) wyciągnęliśmy różniczkowanie po parametrze przed
znak całki.  Poprawność takiego przejścia oznacza spełnienie żądania,  aby rozkład $P\left({y|\Theta}\right)$
był wystarczająco gładki jako funkcja $\Theta$ \cite{Pawitan}. Oznacza to, że $\Theta$ {\it nie
może być brzegową wartością, tzn. istnieje taka funkcja  $g\left(y\right)$ (której całka  $\int_{\cal B}{g\left({y}\right)dy}$
jest skończona),  dla której  w sąsiedztwie prawdziwej wartości
parametru $\Theta$ zachodzi  
$\left|{{\partial  P\left({y|\Theta}\right)} /{\partial\Theta}}\right|\le g\left(y\right)$
gdy 
${{\partial P}/{\partial\Theta}}$
jest traktowana jako funkcja} $y$ \cite{Pawitan}, skąd wynika skończoność całki $ \int_{\cal B} dy \, \partial  P\left({y|\Theta}\right) /{\partial\Theta} $. \\
Zauważmy również, że przy odpowiednim warunku regularności:
\begin{eqnarray}
\label{2 poch po P}
\int_{\cal B}{ dy\, \frac{\partial^{2}}{\partial \Theta^{2}} P\left(y|\Theta \right)} = \frac{\partial^{2}}{\partial\Theta^{2}} \int_{\cal B}{ dy\, P\left(y|\Theta \right)} =0 \; .
\end{eqnarray}
 \\
{\bf  Twierdzenie o wariancji funkcji wynikowej:} Zakładając warunek regularności pozwalający na wyciągnięcie różniczkowania po parametrze przed znak całki, można pokazać, że wariancja (precyzyjnie, macierz wariancji-kowariancji) funkcji wynikowej  $S(\Theta) \equiv S(\widetilde{Y}|\Theta)$ jest równa oczekiwanej IF\footnote{Ponieważ funkcja wynikowa $S(\Theta) \equiv S(\widetilde{Y}|\Theta) = \frac{\partial P(\widetilde{Y}|\Theta)}{\partial \Theta}$ jest $d$-wymiarowym  kolumnowym wektorem  losowym  (\ref{funkcja wynikowa}) z wartością oczekiwaną równą zero, zatem   ${\rm Cov}_{\Theta}\left[ \,S \left( \widetilde{Y}|\Theta \right) \right]$ jest $d\times  d$-wymiarową macierzą kowariancji (współrzędnych) wektora  $S(\widetilde{Y}|\Theta)$. W skrypcie będziemy  stosowali oznaczenie ${\sigma^{2}}_{\!\! \Theta} \,S \left( \Theta \right) \equiv  {\rm Cov}_{ \Theta}\left[ \,S \left( \widetilde{Y}|\Theta \right) \right]$. 
}:
\begin{eqnarray}
\label{var S oraz IF}
{\sigma^{2}}_{\!\! \Theta} \,S \left( \Theta \right) = I_{F}(\Theta) \; .
\end{eqnarray}
Zauważmy, że z powyższego wynika, że $d\times d$-wymiarowa macierz informacji $I_{F}(\Theta)$ jest macierzą kowariancji, co oznacza, że jest ona {\it nieujemnie 
określona}\footnote{Dla dowolnego wektora ${\bf a} = \left(a_{1},...,a_{d} \right)^{T} \in \mathbf{R}^{d}$
oraz macierzy kowariancji 
$C = E \left(\left(Z-E(Z) \right) \left(Z-E(Z) \right)^{T}\ \right)$, gdzie $Z$ jest $d$-wymiarowym 
wektorem losowym, 
zachodzi 
${\bf a}^{T} E\left[\left(Z-E(Z) \right)\left(Z - E(Z) \right)^{T} \right] {\bf a} = {\bf a}^{T} E \left(W  W^{T} \right) {\bf a}$ 
$ = E \left[({\bf a}^{T} W  )({\bf a}^{T} W  )^{T}\right] ={\mathop{\sigma^{2}}}\left({\bf a}^{T} W \right)\ge0$, gdzie
$W=Z-E(Z)\;$ i $E(W)=0$, tzn. macierz $C$ jest nieujemnie określona.
}.  \\
\\
{\bf Dowód} twierdzenia (\ref{var S oraz IF}).
Korzystając z (\ref{znikanie ES})  otrzymujemy:
\begin{eqnarray}
\label{dow varS=I 1}
{\sigma^{2}}_{\Theta} S\left( \Theta \right) &=& \int_{\cal B} d{y} \, P\left(y|\Theta\right) \left( S\left(\Theta\right) - E_{\Theta} S\left(\Theta \right) \right)^{2}   \\
&=& \int{d{y}  \,  P\left({y|\Theta}\right)\left({S\left(\Theta\right)}\right)^{2}} = \int{d{y}\, P\left({y|\Theta}\right) \left({\frac{\partial}{{\partial\Theta}} \ln P\left({y|\Theta}\right)}\right)}^{2} 
\nonumber \\
&=&\int{d{y}\, P\left(y|\Theta\right) \left[\left(\frac{\partial}{\partial\Theta}P\left(y|\Theta\right)\right)/P\left(y|\Theta\right)\right]^{2}} =  \int{dy\,\left[\frac{\partial}{\partial\Theta}P\left(y|\Theta\right)
\right]^{2}/P\left(y|\Theta\right)} \; . \nonumber
\end{eqnarray}
\\
Natomiast korzystając z (\ref{2 poch po P}) otrzymujemy: \begin{eqnarray}
\label{dow varS=I 2}
\!\!\!\!\!\!\!\!\!\!\!\!\!\!\! & & I_{F}(\Theta) = E_{\Theta}\texttt{i\!F}(\Theta) = \int_{\cal B}{d{y} P\left({y|\Theta}\right) \texttt{i\!F}(\Theta)}  = -\int{d{y} P\left({y|\Theta}\right)  \left(\frac{{\partial^{2}}}{{\partial\Theta^{2}}}\ln P\left({y|\Theta}\right)\right)} \nonumber \\
\!\!\!\!\! &=&-\int dy P\left(y|\Theta\right) {\frac{\partial}{\partial\Theta}  }\left[\left(\frac{\partial}{\partial\Theta}P\left(y|\Theta\right)\right)/P\left(y|\Theta\right)\right]  \\
\!\!\!\!\! &=&\int dy P\left(y|\Theta\right) \frac{\left[-\left(\frac{\partial^{2}}{\partial\Theta^{2}}P\left(y|\Theta\right)\right)P\left(y|\Theta\right)+ \left(\frac{\partial}{\partial\Theta}P\left(y|\Theta\right)\right)^{2}\right]}{\left(P\left(y|\Theta\right)\right)^{2}} = \int{ d{y} \, \left[\frac{\partial}{\partial\Theta}P\left(y|\Theta\right)\right]^{2}/P\left(y|\Theta\right)} \; , \;\;\;  \nonumber
\end{eqnarray}
co porównując z (\ref{dow varS=I 1}) kończy  dowód twierdzenia (\ref{var S oraz IF}). \\
\\
{\bf Wniosek}: Dowodząc (\ref{znikanie ES}) okazało się, że przy założeniu warunku reguralności, wartość oczekiwana na przestrzeni próby z funkcji wynikowej zeruje się, tzn. $E_{\Theta}S\left(\Theta\right)=0$ i rezultat ten jest słuszny dla ogólnego przypadku wektorowego.   Z (\ref{var S oraz IF}) wynika również, że macierz wariancji-kowariancji funkcji wynikowej jest równa IF, tzn.  
${\sigma^{2}}_{\Theta} S \left( \Theta \right)= I_{F}(\Theta)$, co jak sprawdzimy poniżej, ma znaczenie dla dodatniej określoności metryki Fishera $g_{ij}$ w teorii pola. Skoro równości (\ref{znikanie ES}) oraz (\ref{var S oraz IF}) zachodzą  dla przypadku wektorowego, zatem są one również słuszne  w przypadku skalarnym. \\
\\
{\bf Ćwiczenie}: Przedstawione dowody dla własności (\ref{znikanie ES}) oraz (\ref{var S oraz IF}) były ogólne. Sprawdzić bezpośrednim rachunkiem, że zachodzą one dla (\ref{fun wyn r normalnego 2 par}) oraz (\ref{oczekiw iF r normalnego 2 par}) w powyższym przykładzie estymacji obu parametrów rozkładu normalnego. \\
\\
{\bf Ważny związek}: Z  (\ref{dow varS=I 1})-(\ref{dow varS=I 2}) widać, że przy spełnieniu wspomnianych własności regularności,  zachodzi:
\begin{eqnarray}
\label{IF 2 poch na kwadrat pierwszej}
\!\!\!\!\!\!\!\! I_{F}(\Theta)  =  E_{\Theta}\texttt{i\!F}(\Theta) = - \int_{\cal B} \! {d{y} P\left({y|\Theta}\right) \left(\frac{{\partial^{2}}}{{\partial\Theta^{2}}}\ln P\left({y|\Theta}\right)\right)} 
= \int_{\cal B} \!{dy P\left(y|\Theta\right) \, \left[\frac{\partial}{\partial \Theta} \ln P\left(y|\Theta\right)\right]^{2} }  . \;\;\;
\end{eqnarray}
 \\
{\bf Uwaga o dodatniej określoności obserwowanej IF}: Niech $g_{ij}$ są elementami $d\times d$ - wymiarowej macierzy $I_{F}(\Theta)$. Z (\ref{IF 2 poch na kwadrat pierwszej}) otrzymujemy dla dowolnego $d$ - wymiarowego wektora $v=(v_{1},v_{2},...,v_{d})^{T}$:
\begin{eqnarray}
\label{iF polokreslona}
v^{T} I_{F}(\Theta) \,v = \! \sum_{i=1}^{d} \sum_{j=1}^{d} v_{i} g_{ij}(\Theta) v_{j} = \!\int_{\cal B} \! dy P(y|\Theta) (  \sum_{i=1}^{d} v_{i} \frac{\partial \ln P(\Theta)}{\partial \theta_{i}} )  ( \sum_{j=1}^{d}  \frac{\partial \ln P(\Theta)}{\partial \theta_{j}} \,v_{j}  ) \geq 0 \; , \;\;\;\;\;
\end{eqnarray}
co oznacza, że oczekiwana informacja Fishera $I_{F}(\Theta)$ jest {\it dodatnio półokreślona} \cite{Pawitan}, jak to zaznaczyliśmy poniżej (\ref{var S oraz IF}). Jednak w  teorii pola  interesuje nas zaostrzenie warunku (\ref{iF polokreslona}) do własności dodatniej określoności. \\ 
\\
Z (\ref{var S oraz IF}) (lub (\ref{IF 2 poch na kwadrat pierwszej})) widać również, że w {\it teorii pola, którą dałoby się 
sformułować dla ciągłych, regularnych, unormowanych rozkładów}, co pociąga za sobą ciągłość  rozkładu funkcji wynikowej na całej przestrzeni próby ${\cal B}$, macierz 
$I_{F}(\Theta)$ jest określona dodatnio. \\
Ponieważ w  Rozdziale~\ref{alfa koneksja} zwrócimy uwagę na fakt, że oczekiwana IF 
określa tzw. {\it metrykę Fishera-Rao} $(g_{ij}):=I_{F}(\Theta)$ na przestrzeni statystycznej ${\cal S}$, zatem:  \\
\\
{\it W teorii pola z ciągłymi, regularnych i unormowanymi rozkładami, metryka Fishera-Rao $g_{ij}$ jest dodatnio określona.}

\section[{Wstęp do geometrii różniczkowej na przestrzeni statystycznej i  $\alpha$-koneksja}]{Wstęp do geometrii różniczkowej na przestrzeni statystycznej i  $\alpha$-koneksja}

\label{alfa koneksja}


Niech zbiorem punktów $\Omega$ reprezentujących konfigurację układu
%
%
będzie przestrzeń próby układu ${\cal B}$. Np. w przypadku próby jednowymiarowej  ${\cal B} \equiv {\cal Y}$, gdzie  ${\cal Y}$ jest zbiorem wszystkich możliwych  wartości ${\bf y}$ zmiennej losowej $Y$. 
Niech $P$ jest  miarą probabilistyczną (prawdopodobieństwem) na  ${\cal B}$.  
Zbiór wszystkich miar na ${\cal B}$ oznaczmy $\Sigma({\cal B})$ i nazwijmy  {\it przestrzenią stanów} modelu.  \\
 \\
{\bf Określenie modelu statystycznego}: Rozważmy podzbiór ${\cal S} \subset \Sigma({\cal B})$, na którym jest zadany układ współrzędnych $(\xi^{i})_{i=1}^{d}$ \cite{Amari Nagaoka book} tak, że ${\cal S}$ jest 
rozmaitością\footnote{
Niech ${\cal S}$ jest zbiorem punktów,  na którym określony jest układ współrzędnych $\phi_{\Xi}: {\cal S} \rightarrow \mathbb{R}^{d}$. {\bf W skrypcie interesują nas tylko  globalne układy współrzędnych}. Wtedy $\phi_{\Xi}$ odwzorowuje każdy punkt $P \in {\cal S}$ w zbiór $d$ liczb rzeczywistych $\phi_{\Xi}(P) \equiv (\xi^{1}(P), \xi^{2}(P),...,\xi^{d}(P))^{T} = ( \xi^{1}, \xi^{2},...,\xi^{d})^{T}  \equiv \Xi$.  \\
\\
Niech istnieje zbiór układów  współrzędnych (czyli atlas) ${\cal A}$ spełniający następujące warunki:\\
1) Każdy element $\phi_{\Xi} = \left[(\xi^{i})_{i=1}^{d}\right] \in {\cal A}$,  jest wzajemnie jednoznacznym odwzorowaniem $\phi_{\Xi}: {\cal S} \rightarrow \mathbb{R}^{d}$ z ${\cal S}$ 
w pewien otwarty podzbiór w $\mathbb{R}^{d}$.\\
2) Dla każdego $\phi_{\Xi} \in {\cal A}$ oraz wzajemnie jednoznacznego odwzorowania $\phi_{\Xi'}$ z ${\cal S}$ w $\mathbb{R}^{d}$, zachodzi \\
równoważność: $(\phi_{\Xi'} = (\xi^{'\,i})_{i=1}^{d} \in {\cal A})$ $\Leftrightarrow$ $(\phi_{\Xi'} \circ \phi_{\Xi}^{-1}$ jest dyfeomorfizmem rzędu $C^{\infty})$.
\begin{eqnarray}
\label{uklad wspolrzednych na S}
Zbi\acute{o}r \; {\cal S} \; z \; tak \; okre\acute{s}lonym \; atlasem \; {\cal A} \; to \; (C^{\infty} \; r\acute{o}\dot{z}niczkowalna) \; rozmaito\acute{s}\acute{c} \; .
\end{eqnarray}
}. 
%
Niech na ${\cal B}$ określony jest $d$~-~wymiarowy model statystyczny (tzn. para $(y, P)$), a precyzyjniej:
\begin{eqnarray}
\label{model statystyczny S}
{\cal S} = \{P_{\Xi} \equiv P(y|\Xi),  \Xi \equiv (\xi^{i})_{i=1}^{d} \in  {V}_{\Xi} \subset \mathbb{R}^{d} \} \; , 
\end{eqnarray}
tzn. rodzina  rozkładów prawdopodobieństwa parametryzowana przez $d$ nielosowych zmiennych o wartościach rzeczywistych $(\xi^{i})_{i=1}^{d}$ należących do przestrzeni parametru ${V}_{\Xi}$, będącej podzbiorem  $\mathbb{R}^{d}$. 
Mówimy, że ${\cal S}$ jest $d$ - wymiarową przestrzenią statystyczną. \\ 
\\
{\bf Notacja}: Ponieważ w danym modelu statystycznym wartość parametru $\Xi$ określa jednoznacznie rozkład prawdopodobieństwa $P$ jako punkt na ${\cal S}$, więc ze względu na wygodę i o ile nie będzie to prowadziło do nieporozumień, sformułowania (punkt) $P_{\Xi} \in {\cal S}$ oraz (punkt) $\Xi \in {\cal S}$ będziemy stosowali zamiennie.  \\  
\\
{\bf Uwaga o niezależności $P$ od parametryzacji}: Oczywiście rozkład prawdopodobieństwa nie zależy od wyboru bazy w przestrzeni statystycznej ${\cal S}$, tzn. gdyby np. $\Theta$ był innym układem współrzędnych (inną parametryzacją) to $P = P_{\Theta} \equiv P(y|\Theta) = P_{\Xi} \equiv P(y|\Xi)$ dla każdego $P \in {\cal S}$.\\
\\
{\bf Określenie macierzy informacyjnej Fishera}: Wprowadźmy oznaczenie: 
\begin{eqnarray}
\label{oznaczenie ln P}
\ell_{\Xi} \equiv l(y|\Xi) \equiv \ln P_{\Xi}  \;  \;\;\;\; {\rm oraz} \;\;\;\; \partial_{i} \equiv \frac{\partial}{\partial \xi^{i}} \;\;\; i=1,2,...,d \; .
\end{eqnarray} 
Dla każdego punktu $P_{\Xi}$,  $d \times d$ - wymiarowa macierz $(g_{ij}(\Xi))$ o elementach:
\begin{eqnarray}
\label{Fisher inf matrix}
\!\!\!\!\!\!\! g_{ij}(\Xi):=  E_{\Xi}(\partial_{i} \ell_{\Xi} \partial_{j} \ell_{\Xi}) =  \! \int_{{\cal B}} \! dy \, P(y|\Xi) \, \partial_{i} l(y|\Xi) \partial_{j} l(y|\Xi) \; , \;\; i,j = 1,2,...,d \; , \;\; \forall\, P_{\Xi} \in {\cal S} \; ,
\end{eqnarray}
jest nazywana {\it macierzą informacyjną Fishera} na  ${\cal S}$ w punkcie $P_{\Xi}$ \cite{Amari Nagaoka book}.  
Wielkość  $E_{\Xi}(.)$ oznacza tutaj wartość oczekiwaną, a całkowanie przebiega po całej przestrzeni próby ${\cal B}$.  \\
Przy założeniu spełnienia warunków regularności (porównaj (\ref{IF 2 poch na kwadrat pierwszej})), macierz  $(g_{ij}(\Xi))$ może zostać zapisana następująco:  
\begin{eqnarray}
\label{Fisher inf matrix plus reg condition} 
g_{ij} = -  E_{\Xi}(\partial_{i} \partial_{j} \ell_{\Xi})  =  -  \int_{{\cal B}} dy \, P(y|\Xi) \, \partial_{i} \partial_{j}  \ln P(y|\Xi) \; , \;\;\; i,j = 1,2,...,d \; , \;\;\; \forall\, P_{\Xi} \in {\cal S} \; .
\end{eqnarray}
\\
%
{\bf Metryka Fishera}: Macierz informacyjna Fishera określa na ${\cal S}$ iloczyn wewnętrzny nazywany {\it metryką Fishera}  $\langle , \rangle$ na ${\cal S}$, definując go w układzie współrzędnych $(\xi^{i})_{i=1}^{d}$ poprzez związek:
\begin{eqnarray}
\label{Fisher metric poprzez macierz informacyjna} 
\langle \partial_{i} , \partial_{j} \rangle := g_{ij} \; , \; \;\;\;\; {\rm gdzie \;\; wektory \;\; bazowe} \;\;\; \partial_{i} \equiv \frac{\partial}{\partial \xi^{i}} \; , \;\;\;\; i=1,2,...,d \; , \;\;\; \forall\, P_{\Xi} \in {\cal S} \; .
\end{eqnarray}
Metryka $g_{ij}$ Fishera-Rao  jest metryką typu 
Riemannowskiego\footnote{
{\bf Określenie przestrzeni Riemannowskiej}: Niech ${\cal S}$ jest rozmaitością i  załóżmy, że dla każdego punktu $P_{\Xi}    \in {\cal S}$  w jej przestrzeni stycznej $T_{P} \equiv T_{P}({\cal S})$  jest określony iloczyn wewnętrzny $\left\langle , \right\rangle_{P}$  taki, że   $\left\langle V, W \right\rangle_{P} \in \mathbb{R}$ oraz posiadający dla dowolnych wektorów $V,W \in T_{P}$ następujące własności:
\begin{eqnarray}
\label{metryka Riemanna} 
& & (\forall \,a,b \in \mathbb{R}) \;\;\;\; \left\langle a V + b W, Z \right\rangle_{P} = a \left\langle V, Z \right\rangle_{P} + b \left\langle  W,  Z \right\rangle_{P} \;\;\;\;\;\;\; \quad {\rm liniowo\acute{s}\acute{c}} \; \\
& &\left\langle V, W \right\rangle_{P} =  \left\langle W, V  \right\rangle_{P}  \;\;\;\;\;\;\;\;\;\; \quad\quad\quad\quad\quad\quad\quad\quad\quad\quad\quad \quad\quad
{\rm symetryczno\acute{s}\acute{c}} \; \\
& & {\rm Je\acute{s}li} \;\;\;\; V \neq 0 \, , \;\;\; {\rm wtedy} \;\;\;  \left\langle V, V \right\rangle_{P} > 0  \;\;\;\;\;\;\;\; \;\;\quad \quad\quad {\rm dodatnia \;okre\acute{s}lono\acute{s}\acute{c}} \; .
\end{eqnarray}
Pierwsze dwie własności oznaczają, że 
$\left\langle , \right\rangle_{P}$ jest  formą dwuliniową. Odwzorowanie 
$g:{\cal S} \ni P \rightarrow \left\langle ,  \right\rangle_{P} $ jest na ${\cal S}$ polem wektorowym  (kowariantnym) rangi 2. Nazywamy je {\it metryką Riemanna na ${\cal S}$, a przestrzeń ${\cal S}$, z tak określoną metryką $g$, nazywamy  przestrzenią Riemannowską} $({\cal S}, \, g)$.\\ 
Gdy dla każdego $P \in {\cal S}$ współrzędne $V^{i}$, $i=1,2,...,d$ dowolnego wektora $V \in T_{P}$ są $C^{\infty}$ (tzn. są analityczne) względem pewnego układu współrzędnych $\Xi$, wtedy metryka Riemanna jest $C^{\infty}$. W skrypcie rozważamy tylko przypadek $C^{\infty}$. \\
Mając metrykę $g$ możemy zdefiniować długość wektora stycznego $V$ przez nią indukowaną:
\begin{eqnarray}
\label{dlugosc z metryki Riemanna} 
\left\| V \right\| = \sqrt{\left\langle V, V \right\rangle_{P}} = \sqrt{\sum_{i,j=1}^{d} g_{ij}(\Xi) V^{i} V^{j}} \; .
\end{eqnarray}
}. 
Warto zauważyć, że na rozmaitości ${\cal S}$ można zdefiniować nieskończoną liczbę metryk Riemannowskich.  Jednak Chentsov pokazał, że  {\it metryka Fishera-Rao jest} wyróżniona spośród wszystkich innych tym, że (z dokładnością do stałego czynnika) jest ona {\it jedyną, która jest redukowana} (w znaczeniu zmniejszania się odległości dowolnych dwóch stanów) {\it w każdym stochastycznym odwzorowaniu}  \cite{Bengtsson_Zyczkowski,Streater}.\\
 \\
{\bf Określenie koneksji afinicznej}: 
Oznacz\-my przez $T_{P}$ przestrzeń styczną do ${\cal S}$ w punkcie $P_{\Xi} \in {\cal S}$. Każdy wektor $V \in T_{P}$ można rozłożyć na wektory bazowe $(\partial_i)_{P}$:
\begin{eqnarray}
\label{V w T Theta}
V = \sum_{i=1}^{d} V^{i} (\partial_{i})_{P} \;\, , \; \;\;\; {\rm gdzie} \;\;\;\; V \in T_{P} \; , \;\;\;\;\; \forall\, P_{\Xi} \in {\cal S} \; ,
\end{eqnarray}
gdzie dolny indeks $_{P}$ oznacza zależność układu współrzędnych $\partial_{i} \equiv \frac{\partial}{\partial \xi^{i}}$, $i=1,2,...,d$, od punktu $P_{\Xi} \in {\cal S}$. \\
\\
Niech $\gamma_{P P'}$ oznacza ścieżkę w ${\cal S}$ łączącą punkty $P$ oraz $P'$. Przyporządkujmy, każdej ścieżce $\gamma_{P P'}$ w ${\cal S}$,  odwzorowanie $\Pi_{\gamma_{P P'}}$:
\begin{eqnarray}
\label{Pi z Tr do Ts}
\gamma_{P P'} \rightarrow \Pi_{\gamma_{P P'}}: T_{P} \rightarrow T_{P'} \; , \;\;\;\;\; \forall\, P_{\Xi} \; {\rm i} \; P'_{\Xi}  \in {\cal S} \; , 
\end{eqnarray}
które przekształca wektory z przestrzeni wektorowej $T_{P}$ do przestrzeni wektorowej $T_{P'}$.\\
\\
%
%
Rozważmy trzy dowolne punkty $P$, $P'$ oraz $P''$ w ${\cal S}$, oraz ścieżki $\gamma_{P  P'}$ z punktu $P$ do $P'$ i   $\gamma_{P' P''}$  z $P'$ do $P''$. 
{\it Mówimy, że przekształcenie $\Pi$ jest koneksją afiniczną jeśli}:
\begin{eqnarray}
\label{Pi jako koneksja afiniczna}
\Pi_{\gamma_{P P''}} = \Pi_{\gamma_{P' P''}} \circ  \Pi_{\gamma_{P P'}} \;\;\;\; {\rm oraz} \;\;\;\; \Pi_{\gamma_{0}} \equiv \Pi_{\gamma_{P P}} = id \; , \;\;\;\;\; \forall\, P_{\Xi} \; {\rm i} \;P'_{\Xi} \in {\cal S} \; ,
\end{eqnarray}
gdzie $id$ jest przekształceniem identycznościowym. \\
\\
Aby określić postać liniowego odwzorowania $\Pi_{P P'}$ pomiędzy $T_{P}$ a $T_{P'}$ musimy, dla każdego $i \in 1, 2,...,d$, określić $\Pi_{P P'}((\partial_i)_{P})$ jako liniową kombinację wektorów bazowych $(\partial_1)_{P'}, (\partial_2)_{P'}, ..., (\partial_{d})_{P'}$ w $T_{P'}$. \\
 \\
{\bf Określenie przesunięcia równoległego}: 
Niech $V_{P}$ będzie wektorem stycznym w $P$. Wektor $V_{P P'}$ nazywamy {\it równoległym przesunięciem wektora} $V_{P}$ z $P$ do  $P'$ wzdłuż krzywej (ścieżki) $\gamma_{P P'}$, wtedy gdy:
\begin{eqnarray}
\label{Pi i rownolegle przesuniecie}
V_{P} \rightarrow  V_{P P'} := \Pi_{\gamma_{P P'}} \, V_{P}\; , \;\;\;\;\; \forall\,  P_{\Xi} \; {\rm i} \;P'_{\Xi} \in {\cal S} \; .
\end{eqnarray}
\\
{\bf Określenie równoległości wektorów}: {\it Dwa wektory $V_{P}$ oraz $V_{P'}$  styczne do ${\cal S}$ w punktach $P$ i $P'$ są równoległe}, jeśli równoległe przesunięcie (określone w (\ref{Pi i rownolegle przesuniecie})) wzdłuż wskazanej krzywej $\gamma$ jednego z nich, powiedzmy $V_{P}$,  do punktu $P'$ ``zaczepienia'' drugiego  z nich, da wektor $V_{P P'}$, który jest {\it proporcjonalny} do $V_{P'}$ (i na odwrót). \\

Koneksja afiniczna pozwala zdefiniować  pochodną kowariantną w następujący sposób. 
Niech $V$ jest dowolnym wektorem w przestrzeni stycznej $T_{P}$, oraz niech $\gamma(t)$, gdzie $t \in \left\langle 0,1 \right\rangle$, oznacza dowolną ścieżkę z $P$ do $P'$ w ${\cal S}$, która wychodzi z $P$ w kierunku $W \in T_{P}$. \\
 \\
{\bf Pochodną kowariantną} (koneksji afinicznej $\Pi$) wektora $V$ w punkcie $P$ i w kierunku $W \in T_{P}$ definiujemy następująco:
\begin{eqnarray}
\label{Pi pochodna kowariantna}
\nabla_{W} V :=   \frac{d}{dt}  \left(\Pi_{\gamma_{P  \gamma(t)}} V \right)_{|_{t=0}} \; , \;\;\; {\rm gdzie} \;\;\;\; \gamma(t=0) = P \;\; , \;\;\;\;\; \forall\, P_{\Xi} \in {\cal S} \; .
\end{eqnarray}
Pojęcie pochodnej kowariantnej pozwala wypowiedzieć się co do równoległości wektorów w punktach $P$ i $P'$.\\
 \\
{\bf Określenie geodezyjnej}: Geodezyjną nazywamy taką krzywą $\gamma$ w ${\cal S}$, której wszystkie wektory styczne są do siebie równoległe. W związku z tym mówimy, że geodezyjna jest {\it samo-równoległą} krzywą w ${\cal S}$. \\
\\
\\
{\bf Współczynniki koneksji}: 
Konkretna analityczna postać współczynników koneksji {\it musi być podana przy ustalonej parametryzacji}\footnote{Niech $\Delta=(\delta^{i})_{i=1}^{d}$ jest innym niż $\Xi=(\xi^{i})_{i=1}^{d}$ układem współrzędnych (inną  parametryzacją) i niech $\tilde{\partial}_{j} \equiv \frac{\partial}{\partial \delta^{j}} = \sum_{i=1}^{d} \frac{\partial \xi^{i}}{\partial \delta^{j}}\, \partial_{i} $. Można pokazać, że współczynniki koneksji w układach współrzędnych $\Delta$ oraz $\Xi$ są związane ze sobą następująco \cite{Amari Nagaoka book}:
\begin{eqnarray}
\label{nowy wspolczynnik koneksji}
\tilde{\Gamma}^{t}_{rs} = \sum_{i, j, \,l=1}^{d} \left(\Gamma^{l}_{ij} \frac{\partial \xi^{i} }{\partial \delta^{r}} \frac{\partial \xi^{j} }{\partial \delta^{s}} + \frac{\partial^{2} \xi^{\,l}}{\partial \delta^{r} \partial \delta^{s}} \right) \frac{\partial \delta^{t}}{\partial \xi^{l}}
\; , \;\;\;\;\;\; t,r,s=1,2,...,d \; .
\end{eqnarray}
Drugi składnik w (\ref{nowy wspolczynnik koneksji}) zależy tylko od postaci transformacji współrzędnych i jest niezależny od koneksji.  Zatem, za wyjątkiem transformacji liniowej, dla której drugi składnik znika,    
współczynniki koneksji nie transformują się przy przejściu do nowego układu współrzędnych jak wielkość tensorowa. Gdyby więc w układzie współrzędnych $\Xi=(\xi^{i})_{i=1}^{d}$ wszystkie  $\Gamma^{l}_{ij} = 0$, to  (za wyjątkiem transformacji liniowej) w nowym układzie współrzędnych $\Delta=(\delta^{i})_{i=1}^{d}$  
nie wszystkie współczynniki $\tilde{\Gamma}^{t}_{rs}$ byłyby również równe zero. \\  
Jednakże postać koneksji afinicznej $\Pi$ jest w nowej parametryzacji taka sama jak (\ref{Pi poprzez partial oraz d theta}), tzn. jest ona współzmiennicza, co oznacza, że:
\begin{eqnarray}
\label{Pi poprzez partial oraz d xi}
\Pi_{P P'}((\tilde{\partial}_{j})_{P}) = (\tilde{\partial}_j)_{P'} - \sum_{i,\,l=1}^{d} d\delta^{i}(\tilde{\Gamma}_{ij}^{l})_{P} (\tilde{\partial}_l)_{P'} \; , \;\;\;{\rm gdzie} \;\;\;\; j=1,2,...,d  \;\;\; {\rm oraz} \;\;\; d\delta^{i} = \delta^{i}(P) - \delta^{i}(P') \;  .
\end{eqnarray}
}
$\Xi \rightarrow P$, tzn. w określonym układzie współrzędnych $\Xi = (\xi^{i})_{i=1}^{d}$. \\ 
Zakładając, że różnica pomiędzy $\Pi_{P P'}((\partial_j)_{P})$ oraz  $(\partial_j)_{P'}$ jest {\it infinitezymalna}, i że może być wyrażona jako liniowa kombinacja {\it różniczek} $d\xi^{1}, d\xi^{2},..., d\xi^{d}$, gdzie:
\begin{eqnarray}
\label{rozniczka theta dla theta}
d\xi^{i} = \xi^{i}(P) - \xi^{i}(P') \; ,  \;\;\;\; i=1,2,...,d \; , \;\;\;\;\;  P_{\Xi} \;{\rm i}\; P'_{\Xi}  \in {\cal S} \; ,
\end{eqnarray}
mamy: 
\begin{eqnarray}
\label{Pi poprzez partial oraz d theta}
\Pi_{P P'}((\partial_{j})_{P}) = (\partial_j)_{P'} - \sum_{i,\,l=1}^{d} d\xi^{i}(\Gamma_{ij}^{l})_{P} (\partial_l)_{P'} \; , \;\;\; j=1,2,...,d  \; , \;\;\;\;\;  P_{\Xi} \;{\rm i}\; P'_{\Xi} \in {\cal S} \; , 
\end{eqnarray}
gdzie $(\Gamma_{ij}^{l})_{P}$, $i,j,l=1,2,...,d$, są $d^{3}$ {\it współczynnikami koneksji} zależącymi od punktu $P$. Koneksję afiniczną $\Pi$, więc i współczynniki koneksji, można (przy ustalonej parametryzacji $\Xi$) określić na różne sposoby. Jeden z nich związany z koneksją Levi-Civita podany jest w (\ref{koneksja Levi-Civita}). Jednak w analizie na przestrzeniach statystycznych szczególnie użyteczna okazała się tzw. $\alpha$-koneksja. \\ 
 \\
{\bf $\alpha$-koneksja}: 
%
{\it W każdym punkcie} $P_{\Xi} \in {\cal S}$, {\it $\alpha$-koneksja} zadaje $d^{3}$ funkcji  $\Gamma^{(\alpha)}_{ij,\,r}\!: \!\Xi \rightarrow \! (\Gamma^{(\alpha)}_{ij,\,r})_{\Xi}$, $i,j,r=1,2,...,d$,  
przyporządkowując mu 
współczynniki koneksji o następującej postaci 
\cite{Amari Nagaoka book}:
\begin{eqnarray}
\label{affine coefficients}
(\Gamma^{(\alpha)}_{ij, \, r})_{\Xi}  \equiv   (\Gamma^{(\alpha)}_{ij, \, r})_{P_{\Xi}} =  E_{\Xi}\left[ \left(
\partial_{i} \partial_{j} \ell_{\Xi} + \frac{1-\alpha}{2}
\partial_{i} \ell_{\Xi} \partial_{j} \ell_{\Xi} \right) \partial_{r} \ell_{\Xi} \right]\; , \;\;\;\;\; \forall\, P_{\Xi} \in {\cal S} \; .
\end{eqnarray}
Koneksja $\alpha$ jest {\it symetryczna}, tzn.:
\begin{eqnarray}
\label{koneksja symetryczna}
\Gamma^{(\alpha)}_{ij,\,l} = \Gamma^{(\alpha)}_{ji,\,l} \; , \;\;\;\;  i,j,l = 1,2,...,d\; , \;\;\;\;\;\;\;\;\;\;\;\;\;\;\;\;\; \forall\, P_{\Xi} \in {\cal S} \; .
\end{eqnarray}
W końcu, jeśli $g^{ij}(P)$ jest $(i,j)$-składową macierzy odwrotnej do macierzy informacyjnej  $(g_{ij}(P))$, to współczynniki $\Gamma^{r\, (\alpha)}_{ij}$ są równe:
\begin{eqnarray}
\label{wspolczynniki koneksji dla g i g-1}
\Gamma^{r\, (\alpha)}_{ij} = \sum_{l=1}^{d} g^{rl} \, \Gamma^{(\alpha)}_{ij,\, l} \; , \;\;\;\;  i,j,r = 1,2,...,d\; , \;\;\;\;\; \forall\, P_{\Xi} \in {\cal S} \; .
\end{eqnarray}
{\bf $\alpha$ - pochodna kowariantna}: Mając metrykę Fishera-Rao $g_{ij}$,  (\ref{Fisher metric poprzez macierz informacyjna}), zdefiniujmy $\alpha$ - pochodną kowariantną $\nabla^{(\alpha)}$ na ${\cal S}$ poprzez  $\alpha$-koneksję afiniczną następująco: 
\begin{eqnarray}
\label{pochodna kowariantna i wspolczynniki koneksji}
\langle \nabla_{\partial_{i}}^{(\alpha)} \partial_{j}, \, 
\partial_{l}\rangle := \Gamma^{(\alpha)}_{ij,\, l} \; , \;\;\;\;  i,j,l = 1,2,...,d\; , \;\;\;\;\; \forall\, P_{\Xi} \in {\cal S} \; .
\end{eqnarray} 
Powyżej, poprzez koneksję afiniczną zdefiniowaliśmy pochodną kowariantną. Ale i odwrotnie, pochodna kowariantna definiuje koneksję \cite{Amari Nagaoka book}. Stąd np. mówimy, że określiliśmy koneksję $\nabla$. \\ 
\\
{\bf Pole wektorowe na ${\cal S}$}: Niech $V: P \rightarrow V_{P}$ jest odwzorowaniem przyporządkowującym każdemu punktowi $P \in {\cal S}$ wektor styczny $V_{P} \in T_{P}({\cal S})$. Odwzorowanie to nazywamy {\it polem wektorowym}. Na przykład, jeśli $(\xi^{i})_{i=1}^{d}$ jest układem współrzędnych, wtedy przyporządkowanie $\frac{\partial}{\partial \xi^{i}}: P \rightarrow \left(\frac{\partial}{\partial \xi^{i}}\right)_{P}$, $i=1,2,...,d$, określa $d\,$ pól wektorowych na ${\cal S}$. \\
Oznaczmy przez $T({\cal S})$  rodzinę wszystkich pól wektorowych klasy $C^{\infty}$ typu $V_{P} = \sum_{i=1}^{d} V_{P}^{i} \left(\partial_{i}\right)_{P} \in T_{P}({\cal S})$ na ${\cal S}$, gdzie $d$ funkcji $V^{i}: P \rightarrow V^{i}_{P}$ nazywamy współrzędnymi pola wektorowego $V$ ze względu na $(\xi^{i})_{i=1}^{d}$.  \\
\\
{\bf Określenie pochodnej kowariantnej pola wektorowego}: Rozważmy  dwa pola wektorowe  $V, W \in T({\cal S})$. Niech w bazie $(\xi^{i})_{i=1}^{d}$ pola te mają postać $V = \sum_{i=1}^{d} V^{i} \partial_{i}$ oraz $W = \sum_{i=1}^{d} W^{i} \partial_{i}$.  Pochodną kowariantną pola $V$ ze względu na $W$ nazywamy pole wektorowe $\nabla_{W} V \in T({\cal S})$, które w bazie  $(\xi^{i})_{i=1}^{d}$ ma postać:
\begin{eqnarray}
\label{pochodna kowariantna V wzgledem W}
\nabla_{W} V = \sum_{i=1}^{d} W^{i} \, \sum_{k=1}^{d} \, \{ \,  \partial_{i} V^{k} + \sum_{j=1}^{d} V^{j} \, \Gamma^{k}_{ij} \, \} \, \partial_{k} \; , \;\;\;\;\; \forall\, P_{\Xi} \in {\cal S} \; .
\end{eqnarray}
Przyjmując $W = \partial_{i}$ oraz $V=\partial_{j}$ łatwo sprawdzić, że związek (\ref{pochodna kowariantna V wzgledem W}) daje:
\begin{eqnarray}
\label{pochodna kowariantna partial wzgledem partial}
\nabla_{\partial_{i}} \partial_{j} =  \sum_{k=1}^{d} \,  \Gamma^{k}_{ij} \, \, \partial_{k} \; , \;\;\;\;\; \forall\, P_{\Xi} \in {\cal S} \; ,
\end{eqnarray}
co po skorzystaniu z (\ref{wspolczynniki koneksji dla g i g-1}) jest zgodne z  określeniem koneksji $\nabla$ w (\ref{pochodna kowariantna i wspolczynniki koneksji}). \\
Można pokazać  \cite{Amari Nagaoka book}, że zachodzi:
\begin{eqnarray}
\label{alfa conection + -}
\nabla^{(\alpha)} = \frac{1 + \alpha}{2} \, \nabla^{(1)} + \frac{1 -
\alpha}{2} \, \nabla^{(-1)}\; , \;\;\;\;\; \forall\, P_{\Xi} \in {\cal S} \; .
\end{eqnarray}
{\bf Uwaga o płaskości ${\cal S}$ ze względu na $\nabla$}: Mówimy, że układ współrzędnych $(\xi^{i})_{i=1}^{d}$ jest {\it afinicznym układem współrzędnych dla koneksji} $\nabla$ gdy zachodzą  (równoważne) warunki:
\begin{eqnarray}
\label{uklad afiniczny dla koneksji}
\nabla_{\partial_{i}} \partial_{j} = 0 \;\;\;\;\;\; {\rm lub \; r\acute{o}wnowa\dot{z}nie} \;\;\;\;\; \Gamma_{ij}^{l} = 0 \;\;\;\; i,j,l=1,2,...,d \; , \;\;\;\;\; \forall\, P_{\Xi} \in {\cal S} \; .
\end{eqnarray}
Jeśli dla zadanej koneksji $\nabla$ odpowiadający jej układ współrzędnych jest afiniczny, to mówimy, że koneksja $\nabla$ jest płaska\footnote{W ogólności, 
znikanie koneksji jest warunkiem wystarczającym lecz niekoniecznym  {\it afinicznej płaskości, która w bardziej fundamentalnym określeniu ma miejsce, gdy tensor krzywizny (Riemanna) $R$ zeruje się na całej rozmaitości $\cal S$}, tzn. zerują się wszystkie jego składowe. 
Składowe tensora krzywizny $R$ w bazie $(\xi^{i})_{i=1}^{d}$ mają  następującą postać: 
\begin{eqnarray}
\label{tensor krzywizny R}
R^{l}_{ijk} \equiv R^{l}_{ijk}(\Gamma) = \partial_{i} \Gamma^{l}_{jk} - \partial_{j} \Gamma^{l}_{ik} + \sum_{s=1}^{d} \Gamma^{l}_{is} \Gamma^{s}_{jk} - \sum_{s=1}^{d} \Gamma^{l}_{js} \Gamma^{s}_{ik} \; , \;\;\;\;\; \forall\, P_{\Xi} \in {\cal S} .
\end{eqnarray}
Gdy na rozmaitości $\cal S$ istnieje globalny układu współrzędnych, wtedy dla koneksji, która jest symetryczna (tak jak np.  $\alpha$-koneksja), znikanie tensora krzywizny pociąga istnienie układu współrzędnych, w którym znika koneksja i oba określenia afinicznej płaskości pokrywają się. \\
} 
lub, że {\it przestrzeń statystyczna ${\cal S}$ jest płaska ze względu na} $\nabla$. \\
Płaskość ${\cal S}$ ze względu na $\nabla$ oznacza, że wszystkie wektory bazowe $\partial_{i} \equiv \frac{\partial}{\partial \xi^{i}}$, $i=1,2,...,d$ są równoległe na całej przestrzeni ${\cal S}$. \\
\\
Modele z koneksją  $\nabla^{(\alpha)}$, które są $\alpha = + 1$ bądź $-1$ płaskie,  odgrywają szczególną rolę w modelowaniu statystycznym  \cite{Pawitan}.\\
\\
{\bf Dwa przykłady rodzin rozkładów prawdopodobieństwa}: Istnieją dwa przypadki rodzin modeli statystycznych, szczególnie istotnych w badaniu podstawowych geometrycznych własności modeli  statystycznych. Pierwsza z nich to rodzina rozkładów eksponentialnych, a druga, rozkładów mieszanych.  \\
\\
{\bf Rodzina modeli eksponentialnych} okazuje się wyjątkowo ważna, nie tylko dla badania własności statystycznych, ale również w związku z jej realizacją w szeregu zagadnieniach  fizycznych. Niech  wielkość próby $N=1$. Niech zmienna losowa $Y$ będzie zmienną typu ciągłego lub dyskretnego. Ogólna postać regularnej rodziny rozkładów  eksponentialnych\footnote{Zlogarytmowane 
modele eksponentialne $\{ \ln p_{\Xi}, \; \Xi \equiv (\xi^{i})_{i=1}^{d} \in  {V}_{\Xi} \subset \mathbb{R}^{d}\}$ są przestrzeniami afinicznymi. Dlatego z punktu widzenia ich geometrycznej charakterystyki odgrywają taką rolę jak proste i powierzchnie w $3$-wymiarowej geometrii Euklidesowej.
} 
jest następująca: 
\begin{eqnarray}
\label{exponential family}
p_{\Xi} \equiv p({\bf y}| \Xi) = \exp \left[ C({\bf y}) +
\sum_{i=1}^{d}
\xi^{i} F_{i}({\bf y})  - \psi(\Xi) \right] 
= \exp \left[ \sum_{i=1}^{d}
\xi^{i} F_{i}({\bf y})  - \psi(\Xi) \right] \, h({\bf y}) \; , 
\end{eqnarray}
gdzie $\xi^{i}$, $\,i=1,2,...,d$, są tzw. {\it parametrami kanonicznymi}, natomiast  $h({\bf y}) = \exp (C({\bf y}))$ jest nieujemną funkcją, która nie zależy od wektorowego parametru $\Xi$. \\
W (\ref{exponential family}) pojawiła się   
$d$-wymiarowa statystyka: 
\begin{eqnarray}
\label{F dla exponential family}
F(Y) = (F_{1}(Y),..., F_{d}(Y))^{T} \equiv (F_{i}(Y))_{i=1}^{d} ,
\end{eqnarray}
nazywana  statystyką {\it kanoniczną}.\\
\\
Całkując obustronnie (\ref{exponential family}) po przestrzeni próby ${\cal B} = {\cal Y}$,  a następnie wykorzystując własność normalizacji $\int_{\cal Y} d {\bf y}\, p({\bf y}| \Xi) = 1$, 
otrzymujemy\footnote{Funkcja $\psi(\Xi)$ okazuje się być tzw. potencjałem transformacji Legendre'a pomiędzy affinicznymi układami współrzędnych modeli eksponentialnych (por. Rozdziały~\ref{Potencjaly ukladow wspolrzednych} oraz \ref{Estymacja w modelach fizycznych na DORC}).}: 
\begin{eqnarray}
\label{psi dla exponential family}
\psi(\Xi) = \ln \int_{\cal Y} d{\bf y} \exp \left[ C({\bf y}) +
\sum_{i=1}^{d}
\xi^{i} F_{i}({\bf y})  \right] \; .
\end{eqnarray}
\\
Modele eksponentialne są $\alpha = 1$ - płaskie, co oznacza, że  \cite{Amari Nagaoka book}:
\begin{eqnarray}
\label{Gamma 1 dla exponential family}
(\Gamma^{(1)}_{ij,\,k})_{\Xi} = 0 \; , \;\;\;\; {\rm dla \; modeli \; eksponentialnych}  \; , \;\;\; \forall\, P_{\Xi} \in {\cal S}  \; .
\end{eqnarray} 
Istotnie, korzystając z oznaczenia wprowadzonego w (\ref{oznaczenie ln P}), otrzymujemy dla modeli zadanych przez (\ref{exponential family}): 
\begin{eqnarray}
\label{partial l w F oraz partial psi dla exponential family}
\frac{\partial l({\bf y}|\Xi)}{\partial \xi^{i}} = F_{i}({\bf y}) - \frac{\partial \psi(\Xi)}{\partial \xi^{i}} \; , 
\end{eqnarray}
skąd (przy okazji) otrzymujemy obserwowaną informację Fishera dla rozkładów eksponentialnych w parametryzacji kanonicznej:
\begin{eqnarray}
\label{partial l dla psi dla exponential family}  
\texttt{i\!F}\left(\Xi \right) = - \frac{\partial^{2} l({\bf y}|\Xi)}{\partial \xi^{j} \partial \xi^{i}} =  \frac{\partial^{2} \psi(\Xi)}{\partial \xi^{j} \partial \xi^{i}} \; .
\end{eqnarray}
Podstawiając (\ref{partial l dla psi dla exponential family}) do (\ref{affine coefficients}), otrzymujemy dla $\alpha=1$  współczynniki $1$-koneksji, równe:
\begin{eqnarray}
\label{affine coefficients dla eksponentialnych}
(\Gamma^{(1)}_{ij, \, r})_{\Xi}  =  E_{\Xi}\left[  \left(\frac{\partial^{2} l(Y|\Xi)}{\partial \xi^{j} \partial \xi^{i}} \right) \frac{\partial l(Y|\Xi)}{\partial \xi_{r}} \right] = -  \frac{\partial^{2} \psi(\Xi)}{\partial \xi^{j} \partial \xi^{i}} E_{\Xi}\left[ \frac{ \partial  l(Y|\Xi)}{ \partial \xi_{r} } \right] = 0 \; , \;\;\; \forall\, P_{\Xi} \in {\cal S}  \; , \;\;\;
\end{eqnarray}
gdzie w ostatniej równości skorzystano z (\ref{znikanie ES}) dla funkcji wynikowej.\\
\\
Ponadto dla modeli eksponentialnych z (\ref{Fisher inf matrix plus reg condition}) oraz (\ref{partial l dla psi dla exponential family}) otrzymujemy następującą postać metryki Fishera-Rao:
\begin{eqnarray}
\label{g dla  exponential family}
g_{ij} = -  E_{\Xi}(\partial_{i} \partial_{j} \ell_{\Xi})  =  \frac{\partial^{2} \psi(\Xi)}{\partial \xi^{j} \partial \xi^{i}} \; .
\end{eqnarray}
Zwróćmy uwagę na równość prawych stron 
(\ref{partial l dla psi dla exponential family}) oraz (\ref{g dla  exponential family}) dla rozkładów eksponentialnych. 
\\
\\
Przykładami modeli z eksponentialnej rodziny rozkładów są:\\
\\
i) {\it Rozkład normalny}, (\ref{rozklad norm theta sigma2}), 
$p\left(Y={\bf y}|\mu, \sigma^{2}\right) =  \frac{1}{\sqrt{2 \pi \, \sigma^2}} \; \exp \left( - {\frac{({\bf y} - \mu)^{2}}{2 \, \sigma^2}} \right)$, ${\bf y} \in \mathbf{R}$, dla którego:
\begin{eqnarray}
\label{rozklad normalny parametry kanoniczne} 
& & C({\bf y}) = 0 \;, \;\;\; F_{1}({\bf y})={\bf y} \; , \;\; F_{2}({\bf y})={\bf y}^{2} \; , \;\; \xi^{1} = \frac{\mu}{\sigma^{2}} \; , \;\; \xi^{2} = - \frac{1}{2 \sigma^{2}} \; ,  \nonumber \\
& & \psi(\xi) = \frac{\mu^{2}}{2 \sigma^{2}} + \ln (\sqrt{2 \pi} \, \sigma) = - \frac{(\xi^{1})^{2}}{4 \xi^{2}} + \frac{1}{2}\ln (- \frac{\pi}{\xi^{2}} ) \; . 
\end{eqnarray}
\\
ii) {\it Rozkład Poissona} (\ref{rozklad Poissona}), $p \left(Y={\bf y}|\mu \right)=\frac{\mu ^{{\bf y}} \exp({-\mu }) }{{\bf y}\, !}\,$, gdzie ${\bf y} = 0,1,...,\infty$, dla którego:
\begin{eqnarray}
\label{rozklad Poissona parametry kanoniczne} 
& & C({\bf y}) = - \ln ({\bf y}!) \;, \;\;\; F({\bf y})={\bf y} \; ,  \;\; \xi = \ln \mu \; , \;\; \psi(\xi) = \mu = \exp \xi  \; . 
\end{eqnarray} 
\\
iii) ({\it Standardowy}) {\it  rozkład eksponentialny}, $p \left(Y={\bf y}|\mu \right)= \mu^{-1} \exp(-{\bf y}/\mu)\,$, gdzie ${\bf y} > 0$, dla którego:
\begin{eqnarray}
\label{rozklad eksponentialny parametry kanoniczne} 
& & C({\bf y}) = 0 \;, \;\;\; F({\bf y})=  {\bf y} \; ,  \;\; \xi = - \frac{1}{\mu} \; , \;\; \psi(\xi) =  \ln \mu = \ln ( -\frac{1}{\xi})  \; . 
\end{eqnarray} 
Modele eksponentialne wyróżnia fakt osiągania 
dolnego ograniczenia nierówności Rao-Cramera \cite{Streater} (porównaj związek (\ref{RC dla 1 N z 1 par oczekiwana IF})). \\
\\
{\bf Wymiar statystyki dostatecznej dla} (\ref{exponential family}):  Dla modeli eksponentialnych zachodzi ważna własność związana z wymiarem statystyki kanonicznej $F(Y)$, (\ref{F dla exponential family}). Rozważmy $N$-elementową próbę $\widetilde{Y} \equiv( Y_{n})_{n=1}^{N}$, dla której każdy  punktowy rozkład ma postać  (\ref{exponential family}). Wtedy funkcja wiarygodności dla próby jest następująca:
\begin{eqnarray}
\label{exponential family N fun wiarygodnosci}
P(y| \Xi) =   \exp \left[ \sum_{i=1}^{d}
\xi^{i} \sum_{n=1}^{N} F_{i}({\bf y}_{n})  - N \, \psi(\Xi) \right] \, \exp \left[ \sum_{n=1}^{N} C({\bf y}_{n}) \right] \; .
\end{eqnarray}
Dostateczna statystyka dla wektorowego parametru oczekiwanego  $(\theta_{i})_{i=1}^{d}$, gdzie $\theta_{i} = E_{\Xi}\left[ F_{i}(Y) \right]$, $i = 1,2,...,d $ (por. (\ref{wartosi oczekiwane eta model eksponentialny}) w Rozdziale~\ref{Estymacja w modelach fizycznych na 
DORC}) ma zatem postać:
\begin{eqnarray}
\label{F dostateczna dla exponential family dla proby N}
\left( \sum_{n=1}^{N} F_{1}({\bf y}_{n}), \sum_{n=1}^{N} F_{2}({\bf y}_{n}), ..., \sum_{n=1}^{N} F_{d}({\bf y}_{n})  \right) \; .
\end{eqnarray}
Jej wymiar jest równy $d$ i jak widać, dla konkretnych reprezentantów ogólnej rodziny eksponentialnej nie zależy on od wymiaru próby $N$. 
Własność ta nie jest spełniona np. dla takich nieeksponentialnych rozkładów jak rozkładu Weibull'a oraz Pareto \cite{Nowak}, dla których wymiar statystyki dostatecznej rośnie wraz z wymiarem próby. %
Inną ważną własnością rozkładów nieeksponentialnych jest to, że dziedzina  (tzn. nośnik) ich funkcji gęstości może zależeć od parametru.  
\\
\\
{\bf Rodzina  mieszanych rozkładów prawdopodobieństwa}:
\begin{eqnarray}
\label{mixture family} 
p_{\Delta} \equiv p({\bf y}|\Delta) 
= C({\bf y}) + \sum_{i=1}^{d} \delta^{i}
F_{i}({\bf y})  \; , 
\end{eqnarray}
gdzie $\delta^{i}$ są tzw. {\it parametrami mieszanymi}. \\
\\
{\bf Ćwiczenie}: Pokazać, że przestrzeń statystyczna rodziny rozkładów mieszanych jest $\alpha = - 1$ - płaska \cite{Amari Nagaoka book}:
\begin{eqnarray}
\label{Gamma 1 dla mieszanej family}
(\Gamma^{(-1)}_{ij,\,k})_{\Xi} = 0 \; , \;\;\;\; {\rm dla \; modeli \; mieszanych}  \; , \;\;\; \forall\, P_{\Xi} \in {\cal S}  \; .
\end{eqnarray} 
\\
{\bf Geometria przestrzeni ${\cal S}$ a parametryzacja}: Jako własności geometryczne przestrzeni ${\cal S}$ przyjmujemy te, które są niezmiennicze ze względu na zmianę parametryzacji. \\
Np. własności geometryczne rodziny modeli eksponentialnych nie zależą od tego czy posłużymy się parametrami kanonicznymi $(\xi^{i})_{i=1}^{d}$ czy oczekiwanymi $\theta_{i} = E_{\Xi}\left[ F_{i}(Y) \right]$, $i = 1,2,...,d $ (por. (\ref{wartosi oczekiwane eta model eksponentialny})).

%
%
%
%

\subsection{Przestrzeń statystyczna dualnie płaska}

\label{Przestrzen dualnie plaska}

Z (\ref{Fisher inf matrix}) oraz (\ref{affine coefficients}) wynika, że:  
\begin{eqnarray}
\label{0 -  koneksja}
\partial_{r}g_{ij} = \Gamma^{(0)}_{ri, \, j} + \Gamma^{(0)}_{rj, \, i}\; , \;\;\;\;\;\;\;\;\;\;\;\;\;\;\;\;\;\; \forall\, P_{\Xi} \in {\cal S} \; , 
\end{eqnarray}
co oznacza, że  $0$-koneksja jest 
metryczna\footnote{{\bf Koneksja metryczna}: 
Poprzez metrykę Riemannowską $g$, której przykładem jest metryka Fishera-Rao, (por. Rozdział~\ref{alfa koneksja}),  można określić 
koneksję $\gamma_{P P'} \rightarrow \Pi_{\gamma_{P P'}}$, która jest metryczna na ${\cal S}$. Z {\it koneksją metryczną} mamy do czynienia gdy:
\begin{eqnarray}
\label{koneksja metryczna}
g_{P'}(\Pi_{\gamma_{P P'}} V,  \; \Pi_{\gamma_{P P'}} W) = g_{P}(V,\, W)\; , \;\;\;\;\; \forall\, P_{\Xi} \;{\rm i} \;P'_{\Xi} \in {\cal S}\; , 
\end{eqnarray}
tzn. gdy {\it iloczyn wewnętrzny jest niezmienniczy ze względu na przesunięcie równoległe} dla wszystkich wektorów stycznych $V$ oraz $W$ i wszystkich ścieżek $\gamma$ z $P$ do $P'$. \\ 
Koneksję metryczną nazywamy {\it koneksją Levi-Civita} (lub  Riemannowską), jeśli jest ona  zarówno metryczna jak i symetryczna. Dla zadanego $g$ koneksja taka  jest określona jednoznacznie jako:
\begin{eqnarray}
\label{koneksja Levi-Civita}
\Gamma_{ij,\,k} =  \frac{1}{2}  \left( \partial_{i}g_{jk} + \partial_{j}g_{ki} - \partial_{k}g_{ij} \right)\; , \;\;\;\;\; \forall\, P_{\Xi} \in {\cal S} \; ,
\end{eqnarray}
gdzie jej {\it symetria} oznacza spełnienie warunku (\ref{koneksja symetryczna}). 
Koneksja (\ref{koneksja Levi-Civita}) spełnia warunek (\ref{0 -  koneksja}). \\
\\
Geodezyjne koneksji Levi-Civita są krzywymi o najmniejszej (mierzonej przez  metrykę $g$) długości, tzn. jej długość pomiędzy punktem  $P$ a $P'$ określona następująco:
\begin{eqnarray}
\label{dlugosc wzdluz gamma}
\left\| \gamma \right\| = \int_{P}^{P'} \left\| \frac{d \gamma}{dt} \right\| dt = \int_{P}^{P'} \sqrt{ \sum_{i,j=1}^{d} g_{ij} \frac{d \gamma^{i}}{dt} \frac{d \gamma^{j}}{dt}} \;dt \; ,
\end{eqnarray}
jest najmniejsza, gdzie $\gamma^{i}(t)$ jest $i$-tą współrzędną punktu 
na krzywej $\gamma(t)$, tzn. $\gamma^{i}(t) := \xi^{i}(\gamma(t))$. 
} 
ze względu na metrykę Fishera-Rao \cite{Amari Nagaoka book}. 
\\
\\
{\it Jednak w ogólności dla $\alpha \neq 0$, $\alpha$-koneksja  afiniczna 
nie jest metryczna,  natomiast spełnia warunek dualności} 
omówiony poniżej. \\
\\
{\bf Koneksje dualne}: Niech na rozmaitości ${\cal S}$ zadana jest pewna metryka Riemannowska $g=\left\langle , \right\rangle$ i dwie koneksje $\nabla$ oraz  $\nabla^{*}$. Metryka ta może być np. metryką $g$ Fishera-Rao. Jeśli dla wszystkich pól wektorowych $V,W,Z\in T({\cal S})$,  zachodzi:
\begin{eqnarray}
\label{def dualnych nabla}
Z\left\langle V, W \right\rangle = \left\langle \nabla_{Z} V, W \right\rangle + \left\langle  V, \nabla^{*}_{Z} W \right\rangle\; , \;\;\;\;\;\;\;\;\;\; \forall\, P_{\Xi} \in {\cal S} \; , 
\end{eqnarray}
wtedy mówimy, że $\nabla$ oraz $\nabla^{*}$ są ze względu ma metrykę $\left\langle , \right\rangle$  {\it dualne} (sprzężone) względem siebie.  W układzie współrzędnych $(\xi^{i})_{i=1}^{d}$ metryka $g$ ma współrzędne $g_{ij}$, a koneksje $\nabla$ oraz $\nabla^{*}$ mają  współczynniki koneksji odpowiednio $\Gamma_{ij,\,r}$ oraz $\Gamma^{*}_{ij, \, r}$. \\
Warunek dualności (\ref{def dualnych nabla}) możemy teraz zapisać 
w postaci\footnote{Wystarczy zapisać (\ref{def dualnych nabla}) dla wektorów bazowych $(\partial_{i})_{i=1}^{d}$, otrzymując: 
\begin{eqnarray}
\label{def dualnych w bazach nabla}
\partial_{r}\left\langle \partial_{i}, \partial_{j} \right\rangle = \left\langle \nabla_{\partial_{r}} \partial_{i}, \partial_{j} \right\rangle + \left\langle  \partial_{i}, \nabla^{*}_{\partial_{r}} \partial_{j} \right\rangle\; , \;\;\;\;\;\;\;\;\;\; \forall\, P_{\Xi} \in {\cal S} \; , 
\end{eqnarray}
skąd wykorzystując (\ref{pochodna kowariantna i wspolczynniki koneksji}) dla koneksji $\nabla$ oraz $\nabla^{*}$, otrzymujemy (\ref{war dualnosci we wspolczynnikach}).
}:
\begin{eqnarray}
\label{war dualnosci we wspolczynnikach}
\partial_{r}g_{ij} = \Gamma_{ri, \, j} + \Gamma^{*}_{rj, \, i}\; , \;\;\;\;\;\;\;\;\;\; \forall\, P_{\Xi} \in {\cal S} \; ,
\end{eqnarray}
będącej uogólnieniem warunku (\ref{0 -  koneksja}) istniejącego dla koneksji metrycznej. \\
Ponadto, koneksja $\nabla^{met} \equiv (\nabla+\nabla^{*})/2$ jest koneksją metryczną, dla  której zachodzi warunek  $\partial_{r}g_{ij} = \Gamma^{met}_{ri, \, j} + \Gamma^{met}_{rj, \, i}$   \cite{Amari Nagaoka book}. 
Można również sprawdzić, że (ze względu na $g$) zachodzi $(\nabla^{*})^{*} = \nabla$.  \\
\\
{\bf Struktura dualna}: Trójkę $(g, \nabla, \nabla^{*}) \equiv ({\cal S}, g, \nabla, \nabla^{*})$ nazywamy {\it strukturą dualną na} ${\cal S}$. 
W ogólności, mając metrykę $g$ oraz koneksję $\nabla$ określoną na ${\cal S}$, koneksja dualna $\nabla^{*}$ jest wyznaczona w sposób jednoznaczny, co jest treścią poniższego twierdzenia.
\\
\\
{\bf Twierdzenie o związku pomiędzy koneksjami dualnymi}: 
Niech $P$ oraz $P'$ są punktami brzegowymi ścieżki $\gamma$, oraz  niech przekształcenia 
$\Pi_{\gamma_{P P'}}$ oraz $\Pi^{*}_{\gamma_{P P'}}$ z $T_{P}(\cal S)$ do $T_{P'}(\cal S)$ opisują {\it równoległe przesunięcie wzdłuż} $\gamma$, odpowiednio ze względu na koneksje afiniczne $\nabla$ oraz $\nabla^{*}$. Wtedy dla wszystkich $V,W \in T_{P}({\cal S})$ zachodzi  \cite{Amari Nagaoka book}:
\begin{eqnarray}
\label{koneksja warunek dualnosci ogolnie}
g_{P'}(\Pi_{\gamma_{P P'}} V, \, \Pi^{*}_{\gamma_{P P'}} W) = g_{P}(V,\, W)\; , \;\;\;\;\;\;\;\;\;\;  \forall\, P_{\Xi}\;{\rm i}\;P'_{\Xi} \in {\cal S} \; .
\end{eqnarray}
Warunek ten jest uogólnieniem warunku (\ref{koneksja metryczna})  istniejącego dla koneksji metrycznej o {\it niezmienniczości iloczynu wewnętrznego ze względu na  przesunięcie równoległe}.  Wyznacza on w sposób jednoznaczny związek pomiędzy $\Pi_{\gamma_{P P'}}$ oraz  $\Pi^{*}_{\gamma_{P P'}}$. \\
\\
{\bf Niezmienniczość iloczynu wewnętrznego dla płaskich koneksji dualnych przy przesunięciu równoległym}: {\it Jeśli}  $\Pi_{\gamma_{P P'}}$ nie zależy na ${\cal S}$ od ścieżki $\gamma$, a tylko od punktów końcowych\footnote{
Mówimy wtedy, że koneksja na ${\cal S}$ jest całkowalna. Gdy przestrzeń ${\cal S}$ jest jedno-spójna, to warunek ten oznacza znikanie tensora krzywizny Riemanna na ${\cal S}$. 
} 
$P$ oraz $ P'$, wtedy 
$\Pi_{\gamma_{P P'}}=\Pi_{P P'}$ na ${\cal S}$.  
Ponieważ przy określonej metryce $g$, koneksja dualna $\nabla^{*}$ jest jednoznacznie wyznaczona dla $\nabla$, zatem również dla koneksji $\nabla^{*}$ jest  na ${\cal S}$ spełniony warunek $\Pi^{*}_{\gamma_{P P'}} = \Pi^{*}_{P P'}$, skąd z (\ref{koneksja warunek dualnosci ogolnie}) przy przesunięciu równoległym otrzymujemy: 
\begin{eqnarray}
\label{koneksja warunek dualnosci ogolnie bez gamma}
g_{P'}(\Pi_{P P'} V, \, \Pi^{*}_{P P'} W) = g_{P}(V,\, W)\; , \;\;\;\;\;\;\;\;\;\;  \forall\, P_{\Xi}\;{\rm i}\;P'_{\Xi} \in {\cal S} \; .
\end{eqnarray}
{\bf Zadanie}: Korzystając z metryki (\ref{g dla  exponential family}) modelu eksponentialnego oraz z (\ref{affine coefficients dla eksponentialnych}) i (\ref{affine coefficients}), sprawdzić  bezpośrednim rachunkiem w parametryzacji kanonicznej $(\xi^{i})_{i=1}^{d}$ warunek (\ref{war dualnosci we wspolczynnikach}), otrzymując: 
\begin{eqnarray}
\label{koneksja warunek dualnosci dla eksponent}
\partial_{r}g_{ij} = \Gamma^{+1}_{ri, \, j} + \Gamma^{-1}_{rj, \, i} = \Gamma^{-1}_{rj, \, i} = \partial_{r} \partial_{i} \partial_{j} \psi (\Xi) \; , \;\;\;\;\;\;\;\;\;\; \forall\, P_{\Xi} \in {\cal S} \; .
\end{eqnarray}

\vspace{2mm}

\subsection{Dualne układy współrzędnych}

\label{Dualne uklady wspolrzednych}

{\bf Określenie dualnie płaskiej przestrzeni}: Mówimy, że $({\cal S}, g, \nabla, \nabla^{*})$ jest {\it dualnie płaską przestrzenią}, jeśli obie koneksje dualne, $\nabla$ oraz $ \nabla^{*}$, są płaskie na ${\cal S}$. Oznacza to, że jeśli koneksja $\nabla$ jest płaska w pewnej bazie $(\xi^{i})_{i=1}^{d}$ to koneksja $\nabla^{*}$ jest płaska w pewnej bazie $(\xi^{*\,i})_{i=1}^{d}$, którą nazywamy bazą dualną do  $(\xi^{i})_{i=1}^{d}$. \\
\\
{\bf $\alpha$ - koneksja dualnie płaska}: \\
Istotność pojęcia $\alpha$-koneksji pojawia się wraz z rozważeniem na przestrzeni statystycznej ${\cal S}$ nie tyle prostej pary  $(g, \nabla^{(\alpha)}) \equiv ({\cal S}, g, \nabla^{(\alpha)})$ ale struktury potrójnej $(g, \nabla^{(\alpha)},$ $\nabla^{(-\alpha)}) \equiv ({\cal S}, g, \nabla^{(\alpha)}, \nabla^{(-\alpha)})$. Powodem jest istnienie poprzez metrykę Fishera-Rao {\it dualności pomiędzy koneksjami}  $\nabla^{(\alpha)}$ oraz $\nabla^{(-\alpha)}$, która okazuje się być ważna przy badaniu modeli statystycznych. 
\\
\\
Podsumowując, {\it dla dowolnego modelu statystycznego ${\cal S}$}, zachodzi \cite{Amari Nagaoka book}:
\begin{eqnarray}
\label{podwojna plaskosc modelu} 
({\cal S} \;\; {\rm jest} \;\; \alpha-p\ell aska) \Leftrightarrow ({\cal S} \;\; {\rm jest} \;\; (-\alpha)-p\ell aska)  \; \; .
\end{eqnarray}
{\bf Przykład}: Model statystyczny  eksponentialny ${\cal S}$ jest  płaski w parametryzacji kanonicznej $(\xi^{i})_{i=1}^{d}$ dla $\alpha=1$.  Istnieje zatem {\it parametryzacja dualna}, w której 
jest on również $\alpha=-1$ płaski.  Np. w Rozdziale~\ref{Estymacja w modelach fizycznych na DORC} okaże się, że dla modeli eksponentialnych spełniających warunek maksymalizacji entropii, bazą dualną do bazy kanonicznej jest baza parametrów oczekiwanych.\\ 
Podobnie jest dla rodziny rozkładów mieszanych, tzn. jest ona jednocześnie $\pm 1$ płaska.
\\
\\
{\bf Określenie dualnych układów współrzędnych}: Zastanówmy się nad ogólną strukturą przestrzeni $({\cal S}, g, \nabla, \nabla^{*})$, która byłaby dualnie {\it płaska}. \\
Z określenia przestrzeni płaskiej ze względu na określoną koneksję (\ref{uklad afiniczny dla koneksji}) oraz z (\ref{podwojna plaskosc modelu}) wynika, że jeśli istnieje układ współrzędnych $\Xi=(\xi^{i})_{i=1}^{d}$ z wektorami bazowymi $\partial_{\xi^i} \equiv \frac{\partial}{\partial \xi^{i}}$, ze względu na który koneksja $\nabla$ jest płaska, tzn. $\nabla_{\partial_{\xi^i}} \partial_{\xi^j} = 0$, $i,j=1,2,...,d$, to istnieje również układ współrzędnych $\Theta=(\theta^{i})_{i=1}^{d}:=(\xi^{*i})_{i=1}^{d}$ z wektorami bazowymi $\partial_{\theta^i} \equiv \frac{\partial}{\partial \theta^{i}}$, ze względu na który koneksja $\nabla^{*}$ jest płaska,  tzn. $\nabla_{\partial_{\theta^j}} \partial_{\theta^l} = 0$, $j,l=1,2,...,d$.\\ 
\\
{\bf Wniosek}: Zatem, gdy pole wektorowe $\partial_{\xi^i}$  jest  $\nabla$- płaskie, więc pole wektorowe $\partial_{\theta^j}$ jest  $\nabla^{*}$-płaskie i z (\ref{koneksja warunek dualnosci ogolnie bez gamma}) {\it wynika stałość} $\,g_{P_{\Xi}}(\partial_{\xi^i}, \partial_{\theta^j})$ na ${\cal S}$. Fakt ten, biorąc pod uwagę wszystkie $d$ stopni swobody zawarte w afinicznym układzie współrzędnych, (\ref{uklad afiniczny dla koneksji}), można zapisać 
jako\footnote{
Niech 
$\Xi = (\xi^{i})_{i=1}^{d}$ będzie układem współrzędnych afinicznych. Wtedy, zgodnie z (\ref{nowy wspolczynnik koneksji}) współczynniki koneksji w układzie współrzędnych $\Theta=(\theta^{i})_{i=1}^{d}$ są równe:
\begin{eqnarray}
\label{gdy Gamma zero w afin uk wsp}
\tilde{\Gamma}^{t}_{rs} = \sum_{i, j, \,l=1}^{d} \left( \frac{\partial^{2} \xi^{\,l}}{\partial \theta^{r} \partial \theta^{s}} \right) \frac{\partial \theta^{t}}{\partial \xi^{l}}
\; , \;\;\;\;\;\; t,r,s=1,2,...,d \; . \nonumber
\end{eqnarray}
Zażądajmy aby nowy układ współrzędnych $\Theta$ był również afiniczny. Koniecznym i wystarczającym warunkiem spełnienia tego żądania jest zerowanie się wszystkich drugich pochodnych $ \frac{\partial^{2} \xi^{\,l}}{\partial \theta^{r} \partial \theta^{s}} = 0\,$, gdzie $l,r,s=1,2,...,d$.  Warunek ten oznacza, że pomiędzy współrzędnymi $(\xi^{i})_{i=1}^{d}$ oraz $(\theta^{i})_{i=1}^{d}$ istnieje {\it afiniczna transformacja}:
\begin{eqnarray}
\label{transformacja affiniczna}
\xi^{\,l} \longrightarrow \theta^{\,l} = \sum_{k=1}^{d} A^{l}_{\; k} \, \xi^{k} + V^{l} \; , \;\;\;\;\;\; l  = 1,2,...,d \;  \; , \;\;\;\;\;\;\;\;\;\; \forall\, P \in {\cal S} , \nonumber
\end{eqnarray}
gdzie $A = (A^{l}_{\; k})  = (\partial \theta^{l}/\partial \xi^{k}) \in GL(d)$ jest niezależną od współrzędnych $d \times d$ - wymiarową nieosobliwą macierzą ogólnej grupy liniowych transformacji, natomiast jest $V = \sum_{l=1}^{d} V^{l} \partial_{\xi^l}$ jest stałym  $d$-wymiarowym wektorem. \\
Metryka $g$ jest tensorem kowariantnym rangi 2, jest to więc forma dwu-liniowa, zatem iloczyn wewnętrzny wektorów układów dualnych $\partial/\partial \xi^{s}$ oraz $\partial/\partial \theta^{r}$,  można zapisać następująco:
\begin{eqnarray}
\label{dzialanie tensora metrycznego}
\left\langle  \partial/\partial \xi^{s} , \; \partial/\partial  \theta^{r}  \right\rangle 
&=& \left\langle  \partial/\partial \xi^{s} , \, \sum_{l=1}^{d} \frac{\partial \xi^{l}}{\partial \theta^{r}} \frac{\partial}{\partial \xi^{l}}  \right\rangle  = \sum_{l=1}^{d} \frac{\partial \xi^{l}}{\partial \theta^{r}}  \left\langle  \partial/\partial \xi^{s} , \; \partial/\partial  \xi^{l}  \right\rangle   \nonumber \\
&=&  \sum_{l=1}^{d} (A^{-1})^{l}_{\;\,r} \, g^{\xi}_{sl}  = \delta_{sr} \;\;\;\;\;\;\;\;\;\;\;\;\;\;\;\;\;\; s, r  = 1,2,...,d  \; , \;\;\;\;\;\;\; \forall\, P \in {\cal S} \, , \;\;\;\;\;\;\;\;\;\;\;\;\;\;  
\end{eqnarray}
gdzie $g^{\xi}_{sl} = \left\langle  \partial/\partial \xi^{s} , \; \partial/\partial  \xi^{l}  \right\rangle$, a ostatnia równość w (\ref{dzialanie tensora metrycznego}) wynika z żądania stałości iloczynu wewnętrznego $\left\langle  \partial/\partial \xi^{s} , \; \partial/\partial  \theta^{r}  \right\rangle $ na ${\cal S}$ dla układów dualnie płaskich zgodnie z (\ref{uklad afiniczny dla koneksji}) oraz z nieosobliwości transformacji afinicznej, tak że układ współrzędnych $\Theta$ ma tyle samo stopni swobody co układ $\Xi$. Zatem otrzymaliśmy warunek (\ref{stalosc il wewn partial i dual partial}) dla układów dualnie płaskich. \\
Przechodząc zgodnie z (\ref{ukl wsp theta kontrawariantne}) od współrzędnych kontrawariantnych $\theta^{r}$ do kowariantnych,  $\theta_{r}$, $r=1,2,..,d$,  
widać, że powyższy związek (\ref{dzialanie tensora metrycznego}) dla wektorów bazowych układów dualnych można zapisać w postaci $\, \left\langle  \partial/\partial \xi^{s} , \; \partial/\partial  \theta_{r}  \right\rangle = \delta_{s}^{r}$ podanej w \cite{Amari Nagaoka book}. \\
{\bf Uwaga}:
Przestrzeń wektorów dualnych $(\partial/\partial  \theta^{r})_{r=1}^{d}$ układu, który jest  afiniczny względem koneksji $\nabla^{*}$ jest izomorficzna z przestrzenią  1-form $({\rm d} \theta^{r})_{r=1}^{d}$ sprzężoną liniowo   
do układu przestrzeni wektorów $(\partial/\partial  \theta^{r})_{r=1}^{d}$. 
Dla 1-form sprzężonych do wektorów bazowych z definicji zachodzi ${\rm d} \theta^{r} (\partial/\partial  \theta^{s}) = \delta^{r}_{s}$, $\,s,r =1,2,...,d$. [1-form ${\rm d} \theta^{r}$, $r =1,2,...,d$, nie należy mylić z przyrostami typu (\ref{rozniczka theta dla theta}) czy jak w (\ref{Pi poprzez partial oraz d xi})].
}: 
\begin{eqnarray}
\label{stalosc il wewn partial i dual partial}
\left\langle \partial_{\xi^i}, \partial_{\theta^j} \right\rangle \equiv g_{P_{\Xi}}\!\left(\frac{\partial}{\partial \xi^{i}}, \, \frac{\partial}{\partial \theta^{j}} \right) = \delta_{i j} \; ,  \; \; \;\;\;\; i,j=1,2,...,d \;\;\;\;\; , \;\;\;\;\; \forall\, P_{\Xi} \in {\cal S} \; ,
\end{eqnarray}
gdzie iloczyn wewnętrzny $\left\langle \cdot, \cdot \right\rangle$ jest wyznaczony w konkretnym układzie współrzędnych, w tym przypadku w $\Xi = (\xi^{i})_{i=1}^{d}$.
Układy współrzędnych $\Xi = (\xi^{i})_{i=1}^{d}$ oraz $\Theta = (\theta^{j})_{j=1}^{d}$ określone na przestrzeni Riemannowskiej $({\cal S}, g)$ i spełniające warunek (\ref{stalosc il wewn partial i dual partial}) nazywamy {\it wzajemnie dualnymi}. Warunek (\ref{stalosc il wewn partial i dual partial}) oznacza {\bf  stałość na ${\cal S}$ iloczynu wewnętrznego dla układów dualnie płaskich}. \\
 \\
W ogólności dla dowolnej przestrzeni Riemannowskiej $({\cal S}, g)$ nie istnieją układy współrzędnych wzajemnie dualne. \\
Jeśli jednak przestrzeń Riemannowska z dualną koneksją $({\cal S}, g, \nabla, \nabla^{*})$ jest dualnie płaska, to taka para układów współrzędnych {\it istnieje}. 
Ale i na odwrót. Jeśli na przestrzeni Riemannowskiej $({\cal S}, g)$ istnieją dwa układy współrzędnych $(\xi^{i})_{i=1}^{d}$ oraz $(\theta^{j})_{j=1}^{d}$  spełniające warunek (\ref{stalosc il wewn partial i dual partial}), wtedy koneksje $\nabla$ oraz $\nabla^{*}$, względem których układy te są afiniczne, są określone, a $({\cal S}, g, \nabla, \nabla^{*})$ jest dualnie płaska. \\
\\
 \\
\\
{\bf Euklidesowy układ współrzędnych}: W przypadku Euklidesowego układu współrzędnych na ${\cal S}$ mamy (z definicji):
\begin{eqnarray}
\label{stalosc il wewn dla Euklidesowego ukl wsp}
\left\langle \partial_{\xi^i}, \partial_{\xi^j} \right\rangle  = \delta_{ij}\; ,  \; \; \;\;\;\; i,j=1,2,...,d \;\;\;\;\; , \;\;\;\;\; \forall\, P_{\Xi} \in {\cal S} \; ,
\end{eqnarray} 
co oznacza, że jest on {\it samo-dualny}.\\
\\
{\bf Uwaga o współczesnym zastosowaniu koneksji dualnych}: Interesującym wydaje się fakt, że pojęcie koneksji 
dualnych ma coraz większe zastosowanie w analizie układów liniowych \cite{Ohara} i szeregów czasowych \cite{Amari Nagaoka book}.  Przykładem może być jej zastosowanie w analizie szeregów czasowych ARMA(p,q) \cite{Brockwell_Machura}, co jest związane z faktem, że zbiór wszystkich szeregów czasowych ARMA(p,q) ma skończoną parametryzację i w związku z tym tworzy on skończenie wymiarową rozmaitość. Aby dokonać analizy porównawczej dwóch szeregów czasowych biorąc pod uwagę problemy ich aproksymacji, estymacji oraz redukcji wymiaru,   analizowanie pojedynczego szeregu czasowego jest niewystarczające i  okazuje się koniecznym rozważanie własności całej przestrzeni tych szeregów wraz z ich strukturą geometryczną \cite{Amari Nagaoka book}.

\subsubsection{Transformacja Legendre'a pomiędzy parametryzacjami dualnymi}

\label{Potencjaly ukladow wspolrzednych}

\vspace{3mm}

Niech $\Xi \equiv (\xi^{i})_{i=1}^{d}$  oraz  $\Theta \equiv (\theta^{i})_{i=1}^{d}$ są wzajemnie dualnymi bazami na ${\cal S}$,  zgodnie z relacją (\ref{stalosc il wewn partial i dual partial}). 
Zdefiniujmy współrzędne metryki $g$ ze względu na układ współrzędnych $(\xi^{i})_{i=1}^{d}$  jako:
\begin{eqnarray}
\label{ukl wsp ze wzgledu na theta}
g^{\xi}_{ij}:= \left\langle \partial_{\xi^i}, \partial_{\xi^j} \right\rangle    \; ,  \; \;\;\;\; i,j=1,2,...,d \;\;\;\;\; , \;\;\;\;\; \forall\, P_{\Xi} \in {\cal S} \; ,
\end{eqnarray} 
a ze względu na układ współrzędnych $(\theta^{j})_{j=1}^{d}$ jako:
\begin{eqnarray}
\label{ukl wsp ze wzgledu na eta dla theta kowariantne}
g^{\theta}_{ij}:= \left\langle \partial_{\theta^i}, \partial_{\theta^j} \right\rangle    \; ,  \; \;\;\;\; i,j=1,2,...,d \;\;\;\;\; , \;\;\;\;\; \forall\, P_{\Theta} \in {\cal S} \; .
\end{eqnarray} 
Przejście od współrzędnych kontrawariantnych $(\theta^{j})_{j=1}^{d}$ w bazie $\partial_{\theta^j} \equiv \partial/\partial\theta^j$ do  kowariantnych $(\theta_{j})_{j=1}^{d}$ w bazie $\partial^{\theta_j}  \equiv \partial/\partial\theta_j$ ma postać: 
\begin{eqnarray}
\label{ukl wsp theta kontrawariantne}
\theta_j := \sum_{k=1}^{d} g^{\theta}_{jk} \, \theta^k \;\;\; {\rm oraz} \;\;\; \partial^{\theta_j} := \sum_{k=1}^{d} g_{\theta}^{jk}  \partial_{\theta^k}   \; ,  \; \;\;\;\; j,k=1,2,...,d \;\;\;\;\; , \;\;\;\;\; \forall\, P_{\Theta} \in {\cal S} \; ,
\end{eqnarray} 
gdzie $g_{\theta}^{jk}$ jest $(j,k)$-składową macierzy odwrotnej do macierzy informacyjnej  $(g^{\theta}_{jk})$. 
Ze względu na  układ współrzędnych $(\theta_{j})_{j=1}^{d}$, współrzędne metryki $g_{\theta}^{ij}$ są równe:
\begin{eqnarray}
\label{ukl wsp ze wzgledu na eta}
g_{\theta}^{ij} = \left\langle \partial^{\theta_i}, \partial^{\theta_j} \right\rangle    \; ,  \; \;\;\;\; i,j=1,2,...,d \;\;\;\;\; , \;\;\;\;\; \forall\, P_{\Theta} \in {\cal S} \; .
\end{eqnarray} 
Podobnie określamy  macierz informacyjną $(g_{\xi}^{jk})$ jako odwrotną do  $(g^{\xi}_{jk})$. \\
\\
Rozważmy transformacje układu współrzędnych:
\begin{eqnarray}
\label{transf bazy dualnej eta w theta}
\partial^{\theta_{j}} \equiv \frac{\partial }{\partial \theta_{j}} = \sum_{i=1}^{d} \frac{\partial \xi^{i} }{\partial \theta_{j}} \frac{\partial}{\partial \xi^{i}} \equiv  \sum_{i=1}^{d} (\partial^{\theta_j} \xi^{i}) \partial_{\xi^i}  \; , \;\;\;\; j=1,2,...,d \;\;\;\;\; , \;\;\;\;\; \forall\, P \in {\cal S} \; 
\end{eqnarray} 
oraz
\begin{eqnarray}
\label{transf bazy dualnej theta w eta}
\!\!\!\!\!\!\! 
\partial_{\xi^i} \equiv \frac{\partial }{\partial \xi^{i}} = \sum_{j=1}^{d} \frac{\partial \theta_{j} }{\partial \xi^{i}} \frac{\partial}{\partial \theta_{j}} \equiv  \sum_{j=1}^{d} (\partial_{\xi^i} \theta_{j}) \partial^{\theta_j} = \sum_{j=1}^{d} (\partial_{\xi^i} \theta^{j}) \partial_{\theta^j}  \; , \;\;\;\; i=1,2,...,d \; , \;\;\;\; \forall\, P \in {\cal S} \; .
\end{eqnarray} 
Zatem po skorzystaniu  z (\ref{transf bazy dualnej eta w theta})-(\ref{transf bazy dualnej theta w eta}) oraz warunku  (\ref{stalosc il wewn partial i dual partial}), można dualne metryki (\ref{ukl wsp ze wzgledu na theta}) oraz (\ref{ukl wsp ze wzgledu na eta})  zapisać następująco:
\begin{eqnarray}
\label{g ij  we wspol przez eta i theta}
g^{\xi}_{ij} = \frac{\partial \theta_{j}}{\partial \xi^{i}} \;\;\;\; {\rm oraz} \;\;\;\; g_{\theta}^{ij} = \frac{\partial \xi^{i}}{\partial \theta_{j}}  \; , \;\;\;\; i,j=1,2,...,d \;\;\;\;\; , \;\;\;\;\; \forall\, P \in {\cal S} \; ,
\end{eqnarray}
co oznacza również, że macierze informacyjne, $I_{F}(\Xi) = (g^{\xi}_{ij})$ w bazie $\Xi \equiv (\xi^{i})_{i=1}^{d}$  oraz $I_{F}(\Theta) = (g_{\theta}^{ij})$ w 
bazie\footnote{W  zależności od kontekstu i aby nie komplikować zapisu,    
przez $\Theta$ będziemy rozumieli  parametr wektorowy we współrzędnych kowariantnych  $(\theta_{i})_{i=1}^{d}$ bądź  kontrawariantnych  $(\theta^{i})_{i=1}^{d}$. Analogicznie postąpimy   dla $\Xi$. 
} 
dualnej $\Theta \equiv (\theta_{i})_{i=1}^{d}$, są względem siebie odwrotne: 
\begin{eqnarray}
\label{macierze informacyjne dualne}
I_{F}(\Xi) = I_{F}^{-1}(\Theta) \; , \;\;\;\;\;\;\;\;\;\;\;\;\;\;\; \forall\, P \in {\cal S} \; .
\end{eqnarray}
\\
Rozważmy z kolei funkcję $\psi:{\cal S} \rightarrow  \mathbb{R}$ oraz następujące cząstkowe równanie różniczkowe:
\begin{eqnarray}
\label{row rozn dla psi i eta}
\partial_{\xi^i} \psi = \theta_{i}  \; , \;\;\;\; i=1,2,...,d \;  \;\;\;\; {\rm tzn.} \;\;\;\;\; d\psi = \sum_{i=1}^{d} \theta_{i} \, d \xi^{i}   \; , \;\;\;\;\;\; \forall\, P_{\Xi} \in {\cal S} \; . 
\end{eqnarray} 
Ze względu na (\ref{g ij  we wspol przez eta i theta}) równanie (\ref{row rozn dla psi i eta}) daje:
\begin{eqnarray}
\label{row rozn 2 rzedu dla psi i eta}
\partial_{\xi^i} \partial_{\xi^j} \psi =g^{\xi}_{ij} \; , \;\;\;\; i,j=1,2,...,d \;\;\;\;\; , \;\;\;\;\; \forall\, P_{\Xi} \in {\cal S} \; .
\end{eqnarray}
Ze względu na dodatnią określoność metryki $g^{\xi}_{ij}$, równanie to oznacza, że druga pochodna $\psi$ tworzy również {\it dodatnio określoną} macierz. Zatem $\psi$ jest {\it ściśle wypukłą} funkcją współrzędnych $\xi^{1},\xi^{2},...,\xi^{d}$, dla każdego $P \in {\cal S}$. 
\\
Podobnie, rozważając  funkcję $\phi:{\cal S} \rightarrow  \mathbb{R}$ oraz cząstkowe równanie różniczkowe:
\begin{eqnarray}
\label{row rozn dla phi i theta}
\partial^{\theta_i} \phi = \xi^{i}  \; , \;\;\;\; i=1,2,...,d \;  \;\;\;\; {\rm tzn.} \;\;\;\;\; d\phi = \sum_{i=1}^{d} \xi^{i} \, d \theta_{i}   \; , \;\;\;\;\; \forall\, P_{\Theta} \in {\cal S} \; .
\end{eqnarray} 
Ze względu na (\ref{g ij  we wspol przez eta i theta}) równanie (\ref{row rozn dla phi i theta}) daje:
\begin{eqnarray}
\label{row rozn 2 rzedu dla phi i theta}
\partial^{\theta_i} \partial^{\theta_j} \phi =g_{\theta}^{ij} \; , \;\;\;\; i,j=1,2,...,d \; , \;\;\;\;\; \forall\, P_{\Theta} \in {\cal S} \; ,
\end{eqnarray}
w związku z czym dodatnia określoność dualnej metryki $g_{\theta}^{ij}$   oznacza, że druga pochodna $\phi$ tworzy  dodatnio określoną macierz. Zatem $\phi$ jest {\it ściśle wypukłą} funkcją współrzędnych $\theta^{1},\theta^{2},...,\theta^{d}$, dla każdego $P \in {\cal S}$. \\ 
\\
{\bf Transformacja Legendre'a}: Powiedzmy, że $\psi$ jest pewnym rozwiązaniem równania (\ref{row rozn 2 rzedu dla psi i eta}). Wtedy po skorzystaniu z  (\ref{g ij  we wspol przez eta i theta}) oraz (\ref{row rozn dla psi i eta}), widać, że od $\psi \equiv \psi(\Xi)$ do $\phi \equiv \phi(\Theta)$ można przejść przez {\it transformację 
Legendre'a}\footnote{Sprawdźmy 
zgodność (\ref{transformacja Legendrea psi w phi}) z warunkami (\ref{row rozn dla psi i eta}) oraz (\ref{row rozn dla phi i theta}). Z (\ref{transformacja Legendrea psi w phi}) otrzymujemy:
\begin{eqnarray}
\label{d phi w theta eta i d psi}
d \phi = \sum_{i=1}^{d} \xi^{i} d\theta_{i} + \sum_{i=1}^{d} d\xi^{i} \theta_{i} - d\psi   \; , \;\;\;\;\; \forall\, P \in {\cal S} \; .
\end{eqnarray}
Korzystając z (\ref{row rozn dla psi i eta}) otrzymujemy $d \phi = \sum_{i=1}^{d} d\xi^{i} \theta_{i}$ jak w  (\ref{row rozn dla phi i theta}). 
}
:
\begin{eqnarray}
\label{transformacja Legendrea psi w phi}
\phi(\Theta) = \sum_{i=1}^{d} \xi^{i} \theta_{i} - \psi(\Xi)   \; , \;\;\;\;\;\; \forall\, P \in {\cal S} \; .
\end{eqnarray}
Podobnie, powiedzmy, że $\phi$ jest pewnym rozwiązaniem równania  (\ref{row rozn 2 rzedu dla phi i theta}). Wtedy poprzez transformację Legendre'a: 
\begin{eqnarray}
\label{transformacja Legendrea phi w psi}
\psi(\Xi) = \sum_{i=1}^{d} \xi^{i} \theta_{i} - \phi(\Theta)   \; , \;\;\;\;\;\; \forall\, P \in {\cal S} \; ,
\end{eqnarray}
można przejść od funkcji $\phi$ do $\psi$. \\
\\ 
{\bf Uwaga}: W ogólności transformacje pomiędzy układami współrzędnych  $\Xi=(\xi^{i})_{i=1}^{d}$ oraz $\Theta=(\theta_{j})_{j=1}^{d}$, które mają postać (\ref{transformacja Legendrea psi w phi}) i (\ref{transformacja Legendrea psi w phi}) nazywamy transformacjami Legendre'a\footnote{Jeśli  $\psi$ oraz $\phi$ są wypukłymi funkcjami na wypukłych przestrzeniach parametrów $V_{\Xi}$ oraz $V_{\Theta}$, gdzie parametry wektorowe mają postać  $\Xi \equiv (\xi^{i})_{i=1}^{d}$ oraz $\Theta \equiv (\theta_{i})_{i=1}^{d}$, to transformacje Legendre'a można sformułować w sposób bardziej ogólny \cite{Amari Nagaoka book}:
\begin{eqnarray}
\label{transformacja Legendrea psi w phi ogolna}
\phi(\Theta) = max_{(\Xi \in V_{\Xi})} \left\{ \sum_{i=1}^{d} \xi^{i} \theta_{i} - \psi(\Xi) \right\}   \; , \;\;\;\;\; \forall\, P \in {\cal S} \; 
\end{eqnarray}
dla $\psi(\Xi)$ będącego rozwiązaniem równania (\ref{row rozn 2 rzedu dla psi i eta}). 
Podobnie, powiedzmy, że $\phi(\Theta)$ jest pewnym rozwiązaiem równania (\ref{row rozn 2 rzedu dla phi i theta}). Wtedy poprzez transformację Legendre'a: 
\begin{eqnarray}
\label{transformacja Legendrea phi w psi ogolna}
\psi(\Xi) = max_{(\Theta \in V_{\Theta})} \left\{ \sum_{i=1}^{d} \xi^{i} \theta_{i} - \phi(\Theta) \right\}   \; , \;\;\;\;\; \forall\, P \in {\cal S} \; ,
\end{eqnarray}
otrzymujemy potencjał $\psi(\Xi)$.
}. \\
\\
{\bf Określenie potencjałów}: Funkcje $\psi$ oraz $\phi$ spełniające odpowiednio warunki  (\ref{row rozn dla psi i eta}) oraz (\ref{row rozn dla phi i theta}), pomiędzy którymi można przejść transformacją Legendre'a (\ref{transformacja Legendrea psi w phi}) lub (\ref{transformacja Legendrea phi w psi}), nazywamy {\it potencjałami} układów współrzędnych (odpowiednio $\Xi$ oraz $\Theta$). \\
\\
Poniżej podamy  twierdzenie podsumowujące powyższe rozważania. \\
 \\
{\bf Twierdzenie} {\it o dualnych układach współrzędnych}: Niech 
$\Xi=(\xi^{i})_{i=1}^{d}$ jest $\nabla$-afinicznym układem współrzędnych na dualnie płaskiej przestrzeni  $({\cal S}, g, \nabla, \nabla^{*})$. Wtedy, ze względu na metrykę $g$, istnieje dualny do $(\xi^{i})_{i=1}^{d}$ układ współrzędnych  $\Theta=(\theta_{i})_{i=1}^{d}$, który jest $\nabla^{*}$-afinicznym układem współrzędnych. Oba te układy współrzędnych są 
związane transformacją Legendre'a zadaną przy potencjałach $\psi(\Xi)$ oraz $\phi(\Theta)$ poprzez związki (\ref{transformacja Legendrea psi w phi}) lub (\ref{transformacja Legendrea phi w psi}). Ponadto współrzędne metryki w tych układach współrzędnych są zadane jako drugie pochodne potencjałów, jak w (\ref{row rozn 2 rzedu dla psi i eta})~oraz~(\ref{row rozn 2 rzedu dla phi i theta}).\\
\\
{\bf Współczynniki koneksji dla układów dualnych}: Na koniec podajmy wyprowadzone z użyciem związku (\ref{war dualnosci we wspolczynnikach}) oraz (\ref{row rozn 2 rzedu dla psi i eta}) postacie współczynników koneksji afinicznej $\Gamma^{\xi\, *}_{ij,\,l}$  \cite{Amari Nagaoka book} (por. (\ref{pochodna kowariantna i wspolczynniki koneksji})):
\begin{eqnarray}
\label{pochodna kow i wspolczynniki koneksji dualne *}
\Gamma^{\xi\, *}_{ij,\, l}:= \langle \nabla_{\partial_{\xi^i}}^{*} \partial_{\xi^j}, \, \partial_{\xi^l}\rangle = \partial_{\xi^i} \partial_{\xi^j} \partial_{\xi^l} \psi(\Xi) \; , \;\;\;\;  i,j,l = 1,2,...,d \; , \;\;\;\;\; \forall\, P_{\Xi} \in {\cal S} \; 
\end{eqnarray}
oraz z użyciem (\ref{war dualnosci we wspolczynnikach}) oraz (\ref{row rozn 2 rzedu dla phi i theta}), współczynniki koneksji afinicznej $\Gamma_{\theta}^{ij,\, l}$:
\begin{eqnarray}
\label{pochodna kow i wspolczynniki koneksji dualne}
\Gamma_{\theta}^{ij,\, l}:= \langle \nabla_{\partial^{\theta_i}} \partial^{\theta_j}, \, \partial^{\theta_l}\rangle = \partial^{\theta_i} \partial^{\theta_j} \partial^{\theta_l} \phi(\Theta) \; , \;\;\;\;  i,j,l = 1,2,...,d \; , \;\;\;\;\; \forall\, P_{\Theta} \in {\cal S} \; ,
\end{eqnarray}
przy czym, ponieważ układy współrzędnych $(\xi^{i})_{i=1}^{d}$ oraz $(\theta_{i})_{i=1}^{d}$ są afiniczne, zatem:
\begin{eqnarray}
\label{pochodna kow i wspolczynniki koneksji dualne zwykle}
\Gamma^{\xi}_{ij,\, l} = \Gamma_{\theta}^{* \, ij,\, l} = 0 , \;\;\;\;  i,j,l = 1,2,...,d \; , \;\;\;\;\; \forall\, P \in {\cal S} \; .
\end{eqnarray}

\subsection{Geometryczne sformułowanie teorii estymacji dla EFI}

\label{Geometryczne sformulowanie teorii estymacji} 

\vspace{2mm}

Dokładne sformułowanie metody ekstremalnej fizycznej informacji (EFI) jest treścią kolejnych rozdziałów skryptu. Poniżej podajemy jedynie jej wstępną charakterystykę z punktu widzenia geometrii różniczkowej na ${\cal S}$.\\
Teoria estymacji w metodzie EFI może być określona geometrycznie w sposób następujący. Załóżmy, że z pewnych powodów  teoretyczny rozkład   $P$ na ${\cal B}$ 
leży na pewnej podprzestrzeni ({\it warstwie}) ${\cal S}_{w} \subseteq {\cal S}$. Warstwa ${\cal S}_{w}$ oraz wymiar ${\cal B}$ nie są z góry określone.   Zakładamy również, że wszystkie rozważane rozkłady, łącznie z empirycznym, leżą w przestrzeni statystycznej ${\cal S}$, która (w przeciwieństwie do estymacji w statystyce  klasycznej \cite{Amari Nagaoka book,Streater}) {\it nie ma znanej postaci metryki}  Fishera-Rao. 
Na podstawie danych, które dały rozkład empiryczny $P_{Obs}$, szukamy punktu należącego do ${\cal S}_{w}$, który spełnia zasady informacyjne i jest szukanym {\it oszacowaniem} $P_{\hat{\,\Xi}}$ rozkładu teoretycznego $P$. 
\\
\\
{\bf Uwaga o estymacji w statystyce klasycznej}: W statystyce klasycznej wyznaczamy krzywą geodezyjną biegnącą przez punkt empiryczny $P_{Obs}$ i szukane oszacowanie $P_{\hat{\,\Xi}}$ stanu układu leżącego na ${\cal S}_{w}$. Ponieważ za wyjątkiem przypadku $\alpha=0$ okazuje się, że $\alpha$-koneksja nie jest metryczna (tzn. nie jest wyprowadzona jedynie z metryki, w naszym przypadku metryki Fishera-Rao), zatem odległość  $P_{Obs}$ od $P_{\hat{\,\Xi}}$ nie jest na ogół najmniejsza z możliwych \cite{Amari Nagaoka book}. 
Geodezyjna ta przecina ${\cal S}_{w}$ w pewnym punkcie należącym do ${\cal S}$, który jest szukanym {\it oszacowaniem} $P_{\hat{\,\Xi}}$ stanu układu. Jak wspomnieliśmy, w statystyce klasycznej istnieje jedno ułatwiające estymację założenie. Otóż znana jest ogólna postać modelu statystycznego ${\cal S}$, zatem znana jest i metryka Fishera-Rao na ${\cal S}$. \\
\\
{\bf Uwaga o estymacji w EFI}: W przeciwieństwie do tego estymacja w metodzie EFI nie może założyć z góry znajomości postaci metryki $g$  Fishera-Rao. Metoda EFI musi wyestymować $g$ i {\it estymacja ta jest dynamiczna}, poprzez konstrukcję odpowiednich  {\it zasad informacyjnych}. \\
{\bf Zasada wariacyjna}: Jedna z tych zasad powinna zapewnić, że po wyestymowaniu metryki Fishera-Rao, znalezione oszacowanie $P_{\hat{\,\Xi}}$ metody EFI będzie również leżeć na geodezyjnej łączącej je z $P_{Obs}$. Stąd pojawia się konieczność wprowadzenia {\it wariacyjnej zasady informacyjnej}.  \\
{\bf Zasada strukturalna}: Druga tzw. {\it strukturalna zasada informacyjna} zapewni, że szukane oszacowanie będzie leżeć w klasie rozwiązań {\it analitycznych} w parametrze $\Xi\,$, w znaczeniu równoważności metrycznej otrzymanego modelu z modelem analitycznym.  
\\
Zatem zasada wariacyjna i strukturalna wyznaczają samospójnie punkt $P_{\hat{\,\Xi}}$ i tym samym wskazują podprzestrzeń statystyczną ${\cal S}_{w}$ . Jednak koneksja afiniczna, dla której układ współrzędnych wzdłuż  krzywej geodezynej łączącej $P_{Obs}$ z $P_{\hat{\,\Xi}}$  jest płaski, nie jest $\alpha$-koneksją Amariego.   
\\
Sformułowaniem i zastosowaniem zasad informacyjnych w estymacji metodą EFI zajmiemy się w kolejnych rozdziałach.

\subsection{Uwaga o rozwinięciu rozkładu w szereg Taylora}

\label{Uwaga o rozwinieciu funkcji w szereg Taylora}

W całej treści skryptu zakładamy, że rozkład prawdopodobieństwa  $P(\Xi)$ (lub jego logarytm $\ln P(\Xi)$), jest wystarczająco gładki, tzn. posiada {\it rozwinięcie w szereg Taylora}  wystarczająco wysokiego rzędu, w każdym punkcie (pod)przestrzeni statystycznej ${\cal S}$  \cite{Dziekuje informacja_2}. Zatem albo jest spełniony warunek analityczności  rozkładu we wszystkich składowych estymowanego parametru $\Xi$,  
albo przynajmniej rozkład prawdopodobieństwa określony w otoczeniu punktu $P \in {\cal S}$ posiada dżet $J_{P}^{r}({\cal S},\text{R})$ wystarczająco wysokiego, choć skoń\-czo\-nego rzędu $r$,  np. wyrazy do drugiego (lub innego określonego, wyższego) rzędu rozwinięcia w szeregu Taylora, podczas gdy wyższe  niż $r$ rzędy rozwinięcia  znikają \cite{Murray_differential geometry and statistics}.   Do rozważań na temat rzędu dżetów powrócimy w  Rozdziale~(\ref{equations of motion}). \\
\\
{\bf Przestrzeń wektorowa dżetów}: Istotną sprawą jest fakt, że zbiór wszystkich $r$-dżetów funkcji w punkcie $P \in {\cal S}$ tworzy skończenie wymiarową przestrzeń wektorową, natomiast ich suma  $J^{r}({\cal S},\text{R}) = \bigcup_{P \in {\cal S}} J_{P}^{r}({\cal S},\text{R})$ jest {\it wiązką wektorową, czyli 
wiązką włóknistą\footnote{{\bf Określenie wiązki włóknistej}: (Różniczkowalna) wiązka włóknista $(E, \pi, {\cal M}, F, G)$ nad ${\cal M}$ składa się z następujących siedmiu elementów \cite{Nakahara}:\\
1. Różniczkowalnej rozmaitości $E$ nazywanej {\it przestrzenią totalną}. \\
2. Różniczkowalnej rozmaitości ${\cal M}$ nazywanej {\it przestrzenią bazową}. \\
3. Różniczkowalnej rozmaitości $F$ nazywanej {\it (typowym) włóknem}. \\
4. Odwzorowania suriektywnego $\pi: E \rightarrow {\cal M}$ nazywanego {\it rzutowaniem}, którego odwrotny obraz  $\pi^{-1}(p) = F_{p} \cong F$ nazywamy włóknem w $p$ gdzie $p \in {\cal M}$. \\
5. Grupy Liego $G$ nazywanej {\it grupą strukturalną}, która działa lewostronnie na $F$. \\
6. Zbioru  otwartych pokryć $\{ U_{i} \}$ rozmaitości ${\cal M}$ z dyfeomorfizmem $\phi_{i}: U_{i} \times F  \rightarrow  \pi^{-1}(U_{i})$ takim, że $\pi \circ \phi_{i}(p, \,f) = p\,$, gdzie $f \in F$. Ponieważ $\phi_{i}^{-1}$ odwzorowuje $\pi^{-1}(U_{i})$ {\it na} {\it iloczyn prosty} $\,U_{i} \times F$ dlatego odwzorowanie $\phi_{i}$ jest nazywane {\it lokalną trywializacją} .\\
7. Wprowadźmy oznaczenie $\phi_{i,p}(f) \equiv \phi_{i}(p, \, f)$. Odwzorowanie $\phi_{i,p}: F \rightarrow F_{p}$ jest dyfeomorfizmem.  Rządamy aby na $U_{i} \bigcap U_{j}\neq \emptyset$ odwzorowanie $t_{ij} \equiv \phi_{i,p}^{-1} \circ \phi_{j,p}: F \rightarrow F$ było elementem grupy $G$. Wtedy $\phi_{i}$ oraz $\phi_{j}$ są związane poprzez gładkie odwzorowanie $t_{ij}: U_{i} \bigcap U_{j} \rightarrow G$ w następujący sposób: $\phi_{j}(p, \, f) = \phi_{i}(p, \, t_{ij}(p) f)$. \\
Odwzorowania $t_{ij}$ są nazwane {\it funkcjami przejścia}. \\
{\bf Oznaczenie}: Czasami na oznaczenie wiązki włóknistej $(E, \pi, {\cal M}, F, G)$ używa się skróconego zapisu $E \stackrel{\pi}{\rightarrow} {\cal M}$ lub nawet tylko $E$. \\
\\
{\bf Przykład}: W powyższych rozważaniach przestrzenią bazową ${\cal  M}$ jest przestrzeń statystyczna ${\cal S}$. Gdy  $T_{P}$ jest  przestrzenią  styczną do  ${\cal S}$ w $P$ a $T_{P}^{*}$ jest przestrzenią  wektorową dualną do $T_{P}$, wtedy typowe włókno $F_{P}$ może być np.  przestrzenią tensorową:
\begin{eqnarray}
\label{tensor q r} \left[T_{P}\right]^{q}_{r} \equiv \underbrace{T_{P} \otimes \cdots \otimes T_{P}}_{q-razy} \otimes \underbrace{T^{*}_{P} \otimes \cdots \otimes T^{*}_{P}}_{r-razy} \; , 
\end{eqnarray}
gdzie $q$ jest indeksem  stopnia kontrawariantnego iloczynu tensorowego  $T_{P}$ a $r$ indeksem stopnia kowariantnego iloczynu tensorowego  $T_{P}^{*}$. Informacja Fishera jest szczególnym przykładem tensora $\left[T_{P}\right]^{0}_{2}$ na ${\cal S}$. Gdy typowe włókno $F$ jest przestrzenią tensorową a przestrzeń wektorowa $T_{P}$ jest $d$ - wymiarowa, to grupa strukturalna $G$ jest w ogólności ogólną  grupą liniowych transformacji $GL(d)$. 
}, 
której włókno jest przestrzenią wektorową nad przestrzenią bazową ${\cal S}$}. \\
Można pokazać, że również $J_{P}^{\infty}({\cal S},\text{R})$ jest  przestrzenią wektorową. Zatem dżety należące do $J_{P}^{\infty}({\cal S},\text{R})$  można dodawać i mnożyć przez liczbę.  Tworzą one  też algebrę co oznacza, że można je mnożyć.  Ważność przestrzeni $J_{P}^{\infty}({\cal S},\text{R})$ ujawnia się przy określeniu rozwinięcia funkcji w szereg Taylora.\\
{\bf Klasy równoważności dżetów}: O funkcjach mówimy, że są w tej samej klasie równoważności dżetów, gdy mają takie samo rozwinięcie Taylora. \\
\\
{\bf Pojęcie odwzorowania Taylora na ${\cal S}$}: Niech $T_{P}^{*}$ jest przestrzenią  wektorową dualną do przestrzeni stycznej $T_{P}$ na przestrzeni statystycznej ${\cal S}$, oraz niech $S^{k}(T^{*}_{P})$ jest przestrzenią wektorową wszystkich 
{\it  symetrycznych\footnote{Odwzorowanie 
nazywamy symetrycznym jeśli jest symetryczne ze względu na permutację zmiennych.
}
wieloliniowych odwzorowań}:
\begin{eqnarray}
\label{odwzorowanie S k}
\underbrace{T_{P} \times \cdots \times T_{P}}_{k-razy} \rightarrow \text{R} \; .
\end{eqnarray}
{\it Rozwinięcie w szereg Taylora} $T$ pewnej funkcji (np. $P(\Xi)$ lub $\ln P(\Xi)$) na ${\cal S}$ określa wtedy odwzorowanie: 
\begin{eqnarray}
\label{odwzorowanie Taylora}
T: J_{P}^{\infty}({\cal S},\text{R}) \rightarrow S(T^{*}_{P}) \;\;\; {\rm gdzie} \;\;\; S(T^{*}_{P}) \equiv \bigoplus_{k \geq 0} S^{k}(T^{*}_{P}) \; ,
\end{eqnarray}
nazywane {\it odwzorowaniem Taylora}. 
Szeregi Taylora spełniają istotną rolę w analizie statystycznej \cite{
Murray_differential geometry and statistics}, o czym przekonamy się  przy wyprowadzeniu podstawowego narzędzia estymacji metody EFI, a mianowicie strukturalnej zasady informacyjnej (por. Rozdział~\ref{structural principle}). 

\newpage

\section[Twierdzenie Rao-Cramera i DORC]{Twierdzenie Rao-Cramera i DORC}

\label{r-c}

Estymatory MNW mają asymptotycznie optymalne własności, tzn. są nieobciążone, zgodne, efektywne  i dostateczne \cite{Nowak}.  Poniższy rozdział poświęcimy efektywności nieobciążonych estymatorów parametru dla dowolnej wielkości próby $N$. \\
\\
{\bf Estymator efektywny}:  
Wartość dolnego ograniczenia na wariancję estymatora, czyli wariancję estymatora efektywnego, podaje poniższe twierdzenie Rao-Cramera. Jego sednem jest stwierdzenie, że osiągnięcie  przez estymator dolnej granicy wariancji podanej w twierdzeniu  
oznacza, że w klasie estymatorów nieobciążonych, które 
spełniają warunek regularności (tzn. mają funkcję rozkładu prawdopodobieństwa nie posiadającą punktów nieciągłości zależnych od estymowanego parametru $\Theta$), nie znajdziemy estymatora z mniejszą wariancją. \\
{\it Estymator efektywny ma więc najmniejszą z możliwych wariancji, jaką możemy uzyskać w procesie estymacji parametru. }

\subsection{Skalarne Twierdzenie Rao-Cramera}

\label{r-c-skalarne}

\textbf{Twierdzenie Rao-Cramera (TRC). Przypadek skalarny}: Niech $F(\widetilde{Y})$ będzie nieobciążonym estymatorem funkcji skalarnego parametru $g\left(\theta\right)$, tzn.: \begin{eqnarray}
\label{E funkcji par skalarnego}
E_{\theta}F\left({\widetilde{Y}}\right)=g\left(\theta\right)
\end{eqnarray} 
oraz niech $I_{F}(\theta)$ będzie informacją Fishera dla parametru $\theta$ wyznaczoną na podstawie próby $\widetilde{Y}$. Zakładając warunki regularności, otrzymujemy: 
\begin{eqnarray}
\label{tw R-C dla funkcji par skalarnego}
{\sigma^{2}}_{\theta}F\left({\widetilde{Y}}\right)\ge
\frac{{\left[{g'\left(\theta\right)}\right]^{2}}}{I_{F}\left(\theta\right)} \; ,
\end{eqnarray}
co jest tezą twierdzenia Rao-Cramera. 
W szczególnym przypadku, gdy $g(\theta)=\theta$, wtedy
z (\ref{tw R-C dla funkcji par skalarnego}) otrzymujemy następującą postać nierówności Rao-Cramera:
\begin{eqnarray}
\label{tw R-C dla par skalarnego}
{\sigma^{2}}_{\theta} F\left({\widetilde{Y}}\right) \ge \frac{{1}}{I_{F}\left(\theta\right)} \; .
\end{eqnarray}
Wielkość: 
\begin{eqnarray}
\label{dolne ogr R-C dla funkcji par skalar}
\frac{\left[{g'\left(\theta\right)}\right]^{2}}{I_{F}\left(\theta\right)} \;\;\; {\rm lub} \;\;\; \frac{1}{I_{F}\left(\theta\right)} \;\;\; {\rm dla} \;\;\; g(\theta) = \theta \;
\end{eqnarray}
nazywana jest dolnym ograniczeniem Rao-Cramera (DORC)\footnote{W jęz. angielskim {\it Cram$\acute{e}$r-Rao lower bound} (CRLB).}. Przypomnijmy, że ponieważ statystyka $F(\widetilde{Y})$ jest estymatorem parametru $\theta$, więc wartości jakie przyjmuje nie zależą od tego parametru. \\ 
\\
{\bf Uwaga}: W przypadku, gdy rozkład zmiennej $Y$ traktowany jako funkcja estymowanego parametru $\theta$ ma dla pewnych wartości tego parametru punkty nieciągłości, wtedy wariancja estymatora parametru $\theta$ występująca po lewej stronie  (\ref{tw R-C dla par skalarnego}) może okazać się mniejsza niż wartość po stronie prawej. Sytuacji nieciągłości  rozkładu w parametrze nie będziemy jednak rozważali. Przeciwnie, {\it zakładamy, że rozkład $P(\Theta)$, i jej logarytm $\ln P(\Theta)$, jest wystarczająco gładki}, tzn. posiada {\it rozwinięcie w szereg Taylora}  wystarczająco wysokiego rzędu, w każdym punkcie (pod)przestrzeni statystycznej ${\cal S}$, jak o tym wspomnieliśmy w Rozdziale~\ref{Uwaga o rozwinieciu funkcji w szereg Taylora}.

\subsubsection{Dowód TRC (wersja dowodu dla przypadku skalarnego)} 

{\it Współczynnik korelacji liniowej Pearsona} dla dwóch zmiennych
losowych $S(\widetilde{Y})$ i $F(\widetilde{Y})$ zdefiniowany jest następująco: 
\begin{eqnarray}
\label{wsp kor Piersona dla S F}
\rho_{\theta}\left({S,F}\right)=\frac{{{\mathop{\rm cov}}_{\theta} \left({S,F}\right)}}{{\sqrt{{\mathop{\sigma_{\theta}^{2}}}\left(S\right)}\sqrt{{\mathop{\sigma_{\theta}^{2}}}\left(F\right)}}} \; .
\end{eqnarray}
Z klasycznej analizy statystycznej wiemy, że  $\rho_{\theta}\left({S,F}\right)\in\left[{-1,1}\right]$, stąd z (\ref{wsp kor Piersona dla S F}) otrzymujemy:
\begin{eqnarray}
\label{nier dla F S i covSF}
{\sigma_{\theta}^{2}}\left(F\right) \ge \frac{{\left|{{\mathop{\rm cov}}_{\theta}\left({S,F}\right)}\right|^{2}}}{{{\mathop{\sigma_{\theta}^{2}}}\left(S\right)}} \; .
\end{eqnarray}
Równość występuje jeżeli współczynnik korelacji liniowej Pearsona jest równy 1, co zachodzi, gdy zmienne $S$ i $F$ są idealnie skorelowane. \\
\\
Niech teraz zmienna losowa $S$ będzie statystyką wynikową $S(\theta) \equiv S(\widetilde{Y}|\theta)$. \\
Pokażmy, że: 
\begin{eqnarray}
\label{postac g'}
g'(\theta) = {\rm cov}_{\theta} \left(S(\theta),F\right) \; . 
\end{eqnarray}
Istotnie, ponieważ $E_{\theta}\left(S(\theta)\right) = 0$, (\ref{znikanie ES}), zatem: 
\begin{eqnarray}
\label{dow covSF = g'theta}
\!\!\!\!\!\!\!\!\! & & 
{\rm cov}_{\theta} \left(S(\theta),F(\widetilde{Y})\right) = E_{\theta}\left(S(\theta) F(\widetilde{Y})\right) = \int_{\cal B}  dy  \, P(y|\theta) S(\theta) F(y) =  \int_{\cal B} dy  P(y|\theta)  \frac{\frac{\partial}{\partial \theta} P(y|\theta)}{P(y|\theta)} F(y) \nonumber \;\;\; \\ 
\!\!\!\!\!\!\!\!\! &=& \int_{\cal B} dy  \frac{\partial}{\partial \theta} P(y|\theta) F(y) 
= \frac{\partial}{\partial \theta} \int_{\cal B} dy \, P(y|\theta) F(y)  = \frac{\partial}{\partial \theta} E_{\theta}F(\widetilde{Y}) = g'(\theta) \; . \;\;\;
\end{eqnarray}
Skoro więc zgodnie z (\ref{var S oraz IF}) zachodzi, $\sigma_{\theta}^{2}S(\theta)=I_{F}(\theta)$, 
więc wstawiając (\ref{dow covSF = g'theta}) do (\ref{nier dla F S i covSF}) otrzymujemy  (\ref{tw R-C dla funkcji par skalarnego}), co kończy dowód TRC.

\subsubsection{Przykład skalarny DORC dla rozkładu normalnego}

\label{DORC dla rozkl norm}

Interesuje nas {\it przypadek  estymacji  skalarnego parametru} $\theta=\mu$ w próbie prostej $\widetilde{Y}$, przy czym zakładamy, że $g(\mu) = \mu$.  
Rozważmy średnią arytmetyczną  $\bar{Y}=\frac{1}{N}\sum_{n=1}^{N} Y_{n}$ (z realizacją $\bar{\bf y}=\frac{1}{N}\sum_{n=1}^{N} {\bf y}_{n}$), która, zakładając jedynie identyczne rozkłady zmiennych $Y_{i}$ próby, jest dla dowolnego rozkładu $p({\bf y})$ zmiennej $Y$, nieobciążonym estymatorem wartości oczekiwanej $\mu \equiv E_{\mu}(Y) = \int_{\cal Y}  d{\bf y}  p({\bf y})\, {\bf y}$, tzn.:  
\begin{eqnarray}
\label{E dla sredniej}
E_{\mu}\left(\bar{Y}\right) = \int_{\cal B}  dy \, \bar{{\bf y}} \, P(y|\mu)  = E_{\mu}(Y) = \mu \; .
\end{eqnarray}
Ponadto  dla próby prostej, z bezpośredniego rachunku otrzymujemy:  \begin{eqnarray}
\label{war dla sredniej y}
{\mathop{\sigma_{\mu}^{2}}}\left({\bar{Y}}\right) = \int_{\cal B} dy \,P(y|\mu)\,  (\bar{{\bf y}}-E(\bar{Y}))^{2}  = \frac{{\sigma^{2}}}{N} \; ,
\end{eqnarray}
gdzie $\sigma^2$ jest wariancją ${\sigma_{\mu}^{2}}(Y)$ zmiennej $Y$. \\
\\
Niech teraz zmienna pierwotna $Y$ ma rozkład normalny $N(\mu, \sigma^2)$. 
Ze związków (\ref{log wiaryg rozklad norm jeden par})  oraz (\ref{rown wiaryg skal})-(\ref{srednia arytmet z MNW}) wiemy, że średnia $\bar{Y}$ jest estymatorem MNW parametru $\mu = E(Y)$, zatem przyjmijmy $F\left(\widetilde{Y}\right) = \hat{\mu}  = \bar{Y}$. Z (\ref{war dla sredniej y}) widzimy więc, że dla zmiennych o rozkładzie normalnym zachodzi:
\begin{eqnarray}
\label{war F dla normalnego}
\sigma_{\mu}^{2}\left(F\right) =  \sigma_{\mu}^{2}\left(\bar{Y}\right) = \frac{\sigma^{2}}{N} \,\; .
\end{eqnarray}
\\
W przypadku {\it rozkładu normalnego} warunek  (\ref{tw R-C dla par skalarnego}) łatwo sprawdzić bezpośrednim rachunkiem. Istotnie, korzystając z (\ref{f wynikowa_1 wym N_1 par}) oraz (\ref{I oczekiwana dla N_parametr mu}), otrzymujemy (por. (\ref{I oczekiwana dla N_parametr mu})): 
\begin{eqnarray}
\label{sprawdzenie RC dla N}
{\mathop{\sigma_{\mu}^{2}}}S\left(\mu\right)={\mathop{\sigma_{\mu}^{2}}}\left({\frac{N}{{\sigma^{2}}}\left({\bar{Y} - \mu}\right)}\right)=\left({\frac{N}{{\sigma^{2}}}}\right)^{2}\frac{{\sigma^{2}}}{N}=\frac{N}{{\sigma^{2}}} = I_{F}(\mu) \; .
\end{eqnarray}
Z (\ref{war F dla normalnego}) oraz (\ref{sprawdzenie RC dla N}) otrzymujemy: 
\begin{eqnarray}
\label{DORC dla rozkl norm wzor}
{\sigma_{\mu}^{2}}({\hat{\mu}})  = \frac{1}{I_F(\mu)} \; ,
\end{eqnarray}
co stanowi DORC (\ref{dolne ogr R-C dla funkcji par skalar}) dla nierówności Rao-Cramera (\ref{tw R-C dla par skalarnego}). Warunek ten otrzymaliśmy już poprzednio dla rozkładu normalnego (por. (\ref{RC dla 1 N z 1 par oczekiwana IF})). Spełnienie go oznacza, że średnia arytmetyczna $\bar{Y}$ jest 
efektywnym\footnote{Tzn. 
pośród estymatorów nieobciążonych parametru $\mu = E_{\mu}(Y)$ i regularnych, posiada najmniejszą z możliwych wariancji. 
} 
estymatorem wartości oczekiwanej zmiennej losowej opisanej rozkładem normalnym $N(\mu, \sigma^2)$. \\
 \\
{\bf Uwaga o rozkładach eksponentialnych}: Rozkład normalny jest szczególnym przypadkiem szerszej klasy rozkładów, które spełniają  warunek DORC. Rozkłady te są tzw. rozkładami eksponentialnymi (\ref{exponential family}) wprowadzonymi w Rozdziale~\ref{alfa koneksja}.  Powyżej, w przypadku rozkładu normalnego, sprawdziliśmy ten fakt bezpośrednim rachunkiem zakładając wpierw typ rozkładu zmiennej $Y$, a potem sprawdzając, że estymator $\hat{\mu}$ parametru $\mu = E_{\mu}(Y)$ osiąga DORC.   \\
%

\subsection{Wieloparametrowe Twierdzenie Rao-Cramera}

\label{r-c-wieloparametrowe} 

\vspace{3mm}

Gdy dokonujemy równoczesnej estymacji $d>1$ parametrów, wtedy funkcja wynikowa $S(\Theta)$ jest $d$-wymiarowym wektorem kolumnowym (\ref{funkcja wynikowa}),  natomiast  obserwowana IF w punkcie ${\hat{\Theta}}$,  czyli $\texttt{i\!F}({\hat{\Theta}})$, oraz wartość oczekiwana z $\texttt{i\!F}(\Theta)$, czyli $I_{F}$ (por.(\ref{infoczekiwana})), są  $d\times d$ wymiarowymi macierzami. \\
\\
{\bf Analogia inflacji  wariancji}: Poniżej pokażemy, że włączenie do analizy dodatkowych parametrów ma (na ogół) wpływ na wartość IF dla interesującego  nas, wyróżnionego parametru.
Sytuacja ta jest analogiczna do problemu inflacji  wariancji estymatora parametru w analizie częstotliwościowej \cite{Kleinbaum}. Poniżej przedstawiona  zostanie odnosząca się do tego problemu wieloparametrowa wersja twierdzenia o dolnym ograniczeniu w nierówności Rao-Cramera (DORC). \\ 
  \\
{\bf Uwaga o wersjach TRC}: Poniżej podamy dwie równoważne  wersje \cite{Amari Nagaoka book} wieloparametrowego Twierdzenia Rao-Cramera (TRC). Pierwsza z nich okaże się być bardzo użyteczna przy wprowadzeniu w  Rozdziale~\ref{Pojecie kanalu informacyjnego} relacji pomiędzy tzw. informacją  Stam'a a pojemnością informacyjną układu. W Rozdziale~\ref{Estymacja w modelach fizycznych na DORC}  przekonamy się, że wersja druga TRC jest użyteczna w sprawdzeniu czy wieloparametrowa estymacja przebiega na DORC. 

\newpage

\subsubsection{Pierwsza wersja wieloparametrowego TRC}

\label{Pierwsza wersja wieloparametrowego TRC}

\textbf{Wieloparametrowe Twierdzenie RC (wersja pierwsza)}: Niech $F(\widetilde{Y})$ będzie  funkcją {\it skalarną} z wartością oczekiwaną:
\begin{eqnarray}
\label{ET dla DORC wielopar}
E_{\Theta}F\left({\widetilde{Y}}\right)=g\left(\Theta \right) \in \mathbf{R} \; 
\end{eqnarray}
oraz $I_{F}(\Theta)$ niech będzie oczekiwaną informacją
Fishera (\ref{infoczekiwana}) dla $\Theta \equiv (\theta_{i})_{i=1}^{d}$ wyznaczoną na przestrzeni próby ${\cal B}$. 
Zachodzi wtedy nierówność:
\begin{eqnarray}
\label{rc wielop}
{\sigma^{2}}_{\!\Theta} \left(F(\widetilde{Y}) \right) \ge {\bf a}^{T} I_{F}^{-1} \left(\Theta \right) {\bf a} \; ,
\end{eqnarray}
gdzie $I_{F}^{-1}$ jest macierzą odwrotną do macierzy informacyjnej  Fishera $I_{F}$, natomiast 
\begin{eqnarray}
\label{alfa} 
{\bf a} = \frac{\partial g\left(\Theta \right)}{\partial \Theta} \,
\; 
\end{eqnarray}
jest $d$-wymiarowym wektorem. \\
\\
{\bf Uwaga}: Warunek (\ref{ET dla DORC wielopar}) oznacza, że skalarna funkcja $F({\widetilde{Y}})$ jest nieobciążonym estymatorem skalarnej funkcji $g\left(\Theta \right)$ wektorowego parametru $\Theta$. 

\subsubsection{Przykład wektorowego DORC}  
\label{Przyklad wektorowego DORC}

Jako ilustrację powyższego wieloparametrowego Twierdzenia RC przedstawimy przykład, przyjmując szczególną postać skalarnej funkcji $g(\Theta)$, o której zakładamy, że jest liniową funkcją składowych $\theta_{i}$ wektora parametrów \cite{Pawitan}:
\begin{eqnarray}
\label{uwad}
g\left(\Theta \right) = {\bf a}^{T} \Theta = \sum_{i=1}^{d} a_{i} \, \theta_{i} \; ,
\end{eqnarray}
gdzie ${\bf a}$ jest pewnym znanym wektorem o stałych składowych $a_{i}$, które  nie zależą od składowych wektora $\Theta$. 
Załóżmy chwilowo, że ${\bf a}^{T}=\left({1,0,\ldots,0}\right)$, tzn. jedynie $a_{1} \neq 0$. Wtedy  z (\ref{uwad}) otrzymujemy $g\left(\Theta \right)=\theta_{1}$, natomiast  (\ref{rc wielop}) w Twierdzeniu RC, ${\sigma^{2}}_{\! \Theta} \left(F \right) \ge {\bf a}^{T} I_{F}^{-1} \left(\Theta \right) {\bf a}$, przyjmuje dla rozważanego nieobciążonego estymatora $F$ parametru $\theta_{1}$, tzn. $E_{\Theta}(F({\widetilde{Y}}))  = \theta_{1}$, postać: 
\begin{eqnarray}
\label{RG dla a1}
{\sigma_{\Theta}^{2}}\left(F\right) \ge \left[{ I_{F}^{-1}\left( \Theta \right)}\right]_{11} =:  I_{F}^{11}\left(\Theta \right) \; ,
\end{eqnarray}
gdzie $I_{F}^{11}{(\Theta)}$ oznacza element (1,1) macierzy  $I_{F}^{-1}\left(\Theta\right)$. Prawa strona nierówności (\ref{RG dla a1}) podaje dolne ograniczenie wariancji estymatora $F$, pod warunkiem, że  $\theta_{1}$ jest wyróżnionym parametrem a wartości pozostałych parametrów {\it nie są znane}. 
Oznaczmy wewnętrzną strukturę $d \times d\,$-wymiarowych macierzy $ I_{F}(\Theta)$ oraz  $I_{F}(\Theta)^{-1}$ następująco:
\begin{eqnarray}
I_{F} \left(\Theta \right) = \left({\begin{array}{cc}
{I_{F 11}} & {I_{F 12}}\\
{I_{F 21}} & {I_{F 22}}\end{array}}\right)\end{eqnarray}
oraz \begin{eqnarray}
I_{F}^{-1}\left(\Theta \right)=\left({\begin{array}{cc}
{I_{F}^{11}} & {I_{F}^{12}}\\
{I_{F}^{21}} & {I_{F}^{22}}\end{array}}\right)\end{eqnarray}
gdzie $I_{F 11}$ oraz  $I_{F}^{11} = \left[{ I_{F}^{-1}\left( \Theta \right)}\right]_{11}$ (zgodnie z oznaczeniem wprowadzonym w (\ref{RG dla a1})) są liczbami,  $I_{F 22}$, $I_{F}^{22}$ są $(d-1) \times (d-1)$-wymiarowymi macierzami, natomiast $(I_{F 12})_{1 \times (d-1)}$, $(I_{F 21})_{(d-1) \times 1}$, $(I_{F}^{12})_{1 \times (d-1)}$, $(I_{F}^{21})_{(d-1) \times 1}$ odpowiednimi wierszowymi bądź kolumnowymi wektorami~o~wymiarze~$(d-1)$. \\
\\
%
{\bf Rozważmy parametr $\theta_{1}$}. Jego informacja Fishera (patrz poniżej {\bf Uwaga o nazwie}) jest równa $I_{F 11}=I_{F 11}\left(\theta_{1} \right)$. {\it Nie oznacza to jednak}, że $\sigma_{\Theta}^{2}\left(F\right)$ oraz $I_{F 11}$ są z sobą automatycznie powiązane  nierównością $\sigma_{\Theta}^{2}\left(F\right) \ge 1/I_{F 11}$, która jest treścią Twierdzenia RC (\ref{tw R-C dla par skalarnego}). Udowodniliśmy ją bowiem tylko dla  przypadku parametru skalarnego,  tzn. gdy tylko jeden parametr jest estymowany, a reszta parametrów jest znana. \\
Określmy relację pomiędzy $(I_{F 11})^{-1}$ oraz $I_{F}^{11}$. 
Oczywiście zachodzi:
\begin{eqnarray}
I_{F}\left( \Theta \right) \cdot I_{F}^{-1} \left( \Theta \right) = \left({\begin{array}{cc}
{I_{F 11}} & {I_{F 12}}\\
{I_{F 21}} & {I_{F 22}}\end{array}}\right)\left({\begin{array}{cc}
{I_{F}^{11}} & {I_{F}^{12}}\\
{I_{F}^{21}} & {I_{F}^{22}}\end{array}}\right)=\left({\begin{array}{cc}
1 & 0 \\
0 & \textbf{1} \end{array}}\right) \; , 
\end{eqnarray}
gdzie $\textbf{1}$ jest $(d-1) \times (d-1)$-wymiarową macierzą jednostkową. \\
Z powyższego mamy: 
\begin{eqnarray}
&(I_{F 11})_{1\times 1}&  \!\!\!\! (I_{F}^{11})_{1\times 1} + (I_{F 12})_{1\times (d-1)} (I_{F}^{21})_{(d-1) \times 1} = 1 \nonumber \\ &\Rightarrow& \;\;\;\; \left( {I_{F}^{11}}\right)^{-1} = I_{F 11} + I_{F 12} I_{F}^{21} \left({I_{F}^{11}} \right)^{-1} \; ,
\end{eqnarray}
\begin{eqnarray}
&(I_{F 21})_{(d-1) \times 1}& \!\!\!\! (I_{F}^{11})_{1\times 1} + (I_{F 22})_{(d-1)\times(d-1)} \; (I_{F}^{21})_{(d-1) \times 1} = (0)_{(d-1)\times 1}  \nonumber \\
&\Rightarrow& \;\;\;\;  I_{F}^{21} = -\left({I_{F 22}}\right)^{-1} I_{F 21} I_{F}^{11}
\end{eqnarray}
skąd otrzymujemy:
\begin{eqnarray}
\label{rownanie macierzowa dla I}
\left({I_{F}^{11}}\right)^{-1} = 
I_{F 11}-I_{F 12}\left({I_{F 22}}\right)^{-1}I_{F 21} \; .
\end{eqnarray}
Ponieważ $I_{F 22}$ jest macierzą informacyjną (dla parametrów $\theta_{2},\theta_{3},...,\theta_{d}$), jest więc ona zgodnie z rozważaniami przedstawionymi poniżej (\ref{iF polokreslona}),  symetryczna i nieujemnie określona. Symetryczna i nieujemnie określona jest zatem $(I_{F 22})^{-1}$. Ponieważ z symetrii macierzy $I_{F}$ wynika $I_{F 12}=(I_{F 21})^{T}$, zatem ostatecznie forma kwadratowa 
$I_{F 12} \left({I_{F 22}} \right)^{-1} I_{F 21} \geq 0$, stąd z (\ref{rownanie macierzowa dla I}) otrzymujemy:
\begin{eqnarray}
\label{porownanie I11 z I11do-1}
(I^{F 11})^{-1} \le I_{F 11} \;\;  \Rightarrow \;\;  I^{F 11} \geq \frac{1}{I_{F 11}} \; ,
\end{eqnarray}
co zgodnie z (\ref{RG dla a1}) oznacza, że:
\begin{eqnarray}
\label{porownanie sigma I11 z I11do-1}
\sigma_{\Theta}^{2}\left(F\right) \geq   I_{F}^{11} \geq \frac{1}{ I_{F 11}} \; .
\end{eqnarray}
{\bf Wniosek}: Zatem widzimy, że $I_{F}^{11}$ daje {\it silniejsze ograniczenie}  niż $(I_{F 11})^{-1}$. Tzn. w przypadku estymacji wieloparametrowej należy zastosować związek $\sigma^{2}\left(F\right) \geq   I_{F}^{11}$, (\ref{RG dla a1}), gdyż to właśnie on jest właściwy na podstawie wieloparametrowego Twierdzenia RC. Zastosowanie $\sigma_{\Theta}^{2}\left(F\right) \geq  1/I_{F 11}$, tak jak byśmy mieli do czynienia z przypadkiem skalarnym, może błędnie zaniżyć wartość dolnego ograniczenia na $\sigma_{\Theta}^{2}\left(F\right)$. \\
  \\
{\bf Uwaga o  nazwie $I_{F 11}$}: W ``statycznie'' ukierunkowanej analizie statystycznej wielkość  $(I_{F}^{11})^{-1}$ jest 
interpretowana jako informacja Fishera dla $\theta_{1}$ -- {\bf jednak w  treści skryptu odstąpimy od tej nazwy}. Okazuje się, że w  analizie ukierunkowanej na estymację ``dynamiczną'', tzn. generującą równania różniczkowe dla rozkładów, bardziej użyteczne jest nazwać  $I_{F}^{11}$ po prostu {\it dolnym  ograniczeniem RC na wariancję estymatora  parametru} $\theta_{1}$ w sytuacji gdy pozostałe parametry są nieznane (tzn. trzeba je estymować z próby  równocześnie z $\theta_{1}$). {\it Natomiast $I_{F 11}$ będziemy nazywali, zgodnie z tym jak to uczyniliśmy, informacją Fishera parametru $\theta_{1}$} i to niezależnie od tego czy inne parametry są równocześnie estymowane, czy też nie. \\
  \\
{\bf Podsumowanie na temat zaniżenia DORC}: Należy pamiętać, że estymując parametr $\theta_{1}$ należy być świadomym faktu występowania równoczesnej estymacji innych parametrów, gdyż wstawienie wartości $I_{F 11}$ do nierówności RC może w przypadku estymacji wieloparametrowej doprowadzić do zaniżenia wartości dolnego ograniczenia wariancji tego parametru. \\
  \\
{\bf Przypadek ``pseudo-skalarny''}: Istnieje jednak pewien wyjątek spowodowany dokładnym zerowaniem się  $I_{F 12}$ dla dowolnego $N$. Wtedy  z  (\ref{rownanie macierzowa dla I}) wynika, że wzrost wariancji estymatora parametru związany z dodaniem nowych parametrów o nieznanych wartościach byłby równy zeru. Tak też  było w rozważanym wcześniej  przykładzie   rozkładu normalnego $N\left({\mu,\sigma}\right)$ (porównaj (\ref{oczekiw iF r normalnego 2 par}) z (\ref{I oczekiwana dla N_parametr mu})).\\ 
Podobnie, taki szczególny przypadek zachodzi, 
gdy wieloparametrowym  rozkładem prawdopodobieństwa jest  wiarygodność $N$-wy\-mia\-rowej próby $P(\theta_1,\theta_2,,...,\theta_{N}) = \prod_{n=1}^{N} p_{\theta_{n}}$, gdzie każdy estymowany parametr $\theta_n$ określa tylko jeden  punktowy rozkład $p_{\theta_{n}}$. Wtedy macierz informacyjna Fishera $I_{F}$ jest diagonalna i zachodzi  $I_{F nn}=(I_{F}^{nn})^{-1}$, a  w miejsce (\ref{porownanie sigma I11 z I11do-1}) otrzymujemy dla każdego parametru $\theta_{n}$:
\begin{eqnarray}
\label{I11 rowna sie I11do-1 przypadek diagonalny}
\sigma_{\Theta}^{2}\left(F_{n}\right) \geq   I_{F}^{nn} = \frac{1}{ I_{F nn}} \; , \;\;\; n=1,2,...,N \;, \;\;\;\; {\rm gdy} \;\;\; I_{F} \; \; {\rm  diagonalne}  \; ,
\end{eqnarray}
jako szczególny przypadek nierówności RC, gdzie $F_{n}$ jest estymatorem $\hat{\theta}_{n}$ parametru $\theta_{n}$.



\subsubsection{Druga wersja wieloparametrowego TRC}

\label{Druga wersja wieloparametrowego TRC}

Niech $\hat{\Theta} \equiv \hat{\Theta}(\widetilde{Y}) = (\hat{\theta}_{i}(\widetilde{Y}))_{i=1}^{d}$ jest nieobciążonym estymatorem parametru  $\Theta~=~(\theta_{i})_{i=1}^{d}$, co oznacza, że zachodzi:
\begin{eqnarray}
\label{wektorowy nieobciazany estymator}
\Theta = E_{\Theta}\left[ \hat{\Theta}(\widetilde{Y}) \right] \; , \;\;\;\;\; \forall \, P_{\Theta} \in {\cal S} \; .
\end{eqnarray}
{\bf (Oczekiwaną) macierzą kowariancji}\footnote{Czyli tzw. macierzą oczekiwanego błędu kwadratowego.
}
$V_{\Theta}[ \hat{\Theta} ]$  nieobciążonego estymatora $\hat{\Theta}$ w bazie $\Theta$ nazywamy $d \times d$ - wymiarową macierz o elementach: 
\begin{eqnarray}
\label{oczekiwana macierz kowariancji}
V_{\Theta \, ij} [ \hat{\Theta} ]  := E_{\Theta}\left[ (\hat{\theta}_{i}(\widetilde{Y}) - \theta_{i})\; (\hat{\theta}_{j}(\widetilde{Y}) - \theta_{j}) \right] \; , \;\;\;\;\; i,j = 1,2,...,d, \;\;\;\;\; \forall \, P_{\Theta} \in {\cal S} \; .
\end{eqnarray}
Zachodzi następujące twierdzenie.\\
 \\
{\bf Wieloparametrowe Twierdzenie RC (wersja druga)}: Macierz kowariancji $V_{\Theta}[ \hat{\Theta} ]$ nieobciążonego estymatora $\hat{\Theta}$ spełnia nierówność \cite{Amari Nagaoka book}:
\begin{eqnarray}
\label{twierdzenie RC wersja 2}
V_{\Theta}[ \hat{\Theta} ] \geq I_{F}^{-1}(\Theta) \;\;\;\;\;\;\;\;\;\; \forall \, P_{\Theta} \in {\cal S} \; ,
\end{eqnarray}
co oznacza, że macierz $V_{\Theta}[ \hat{\Theta} ] - I_{F}^{-1}(\Theta)$ jest dodatnio półokreślona. \\
\\
{\bf Estymator efektywny}: Nieobciążony estymator $\hat{\Theta}$ spełniający równość w nierówności (\ref{twierdzenie RC wersja 2}):
\begin{eqnarray}
\label{DORC wersja 2 TRC}
V_{\Theta}[ \hat{\Theta} ] = I_{F}^{-1}(\Theta)  \;\;\;\;\;\;\;\;\;\; \forall \, P_{\Theta} \in {\cal S} \; .
\end{eqnarray}
nazywamy estymatorem {\it efektywnym} parametru $\Theta$. 

\newpage

\section[Entropia informacyjna Shannona i entropia względna]{Entropia informacyjna Shannona i entropia względna}

\label{shan} 

W rozdziale tym omówimy pojęcie, które podaje {\it globalną charakterystykę} pojedynczego rozkładu prawdopodobieństwa, tzn. entropię Shannona. Dokładniejsze omówienie własności entropii Shannona można znaleźć w \cite{Bengtsson_Zyczkowski}. Z treści poniższych przykładów wynika jaki jest rozmiar próby $N$. \\
\\
{\bf Entropia Shannona}: Niech $P(\omega)$ będzie rozkładem prawdopodobieństwa\footnote{{\bf Przestrzeń 
stanów modelu}: Mówimy, że na przestrzeni zdarzeń  $\Omega$ została określona funkcja $\omega \rightarrow P(\omega)$ spełniająca warunki,  
$P(\omega) \ge 0$  oraz $ \sum \limits_{\omega}P(\omega) = 1$,
nazywana wtedy miarą probabilistyczną.  {\it Zbiór wszystkich miar probabilistycznych określonych na $\Omega$ tworzy  przestrzeń  stanów modelu}.
} 
określonym na przestrzeni zdarzeń $\Omega$, gdzie $\omega$ jest puntem w $\Omega$.  
Jeśli przestrzeń zdarzeń $\Omega$ jest dyskretna, to {\it informacyjna entropia Shannona} jest zdefiniowana następująco:
\begin{eqnarray}
\label{shannon dla omega}
S_{H}\left(P\right) = - k \sum\limits_{\omega \in \,\Omega} {P (\omega) \ln P (\omega)} \; ,
\end{eqnarray}
gdzie $k$ jest liczbą 
dodatnią\footnote{W przypadku  
statystycznej entropii fizycznej $\aleph$ może być np. {\it liczbą konfiguracji określonej liczy molekuł przy zadanej energii całkowitej układu}. Wtedy $k$ jest utożsamiane ze stałą Boltzmann'a $k_{B}$. Dla układu określonego w przestrzeni ciągłej $\mathbb{R}^{3}$ liczba konfiguracji  jest nieskończona. Gdyby ograniczyć się do skończonej podprzestrzeni i podzielić ją na  komórki o skończonej wielkości, i podobnie uczynić w przestrzeni pędowej, to liczba możliwych konfiguracji układu byłaby skończona a jego entropia mogłaby być policzona. Jednakże poprawny rachunek entropii wymaga wtedy utożsamienia konfiguracji powiedzmy $n$ cząstek różniących się jedynie ich permutacją, w ramach jednej klasy równoważności.  Na fakt, że właściwa przestrzeń próby ma w tym przypadku nie $\aleph$ lecz $\aleph/n!$ punktów zwrócił uwagę Gibbs, a otrzymaną przestrzeń próby nazywa się przestrzenią próby Gibbsa. Np. dla 1 $cm^{3}$ cieczy w zwykłych warunkach $n \approx 10^{23}$. Problem ten nie będzie rozważany dalej w niniejszym skrypcie.}, 
którą przyjmiemy dalej jako równą 1. \\
\\
{\bf Rozkład prawdopodobieństwa jako element sympleksu rozkładów}:  Niech $\Omega$ jest rozpięta przez skończoną liczbę $\aleph$ elementów będących  możliwymi wynikami  doświadczenia, w rezultacie  którego otrzymujemy wartości ${\bf y}_{i}$, $i=1,2,...,\aleph$, zmiennej losowej $Y$. Rozkład prawdopodobieństwa $P$ jest wtedy reprezentowany przez wektor $\vec{p} = (p_{i})_{i=1}^{\aleph}\,$,  należący do
{\it sympleksu rozkładów prawdopodobieństwa} \cite{Bengtsson_Zyczkowski}, {\it tzn. jego  $\aleph$ składowych spełnia warunki}:
\begin{eqnarray}
\label{rozklad prawdopod}
p_{i} \ge 0 \;\;\;\;  {\rm oraz} \;\;\;\;  \sum \limits_{i=1}^{\aleph} p_{i} = 1 \; .
\end{eqnarray}
{\it m-Sympleks rozkładów prawdopodobieństwa, czyli zbiór}  $\Delta^{m}=\{\left(p_{1},p_{2},...,p_{m},p_{m+1} \right)$
$\in \mathbf{R}^{m+1} ;$ $p_{j} \ge 0, \; j=1,...,m+1$ gdzie $m+1 \leq \aleph$, dla  $\sum_{j=1}^{m+1}{p_{j}} \le  1 \}$ {\it jest zbiorem wypukłym}, tzn. każdy punkt $\vec{p}_{m} \in \Delta^{m}$ można przedstawić jako $\vec{p}_{m} = \sum_{i=1}^{m+1} \lambda_{i} p_{i}$, gdzie $\sum_{i=1}^{m+1} \lambda_{i} = 1 \,$.\\
\\
Entropia Shannona (\ref{shannon dla omega}) rozkładu (\ref{rozklad prawdopod}) ma postać: 
\begin{eqnarray}
\label{shannon}
S_{H}\left(P\right) = - \sum\limits_{i=1}^{\aleph}{p_{i}\ln p_{i}} \; .
\end{eqnarray}
Zapis $S_{H}(P)$, gdzie w argumencie pominięto oznaczenie zmiennej losowej $Y$ podkreśla, że {\it jedyną rozważaną  przez nas cechą zmiennej losowej jest jej rozkład prawdopodobieństwa} $P$. 

\subsection{Interpretacja entropii Shannona}

Maksymalna możliwa wartość entropii 
Shannona\footnote{Ze 
względu na warunek $\sum\limits_{i=1}^{\aleph} p_{i} = 1$ maksymalizujemy funkcję $S_{H\,war}\left(P\right) = S_{H}\left(P\right) - \lambda \left( 1 - \sum\limits_{i=1}^{\aleph}  p_{i} \right) \, $, licząc pochodną po $p_{j}$, $j=1,2,...,\aleph\,$, gdzie $\lambda$ jest czynnikiem Lagrange'a.
} 
wynosi $\ln \aleph$ i jest osiągnięta, gdy wszystkie wyniki są równo prawdopodobne
($p_{i}=1/\aleph$), tzn. gdy stan układu jest maksymalnie zmieszany.
\\ 
\\
{\bf Układ w stanie czystym}: Gdy jeden z wyników jest pewny, wtedy tylko  jedna, odpowiadająca mu  współrzędna wektora $\vec{p}$ jest równa jeden, a pozostałe są równe 0. Mówimy wtedy, że układ znajduje się w stanie czystym, a  odpowiadająca mu  wartość entropii Shannona jest minimalna i równa zero. \\
\\
{\bf Entropia jako miara niepewności wyniku eksperymentu}: Z powyższych przykładów można wnioskować, że entropię Shannona można interpretować jako {\it miarę niepewności otrzymania wyniku eksperymentu}  będącego realizacją 
rozkładu prawdopodobieństwa $P$ lub inaczej, 
jako wielkość informacji koniecznej {\it do określenia} wyniku,
który może się pojawić w rezultacie przeprowadzenia eksperymentu na układzie. 
\\
Podstawowe własności entropii Shanonna można znaleźć w \cite{Bengtsson_Zyczkowski}. \\

\subsection{Przypadek ciągłego rozkładu prawdopodobieństwa}

\label{Przypadek ciąglego rozkladu prawdopodobienstwa}

\vspace{4mm}

Przejście z dyskretnego do ciągłego rozkładu
prawdopodobieństwa  polega na zastąpieniu  sumowania w (\ref{shannon}) całkowaniem po całym zakresie zmienności zmiennej losowej $Y$. W ten sposób otrzymujemy Boltzmanowską postać entropii Shannona:
\begin{eqnarray}
\label{eboltmana}
S_{H}(P) = - \int\limits_{-\infty}^{+\infty}{d{\bf y}\, P\left({\bf y}\right) \ln P\left({\bf y}\right)} \; .
\end{eqnarray}
Jednakże dla pewnych funkcji rozkładu $P({\bf y})$, całka (\ref{eboltmana}) może nie być określona. \\
\\
{\bf Czysty stan klasyczny}: Jako ilustrację powyższego stwierdzenia rozważmy sytuację \cite{Bengtsson_Zyczkowski}, gdy rozkład $P({\bf y})$ przyjmuje w przedziale $[0,t]$ wartość
$t^{-1}$ i zero wszędzie poza tym przedziałem. Wtedy jego entropia $S_{H}(P)$ jest równa $\ln t$ i dla $t \rightarrow 0$ dąży do {\it minus nieskończoności}. 
Procedura ta odpowiada przejściu do punktowego, {\it czystego 
stanu klasycznego} opisanego dystrybucją delta Diraca, dla której $S_{H}(P) = - \infty$. \\
Zatem przyjmując poziom zerowy entropii jako punkt odniesienia, dokładne określenie stanu opisanego deltą Diraca (której fizycznie mógłby odpowiadać nieskończony skok w gęstości rozkładu substancji cząstki) wymaga dostarczenia nieskończonej ilości informacji o układzie. Do sprawy powrócimy w jednym z kolejnych rozdziałów. \\
\\
{\bf Problem transformacyjny $S_{H}(P)$}:  Definicja (\ref{eboltmana}) ma pewien formalny minus, związany z  brakiem  porządnych własności transformacyjnych entropii Shannona. Omówimy go poniżej. \\
\\
Ze względu na unormowanie prawdopodobieństwa do jedności, gęstość  rozkładu prawdopodobieństwa przekształca się przy transformacji  układu współrzędnych tak jak odwrotność objętości. \\
\\
\\
{\bf Przykład zmiennej jednowymiarowej}: Z unormowania   $\int_{-\infty}^{+\infty} d{\bf y}\, P\left( {\bf y} \right) = 1$ wynika, że $P({\bf y})$ musi transformować się tak, jak $1/{\bf y}$.  
\\
Rozważmy transformację układu współrzędnych ${\bf y} \rightarrow {\bf y}'$. 
Różniczka zmiennej $Y$ transformuje się wtedy zgodnie z $\,d{\bf y}' = J(\frac{{\bf y}}{{\bf y}'}) d{\bf y}$,  gdzie $J$ jest jaco\-bia\-nem transformacji, natomiast rozkład prawdopodobieństwa transformuje się następująco:  $\,P'({\bf y}') = \,$  $J^{-1}(\frac{{\bf y}}{{\bf y}'})P({\bf y})$.  
Zatem tak jak to powinno być, unormowanie rozkładu w transformacji pozostaje niezmiennicze, tzn.:
\begin{eqnarray}
\label{niezmienniczosc unormowania}
\int\limits_{-\infty}^{+\infty} d{\bf y}\, P\left({\bf y}\right) =  \int\limits_{-\infty}^{+\infty} d{\bf y}'\, P'\left({\bf y}'\right) = 1 \; . 
\end{eqnarray}
Rozważmy teraz entropię Shannona układu określoną dla rozkładu ciągłego jak w (\ref{eboltmana}), $S_{H}(P)$ $=-\int{d{\bf y}\, P\left({\bf y}\right) \ln P\left({\bf y}\right)}$. Jak to zauważyliśmy powyżej, entropia układu jest miarą nieuporządkowania w układzie, 
bądź informacji potrzebnej do określenia wyniku eksperymentu. 
Zatem również i ona powinna być niezmiennicza przy rozważanej transformacji. Niestety, chociaż miara probabilistyczna pozostaje niezmiennicza, to ponieważ $\ln P\left({\bf y}\right) \neq \ln P'\left({\bf y}'\right)$ zatem  $S_{H}(P) \neq S_{H}(P')$. \\
\\
Tak więc {\it logarytm z gęstości rozkładu prawdopodobieństwa nie jest niezmienniczy  przy transformacji układu współrzędnych} i w konsekwencji otrzymujemy następujący wniosek, słuszny również w przypadku rozkładu prawdopodobieństwa wielowymiarowej zmiennej losowej $Y$.
\\
\\
{\bf Wniosek}: Entropia Shannona (\ref{eboltmana}) nie jest niezmiennicza przy transformacji układu współrzędnych przestrzeni bazowej ${\cal B}$. \\
\\
Z drugiej strony,  ze względu na wyjątkowe pośród innych entropii własności entropii Shannona dla rozkładu dyskretnego \cite{Bengtsson_Zyczkowski}, zrezygnowanie z jej ciągłej granicy  (\ref{eboltmana}) mogłoby się okazać decyzją chybioną. Również  jej związek z informacją Fishera omówiony dalej, przekonuje o istotności pojęcia entropii Shannona w jej formie ciągłej.  \\
{\bf Entropia względna jako rozwiązanie problemu transformacji}: Proste rozwiązanie zaistniałego problemu polega na zaobserwowaniu, że ponieważ iloraz dwóch gęstości $P({\bf y})$ oraz $P_{ref}({\bf y})$ transformuje się jak skalar, tzn.: 
\begin{eqnarray}
\label{iloraz rozkladow w transformacji}
\frac{P'({\bf y}')}{P_{ref}'({\bf y}')} = \frac{J^{-1}(\frac{{\bf y}}{{\bf y}'}) P({\bf y})}{J^{-1}(\frac{{\bf y}}{{\bf y}'}) P_{ref}({\bf y})}  = \frac{P({\bf y})}{P_{ref}({\bf y})} \; , 
\end{eqnarray}
gdzie $P_{ref}({\bf y})$ występuje jako pewien rozkład referencyjny, zatem wielkość nazywana {\it entropią względ\-ną}: 
\begin{eqnarray}
\label{entropia wzgledna roz ciagle}
S_{H}(P|P_{ref}) \equiv  \int\limits _{-\infty}^{+\infty}{d{\bf y}\, P\left({\bf y}\right) \ln \frac{{P\left({\bf y}\right)}}{{P_{ref}\left({\bf y}\right)}}} \; ,
\end{eqnarray}
{\it posiada już własność niezmienniczości}: 
\begin{eqnarray}
\label{entropia wzgledna roz ciagle  niezm}
S_{H}(P'|P'_{ref})  = S_{H}(P|P_{ref}) \; 
\end{eqnarray}
{\it przy transformacji układu współrzędnych}.  \\
\\
\\
{\bf Entropia względna jako wartość oczekiwana}: Zwrócmy uwagę, że entropia względna jest wartością oczekiwaną  logarytmu dwóch rozkładów, $P({\bf y})$ oraz $P_{ref}({\bf y})$,  zmiennej $Y$:
\begin{eqnarray}
\label{entropia wzgledna jako wartosc oczekiwana ilorazu}
S_{H}(P|P_{ref}) = E_{P}\left( \ln \frac{{P\left(Y\right)}}{{P_{ref}\left(Y\right)}} \right)\; ,
\end{eqnarray}
wyznaczoną przy założeniu, że zmienna losowa $Y$ ma rozkład $P$. \\
\\
{\bf Entropia względna a analiza doboru modelu}: W ten sposób problem logarytmu ilorazu funkcji wiarygodności (czy w szczególności dewiancji) wykorzystywanego w analizie braku dopasowania modelu (por. Rozdziały~\ref{Wnioskowanie w MNW}-\ref{regresja klasyczna}),  powrócił w postaci konieczności wprowadzenia entropii względnej. Istotnie, wiemy, że pojęcie logarytmu ilorazu rozkładów okazało się już   użyteczne w porównywaniu modeli  statystycznych i wyborze  modelu bardziej ``wiarygodnego''. Wybór modelu powinien być   niezmienniczy ze względu na transformację układu współrzędnych przestrzeni bazowej ${\cal B}$. Logarytm ilorazu rozkładów posiada żądaną własność. Jego wartością oczekiwaną, która jest entropią względną, zajmiemy się w Rozdziale~(\ref{wzg}). Przez wzgląd na związek IF z entropią względną dla rozkładów różniących się infinitezymalnie mało, jej  pojęcie  będzie nam towarzyszyło do końca skryptu. \\
\\
{\bf Nazwy entropii względnej}: Entropię względną (\ref{entropia wzgledna roz ciagle}) nazywana się również entropią Kullbacka-Leiblera (KL) lub  dywergencją informacji.

\subsection{Entropia względna jako miara odległości}

\label{wzg} 

Rozważmy  eksperyment, którego wyniki są generowane z pewnego określonego, chociaż nieznanego rozkładu  prawdopodobieństwa $P_{ref}$ należącego do obszaru ${\cal O}$. Obszar ${\cal O}$ jest nie posiadającym izolowanych punktów zbiorem  rozkładów prawdopodobieństwa, który  jest  przestrzenią  metryczną zupełną. Oznacza to, że na ${\cal O}$ można określić odległość oraz każdy ciąg Cauchy'ego ma granicę należącą do ${\cal O}$. \\ 
\\
{\bf Twierdzenie Sanova} \cite{Sanov}:
Jeśli mamy $N$ - wymiarową próbkę niezależnych pomiarów pochodzących z rozkładu  prawdopodobieństwa $P_{ref}$ pewnej zmiennej losowej, to prawdopodobieństwo $Pr$, że empiryczny rozkład (częstości) $\hat{{\cal P}}$ wpadnie w obszar ${\cal O}$, spełnia asymptotycznie związek: 
\begin{eqnarray}
\label{Sanov dokladnie}
\lim_{N \rightarrow \infty } \frac{1}{N} \ln Pr \left\{ \hat{{\cal P}} \in {\cal O} \right\}  = - \beta \;\;\; {\rm gdzie} \;\;\;\; \beta = \inf_{P_{\cal O} \in \, {\cal O}} S_{H} \left( P_{\cal O}|P_{ref} \right) \; , 
\end{eqnarray}
który w przybliżeniu można zapisać następująco:
\begin{eqnarray}
\label{Sanov}
Pr(\hat{{\cal P}} \in {\cal O}) \sim e^{- N S_{H} \left( {P_{\cal O}|P_{ref}} \right)} \; ,
\end{eqnarray}
gdzie $P_{\cal O}$ jest rozkładem należącym do ${\cal O}$ różnym od 
$P_{ref}$ z najmniejszą wartością entropii względnej  $S_{H} \left( P_{\cal O}|P_{ref} \right)$. 
Rozkład $P_{\cal O}$ uznajemy za rozkład wyestymowany na podstawie empirycznego rozkładu  częstości $\hat{{\cal P}}$ otrzymanego w obserwacji. \\
\\
{\bf Wniosek}: Zauważmy, że Twierdzenie Sanova jest rodzajem  prawa wielkich liczb, zgodnie z którym dla wielkości próby $N$ dążącej do nieskończoności, {\it prawdopodobieństwo zaobserwowania rozkładu częstości $\hat{{\cal P}}$ należacego do ${\cal O}$  różnego od prawdziwego  rozkładu} $P_{ref}$ (tzn. tego który generował wyniki eksperymentu), {\it dąży do zera}. \\
Fakt ten wyraża właśnie relacja (\ref{Sanov}), a ponieważ $S_{H} \left(P_{\cal O}|P_{ref} \right)$ jest w jej wykładniku,  zatem {\it entropia względna określa tempo w jakim prawdopodobieństwo $Pr(\hat{{\cal P}})$ dąży do zera wraz ze wzrostem~$N$}. \\  
{\bf Entropia względna dla rozkładów dyskretnych}: Jeśli $P_{ref} \equiv (p^{i})_{i=1}^{\aleph}$ jest dyskretnym  rozkładem prawdopodobieństwa o $\aleph$ możliwych wynikach, wtedy rozkład $P_{\cal O} \equiv (p^{i}_{\cal O})_{i=1}^{\aleph} \in {\cal O}$ jest też dyskretnym rozkładem o $\aleph$ wynikach, a  entropia względna  $S_{H} \left( P_{\cal O}|P_{ref} \right)$ ma postać:
\begin{eqnarray}
\label{entropia wzgl dyskret}
S_{H} \left( {P_{\cal O}|P_{ref}} \right) =  \sum\limits_{i=1}^{\aleph}{p_{\cal O}^{i}\ln\frac{{p_{\cal O}^{\, i}}}{{p^{\, i}}}} \; . 
\end{eqnarray}
\\
Dla  rozkładów ciągłych entropia względna została określona w (\ref{entropia wzgledna roz ciagle}). Przekonamy się, że entropia względna jest miarą określającą jak bardzo dwa rozkłady różnią się od siebie. \\
\\
{\bf Przykład}: W celu ilustracji twierdzenie Sanova załóżmy, że  przeprowadzamy doświadczenie rzutu niesymetryczną 
monetą z wynikami orzeł, reszka, zatem $\aleph=2$. Rozkład teoretyczny $P_{ref}$ jest więc zero-jedynkowy. {\it Natomiast w wyniku pobrania  $N$-elementowej próbki dokonujemy jego estymacji 
na podstawie rozkładu empirycznego $\hat{{\cal P}}$ częstości pojawienia się wyników} orzeł lub reszka. Zatem: 
\begin{eqnarray}
\label{zerojeden oraz empir}
P_{ref} \equiv (p^{i})_{i=1}^{\aleph=2} = (p, 1-p) \;\;\; {\rm oraz} \;\;\; \hat{{\cal P}} \equiv (\hat{p}^{\,i})_{i=1}^{\aleph=2} = \left( \frac{m}{N}, \,1 - \frac{m}{N} \right) \; . 
\end{eqnarray}
Twierdzenie Bernoulliego mówi, że prawdopodobieństwo pojawienia się wyniku orzeł z częstością $m/N$ w $N$-losowaniach wynosi:    
\begin{eqnarray}
\label{bernoul}
Pr(\hat{{\cal P}}) \equiv Pr\left({\frac{m}{N}}\right) = \left(\begin{array}{l}
N\\
m\end{array}\right) p^{m}\left({1-p}\right)^{N-m} \; .
\end{eqnarray}
Biorąc logarytm naturalny obu stron (\ref{bernoul}), następnie stosując słuszne dla dużego $n$ przybliżenie Stirlinga, $\ln n!\approx n\ln n-n$, dla każdej silni w wyrażeniu $\left(\begin{array}{l}
N\\
m\end{array}\right)$, i w końcu biorąc eksponentę obu stron, można otrzymać  \cite{Bengtsson_Zyczkowski}:
\begin{eqnarray}
\label{pew}
Pr(\hat{{\cal P}}) \approx e^{-N S_{H}\left( \hat{{\cal P}}|P_{ref} \right)} \; ,
\end{eqnarray}
gdzie 
\begin{eqnarray}
\label{bernoul_entropia wzgl}
S_{H}\left( \hat{{\cal P}}|P_{ref} \right) &=&  \left[{\frac{m}{N}\left({\ln\frac{m}{N} - \ln p}\right) + \left({1-\frac{m}{N}}\right)\left({\ln\left({1-\frac{m}{N}}\right)- \ln\left({1-p}\right)}\right)}\right] \nonumber \\
&=&  \sum\limits_{i=1}^{\aleph=2}{\hat{p}^{\,i}\ln\frac{{\hat{p}^{\,i}}}{{p^{\, i}}}} \; .
\end{eqnarray}
W ostatnej równości skorzystano z (\ref{zerojeden oraz empir}) otrzymując entropię względną (\ref{entropia wzgl dyskret}) dla przypadku liczby wyników $\aleph=2$. \\
\\
{\bf Podsumowanie}: W powyższym przykładzie entropia względna $S_{H}\left( \hat{{\cal P}}|P_{ref} \right)$ pojawiła się jako pojęcie wtórne, wynikające z wyznaczenia prawdopodobieństwa otrzymania w eksperymencie empirycznego rozkładu częstości $\hat{{\cal P}} \in {\cal O}$ jako oszacowania rozkładu $P_{ref}$. Fakt ten oznacza, że {\it entropia względna}  nie jest tworem sztucznym, wprowadzonym do teorii jedynie dla wygody jako ``jakaś'' miara odległości pomiędzy rozkładami, lecz, że {\it jest właściwą dla  przestrzeni statystycznej rozkładów miarą probabilistyczną tej odległości}, tzn. ``dywergencją informacji'' pomiędzy rozkładami.  Okaże się też, że spośród innych miar odległości jest ona wyróżniona poprzez jej zwiazek ze znaną już nam informacją Fishera. \\
\\
W końcu podamy twierdzenie o dodatniości entropii względnej, które jeszcze bardziej przybliży nas do zrozumienia entropii jako miary odległości pomiędzy rozkładami, rozwijanego w treści następnego rozdziału. \\
\\
{\bf Twierdzenie (Nierówność informacyjna)}: 
Jeśli  $P({\bf y})$ oraz $P_{ref}({\bf y})$ są dwoma rozkładami gęstości prawdopodobieństwa, wtedy entropia względna spełnia następującą nierówność:
\begin{eqnarray}
\label{nierownosc informacyjna}
S_{H}\left(P({\bf y})|P_{ref}({\bf y}) \right) \geq 0 \; , 
\end{eqnarray}
i nierówność ta jest ostra za wyjątkiem przypadku, gdy $P({\bf y}) = P_{ref}({\bf y})$ \cite{Pawitan}. 


\section[Geometria przestrzeni rozkładów prawdopodobieństwa i metryka Rao-Fishera]{Geometria przestrzeni rozkładów prawdopodobieństwa i metryka Rao-Fishera}

\label{geometria i metryka Fishera-Rao}

Pojęcie metryki Fishera wprowadziliśmy już w  Rozdziale~\ref{alfa koneksja}. Do ujęcia tam przedstawionego dojdziemy jeszcze raz, wychodząc od pojęcia entropii względnej.\\ 
\\
Rozważmy dwa rozkłady prawdopodobieństwa:  $P=(p^{i})_{i=1}^{\aleph}$ oraz 
$P'=P+dP = (p'^{i}=p^{i}+dp^{i})_{i=1}^{\aleph}$,  różniące się {\it infinitezymalnie mało}, przy czym $p_{i}\neq 0$ dla każdego $i$. 
Rozkłady $P$ oraz $P'$ spełniają warunek unormowania  $\sum_{i=1}^{\aleph} p^{i} = 1$ oraz
$\sum_{i=1}^{\aleph} p^{i} + dp^{i} = 1$, skąd: 
\begin{eqnarray}
\label{suma dp zero}
\sum\limits_{i=1}^{\aleph} dp^{i} = 0 \; .
\end{eqnarray}
Ponieważ $dp^{i}/p^{i}$ jest wielkością infinitezymalnie małą, zatem  z rozwinięcia:
\begin{eqnarray}
\label{rozwinięcie} 
\ln ( 1 + \frac{dp^{i}}{p^{i}} ) = \frac{dp^{i}}{p^{i}} - 1/2  (\frac{dp^{i}}{p^{i}})^{2} + ... 
\end{eqnarray}
otrzymujemy, że entropia KL rozkładów $P$ oraz  $P'$ wynosi:
\begin{eqnarray}
\label{S dla P oraz P+dP}
S_{H} \left( P|P+dP \right) = \sum_{i=1}^{\aleph} p^{i} \ln \frac{p^{i}}{p^{i}+dp^{i}} = - \sum_{i=1}^{\aleph} p^{i} \ln \frac{p^{i}+dp^{i}}{p^{i}} \approx \frac{1}{2} \sum_{i=1}^{\aleph} \frac{dp^{i} dp^{i}}{p^{i}} \; .
\end{eqnarray}
Ostatnia postać $S_{H} \left( P|P+dP \right)$ sugeruje, że entropia  KL określa w naturalny sposób na przestrzeni statystycznej ${\cal S}$ infinitezymalny kwadrat odległości pomiędzy rozważanymi rozkładami, co onaczałoby również  {\it lokalne} określenie na przestrzeni statystycznej ${\cal S}$ pewnej metryki. Do jej związku z metryką Rao-Fisher'a wprowadzoną w Rozdziale~\ref{alfa koneksja} powrócimy później. \\
\\
{\bf Infinitezymalny interwał w ${\cal S}$}: Niech $d\vec{p}=\left(dp^{1},....,dp^{\aleph}\right)$ jest infinitezymalnym 
wektorem w przestrzeni rozkładów prawdopodobieństwa, spełniającym warunek (\ref{suma dp zero}). Wprowadzając na przestrzeni statystycznej ${\cal S}$ aparat matematyczny geometrii różniczkowej \cite{Amari Nagaoka book}, zapiszmy kwadrat różniczkowego interwału w tej przestrzeni następująco: \begin{eqnarray} 
\label{ds2 poprzez P}  d s^{2} = \sum_{i,\, j =1}^{\aleph} g_{i j} \, 
dp^{i}\,dp^{j} \; .
\end{eqnarray}
W celu uzgodnienia (\ref{S dla P oraz P+dP}) z (\ref{ds2 poprzez P}) wprowadźmy na ${\cal S}$ metrykę: 
\begin{eqnarray} 
\label{metryka Rao-Fishera z KL}
g_{i j} =  \frac{\delta_{ij}}{p^{i}} \; .
\end{eqnarray}
Związek (\ref{ds2 poprzez P}) oznacza, że {\it liczba możliwych wyników $\aleph$ określa wymiar przestrzeni} ${\cal S}$ oraz, że entropia względna KL definuje dla infinitezymalnie bliskich rozkładów symetryczną metrykę na przestrzeni statystycznej ${\cal S}$, pozwalającą mierzyć odległość pomiędzy tymi rozkładami. Zastępując prawdopodobieństwa $p^{i}$ częstościami, metryka ta staje się statystyką związana z obserwowaną informacyjną Fishera  
wprowadzoną  poprzednio. \\
  \\
Powyżej stan układu określony był w reprezentacji rozkładu prawdopodobieństwa zmiennej losowej, tzn. określony był przez podanie rozkładu prawdopodobieństwa.  Dwóm infinitezymalnie blisko leżącym stanom opowiadały rozkłady $P$ oraz $P'$. \\
\\
{\bf Reprezentacja amplitudowa ${\cal S}$}: Gdy interesują nas czysto geometryczne własności metryki Fishera, wtedy wygodne jest użycie innej reprezentacji do opisu stanu układu, określonej następująco. Niech  $Q \equiv (q^{i})_{i=1}^{\aleph}$ oraz $Q' \equiv (q'^{i})_{i=1}^{\aleph} = (q^{i} + d q^{i})_{i=1}^{\aleph}$ opisują te same co poprzednio, dwa infinitezymalnie blisko leżące stany układu  tyle, że zarządajmy, aby infinitezymalny kwadrat interwału pomiędzy nimi (\ref{ds2 poprzez P}) był równy: 
\begin{eqnarray} 
\label{ds2 przy amplitudzie}
d s^{2} = 4 \sum_{i=1}^{\aleph} dq^{i}\,dq^{i} \; .
\end{eqnarray} 
Porównując formułę (\ref{ds2 przy amplitudzie}) z (\ref{ds2 poprzez P}) zauważamy, że zgadzają się one z sobą o ile $Q = (q^{i})_{1}^{{\aleph}}$ oraz $P =
(p^{i})_{1}^{{\aleph}}$ są powiązane związkiem:
\begin{eqnarray} 
\label{poch amplituda a poch rozklad}
d q^{i} = \frac{d p^{i}}{2 \sqrt{p^{i}}}  \;\;\; i = 1, ..., {\aleph} \; ,
\end{eqnarray}
co zachodzi wtedy gdy 
\begin{eqnarray} 
\label{amplituda a rozklad}
q^{i} = \sqrt{p^{i}} \;\;\; i = 1, ..., {\aleph} \; .
\end{eqnarray} 
Wielkości $\; q^{i}$ nazywamy {\it amplitudami 
układu}\footnote{  
Fisher korzystał z amplitud prawdopodobieństwa niezależnie od ich  pojawienia się w mechanice kwantowej.
}. 
Definują one na ${\cal S}$ 
nowe współrzędne, dla których $q^{i} \ge 0$ dla każdego
$i$. \\
\\
{\bf ${\cal S}$ z geometrią jednostkowej sfery}: Otrzymana w bazie $q^{i}$ geometria ${\cal S}$ jest geometrią jednostkowej sfery, tzn. 
ze względu na unormowanie rozkładu prawdopodobieństwa do jedności, amplitudy 
spełniają następujący warunek unormowania na promieniu jednostkowym: 
\begin{eqnarray} 
\label{unormowanie amplitud}
\;\;\sum_{i=1}^{\aleph} p^{i} =
\sum_{i=1}^{\aleph} q^{i}\,q^{i} = 1 \; .
\end{eqnarray}  
Na sferze tej możemy określić odległość geodezyjną $D_{Bhatt}$, tzw. odległość Bhattacharyya'  
pomiędzy dwoma   rozkładami prawdopodobieństwa $P$ oraz $P'$, jako długość kątową  
li\-czo\-ną wzdłuż koła wielkiego pomiędzy dwoma wektorami $Q$ oraz $Q'$ o składowych będących 
amplitudami $q^{i}=\sqrt{p^{i}}$ oraz $q^{'i}=\sqrt{p^{'i}}$. \\
\\
{\bf Odległości Bhattacharyya}: Kwadrat infinitezymalnego interwału (\ref{ds2 przy amplitudzie}) jest przykładem odległości Bhattacharyya, którą ogólnie określamy nastepująco: Jeśli $P=(p^{i})$ oraz $P'=(p'^{i})$ są rozkładami prawdopodobieństwa, wtedy odległość Bhattacharyya pomiędzy $Q$ oraz $Q'$ jest iloczynem wewnętrznym określonym na przestrzeni statystycznej ${\cal S}$ następująco:
\begin{eqnarray} 
\label{odleglosc Bhattacharyya}
\cos D_{Bhatt} = \sum_{i=1}^{\aleph} q^{i}\,q'^{i} = \sum_{i=1}^{\aleph} \sqrt{p^{i}\,p'^{i}}
\equiv B(p,p') \; .
\end{eqnarray} 
{\it Statystyczna odległością Bhattacharyya wygląda więc jak iloczyn wewnętrzny mechaniki kwantowej}. 
W kolejnych rozdziałach przekonamy się, że nie jest to błędne skojarzenie. \\
\\
{\bf Hessian entropii Shannona}: Zauważmy, że metryka $g_{ij}$ jest  Hessianem (tzn. macierzą drugich pochodnych) entropii Shannona: 
\begin{eqnarray}
\label{hesian z S}
g_{ij} = - \partial_{i}\partial_{j}S_{H}\left(p\right) = \frac{\partial}{\partial p^{i}}\frac{\partial}{\partial p^{j}}\sum\limits _{k=1}^{\aleph}{p^{k}\ln p^{k}} = \frac{\delta^{ij}}{p^{j}}\,
\geq 0 \, ,
\end{eqnarray}
zgodnie z (\ref{metryka Rao-Fishera z KL}). Powyższy związek oznacza, że fakt wklęsłości entropii 
Shannona \cite{Bengtsson_Zyczkowski} daje dodatnią określoność metryki  $g_{ij}$ na ${\cal S}$. \\
\\
{\bf Metryka indukowana z $g_{ij}$}: Uzasadnijmy fakt nazwania czasami metryki $g_{ij}$, (\ref{metryka Rao-Fishera z KL}), metryką Rao-Fishera.\\
\\  
Załóżmy, że interesuje nas pewna podprzestrzeń przestrzeni  rozkładów prawdopodobieństwa ${\cal S}$. 
Wprowadźmy na niej układ współrzędnych $\theta^{a}$. Korzystając z (\ref{metryka Rao-Fishera z KL}) widać, że {\it metryka $g_{ij} = \frac{\delta^{ij}}{p^{j}}$ 
indukuje w tej podprzestrzeni ${\cal S}$ metrykę}: 
\begin{eqnarray}
\label{metryka Rao-Fishera w ukladzie wsp}
\!\!\! g_{ab} &=& \sum\limits _{i,j=1}^{\aleph} {\frac{{\partial p^{i}}}{{\partial\theta^{a}}} \frac{{\partial p^{j}}}{{\partial\theta^{b}}}} \, g_{ij} 
= \sum\limits_{i=1}^{\aleph}{\frac{{\partial_{a}p^{i} \partial_{b} p^{i}}}{{p^{i}}}} = \sum\limits_{i=1}^{\aleph} p^{i} \, \partial_{a} \ln p^{i} \partial_{b} \ln p^{i} \nonumber \\
&\equiv&  E \left( \partial_{a} \ell_{\Theta} \, \partial_{b} \ell_{\Theta} \right) = E_{\Theta} \texttt{i\!F}\left(\Theta \right) =  I_{F} \left(\Theta \right) \; .
\end{eqnarray}
gdzie skorzystano z zapisu $\ell_{\Theta} \equiv l\left(y|\Theta \right) = \ln P\left(\Theta\right) = (\ln p^{i}\left(\Theta\right))_{i=1}^{\aleph}$ oraz  $\partial_{a} \equiv{\partial }/{\partial\theta^{a}}$,  (\ref{oznaczenie ln P}).\\
\\
W Rozdziale~\ref{alfa koneksja} we wzorze (\ref{Fisher inf matrix}) zdefiniowaliśmy metrykę Rao-Fishera jako $g_{ab} = E_{\vartheta}(\partial_{a} \ell_{\Theta} \,  \partial_{b} \ell_{\Theta})$. Tak więc ostateczne otrzymaliśmy zgodność nazwania metryki (\ref{metryka Rao-Fishera z KL}) metryką  Rao-Fishera. {\it Metryka $g_{ij}$ jest jednak wielkością  obserwowaną a nie oczekiwaną, tak jak metryka  Rao-Fishera $g_{ab}$}. \\
\\
{\bf Metryka Roa-Fishera zapisana w amplitudach}: Korzystając ze związku $q^{i} = \sqrt{p^{i}}$, (\ref{amplituda a rozklad}), pomiędzy prawdopodobieństwami i amplitudami,  metrykę Rao-Fishera (\ref{metryka Rao-Fishera w ukladzie wsp}) w reprezentacji amplitudowej można zapisać następująco:
\begin{eqnarray}
\label{metryka Rao-Fishera w ukladzie wsp - amplitudy}
g_{ab} =  4 \sum\limits_{i=1}^{\aleph} \frac{\partial q^{i}}{\partial \theta^{a}} \frac{\partial q^{i}}{\partial \theta^{b}} \; .
\end{eqnarray}

\newpage

{\bf Przejście do zmiennej ciągłej}: Gdy liczba parametrów $\theta^{a}$ (więc i wektorów bazowych~$\vec{\theta}^{a}$) rozpinających osie układu współrzędnych rozważanej  podprzestrzeni statystycznej jest  skończona, wtedy można dokonać następującego uogólnienia metryki na przypadek ciągłego rozkładu
prawdopodobieństwa $P(y)$ zmiennej losowej $Y$:
\begin{eqnarray}
\label{metryka Rao-Fishera w ukl wsp dla r ciaglego}
g_{ab} = \int_{\cal Y}{d{\bf y}}\frac{{\partial_{a} P\;\partial_{b} P}}{P} = \int_{\cal Y}{d{\bf y}\; P\; \partial_{a} \ell_{\Theta} \, \partial_{b} \ell_{\Theta}} \equiv  E \left( \partial_{a} \ell_{\Theta} \, \partial_{b} \ell_{\Theta} \right) =  E_{\Theta} \texttt{i\!F}\left(\Theta \right) =  I_{F} \left(\Theta \right)  \; ,
\end{eqnarray}
gdzie, ponieważ przestrzeń zdarzeń ciągłego rozkładu  prawdopodobieństwa jest nieskończenie wymiarowa, w miejsce sumowania po $i=1,...,\aleph$ pojawiło się całkowanie po wartościach ${\bf y} \in {\cal Y}$ zmiennej losowej $Y$. 
Otrzymana postać metryki jest jawnie niezmiennicza ze względu na zmianę układu współrzędnych ${\bf y} \rightarrow {\bf y}'$ w przestrzeni bazowej ${\cal Y}$.
\\
Na koniec, dokonując reparametryzacji i przechodząc do amplitud $q({\bf y}|\Theta) = {\sqrt P({\bf y}|\Theta)}\, $, można metrykę (\ref{metryka Rao-Fishera w ukl wsp dla r ciaglego}) dla przypadku rozkładu ciągłego zapisać następująco:
\begin{eqnarray}
\label{metryka Rao-Fishera w ukl wsp dla r ciaglego - amplitudy}
g_{ab} = 4 \int_{\cal Y}{d{\bf y}} \, \partial_{a} q({\bf y}|\Theta) \;\partial_{b} q({\bf y}|\Theta)
 \; ,
\end{eqnarray}
gdzie jawnie zaznaczono zależność amplitudy rozkładu od parametru $\Theta$.\\
\\
{\bf Przypadek funkcji wiarygodności}: Powyższa analiza dotyczyła przypadku próby $N=1$ - wymiarowej. Gdyby rozkładem  prawdopodobieństwa $P$ była funkcja wiarygodności próby, wtedy w miejscu zmiennej losowej $Y$ pojawiłaby się próba $\widetilde{Y} \equiv( Y_{n})_{n=1}^{N}$, a w miejscu przestrzeni ${\cal Y}$, przestrzeń próby ${\cal B}$. Poza tym,  rozważania w obecnym  rozdziale pozostałyby takie same \cite{Amari Nagaoka book}.\\ 
\\
Omówiną sytuację odległości na przestrzeni statystycznej przedstawia graficznie poniższy rysunek dla liczby parametrów $d=2$.\\
\includegraphics[height=9.5cm]{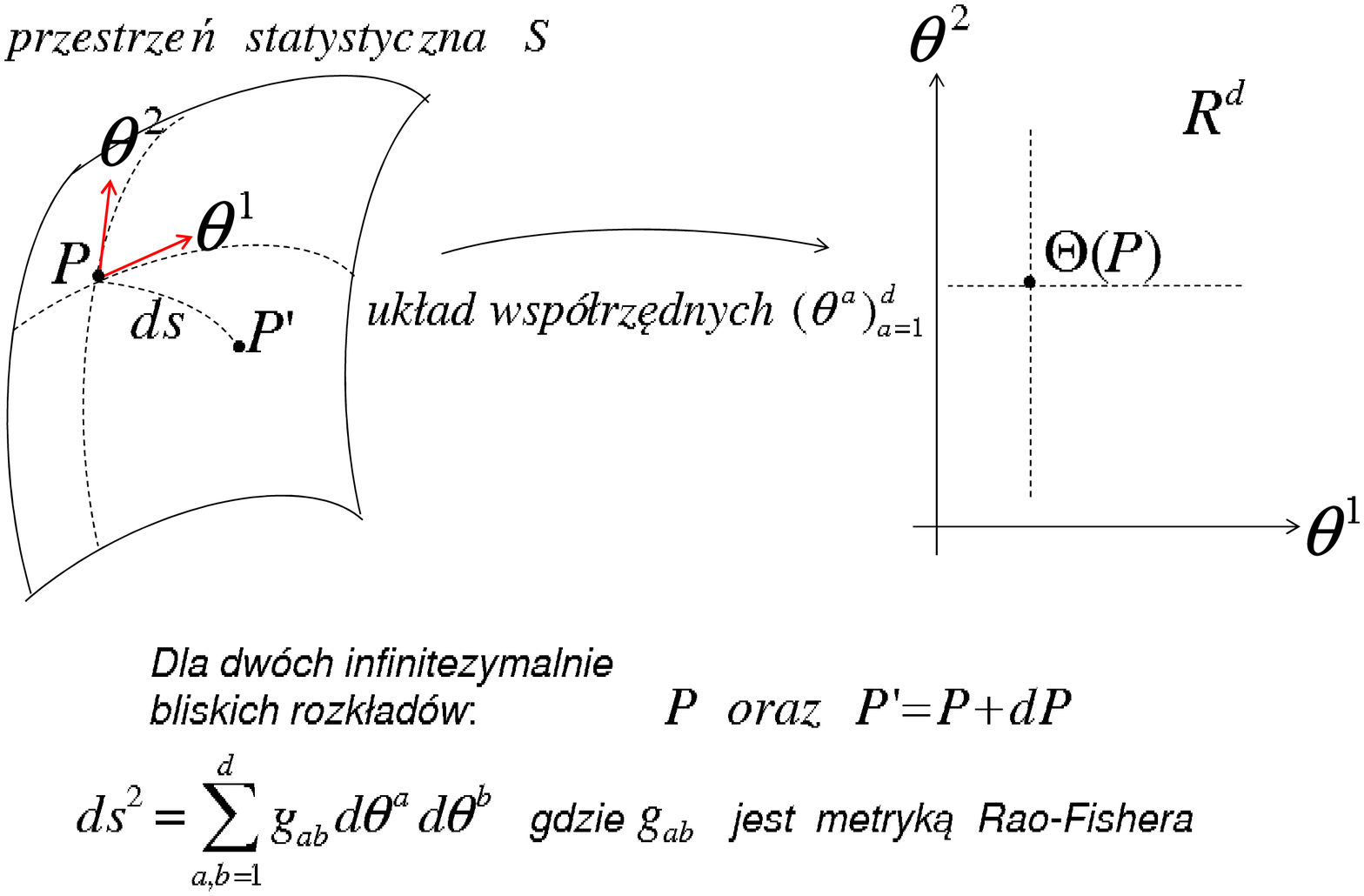}\\

\newpage

{\bf Przykład}: Wyznaczyć kwadrat infinitezymalnego interwału dla dwóch stanów posiadających rozkład normalny $N(\mu, \sigma)$. 
Rozkład normalny ma postać: 
\begin{eqnarray}
\label{rozklad 2-wym norm przyklad}
P\left({\bf y},\mu,\sigma\right)=\frac{1}{{\sqrt{2\pi}\sigma}}e^{-\frac{{\left({{\bf y} - \mu}\right)^{2}}}{{2\sigma^{2}}}} \; ,
\end{eqnarray}
gdzie wektor parametrów $\Theta \equiv (\theta^{a})_{a=1}^{2} = (\mu, \sigma)^{T}$. 
Macierz informacyjna Fishera dla rozkładu normalnego $N(\mu, \sigma)$ ma wyznaczoną poprzednio postać (\ref{oczekiw iF r normalnego 2 par}). Zatem metryka Rao-Fishera na 2-wymiarowej przestrzeni
normalnych rozkładów prawdopodobieństwa z układem współrzędnych $\mu, \sigma$ ma (dla próby $N=1$), postać:
\begin{eqnarray}
\label{oczekiw iF r normalnego 2 par dla N=1}
(g_{ab}) =  I_{F} \left(\Theta \right) =  E_{\Theta} \texttt{i\!F}\left(\Theta \right) =  \left({\begin{array}{cc}
{\frac{1}{{ \sigma^{2}}}} & 0\\
0 & {\frac{2}{{ \sigma^{2}}}} \end{array}}\right)  \; ,
\end{eqnarray}
tzn. składowe metryki (\ref{metryka Rao-Fishera w ukl wsp dla r ciaglego}) są równe:
\begin{eqnarray}
\label{metryka w przykladzie}
g_{\mu\mu} = 
\frac{1}{{\sigma^{2}}} \; , \;\;\; 
g_{\mu\sigma} = g_{\sigma \mu}  
= 0 \; , \;\;\; 
g_{\sigma\sigma} = 
\frac{2}{{\sigma^{2}}} \; .
\end{eqnarray}
Zatem otrzymany kwadrat infinitezymalnego interwału na 2-wymiarowej (pod)przestrzeni statystycznej ${\cal S}$ wynosi \cite{Bengtsson_Zyczkowski}:
\begin{eqnarray}
\label{ds2 dla r norm}
ds^{2} = g_{\mu\mu} d\mu^{2} + g_{\mu\sigma}d\mu d\sigma + g_{\sigma\sigma}d\sigma^{2} = \frac{1}{{\sigma^{2}}}\left({d\mu^{2} + 2d\sigma^{2}}\right) \; .
\end{eqnarray}
Odpowiada on metrce Poincarégo ze stałą ujemną krzywizną. W końcu, z postaci rozkładu (\ref{rozklad 2-wym norm przyklad}) otrzymujemy również bazę w przestrzeni stycznej do ${\cal S}$: 
\begin{eqnarray}
\label{baza w przestrzeni stycz r norm}
\partial_{\mu} \ln P \left({\bf y}|(\mu,\sigma)\right) = \frac{{\bf y} - \mu}{\sigma^{2}} \; ,\;\;\;\; 
\partial_{\sigma} \ln P\left({\bf y}|(\mu,\sigma)\right) = \frac{\left({\bf y} - \mu \right)^{2}}{\sigma^{3}}-\frac{1}{\sigma} \; .
\end{eqnarray}
{\bf Wniosek z przykładu}: {\it Na osi $\mu$, na której $\sigma =0$ leżą punkty odpowiadające  klasycznym stanom czystym i są one zgodnie z (\ref{ds2 dla r norm}) nieskończenie 
daleko odległe od dowolnego punktu we wnętrzu górnej półpłaszczyzny} $\sigma > 0$. W konsekwencji oznacza to, że wynik pewny, któremu odpowiada rozkład klasyczny z  $\sigma=0$, jest łatwy do odróżnienia od każdego  innego. 

\section[Informacja Fishera]{Informacja Fishera}

\subsection{Informacja Fishera jako entropia}
\label{Informacja Fishera jako entropia}

W powyższych rozważaniach informacja Fishera $I_{F}$ pojawiła się   poprzez nierówność Rao-Cra\-mera (\ref{tw R-C dla par skalarnego}),  jako wielkość określająca graniczną dobroć procedury estymacyjnej parametru rozkładu, tzn. o ile estymator efektywny  istnieje, to informacja Fishera określa minimalną wartość jego wariancji. 
Ponieważ im informacja Fishera mniejsza, tym to graniczne  oszacowanie parametru gorsze, zatem jest ona również miarą stopnia nieuporządkowania układu określonego rozkładem prawdopodobieństwa. Takie zrozumienie informacji Fishera odnosi się do typu analizowanego rozkładu prawdopodobieństwa  i jako miara  nieuporządkowania rozkładu oznacza brak przewidywalności procedury estymacyjnej (spróbuj pomyśleć o sensie oszacowania wartości oczekiwanej rozkładu jednorodnego).\\
Poniżej uzasadnimy stwierdzenie, że informacja Fishera okazuje się być  proporcjonalna do  entropii Kullbacka-Leiblera rozkładów różniących się infinitezymalnie mało {\it w parametrze} rozkładu  \cite{Frieden}. \\
 \\
Szukany związek informacji Fishera z entropią Kullbacka-Leiblera pokażemy w trzech krokach.  Rozważmy $N$-wymiarową próbę $Y_{1}, Y_{2}, ..., Y_{N}$, gdzie każda ze zmiennych losowych $Y_{n}$ jest określona na przestrzeni ${\cal Y}$ i  posiada  rozkład prawdopodobieństwa $p\left({\bf y}_{n}|\theta \right)$.  Przyjmijmy, dla uproszczenia rozważań, że  $\Theta = \theta$ jest parametrem skalarnym. \\
\\
{\bf (1)} Funkcja wiarygodności  $P(y\,|\theta)$ ma postać: 
\begin{eqnarray}
\label{fun wiaryg jeden par}
P\left({y|\theta}\right) = \prod\limits_{n=1}^{N} {p_{n}\left({{\bf y}_{n}|\theta}\right)} \; ,
\end{eqnarray}
gdzie $y\equiv({\bf y}_{n})_{n=1}^{N}$ jest realizacją  próby.  \\
Zauważmy, że informację Fishera parametru $\theta$ określoną w (\ref{infoczekiwana}) i (\ref{I jako krzywizna dla P}):
\begin{eqnarray}
\label{inf I jeden parametr - 2 pochodna}
I_{F} = -\int_{\cal B}{dy\, P\left({y|\theta}\right)\frac{{\partial^{2}\ln P \left({y|\theta}\right)}}{{\partial\theta^{2}}}}  \; ,
\end{eqnarray}
gdzie ${\cal B}$ jest przestrzenią próby, a $dy \equiv d^{N}{\bf y} = d{\bf y}_{1} d{\bf y}_{2} ... d{\bf y}_{N}$, można zapisać  następująco:
\begin{eqnarray}
\label{inf I jeden parametr - kwadrat 1 pochodnej}
I_{F} = \int_{\cal B}{dy\, P(y\,|\theta)}\left({\frac{{\partial \ln  P(y\,|\theta)}}{{\partial\theta}}}\right)^{2} \; .
\end{eqnarray}
\\
Istotnie, postać (\ref{inf I jeden parametr - kwadrat 1 pochodnej}) otrzymujemy z (\ref{inf I jeden parametr - 2 pochodna}) po skorzystaniu z:
\begin{eqnarray}
\frac{\partial}{\partial\theta}\left({\frac{{\partial\ln P}}{{\partial\theta}}}\right)=\frac{\partial}{\partial\theta}\left({\frac{1}{P}\frac{{\partial P}}{{\partial\theta}}}\right)=-\frac{1}{{P^{2}}}\left({\frac{{\partial P}}{{\partial\theta}}}\right)^{2}+\frac{1}{P}\left({\frac{{\partial^{2}P}}{{\partial\theta^{2}}}}\right) \; 
\end{eqnarray}
oraz warunku reguralności pozwalającego na wyłączenie różniczkowania po parametrze przed całkę (por. Rozdział~\ref{E i var funkcji wynikowej}):
\begin{eqnarray}
\int_{\cal B}{dy\left({\frac{{\partial^{2}P}}{{\partial\theta^{2}}}}\right)} = \frac{{\partial^{2}}}{{\partial\theta^{2}}}\int_{\cal B}{dy\, P} = \frac{{\partial^{2}}}{{\partial\theta^{2}}} 1 = 0 \; .
\end{eqnarray}
\\
{\bf (2) Zadanie}: Pokazać, że zachodzi następujący {\it rozkład informacji Fishera} parametru $\theta$: 
\begin{eqnarray}
\label{I dla pn jeden parametr}
I_{F} = \sum_{n=1}^{N}{I_{F n}} \;\;\;\; {\rm gdzie} \;\;\;\; I_{F n} = {\int_{\cal Y}{d{\bf y}_{n}{p_{n}\left({{\bf y}_{n}|\theta}\right)}\left({\frac{{\partial\ln p_{n}\left({{\bf y}_{n}|\theta}\right)}}{{\partial\theta}}}\right)^{2}}} \; ,
\end{eqnarray} 
na składowe informacje $I_{F n}$, zależne jedynie od rozkładów punktowych $p_{n}({\bf y}|\theta)$. \\
\\
{\bf Rozwiązanie}: Po skorzystaniu z (\ref{fun wiaryg jeden par}) zauważamy, że:
\begin{eqnarray}
\label{poch lnP}
\ln P\left({y;\theta}\right) = \sum_{n=1}^{N}{\ln p_{n}} \;, \;\;\; {\rm zatem} \;\;\;\;\; \frac{{\partial\ln P}}{{\partial\theta}}=\sum_{n=1}^{N}{\frac{1}{{p_{n}}}}\frac{{\partial p_{n}}}{{\partial\theta}} \; 
\end{eqnarray}
i podnosząc ostatnie wyrażenie do kwadratu otrzymujemy: 
\begin{eqnarray}
\label{kwadrat poch lnP}
\left({\frac{{\partial\ln P}}{{\partial\theta}}}\right)^{2} = \sum_{{\scriptstyle {{n,m=1}\hfill\atop {\scriptstyle m\ne n\hfill}}}}^{N}{\frac{1}{{p_{m}}}\frac{1}{{p_{n}}}}\frac{{\partial p_{m}}}{{\partial\theta}}\frac{{\partial p_{n}}}{{\partial\theta}} + \sum_{n=1}^{N}{\frac{1}{{p_{n}^{2}}}}\left({\frac{{\partial p_{n}}}{{\partial\theta}}}\right)^{2} \; .
\end{eqnarray}
Wstawiając (\ref{fun wiaryg jeden par}) oraz (\ref{kwadrat poch lnP}) do  (\ref{inf I jeden parametr - kwadrat 1 pochodnej}) otrzymujemy: 
\begin{eqnarray}
\label{I przed skorzstaniem z rozkl brzegowych}
I_{F} = \int_{\cal B}{dy\prod_{k=1}^{N}{p_{k}}} {\sum_{{\scriptstyle {{n,m=1}\hfill\atop {\scriptstyle m\ne n\hfill}}}}^{N}{\frac{1}{{p_{m}}}\frac{1}{{p_{n}}}}\frac{{\partial p_{m}}}{{\partial\theta}}\frac{{\partial p_{n}}}{{\partial\theta}} + \int_{\cal B}{dy\prod_{k=1}^{N}{p_{k}}} \sum_{n=1}^{N}{\frac{1}{{p_{n}^{2}}}}\left({\frac{{\partial p_{n}}}{{\partial\theta}}}\right)^{2}}  \; .
\end{eqnarray}
Ze względu na warunek normalizacji rozkładów brzegowych:
\begin{eqnarray}
\label{normalizacja rozkladow brzegowych}
\int_{\cal Y} d{\bf y}_{n} \, p_{n}({\bf y}_{n}|\theta) = 1 \; ,
\end{eqnarray}
w (\ref{I przed skorzstaniem z rozkl brzegowych}) z iloczynu $\prod_{k=1}^{N}{p_{k}}$ pozostaje w pierwszym składniku  jedynie $p_{m}p_{n}$ (dla $k\neq m$ oraz $k\neq n$), natomiast w drugim składniku pozostaje jedynie $p_{n}$ (dla $k\neq n$), tzn.:
\begin{eqnarray}
\label{upr}
I_{F} = \sum_{{\scriptstyle {{n,m=1} \hfill \atop {\scriptstyle m\ne n\hfill}}}}^{N}{\int_{\cal Y} {\int_{\cal Y}{d{\bf y}_{m}d{\bf y}_{n}\frac{{\partial p_{m}}}{{\partial\theta}}\frac{{\partial p_{n}}}{{\partial\theta}}}}} + \sum_{n=1}^{N}{\int_{\cal Y}{d{\bf y}_{n}\frac{1}{{p_{n}}}\left({\frac{{\partial p_{n}}}{{\partial\theta}}}\right)^{2}}} \; .
\end{eqnarray}
W końcu, ze względu na:
\begin{eqnarray}
\label{poch normalizacji}
\int_{\cal Y}{d{\bf y}_{n}\frac{{\partial p_{n}}}{{\partial\theta}}} = \frac{\partial}{{\partial\theta}}\int_{\cal Y}{d{\bf y}_{n}p_{n}}=\frac{\partial}{{\partial\theta}} 1 = 0 \; ,
\end{eqnarray}
pierwszy składnik w (\ref{upr}) zeruje się i pozostaje jedynie drugi,  zatem: 
\begin{eqnarray}
\label{postac I koncowa w pn}
I_{F} = \sum_{n=1}^{N}{\int_{\cal Y}{d{\bf y}_{n}\frac{1}{{p_{n}}}\left({\frac{{\partial p_{n}}}{{\partial\theta}}}\right)^{2}}} = \sum_{n=1}^{N} \int_{\cal Y} d{\bf y}_{n}p_{n}\left({\frac{{\partial\ln p_{n}}}{{\partial\theta}}}\right)^{2} \; .
\end{eqnarray}
Ostatecznie otrzymujemy więc szukany rozkład  informacji Fishera (\ref{I dla pn jeden parametr}):
\begin{eqnarray}
I_{F} = \sum_{n=1}^{N}{I_{F n}} \;\;\;\; {\rm gdzie} \;\;\;\; I_{F n} = {\int_{\cal Y}{d{\bf y}_{n}{p_{n}\left({{\bf y}_{n}|\theta}\right)}\left({\frac{{\partial\ln p_{n}\left({{\bf y}_{n}|\theta}\right)}}{{\partial\theta}}}\right)^{2}}} \; . \nonumber
\end{eqnarray} \\
\\
{\bf (3) Zadanie}: Znaleźć związek informacji Fishera z entropią Kullbacka-Leiblera rozkładów $P \left({y|\theta}\right)$ oraz $P\left({y|\theta + \Delta \theta}\right)$ różniących się infinitezymalnie mało w parametrze $\theta$.\\
\\
\\
{\bf Rozwiązanie}:  Zastąpmy całkę w (\ref{I dla pn jeden parametr}) dla $I_{F n}$ sumą Riemanna: 
\begin{eqnarray}
\label{dys}
I_{F n} &=& {\sum\limits _{k}{\Delta {\bf y}_{nk}\frac{1}{{p_{n}\left({{\bf y}_{nk}|\theta}\right)}}\left[{\frac{{p_{n}\left({{\bf y}_{nk}|\theta+\Delta\theta}\right) - p_{n}\left({{\bf y}_{nk}|\theta}\right)}}{{\Delta\theta}}}\right]^{2}}} \nonumber \\
&=& 
\left({\Delta\theta}\right)^{-2}  \Delta {\bf y}_{n} \sum\limits _{k}{\, p_{n} \left({{\bf y}_{nk}|\theta}\right)\left[{\frac{{p_{n}\left({{\bf y}_{nk}|\theta + \Delta\theta}\right)}}{{p_{n}\left({{\bf y}_{nk}|\theta}\right)}} - 1}\right]^{2}} \; .
\end{eqnarray}
Powyższa zamiana całkowania na sumę jest wprowadzona dla wygody i jest ścisłym przejściem w granicy $\Delta {\bf y}_{nk} 
{\scriptstyle {\;\; \longrightarrow   \hfill \atop {\scriptstyle k \rightarrow   \infty  \hfill}}} 0$.  W drugiej równości w (\ref{dys}) przyjęto równe przyrosty $\Delta {\bf y}_{nk} = \Delta {\bf y}_{n}$ dla każdego $k$, co w tej  granicy nie zmienia wyniku. Natomiast przejście dla pochodnej po $\theta$ pod całką $I_{F n}$ w (\ref{I dla pn jeden parametr}) dokonane  w (\ref{dys}) jest słuszne w granicy $\Delta\theta \rightarrow 0$. \\ 
\\
W granicy $\Delta\theta \to 0$ każde z powyższych wyrażeń  ${{p_{n}\left({{\bf y}_{nk}|\theta + \Delta\theta}\right)}\mathord{\left/{\vphantom{{p_{n}\left({{\bf y}_{nk}|\theta + \Delta\theta}\right)}{p_{n}\left({{\bf y}_{nk}|\theta}\right)}}}\right.\kern -\nulldelimiterspace}{p_{n}\left({{\bf y}_{nk}|\theta}\right)}}$
dąży do 1.  Wtedy każda z wielkości: 
\begin{eqnarray}
\label{delta}
\delta_{\Delta \theta}^{n} \equiv \frac{{p_{n}\left({{\bf y}_{nk}|\, \theta + \Delta \theta}\right)}}{{p_{n}\left({{\bf y}_{nk}| \theta}\right)}} - 1 
\end{eqnarray}
staje się mała i rozwijając funkcję logarytmu do wyrazu drugiego rzędu: 
\begin{eqnarray}
{\ln \left({1 + \delta_{\Delta \theta}^{n}}\right) \approx \delta_{\Delta \theta}^{n} - {{(\delta_{\Delta \theta}^{n})^{2}} \mathord{\left/{\vphantom{{(\delta_{\Delta \theta}^{n})^{2}}2}} \right. \kern -\nulldelimiterspace}2}} \; ,
\end{eqnarray}
otrzymujemy:
\begin{eqnarray}
\label{nu2}
{(\delta_{\Delta \theta}^{n})^{2} \approx 2 \left[{\delta_{\Delta \theta}^{n} - \ln \left({1 + \delta_{\Delta \theta}^{n}}\right)}\right]} \; .
\end{eqnarray}
Korzystając z (\ref{delta}) oraz (\ref{nu2}) wyrażenie (\ref{dys})
można zapisać następująco:
\begin{eqnarray}
\label{I trzy skladniki}
I_{F n} &=& -2 \left({\Delta\theta}\right)^{-2}  \Delta {\bf y}_{n} \left[ \sum\limits _{k}{p_{n}\left({{\bf y}_{nk}|\, \theta}\right) \; \ln\left({\frac{{p_{n}\left({{\bf y}_{nk}|\, 
\theta + \Delta \theta} \right)}}{{p\left({{\bf y}_{k}|\theta}\right)}}}\right)} \right. \nonumber \\
&-& \left.  \sum\limits_{k} {p_{n}\left({{\bf y}_{nk}|\theta + \Delta \theta}\right)} +  \sum\limits_{k}{p_{n} \left({{\bf y}_{nk}|\theta} \right)} \right] \; , \;\;\; {\rm dla} \;\;\; \Delta \theta \rightarrow 0 \; .
\end{eqnarray}
Ze względu na warunek normalizacji zachodzi $\sum\limits_{k} {p_{n}\left({{\bf y}_{nk}|\theta + \Delta \theta}\right)} = \sum\limits_{k}{p_{n} \left({{\bf y}_{nk}|\theta} \right)} = 1$, zatem dwie ostatnie sumy po $k$ w nawiasie kwadratowym znoszą
się wzajemnie i (\ref{I trzy skladniki}) redukuje się do postaci:
\begin{eqnarray}
\label{I porownanie z Sn}
I_{Fn} &=& -2\left({\Delta\theta}\right)^{-2}   \sum\limits_{k} \Delta {\bf y}_{n} \, {p_{n}\left({{\bf y}_{nk}|\theta}\right) \ln\left({\frac{{p_{n}\left({{\bf y}_{nk}|\, \theta + \Delta \theta}\right)}}{{p_{n}\left({{\bf y}_{nk}|\theta}\right)}}}\right)} \nonumber \\
&=& - 2 \left({\Delta\theta}\right)^{-2}  \int_{\cal Y} d{\bf y}_{n}\; p_{n} \left({\bf y}_{n}|\theta\right) \ln\left({\frac{{p_{n}\left({{\bf y}_{n}|\theta + \Delta\theta}\right)}}{{p_{n}\left({{\bf y}_{n}|\theta}\right)}}}\right) \nonumber \\
&=&  2 \left({\Delta\theta}\right)^{-2}  S_{H}\left[p_{n} \left({{\bf y}_{n}|\theta}\right)| p_{n}\left({{\bf y}_{n}|\theta + \Delta \theta}\right)\right]  \; , \;\;\;\;\; {\rm dla} \;\;\; \Delta \theta \rightarrow 0 \; ,
\end{eqnarray}
gdzie w ostatnim przejściu skorzystaliśmy z (\ref{entropia wzgledna roz ciagle}). 
Postać $I_{F n}$ w drugiej lini w (\ref{I porownanie z Sn}) można również, wykorzystując  unormowanie rozkładów $p_{m}( {{\bf y}_{m}|\theta})$, zapisać następująco: 
\begin{eqnarray}
\label{I porownanie z S druga linia}
\!\!\!\!\!\!\!\!\!\!\! I_{F n} = - 2 \left({\Delta\theta}\right)^{-2} \!\!
\prod_{{\scriptstyle {{m=1} \hfill \atop {\scriptstyle m \ne n\hfill}}}}^{N} \!\! \int_{\cal Y} \!\! {d{\bf y}_{m} {p_{m}( {{\bf y}_{m}|\theta} )}} 
 \!\! \int_{\cal Y} \!\! d{\bf y}_{n}\; p_{n} \left({\bf y}_{n}|\theta\right) \ln \! \left({\frac{{p_{n}\left({{\bf y}_{n}|\theta + \Delta\theta}\right)}}{{p_{n}\left({{\bf y}_{n}|\theta}\right)}}} \right)  , \;\, {\rm dla} \;\, \Delta \theta \rightarrow 0   , \;\;\;\;\,
\end{eqnarray}
skąd po skorzystaniu z $I_{F} = \sum_{n=1}^{N}{I_{F n}} $, (\ref{I dla pn jeden parametr}), otrzymujemy:
\begin{eqnarray}
\label{I porownanie z S}
\!\!\!\!\! I_{F} = \sum_{n=1}^{N}{I_{F n}} &=& - 2 \left({\Delta\theta}\right)^{-2} \sum_{n=1}^{N}
\prod_{{\scriptstyle {{m=1} \hfill \atop {\scriptstyle m \ne n\hfill}}}}^{N} \int_{\cal Y} {d{\bf y}_{m} {p_{m}\left({{\bf y}_{m}|\theta}\right)}} 
 \int_{\cal Y} d{\bf y}_{n}\; p_{n} \left({\bf y}_{n}|\theta\right) \ln\left({\frac{{p_{n}\left({{\bf y}_{n}|\theta + \Delta\theta}\right)}}{{p_{n}\left({{\bf y}_{n}|\theta}\right)}}}\right) \nonumber \\
&=& - 2 \left({\Delta\theta}\right)^{-2} 
\int_{\cal B} {dy \prod_{m=1}^{N} {p_{m}\left({{\bf y}_{m}|\theta}\right)}} 
 \ln \left({\frac{\prod_{n=1}^{N}{p_{n}\left({{\bf y}_{n}|\theta + \Delta\theta}\right)}}{{\prod_{n=1}^{N} p_{n}\left({{\bf y}_{n}|\theta}\right)}}}\right) \nonumber \\
&=&  2 \left({\Delta\theta}\right)^{-2}  S_{H}\left[P \left({y|\theta}\right)| P\left({y|\theta + \Delta \theta}\right)\right]  \; , \;\;\;\;\; {\rm dla} \;\;\; \Delta \theta \rightarrow 0 \; .
\end{eqnarray}
Tak więc (\ref{I porownanie z S}) daje  wsparcie dla intuicji wspomnianej na początku obecnego Rozdziału, która mówi, że skoro entropia jest miarą nieuporządkowania układu to informacja Fishera jest również miarą jego nieuporządkowania. \\
\\
{\bf Uwaga o lokalności związku IF z KL}: O własnościach globalnych rozkładu wypowiada się entropia Shannona, natomiast wniosek wynikający z (\ref{I porownanie z S}) ma sens tylko dla entropii względnej.
Zatem związek ten może być co najwyżej sygnałem, że  niektóre  własności układu związane z entropią Shannona mogą być ujęte w języku informacji Fishera. Na niektóre sytuacje, w których ma to miejsce zwrócimy uwagę w przyszłości. \\
\\
{\bf Rozkład entropii KL dla rozkładów punktowych}: \\
Ze względu na $I_{F} = \sum_{n=1}^{N}{I_{F n}} $, (\ref{I dla pn jeden parametr}),  z po\-równania (\ref{I porownanie z S}) z (\ref{I porownanie z Sn})  wynika dodatkowo, że  
entropia względna rozkładów $P \left({y|\theta}\right)$ oraz $P\left({y|\theta + \Delta \theta}\right)$ jest sumą entropii  względnych odpowiadających im rozkładów punktowych $p_{n} \left({{\bf y}_{n}|\theta}\right)$ oraz $p_{n}({{\bf y}_{n}| \theta + \Delta \theta})$: 
\begin{eqnarray}
\label{S jako suma Sn delta th inf}  
\!\!\!\!\!\!\!\! S_{H}\left[P \left({y|\theta}\right)| P\left({y|\theta + \Delta \theta}\right)\right] = \sum_{n=1}^{N} S_{H}\left[p_{n} \left({{\bf y}_{n}|\theta}\right)| \, p_{n}\left({{\bf y}_{n}|\theta + \Delta \theta}\right)\right] 
\;\;\;\;\; {\rm dla \;\; dowolnego} \;\;\; \Delta \theta 
\; . \;\;\;\;\;\;
\end{eqnarray}
Ponieważ warunek $\Delta \theta \rightarrow 0$ nie był wykorzystywany w otrzymaniu (\ref{S jako suma Sn delta th inf}) z porównania (\ref{I porownanie z Sn}) oraz (\ref{I porownanie z S}), zatem związek ten jest słuszny dla dowolnego $\Delta \theta$. 
Natomiast należy pamiętać, że rozkłady punktowe w definicji funkcji wiarygodności są {\it nieskorelowane} dla różnych $n$, dlatego też  związek  (\ref{S jako suma Sn delta th inf}) jest słuszny tylko w tym przypadku. \\

\section[Pojęcie kanału informacyjnego]{Pojęcie kanału informacyjnego}

\label{Pojecie kanalu informacyjnego}

\vspace{3mm}

Niech pierwotna zmienna  losowa $Y$ przyjmuje  wartości wektorowe ${\bf y} \in {\cal Y}$. Wektor ${\bf y}$ może być np. wektorem położenia. 
Zatem, aby wprowadzony opis był wystarczająco ogólny, wartości ${\bf y} \equiv ({\bf y}^{\nu})$ mogą posiadać np. indeks wektorowy $\nu$. Wartości te są realizowane zgodnie z łącznym rozkładem $p({\bf y}|\Theta)$ właściwym dla badanego układu. \\
Rozważmy $N$-wymiarową próbę. 
Oznaczmy przez $y \equiv ({\bf y}_{n})_{n=1}^{N} = ({\bf y}_{1}, ..., {\bf y}_{N})$ dane będące  realizacjami próby $\widetilde{Y} = (Y_{1}, Y_{2}, ..., Y_{N})$ dla pierwotnej zmiennej $Y$, gdzie ${\bf y}_{n} \equiv ({\bf y}_{n}^{\nu})$ oznacza $n$-tą wektorową obserwację w próbce ($n=1,2,...,N$). Rozkład łączny  próby jest określony przez $P\left(y|\Theta\right)$.\\
\\
{\bf Dodatkowy indeks parametru}: Podobnie, również parametry  rozkładu mogą mieć dodatkowy indeks. Niech indeks $\alpha$ określa pewną dodatkową współrzędną wektorową parametru ${\theta}_{i}$, gdzie jak w poprzednich rozdziałach $i=1,2,...,d$. Zatem wektor parametrów ma teraz postać:
\begin{eqnarray}
\label{parametr wektorowy}
\Theta = \left({{\theta}_{1},{\theta}_{2},...,{\theta}_{d}}\right) \; , \;\;\; {\rm gdzie} \;\;\;\; {\theta}_{i}=\left({\theta_{i}^{\alpha}} \right) \; ,  \;\;\;\; \alpha = 1,2,... \; .
\end{eqnarray}
Wariancja estymatora $\hat{\theta}_{i}^{\alpha}$ parametru $\theta_{i}^{\alpha}$ 
ma postać: 
\begin{eqnarray}
\label{wariancja estymatora theta i alfa}
{\sigma^{2}} \,( \hat{\theta}_{i}^{\alpha}) = \int_{\cal B} dy \, P\left(y|\Theta\right) \left(\hat{\theta}_{i}^{\alpha}\left(y\right)-\theta_{i}^{\alpha}\right)^{2} \; ,
\end{eqnarray}
gdzie $\hat{\theta}_{i}^{\alpha}\left({y}\right)$ jest estymatorem parametru $\theta_{i}^{\alpha}\,$, a całkowanie  przebiega po całej przestrzeni próby ${\cal B}$, tzn. po wszystkich możliwych realizacjach $y$. \\
Według  Rozdziału~\ref{Przyklad wektorowego DORC}, dla każdego  wyróżnionego parametru $\theta_{i}^{\alpha}$ wariancja ${\sigma^{2}} \, (\hat{\theta}_{i}^{\alpha})$  jego estymatora (\ref{wariancja estymatora theta i alfa}) jest związana z informacją Fishera $I_{F i \alpha}$ parametru  $\theta_{i}^{\alpha}$ poprzez nierówność (\ref{porownanie sigma I11 z I11do-1}):  
\begin{eqnarray}
\label{krzy3}
{\sigma^{2}}\,(\hat{\theta}_{i}^{\alpha}) \ge {I_{F}^{i \alpha}} \ge \frac{1}{I_{F i \alpha}} \; ,
\end{eqnarray}
gdzie $I_{F}^{i \alpha} \equiv I_{F}^{i\alpha, \,i\alpha}$ jest dolnym ograniczeniem RC dla parametru ${\theta}_{i}^{\alpha}$ w przypadku wieloparametrowym, natomiast: 
\begin{eqnarray}
\label{pojemnosc jeden kanal inform Fishera}
I_{F i \alpha} \equiv I_{F i\alpha, \,i\alpha} \; 
\end{eqnarray}
jest pojemnością informacyjną w {\it pojedynczym  kanale informacyjnym} $(i, \alpha)$ czyli {\it informacją Fishera dla parametru} ${\theta}_{i}^{\alpha}$.  Zgodnie z (\ref{inf I jeden parametr - kwadrat 1 pochodnej}) jest ona równa\footnote{Porównaj przejście od (\ref{inf I jeden parametr - 2 pochodna}) do (\ref{inf I jeden parametr - kwadrat 1 pochodnej}).
}: 
\begin{eqnarray}
\label{krzy4}
I_{F i \alpha} \equiv I_{F} \left(\theta_{i \alpha} \right) = \int_{\cal B} {dy \, P\left(y|\Theta\right)} \left({\frac{{\partial \ln P\left(y|\Theta\right)}}{{\partial \theta_{i}^{\alpha}}}}\right)^{2} \; .
\end{eqnarray}
\\
{\bf Informacja Stama}: Wielkość $\frac{1}{{\sigma^{2}} \, (\hat{\theta}_{i}^{\alpha})}$
odnosi się do 
pojedynczego  kanału $\left(i, \alpha\right)$. 
Sumując  
ją po indeksach  $\alpha$ oraz $i$ otrzymujemy tzw. informację Stama $I_{S}$ \cite{Stam}:
\begin{eqnarray}
\label{informacja Stama}
0 \leq  I_{S} \equiv \sum\limits_{i} {\sum\limits_{\alpha} \frac{1}{{\sigma^{2}}\,(\hat{\theta}_{i}^{\alpha})}} \; ,
\end{eqnarray}
która jest skalarną miarą jakości jednoczesnej estymacji we wszystkich kanałach informacyjnych. {\it Informacja Stama jest z definicji zawsze wielkością nieujemną.}\\
\\
{\bf Pojemność informacyjna $I$}: W końcu, sumując lewą i prawą stronę (\ref{krzy3}) po indeksach  $\alpha$ oraz $i$ otrzymujemy następującą nierówność dla $I_{S}$: 
\begin{eqnarray}
\label{informacja Stama vs pojemnosc informacyjna}
0 \leq I_{S} \equiv \sum\limits_{i} \sum\limits_{\alpha} \frac{1}{{\sigma^{2}}\,(\hat{\theta}_{i}^{\alpha})} \le \sum\limits_{i} {\sum\limits_{\alpha}{I_{F i \alpha}}} =: I \equiv C  \; ,
\end{eqnarray}
gdzie $C$, oznaczana dalej jako $I$, nazywana jest {\it pojemnością informacyjną układu}. Zgodnie z (\ref{informacja Stama vs pojemnosc informacyjna}) i (\ref{krzy4}) jest ona równa: 
\begin{eqnarray}
\label{pojemnosc C}
I = \sum\limits_{i}{\sum\limits _{\alpha}{I_{F i \alpha}}}=\sum\limits_{i}{\int_{\cal B} {dy \; P\left(y|\Theta\right) \sum\limits _{\alpha}{\left(\frac{\partial\ln P\left(y|\Theta\right)}{\partial\theta_{i}^{\alpha}}\right)^{2}}}} \; .
\end{eqnarray}
Jak się okaże, pojemność informacyja $I$ jest najważniejszym  pojęciem statystyki leżącym u podstaw np. członów kinetycznych różnych modeli teorii pola. Jest ona uogólnieniem pojęcia informacji Fishera dla przypadku pojedynczego, skalarnego parametru na przypadek wielopa\-ra\-me\-trowy. 
 
\subsection{Pojemność informacyjna dla zmiennej losowej położenia}

\label{Poj inform zmiennej los poloz}

Zawęźmy obszar analizy do szczególnego przypadku, gdy interesującym nas oczekiwanym parametrem jest wartość oczekiwana  zmiennej położenia układu $Y$: 
\begin{eqnarray}
\label{wartosc oczekiwana EY}
\theta \equiv E(Y)= (\theta^{\nu}) \; , \;\;\;\; {\rm gdzie} \;\;\;\;\; \theta^{\nu} = \int_{\cal Y} d {\bf y}\, p({\bf y})\,  {\bf y}^{\nu} \; .
\end{eqnarray}
Wtedy $N$-wymiarowa próbka $y \equiv ({\bf y}_{n})_{n=1}^{N} = ({\bf y}_{1}, ..., {\bf y}_{N})$ 
jest  realizacją próby $\widetilde{Y}$ dla {\it położeń układu}, 
%
%
a wartość oczekiwana  $\theta_{n}$ położenia układu w $n$-tym punkcie (tzn. pomiarze)  próby wynosi:
\begin{eqnarray}
\label{wartosc oczekiwana EYn}
\theta_{n} \equiv E(Y_{n}) =  (\theta_{n}^{\nu}) \; , \;\;\; {\rm gdzie} \;\;\;\;  \theta_{n}^{\nu} = \int_{\cal B} dy \, P(y|\Theta) \, {\bf y}_{n}^{\nu}  \; .
\end{eqnarray}
{\bf Liczba  oczekiwanych parametrów}: Gdy, jak to ma miejsce w rozważanym przypadku, jedynym parametrem rozkładu, który nas interesuje jest wartość oczekiwana położenia $\theta_{n} \equiv (\theta_{n}^{\nu})$, gdzie $n =1,2,...,N$ jest indeksem próby, wtedy parametr wektorowy $\Theta=(\theta_{n})_{n=1}^{N}$. 
Zatem liczba parametrów $\theta_{n}$ pokryła się z wymiarem próby $N$, a indeksy parametru $\theta_{i}^{\alpha}$ są następujące: $i \equiv n$, gdzie $n=1,2,...,N$, oraz $\alpha \equiv \nu$, gdzie $\nu$ jest indeksem wektorowym współrzędnej ${\bf y}_{n}^{\nu}$. Oznacza to, że wymiar parametru $\Theta$ jest taki sam jak wymiar przestrzeni próby ${\cal B}$. \\
  \\
{\bf Współrzędne kowariantne i kontrawariantne}: Rozważania obecnego Rozdziału jak i innych części skryptu są związane z analizą przeprowadzaną w czasoprzestrzeni Minowskiego. Dlatego koniecznym okazuje się rozróżnienie pomiędzy współrzędnymi kowariantnymi ${\bf y}_{n \nu}$ i kontrawariantnymi ${\bf y}_{n}^{\, \mu}$. 
Związek pomiędzy nimi, tak dla wartości losowego wektora położenia jak i dla odpowiednich wartości oczekiwanych, jest następujący:
\begin{eqnarray}
\label{wsp theta kontra i kowariantne}
{\bf y}_{n \nu} = \sum_{\mu=0}^{3} \eta_{\nu \mu} \, {\bf y}_{n}^{\, \mu} \; , \;\;\;\;\;\; \theta_{n \nu} = \sum_{\mu=0}^{3} \eta_{\nu \mu} \, \theta_{n}^{\, \mu} \; , 
\end{eqnarray}
gdzie $(\eta_{\nu \mu})$ jest tensorem metrycznym przestrzeni   ${\cal Y}$. W przypadku wektorowego indeksu Minkowskiego $\nu = 0,1,2,3,...$ przyjmujemy następującą postać tensora metrycznego:
\begin{eqnarray}
\label{metryka M}
\eta_{\nu\mu}=\left({\begin{array}{ccccc}
{1} & 0 & 0 & 0 & 0\\
0 & {-1} & 0 & 0 & 0\\
0 & 0 & {-1} & 0 & 0\\
0 & 0 & 0 & {-1} & 0\\
0 & 0 & 0 &   0 & \ddots
\end{array}}\right)
\end{eqnarray}
lub w skrócie $(\eta_{\nu \mu}) = diag(1,-1,-1,-1,...)$. Symbol ``diag'' oznacza  macierz diagonalną  z niezerowymi elementami na przekątnej głównej oraz zerami poza nią. Natomiast dla Euklidesowego indeksu wektorowego $\nu = 1,2,3,...\,$,  tensor metryczny ma postać:
\begin{eqnarray}
\label{metryka E}
(\eta_{\nu \mu}) = diag(1,1,1,...) \;\, .
\end{eqnarray}
{\bf Założenie o niezależności $p_{n}$ od $\theta_{m}$ dla $m \neq n$}: W rozważaniach niniejszego skryptu zmienne $Y_{n}$ próby $\widetilde{Y}$ są niezależne (tzn. zakładamy, że pomiary dla $m \neq n$ są w próbie  niezależne). Oznacza to również, że wartość oczekiwana położenia $\theta_{m}  =  \int dy \, P(y|\Theta) \, {\bf y}_{m} $, nie ma wpływu na rozkład $p_{n}({\bf y}_{n}|\theta_{n})$ dla  indeksu próby $m \neq n$. Wtedy dane są generowane zgodnie z punktowymi rozkładami spełniającymi warunek: 
\begin{eqnarray}
\label{rozklady punktowe polozenia}
p_{n}({\bf y}_{n}|\Theta) = p_{n}({\bf y}_{n}|{\theta}_{n}) \; , \;\;\;\; {\rm gdzie} \;\;\; n=1,...,N \; ,
\end{eqnarray}
a   wiarygodność próby jest iloczynem:
\begin{eqnarray}
\label{funkcja wiaryg param wektorowy}
P\left(y|\Theta\right) = \prod_{n=1}^{N} p_{n}({\bf y}_{n}|\theta_{n}) \; . 
\end{eqnarray}
%
{\bf Pojemność informacyjna kanału}: Dla parametru położenia (czaso)przestrzennego, pojemność kanału informacyjnego $I$ dla układu, zdefiniowana ogólnie w (\ref{pojemnosc C}), przyjmuje 
postać\footnote{Lokalne 
własności funkcji wiarygodności $P(y|\Theta)$ opisuje obserwowana macierz informacji Fishera:
\begin{eqnarray}
\label{observed IF empirical statistics} \texttt{i\!F} = \left(-
\frac{\partial^{2} ln P(\Theta)}{\partial \theta_{n'}^{\nu}
\partial \theta_{n}^{\mu}}\right) \, ,
\end{eqnarray}
która pozostaje symetryczna i dodatnio określona.
Jak wiemy (por. Rozdział~\ref{alfa koneksja}) jej wartość oczekiwana na ${\cal B}$ zadaje geometryczną strukturę nazywaną metryką Rao-Fishera na przestrzeni statystycznej $S$  
\cite{Amari Nagaoka book}. 
}:
\begin{eqnarray}
\label{pojemnosc informacyjna Minkowskiego}
I = \sum\limits_{n=1}^{N}{I_{F n}} \; ,
\end{eqnarray}
gdzie 
\begin{eqnarray}
\label{pojemnosc C dla polozenia}
I_{F n} &\equiv& I_{F n} \left(\theta_{n} \right) = \int_{\cal B} {dy \, P\left(y|\Theta\right)} \, \left( \, \nabla_{\theta_{n}} \ln P\left(y|\Theta\right) \cdot \nabla_{\theta_{n}} \ln P\left(y|\Theta\right) \, \right) \nonumber \\
& = & 
{\int_{\cal B}{dy \; P\left(y|\Theta\right) \sum\limits_{\nu, \, \mu = (0),1,2,...} \eta^{\nu \mu} {\left(\frac{\partial\ln P\left(y|\Theta\right)}{\partial\theta_{n}^{\nu}} \; \frac{\partial \ln P \left(y|\Theta\right)}{\partial\theta_{n}^{\mu}} \right)}}}
\; .
\end{eqnarray}
Tensor $(\eta^{\nu \mu}) = diag(1,-1,-1,-1)$ jest tensorem dualnym do $(\eta_{\nu \mu})$, tzn. $\sum_{\mu=0}^{3} \eta_{\nu \mu} \eta^{\gamma \mu}$ $= \delta_{\nu}^{\gamma}$, przy czym    $\delta_{\nu}^{\gamma}$ jest deltą Kroneckera, a  
$\nabla_{\theta_{n}} \equiv \frac{\partial}{\partial \theta_{n}} = \! \sum_{\nu} \frac{\partial}{\partial \theta_{n}^{\nu}} \, d {\bf y}_{n}^{\nu} \, $. W związku (\ref{pojemnosc C dla polozenia}) ``$\cdot$'' oznacza iloczyn wewnętrzny zdefiniowany przez 
tensor metryczny $(\eta^{\nu \mu})$.
\\
\\
{\bf Uwaga o niefaktoryzowalności czasoprzestrzennych indeksów położenia}: $\;\;$ W pomiarze wybranej $\nu$-tej współrzędnej położenia nie można  wykluczyć odchyleń (fluktuacji) wartości współrzędnych do niej ortogonalnych. 
Oznacza to, że wartość oczekiwana $\nu$-tej współrzędnej położenia nie jest w (\ref{wartosc oczekiwana EY}) liczona z jakiegoś rozkładu typu $p({\bf y}^{\nu})$, lecz musi być liczona z łącznego rozkładu $p({\bf y})$ dla wszystkich współrzędnych ${\bf y}^{\nu}$. 
W konsekwencji, w przypadku zmiennych położenia przestrzennego i ich parametrów naturalnych określonych w (\ref{wartosc oczekiwana EY}),  całkowanie w (\ref{pojemnosc C dla polozenia}) nie może zostać sfaktoryzowane ze względu na współrzędną wektorową $\nu$. \\
Zgodnie z powyższą uwagą, wariancja  
estymatora $\hat{\theta}_{n} \left({y}\right)$ parametru  $\theta_{n}$ 
powinna przyjąć postać: 
\begin{eqnarray}
\label{wariancja estymatora theta i alfa}
\;\;\;\;\;\;\; {\sigma^{2}} \,( \hat{\theta}_{n}) &=& \int_{\cal B} dy \, P\left(y|\Theta\right) \left(\hat{\theta}_{n} \left(y\right) - \theta_{n} \right) \cdot \left(\hat{\theta}_{n}  \left(y\right)-\theta_{n} \right) \\
&=& \int_{\cal B} dy \, P\left(y|\Theta\right) \sum\limits_{\nu, \, \mu = (0),1,2,...} \eta_{\nu \mu} \left(\hat{\theta}_{n}^{\nu} \left(y\right) - \theta_{n}^{\nu} \right) \, \left(\hat{\theta}_{n}^{\mu}  \left(y\right) - \theta_{n}^{\mu} \right) 
\; . \nonumber
\end{eqnarray}
Rozważania przedstawione na końcu Rozdziału~\ref{Przyklad wektorowego DORC} oznaczają, że ze względu na (\ref{rozklady punktowe polozenia}), dla każdego  wyróżnionego parametru $\theta_{n}\,$ wariancja ${\sigma^{2}} \, (\hat{\theta}_{n})$  jego estymatora (\ref{wariancja estymatora theta i alfa}) w 
jego kanale informacyjnym jest związana z informacją Fishera 
$I_{F n} = I_{F n}(\theta_{n})$  
parametru  $\theta_{n}$ poprzez nierówność (\ref{I11 rowna sie I11do-1 przypadek diagonalny}):  
\begin{eqnarray}
\label{Rao Cram uogolnienie}
\frac{1}{{\sigma^{2}}\,\left(\hat{\theta}_{n} \right)} \leq  \frac{1}{I_{F}^{n}} \leq  I_{F n} \;\;\; {\rm gdzie} \;\;\; n=1,2,...,N \; ,
\end{eqnarray}
będącą 
uogólnieniem nierówności  informacyjnej  Rao-Cramera (\ref{tw R-C dla par skalarnego}), 
gdzie  $I^{n}_{F}$ jest DORC dla parameteru ${\theta}_{n}$. \\
%
\\
{\bf Uwaga o zmiennych Fisher'owskich}: Faktu {\it zależności statystycznej} zmiennych  położenia przestrzennego dla różnych indeksów ${\nu}$ nie należy mylić z posiadaną przez nie  {\it niezależnością analityczną}, która oznacza, że zmienne $Y$ są tzw. zmiennymi Fisher'owskimi, dla których: 
\begin{eqnarray}
\label{zmienne Fisherowskie}
\frac{\partial {\bf y}^{\nu}}{\partial {\bf y}^{\mu}} = \delta^{\nu}_{\mu}  \; .
\end{eqnarray}
\\
%
%
%
%
{\bf Związek pomiędzy informacją Stama i pojemnością $I$}:  
Wielkość $\frac{1}{{\sigma^{2}} \, (\hat{\theta}_{n})}$ 
odnosi się do estymacji w $n$-tym kanale informacyjnym. 
Sumując po 
indeksie $n$ otrzymujemy {\it informację Stama} $I_{S}$ \cite{Stam,Frieden}:
\begin{eqnarray}
\label{informacja Stama Minkowskiego}
0 \leq  I_{S} \equiv \sum\limits_{n=1}^{N} \frac{1}{{\sigma^{2}}\,(\hat{\theta}_{n})} =: \sum\limits_{n=1}^{N} I_{S n} \;\, .
\end{eqnarray}
%
%
Ponieważ $\theta_{n} = (\theta_{n}^{\nu})$ jest parametrem wektorowym, zatem $I_{S n} = \frac{1}{{\sigma^{2}}\,(\hat{\theta}_{n})}$ jest informacją Stama  czasoprzestrzennych kanałów dla $n$-tego pomiaru w 
próbie\footnote{Uwzględnienie 
metryki Minkowskiego w definicji informacji Stama można zrozumić również jako konsekwencję ogólnego wskazania przy liczeniu średniej kwadratowej wielkości mierzalnej w dowolnej metryce Euklidesowej. W sytuacji gdy obok indeksów przestrzennych $x_{i}$, $i=1,2,3$, występuje indeks czasowy $t$, należy w rachunkach w czterowymiarowej czasoprzestrzeni Euklidesowej uwzględnić we współrzędnych przestrzennych jednostkę urojoną $i$, {\it łącznie z uwzględnieniem tego faktu w prawie propagacji błędów}.  W związku z tym  w oryginalnej analizie EFI Friedena-Soffera, zmienna losowa czterowektora położenia oraz jej wartość oczekiwana mają  odpowiednio postać $(Y_{0} = \,c\,T,  i\, \vec{Y})$ oraz  $(\theta_{0} = \,c \,E_{\Theta}(T), \,\vec{\theta} = i \, E(\vec{Y}))$, co nie zmienia rezultatów analizy zawartej w skrypcie, odnoszącej się do równań mechaniki falowej oraz termodynamiki. Nie zmienia to również rezultatów analizy relatywisycznej mechaniki kwantowej  \cite{Sakurai 2}. 
Jednakże opis wykorzystujący metrykę Minkowskiego wydaje się autorowi skryptu korzystniejszy z punktu widzenia zrozumienia  konstrukcji niepodzielnego ekperymantalnie kanału informacyjnego (por. Uwaga o indeksie próby). 
}. 
\\ 
W końcu, sumując lewą i prawą stronę w (\ref{Rao Cram uogolnienie}) względem indeksu $n$  i biorąc pod uwagę  (\ref{informacja Stama}),  zauważamy, że   $I_{S}$ spełnia nierówność: 
\begin{eqnarray}
\label{informacja Stama vs pojemnosc informacyjna Minkowskiego}
0 \leq I_{S} \equiv \sum\limits_{n=1}^{N} I_{S n} \le \sum\limits_{n=1}^{N} {I_{F n}} = I   
\; ,
\end{eqnarray}
gdzie $I$ jest pojemnością kanału informacyjnego (\ref{pojemnosc informacyjna Minkowskiego}). 
Nierówność (\ref{informacja Stama vs pojemnosc informacyjna Minkowskiego}) jest minimalnym uogólnieniem ``jednokanałowej'' nierówności Rao-Cram{\'e}r'a  (\ref{Rao Cram uogolnienie}), potrzebnym z punktu widzenia przeprowadzanego pomiaru.
%
\\
\\
Z punktu widzenia modelowania fizycznego, pojemność kanału informacyjnego  $I$ jest najważniejszym pojęciem statystycznym,  
leżącym u podstaw 
członów kinematycznych \cite{Frieden} różnych modeli teorii pola. %
%
%
%
%
Zgodnie z (\ref{informacja Stama}) okazało się, że zarówno dla  metryki Euklidesowej (\ref{metryka E}) jak i metryki Minkowskiego  (\ref{metryka M}), estymacja jest wykonywana dla dodatniej informacji $I_{S}$. 
Zatem z  (\ref{informacja Stama vs pojemnosc informacyjna}) wynika, że  $I$ jest również nieujemna. 
W Rozdziale~\ref{Informacja Fouriera} według (\ref{I by Fourier to m}) 
okaże się, że  $I$ 
jest nieujemnie zdefiniowana dla teorii pola dla cząstek, które mają nieujemny kwadrat masy \cite{dziekuje za neutron}.  
Chociaż w każdym szczególnym modelu teorii pola  
z przestrzenią Minkowskiego fakt ten powinien zostać sprawdzony, to z punktu widzenia teorii estymacji jest jasne, że:
\begin{eqnarray}
\label{przyczynowosc}
{\sigma^{2}}\,(\hat{\theta}_{n}) \geq 0 \; .
\end{eqnarray} 
Sytuacja ta ma zawsze miejsce dla {\it procesów przyczynowych}. \\
\\
\\
{\bf Uwaga o indeksie próby i kanale pomiarowym}: Indeks próby $n$ jest najmniejszym   indeksem kanału informacyjnego, w którym dokonywany jest pomiar. Tzn. gdyby indeks próby można było dodatkowo ``zaindeksować'', np. indeksem czasoprzestrzennym, wyznaczając podkanały, to i tak nie można by dokonać pomiaru tylko w jednym z tak wyznaczonych  podkanałów (nie dokonując go równocześnie w pozostałych podkanałach posiadających indeks próby $n$). {\it Kanał  niepodzielny z punktu wiedzenia eksperymentu nazwijmy kanałem pomiarowym}. \\
\\
%
{\bf Uwaga o analizie we fragmencie kanału pomiarowego}: W przypadku ograniczenia analizy do fragmentu kanału pomiarowego należy się upewnić, czy pozostała w analizie część informacji Stama ma wartość dodatnią. Np. w przypadku zaniedbania czasowo zaindeksowanej części  czasoprzestrzennego kanału pomiarowego, otrzymana nierówność Stama dla  składowych przestrzennych ma postać:
\begin{eqnarray}
\label{nierownosc informacja Stama dla podkanalow przestrzennych}
\!\!\! 0 \leq  I_{S} \! &=& \! \sum_{n=1}^{N} I_{S n} \nonumber \\
& & \leq \sum_{n=1}^{N} {\int{d\vec{y} \, P\left(\vec{y}|\vec{\Theta}\right) \sum \limits_{i=1}^{3}{\frac{\partial\ln P\left(\vec{y}|\vec{\Theta}\right)}{\partial\theta_{n i}} \; \frac{\partial\ln P\left(\vec{y}|\vec{\Theta}\right)}{\partial\theta_{n i}}}}} =: \sum_{n=1}^{N} I_{F n} = I\; ,
\end{eqnarray}
gdzie znak wektora oznacza, że analiza zarówno w przestrzeni próby  jak i przestrzeni parametrów została obcięta do części przestrzennej zmiennych losowych i parametrów. \\
%
\\
{\bf Uwaga o symetrii}: Z punktu widzenia pomiaru, błąd estymacji 
składowych wchodzących jednocześnie w wyznaczenie {\it długości} czterowektora w $n$-tym kanale, jest niezależny od układu współrzędnych w przestrzeni Minkowskiego.  Dlatego wielkość $I_{S n} =  1/{\sigma^{2}}\,(\hat{\theta}_{n})$ określona poprzez   
(\ref{informacja Stama Minkowskiego}) oraz (\ref{wariancja estymatora theta i alfa}) jest dla tensora metrycznego  (\ref{metryka M}) niezmiennicza ze względu na transformację Lorentz'a (pchnięcia i obroty). W przypadku tensora metrycznego (\ref{metryka E}) jest ona niezmiennicza ze względu na transformacje Galileusza. \\
Jeśli chodzi o pojemność informacyjną  $I$, to w przypadku niezależności pomiarów w próbie, jest ona również niezmiennicza ze względu na transformację Lorentz'a  w przestrzeni z metryką Minkowskiego (czy transformację Galileusza w przestrzeni Euklidesowej), o ile niezmiennicze jest każde $I_{n}$.\footnote{Pojemność informacyjna: 
\begin{eqnarray}
\label{one channel Fisher_information - I z iF}
I_{n} = \int_{\cal B} d y \, P(y|\Theta) \, \sum_{\nu=0}^{3} \texttt{i\!F}_{n n \nu}^{\;\;\;\;\, \nu} \; 
\end{eqnarray}
jest niezmiennicza ze względu na gładkie odwracalne odwzorowania $Y \rightarrow  X$, gdzie $X$ jest nową zmienną \cite{Streater}. Jest ona również niezmiennicza ze względu na odbicia przestrzenne i czasowe.
}
Warunki niezmienniczości $I_{Sn}$ oraz $I$ schodzą się,  
gdy w nierówności Rao-Cramera (\ref{Rao Cram uogolnienie}) osiągana jest równość. \\
Nierówność Rao-Cramera okazuje się niezmiennicza ze względu na podstawowe transformacje 
\cite{Frieden,PPSV}. Istotnie, zgodnie z powyższymi rozważaniami, właściwym pomiarem niezależnym od przyjętego układu współrzędnych jest pomiar kwadratu długości $\sum_{\nu=0}^{3} {\bf y}_{\nu} {\bf y}^{\nu}$, a nie pojedynczej współrzędnej ${\bf y}^{\nu}$.  
Więcej na temat niezmienniczości DORC 
ze względu na przesunięcie, odbicie przestrzenne,
obroty i transformację affiniczną oraz transformacje unitarne można znaleźć w \cite{PPSV}. \\
\\
{\bf Kryterium minimalizacji $I$ ze względu na $N$}: Na koniec zauważmy, że sumowanie w (\ref{pojemnosc C dla polozenia})
przebiega od $n=1$ do $n=N$. Każdy $n$-ty wyraz w sumie wnosi analityczny wkład jako stopień swobody dla $I$. O ile dodane stopnie swobody nie  wpływają na już istniejące, to ponieważ każdy, cały  wyraz w sumie po $n$ jest nieujemny, to informacja $I$ ma tendencje do wzrostu wraz ze wzrostem $N$. Kryterium minimalizacji $I$ ze względu na $N$  posłużyło Friedenowi i Sofferowi jako dodatkowy warunek przy konstrukcji np. równań ruchu. Nie znaczy to, że modele z większym $N$ zostały automatycznie wykluczone, tylko że  im większe jest $N$ tym więcej stopni swobody wchodzi do opisu obserwowanego zjawiska i opisywane zjawisko jest bardziej złożone. Zagadnienie to omówimy w przykładach, w dalszej części skryptu. 
%
%
%
%

\section[Pomiar układu w podejściu Friedena-Soffera]{Pomiar układu w podejściu Friedena-Soffera}

\label{Podstawowe zalozenie Friedena-Soffera}

Jak w Rozdziale~\ref{Poj inform zmiennej los poloz}, rozważmy zmienną losową $Y$ położenia układu, przyjmującą wartość ${\bf y}$, która jest punktem zbioru  ${\cal Y}$. 
Może to być punkt czasoprzestrzenny przestrzeni Minkowskiego, co ma miejsce w rozważaniach związanych z opisem układu np. w mechanice falowej. Wartości ${\bf y} \equiv ({\bf y}^{\nu})_{\nu=0}^{3} \in {\cal Y} \equiv \mathbb{R}^{4}$ są realizowane zgodnie z rozkładem  $p({\bf y})$ właściwym dla układu\footnote{Przy wyprowadzaniu równań generujących rozkład w fizyce statystycznej ${\bf y}$ może być np. wartością energii $\epsilon$ układu \cite{Frieden} i wtedy ${\bf y} \equiv \epsilon \in {\cal Y} \equiv \mathbb{R}$.}. \\
 \\
{\bf Podstawowe założenie fizyczne podejścia Friedena-Soffera}:  Niech dane $y \equiv ({\bf y}_{n})_{n=1}^{N} = ({\bf y}_{1}, ..., {\bf y}_{N})$ są  realizacjami próby dla położeń układu, gdzie ${\bf y}_{n} \equiv ({\bf y}_{n}^{\nu})_{\nu=0}^{3}$.
{\it Zgodnie z założeniem zaproponowanym przez Frieden'a i Soffer'a} \cite{Frieden}{\it, ich zebranie następuje przez sam układ w zgodzie z rozkładami gęstości prawdopodobieństwa, $p_{n}({\bf y}_{n}|\theta_{n})$, gdzie $n=1,...,N$}.  \\
\\
Treść powyższego założenia fizycznego można wypowiedzieć następująco: {\it Układ próbkuje dostępną mu czasoprzestrzeń, ``zbierając dane i dokonując analizy statystycznej'', zgodnie z zasadami informacyjnymi} (wprowadzonymi w Rozdziale~\ref{Zasady informacyjne}). \\
\\
{\bf Przestrzenie statystyczne próby ${\cal S}$, punktowa ${\cal S}_{4}$ oraz ${\cal S}_{N \times 4} \;$}: Niech rozważana przestrzeń jest czasoprzestrzenią Minkowskiego.  Wtedy każdy z rozkładów $p_{n}({\bf y}_{n}|\theta_{n})$ jest punktem modelu statystycznego  ${\cal S}_{4} = \{p_{n}({\bf y}_{n}|\theta_{n})\}$  parametryzowanego przez naturalny parametr, tzn. przez wartość oczekiwaną $\theta_{n} \equiv (\theta_{n}^{\nu})_{\nu=0}^{3} = E(Y_{n})$, jak w  (\ref{wartosc oczekiwana EYn}). 
Zbiór wartości $d = 4 \times N$ - wymiarowego parametru  $\Theta=(\theta_{n})_{n=1}^{N}$ tworzy współrzędne dla łącznego rozkładu  
$P(y\,|\Theta)$, będącego punktem na $d=4 \times N$-wymiarowej rozmaitości, 
która jest (pod)przestrzenią statystyczną  ${\cal S} \subset \Sigma({\cal B})$ \cite{Amari Nagaoka book} (por. (\ref{model statystyczny S})): 
\begin{eqnarray}
\label{model statystyczny S z Theta}
{\cal S} = \{P_{\Theta} \equiv P(y\,|\Theta),  \Theta \equiv (\theta_{n})_{n=1}^{N} \in  {V}_{\Theta} \subset \Re^{d} \} \; , 
\end{eqnarray}
określoną na przestrzeni próby ${\cal B}$. Jak wiemy z Rozdziału~\ref{alfa koneksja}, rozmaitość ${\cal S}$ jest rodziną rozkładów prawdopodobieństwa  parametryzowaną przez rzeczywistą, nie losową zmienną  wektorową 
$\Theta \equiv (\theta_{n})_{n=1}^{N} \in {V}_{\Theta}$, 
która w rozważanym przypadku parametru położenia, tworzy $N \times 4$-wymiarowy lokalny układ współrzędnych. 
%
%
%
%
Zatem, ponieważ próba $\widetilde{Y}$ jest $N \times 4$-wymiarową zmienną losową, więc 
wymiary przestrzeni próby ${\cal B}$ i wektorowego parametru  $\Theta \equiv (\theta_{n}^{\nu})_{n=1}^{N}$ są takie same\footnote{Jednakże przypomnijmy, że w ogólnym przypadku estymacji, wymiar wektora parametrów  $\Theta \equiv (\theta_{i})_{i=1}^{d}$ oraz wektora próby $y \equiv ({\bf y}_{n})_{n=1}^{N}$ może być inny. 
}.  \\
W przyszłości okaże się, że z powodu  związku (\ref{rozklady punktowe polozenia}) 
analiza na przestrzeni statystycznej określonej przez (\ref{model statystyczny S z Theta}) z parametrami $(\theta_{n}^{\nu})_{n=1}^{N}$ tworzącymi   $N \times 4$-wymiarowy lokalny układ współrzędnych, efektywnie redukuje się do analizy na 
${\cal S}_{N \times 4}  \equiv \{\bigoplus_{n=1}^{N} p_{n}(y\,|\theta_{n}) \}$. 
Jednakże, ponieważ wartości parametru  $\theta_{n}$ mogą się zmieniać od jednego punktu $n$ próby do innego punktu $n'$, zatem 
nie może być ona w ogólności sprowadzona do analizy na  ${\cal S}$ poprzez samo przeskalowanie  metryki Riemanna oraz (dualnej) koneksji na ${\cal S}_{4}$  przez czynnik $N$, jak to ma miejsce we wnioskowaniu pojawiającym się w statystyce klasycznej  \cite{Amari Nagaoka book}. 
W przyszłości liczbę $N$ parametrów $\theta_{n}$, będącą wymiarem próby $y \equiv ({\bf y}_{n})_{n=1}^{N}$,  będziemy nazywali rangą  pola. \\
\\
{\bf Uwaga o podmiocie}: Jednak pamiętajmy, że estymacji dokonuje tylko człowiek, zatem na metodę EFI należy patrzeć tylko jak na {\it pewien  model analizy statystycznej}.\\
\\
Interesująca nas  statystyczna procedura estymacyjna dotyczy wnioskowania o $p_{n}({\bf y}_{n}|\theta_{n})$ na podstawie danych $y$ z wykorzystaniem funkcji wiarygodności $P(y|\Theta)$. 
Załóżmy, że dane zbierane przez układ $y = \left({\bf y}_1,{\bf y}_2,...,{\bf y}_N\right)$ są uzyskane niezależnie tak, że łączny rozkład prawdopodobieństwa dla próby faktoryzuje się na rozkłady brzegowe:
\begin{eqnarray}
P(\Theta) \equiv P\left(y|\Theta \right) = \prod\limits_{n=1}^N {p_n \left({\bf y}_n|\Theta \right)} = \prod\limits_{n=1}^N {p_n\left(y_n| \theta_n \right)} \; ,
\end{eqnarray}
gdzie w ostatniej równości skorzystano z założenia, że parametr $\theta_{m}$ dla $m \neq n$ nie ma wpływu na rozkład zmiennej $Y_n$.\\
\\
{\bf Wstępne określenie postaci kinematycznej $I$}: Centralna część pracy Frieden'a i Soffer'a jest związana  z przejściem od pojemności informacyjnej  $I$ zadanej równaniem (\ref{pojemnosc C dla polozenia}) oraz (\ref{pojemnosc informacyjna Minkowskiego}):
\begin{eqnarray}
\label{pojemnosc C dla polozenia - powtorka wzoru}
I = \sum\limits_{n=1}^{N} I_{n} = \sum\limits_{n=1}^{N}{\int_{\cal B}{dy \; P\left(y|\Theta\right)\sum\limits _{\nu=0}^{3}{\left(\frac{\partial\ln P\left(y|\Theta\right)}{\partial\theta_{n \nu}} \; \frac{\partial\ln P\left(y|\Theta\right)}{\partial\theta_{n}^{\nu}} \right)}}} \;  ,
\end{eqnarray}
gdzie $d y:= d^{4}{\bf y}_{1}...d^{4}{\bf y}_{N}\,$ oraz $d^{4}{\bf y}_{n}=d {\bf y}_{n}^{0} d {\bf y}_{n}^{1} d {\bf y}_{n}^{2} d {\bf y}_{n}^{3} \,$, do tzw. postaci kinematycznej wykorzystywanej w teorii pola oraz fizyce statystycznej. \\
  \\
%
%
Rachunek analogiczny jaki doprowadził z (\ref{inf I jeden parametr - kwadrat 1 pochodnej})  do (\ref{I dla pn jeden parametr}) wygląda  teraz w skrócie następująco. 
Przekształćmy  pochodną $\ln P$ w (\ref{pojemnosc C dla polozenia - powtorka wzoru}) do postaci:
\begin{eqnarray}
\frac{\partial \ln P \left(y|\Theta \right)}{\partial \theta_{n\nu}} = \frac{\partial}{\partial \theta_{n\nu}}\sum\limits_{n=1}^N {\ln p_{n} \left({\bf y}_n|\theta_n \right)} = \sum\limits_{n=1}^N {\frac{1}{p_n \left({\bf y}_n|\theta_n \right)} \frac{\partial{p_n \left({\bf y}_n|\theta_n \right)} }{\partial \theta_{n\nu}}} \; .
\end{eqnarray}
Pamiętając o unormowaniu każdego z rozkładów brzegowych, $\int_{\cal Y} d^{4}{\bf y_{n}} $ $ p_n \left({\bf y}_n|\theta_n \right) =  1$,  otrzymujemy postać pojemności informacyjnej: 
\begin{eqnarray}
\label{postac I bez log p po theta}
I = \sum_{n=1}^N {\int_{\cal Y} d^{4}{\bf y}_n \frac{1}{{p_{n} \left(  {\bf y}_n|\theta_{n}  \right)}} \sum\limits_{\nu=0}^{3}  {\left( {\frac{{\partial p_{n} \left(  {\bf y}_n|\theta_{n}  \right)}}{{\partial \theta_{n \nu} }}}  {\frac{{\partial p_{n} \left(  {\bf y}_n|\theta_{n}  \right)}}{{\partial \theta_{n}^{ \nu} }}} \right) } } \; ,
\end{eqnarray}  
będącą uogólnieniem (\ref{I dla pn jeden parametr}). \\
\\
W końcu przejdźmy do  amplitud $q_{n}\left( {\bf y}_n|\theta_{n}  \right)$ określonych jak w (\ref{amplituda a rozklad}): 
\begin{eqnarray}
p_{n} \left(  {\bf y}_n|\theta_{n}  \right) = q_{n}^{2}\left(  {\bf y}_n|\theta_{n}  \right) \; .
\end{eqnarray}
Proste rachunki dają: 
\begin{eqnarray}
\label{potrz}
I = 4 \sum\limits_{n=1}^N \int_{\cal Y} {d^{4}{\bf y}_{n} \sum\limits_{\nu=0}^{3}  {\left( {\frac{{\partial q_{n} \left(  {\bf y}_n|\theta_{n}  \right)}}{{\partial \theta_{n \nu} }}}  {\frac{{\partial q_{n} \left(  {\bf y}_n|\theta_{n}  \right)}}{{\partial \theta_{n}^{ \nu} }}} \right) } } \; , 
\end{eqnarray}
czyli prawie kluczową postać pojemności informacyjnej dla rachunku metody EFI Friedena-Soffera. Jedyne co trzeba jeszcze zrobić, to przejść od  przestrzeni statystycznej ${\cal S}$ z bazą 
$(\theta_{n})_{n=1}^{N}$ dla reprezentacji amplitud danych pomiarowych ${\bf y}_{n} \in {\cal Y}$, do przestrzeni amplitud  przesunięć  ${\bf x}_{n} := {\bf y}_{n} - \theta_{n}$  określonych na przestrzeni bazowej ${\cal X}$.
Poświęcimy temu zagadnienu Rozdział~\ref{The kinematical form of the Fisher information}.

\subsection{Przykład: Estymacja w fizycznych modelach  eksponentialnych}

\label{Estymacja w modelach fizycznych na DORC}

W Rozdziale~\ref{Dualne uklady wspolrzednych} wprowadzone zostało pojęcie dualnych affinicznych układów współrzędnych na przestrzeni statystycznej ${\cal S}$. Obecny rozdział poświęcony jest zastosowaniu modeli eksponentialnych (\ref{exponential family}) i szczególnej roli {\it parametrów dualnych} związanych z koneksją $\nabla^{(-1)}$. Rodzina modeli eksponentialnych wykorzystywana jest w teorii estymacji szeregu  zagadnień  fizycznych. Jej najbardziej znaną realizacją jest estymacja metodą maksymalnej entropii, 
sformułowana w poniższym twierdzeniu dla wymiaru próby $N=1$. \\
\\
Niech:
\begin{eqnarray}
\label{entropia w dowodzie o maks entropii}
S_{H}(p) = - \int_{\cal Y} d{\bf y}  p({\bf y}|\Xi) \ln p({\bf y}|\Xi) \; ,  \;\;\;\;\; \forall\, p_{\Xi} \in {\cal S} \; 
\end{eqnarray}
jest entropią Shannona stanu układu zadanego rozkładem $p({\bf y}|\Xi)$ z parametrami $\Xi$, a $F_{i}(Y)$, $i=1,2,...,d$, układem niezależnych zmiennych losowych o określonych wartościach oczekiwanych: 
\begin{eqnarray}
\label{wartosi oczekiwane eta model eksponentialny}
\theta_{i} =  \theta_{i}(\Xi) = E_{\Xi}\left[ F_{i}(Y) \right] =   \int_{\cal Y} d{\bf y} p({\bf y}|\Xi) F_{i}({\bf y}) \; , \;\;\;\; i = 1,2,...,d \;\;\;\;\; , \;\;\;\;\; \forall\, p_{\Xi} \in {\cal S} \; .
\end{eqnarray} 
Z (\ref{wartosi oczekiwane eta model eksponentialny}) widać, że $F_{i}(Y)$ są {\it nieobciążonymi estymatorami parametrów} $\theta_{i}$, tzn.:
\begin{eqnarray}
\label{Fi jako estymatory wartosi oczekiwane eta model eksponentialny}
\hat{\theta}_{i} =  F_{i}(Y) \; , \;\;\;\; i=1,2,...,d \; .
\end{eqnarray} 
{\bf Twierdzenie o stanie z maksymalną entropią (TME)}.  Istnieje jednoznacznie określony  unormowany stan układu posiadający {\it maksymalną entropię}, zadany  następująco: 
\begin{eqnarray}
\label{model eksponentialny dla stanu maks entropii}
p_{\Xi} \equiv p({\bf y}|\Xi) = Z^{-1} \exp \left(  \sum_{i=1}^{d} \xi^{i} F_{i}({\bf y}) \right) \; ,  \;\;\;\;\; \forall\, p_{\Xi} \in {\cal S} \; 
\end{eqnarray}
w bazie kanonicznej $\Xi \equiv (\xi^{i})_{i=1}^{d}$ modelu  eksponentialnego (\ref{exponential family}), 
gdzie stała normalizacyjna $Z$ jest tzw. {\it funkcją partycji}  (podziału) \cite{Jurek Dajka kwantowe metody opisu}: 
\begin{eqnarray}
\label{model eksponentialny funkcja podzialu}
Z \equiv Z(\Xi) = \int_{\cal Y} d{\bf y} \exp \left(  \sum_{i=1}^{d} \xi^{i} F_{i}({\bf y}) \right) \; .
\end{eqnarray}\\
{\bf Dowód}: Zgodnie z {\it twierdzeniem Lagrange'a} wiemy, że maksimum warunkowe dla funkcji $S_{H}(p)$, przy dodatkowym warunku normalizacyjnym: 
\begin{eqnarray}
\label{normalizacja dla rozkl eksp}
\int_{\cal Y} d{\bf y} \,p({\bf y}|\Xi) = 1 \; 
\end{eqnarray}
oraz $d$  warunkach związanych z zadaniem wartości oczekiwanych (\ref{wartosi oczekiwane eta model eksponentialny}), jest równoważne wyznaczeniu bezwarunkowego ekstremum funkcji: 
\begin{eqnarray}
\label{funkcja S warunkowa z norm i d war}
S_{war}(p)\! &=& \! \int_{\cal Y} d{\bf y} s_{war}(p) := S_{H}(p) -   \xi^{0} \left (1 - \int_{\cal Y} d{\bf y} \,p({\bf y}|\Xi) \right) -  \sum_{i=1}^{d} \xi^{i} \left( \theta_{i} - \int_{\cal Y} d{\bf y} \, p({\bf y}|\Xi) F_{i}({\bf y}) \right) \nonumber \\
&=&  \! \int_{\cal Y} d{\bf y} \, p({\bf y}|\Xi) \left( - \ln p({\bf y}|\Xi) +  \xi^{0} + \sum_{i=1}^{d} \xi^{i}  F_{i}({\bf y}) \right) -  \xi^{0} -   \sum_{i=1}^{d} \xi^{i} \theta_{i} \; ,
\end{eqnarray}
ze względu na wariację $p({\bf y}|\Xi)$, 
gdzie $\xi^{0}$ oraz $(\xi^{i})_{i=1}^{d}$ są czynnikami Lagrange'a. 
Ponieważ $S_{H}(p)$ jest ścisle wklęsła \cite{Bengtsson_Zyczkowski}, zatem otrzymujemy {\it maksimum, które jest wyznaczone jednoznacznie}. \\
\\
Warunek ekstremizacji funkcjonału\footnote{Funkcjonał, w tym przypadku $S_{war}(p)$, jest liczbą, której wartość zależy od funkcji $p({\bf y}|\Xi)$. 
} 
$S_{war}(p)$ ze względu na $p({\bf y}|\Xi)$, tzn. $\delta_{(p)} S_{war} = 0$, prowadzi do równania Eulera-Lagrange'a: 
\begin{eqnarray}
\label{EL eq dla entropii}
\frac{\partial }{\partial {\bf y}} \left({\frac{{\partial s_{war}}}{{\partial \left(  \frac{\partial p({\bf y}|\Xi)}{\partial {\bf y}}  \right) }}}\right) = \frac{{\partial s_{war}}}{{\partial p({\bf y}|\Xi)}}  \;\; ,
\end{eqnarray}
gdzie zgodnie z (\ref{funkcja S warunkowa z norm i d war}) postać funkcji podcałkowej $s_{war}$ wynosi: 
\begin{eqnarray}
\label{postac s_war}
s_{war} = p({\bf y}|\Xi) \left( - \ln p({\bf y}|\Xi) +  \xi^{0} + \sum_{i=1}^{d} \xi^{i}  F_{i}({\bf y}) \right) \; .
\end{eqnarray}
Podstawiając $s_{war}$ do (\ref{EL eq dla entropii}) otrzymujemy:
\begin{eqnarray}
\label{wariacja S warunkowanego}
- \ln p({\bf y}|\Xi)  + \xi^{0} + \sum_{i=1}^{d} \xi^{i} F_{i}({\bf y})  - 1 = 0 \; .
\end{eqnarray}
Równanie (\ref{wariacja S warunkowanego}) daje szukaną postać rozkładu $p({\bf y}|\Xi)$ maksymalizującego entropię $S_{H}(p)$:
\begin{eqnarray}
\label{rozwiazanie dla stanu maks entropii}
p({\bf y}|\Xi) = A \, \exp \left( \sum_{i=1}^{d} \xi^{i} F_{i}({\bf y}) \right) \; , \;\;\;\; {\rm gdzie} \;\;\;\; A = 1/\exp(1 - \xi^{0}) = const. \;\; , 
\end{eqnarray}
który jak to widać z (\ref{exponential family}) jest typu eksponentialnego z parametrami kanonicznymi $\xi^{i}$,  $i=1,2,...,d$  oraz  $C({\bf y}) = 0$. Ponadto z warunku normalizacji (\ref{normalizacja dla rozkl eksp}) otrzymujemy $A = Z^{-1}$,  gdzie $Z$ jest funkcją partycji (\ref{model eksponentialny funkcja podzialu}). c.n.d. \\
 \\
{\bf Parametry dualne}. Po rozpoznaniu, że  model maksymalizujący  entropię stanu układu jest modelem eksponentialnym  (\ref{model eksponentialny dla stanu maks entropii}) w parametryzacji kanonicznej $\Xi$,  możemy (\ref{model eksponentialny dla stanu maks entropii}) zapisać w postaci (\ref{exponential family}):
\begin{eqnarray}
\label{exponential family dla maks entrop}
p_{\Xi} \equiv p({\bf y}| \Xi) = \exp \left[
\sum_{i=1}^{d}
\xi^{i} F_{i}({\bf y})  - \psi(\Xi) \right] \; ,  \;\;\;\;\; \forall\, p_{\Xi} \in {\cal S} \; ,
\end{eqnarray}
gdzie 
\begin{eqnarray}
\label{Z dla exponential family dla maks entrop}
Z(\Xi) = \exp \left[\psi(\Xi) \right] \; .
\end{eqnarray}
Ponieważ   zgodnie z (\ref{psi dla exponential family}),   $\psi(\Xi) = \ln \int_{\cal Y} d{\bf y} \exp \left[ 
\sum_{i=1}^{d}
\xi^{i} F_{i}({\bf y})  \right] $,  zatem z (\ref{wartosi oczekiwane eta model eksponentialny}) oraz wykorzystując (\ref{exponential family dla maks entrop}),  otrzymujemy:
\begin{eqnarray}
\label{eta dla rozkl eksponentialnego entropia}
\theta_{i} = \partial_{\xi^i} \psi(\Xi) \; , \;\;\;\; i=1,2,...,d \;\;\;\;\; , \;\;\;\;\; \forall\, p_{\Xi} \in {\cal S} \; ,
\end{eqnarray}
gdzie skorzystano z oznaczenia $\partial_{\xi^i} \equiv \partial/\partial \xi^{i}$. \\
Korzystając z $\partial_{\xi^j} \partial_{\xi^i} \ln p({\bf y}|\Xi) =  - \partial_{\xi^j} \partial_{\xi^i} \psi(\Xi)$, (\ref{partial l dla psi dla exponential family}),  oraz   $g^{\xi}_{ij} = -  E_{\Xi}(\partial_{\xi^i} \partial_{\xi^j} \ln p({\bf y}|\Xi))$, (\ref{Fisher inf matrix plus reg condition}),  otrzymujemy również: 
\begin{eqnarray}
\label{row rozn 2 rzedu dla psi i eta dla eksponential}
g^{\xi}_{ij} = \partial_{\xi^i} \partial_{\xi^j} \psi(\Xi) \; , \;\;\;\; i,j=1,2,...,d \; , \;\;\;\; i,j=1,2,...,d \;\;\;\;\; , \;\;\;\;\; \forall\, p \in {\cal S} \; .
\end{eqnarray}
\\
{\bf Dualne układy modelu maksymalizującego entropię}: 
Z (\ref{Gamma 1 dla exponential family}) wiemy, że układ współrzędnych $(\xi^{i})_{i=1}^{d}$ jest $\alpha = 1$ - afinicznym układem współrzędnych modelu eksponentialnego. Zatem z (\ref{eta dla rozkl eksponentialnego entropia}) oraz (\ref{row rozn 2 rzedu dla psi i eta dla eksponential}) wynika, że $(\theta_{i})_{i=1}^{d}$ {\it jest 
$\alpha = (- 1)$ - afinicznym układem współrzędnych modelu eksponentialnego dualnym do} $(\xi^{i})_{i=1}^{d}$. \\
Parametry $\theta_{i}$,  $i=1,2,...,d$, nazywamy z przyczyn podanych powyżej  {\it parametrami dualnymi} do $(\xi^{i})_{i=1}^{d}$ lub {\it parametrami oczekiwanymi} modelu statystycznego ${\cal S}$. \\
\\
{\bf Estymacja na DORC}: Korzystając z definicji (\ref{Fisher inf matrix}) metryki Rao-Fishera, z postaci (\ref{exponential family dla maks entrop}) rozkładu eksponentialnego oraz z (\ref{eta dla rozkl eksponentialnego entropia}), otrzymujemy postać macierzy informacyjnej $I_{F}$ w parametryzacji kanonicznej $\Xi$:
\begin{eqnarray}
\label{Fisher inf matrix eksonent dla max entrop dla Theta}
\!\!\!\!\!\! I_{F\, ij}(\Xi) \equiv g^{\xi}_{ij} &=&  E_{\Xi}\left[ \partial_{\xi^i} \ln p(Y| \Xi) \; \partial_{\xi^j} \ln p(Y| \Xi) \right] \nonumber \\
&=& E_{\Xi}\left[ (F_{i}(Y)  - \theta_{i}) (F_{j}(Y)  - \theta_{j})  \right]   \; , \;\;\; i,j = 1,2,...,d \; , \;\; \forall\, p_{\Xi} \in {\cal S} \; . \;\;\;\;\;\; 
\end{eqnarray}
Niech $\hat{\Theta} \equiv ({\hat{\theta}}_{i})_{i=1}^{d}$ są estymatorami parametrów $\Theta \equiv (\theta_{i})_{i=1}^{d}$ bazy dualnej do $\Xi$. Ponieważ z (\ref{Fi jako estymatory wartosi oczekiwane eta model eksponentialny}) mamy $\hat{\theta}_{i} =  F_{i}(Y)$, $i=1,2,...,d$, zatem 
po prawej stronie (\ref{Fisher inf matrix eksonent dla max entrop dla Theta}) stoją elementy macierzy  kowariancji $V_{\Xi}(\hat{\Theta})$ estymatorów $\hat{\Theta}$: 
\begin{eqnarray}
\label{macierz kowar estymatorow Eta eksonent dla max entrop}
V_{\Xi \, ij}(\hat{\Theta}) =  E_{\Xi}\left[ (\hat{\theta}_{i}  - \theta_{i}) (\hat{\theta}_{j}  - \theta_{j})  \right]   \; , \;\;\;\; i,j = 1,2,...,d \; , \;\; \forall\, p_{\Xi} \in {\cal S} \; .
\end{eqnarray}
Równość (\ref{Fisher inf matrix eksonent dla max entrop dla Theta}) można więc zapisać następująco:
\begin{eqnarray}
\label{rownosc Fisher Theta inf i macierzy kowariancji maks entrop}
V_{\Xi}(\hat{\Theta}) = I_{F}(\Xi)
  \; , \;\;\;\;\;\;\;\;\;\;\; \forall\, p_{\Xi} \in {\cal S} \; .
\end{eqnarray}
Ponieważ macierze informacyjne w bazach dualnych są względem siebie odwrotne, tzn. $I_{F}(\Xi) = I_{F}^{-1}(\Theta)$, (\ref{macierze informacyjne dualne}), oraz macierz kowariancji określonych zmiennych losowych (w tym przypadku $\hat{\Theta}$) nie zależy od bazy w ${\cal S}$:
\begin{eqnarray}
\label{macierz kowariancji V dla Theta w roznych bazach}
V_{\Theta}(\hat{\Theta}) = V_{\Xi}(\hat{\Theta})
  \; , \;\;\;\;\;\;\;\;\;\;\; \forall\, p \in {\cal S} \; ,
\end{eqnarray}
więc  z (\ref{rownosc Fisher Theta inf i macierzy kowariancji maks entrop}) otrzymujemy następujący związek: 
\begin{eqnarray}
\label{rownosc odwrot Fisher Eta inf i m kowariancji maks entrop}
V_{\Theta}(\hat{\Theta}) = I_{F}^{-1}(\Theta)
  \; , \;\;\;\;\;\;\;\;\;\;\; \forall\, p_{\Theta} \in {\cal S} \; .
\end{eqnarray}
\\
{\bf Wniosek}: Ze względu na Twierdzenie Rao-Cramera (\ref{twierdzenie RC wersja 2}) powyższy warunek  oznacza, że estymacja parametrów dualnych jest dla modeli eksponentialnych,  spełniających warunek maksymalnej entropii dokonywana na DORC. \\
\\
\\
{\bf Przykład dualnego układu współrzędnych. Rozkład normalny}: Z Rozdziału~\ref{alfa koneksja}, wzór (\ref{rozklad normalny parametry kanoniczne}), wiemy, że rozkład normalny jest typem modelu eksponentialnego z $C({\bf y}) = 0$. Oznacza to, że może się on pojawić jako rezultat estymacji spełniającej założenia  
TME. Korzystając z postaci rozkładu normalnego oraz z (\ref{wartosi oczekiwane eta model eksponentialny}) i (\ref{rozklad normalny parametry kanoniczne}) otrzymujemy dualne parametry tego modelu:
\begin{eqnarray}
\label{parametry dualne modelu normalnego}
\theta_{1} &\equiv&  \theta_{1}(\xi^{1},\xi^{2}) = \int_{\cal Y} d{\bf y} p({\bf y}|\Xi) {\bf y} =  - \frac{\xi^{1}}{2 \xi^{2}} = \mu \; , \nonumber \\
\theta_{2} &\equiv&  \theta_{2}(\xi^{1},\xi^{2}) = \int_{\cal Y} d{\bf y} p({\bf y}|\Xi) {\bf y}^{2}  = \frac{(\xi^{1})^{2} - 2 \xi^{2} }{4 (\xi^{2})^{2}}  = \mu^{2} + \sigma^{2} \; .
\end{eqnarray}
Ponadto estymatory $\hat{\theta}_{1} = F_{1}(Y) = Y$ oraz $\hat{\theta}_{1} = F_{1}(Y) = Y^{2}$ są niezależne. 
Sprawdźmy, że zachodzi warunek 
konieczny ich niezależności, a mianowicie brak korelacji: 
\begin{eqnarray}
\label{niezaleznosc estymatorow F  modelu normalnego}
& & E_{\Xi}\left[ (\hat{\theta}_{1} - \theta_{1}) \, (\hat{\theta}_{2} - \theta_{2}) \right] = \int_{\cal Y} d{\bf y} \, p({\bf y}|\Xi) \,  (F_{1}({\bf y}) - \theta_{1}) \, (F_{2}({\bf y}) - \theta_{2})  \nonumber \\
& & = \frac{1}{\sqrt{2 \pi \, \sigma^2}}
\int_{\cal Y} d{\bf y}   \, e^{\frac{({\bf y} - \mu)^{2}}{2 \, \sigma^2}}  \left( {\bf y} - \mu \right) \, [ {\bf y}^{2} - (\mu^{2} + \sigma^{2}) ]   \\
& & = \frac{1}{\sqrt{2 \pi \, \sigma^2}} \left(
\int_{\cal Y} d{\bf y}   \, e^{\frac{({\bf y} - \mu)^{2}}{2 \, \sigma^2}}\; ({\bf y} - \mu) \, [ {\bf y}^{2} - \mu^{2} ]  
 - \sigma^{2} 
\int_{\cal Y} d{\bf y}   \, e^{\frac{({\bf y} - \mu)^{2}}{2 \, \sigma^2}} \; ({\bf y} - \mu)  \right)  = 0 \nonumber
\; ,
\end{eqnarray}
gdzie w drugiej linii skorzystano z postaci (\ref{rozklad norm theta sigma2}) rozkładu normalnego,  a w ostatniej z zerowania się obu całek z  osobna. \\
\\
{\bf Wniosek}: Z powyższych ogólnych rozważań wnioskujemy więc, że estymacja z wieloparametrowym rozkładem normalnym spełnia DORC. Poprzednio w  Rozdziale~\ref{iF oraz I_definicje}, w wyniku bezpośredniego  rachunku dla jednoparametrowej estymacji wartości oczekiwanej $\mu$, otrzymaliśmy w (\ref{RC dla 1 N z 1 par oczekiwana IF}) ten sam wynik.    \\
Jednakże dla rozkładu normalnego, w którym chcielibyśmy dokonać jednoczesnej estymacji  para\-metrów $\mu$ oraz $\sigma^{2}$, pojawiłby się problem z zastosowaniem TRC  wynikający z faktu, że dla skończonego wymiaru próby $N$ estymator $\hat{\sigma^{2}}$ jest obciążony. Obecnie wiemy, że parametrami oczekiwanymi, którymi należy się posłużyć aby zastosować TR i przekonać się, że model normalny spełnia DORC są $\mu$ oraz suma $\mu^{2} + \sigma^{2}$ (zamiast $\sigma^{2}$). Chociaż powyższy rachunek został przeprowadzony dla $N=1$, jednak wniosek dla dowolnego $N$ nie ulega zmianie. 
\\
\\
{\bf Przykład dualnego układu współrzędnych. Rozkład standardowy eksponentialny}: TME ma swoją reprezentację w fizyce statystycznej. Otóż stan w równowadze termicznej, który maksymalizuje termodynamiczną entropię Boltzmanna $S_{B}(p) := k_{B} \,S_{H}(p)$, gdzie $k_{B}$ jest dodatnią stałą Boltzmanna, posiada przy warunku $E_{\xi}[E] = \bar{\epsilon}$ nałożonym na wartość oczekiwaną (zmiennej losowej) energii $E$ cząstki gazu, rozkład Boltzmanna:  
\begin{eqnarray}
\label{rozklad Boltzmanna z max entropii}
p(\epsilon|\xi) = \frac{1}{Z} \, e^{-  \frac{2 \, \epsilon}{3 \,k_{B} T}}
\; ,
\end{eqnarray}
gdzie $\epsilon$ jest realizacją $E$, a $T$ jest (entropijną) temperaturą (por. Dodatek~\ref{Wyprowadzenie drugiej zasady termodynamiki}). Rozkład (\ref{rozklad Boltzmanna z max entropii}) jest standardowym rozkładem eksponentialnym  z $C(\epsilon) = 0$, (\ref{rozklad eksponentialny parametry kanoniczne}), dla wymiaru próby $N=1$. Zgodnie z oznaczeniami wprowadzonymi w  (\ref{rozklad eksponentialny parametry kanoniczne}), jeden parametr kanoniczny $\xi$, jedna funkcja $\hat{\theta}_{\epsilon} = F(E)$ będąca estymatorem parametru oczekiwanego $\theta_{\epsilon}$ oraz potencjał $\psi(\xi)$ układu współrzędnych, mają postać:
\begin{eqnarray}
\label{rozklad eksponentialny parametry kanoniczne 2} 
F(E)=  E \; ,  \;\; \xi = - \frac{2}{3 k_{B} T} \; , \;\; \theta_{\epsilon} = \bar{\epsilon} \; , \;\; \psi(\xi) = \ln ( -\frac{1}{\xi}) = \ln Z  \; . 
\end{eqnarray} 
\\
Do rozkładu Boltzmanna powrócimy w Rozdziale~\ref{rozdz.energia}, gdzie wyprowadzimy go odwołując się do wspomnianych w Rozdziale~\ref{Geometryczne sformulowanie teorii estymacji} zasad informacyjnych.  

\newpage

{\bf Kolejne przykłady zastosowania estymacji eksponentialnej z bazą dualną  $\Theta$}: 
Estymacja tego typu znajduje swoje zastosowanie wtedy, gdy mikrostan układu jest po krótkim okresie czasu zastąpiony  makrostanem, co oznacza estymację stanu układu poprzez pewien oszacowujący stan otrzymany metodą maksymalnej entropii na rozmaitości modelu eksponentialnego (np. dla wolno  zmieniających się) zmiennych makroskopowych \cite{Streater}. Przykładami realizacji tej procedury estymacyjnej są: \\ 
\\
{\bf - Metoda analizy nieliniowej dynamiki} Kossakowskiego i Ingardena, którzy zrealizowali powyższą  procedurę, dokonując  ciągłego rzutowania mikrostanu układu na łatwiejsze w opisie makrostany układu, leżące na rozmaitości stanów eksponentialnych. Z   zaproponowanej analizy statystycznej  wynikła możliwość realizacji  nieliniowej dynamiki układu, opisanej  jako konsekwencja optymalnej {\it estymacji} {\bf stanu układu}, pojawiającego się po upływie każdego kolejnego odstępu  czasu (w którym dynamika układu przebiegała w sposób liniowy), {\bf stanem} {\it leżącym na rozmaitości eksponentialnej} \cite{Streater}. \\
\\
{\bf - Model Onsagera} realizujący tego typu estymację 
w badaniu zjawiska przepływów energii lub masy,  w sytuacji, gdy są one liniowymi funkcjami bodźców (pełniących rolę parametrów) wywołujących taki przepływ. Teoria Onsagera ma zastosowanie do zjawisk mających  charakter {\it procesów quasistatystycznych}. Zatem stosuje się ona do sytuacji, gdy materiał, w którym zachodzi zjawisko jest w lokalnej równowadze, tzn. związki zachodzące lokalnie i w tej samej chwili czasu pomiędzy własnościami cieplnymi i mechanicznymi materiału są takie same, jak dla jednorodnego układu znajdującego się  w równowadze termodynamicznej.  W ramach jego teorii sformułowano zasady wariacyjne dla opisu liniowej termodynamiki  procesów nieodwracalnych. \\
\\
{\bf Podsumowanie}: Z powyższej analizy wynika, że metoda maksymalnej entropii dla nieobciążonych estymatorów $\hat{\theta}_{i} = F_{i}(Y)$, $i=1,2,...d$, wykorzystuje $\alpha = (-1)$ -- płaską bazę $\Theta$ dualną  do $\alpha = (+1)$ -- płaskiej bazy kanonicznej $\Xi$ rozkładu eksponentialnego z $C({\bf y}) = 0$. Ze względu na  płaskość modelu {\it eksponentialnego} w bazie kanonicznej $\Xi$, (\ref{Gamma 1 dla exponential family}), macierz informacyjna $I_{F}$ w bazie dualnej $\Theta$ jest odwrotna do $I_{F}$  w bazie $\Xi$, skąd w (\ref{rownosc odwrot Fisher Eta inf i m kowariancji maks entrop}) przekonaliśmy się, że estymacja w bazie koneksji affinicznej  $\nabla^{(-1)}$ przebiega na DORC. Oznacza to, że również dla wektorowego parametru oczekiwanego $\Theta$ jego estymacja jest  {\it efektywna} w klasie entropijnych modeli eksponentialnych. \\
Jednak idąc dalej, dokładniejsza niż to wynika z TRC, {\it modelowa estymacja}, tzn. związana z konstrukcją nieobciążonych estymatorów parametrów, nie jest możliwa. 
\\
\\
{\bf O tym co w kolejnej części skryptu}: W kolejnej części skryptu zajmiemy się estymacją związaną z zasadami informacyjnymi nałożonymi na tzw. {\it fizyczną informację układu}.

\chapter[Zasady informacyjne]{Zasady informacyjne}

\label{Zasady informacyjne}

\section[Estymacja w statystyce klasycznej a estymacja fizyczna. Postawienie problemu]{Estymacja w statystyce klasycznej a estymacja fizyczna. Postawienie problemu}

\label{physical estim}

W dotychczasowej analizie przedstawiona została MNW w statystyce. Polega ona na estymacji parametrów pewnego zadanego rozkładu. Na przykład w analizie regresji na podstawie pewnej wcześniejszej wiedzy na temat zachowania się zmiennej objaśnianej, zakresu wartości jakie może przyjmować oraz jej charakteru (ciągła czy dyskretna) postulujemy warunkowy rozkład i model regresji, a  następnie konstruujemy funkcję wiarygodności, którą maksymalizując 
otrzymujemy estymatory parametrów strukturalnych modelu. Opracowanie skutecznego algorytmu znajdowania estymatorów MNW oraz ich odchyleń standardowych jest centralnym problemem np. w rutynowych aplikacjach służących do analizy  uogólnionych regresyjnych  modeli liniowych. 
W analizie tej najważniejszym wykorzystywanym  algorytmem jest ogólny algorytm metody iteracyjnie ważonych najmniejszych kwadratów, a jedną z jego głównych analitycznych procedur jest procedura Newton-Raphson'a \cite{Pawitan,Mroz}. \\
Niech parametr wektorowy $\Theta \equiv (\theta_{n})_{n=1}^{N}$ jest zbiorem wartości oczekiwanych zmiennej losowej położenia układu w $N$ pomiarach, jak to przyjęliśmy w Rozdziale~\ref{Pojecie kanalu informacyjnego}.  Przypomnijmy więc, że MNW jest wtedy  skoncentrowana na układzie $N$ równań wiarygodności (\ref{rown wiaryg}): 
\begin{eqnarray}
S(\Theta)\mid_{\Theta=\hat{\Theta}}\;\equiv\frac{\partial}{\partial\Theta} \ln P(\Theta)\mid_{\Theta=\hat{\Theta}}=0\;, \nonumber
\end{eqnarray}
których rozwiązanie daje $N$ elementowy zbiór $\hat{\Theta}\equiv(\hat{\theta}_{n})_{n=1}^{N}$ estymatorów parametrów. Tzn. układ  równań wiarygodności tworzy $N$ warunków na estymatory parametrów, które maksymalizują wiarygodność próbki. \\
Estymacja w fizyce musi się rozpocząć na wcześniejszym etapie. Wychodząc od zasad informacyjnych, którym poświęcony będzie
kolejny rozdział, estymujemy odpowiednie dla opisywanego zagadnienia fizycznego równania ruchu, których rozwiązanie  daje  odpowiedni rozkład wraz z parametrami. Tak więc zastosowanie zasad informacyjnych nałożonych na funkcję wiarygodności zamiast MNW stanowi o podstawowej różnicy pomiędzy analizą statystyczną wykorzystywaną w konstrukcji modeli fizycznych, a statystyką klasyczną. Oczywiście oznacza to, że informacja Fishera zdefiniowana poprzednio na przestrzeni statystycznej ${\cal S}$ musi zostać związana z bazową przestrzenią  ${\cal Y}$ przestrzeni próby, tak aby można ją wykorzystać do konstrukcji równań ruchu.

\subsection{Strukturalna zasada informacyjna. Metoda EFI}

\label{structural principle}

Poniższe rozważania prezentują analizę, leżącą u podstaw strukturalnej zasady informacyjnej \cite{Dziekuje informacja_2}. Ta zaś leży u podstaw metody  estymacji  statystycznej EFI zaproponowanej przez Friedena i Soffera~\cite{Frieden}.  \\
Niech $V_{\Theta}$ jest przestrzenią parametru $\Theta$, tzn. $\Theta \in V_{\Theta}$. Wtedy logarytm funkcji wiarygodności $\ln  P: V_{\Theta} \rightarrow \text{R}$  jest funkcją określoną na przestrzeni $V_{\Theta}$ o wartościach w zbiorze liczb rzeczywistych $\text{R}$. Niech $\tilde{\Theta} \equiv (\tilde{\theta}_{n})_{n=1}^{N} \in V_{\Theta}$ jest inną wartością parametru lub wartością estymatora $\hat{\Theta}$ parametru $\Theta$.
Rozwińmy w punkcie $\tilde{\Theta}$ funkcję $\ln P(\tilde{\Theta})$ w szereg Taylora
wokół prawdziwej wartości  $\Theta$:  
\begin{eqnarray}
\label{rozwiniecie w szereg Taylora}
\ln\frac{P(\tilde{\Theta})}{P(\Theta)} =   \sum_{n=1}^{N}\frac{\partial \ln P(\Theta)}{\partial\theta_{n}}(\tilde{\theta}_{n}-\theta_{n})  + \frac{1}{2} \!\!\sum_{n,n'=1}^{N} \!\frac{\partial^{2} \ln P(\Theta)}{\partial\theta_{n'}\partial\theta_{n}} \, (\tilde{\theta}_{n}- \theta_{n})(\tilde{\theta}_{n'}-\theta_{n'}) + R_{3} \; ,
\end{eqnarray}
gdzie użyto oznaczenia  $\frac{\partial P(\Theta)}{\partial\theta_{n}} \equiv \frac{\partial P(\widetilde{\Theta})}{\partial \tilde{\theta}_{n}}\mid_{\widetilde{\Theta} = \Theta}$, oraz podobnie dla wyższych rzędów rozwinięcia, a $R_{3}$ jest resztą rozwinięcia trzeciego rzędu. \\
\\
{\bf Znaczenie zasady obserwowanej}: Wszystkie człony w (\ref{rozwiniecie w szereg Taylora}) są statystykami na przestrzeni próby ${\cal B}$, więc tak jak i układ równań wiarygodności, równanie (\ref{rozwiniecie w szereg Taylora})  jest określone na poziomie obserwowanym.  Jest ono żądaniem analityczności (logarytmu) funkcji wiarygodności na przestrzeni statystycznej ${\cal S}$, co   stanowi punkt wyjścia dla konstrukcji obserwowanej, różniczkowej,  strukturalnej zasady informacyjnej (\ref{micro form of information eq}), słusznej niezależnie od wyprowadzonej w (\ref{expected form of information eq}) postaci całkowej. Postać obserwowana wraz z zasadą  wariacyjną (\ref{var K}) jest, obok postaci oczekiwanej (\ref{expected form of information eq}), podstawą estymacji EFI równań ruchu teorii pola, lub równań generujących rozkład. 
\\
\\
Zdefiniujmy obserwowaną strukturę układu $\texttt{t\!F}$ w następujący sposób: 
\begin{eqnarray}
\label{structure T}
\texttt{t\!F} \equiv \ln \frac{P(\tilde{\Theta})}{P(\Theta)} - R_{3} \; .
\end{eqnarray}
Na poziomie obserwowanym (nazywanym czasami mikroskopowym) możemy rozwinięcie Taylora (\ref{rozwiniecie w szereg Taylora}) zapisać następująco: 
\begin{eqnarray}
\label{micro structure eq}
\Delta_{LHS} \equiv \sum_{n=1}^{N}2\, \frac{\partial \ln P}{\partial\theta_{n}}(\tilde{\theta}_{n} - \theta_{n}) - \sum_{n=1}^{N}2\, \frac{\texttt{t\!F}}{N} \, = \!\! \sum_{n,n'=1}^{N} \texttt{i\!F}_{nn'} \,(\tilde{\theta}_{n} - \theta_{n})(\tilde{\theta}_{n'}-\theta_{n'}) \equiv \Delta_{RHS} \; ,
\end{eqnarray}
gdzie $\texttt{i\!F}$ jest znaną już z 
(\ref{I jako krzywizna dla P})-(\ref{I obserwowana}) obserwowaną macierzą informacyjną Fishera: 
\begin{eqnarray}
\label{observed IF}
\texttt{i\!F} \equiv \left(-\frac{\partial^{2} \ln P(\Theta)}{\partial \theta_{n'} \partial\theta_{n}}\right) = \left(-\frac{\partial^{2} \ln P(\tilde{\Theta})}{\partial \tilde{\theta}_{n'} \partial \tilde{\theta}_{n}}\right)_{|_{\widetilde{\Theta} = \Theta}} \, ,
\end{eqnarray}
która jako macierz odwrotna do macierzy kowariancji, 
jest symetryczna i dodatnio określona\footnote{Co oznacza, że  zakładamy, że funkcja $\ln P$ jest wypukła w otoczeniu   
prawdziwej wartości parametru  $\Theta$.} (Rozdział~\ref{E i var funkcji wynikowej}).  Oznacza to, że istnieje 
ortogonalna macierz $U$ taka, że $\Delta_{RHS}$ występujące w (\ref{micro structure eq}), a zatem również $\Delta_{LHS}$, może być zapisane w tzw. postaci normalnej \cite{kompendium matematyki}:
\begin{eqnarray}
\label{normal form}
\Delta_{LHS} =  \sum_{n=1}^{N}m_{n}\tilde{\upsilon}_{n}^{2} =  \sum_{n,n'=1}^{N}  \texttt{i\!F}_{nn'}  \,(\tilde{\theta}_{n} - \theta_{n})(\tilde{\theta}_{n'} - \theta_{n'}) \equiv \Delta_{RHS} 
 \, ,
\end{eqnarray}
gdzie $\tilde{\upsilon}_{n}$ są pewnymi funkcjami 
$\tilde{\theta}_{n}$, a  $m_{n}$ są elementami dodatnio określonej macierzy $\texttt{m\!F}$ (otrzymanymi dla $\Delta_{LHS}$), która z powodu równości  (\ref{normal form}) musi być równa macierzy diagonalnej otrzymanej dla $\Delta_{RHS}$, tzn.:
\begin{eqnarray}
\label{form of M}
\texttt{m\!F} = D^{T}  U^{T}\, \texttt{i\!F} \, U \, D \;\; . 
\end{eqnarray}
Macierz $D$ jest diagonalną macierzą skalującą o elementach $d_{n}\equiv\sqrt{\frac{m_{n}}{\lambda_{n}}}$, gdzie $\lambda_{n}$ są wartościami własnymi macierzy $\texttt{i\!F}$. \\
\\
Związek (\ref{form of M}) można zapisać w postaci ważnego strukturalnego równania macierzowego będącego bezpośrednią konsekwencją analityczności logarytmu funkcji wiarygodności na przestrzeni statystycznej ${\cal S}$ oraz postaci normalnej formy kwadratowej (\ref{normal form}):
\begin{eqnarray}
\label{micro form of information eq macierzowe}
\texttt{q\!F} + \texttt{i\!F} = 0 \; , 
\end{eqnarray}
gdzie 
\begin{eqnarray}
\label{micro form of qF}
\texttt{q\!F} = -U \, (D^{T})^{-1} \, \texttt{m\!F} \, D^{-1} \, U^{T} \, ,
\end{eqnarray}
nazwijmy {\it obserwowaną macierzą struktury}.  \\
\\
{\bf Dwa proste przypadki $\texttt{q\!F}$}: Istnieją  dwa szczególne przypadki, które prowadzą do prostych realizacji fizycznych. \\
Pierwszy z nich  związany jest z założeniem, że rozkład jest {\it reguralny} \cite{Pawitan}. Wtedy, zakładając dodatkowo, że dla wszystkich $n=1,...,N$ zachodzi $\frac{\partial lnP}{\partial\theta_{n}}=0$, z równania (\ref{micro structure eq}) widzimy, że:
\begin{eqnarray}
\label{M for logL zero}
\texttt{m\!F} = (2\,\delta_{nn'})\;,\;\;\;\tilde{\upsilon}_{n} = \sqrt{\frac{\texttt{t\!F}}{N}} \;  \;\;\; {\rm oraz} \;\;\; 
d_{n}=\sqrt{2/\lambda_{n}} \; .
\end{eqnarray}
Natomiast drugi przypadek związany jest z założeniem, że 
$\texttt{t\!F}=0$ i wtedy z (\ref{micro structure eq}) otrzymuje się postać ``równania master'' (porównaj dalej (\ref{L master oczekiwana})). W przypadku tym: 
\begin{eqnarray}
\label{M for T zero}
\texttt{m\!F} = diag\left(2\,\frac{\partial \ln P}{\partial\theta_{n}}\right) \, ,\;\;\tilde{\upsilon}_{n} = \sqrt{\tilde{\theta}_{n} - \theta_{n}} \, , \;\;\texttt{t\!F} = 0 \;  \;\;\; {\rm oraz} \;\;\; d_{n}=\sqrt{2\,\frac{\partial lnP}{\partial\theta_{n}}/\lambda_{n}} \; ,
\end{eqnarray}
co oznacza, że nie istnieje złożona struktura układu. \\
\\
{\bf Obserwowana strukturalna zasada informacyjna}: Sumując wszystkie elementy zarówno obserwowanej macierzy informacyjnej Fishera $\texttt{i\!F}$ jak i obserwowanej macierzy struktury $\texttt{q\!F}$, równanie macierzowe (\ref{micro form of information eq macierzowe}) prowadzi do {\it obserwowanej  strukturalnej zasady informacyjnej} Frieden'a: 
\begin{eqnarray}
\label{micro form of information eq}
\sum_{n,n'=1}^{N}(\texttt{i\!F})_{nn'} + \sum_{n,n'=1}^{N}(\texttt{q\!F})_{nn'}  = 0 \; .
\end{eqnarray}
  \\
{\bf Znaczenie analityczności $P$ oraz postaci $\texttt{i\!F}$} dla EFI: Analityczność funkcji wiarygodności na przestrzeni statystycznej ${\cal S}$, wyrażona istnieniem rozwinięcia w szereg Taylora (\ref{rozwiniecie w szereg Taylora}) oraz symetryczność i dodatnia określoność obserwowanej macierzy  informacyjnej Fishera (\ref{observed IF}) są zasadniczymi  warunkami, które czynią analizę  Friedena-Soffera w ogóle możliwą. Okazuje się jednak, że w ogólności, dla otrzymania równań EFI należy odwołać się dodatkowo do  wprowadzonej poniżej całkowej zasady strukturalnej. 
\\
\\
{\bf Całkowa strukturalna zasada informacyjna}: 
Całkując  obie strony równania (\ref{micro form of information eq}) po całej przestrzeni próby ${\cal B}$ (lub na jej podprzestrzeni) z miarą  $d y\, P(\Theta)$,  gdzie jak zwykle stosujemy oznaczenie $dy \equiv d^{N}{\bf y}$, otrzymujemy całkową postać {\it informacyjnej zasady strukturalnej}:
\begin{eqnarray}
\label{expected form of information eq}
Q + I = 0 \; ,
\end{eqnarray}
gdzie $I$ jest uogólnieniem pojemności informacyjnej Fishera (\ref{pojemnosc C}) (por. (\ref{IF 2 poch na kwadrat pierwszej})): 
\begin{eqnarray}
\label{iF and I}
I = \int_{\cal B} d y\, P(\Theta) \; \sum_{n,n'=1}^{N}(\texttt{i\!F})_{nn'} \; , 
\end{eqnarray}
natomiast $Q$ jest informacją strukturalną  ($SI$): 
\begin{eqnarray}
\label{qF and Q}
Q = \int_{\cal B} d y\, P(\Theta) \; \sum_{n,n'=1}^{N}(\texttt{q\!F})_{nn'} \; .
\end{eqnarray}
Pierwotnie, w innej, informatycznej formie i  interpretacji, zasada (\ref{expected form of information eq}) została zapostulowana  w \cite{Frieden}. Powyższa, fizyczna postać zasady strukturalnej (\ref{expected form of information eq}) została zapostulowana w \cite{Dziekuje informacja_1}, a następnie wyprowadzona, jak to przedstawiono powyżej w \cite{Dziekuje informacja_2}. \\
\\
Obserwowana zasada informacyjna (\ref{micro form of information eq}) jest równaniem strukturalnym współczesnych modeli fizycznych wyprowadzanych metodą EFI. Natomiast użyteczność oczekiwanej strukturalnej zasady informacyjnej (\ref{expected form of information eq}) okaże się być jasna przy, po pierwsze określeniu zmodyfikowanej obserwowanej zasady strukturalnej, po drugie, przy definicji całkowitej fizycznej informacji (\ref{physical K}) oraz po trzecie,  przy sformułowaniu informacyjnej zasady wariacyjnej (\ref{var K}). Oczekiwana zasada strukturalna jako taka, tzn. w postaci całkowej (\ref{expected form of information eq}),  nie jest rozwiązywana jednocześnie z zasadą wariacyjną, co jest czasami jej przypisywane. \\

\subsubsection{Całka rozwinięcia Taylora}

\label{postac calkowa Taylora}

Scałkujmy (\ref{rozwiniecie w szereg Taylora}) 
na na całej przestrzeni próby ${\cal B}$ (lub na jej podprzestrzeni) 
z miarą  $d y\, P(\Theta)$.  W wyniku otrzymujemy pewną całkową formę  strukturalnego równania estymacji modeli:
\begin{eqnarray}
\label{Freiden like equation}
&  & \!\!\!\int_{\cal B}\!\! d y P(\Theta) \left(\ln\frac{P(\tilde{\Theta})}{P(\Theta)}- R_{3} -  \sum_{n=1}^{N}\frac{\partial \ln P(\Theta)}{\partial\theta_{n}}(\tilde{\theta}_{n}-\theta_{n}) \right) \nonumber \\
 &  & \!\! = \frac{1}{2}\!\int_{\cal B}\!\! d y\, P(\Theta) \!\!\sum_{n,n'=1}^{N} \!\frac{\partial^{2} \ln P(\Theta)}{\partial\theta_{n'}\partial\theta_{n}} \, (\tilde{\theta}_{n}- \theta_{n})(\tilde{\theta}_{n'}-\theta_{n'}) \; .
\end{eqnarray}
Wyrażenie po lewej stronie (\ref{Freiden like equation}) ma postać zmodyfikowanej entropii względnej. \\
Następnie, definując  $\widetilde{{\cal Q}}$ jako: 
\begin{eqnarray}
\label{structure Q}
\widetilde{{\cal Q}} = \int_{\cal B}\, d y\, P(\Theta)\,\left(\texttt{t\!F} - \sum_{n=1}^{N}\frac{\partial \ln P}{\partial\theta_{n}}(\tilde{\theta}_{n}-\theta_{n})\right) \, 
\end{eqnarray}
otrzymujemy równanie będące całkową formą strukturalnej zasady informacyjnej: 
\begin{eqnarray}
\label{structure eq}
-  \widetilde{{\cal Q}} = \widetilde{I} \equiv  \! \frac{1}{2} \! \int_{\cal B}\!\! d y \, P(\Theta) \!\! \sum_{n,n'=1}^{N} \! \left( - \frac{\partial^{2} \ln P}{\partial\theta_{n'} \partial\theta_{n}} \right) (\tilde{\theta}_{n}-\theta_{n})(\tilde{\theta}_{n'}- \theta_{n'}) \; .  
\end{eqnarray}
\\
{\bf Uwaga}: Równanie (\ref{structure eq}) jest wtórne wobec bardziej fundamentalnego  równania strukturalnego (\ref{micro form of information eq}) słusznego na poziomie obserwowanym, tzn. pod całką. Chociaż równanie (\ref{structure eq}) nie jest bezpośrednio wykorzystywane w metodzie EFI, to jest ono stosowane do badania własności nieobciążonych estymatorów $\tilde{\Theta}$ parametrów $\Theta$ \cite{Murray_differential geometry and statistics}. Zagadnienie to wykracza poza zakres skryptu.

\subsubsection{$I$ oraz $Q$ dla parami niezależnych zmiennych położeniowych próby}

\label{zmienne Yn niezalezne}

Rozważmy jeszcze  postać $SI$ wyrażoną w amplitudach w szczególnym przypadku zmiennych $Y_{n}$ parami niezależnych. W takim przypadku amplituda $q_{n}$  nie zależy od ${\bf y}_{n}$ dla $n' \neq n$, czyli ma postać $q_{n}({\bf y}_{n})$, natomiast  $(\texttt{i\!F})$ jest  diagonalna, tzn. ma postać:
\begin{eqnarray}
\label{iF diagonalne}
(\texttt{i\!F})_{nn'} = \delta_{nn'} \texttt{i\!F}_{nn} \equiv \texttt{i\!F}_{n}  \; ,
\end{eqnarray}
gdzie $\delta_{nn'}$ jest deltą Kroneckera. 
W takim razie,  zgodnie z (\ref{form of M}) oraz (\ref{micro form of qF}) obserwowana macierz strukturalna jest diagonalna i  jej ogólna postać jest następująca:
\begin{eqnarray}
\label{qF diagonalne}
(\texttt{q\!F})_{nn'} = \delta_{nn'} \; \texttt{q\!F}_{nn}\left( q_{n}({\bf y}_{n}), q_{n}^{(r)}({\bf y}_{n}) \right) \equiv \texttt{q\!F}_{n}\left( q_{n}({\bf y}_{n}) \right)\; ,
\end{eqnarray}
tzn. nie zależy od amplitud $q_{n'}({\bf y}_{n'})$ i jej pochodnych dla $n' \neq n$. Powyżej $q_{n}^{(r)}({\bf y}_{n})$ oznaczają pochodne  rzędu $r=1,2,... \,$. Zobaczymy, że dla teorii pola w $\texttt{q\!F}_{n}$ pojawią się pochodne co najwyżej pierwszego rzędu. Fakt ten wynika stąd, że swobodne pola rangi $N$, z którymi będziemy mieli do czynienia, będą  spełniały równanie Kleina-Gordona. \\
\\
{\bf Uwaga}: Oznaczenie $\texttt{q\!F}_{n}$,  jak również  jawne  zaznaczenie w argumencie obserwowanej $SI$ tylko amplitudy $q_{n}({\bf y}_{n})$, będą stosowane w dalszej części skryptu. \\
 \\
Wykorzystując (\ref{iF diagonalne}) oraz (\ref{qF diagonalne}), pojemność informacyjna (\ref{iF and I}) przyjmuje w rozważanym przypadku postać: 
\begin{eqnarray}
\label{I dla niezaleznych Yn}
I = \int_{\cal B} d y\, \textit{i} = \int_{\cal B} d y\, P(\Theta) \; \sum_{n=1}^{N} \texttt{i\!F}_{n} \; , 
\end{eqnarray}
natomiast informacja strukturalna (\ref{qF and Q}) jest następująca: 
\begin{eqnarray}
\label{Q dla niezaleznych Yn}
Q = \int_{\cal B} d y\, \textit{q} = \int_{\cal B} d y\, P(\Theta) \; \sum_{n=1}^{N} \texttt{q\!F}_{n}( q_{n}({\bf y}_{n})) \; .
\end{eqnarray}
Powyżej $\textit{i}$ jest {\it gęstością pojemności informacyjnej}:
\begin{eqnarray}
\label{gestosc i dla niezaleznych Yn}
\textit{i} :=  P(\Theta) \; \sum_{n=1}^{N} \texttt{i\!F}_{n} \; ,
\end{eqnarray}
natomiast $\textit{q}$ jest {\it gęstością informacji strukturalnej}: 
\begin{eqnarray}
\label{gestosc q dla niezaleznych Yn}
\textit{q} :=  P(\Theta) \; \sum_{n=1}^{N} \texttt{q\!F}_{n}( q_{n}({\bf y}_{n})) \; .
\end{eqnarray}
{\bf Obserwowana zasada strukturalna zapisana w gęstościach}: Zarówno $\textit{i}$ jak i $\textit{q}$ są określone na poziomie obserwowanym. Zatem korzystając z (\ref{iF diagonalne}) oraz (\ref{qF diagonalne}), możemy {\it obserwowaną} informacyjną zasadę strukturalną (\ref{micro form of information eq}) zapisać w postaci:
\begin{eqnarray}
\label{obserwowana zas strukt z P}
\textit{i} + \textit{q} = 0\; .
\end{eqnarray}
Zasada ta, a raczej jej zmodyfikowana wersja, jest obok wariacyjnej zasady informacyjnej, 
wykorzystywana  w celu otrzymania równań ruchu (bądź równań generujących rozkład) metody EFI. Zarówno zmodyfikowana  obserwowana zasada strukturalna jak i zasada wariacyjna są określone poniżej.\\
\\
{\bf Uwaga}: W treści skryptu gęstość pojemności informacyjnej $\textit{i}$ jest zawsze związana z postacią (\ref{observed IF}) 
obserwowanej informacji Fishera $\texttt{i\!F}$. 
\\
\\
Na koniec zauważmy, że ze względu na unormowanie rozkładów brzegowych $\int d^{4}{\bf y}_{n} \, p_{n}({\bf y}_{n}|\theta_{n}) =1$, postać $Q$ podaną w (\ref{Q dla niezaleznych Yn})  można zapisać następująco:
\begin{eqnarray}
\label{Q dla niezaleznych Yn w d4y}
Q = \sum_{n=1}^{N} \int d^{4}{\bf y}_{n} \, p_{n}({\bf y}_{n}|\theta_{n}) \, \texttt{q\!F}_{n}\,( q_{n}({\bf y}_{n})) \; .
\end{eqnarray}
Ważna kinematyczna postać $I$ zostanie wprowadzona w Rozdziale~\ref{The kinematical form of the Fisher information}, natomiast postacie $Q$ będą pojawiały się w toku rozwiązywania konkretnych fizycznych problemów.

\section[Przepływ informacji]{Przepływ informacji}

\label{information transfer}

Informacja Fishera $I_{F}$ jest infinitezymalnym typem entropii Kulback-Leibler'a (Rozdział~\ref{Informacja Fishera jako entropia}) wzór (\ref{I porownanie z S}). W statystycznej estymacji KL służy jako narzędzie analizy wyboru modelu \cite{Brockwell_Machura,Zajac}, o czym możemy się przekonać, zauważając, że jest ona związana z wartością oczekiwaną statystki ilorazu wiarygodności (\ref{statystyka ilorazu wiaryg}), wprowadzonej w Rozdziale~\ref{Analiza regresji Poissona}, właśnie w celu porównywania wiarygodności modeli. 
Chociażby z tego powodu, pojawia się przypuszczenie, że pojemność informacyjna $I$ mogłaby, po nałożeniu, jak się okazuje strukturalnej i wariacyjnej  zasady informacyjnej \cite{Frieden,Dziekuje informacja_1}, stać się podstawą równań ruchu (lub równań generujących rozkład) układu fizycznego. Równania te miałyby być najlepsze z punktu widzenia zasad informacyjnych, co jest sednem metody EFI Friedena-Soffera. \\
\\
Zgodnie z Rozdziałem~\ref{Podstawowe zalozenie Friedena-Soffera},  główna statystyczna myśl stojąca za metodą EFI jest następująca:  próbkowanie czasoprzestrzeni następuje przez sam układ nawet wtedy, gdy on sam nie jest poddany rzeczywistemu pomiarowi. Sprawę należałoby rozumieć tak, że układ dokonuje  próbkowania czasoprzestrzeni używając charakterystycznego, swojego własnego pola (i związanej z nim amplitudy) rangi $N$, która jest wymiarem próby, próbkując swoimi kinematycznymi  ``Fisherowskimi'' stopniami swobody przestrzeń położeń jemu dostępną. Przejście od postaci statystycznej pojemności informacyjnej (\ref{pojemnosc C dla polozenia}) do jej reprezentacji kinematycznej zostanie omówione poniżej w Rozdziale~\ref{The kinematical form of the Fisher information}. \\
  \\
Rozważmy następujący, informacyjny schemat układu. Zanim nastąpi pomiar, którego dokonuje sam układ, ma on  pojemność informacyjną $I$ zawartą w swoich kinematycznych stopniach swobody oraz  informację strukturalną $Q$ układu zawartą w swoich strukturalnych stopniach swobody, jak to przedstawiono symbolicznie na poniższym Rysunku. 
\\
\begin{figure}[htb]
\label{przeplyw informacji w ukladzie}
\includegraphics[clip,width=0.47\textwidth,height=2cm]{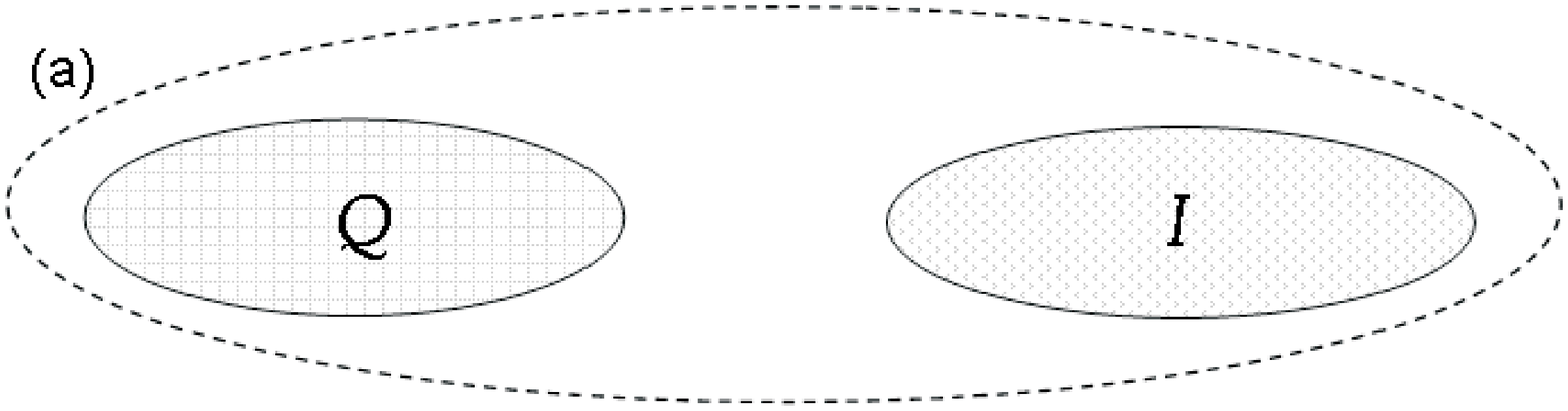} 
\hfill{}
\includegraphics[clip,width=0.47\textwidth,height=2.1cm]{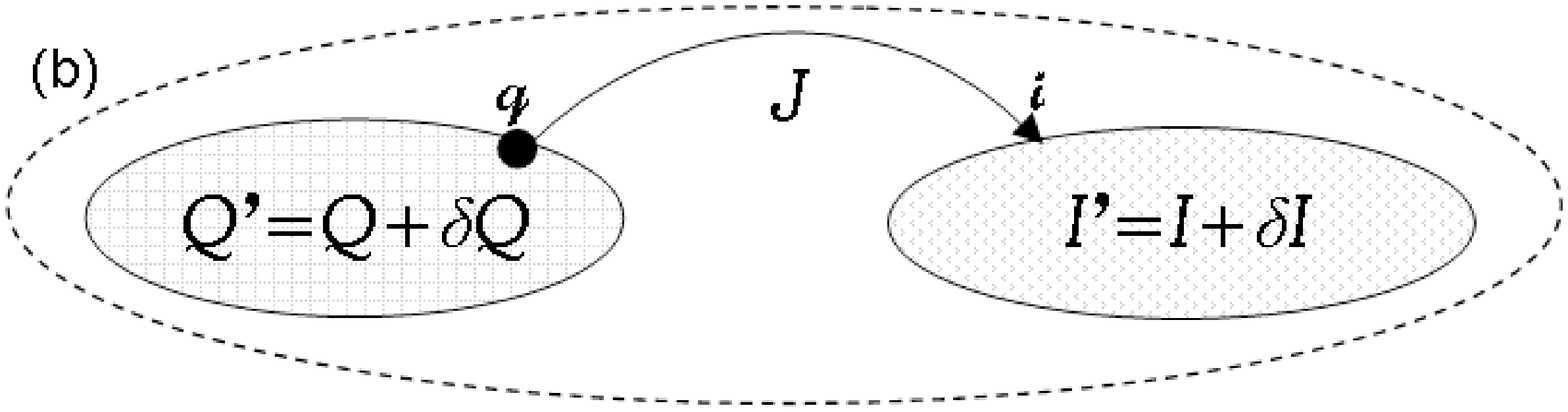}
\caption{Panel: (a)~Układ przed pomiarem : $Q$ jest $SI$ układu zawartą w strukturalnych stopniach swobody, a  $I$
jest pojemnością informacyjną układu zawartą w kinematycznych stopniach swobody. 
(b)~Układ po pomiarze: $Q'$ jest $SI$, a  $I'$ jest  pojemnością informacyjną układu po pomiarze. Ponieważ transfer informacji ($TI$) w pomiarze przebiega z $J\geq0$ zatem   $\delta Q=Q'-Q\leq0$ oraz $\delta I=I'-I\geq0$. W pomiarze idealnym  $\delta I=-\delta Q$.}
\end{figure}\\
``W chwili włączenia'' pomiaru, podczas którego 
transfer informacji ($TI$) przebiega zgodnie z następującymi zasadami (Rysunek~3.1):
\begin{eqnarray}
\label{delta Q and I}
J \geq 0 \; , \;\;\; {\rm zatem}\;\;\; \delta I=I' - I \geq 0 \; ,\; \; \delta Q = Q'  - Q \leq 0  \; ,
\end{eqnarray}
gdzie $I'$, $Q'$ są odpowiednio $IF$ oraz $SI$ układu po pomiarze,
natomiast $J$ jest dokonanym transferem informacji ($TI$).\\
Postulujemy, że w pomiarze  $TI$ ``w punkcie $\emph{q}$'' jest idealny, co oznacza, że:
\begin{eqnarray}
\label{delta Q and I w idealnym pomiarze}
Q = Q' + J = Q + \delta Q + J \, , \;\; {\rm zatem} \;\;\; \delta Q = - J \; . 
\end{eqnarray}
Oznacza to, że ``w punkcie $\emph{q}$'' przekazana jest cała zmiana $SI$. \\
\\
Z drugiej strony ``w punkcie $\emph{i}$'' zasada związana z $TI$ jest następująca:
\begin{eqnarray}
\label{TI w punkcie i}
I' \leq I + J \; \;\; {\rm zatem}  \;\;\; 0 \leq \delta I = I' - I \leq J \; .
\end{eqnarray}
Dlatego 
\begin{eqnarray}
\label{J > 0 i zwiazek delta dla I oraz Q}
{\rm poniewa\dot{z}} \;\;\;\; J \geq 0 \; , \;\;\; {\rm zatem} \;\;\; |\delta I| \leq |\delta Q| \; , 
\end{eqnarray}
co jest rozsądnym resultatem, gdyż w pomiarze może nastąpić utrata informacji. Gdyby  ``w punkcie $\emph{i}$'' $TI$ był idealny, wtedy cały pomiar byłby idealny, tzn.:
\begin{eqnarray}
\label{caly pomiar idealny}
\delta Q = -\delta I \;\; \Leftrightarrow  \;\;\; {\rm pomiar \;\; idealny} \; .
\end{eqnarray}
\\
W \cite{Dziekuje informacja_1,Mroziakiewicz} zostało zapostulowane istnienie nieujemnej  addytywnej całkowitej (totalnej) fizycznej informacji ($TFI$): 
\begin{eqnarray}
\label{physical K}
K = I + Q  \geq 0 \; .
\end{eqnarray}
Wybór intuicyjnego warunku  $K \geq 0$ \cite{Mroziakiewicz}  jest związany ze {\it strukturalną zasadą  informacyjną} zapisaną w postaci obserwowanej:
\begin{eqnarray}
\label{condition from K obserwowana} \sum_{n,n'=1}^{N}(\texttt{i\!F})_{nn'} + \kappa \!\! \sum_{n,n'=1}^{N}(\texttt{q\!F})_{nn'}  = 0 \; 
\end{eqnarray}
lub oczekiwanej:
\begin{eqnarray}
\label{condition from K} 
I + \kappa \, Q  = 0 \; .
\end{eqnarray}
Dla szczególnego przypadku $\kappa = 1$, w Rozdziale~\ref{structural principle} została wyprowadzona  \cite{Dziekuje informacja_2}  postać obserwowana zasady strukturalnej (\ref{micro form of information eq}):
\begin{eqnarray}
\label{ideal condition from K obserwowana}
\sum_{n,n'=1}^{N}(\texttt{i\!F})_{nn'} +   \sum_{n,n'=1}^{N}(\texttt{q\!F})_{nn'}  = 0 \;\;\; {\rm dla}\;\;\; \kappa = 1 \; , 
\end{eqnarray} 
oraz jej oczekiwany odpowiednik (\ref{expected form of information eq}): 
\begin{eqnarray}
\label{ideal condition from K}
I + Q = 0 \;\;\; {\rm dla}\;\;\; \kappa = 1 \; .
\end{eqnarray} 
Współczynnik $\kappa$ został nazwany w \cite{Frieden} współczynnikiem efektywności. W praktyce przyjmuje on dwie możliwe wartości \cite{Frieden}:
\begin{eqnarray}
\label{wartosc kappa}
\kappa = 1 \;\; \vee \;\; \frac{1}{2} \; .
\end{eqnarray}
Jego znaczenie zostanie omówione w Rozdziale~\ref{Kryteria informacyjne w teorii pola}.  
W przypadku określonym w (\ref{ideal condition from K}), otrzymujemy całkowitą fizyczną informację  $K$ równą:
\begin{eqnarray}
\label{ideal K}
K = I + Q = 0 \;\;\; {\rm dla} \;\;\; \kappa = 1 \; . 
\end{eqnarray}
W końcu zauważmy, że w zgodzie z zapostulowanym zachowaniem się układu w pomiarze, otrzymaliśmy z  warunków (\ref{delta Q and I w idealnym pomiarze}) i (\ref{TI w punkcie i}) nierówność $\delta I\leq J = - \delta Q$, z czego wynika, że:  
\begin{eqnarray}
K' = I' + Q'  \leq  (I + J) + (Q - J) = I + Q = K \; \Rightarrow  \;\;\; K' \leq K \; .
\end{eqnarray}
Dla pomiaru idealnego (\ref{caly pomiar idealny})  otrzymaliśmy $\delta I=-\delta Q$ skąd $K'=K$, co oznacza, że informacja fizyczna $TFI$ pozostaje w tym przypadku niezmieniona. Jeśli pomiar idealny byłby wykonany na poziomie próbkowania czasoprzestrzeni przez sam układ, wtedy warunek ten mógłby prowadzić do wariacyjnej zasady informacyjnej (\ref{var K}), tzn.:
\begin{eqnarray}
\label{zasada wariacyjna}
\delta I = - \delta Q \;\; \Rightarrow \;\; \delta(I + Q) = 0 \;  .
\end{eqnarray}
Chociaż rozumowanie powyższe wydaje się być rozsądne, jednak ściśle mówiąc słuszność przyjęcia zasad informacyjnych,  strukturalnej oraz wariacyjnej, powinno wynikać z dwóch rzeczy. Po pierwsze z ich wyprowadzenia, a po drugie z ich użyteczności. 
Wyprowadzenie zasady strukturalnej (dla $\kappa=1$) zostało pokazane w Rozdziale~\ref{structural principle}. \\
Natomiast powyższe wnioskowanie, które doprowadziło do warunku (\ref{zasada wariacyjna}) oraz sama implikacja wewnątrz niego, może służyć jedynie jako przesłanka słuszności zasady wariacyjnej. 
W Rozdziale~\ref{Geometryczne sformulowanie teorii estymacji}  stwierdziliśmy, że jej słuszność wynika z żądania aby rozkład empiryczny oraz rozkład wyestymowany metodą EFI, leżały na wspólnej geodezyjnej w przestrzeni statystycznej ${\cal S}$. \\
%
Co do użyteczności zasady wariacyjnej w metodzie EFI, to jest ona  oczywista, bowiem prowadzi ona do owocnego w zastosowaniach równania Eulera-Lagrange'a. \\
\\
{\bf Analityczny przypadek układu równań informacyjnych metody EFI}: {\it Obserwowana zasada strukturalna} zapisana w gęstościach   (\ref{obserwowana zas strukt z P}), ale  uwzględniająca postać  obserwowanej zasady strukturalnej z $\kappa$ (\ref{condition from K obserwowana}),  jest następująca:
\begin{eqnarray}
\label{obserwowana zas strukt z P i z kappa}
\textit{i} + \kappa \, \textit{q} = 0\; .
\end{eqnarray}
Drugą zasadą informacyjną jest {\it zasada wariacyjna} (skalarna). Ma ona postać \cite{Dziekuje informacja_1}:
\begin{eqnarray}
\label{var K}
\delta K = \delta(I + Q) = 0 \; \, \Rightarrow \; \, K = I + Q \;\;{\rm jest\; ekstremalne} \; .
\end{eqnarray}
\\
{\bf Warunek geometrycznej struktury na ${\cal S}$}: Kolejnym warunkiem narzuconym na rozwiązania metody EFI, a oczywistym od początku analizy, jest warunek normalizacji i reguralności rozkładu prawdopodobieństwa. Warunek ten oznacza możliwość przejścia, podanego w (\ref{IF 2 poch na kwadrat pierwszej}) i (\ref{Fisher inf matrix}), od pierwotnej postaci obserwowanej informacji Fishera (\ref{observed IF}) do postaci potrzebnej dla zdefiniowania przestrzeni statystycznej  ${\cal S}$ jako przestrzeni metrycznej z metryką Rao-Fishera (\ref{Fisher inf matrix plus reg condition}) i  $\alpha$-koneksją (\ref{affine coefficients}) (por. Rozdział~\ref{alfa koneksja}). Obie postacie obserwowanej informacji Fishera, pierwsza  $\texttt{i\!F}$, (\ref{observed IF}), która jest pierwotną formą z punktu widzenia analityczności funkcji wiarygodności oraz druga, {\it metryczna}  $\widetilde{\texttt{i\!F}}$, (\ref{observed IF Amari}), która jest istotna dla geometrycznej analizy modelu, są równoważne tylko na poziomie oczekiwanym, tzn. pod całką (por. Rozdział~\ref{IF 2 poch na kwadrat pierwszej}).  Powyższe rozważania zostaną zilustrowane przykładami zawartymi w dalszej części skryptu. \\
\\
{\bf Podstawowy układ równań informacyjnych EFI i zmodyfikowane równanie strukturalne}:  Aby wyjaśnić powyższy problem na wstępnym,  symbolicznym poziomie, wprowadźmy zmodyfikowane równanie strukturalne, uwzględniające również współczynnik $\kappa$ występujący w (\ref{obserwowana zas strukt z P i z kappa}). Niech  $\widetilde{\texttt{i\!F}}$ jest kwadratową postacią obserwowanej informacji Fishera (\ref{observed IF Amari})
tak, że odpowiadająca jej gęstość pojemności informacyjnej:
\begin{eqnarray}
\label{gestosc i Amarii}
\widetilde{\textit{i}} :=  P(\Theta)  \sum_{n,n'=1}^{N} \widetilde{\texttt{i\!F}}_{n n'} \; ,
\end{eqnarray}
daje na poziomie oczekiwanym pojemność informacyjną $I = \int_{\cal B} \! dy \,  \widetilde{\textit{i}}$ $= \int_{\cal B} \! dy \,  \textit{i}\,$. \\
Wprowadźmy zamiast (\ref{obserwowana zas strukt z P i z kappa}) {\it zmodyfikowaną obserwowaną zasadę strukturalną} zapisaną w następujący sposób: 
\begin{eqnarray}
\label{zmodyfikowana obserwowana zas strukt z P i z kappa}
\widetilde{\textit{i'}} + \widetilde{\mathbf{C}} + \kappa \, \textit{q} = 0 \; , \;\;\; {\rm przy \; czym} \;\;\; I = \int_{\cal B} \! dy \,  \widetilde{\textit{i}} = \int_{\cal B} \! dy \,  (\widetilde{\textit{i'}} + \widetilde{\mathbf{C}} ) \; , 
\end{eqnarray}
gdzie $ \widetilde{\mathbf{C}}$ jest pochodną zupełną, która wynika z całkowania przez części całki $I = \int_{\cal B} \! dy \,  \widetilde{\textit{i}}$. Ponieważ $I = \int_{\cal B} \! dy \, \widetilde{\textit{i}}\,$ zatem  informacyjna  zasada wariacyjna ma postać: 
\begin{eqnarray}
\label{var K rozpisana}
\delta(I + Q) = \delta \! \int_{\cal B} \! dy \, ( \widetilde{\textit{i}} +  \textit{q} ) = 0  \; . 
\end{eqnarray}
Rozwiązanie równań (\ref{zmodyfikowana obserwowana zas strukt z P i z kappa}) oraz (\ref{var K rozpisana}) jest równoważne rozwiązaniu równań (\ref{obserwowana zas strukt z P i z kappa}) oraz (\ref{var K}) co najmniej pod całką, tzn.: 
\begin{eqnarray}
\label{rownowaznosc strukt i zmodyfikowanego strukt}
\int_{\cal B} \! dy \, ( \widetilde{\textit{i'}}  + \widetilde{\mathbf{C}} + \kappa \, \textit{q} )  = 0 \; \;\; \Leftrightarrow \;\;\;  \int_{\cal B} \! dy \, ( \textit{i} + \kappa \, \textit{q} )  = 0 \; . 
\end{eqnarray}
Powyższą symboliczną konstrukcję (\ref{zmodyfikowana obserwowana zas strukt z P i z kappa}) zaprezentujemy na przykładach w dalszej części skryptu (Rozdział~\ref{Przyklady}).  
Jej zrozumienie jest następujące: Modele "`nie do końca równoważne"' pod względem analitycznym są, z dokładnością do wycałkowania $I$ przez części, równoważne pod względem metrycznym. To znaczy, istnieje pewien związek pomiędzy ich różniczkowalnością, a mianowicie wszystkie one są metrycznie (a więc na poziomie całkowym) równoważne modelowi analitycznemu, tzn. posiadającemu rozwinięcie w szereg Taylora.   \\
\\
{\bf Podsumowanie}.  Należy podkreślić, że równanie całkowe (\ref{ideal K}) oraz zasada wariacyjna (\ref{var K}) {\it nie} tworzą pary równań metody EFI rozwiązywanych samospójnie. 
\\
Natomiast obie zasady, obserwowana zmodyfikowana zasada strukturalna (\ref{zmodyfikowana obserwowana zas strukt z P i z kappa}) oraz zasada wariacyjna (\ref{var K rozpisana})  są podstawą metody estymacyjnej EFI. 
Tworzą one układ  dwóch równań różniczkowych dla wprowadzonych w Rozdziale~\ref{geometria i metryka Fishera-Rao} amplitud  układu (\ref{amplituda a rozklad}). 
Układ ten może być zgodny, dając samospójne  rozwiązanie dla amplitud \cite{Frieden} i prowadząc przy $\kappa=1$ lub $1/2$ do dobrze znanych modeli teorii pola (Rozdział~\ref{Kryteria informacyjne w teorii pola}) lub modeli fizyki statystycznej (Rozdział~\ref{Przyklady}). Ponadto, strukturalna (wewnętrzna) zasada informacyjna (\ref{condition from K}) \cite{Dziekuje informacja_1} jest operacyjnie równoważna  zapostulowanej przez Frieden'a \cite{Frieden}, więc jako wyprowadzona powinna mieć  przynajmniej  taką samą moc przewidywania jak i ona. 
Wiele z podstawowych modeli zostało już wyliczonych \cite{Frieden}, jednak ich ponowne przeliczenie \cite{Mroziakiewicz} przy powyżej podanej interpretacji informacji  fizycznej $K$ może dać lepsze zrozumienie samej metody EFI i jej związku z istniejącym już modelowaniem zjawisk w fizyce oraz jej ograniczeń. \\
\\
{\bf Zasada ekwipartycji entropii względnej}: W końcu, w strukturalnej zasadzie informacyjnej ciekawe jest  również to, że stanowi ona warunek zerowego podziału dla $TFI$, który jest dawno poszukiwanym warunkiem zasady ekwipartycji entropii (w tym przypadku infinitezymalnej entropii względnej).\\
 \\
{\bf Uwaga o podejściu Friedena}: Wspomnieliśmy o tym, że pomysł metody EFI pochodzi od Friedena. Jednak mówiąc w skrócie, Frieden i Soffer \cite{Frieden} podeszli inaczej do informacji strukturalnej. W \cite{Frieden} wprowadzono tzw. informację związaną J, która ma interpretację informacji zawartej w układzie przed pomiarem. Chociaż, aksjomaty Frieden'a są równoważne powyższym warunkom (\ref{condition from K}) oraz (\ref{var K}),  o ile ${\rm J} = -Q$, to jednakże różnica pomiędzy podejściami jest widoczna. A mianowicie, o ile w podejściu Friedena-Soffera układ doświadcza transferu informacji ${\rm J} \rightarrow I$, mając w każdej chwili czasu tylko jeden z tych typów informacji, o tyle w naszym podejściu system jest charakteryzowany jednocześnie przez $I$ oraz $Q$ w każdej chwili czasu. \\
\\
{\bf Uwaga o podobieństwie EFI i teorii Jaynes'a}: Metoda  EFI zaproponowana przez  Frie\-dena i Soffera  \cite{Frieden} jest konsekwencją postulatu podobnego do zasady Jaynes’a\footnote{E.~Jaynes, 
Information Theory and Statistical Mechanics, 
\textit{Phys.Rev.} \textbf{106},  620–630 (1957). 
E.~Jaynes,  
Information Theory and Statistical Mechanics. II, 
\textit{Phys.Rev.} \textbf{108}, 171–190 (1957). }. 
Mianowicie podobieństwo obu teorii leży w tym, że poprzez zasadę  wariacyjną wiążą one strukturalne (Boltzmann'owskie) stopnie swobody z  kinematycznymi (Shannona) stopniami swobody\footnote{Metod tych nie należy jednak utożsamiać. Należy  pamiętać, że macierz informacyjna Fishera, wykorzystywana w EFI, oraz entropia Shannona, wykorzystywana w podejściu Jaynes'a, są różnymi pojęciami. Macierz informacyjna Fishera jest Hessianem (\ref{hesian z S}) entropii Shannona.
}. \\
Według podejścia Jaynes'a, maksymalizacja entropii Shannona  
względem prawdopodobieństw mikro-stanu układu, posiadającego znane  własności, np. ustaloną energię, umożliwia identyfikację  termodynamicznej entropii Boltzmanna jako zmaksymalizowanej entropii   Shannona, a następnie na konstrukcję funkcji stanu, np. energii swobodnej.

\section[Kinematyczna postać informacji Fishera]{Kinetyczna postać informacji Fishera}

\label{The kinematical form of the Fisher information}


Centralna część pracy Frieden'a i Soffer'a związana jest z transformacją postaci pojemności informacyjnej  $I$ zadanej równaniem (\ref{pojemnosc C dla polozenia}) oraz (\ref{pojemnosc informacyjna Minkowskiego}) do tzw. postaci kinematycznej wykorzystywanej w teorii pola oraz fizyce statystycznej. 
W obecnym rozdziale zaprezentujemy podstawowe założenia, które 
doprowadziły do konstrukcji kinematycznego członu (całki) działania dla czterowymiarowych modeli teorii pola. 
%
Przejście to ma następującą postać \cite{Frieden}.  \\
\\
Zgodnie z podstawowym założeniem Friedena-Soffera, $N$-wymiarowa próbka ${\bf y}_{n} \equiv ({\bf y}_{n}^{\nu})$ jest pobierana przez układ posiadający rozkład $p_{n}({\bf y}_{n})$, gdzie obok indeksu próby $n=1,2,...,N$ wprowadzono indeks (czaso)przestrzenny $\nu = (0),1,2,3$. 
Zgodnie z Rozdziałem~\ref{geometria i metryka Fishera-Rao}, wzór (\ref{amplituda a rozklad}), metryka Fishera na (pod)rozmaitości ${\cal S}$ prowadzi w naturalny sposób do pojęcia rzeczywistej amplitudy $q({\bf y}_{n}|\theta_{n}) \equiv \sqrt{p({\bf y}_{n}|\theta_{n}})$ pola układu. Od razu skorzystano też z zapisu, który sugeruje niezależność rozkładu dla $Y_{n}$ od $\theta_{m}$, gdy $m \neq n$.  \\
\\
Jak przedstawiliśmy w Rozdziale~\ref{Podstawowe zalozenie Friedena-Soffera} pojemność informacyjna (\ref{pojemnosc C dla polozenia - powtorka wzoru}) może zostać zapisana jako (\ref{potrz}): 
\begin{eqnarray}
\label{Fisher_information with q dla przejscia}
I \equiv  I(\Theta)  = 4 \sum\limits_{n=1}^N \int_{\cal Y} {d^{4}{\bf y}_{n} \sum\limits_{\nu=0}^{3}  {\left( {\frac{{\partial q_{n} \left(  {\bf y}_n|\theta_{n}  \right)}}{{\partial \theta_{n \nu} }}}  {\frac{{\partial q_{n} \left(  {\bf y}_n|\theta_{n}  \right)}}{{\partial \theta_{n}^{ \nu} }}} \right) } } \; .
\end{eqnarray}
\\
\\
{\bf Addytywny rozkład położeń i reguła łańcuchowa}: Niech ${\bf x}_{n} \equiv({\bf x}^{\nu}_{n})$ są przesunięciami (np. addytywnymi fluktuacjami) danych ${\bf y}_{n} \equiv ({\bf y}_{n}^{\nu})$ od ich wartości oczekiwanych $\theta^{\nu}_{n}$, tzn.:
\begin{eqnarray}  
\label{parameters separation}
{\bf y}_{n}^{\nu} = \theta^{\nu}_{n} + {\bf x}^{\nu}_{n} \; .
\end{eqnarray}
Przesunięcia ${\bf x}^{\nu}_{n}$ są zmiennymi Fisher'owskimi, spełniając warunek   $\frac{\partial {\bf x}^{\nu}}{\partial {\bf x}^{\mu}} = \delta^{\nu}_{\mu}$, (\ref{zmienne Fisherowskie}).\\
\\
Odwołując się do ``reguły łańcuchowej'' dla pochodnej:
\begin{eqnarray}
\label{chain rule}
\frac{\partial}{\partial {\bf \theta_{n}^{\nu}}} =  \frac{\partial ({\bf y_{n}^{\nu}} -
\theta_{n}^{\nu})}{\partial \theta_{n}^{\nu}} \, \frac{\partial}{\partial ({\bf
y_{n}^{\nu}} - \theta_{n}^{\nu})}  = - \; \frac{\partial}{\partial
({\bf y_{n}^{\nu}} - \theta_{n}^{\nu})} = - \; \frac{\partial}{\partial {\bf x_{n}^{\nu}}} \; 
\end{eqnarray}
oraz uwzględniając $d^{4}{\bf x}_{n}=d^{4}{\bf y}_{n}$, co wynika z tego, że parametry $\theta_{n}$ są stałymi, 
możemy przejść od postaci statystycznej  (\ref{Fisher_information with q dla przejscia}) do {\it postaci kinematycznej $IF$}~: 
\begin{eqnarray}  
\label{Fisher_information-kinetic form}
I  =  4  \sum_{n=1}^{N} \int_{{\cal X}_{n}} \!\! d^{4}{\bf x}_{n}  \sum_{\nu} \frac{\partial q_{n}({\bf x}_{n})}{\partial {\bf x}_{n \nu}}
\frac{\partial q_{n}({\bf x}_{n})}{\partial {\bf x}^{\nu}_{n}} \; ,
\end{eqnarray}
gdzie $d^{4}{\bf x}_{n}=d {\bf x}_{n}^{0} d {\bf x}_{n}^{1} d {\bf x}_{n}^{2} d {\bf x}_{n}^{3} $. W (\ref{Fisher_information-kinetic form}) wprowadzono oznaczenie:
\begin{eqnarray}  
\label{zapis dla qn w xn}
q_{n}({\bf x}_{n}) \equiv q_{n}({\bf x}_{n}+\theta_{n}|\theta_{n}) = q_{n}({\bf y}_{n}|\theta_{n}) \; ,
\end{eqnarray}
{\it pozostawiając całą informację o $\theta_{n}$ w indeksie $n$ amplitudy $q_{n}({\bf x}_{n})$}.\\
\\
{\bf Kinematyczna postać IF dla $q_{n}$}: Zakładając, że zakres zmienności wszystkich ${\bf x}_{n}^{\nu}$ jest dla każdego $n$ taki sam, możemy pominąć indeks $n$ przy tej zmiennej (ale nie przy amplitudzie $q_{n}$), otrzymując postać: 
\begin{eqnarray}  
\label{Fisher_information-kinetic form bez n}
I  =  4  \sum_{n=1}^{N} \int_{\cal X} \!\! d^{4}{\bf x}  \sum_{\nu} \frac{\partial q_{n}({\bf x})}{\partial {\bf x}_{\nu}}
\frac{\partial q_{n}({\bf x})}{\partial {\bf x}^{\nu}} \; ,
\end{eqnarray}
którą wykorzystamy przy wyprowadzeniu równań generujących fizyki statystycznej \cite{Frieden}, ale która  została również  wykorzystana do wyprowadzenia elektrodynamiki Maxwella metodą EFI  \cite{Frieden}. \\
  \\
{\bf Uwaga}: {\it Wymiar próby $N$ jest  rangą pola układu zdefiniowanego jako zbiór  amplitud $\left(q_{n}({\bf x}_{n})\right)_{n=1}^{N}$}. \\
 \\
W Rozdziale~\ref{structural principle} pokazaliśmy, że strukturalna zasada informacyjna $I + Q =0$ jest artefaktem istnienia rozwinięcia $\ln P(\tilde{\Theta})$ w szereg Taylora\footnote{W  \cite{Dziekuje informacja_2} była użyta miara $d^{N}{\bf x} \, P(\Theta)$ zamiast $d^{N}{\bf y} \,
P(\Theta)$. Nie zmienia to jednak dowodu strukturalnej zasady informacyjnej, lecz poszerza jego zastosowanie na sytuacje, które {\it nie posiadają niezmienniczości przesunięcia},  założenia nie wykorzystywano w dowodzie.} wokół prawdziwej wartości parametru $\Theta$. Obecnie znamy już ogólną postać kinematyczną $I$ części pomiarowej zasady strukturalnej. W metodzie EFI, jej część strukturalna $Q$ ma postać zależną od np. fizycznych więzów nałożonych na układ. Zagadnieniem tym zajmiemy się w kolejnych Rozdziałach~\ref{Kryteria informacyjne w teorii pola} oraz  \ref{Przyklady}.\\
  \\
{\bf Amplitudy zespolone}: Kolejnym założeniem jest konstrukcja {\it składowych funkcji falowej} składanych z amplitud\footnote{Amplitudy $q_{n}$ są w przypadku rozkładów ciągłych  związane z $p_{n}$, które są {\it gęstościami}  prawdopodobieństw.} 
w następujący sposób \cite{Frieden}: 
\begin{eqnarray}
\label{amplitudapsi dla roznych xn}
\psi_{n}({\bf x}_{2n-1},{\bf x}_{2n}) \equiv \frac{1}{{\sqrt{N}}}\left( q_{2n-1}({\bf x}_{2n-1}) + i \, q_{2n}({\bf x}_{2n}) \right)\; , \quad\quad n=1,...,{N/2} \; .
\end{eqnarray}
Powyższa postać jest uogólnieniem konstrukcji Friedena, który  tworząc funkcję falową układu złożył $n$-tą {\it składową funkcji falowej} z amplitud w następujący sposób \cite{Frieden}:
\begin{eqnarray}
\label{amplitudapsi}
\psi_{n}({\bf x}) \equiv \frac{1}{{\sqrt{N}}}\left( q_{2n-1}({\bf x}) + i \, q_{2n}({\bf x}) \right)\; , \quad\quad n=1,...,{N/2} \; .
\end{eqnarray}
Dokładniej mówiąc, aby posłużenie się funkcją falową (\ref{amplitudapsi}) miało sens, musi przynajmniej pod całką zachodzić równoważność zmiennych: 
\begin{eqnarray}
\label{x n rownowaznosc}
{\bf x}_{n} \equiv {\bf x} \;\;\; {\rm dla \;\;  wszystkich} \;\;\; n =1,2,...,N \; .
\end{eqnarray}
Założenie to całkiem wystarcza przy liczeniu wartości oczekiwanych oraz prawdopodobieństw.\\
Przy założeniu postaci (\ref{amplitudapsi}) dla $n$-tej składowej\footnote{O ile nie będzie to prowadziło do nieporozumień,  będziemy pominijali słowo ``składowa''.}, postać {\it funkcji falowej}~~Friedena jest następująca:
\begin{eqnarray}
\label{psi zespolona}
\psi({\bf x}) \equiv (\psi_{n}\left({\bf x})\right)_{n=1}^{N/2} \; .
\end{eqnarray}
Zbiór $N/2$ składowych funkcji falowych $\psi_{n}$ nazwijmy {\it funkcją falową układu rangi $N$}. \\
%
Zauważmy, że zachodzą następujące równości:
\begin{eqnarray}
\label{analizaq dla q}
\!\sum\limits _{n=1}^{N}{q_{n}^{2}}= \left({q_{1}^{2}+q_{3}^{2}+...+q_{N-1}^{2}}\right)+\left({q_{2}^{2}+q_{4}^{2}+...+q_{N}^{2}}\right) = \sum\limits_{n=1}^{{N/2}}{\left({q_{2n-1}}\right)^{2}+\left({q_{2n}}\right)^{2}} \; 
\end{eqnarray}
i analogicznie:
\begin{eqnarray}
\label{analizaq}
\sum\limits_{n=1}^{N} \frac{\partial q_{n}}{\partial {\bf x}_{n \nu}} \frac{\partial q_{n}}{\partial {\bf x}_{n}^{\nu}}  = \sum\limits_{n=1}^{N/2} \left(\frac{\partial q_{2n-1}}{\partial {\bf x}_{n \nu}} \frac{\partial q_{2n-1}}{\partial {\bf x}_{n}^{\nu}} + \frac{\partial q_{2n}}{\partial {\bf x}_{n \nu}} \frac{\partial q_{2n}}{\partial {\bf x}_{n}^{\nu}} \right)  \; .
\end{eqnarray}
{\it Zakładając dla wszystkich poniższych rozważań słuszność (\ref{x n rownowaznosc}), przynajmniej pod całką, oraz postać   funkcji falowych (\ref{amplitudapsi})},  dokonajmy  następującego ciągu przekształceń  dla  (\ref{Fisher_information-kinetic form bez n}):
\begin{eqnarray}
\label{krok zamiany I dla q na psi}
I &=& 4 \sum\limits_{n=1}^{N} \int_{\cal X} d^{4}{\bf x} 
\sum\limits_{\nu} \frac{\partial q_{n}}{\partial {\bf x}_{\nu}} \frac{\partial q_{n}}{\partial {\bf x}^{\nu}}  = 
4 \sum\limits_{n=1}^{N/2} \int_{\cal X} d^{4}{\bf x} \sum\limits_{\nu} \left[ \frac{\partial q_{2n-1}}{\partial {\bf x}_{\nu}} \frac{\partial q_{2n-1}}{\partial {\bf x}^{\nu}} +  \frac{\partial q_{2n}}{\partial {\bf x}_{\nu}} \frac{\partial q_{2n}}{\partial {\bf x}^{\nu}} 
\right] \nonumber \\ 
&=& 4 N \sum\limits_{n=1}^{{N/2}} \int_{\cal X} d^{4}{\bf x}
\sum\limits_{\nu} \frac{1}{{\sqrt{N}}} \frac{\partial\left({q_{2n-1} -  i q_{2n}} \right)}{\partial {\bf x}_{\nu}} \frac{1}{\sqrt{N}} \frac{\partial\left({q_{2n-1}  + iq_{2n}}\right)}{\partial {\bf x}^{\nu}} \; ,
\end{eqnarray}
gdzie w ostatnim przejściu skorzystano z przekształcenia typu: 
\begin{eqnarray}
\label{q2 na q q}
q_{2n-1}^{k} \,q_{2n-1}^{k} + q_{2n}^{k}\, q_{2n}^{k}  = (q_{2n-1}^{k} -  i \,q_{2n}^{k})( q_{2n-1}^{k} +  i \,q_{2n}^{k})  \; , 
\end{eqnarray}
z indeksem $k$ oznaczającym pochodną rzędu $k=0,1,...\,$. \\
\\
{\bf Kinematyczna postać IF dla $\psi$}:  Odwołując sią do definicji (\ref{amplitudapsi}) funkcji falowej, otrzymujemy:
\begin{eqnarray}
\label{inf F z psi}
I = 4 N \sum\limits_{n=1}^{{N/2}} \int_{\cal X} d^{4}{\bf x} 
\sum\limits_{\nu}{\frac{\partial \psi_{n}^{*}({\bf x})}{\partial {\bf  x}_{\nu}}}{\frac{\partial \psi_{n}({\bf x})}{\partial {\bf x}^{\nu}}} \; .
\end{eqnarray}
Pojemnośc informacyjna (\ref{inf F z psi}) ma typową postać  np. dla relatywistycznej mechaniki falowej, odpowiadającą części kinetycznej całki działania. Dlatego właśnie oczekiwaną informację Fishera nazwał  Frieden {\it informacją kinetyczną}. W  \cite{Frieden} użyto jej  do wyprowadzenia równań Kleina-Gordona oraz Diraca  metodą EFI  \cite{Frieden}. \\
 \\ 
{\bf Rozkładu prawdopodobieństwa przesunięcia w układzie}: Korzystając z twierdzenia o prawdopodobieństwie całkowitym, {\it gęstość} rozkładu prawdopodobieństwa przesunięcia (lub fluktuacji) w układzie może być zapisana następująco \cite{Frieden}: 
\begin{eqnarray}
\label{p jako suma po qn2 przez N}
p\left({\bf x}\right) &=& \sum_{n=1}^{N} p\left({\bf x}|{\theta}_{n} \right) r\left({\theta}_{n}\right) = \sum_{n=1}^{N} {p_{n}\left( {\bf x}_{n}|{\theta}_{n}\right) r\left({\theta}_{n}\right)} = \frac{1}{N} \sum_{n=1}^{N} q_{n}^{2} \left({\bf x}_{n}|{\theta}_{n}\right) \nonumber \\
&=& \frac{1}{N} \sum_{n=1}^{N} q_{n}^{2} \left({\bf x}\right) \; ,
\end{eqnarray}
gdzie skorzystano z założenia, że $n$-ta wartość oczekiwana ${\theta}_{n}$ nie ma dla $m \neq n$  wpływu na rozkład przesunięcia ${\bf x}_{m}$ oraz jak zwykle z postaci amplitudy  $q_{n}^{2} = p_{n}$.  Prawdopodobieństwo $p_{n}$ jest prawdopodobieństwem pojawienia się wartości ${\bf x}_{n}$ zmiennej losowej przesunięcia (lub fluktuacji) z rozkładu generowanego z parametrem  ${\theta}_{n}$, tzn. ma ono interpretację prawdopodobieństwa warunkowego  $p_{n}\left({\bf x}_{n}|{\theta}_{n}\right)$. Funkcję  
$r\left({\theta}_{n}\right) = \frac{1}{N}$ można nazwać funkcją ``niewiedzy'', gdyż jej postać jest odzwierciedleniem całkowitego braku wiedzy odnośnie tego, która z $N$ możliwych wartości ${\theta}_{n}$ pojawi się w konkretnym $n$-tym z $N$ eksperymentów próby.  \\ 
{\bf Postać rozkładu  dla $\psi$}: W końcu, korzystając z (\ref{amplitudapsi}), (\ref{analizaq}), (\ref{q2 na q q}) oraz (\ref{p jako suma po qn2 przez N}) widać, że: 
\begin{eqnarray}
\label{prawdpsi}
p\left({\bf x}\right) = \sum_{n=1}^{N/2} {\psi_{n}^{*}\left({\bf x}\right) \psi_{n}}\left({\bf x}\right) \; 
\end{eqnarray}
jest gęstością rozkładu prawdopodobieństwa przesunięcia (lub fluktuacji) ${\bf x}$ w układzie opisanym funkcją falową (\ref{psi zespolona}).\\
\\
{\bf Uwaga o różnicy z podejściem Friedena}: W całym powyższym wyprowadzeniu nie użyliśmy podstawowego założenia Friedena-Soffera  o  {\it niezmienniczości rozkładu ze względu na przesunięcie}, tzn.: 
\begin{eqnarray}
\label{shift inv property}
p_{n} ({\bf x}_{n}) = p_{x_{n}} ({\bf x}_{n}|\theta_{n}) = p_{n} ({\bf y}_{n}|\theta_{n}) \; , \;\;\; {\rm gdzie} \;\;\; {\bf x}_{n}^{\nu} \equiv {\bf y}_{n}^{\nu} - \theta_{n}^{\nu} \; ,
\end{eqnarray}
gdzie ${\bf y}_{n} \equiv ({\bf y}_{n}^{\nu})$, ${\bf x}_{n} \equiv ({\bf x}_{n}^{\nu})$ oraz  $\theta_{n} \equiv (\theta_{n}^{\nu})$. Założenie to nie jest potrzebne przy wyprowadzeniu postaci (\ref{Fisher_information-kinetic form}) pojemności informacyjnej.\\ {\bf Uwaga}: Co więcej, {\it informacja o $\theta_{n}$ musi pozostać w rozkładzie $p_{n}$ oraz jego amplitudzie $q_{n}$. Wcześniej umówiliśmy się, że indeks $n$ zawiera tą informację}. Po umiejscowieniu informacji o $\theta_{n}$ w indeksie $n$ można, w razie potrzeby wynikającej np. z fizyki zjawiska, zażądać  dodatkowo niezmienniczości ze względu na przesunięcie. 


\subsection{Postać kinematyczna pojemności zapisana w prawdopodobieństwie}

\label{Postac kinematyczna pojemnosci zapisana w prawdopodobienstwie}

Poniżej podamy postać kinematyczną pojemności zapisaną w  (punktowych) prawdopodobieństwach próby. Postać ta jest bardziej pierwotna niż (\ref{Fisher_information with q dla przejscia}), chociaż w treści skryptu wykorzystywana jedynie w Dodatku. \\
\\
Puntem wyjścia jest pojemność (\ref{postac I bez log p po theta}):
\begin{eqnarray}
I \equiv  I(\Theta) = \sum_{n=1}^N {\int_{\cal Y} d^{4}{\bf y}_n \frac{1}{{p_{n} \left(  {\bf y}_n|\theta_{n}  \right)}} \sum\limits_{\nu=0}^{3}  {\left( {\frac{{\partial p_{n} \left(  {\bf y}_n|\theta_{n}  \right)}}{{\partial \theta_{n \nu} }}}  {\frac{{\partial p_{n} \left(  {\bf y}_n|\theta_{n}  \right)}}{{\partial \theta_{n}^{ \nu} }}} \right) } } \; . \nonumber
\end{eqnarray}  
Korzystając z przejścia do addytywnych przesunięć ${\bf x}_{n} \equiv({\bf x}^{\nu}_{n})$ ,(\ref{parameters separation}), oraz z ``reguły łańcuchowej'' (\ref{chain rule}) dla pochodnej, otrzymujemy (podobnie do (\ref{Fisher_information-kinetic form})) następującą {\it kinematyczną postać pojemności informacyjnej}, wyrażoną w prawdopodobieństwach:
\begin{eqnarray}
\label{postac I dla p po x}
I = \sum_{n=1}^N {\int_{{\cal X}_{n}} d^{4}{\bf x}_{n} \frac{1}{{p_{n} \left(  {\bf x}_{n}  \right)}} \sum\limits_{\nu=0}^{3}  {\left( {\frac{{\partial p_{n} \left(  {\bf x}_{n}  \right)}}{{\partial {\bf x}_{n \nu} }}}  {\frac{{\partial p_{n} \left(  {\bf x}_{n}  \right)}}{{\partial {\bf x}_{n}^{ \nu} }}} \right) } } \; , 
\end{eqnarray}  
gdzie,  podobnie jak poprzednio dla amplitud, pozostawiliśmy  całą informację o {\it  $\theta_{n}$ w indeksie $n$ rozkładu $p_{n}({\bf x}_{n})$}. \\
\\
{\bf Postać kinematyczna $I$ zapisana w prawdopodobieństwie}: W końcu,  zakładając, że zakres zmienności wszystkich ${\bf x}_{n}^{\nu}$ jest dla każdego $n$ taki sam,  pomijamy  indeks $n$ przy tej zmiennej (ale nie przy rozkładzie $p_{n}$), otrzymując:
\begin{eqnarray}
\label{postac I dla p po x bez n}
I = \sum_{n=1}^N {\int_{\cal X} d^{4}{\bf x} \frac{1}{{p_{n} \left(  {\bf x}  \right)}} \sum\limits_{\nu=0}^{3}  {\left( {\frac{{\partial p_{n} \left(  {\bf x}  \right)}}{{\partial {\bf x}_{\nu} }}}  {\frac{{\partial p_{n} \left(  {\bf x}  \right)}}{{\partial {\bf x}^{ \nu} }}} \right) } } \; . 
\end{eqnarray}  
Postać tą wykorzystamy w Dodatku jako pierwotną przy wyprowadzeniu elektrodynamiki Max\-we\-lla, granicy słabego pola w teorii grawitacji oraz twierdzenia $I$ fizyki statystycznej.

\section[Równania master]{Równania master}

\label{master eq}

Podejdźmy nieco inaczej niż w Rozdziale~\ref{structural principle} do problemu estymacji. Rozwińmy $P(\tilde{\Theta})$ w szereg Taylora  wokół prawdziwej wartości parametru $\Theta$ i wycałkujmy po całej przestrzeni próby ${\cal B}$, otrzymując:
\begin{eqnarray}
\label{rozw w szereg T dla P}
 &  & \!\!\!\!\!\!\!\!\int_{\cal B}\!\! d y \!\left(P(\tilde{\Theta})-P(\Theta)\right) = \! \int_{\cal B} \! d y \!\left(\sum_{n=1}^{N}\!\frac{\partial P(\Theta)}{\partial\theta_{n}}(\tilde{\theta}_{n}-\theta_{n})\right.\nonumber \\
 & + & \left.\frac{1}{2}\sum_{n,n'=1}^{N} \frac{\partial^{2}P(\Theta)}
{\partial\theta_{n'}\partial\theta_{n}}(\tilde{\theta}_{n}-\theta_{n})(\tilde{\theta}_{n'}-\theta_{n'})+\cdots\right) \; ,
 \end{eqnarray}
gdzie użyto oznaczenia  $\frac{\partial P(\Theta)}{\partial\theta_{n}} \equiv \frac{\partial P(\widetilde{\Theta})}{\partial \tilde{\theta}_{n}}\mid_{\widetilde{\Theta} = \Theta}$ oraz podobnie dla wyższych rzędów rozwinięcia.
Ponieważ całkowanie zostaje wykonane po całej przestrzeni próby ${\cal B}$, zatem biorąc pod uwagę warunek normalizacji  $\int_{\cal B} d y \, P(\Theta) = \int_{\cal B} d y \, P(\tilde{\Theta})=1$
widzimy, że lewa strona równania (\ref{rozw w szereg T dla P}) jest równa zero. Pomijając człony wyższego rzędu\footnote{Co jest słuszne 
nawet na poziomie gęstości o ile tylko $P(\tilde{\Theta})$
$\in$ ${\cal S}$ nie posiada  w $\Theta$ wyższych dżetów niż drugiego rzędu.}, otrzymujemy:
\begin{eqnarray}
\label{L expand macroscop bez P P}
\int_{\cal B} \! d y \!\left(\sum_{n=1}^{N}\!\frac{\partial P(\Theta)}{\partial\theta_{n}}(\tilde{\theta}_{n}-\theta_{n}) 
 + \frac{1}{2}\sum_{n,n'=1}^{N} \frac{\partial^{2}P(\Theta)}
{\partial\theta_{n'}\partial\theta_{n}}(\tilde{\theta}_{n}-\theta_{n})(\tilde{\theta}_{n'}-\theta_{n'}) \right)  = 0 \; .
\end{eqnarray}
Dla estymatorów $\tilde{\Theta}$  lokalnie nieobciążonych 
\cite{Amari Nagaoka book} 
równanie (\ref{L expand macroscop bez P P}) przyjmuje dla konkretnych $n$ oraz $n'$ następującą postać {\bf równania master}: 
\begin{eqnarray}
\label{L master oczekiwana}
\int_{\cal B} \! d y  \; \frac{\partial^{2}P(\Theta)}
{\partial\theta_{n'}\partial\theta_{n}}\,(\tilde{\theta}_{n}-\theta_{n})(\tilde{\theta}_{n'}-\theta_{n'})   = 0  \; , \;\;\;\;\;\; n, n' =1,2,...,N \; .
\end{eqnarray}
%
Gdy parametr  $\theta_{n}^{\nu}$ ma index Minkowskiego 
$\nu$, wtedy można pokazać, że wykorzystując  $P=\prod_{n=1}^{N}p_{n}({\bf y}_{n})$ 
w (\ref{L master oczekiwana}) otrzymujemy, po przejściu do zmiennych Fisherowskich (porównaj (\ref{zmienne Fisherowskie})),  równanie mające w granicy $\tilde{\theta}_{n} \rightarrow \theta_{n}$ następującą postać  {\bf obserwowaną  równania master}:
\begin{eqnarray}
\label{conservation flow eq}
\frac{\partial p_{n}({\bf y}_{n})}{\partial t_{n}} + \sum_{i=1}^{3}\frac{\partial\, p_{n}({\bf y}_{n})}{\partial {\bf y}_{n}^{i}}\, v_{n}^{i}=0\;,\;\;\; \;\;\; {\rm gdzie} \;\;\;\; v_{n}^{i} = \lim_{\tilde{\theta}_{n} \rightarrow \theta_{n}} \hat{v}_{n}^{i}\equiv\frac{\tilde{\theta}_{n}^{i} - \theta_{n}^{i}}{\tilde{\theta}_{n}^{0} - \theta_{n}^{0}}  \; , \;\;\; n=1,2,...,N \; , \;\;\;
\end{eqnarray}
będącego typem  {\bf równania ciągłości strumienia}, 
gdzie  $t_{n}\equiv {\bf y}_{n}^{0}$. W (\ref{conservation flow eq})  $\, \theta_{n}^{i}$ oraz $\theta_{n}^{0}$ są odpowiednio wartościami  oczekiwanymi położenia oraz czasu układu. 

\section[Podsumowanie rozważań]{Podsumowanie rozważań}

\label{Podsumowanie rozwazan}

Podstawowym przesłaniem wyniesionym z metody estymacyjnej Friedena-Soffera jest to, że $TFI$ jest poprzednikiem Lagrangianu układu \cite{Frieden}. Temat ten rozwiniemy w kolejnym rozdziale. Pewnym minusem teorii Friedena-Soffera mogła wydawać się konieczność zapostulowania nowych zasad informacyjnych. 
Co prawda z puntu widzenia fenomenologii skuteczność tych zasad w wyprowadzeniu dużej liczby modeli użytecznych do opisu zjawisk wydaje się być całkiem satysfakcjonująca, jednak wyprowadzenie tych zasad przesunęłoby   teorię do obszaru bardziej podstawowego. Pozwoliłoby to zarówno na podanie jej przyszłych ograniczeń fenomenologicznych jak i jej możliwych teoretycznych uogólnień. \\
W tym kontekstcie, tą właśnie rolę spełnia wyprowadzenie strukturalnej zasady informacyjnej jako konsekwencji analityczności logarytmu funkcji wiarygodności w otoczeniu prawdziwej wartości parametru $\Theta$ oraz wskazanie geometrycznego znaczenia informacyjnej zasady wariacyjnej, 
która leżu u podstaw zasady ekstremizacji działania fizycznego. W obecnym rozdziale zwrócono uwagę, że u podstaw informacyjnego zrozumienia zasady wariacyjnej 
może leżeć idea idealnego pomiaru \cite{Dziekuje informacja_2}, przy której wariacja pojemności informacyjnej $I$ jest równa (z wyjątkiem znaku) wariacji informacji strukturalnej $Q$. \\
W powyższym rozdziale wyprowadzono też równanie master (\ref{L master oczekiwana}) dla funkcji wiarygodności, które prowadzi do   równania ciągłości strumienia dla punktowego rozkładu w próbie  (\ref{conservation flow eq}). Ciekawe jest to, że równanie master pojawia się z rozwinięcia  funkcji wiarygodności w szereg Taylora wokół prawdziwej wartości parametru $\Theta$. Siłą rzeczy (por. (\ref{rozw w szereg T dla P})) nie pojawia się więc w nich (logarytmiczna) część nieliniowa
struktury układu. Ta gałaź uogólnienia MNW  w klasycznej statystycznej estymacji leży bliżej teorii procesów stochastycznych \cite{Sobczyk_Luczka} niż EFI.  \\
\\
Wyprowadzając  w  Rozdziale~\ref{structural principle} strukturalną zasadę informacyjną \cite{Dziekuje informacja_2} wykazano, że  metoda  Frie\-dena-Soffera  jest pewną modyfikacją MNW, pozwalającą,  jak się okaże, na nieparametryczną estymację równań ruchu teorii pola lub równań generujących rozkład fizyki statystycznej \cite{Frieden}. Wiele z tych równań otrzymano już w  \cite{Frieden}  zgodnie z informatycznym zrozumieniem Friedena-Soffera wspomnianym na końcu Rozdziału~\ref{information transfer}. W  \cite{Mroziakiewicz} wyprowadzenia te zostały sprawdzone dla przyjętej w obecnym skrypcie fizycznej postaci zasad informacyjnych \cite{Dziekuje informacja_1}. \\
Jednakże dopiero wyprowadzenie strukturalnej zasady informacyjnej pozwala na faktoryzację z obserwowanej $SI$ części, będącej miarą probabilistyczną 
i w związku z tym  na prawidłowe umieszczenie rozkładów spełniających równania różniczkowe metody EFI w odpowiednich podprzestrzeniach przestrzeni statystycznej.
Dlatego omówieniu bądź przeliczeniu niektórych rozwiązań EFI z uwzględnieniem tego faktu poświęcimy dwa następne rozdziały.  \\
Należy jednak podkreślić, że Frieden, Soffer i ich współpracownicy Plastino i Plastino,  podali metodę rozwiązania  układu (różniczkowych) zasad informacyjnych dla problemu EFI, która jest bardzo skuteczna, gdyż poza warunkami brzegowymi i ewentualnymi równaniami ciągłości nie jest  ograniczona przez żadną konkretną postać rozkładu. Metodę tą wykorzystamy w dalszym ciągu analizy.


\chapter[Kryteria informacyjne w  teorii pola]{Kryteria informacyjne w  teorii pola}

\label{Kryteria informacyjne w teorii pola}

{\it Główne estymacyjne przesłanie metody} EFI. Jak  stwierdziliśmy poprzednio, ponieważ podstawowa myśl stojąca za metodą 
EFI jest następująca:
Skoro $IF$ jest infinitezymalnym typem entropii względnej Kulback-Leibler'a, która służy do 
statystycznego wyboru pomiędzy zaproponowanymi  {\it ręcznie} modelami, zatem po dodatkowym {\it ręcznym}, aczkolwiek uzasadnionym, nałożeniu  różniczkowych zasad strukturalnych na układ, staje się ona  metodą estymującą  równania ruchu i ich wyboru drogą wymogu spełnienia 
zasad\footnote{W tym punkcie, działa EFI w stosunku do estymowanego równania ruchu  jak MNW w stosunku do estymowanego  parametru dla rozkładu znanego 
typu.
}  
informacyjnych.  Bądź,  jeśli ktoś woli, metoda EFI jest metodą estymującą rokłady, które są rozwiązaniami tych równań. Jest więc to metoda estymacji nieparametrycznej. Wspomniane równania to np. równania ruchu teorii pola bądź równania generujące rozkłady fizyki statystycznej. 


\section[Informacja Fishera i klasyfikacja modeli]{Informacja Fishera i klasyfikacja modeli 
}

\label{Informacja Fishera i klasyfikacja modeli}


Obecny rozdział poświęcony jest głównie przedstawieniu wstępnej klasyfikacji modeli fizycznych ze względu na skończoność (bądź nieskończoność) pojemności informacyjnej $I$. 
Ponadto, poniższe rozważania dla modeli ze skończonym $I$ dotyczą  wyłącznie modeli metody EFI. Kolejna, bardziej  szczegółowa klasyfikacja  pozwala sklasyfikować modele ze względu na wielkość próby $N$.\\
Jak pokażemy mechanika klasyczna posiada nieskończoną pojemność informacyjną $I$. Ściśle mówiąc, mechanika klasyczna jest teorią z  symplektyczną strukturą  rozmaitości  i nie posiada struktury  statystycznej. Czasami jednak słyszy się stwierdzenie, że jest ona stochastyczną granicą mechaniki kwantowej. Ale i na odwrót, według von Neumann'a \cite{Neumann} teoria kwantowa jest niespójna z istnieniem zespołów nie posiadających rozmycia (rozproszenia). W związku z tym, dość powszechnie uważa się, że występowanie  
odstępstw od klasycznego zachowania się układów  można uchwycić jedynie na poziomie statystycznym \cite{Peres}. 
  \\
Poniżej udowodnimy twierdzenie klasycznej statystyki mówiące o  niemożliwości wyprowadzenia mechaniki falowej\footnote{Wspólnotę mechaniki falowej z kwantową ograniczymy do typów równań różniczkowych zagadnienia Sturm'a-Liouville'a oraz zasady nieoznaczoności Heisenberga. Dowód przeprowadzony w ramach mechaniki falowej w obszarze wspólnym dla obu teorii falowych, uznajemy co najwyżej jako przesłankę jego słuszności w mechanice kwantowej.
} 
metody EFI z mechaniki klasycznej. W tym celu wykorzystamy statystyczne pojęcie pojemności informacyjnej, które jest  narzędziem dla dwóch sprzężonych z sobą zagadnień, a mianowicie powyżej wspomnianego statystyczego dowodu o niewyprowadzalności mechaniki kwantowej z klasycznej i związanego z nim problemu konsystencji samospójnego  formalizmu. Ostatni fakt wykorzystywany jest  w takich gałęziach badań fizycznych jak nadprzewodnictwo \cite{superconductivity}, fizyka atomowa i cząstek elementarmych \cite{bib B-K-1} oraz astrofizyka \cite{Bednarek}.

\subsection{Podział modeli ze względu na $N$ oraz kategorie $I$}

\label{rozdzielnosci mech fal i klas}

Jak dotąd nie odnieśliśmy się do wartości $N$ wymiaru próby. Pierwsza klasyfikacja związana z $N$ jest ogólna. Tzn. pokażemy, że modele należą do dwóch różnych, ogólnych kategorii z różną wartością $N$. Pierwsza z nich posiada skończoną wartość $N$ i jest związana ze skończoną wartością $I$.  Obejmuje ona modele mechaniki falowej i klasycznych teorii pola, gdyż jedno skończenie wymiarowe, polowe rozwiązanie równań ruchu określa ewolucję układu wraz z pełnym określeniem  jego struktury w przestrzeni i czasie. 
Natomiast mechanika klasyczna należy do drugiej kategorii z nieskończonym $N$, 
gdyż rozwiązanie równania ruchu nie określa struktury cząstki, która 
musi być niezależnie od tego równania określona poprzez zdefiniowanie, w każdym punkcie toru cząstki, jej punktowej struktury (np. poprzez dystrybucję $\delta$-Diraca). Mechanika klasyczna okazuje się posiadać nieskończoną wartość~$I$. 
%
%

\subsubsection{Dowód podziału na dwie kategorie $I$} 

\label{kategorie I - dowod}

Zobrazujmy powyższe słowa następującą analizą. Dla uproszczenia rozważmy układ jednowymiarowy w położeniu\footnote{Wielowymiarowość czasoprzestrzenna może, w kontekście obecnych rozważań,  zmienić co najwyżej znak pojemności informacyjnej.}. Załóżmy wpierw, że układ jest opisany przez nieosobliwą dystrybucję.  Wtedy dla $N\rightarrow\infty$ pojemność informacyjna $I$, (\ref{I dla pn jeden parametr}), rozbiega się do nieskończoności. Taka sama sytuacja zachodzi jednak dla każdej {\it osobliwej} dystrybucji jak np. dystrybucja $\delta$-Diraca. Sprawdźmy, że tak jest istotnie. Rozważmy punktową cząstkę swobodną, dla uproszczenia w spoczynku, w położeniu $\theta$, oraz  $\delta$-Diracowski ciąg funkcji, np. ciąg funkcji Gaussa:
\begin{eqnarray}
\label{ciag Gaussa}
\left\{\delta_{k}(y_{n}|\theta) = \frac{k}{\sqrt{\pi}}\, \exp(-k^{2}(y_{n}-\theta)^{2})\right\} \; , \;\;\; {\rm gdzie} \;\;\; k = 1,2,3,... \;\; .
\end{eqnarray}   
Wtedy, ponieważ dla określonego indeksu $k$  ciągu (\ref{ciag Gaussa}), pojemność  informacyjna (\ref{I dla pn jeden parametr}): 
\begin{eqnarray}
\label{pojemnosc I dla ciag Gaussa}
I_{k} = \sum_{n=1}^{N} {\int{dy_{n} {\delta_{k}(y_{n}|\theta)}\left({\frac{{\partial \ln \delta_{k}(y_{n}|\theta)}}{{\partial\theta}}}\right)^{2}}} \; , \;\;\; {\rm gdzie} \;\;\; k = 1,2,3,... \;\; 
\end{eqnarray}   
jest równa:
\begin{eqnarray}
\label{wartosc pojemnosci I_k dla ciag Gaussa}
I_{k} = \frac{N}{\sigma_{k}^{2}} \; , \;\;\; {\rm dla} \;\;\; k = 1,2,3,... \;\; ,
\end{eqnarray}     
gdzie  $\sigma_{k}^{2}=\frac{1}{2k^{2}}$ opisuje wariancję położenia cząstki dla $k$-tego elementu ciągu (\ref{ciag Gaussa}), więc widzimy, że  $I_{k}$ rozbiega się do nieskończoności dla $N \rightarrow \infty\,$ i nawet jeszcze mocniej, gdy dodatkowo $k  \rightarrow \infty$. \\
Posumowując, dla $N \rightarrow \infty$ pojemność informacyjna $I$ nie istnieje, obojętnie z jaką dystrybucją mielibyśmy do czynienia. \\
Istnieją więc {\it  dwie, powyżej wymienione, rozdzielne kategorie modeli odnoszące się do wymiaru $N$ próby}. 
Tzn. dla jednych, takich jak mechanika falowa i teorie pola, $N$ oraz $I$ są  skończone,  
podczas gdy mechanika klasyczna tworzy osobną klasę z nieskończonym $N$ oraz $I$.  
To kończy dowód \cite{Dziekuje informacja_1} o 
niewyprowadzalności modeli falowych i teorio-polowych z mechaniki klasycznej.  \\
  \\
Powyższy dowód nie obejmuje możliwości wyprowadzenia mechaniki falowej (czy też teorii kwantowych) z klasycznej teorii pola bądź samospójnej teorii pola 
\cite{bib B-K-1,Dziekuje_Jacek_nova_2,Dziekuje_Jacek_nova_1}. \\
{\bf Uwaga}: Oznacza to, że mechanika klasyczna nie ma skończonego statystycznego 
pochodzenia\footnote{Frieden co prawda wyprowadził  również mechanikę klasyczną z mechaniki falowej, ale jedynie jako graniczny przypadek  $\hbar \rightarrow 0$, a wartość $N$ w tym wyprowadzeniu jest nieistotna  \cite{Frieden}.
}, 
chyba, że tak jak w (\ref{ciag Gaussa}) wprowadzi się nieskończoną 
liczbę statystycznych parametrów, co jednak pociąga za sobą nieskończoność pojemności informacyjnej $I$.


\subsection{Podział modeli ze skończonym $I$ na podklasy z różnym $N$}

\label{podzial modeli ze wzgledu na N}

Jak już  wspomninaliśmy, Frieden i Soffer \cite{Frieden} wyprowadzili modele falowe  posługując się pojęciem pojemności informacyjnej $I$ oraz zasadami informacyjnymi estymacyjnej metody EFI. Rozwiniemy ten temat w dalszej części obecnego rozdziału. Na razie zauważmy, że stosując 
zasadniczo\footnote{Wyjątkiem jest równanie Kleina-Gordona, w wyprowadzeniu którego odwołujemy się jedynie do zasady wariacyjnej, natomiast struktura układu jest narzucona z góry (por. (\ref{Q for N scalar}).)
} 
jednocześnie obie zasady informacyjne, strukturalną $\widetilde{\textit{i'}} + \widetilde{\mathbf{C}} + \kappa \, \textit{q} = 0$, (\ref{zmodyfikowana obserwowana zas strukt z P i z kappa}), oraz wariacyjną $\delta(I + Q) = 0$, (\ref{var K rozpisana}),  oraz uwzględniając odpowiednie fizyczne więzy (wyrażone narzuceniem  na układ np. równania ciągłości, symetrii oraz warunków brzegowych),  
otrzymujemy zróżnicowanie ze  względu na $N$ modeli posiadających skończone wartości $N$ oraz $I$. I tak równanie  Kleina-Gordona oraz równanie Schr{ö}dingera jako jego nierelatywistyczna granica (por. Dodatek~\ref{Rownanie Schrodingera}) posiadają rangę pola $N=2$, równanie Diraca posiada $N=8$, równania Maxwell'a posiadają $N=4$, a teoria grawitacji,  zasadniczo bardziej w ujęciu Logunova  \cite{Denisov-Logunov} niż ogólnej teorii względności, posiada $N=10$ (Dodatek~\ref{general relativity case}).

\subsection{Konkluzje i konsekwencje podziału modeli na  kategorie  $I$}

\label{konkluzje o dwoch kategoriach I}

Powyżej otrzymaliśmy rezultat mówiący, że wszystkie modele opisane strukturalną zasadą informacyjną należą do kategorii skończonej wartości pojemności informacyjnej $I$ oraz, że  mechanika klasyczna należy do kategorii nieskończonego $I$. Zatem w ramach zagadnień rozważanych w skrypcie, granica nie leży pomiędzy tym co micro a makro, ale przebiega pomiędzy teoriami, które mają pochodzenie statystyczne  oraz tymi, które  mają pochodzenie klasyczno-mechaniczne. Albo lepiej, pomiędzy tym co ma pochodzenie falowe lub szerzej, teorio-polowe, oraz tym co ma pochodzenie ściśle punktowe. \\
Ponieważ w konstrukcji modeli klasycznej teorii pola oraz mechaniki falowej, użyty jest ten sam statystyczny formalizm informacji Fishera, dlatego jest ona również właściwym narzędziem w konstrukcji samospójnych teorii pola \cite{bib B-K-1},  łącząc modele mechaniki falowej i klasycznej teorii pola w jeden, logicznie spójny aparat matematyczny.  \\
Jak wiemy, aby otrzymać jakąkolwiek teorię pola, metoda  EFI używa dwóch nowych zasad, wariacyjnej (\ref{var K rozpisana}), która minimalizuje całkowitą fizyczną informację układu oraz obserwowanej  (\ref{zmodyfikowana obserwowana zas strukt z P i z kappa}) i oczekiwanej (\ref{rownowaznosc strukt i zmodyfikowanego strukt}) zasady strukturalnej, która 
tą informację zeruje. Frieden i Soffer \cite{Frieden} zwrócili uwagę, że pojęcie informacji poprzedza pojęcie fizycznego działania, 
a wprowadzony formalizm można słusznie nazwać podejściem Friedena do równań ruchu. Sporo też na tej drodze konstrukcji modeli fizycznych już zrobiono. Jednakże  liczne zagadnienia, ze względu na odmienne niż w  \cite{Frieden} zrozumienie zasady strukturalnej (patrz \cite{Dziekuje informacja_1,Mroziakiewicz,Dziekuje informacja_2} oraz obecny skrypt),  wymagają ponownego zinterpretowania i zrozumienia. Ciągle na ogólne opracowanie czeka wprowadzenie do formalizmu informacyjnych poprzedników źródeł oraz lepsze zrozumienie fizyki leżącej u podstaw znaczenia wymiaru próby $N$. Poniższe rozważania służą usystematyzowaniu istniejącego już statystycznego aparatu pojęciowego informacji kinetycznej i strukturalnej metody EFI oraz lepszemu opisowi związku informacji fizycznej z całką działania.

\section[Równania różniczkowe metody EFI]{Równania różniczkowe metody EFI 
}

\label{equations of motion}


Kolejna część obecnego rozdziału poświęcona jest omówieniu  rozwiązań zasad informacyjnych metodą EFI dla modeli mechaniki falowej i teorii pola \cite{Dziekuje za models building}. Punktem wyjścia jest pojemność informacyjna $I$ w jej 
kinematycznych postaciach  
(\ref{Fisher_information-kinetic form bez n}):
\begin{eqnarray}
I  =  4  \sum_{n=1}^{N} \int_{\cal X} \!\! d^{4}{\bf x}  \sum_{\nu=0}^{3} \frac{\partial q_{n}({\bf x})}{\partial {\bf x}_{\nu}}
\frac{\partial q_{n}({\bf x})}{\partial {\bf x}^{\nu}} \;  \nonumber
\end{eqnarray}
bądź  (\ref{inf F z psi}):
\begin{eqnarray}
I = 4 N \sum\limits_{n=1}^{{N/2}} \int_{\cal X} d^{4}{\bf x} 
\sum\limits_{\nu=0}^{3} {\frac{\partial\psi_{n}^{*}({\bf x})}{\partial {\bf  x}_{\nu}}}{\frac{\partial \psi_{n}({\bf x})}{\partial {\bf x}^{\nu}}} \; , \nonumber
\end{eqnarray}
wyprowadzonych w Rozdziale~\ref{The kinematical form of the Fisher information}, gdzie ${\bf x}^{\nu}_{n}$ 
są zgodnie z (\ref{parameters separation}) przesunięciami wartości pomiarowych położenia zebranymi przez układ od ich wartości oczekiwanych. 
Wyprowadzenie (\ref{Fisher_information-kinetic form bez n}) oraz (\ref{inf F z psi}) zostało zaprezentowane w  Rozdziale~\ref{The kinematical form of the Fisher information} i nie zakłada ono (w przeciwieństwie do orginalnego wyprowadzenia Friedena-Soffera) konieczności istnienia niezmienniczości przesunięcia rozkładów prawdopodobieństwa. 
Pewne informacje na temat niezmienniczości Lorentzowskiej pojemności informacyjnej $I$ zostały podane  w Rozdziale~\ref{Poj inform zmiennej los poloz}. \\ 
Uogólnienia powyższych kinematycznych postaci na przypadek występowania w układzie pól cechowania omówimy w dalszej części rozdziału. 

\subsection{Ogólna postać funkcji gęstości TFI oraz obserwowane zasady informacyjne}

\label{ogolna postac TFI i zasad obserwowanych}

Wyprowadzenie strukturalnej zasady informacyjnej zostało przedstawione w  Rozdziale~\ref{structural principle}.  Odwołuje się ono do pełnych  danych pomiarowych $({\bf y}_{n})_{n=1}^{N}$, ale jego postać dla przesunięć $({\bf x}_{n})_{n=1}^{N}$ jest dokładnie taka sama  \cite{Dziekuje informacja_2}. Tak więc, poniżej stosowane zasady informacyjne, strukturalna oraz wariacyjna, będą odwoływały się do miary probabilistycznej $ d{\bf x} \, p_{n}(\bf x)$ określonej na przestrzeni przesunięć ${\bf x} \in {\cal X}$ jako przestrzeni bazowej, gdzie ${\cal X}$ jest czasoprzestrzenią Minkowskiego $R^{4}$. \\
\\
Przystąpmy do przedstawienia konstrukcji mechaniki falowej i teorii pola zgodnie z metodą EFI. Według  równania  (\ref{physical K}) TPI została określona jako  $K=Q+I$. 
Ponieważ przesunięcie ${\bf x}_{n}$ nie zależy od parametru $\theta_{m}$ dla $m\neq n$ oraz zakres całkowania dla wszystkich ${\bf x}_{n}$ jest taki sam,  dlatego  $I$ 
redukuje się do diagonalnych postaci (\ref{Fisher_information-kinetic form bez n}) bądź (\ref{inf F z psi}), a  $Q$ do postaci:
\begin{eqnarray}
\label{Q diag z q2}
Q = \sum_{n=1}^{N}\int_{\cal X} d^{4}{\bf x}\, q_{n}^{2}({\bf x})\,\texttt{q\!F}_{n}(q_{n}({\bf x})) \; ,
\end{eqnarray}
zgodnie z oznaczeniem w (\ref{Q dla niezaleznych Yn w d4y}), bądź w przypadku pola $\psi({\bf x})$, (\ref{psi zespolona}), do ogólnej (jak zwykle rzeczywistej) postaci:
\begin{eqnarray}
\label{Q diag z psi2}
Q \equiv Q_{\psi} =  \int_{\cal X} d^{4}{\bf x} \sum_{n,n'=1}^{N/2} \, \psi_{n}^{*}({\bf x}) \psi_{n'}({\bf x}) \, \texttt{q\!F}_{nn'}^{\psi}(\psi({\bf x}), \psi^{*}({\bf x}), \psi^{(l)}({\bf x}), \psi^{*(l)}({\bf x}))  \; ,
\end{eqnarray}  
przy czym cała funkcja podcałkowa jest wielomianem pól $\psi({\bf x})$ oraz  $\psi^{*}({\bf x})$, stopnia nie mniejszego niż 2, oraz ich pochodnych rzędu $l = 1,2,...\;$ (por. (\ref{qF diagonalne})), natomiast $\texttt{q\!F}^{\psi}_{nn'}$ jest pewną obserwowaną (w ogólności zespoloną) informacją strukturalną układu.  Konkretną, jak sie okazuje prostą  postać $Q$ dla przypadku pól skalarnych Kleina-Gordona oraz pola Diraca omówimy poniżej. \\
  \\
{\bf Gęstość TFI}: Korzystając z (\ref{Fisher_information-kinetic form bez n}),  (\ref{Q diag z q2}) oraz (\ref{physical K}), możemy zapisać TFI w postaci: 
\begin{eqnarray}
\label{TPI diag}
\mathbb{S} \equiv K = \int_{\cal X} d^{4}{\bf x}\, k\;,
\end{eqnarray}
gdzie dla pola opisanego amplitudami $q_{n}$:
\begin{eqnarray}
\label{k form}
k = 4\sum_{n=1}^{N}\left[\;\sum_{\nu=0}^{3}\frac{\partial q_{n}({\bf x})}{\partial {\bf x}_{\nu}} \frac{\partial q_{n}({\bf x})}{\partial {\bf x}^{\nu}} \, + \, \frac{1}{4}\, q_{n}^{2}({\bf x})\,\texttt{q\!F}_{n}(q_{n}({\bf x}))\right] \; .
\end{eqnarray}
natomiast dla pola opisanego amplitudami $\psi_{n}$:
\begin{eqnarray}
\label{k form dla psi}
k &=& 4 N \sum_{n, n'=1}^{N/2}\left[\;\sum_{\nu=0}^{3} \delta_{nn'}{\frac{\partial\psi_{n}^{*}({\bf x})}{\partial {\bf  x}_{\nu}}}{\frac{\partial \psi_{n'}({\bf x})}{\partial {\bf x}^{\nu}}} \, \right. \nonumber \\
&+& \left. \, \frac{1}{4}\, \, \psi_{n}^{*}({\bf x}) \psi_{n'}({\bf x}) \, \texttt{q\!F}_{nn'}^{\psi}(\psi({\bf x}), \psi^{*}({\bf x}), \psi^{(l)}({\bf x}), \psi^{*(l)}({\bf x})) \right] \; .
\end{eqnarray}
Równoważność  w (\ref{TPI diag}) sugeruje, że $K$ {\it pełni funkcję statystycznego poprzednika} ({\it całki}) {\it działania} $\mathbb{S}$, natomiast $k$, będące {\it funkcją gęstości} TFI,  jest statystycznym poprzednikiem {\it gęstości Lagrangianu} ${\cal L}$. Sprawie tej poświęcimy jeden z poniższych rozdziałów. \\
  \\
{\bf Uwaga o sformułowaniu Lagrange'a i rząd dżetów funkcji wiarygodności}: W dalszej części skryptu założymy, że obserwowana informacja strukturalna nie zawiera pochodnych pól rzędu wyższego niż $l=1$. Założenie to ma charakter fizyczny. Oznacza ono, że jeśli współrzędne uogólnione (u nas amplitudy) oraz prędkości uogólnione (u nas pochodne amplitud) układu są zadane w pewnej chwili czasu, to ewolucja układu jest całkowicie określona, o ile równania ruchu są 2-giego rzędu. 
Odpowiada to sformułowaniu Lagrange'a wykorzystywanemu w badaniu dynamicznych i termodynamicznych własności układów. \\
Fakt ten z punktu widzenia statystycznego oznacza, że interesują nas tylko takie (pod)przestrzenie statystyczne, dla których 
wszystkie możliwe logarytmy funkcji wiarygodności posiadają $r$-jety $J_{p}^{\,r}({\cal S},\text{R})$ w otoczeniu $U_{p}$ punktu $p \equiv P(\Theta)$ $\in$ ${\cal S}$  rzędu $r \leq 2$  (por. Rozdział~\ref{r-c}). 
Jest to istotne z punktu widzenia obserwowanej IF (\ref{observed IF}) zdefiniowanej pierwotnie poprzez drugie pochodne logarytmu funkcji wiarygodności po parametrach, co z kolei uwożliwia konstrukcję strukturalnej zasady informacyjnej (Rozdział~\ref{structural principle}), która jest równaniem  metody EFI. \\
Jednocześnie oczekiwana IF (\ref{iF and I}) wchodzi  w nierówność Rao-Cramera, której pewną postacią jest, po dokonaniu w informacji Fishera transformacji Fouriera do przestrzeni pędowej, zasada nieoznaczoności Heisenberga  (Dodatek~\ref{Zasada nieoznaczonosci Heisenberga}). Zatem fakt występowania w funkcji Lagrange'a kwadratu pierwszych pochodnych byłby (z tego punktu widzenia) artefaktem konieczności wykonania przez układ estymacji jego czasoprzestrzennych położeń, która to estymacja posiada dolne ogranicze Rao-Cramera na dokładność jednoczesnej estymacji położenia oraz prędkości. \\
\\
\\
{\bf Warunek stałości metryki Rao-Fishera}: Odpowiedzmy jeszcze na pytanie, co z punktu widzenia {\it statystycznego}  oznacza niewystępowanie w rozwinięciu $\ln P(\tilde{\Theta})$ w szereg Taylora (\ref{Freiden like equation}) wyrazów rzędu wyższego  niż drugi (tzn. brak jetów o rzędzie $r>2$). Sytuacja ta ma miejsce, gdy obserwowana informacja  Fishera $\texttt{i\!F}$, (\ref{observed IF}), nie zależy od parametru $\Theta \in V_{\Theta}$, gdzie $V_{\Theta}$ jest przestrzenią parametru $\Theta$. Wtedy bowiem jej pochodne po parametrze $\Theta=(\theta_{n})$ są w równe zero dla każdego punktu $p'=P(\Theta')$ w otoczeniu $U_{p}$. Zatem:
\begin{eqnarray}
\label{observed IF wynika nie jet}
\!\!\!\!\!\!\!\! \frac{\partial }{\partial \Theta'} \texttt{i\!F}\left|_{p'\in \,U_{p}} \right. = \frac{\partial }{\partial \Theta'} \left(-\frac{\partial^{2} \ln P(\Theta')}{\partial \Theta'^{\,2}}\right)\left|_{p'\in \,U_{p}} \right. = 0 \; \Rightarrow \; j_{p'}^{\,r}({\cal S},\text{R}) - j_{p'}^{\,2}({\cal S},\text{R})=0 \; {\rm dla} \; r>2 \,, 
\end{eqnarray}
gdzie $j_{p}^{\,r}({\cal S},\text{R})$ jest dowolnym elementem przestrzeni jetów $J_{p}^{\,r}({\cal S},\text{R})$. Lewa strona powyższej implikacji oznacza, że: 
\begin{eqnarray}
\label{observed IF stala}
(\texttt{i\!F})_{nn'}\left|_{p'\in \,U_{p}} \right. = const. \;\;\; {\rm na } \;\;\; U_{p} \; ,
\end{eqnarray}
tzn. obserwowana informacja Fishera $\texttt{i\!F}$ nie zależy w $U_{p}$ od parametru $\Theta$, z czego wynika {\it nie\-zależność oczekiwanej} IF, {\it czyli metryki Rao-Fishera},  {\it od parametru} $\Theta$ w $U_{p}$. \\
%
Takie zachowanie się metryki Rao-Fishera ma następującą ciekawą  konsekwencję. Otóż w Roz\-dziale \ref{Informacja strukturalna EPR}  okaże się, że fakt {\it stałości metryki Rao-Fishera} (\ref{metryka Rao-Fishera dla EPR}) jest odpowiedzialny za otrzymanie w ramach metody EFI znanych formuł mechaniki kwantowej (\ref{wynikEPR}), opisujących  splątanie w problemie EPR-Bohm'a. 
%
\subsubsection{Postać obserwowana zasad informacyjnych} 

\label{obserw zasady inform}

Nałóżmy na układ informacyjną obserwowaną zasadę strukturalną $\widetilde{\textit{i'}} + \widetilde{\mathbf{C}} + \kappa \, \textit{q} = 0$, (\ref{zmodyfikowana obserwowana zas strukt z P i z kappa}),  oraz informacyjną zasadę  wariacyjną $\delta(I + Q)=0$, (\ref{var K rozpisana}). \\  
\\
{\bf Przypadek z amplitudą $q$}: 
Ze zmodyfikowanej obserwowanej zasady strukturalnej  (\ref{zmodyfikowana obserwowana zas strukt z P i z kappa}), biorąc pod uwagę wcześniejsze przejścia pomiędzy (\ref{IF 2 poch na kwadrat pierwszej}), (\ref{pojemnosc C dla polozenia - powtorka wzoru}),  (\ref{potrz}) oraz (\ref{Fisher_information-kinetic form bez n}), wynika warunek zerowy:
\begin{eqnarray}
\label{eq zero}
\sum_{\nu=0}^{3}\frac{\partial q_{n}({\bf x})}{\partial {\bf x}_{\nu}} \frac{\partial q_{n}({\bf x})}{\partial {\bf x}^{\nu}}  + \,\frac{\kappa}{4}\, q_{n}^{2}({\bf x})\,\texttt{q\!F}_{n}(q_{n}({\bf x}))=0 \; , \;\;\; n =1,2,...,N \; .
\end{eqnarray}
Natomiast  z zasady wariacyjnej (\ref{var K rozpisana}) wynika układ równań Eulera-Lagrange'a: 
\begin{eqnarray}
\label{EL eq}
\sum_{\nu=0}^{3} \frac{\partial }{\partial {\bf x^{\nu}}} \left({\frac{{\partial k}}{{\partial (  \frac{\partial q_{n}({\bf x})}{\partial {\bf x}^{\nu}}  ) }}}\right) = \frac{{\partial k}}{{\partial q_{n}({\bf x})}}  \; , \;\;\; n =1,2,...,N \; ,
\end{eqnarray}
gdzie $k$ zostało podane w (\ref{k form}). \\
\\
{\bf Przypadek z  amplitudą $\psi$}: 
Odpowiednia dla pola $\psi$ obserwowana, zmodyfikowana postać zasady strukturalnej (\ref{zmodyfikowana obserwowana zas strukt z P i z kappa}), która bierze pod uwagę (\ref{inf F z psi}),    
jest następująca:
\begin{eqnarray}
\label{strukt row dla psi}
&&\sum_{\nu=0}^{3} {\frac{\partial\psi_{n}^{*}({\bf x})}{\partial {\bf  x}_{\nu}}}{\frac{\partial \psi_{n}({\bf x})}{\partial {\bf x}^{\nu}}}    \\
&+& \frac{\kappa}{4}\, \sum_{n'=1}^{N/2}  \, \psi_{n}^{*}({\bf x}) \psi_{n'}({\bf x}) \, \texttt{q\!F}_{nn'}^{\psi}(\psi({\bf x}),\psi({\bf x}), \frac{\partial \psi({\bf x})}{\partial {\bf x}},  \frac{\partial \psi^{*}({\bf x})}{\partial {\bf x}} ) = 0 \; , \;\;\;\;\;\; n=1,2,...,N/2 \; . \nonumber
\end{eqnarray}
Natomiast układ równań Eulera-Lagrang'a wynikający z zasady wariacyjnej (\ref{var K rozpisana}) jest następujący:
\begin{eqnarray}
\label{EL eq dla psi}
\sum_{\nu=0}^{3} \frac{\partial }{\partial {\bf x^{\nu}}} \left({\frac{{\partial k}}{{\partial (  \frac{ \partial \psi_{n}^{*}({\bf x})}{\partial {\bf x}^{\nu}}  ) }}}\right) = \frac{{\partial k}}{{\partial \psi_{n}^{*}({\bf x})}}  \; , \;\;\; n =1,2,...,N/2 \; ,
\end{eqnarray}
gdzie $k$ zostało podane w (\ref{k form dla psi}).\\ 
Zauważmy, że powyższa postać $k$, (\ref{k form dla psi}), jest na tyle ogólna, że aby zobaczyć działanie metody EFI wynikające z równań (\ref{strukt row dla psi}) i (\ref{EL eq dla psi}), należy podać konkretną postać $k$, dla każdego zagadnienia z polem typu $\psi$  z osobna. 
  \\
Natomiast, jak się przekonamy, równania (\ref{eq zero}) oraz (\ref{EL eq}) z amplitudami $q_{n}$ są już zapisane w postaci bliskiej ich  bezpośredniego użycia i otrzymania 
jawnej postaci  $\texttt{q\!F}_{n}$ oraz rozwiązań metody EFI, czyli odpowiednich fizycznych równań ruchu (bądź równań generujących, por. Rozdział~\ref{Przyklady}) dla amplitud  $q_{n}({\bf x})$. \\
%
%
\\
Let us summarize this Section. Postacie kinetyczne   (\ref{Fisher_information-kinetic form bez n}) oraz (\ref{inf F z psi}), oparte o informację Fishera  (\ref{observed IF Amari}), są  wykorzystywane do konstrukcji równań ruchu (lub równań generujących rozkład) modeli fizycznych. Występują one w (\ref{eq zero})-(\ref{EL eq dla psi}). Natomiast pierwotna postać pojemności $I$ ma swój początek w (\ref{I dla niezaleznych Yn}) oraz w (\ref{observed IF}) i (\ref{iF diagonalne}).  
Zgodnie z (\ref{rownowaznosc strukt i zmodyfikowanego strukt}),  postacie te są równoważne  na poziomie oczekiwanym. 
\\
Przy samospójnym rozwiązywaniu strukturalnej i wariacyjnej zasady informacyjnej metody EFI, wykorzystywana jest postać obserwowana zasady strukturalnej (\ref{zmodyfikowana obserwowana zas strukt z P i z kappa}), wynikająca z żądania analityczności logarytmu funkcji wiarygodności oraz metryczności przestrzeni statystycznej ${\cal S}$ 
(por. Rozdział~\ref{structural principle}). 
Natomiast  oczekiwana strukturalna zasada informacyjna  (\ref{expected form of information eq}) jest narzędziem pomocniczym w definicji całkowitej fizycznej informacji $K$, (\ref{physical K}), oraz informacyjnej zasady wariacyjnej (\ref{var K rozpisana}). 
\\
\\
Poniżej przekonamy się, że wszystkie modelowe różnice leżą po stronie   $\texttt{q\!F}_{n}$, której postać zależy od konkretnego fizycznego scenariusza, włączając w to  symetrie oraz warunki brzegowe. \\
Po raz pierwszy równania (\ref{eq zero}) oraz (\ref{EL eq})  otrzymali Frieden i Soffer \cite{Frieden}.  Jednak powyższa ich forma uwzględnia inną interpretację $Q$ (jako obecnej stale podczas ewolucji układu) oraz faktoryzację probabilistycznego czynnika $q_{n}^{2}({\bf x})$ zawartego obligatoryjnie w mierze 
całkowej\footnote{Również  wyprowadzenie zasady  strukturalnej (\ref{eq zero}) czyni ją mniej fundamentalną niż w sformułowaniu Friedena-Soffera \cite{Frieden}, a tym co staje się fundamentalne jest funkcja wiarygodności $P(\Theta)$ próbki, czyli łączna gęstość prawdopodobieństwa jej (niewidocznej dla badacza) realizacji.
}. 

\section[TFI oraz specyficzne formy $Q$ w teorii pola]{TFI oraz specyficzne formy $Q$ w teorii pola}

\label{structural inf}


Poniżej zebrano i rozwinięto wyniki EFI otrzymane poprzednio w \cite{Frieden}. Jednakże zapisano je, szczególnie dla $Q$, w otwartej formie z punktu widzenia analizy porównawczej modeli \cite{Dziekuje za models building}. 
Metoda EFI prowadzi do takiego sformułowania metody teorii pola, która jest zgodna z dzisiejszym opisem mechaniki falowej dla szerokiej klasy 
struktur. \\
%
%
Jednakże, czasami metoda EFI wraz z całym towarzyszącym jej statystycznym aparatem pojęciowym może doprowadzić do korekty  istniejącej analizy. Z sytuacją taką możemy mieć do czynienia np. w przypadku sformułowania zasady nieoznaczoności Heisenberga dla pola {\it świetlnego}. Otóż w świetle nowych eksperymentów, w wyniku których otrzymano {\it za wąski impuls świetlny w częstotliwości}   \cite{Roychoudhuri}, standardowa Fourierowska podstawa zasady nieoznaczoności jest ostatnio kwestionowana. Wytłumaczenie istoty nierówności Heisenberga w oparciu o nierówność Rao-Cramera wraz z rozróżnieniem pomiędzy estymacją parametru w przypadku skalarnym \cite{Frieden} (por. Dodatek~\ref{Zasada nieoznaczonosci Heisenberga}) i wektorowym, dla którego zachodzi ciąg nierówności (\ref{porownanie sigma I11 z I11do-1}), może okazać się kluczem do zrozumienia pytań,  narastających  na skutek nowych eksperymentów ze światłem. \\
  \\
W metodzie EFI, informacja strukturalna $Q$ musi zostać wyprowadzona z użyciem zasady strukturalnej, wariacyjnej i czasami pewnych dodatkowych warunków symetrii,  biorących pod uwagę specyficzny fizyczny scenariusz teorii. 
Szeroki opis metod stosowanych przy rozwiązywaniu równań (\ref{EL eq}) oraz (\ref{eq zero})  można znaleźć w  \cite{Frieden}. 
  \\
Jednak poniższe  rozważania powinny okazać się pomocne w zrozumieniu  metody, szczególnie dla układu zasad informacyjnych (\ref{strukt row dla psi}) oraz (\ref{EL eq dla psi})  dla pola $\psi$. Poniżej zostanie pokazane jak możliwe rozwiązania zasad informacyjnych, strukturalnej i wariacyjnej,  przewiduje pojawienia się trzech typów pól: $N$-skalarów \cite{Dziekuje za models building}, fermionów oraz bozonów \cite{Frieden}. \\

\subsection{Informacja Fouriera}

\label{Informacja Fouriera}

Rozważmy cząstkę jako układ  opisany polem rangi $N$ poprzez zbiór zespolonych funkcji falowych $\psi_{n}({\bf x})$, 
$n=1,2,...,N/2$, określonych w czasoprzestrzeni ${\cal X}$ położeń  ${\bf x} \equiv ({\bf x}^{\mu})_{\mu=0}^{\, 3} = (c t,\, {\bf x}^{1},{\bf x}^{2},{\bf x}^{3})$, 
zgodnie z konstrukcją  (\ref{amplitudapsi}) oraz (\ref{x n rownowaznosc}) z Rozdziału~\ref{The kinematical form of the Fisher information}. 
Ich transformaty Fouriera $\phi_{n}({\bf p})$ w  sprzężonej do przestrzeni przesunięć ${\cal X}$, energety\-czno-pędowej przestrzeni  ${\cal P}$ czteropędów ${\bf p} \equiv (\wp^{\mu})_{\mu=0}^{\,3} = (\frac{E}{c},\, \wp^{1},\wp^{2},\wp^{3})$, mają postać: 
\begin{eqnarray}
\label{Fourier transf}
\phi_{n}({\bf p}) = \frac{1}{(2\pi\hbar)^{2}} \int_{\cal X} d^{4} {\bf x} \; \psi_{n}({\bf x}) \, e^{i\,(\,\sum_{\nu=0}^{3}{\bf x}^{\nu} \wp_{\nu})/\hbar} \; ,
\end{eqnarray}
gdzie $\sum_{\nu=0}^{3} {\bf x}^{\nu} \wp_{\nu} = E t - \sum_{l=1}^{3} {\bf x}^{l} \wp^{l}$, a $\hbar$ jest stałą Plancka. \\
  \\
{\bf Uwaga}: Transformacja  Fouriera jest {\it unitarną transformacją  zachowującą miarę} na przestrzeni $L^{2}$ funkcji całkowalnych z kwadratem, tzn.:
\begin{eqnarray}
\label{miara zachowana}
\int_{\cal X} d^{4}{\bf x}\,\psi_{n}^{*}({\bf x})\,\psi_{m}({\bf x}) = \int_{\cal P} d^{4}{\bf p}\,\phi_{n}^{*}({\bf p})\,\phi_{m}({\bf p})  \; ,
\end{eqnarray}
zatem wykorzystując warunek normalizacji prawdopodobieństwa (\ref{prawdpsi})  
otrzymujemy\footnote{Równość:
\begin{eqnarray}
\label{tw Parsevala}
\int_{\cal X} d^{4}{\bf x}\,\psi_{n}^{*}({\bf x})\,\psi_{n}({\bf x}) = \int_{\cal P} d^{4}{\bf p}\,\phi_{n}^{*}({\bf p})\,\phi_{n}({\bf p})  \; ,
\end{eqnarray}
jest treścią twierdzenia Parseval'a.
}: 
\begin{eqnarray}
\label{norm condition}
\sum_{n=1}^{N/2} \int_{\cal X} d^{4}{\bf x} \,|\psi_{n}({\bf x})|^{2} = \sum_{n=1}^{N/2}\int_{\cal P} d^{4} {\bf p}\,|\phi_{n}({\bf p})|^{2} = 1 \; ,
\end{eqnarray}
gdzie $|\psi_{n}({\bf x})|^{2} \equiv \psi_{n}^{*}({\bf x})\,\psi_{n}({\bf x})$ oraz $|\phi_{n}({\bf p})|^{2} \equiv \phi_{n}^{*}({\bf p})\,\phi_{n}({\bf p})$.
Korzystając z (\ref{Fourier transf}) możemy zapisać $I$ podane wzorem (\ref{inf F z psi}) w następujący sposób:
\begin{eqnarray}
\label{I by Fourier tr}
I\left[\psi({\bf x})\right] = I\left[\phi({\bf p})\right] = \frac{4N}{\hbar^{2}} \int_{\cal P} d^{4} {\bf p} \sum_{n=1}^{N/2} |\phi_{n}({\bf p})|^{2}\,(\frac{E^{2}}{c^{2}} - \vec{\wp}^{\,2}\,) \, ,
\end{eqnarray}
gdzie $\vec{\wp}^{\,2} = \sum_{k=1}^{3}\wp_{k} \wp^{k}\,$.\\
  \\
{\bf Określenie kwadratu masy cząstki}: Ponieważ $I$ jest z definicji sumą po wartościach oczekiwanych (por. (\ref{krzy4}) i (\ref{pojemnosc C})), dlatego kwadrat {\it masy} cząstki zdefiniowany jako \cite{Frieden}:
\begin{eqnarray}
\label{m E p}
m^{2} := \frac{1}{c^{2}} \int_{\cal P}  d^{4} {\bf p} \sum_{n=1}^{N/2} |\phi_{n}({\bf p})|^{2} \,(\frac{E^{2}}{c^{2}}-\vec{\wp}^{\,2}\,) \;
\end{eqnarray}
jest stałą niezależnie od statystycznych fluktuacji energii $E$ oraz pędu $\vec{\wp}$, tzn. przynajmniej wtedy, gdy całkowanie jest wykonane (czyli jako średnia). 
Tak więc, dla cząstki swobodnej możemy (\ref{I by Fourier tr}) zapisać następująco: 
\begin{eqnarray}
\label{I by Fourier to m}
I\left[\psi({\bf x})\right] = I\left[\phi({\bf p})\right] = 4N(\frac{m\, c}{\hbar})^{2} = const. \;\; .
\end{eqnarray}
{\bf Informacja Fouriera z $\psi$}: Powyższy warunek oznacza, że: 
\begin{eqnarray}
\label{free field eq all 1}
K_{F} \equiv I\left[\psi({\bf x}^{\mu})\right] - I\left[\phi(p^{\mu})\right]=0 \; , 
\end{eqnarray} 
co korzystając ze stałości  $4N(\frac{m\, c}{\hbar})^{2}$ oraz  (\ref{norm condition}) można zapisać jako warunek spełniony przez pole swobodne rangi $N$:
\begin{eqnarray}
\label{free field eq all 2}
K_{F} = 4 \, N \int_{\cal X} d^{4}{\bf x} \sum_{n=1}^{N/2} \left[\;\sum_{\nu=0}^{3}\left( \frac{\partial\psi_{n}}{\partial {\bf x}_{\nu}} \right)^{\!\!*}\; \frac{\partial\psi_{n}}{\partial {\bf x}^{\nu}}-(\frac{m\, c}{\hbar})^{2}\psi_{n}^{*}\,\psi_{n}\right]  = 0 \; .
\end{eqnarray}
Wielkość $K_{F}$ definiuje tzw. {\it informację Fouriera} ($F$), a  $k_{F}$ jej gęstość:
\begin{eqnarray}
\label{gestosc kF inf Fouriera}
k_{F} = 4 \, N \sum_{n=1}^{N/2} \left[\;\sum_{\nu=0}^{3}\left( \frac{\partial\psi_{n}}{\partial {\bf x}_{\nu}} \right)^{\!\!*}\; \frac{\partial\psi_{n}}{\partial {\bf x}^{\nu}}-(\frac{m\, c}{\hbar})^{2}\psi_{n}^{*}\,\psi_{n}\right]  \; .
\end{eqnarray}
\\
Pomijając fakt, że z powyższych rachunków  $m^2$ wyłania się jako średnia,  
równanie (\ref{free field eq all 1}), a zatem (\ref{free field eq all 2}), jest odbiciem twierdzenia Parseval's (\ref{tw Parsevala}) i jako takie jest ono zdaniem tautologicznym. Fakt ten oznacza, że 
transformacja Fouriera odzwierciedla jedynie zmianę bazy w przestrzeni statystycznej ${\cal S}$. Dlatego też, sam z siebie,  warunek (\ref{free field eq all 2}) nie nakłada żadnego dodatkowego więzu na układ, {\it chyba, że  $4N(\frac{m\, c}{\hbar})^{2}$ jest zadana jako  informacja strukturala  $Q$ układu}. Tylko wtedy (\ref{free field eq all 2}) staje się informacyjną zasadą strukturalną dla układu definującego szczególny typ pola omawianego w Rozdziale~\ref{Klein-Gordon scalars}.

\subsubsection{Informacja Fouriera dla amplitudy rzeczywistej}

\label{Informacja Fouriera dla amplitudy rzeczywistej}

Rozważmy z kolei cząstkę, jako układ opisany polem rangi $N$  określonym   poprzez zbiór rzeczywistych amplitud $q_{n}({\bf x})$, $n=1,2,...,N$, na  czasoprzestrzeni przesunięć ${\bf x} \equiv ({\bf x}^{\mu})_{\mu=0}^{\, 3} \in {\cal X}$  
i posiadającą pojemność kanału informacyjnego jak w   (\ref{Fisher_information-kinetic form bez n}). 
Zespolone transformaty Fouriera $\tilde{q}_{n}({\bf p})$ rzeczywistych  funkcji $q_{n}({\bf x})$, gdzie ${\bf p} \equiv (\wp^{\mu})_{\mu=0}^{\,3} \in {\cal P}$  
jest czteropędem, mają postać: 
\begin{eqnarray}
\label{Fourier transf dla q}
\tilde{q}_{n}({\bf p}) = \frac{1}{(2\pi\hbar)^{2}} \int_{\cal X} d^{4} {\bf x} \; q_{n}({\bf x}) \, e^{i\,(\,\sum_{\nu=0}^{3} {\bf x}^{\nu} \wp_{\nu})/\hbar} \; .
\end{eqnarray}
%
%
%
%
Podobnie jak w (\ref{miara zachowana}), transformacja Fouriera spełnia związek: 
\begin{eqnarray}
\label{miara zachowana dla q}
\int_{\cal X} d^{4}{\bf x}\,q_{n}^{*}({\bf x})\,q_{m}({\bf x}) = \int_{\cal P} d^{4}{\bf p}\,\tilde{q}_{n}^{\,*}({\bf p})\,\tilde{q}_{m}({\bf p})  \; .
\end{eqnarray}
Wykorzystując warunek unormowania prawdopodobieństwa:
\begin{eqnarray}
\label{norm condition dla q}
\frac{1}{N}\sum_{n=1}^{N} \int d^{4}{\bf x} \, q_{n}^{2}({\bf x}) = 1 \; , 
\end{eqnarray} 
%
otrzymujemy, jako konsekwencję twierdzenia Parseval'a: 
\begin{eqnarray}
\label{norm condition dla q i tw Parseval}
\frac{1}{N}\sum_{n=1}^{N} \int_{\cal X} d^{4}{\bf x} \, q_{n}^{2}({\bf x}) = \frac{1}{N} \sum_{n=1}^{N} \int_{\cal P} d^{4} {\bf p}\,| \tilde{q}_{n}({\bf p})|^{2} = 1 \; ,
\end{eqnarray}
gdzie $|q_{n}({\bf x})|^{2} \equiv q_{n}^{2}({\bf x})$ i $| \tilde{q}_{n}({\bf p})|^{2} \equiv \tilde{q}_{n}^{*}({\bf p})\,\tilde{q}_{n}({\bf p})$. \\
\\
{\bf Informacja Fouriera dla $q$}: Podobne rachunki jak wykonane poprzednio dla zespolonego pola rangi $N$, prowadzą 
w przypadku pola określonego poprzez zbiór $N$ rzeczywistych amplitud $q_{n}({\bf x})$ i dla pojemności informacyjnej kanału $I$,  (\ref{Fisher_information-kinetic form bez n}), do następującej jego postaci: 
\begin{eqnarray}
\label{I by Fourier tr dla q}
I\left[q({\bf x})\right] = I\left[\tilde{q}({\bf p})\right] = \frac{4}{\hbar^{2}} \int_{\cal P} d^{4} {\bf p} \sum_{n=1}^{N} |\tilde{q}_{n}({\bf p})|^{2}\,(\frac{E^{2}}{c^{2}} - \vec{\wp}^{\,2}\,) \, ,
\end{eqnarray}
gdzie  $\vec{\wp}^{\,2} = \sum_{k=1}^{3}\wp_{k} \wp^{k}\,$.
\\
  \\
{\bf Określenie kwadratu masy cząstki}: Zatem podobnie jak dla pola zespolonego, również dla pola rzeczywistego, kwadrat masy cząstki zdefiniowany następująco:
\begin{eqnarray}
\label{m E p dla q} 
m^{2} := \frac{1}{N\, c^{2}} \int_{\cal P}  d^{4} {\bf p} \sum_{n=1}^{N} |\tilde{q}_{n}({\bf p})|^{2} \,(\frac{E^{2}}{c^{2}}-\vec{\wp}^{\,2}\,) \;
\end{eqnarray}
jest stałą, która nie zależy od statystycznych fluktuacji energii $E$ oraz pędu $\vec{\wp}$. Tak więc i dla cząstki opisanej polem rzeczywistym rangi $N$, możemy (porównaj (\ref{I by Fourier to m})) zapisać (\ref{I by Fourier tr dla q}) w postaci: 
\begin{eqnarray}
\label{I by Fourier to m dla q}
I\left[q({\bf x})\right] = I\left[\tilde{q}({\bf p})\right] = 4 N \, (\frac{m\, c}{\hbar})^{2} = const. \;\; ,
\end{eqnarray}
co oznacza, że warunek (\ref{informacja Stama vs pojemnosc informacyjna}), $I \geq 0$, pociąga za sobą w zgodzie z  (\ref{I by Fourier to m dla q}) warunek $m^{2} \geq 0$, mówiący o nieobecność tachionów w teorii \cite{dziekuje za neutron}, a wynikający z jej przyczynowości, zgodnie z (\ref{przyczynowosc}) oraz (\ref{informacja Stama vs pojemnosc informacyjna Minkowskiego}). \\
Zauważmy, że w zgodzie z (\ref{m E p dla q}), zerowanie się masy cząstki byłoby niemożliwe dla czasoprzestrzeni z metryką Euklidesową (\ref{metryka E}).  \\
\\
{\bf Informacja Fouriera dla pola typu $q$}: Związek (\ref{I by Fourier to m dla q}) pozwala na zapisanie informacji Fouriera, w przypadku rzeczywistego swobodnego pola rangi $N$, w następującej postaci: 
\begin{eqnarray}
\label{free field eq all 2 real amplitudes}
K_{F} = 4\int_{\cal X} d^{4}{\bf x}\sum_{n=1}^{N}  \left[\;\sum_{\nu=0}^{3}\frac{\partial q_{n}}{\partial {\bf x}_{\nu}} \frac{\partial q_{n}}{\partial {\bf x}^{\nu}} -  \,(\frac{m\, c}{\hbar})^{2}q_{n}^{\,2}\right]=0 \; .
\end{eqnarray}
Wnioski płynące z  zastosowania powyższej postaci informacji Fouriera w przypadku pola bezmasowego, można znaleźć w 
Dodatku~\ref{Maxwell field}.  

%
%
%
%

\subsection{Skalary Kleina-Gordona}

\label{Klein-Gordon scalars}

Rozważmy pole skalarne, którego TFI oznaczmy jako $K_{S}$. 
Równanie ruchu Kleina-Gordona dla swobodnego pola skalarnego rangi $N$ wynika z wariacyjnej zasady informacyjnej (\ref{var K rozpisana}): 
\begin{eqnarray}
\label{wariacyjna zas dla pola skalarnego}
\delta_{(\psi^{*})}K_{S} = 0 \; ,
\end{eqnarray}
gdzie na $k=k_{S}$, określone ogólnie związkiem (\ref{k form dla psi}), nałożony jest dodatkowy warunek, wynikający z następującej postaci informacji strukturalnej $Q$: 
\begin{eqnarray}
\label{Q for N scalar}
Q \left[\phi({\bf p})\right] = Q_{S}  = \int_{\cal X} d^{4}{\bf x} \, \textit{q}_{S} \equiv - \, 4\, N(\frac{m\, c}{\hbar})^{2} \; .
\end{eqnarray}
Dla swobodnego skalarnego pola Kleina-Gordona, TFI jest równa jego informacji Fouriera (\ref{free field eq all 2}), tzn. $K=K_{S}=K_{F}$. Powyżej $\textit{q}_{S}$ jest gęstością informacji strukturalnej dla pola skalarnego: 
\begin{eqnarray}
\label{gestosc inf strukt dla pola skalarnego}
\textit{q}_{S} = - 4 \, N (\frac{m\, c}{\hbar})^{2} \sum_{n=1}^{N/2} \psi_{n}^{*}\,\psi_{n}   \; .
\end{eqnarray}
Zatem z wariacyjnej zasady informacyjnej $\delta_{(\psi^{*})}K_{S} = 0$ 
wynika $N/2$ równań Eulera-Lagrange'a (\ref{EL eq dla psi}), które dla gęstości  TFI równej $k = k_{S} = k_{F}$, (\ref{gestosc kF inf Fouriera}), prowadzą do $N/2$ równań Kleina-Gordona\footnote{Patrz równanie (\ref{row KL dla swobodnego}) w przypisie.}. 
Ich wyprowadzenie można znaleźć w \cite{Frieden,Mroziakiewicz}.\\ 
Uzasadnienie faktu, że pole $(\psi_{n})$ z gęstością informacji strukturalnej $\textit{q}_{S}$ zadaną przez (\ref{gestosc inf strukt dla pola skalarnego}) jest polem skalarnym, wymaga rozważań związanych z badaniem reprezentacji transformacji izometrii pojemności kanału informacyjnego $I$, co odkładamy do Rozdziału~\ref{Dirac field}.\\
\\
{\bf Niezmienniczość Fouriera zasady strukturalnej}: Ponieważ dla (\ref{Q for N scalar}) warunek 
(\ref{free field eq all 2}) stanowi  oczekiwaną strukturalną zasadę informacyjną $I + Q  = 0$, (\ref{expected form of information eq}),  
zatem dla swobodnego pola skalarnego transformacja Fouriera 
jest transformacją  unitarną, ze względu na którą warunek (\ref{free field eq all 2})  pozostaje niezmienniczy. \\
\\
{\bf Masa układu a Fourierowskie splątanie}: Powyższy fakt  
oznacza, że transformacja Fouriera tworzy rodzaj samosplątania pomiędzy reprezentacją położeniową a pędową  realizowanych wartości zmiennych układu występujących 
w $I$ \cite{Frieden},  
a wyprowadzenie oczekiwanej zasady strukturalnej $I+Q = 0$, (por. (\ref{expected form of information eq})), jako  konsekwencji 
analityczności logarytmu funkcji wiarygodności \cite{Dziekuje informacja_2}, 
wyjaśnia je jako splątanie pędowych stopni swobody układu 
spowodowane jego masą  (\ref{m E p}). \\
\\
{\bf Uwaga}: Podkreślmy, że w przypadku swobodnego pola skalarnego, 
oczekiwana zasada structuralna, $I + Q  = 0$, jest jedynie odbiciem warunku  (\ref{Q for N scalar}). 
Jest on warunkiem brzegowym i nie jest on rozwiązywany samo-spójnie wraz z wariacyjną zasadą informacyjną (\ref{wariacyjna zas dla pola skalarnego}). 
\\
  \\
{\bf Typy pól skalarnych}: Podajmy wynikające z powyższej analizy metody EFI dwa typy pól skalarnych:\\
  \\
{\it Zwykłe (naładowane) pole skalarne}: Pole skalarne mające rangę $N=2$ ma tylko jedną składową zespoloną, tzn.  $\psi\equiv(\psi_{n})=\psi_{1}$  \cite{Frieden}. W przypadku tym informacja strukturalna (\ref{Q for N scalar}) jest równa  $Q_{S} = - 8(\frac{m\, c}{\hbar})^{2}$. Naładowane pole skalarne Higgsa $H^{+}$ \cite{ASSzZZ} mogłoby teoretycznie być jego przykładem.\\
  \\ 
{\it $N$-skalary}: W przypadku $n>1$ składowe $\psi_{n}$ podlegają ewolucji opisanej przez $n=N/2$ nie sprzężonych równań Kleina-Gordona z dwoma dodatkowymi więzami. Pierwszy z nich oznacza, że wszystkie pola $\psi_{n}$ mają taką samą masę $m$, a drugim jest warunek normalizacji (\ref{norm condition}).
%
%
Informacja strukturalna $Q$ takiego układu jest określona przez ogólną postać (\ref{Q for N scalar}) dla pola skalarnego rangi $N$. Owe skalarne pola Kleina-Gordona rangi $N$  nazwijmy {\it $N$-skalarami} \cite{Dziekuje za models building}. Są one teoretycznie realizowane w ramach tzw. $\sigma$-modeli teorii pola  \cite{sigma-field theory,Dziekuje za models building}. \\
%
\\
W kolejnym rozdziale omówimy postać TFI oraz $Q$ dla równania Diraca. 
Podstawowe fakty metody EFI dla pól cechowania w elektrodynamice Maxwella \cite{Frieden} omówione są w  Dodatku~\ref{Maxwell field}. Również w Dodatku~\ref{general relativity case} zamieszczona jest postać $Q$ w  teorii grawitacji \cite{Frieden}. W opracowaniu jest postać $Q$ dla pól nieabelowych \cite{Dziekuje za models building}. 

\subsection{TFI równania Kleina-Gordona dla pól rangi $N$}

\label{TPI of the Klein-Gordon equation}

Rozdział ten poświęcony jest konstrukcji równania Diraca metodą EFI, z uwzględnieniem pól cechowania\footnote{Zgodnie z uwagą uczynioną powyżej, skrót rezultatów metody EFI dla samych pól cechowania w elektrodynamice Maxwella znajduje się w Dodatku~\ref{Maxwell field}.}. Szczególną uwagę zwrócono na problem kwadratury TFI pola Kleina-Gordona \cite{Frieden,Dziekuje za models building}.

\subsubsection{Wstępna foliacja  ${\cal S}$ oraz pochodna kowariantna. Ogólny zarys problemu}

\label{foliation of S}


{\bf Wybór przestrzeni bazowej członu kinetycznego}: Wyjściowa struktura modelu EFI opierała się o analizę wartości oczekiwanej informacji fizycznej $K = I + Q$ 
na przestrzeni bazowej próby ${\cal B} = {\cal Y}_{1} \times {\cal Y}_{1} \times ... \times {\cal Y}_{N}$. Niezależne zmienne losowe $Y_{n}$, $n=1,2,...,N\,$, przyjmowały wartości ${\bf y}_{n} \in {\cal Y}_{n}$. Po przejściu od pełnych danych $y \equiv ({\bf y}_{n})_{n=1}^{N} \in {\cal B}$ do niezależnych zmiennych losowych przesunięć $X_{n} = Y_{n} - \theta_{n}$, $n=1,2,...,N\,$, oraz utożsamieniu zbiorów wartości tych zmiennych losowych, tzn. przyjęciu, że ${\cal X}_{n=1} \equiv {\cal X}$, $n=1,2,...,N\,$,  zostały skonstruowane kinematyczne postacie pojemności informacyjnej (\ref{Fisher_information-kinetic form bez n}) oraz (\ref{inf F z psi}). {\it Zatem przestrzenią bazową członów kinetycznych jest zbiór przesunięć} ${\cal X}$.\\
\\
{\bf Estymacja na włóknach}: Zatem model EFI, z którego wyłoni się model teorii pola jest budowany na {\it przestrzeni bazowej}  
${\cal X}$,  będącej w rozważanych przez nas przypadkach czasoprzestrzenią Minkowskiego ${\cal X} \equiv R^{4}$. 
Pojawiła się ona jako konsekwencja 
transformacji modelu statystycznego ${\cal S}$ z określonego w przestrzeni parametru $V_{\Theta} \equiv R^{4}$ do przestrzeni przesunięć  ${\cal X}$.  Jednakże jednocześnie, tak przed jak i po jego przedefiniowaniu, model statystyczny pozostaje zdefiniowany ponad przestrzenią bazową  ${\cal B}$, która jest oryginalną  przestrzenią próby. 
Następnie, w celu uczynienia kinematycznej postaci pojemności informacyjnej $I$ niezmienniczą ze względu na lokalne transformacje cechowania, musimy, poprzez zdefiniowanie pochodnej kowariantnej na przestrzeni bazowej ${\cal X}$, zapisać $I$ w postaci współzmienniczej. Z kolei, zdefiniowanie tej  pochodnej kowariantnej oznacza konieczność podania układu współrzędnych na przestrzeni statystycznej ${\cal S}$, co wynika z tego, że pola cechowania są (jak się okaże) amplitudami typu $q_{n}$ (por. Dodatek~\ref{Maxwell field}).  \\
Zatem wprowadzenie układu współrzędnych na  ${\cal S}$ nie jest  zadaniem trywialnym \cite{Amari Nagaoka book}. Sytuacja ta wynika z konieczności wykonania analizy EFI dla pól rangi $N$, które z góry przekształcają się zgodnie z  transformacjami, których parametry mogą zależeć lokalnie od położenia w przestrzeni bazowej ${\cal X}$. Zatem na ${\cal X}$ określana jest {\it strukturalna grupa symetrii} $G$ wspomnianej powyżej transformacji, która w teorii pola jest grupą Liego pola cechowania. Konstruuje się więc główną wiązkę włóknistą $E({\cal X},G)\,$,  a fizyczną estymację przeprowadza się na włóknach\footnote{
{\bf Uwaga o głównej wiązce (włóknistej)}: Mówiąc zwięźle, główna wiązka to taka wiązka włóknista (określona w Rozdziale~\ref{Uwaga o rozwinieciu funkcji w szereg Taylora}), której włókno jest strukturalną grupą symetrii $G$. \\ 
Podsumujmy jednak całą dotychczasową informację na temat głównej wiązki włóknistej precyzyjnie: 
Mając rozmaitość ${\cal X}$ oraz grupę Liego $G$, główna wiązka włóknista $E({\cal X},G)\,$ jest rozmaitością taką, że:\\
1. Grupa $G$ działa na $E$ w sposób różniczkowalny i bez punktów stałych.\\
2. Przestrzeń bazowa ${\cal X}=E/G$, tzn. ${\cal X}$ jest przestrzenią ilorazową $E$ względem $G$, oraz  istnieje różniczkowalne odwzorowanie (nazywane rzutowaniem) $\pi: E \rightarrow {\cal X}$. 
\\
3. Dla każdej mapy $\{ U_{i} \}$ w atlasie dla ${\cal X}$, istnieje różniczkowalne i odwracalne odwzorowanie $\phi_{j}: \pi^{-1}(U_{j}) \rightarrow U_{j} \times G $ zadane przez $E \rightarrow (\pi({\cal P}), f({\cal P}))$ w każdym punkcie  ${\cal P} \in E$,  gdzie $f: \pi^{-1}(U_{j}) \rightarrow G$ spełnia warunek $f(g \,{\cal P}) = g \, f({\cal P})$, dla każdego $g \in G$. \\
Obraz $\pi^{-1}$, czyli $U_{j} \times G$, jest nazywany włóknem. Zatem każde włókno niesie z sobą kopię grupy 
strukturalnej $G$.
}.\\ 
%
%
%
%
%
Zauważmy, że skoro kwadrat pola cechowania jest elementem ${\cal S}$, więc dla określonej algebry grupy cechowania, wybór cechowania dokonuje częściowej foliacji\footnote{Jako, że baza dla pozostałych pól nie jest jeszcze wybrana. 
} 
przestrzeni ${\cal S}$ na  pierwsze warstwy. 
Dopiero w tym momencie, zasady informacyjne umożliwiają dokonanie  wyboru kolejnych foliacji przestrzeni ${\cal S}$ na warstwy związane z wszystkimi szczególnymi reprezentacjami grupy $G$ \cite{Dziekuje za models building}, których wymiar jest ściśle związany z rangą pól $N$ \cite{Frieden}.  \\
\\
{\bf Sens powyższego rozważania} ujmijmy tak: Cała procedura EFI musi nie tylko od początku wybrać typ amplitudy (tzn. $q_{n}$ lub $\psi_{n}$), określić zasady informacyjne i warunki brzegowe, ale musi być  wykonana od początku we właściwym układzie współrzędnych, który  uwzględnia istnienie  strukturalnej grupy symetrii $G$ oraz związanych z nią pól cechowania. 
Pole cechowania umożliwia bowiem wybór podprzestrzeni $T_{{\cal P}}H$ przestrzeni stycznej $T_{{\cal P}}E$ wiązki głównej $E\equiv E({\cal X},G)$ w każdym jej punkcie ${\cal P}$, tak, że  $T_{{\cal P}}E=T_{{\cal P}}G\otimes T_{{\cal P}}H$, przy czym baza na $T_{{\cal P}}H$ może być zdefiniowana jako liniowa kombinacja czterowymiarowej bazy:
\begin{eqnarray}
\label{baza kowariantna}
(\partial_{\mu}) \equiv \left(\frac{\partial}{\partial {\bf x}^{\mu}} \right) \equiv \left( \frac{\partial}{\partial (c\,t)} \,, \, \frac{\partial}{\partial {\bf x}^{1}}, \frac{\partial}{\partial {\bf x}^{2}}, \frac{\partial}{\partial {\bf x}^{3}} \right) \equiv \left( \frac{\partial}{\partial (c\,t)} \,, \, \vec{\nabla} \right) 
\end{eqnarray}
oraz generatorów $D_{G}$ infinitezymalnych transformacji grupy $G$.\\ 
\\
{\bf Przykład}: W przypadku pól cechowania $A_{\mu}$ grupy $G=U(1)$ pochodna kowariantna ma postać:
\begin{eqnarray}
\label{poch kowariantna na H}
D_{\mu}\equiv(D_{0}, D_{l})=\partial_{\mu} - i \, \frac{e}{c\,\hbar} \, A_{\mu} \, ,
\end{eqnarray}
gdzie $e$ jest ładunkiem elektronu. Zatem, wprowadzając pochodną kowariantną do zasad informacyjnych i stosując metodę EFI, otrzymujemy równania ruchu, które rozwiązując dają bazę na znalezionych (pod)rozmaitościach  przestrzeni statystycznej ${\cal S}$. \\
  \\
Jest wiele sposobów, na które współzmiennicza postać pojemności informacyjnej $I$ może być odczytana. Poniżej skoncentrujemy się na dwóch z nich, jednej dla pól skalarnych rangi $N$ (tzn. $N$-skalarów) oraz drugiej, dla pól fermionowych rangi $N$.

\subsubsection{Równanie ruchu Diraca dla pola swobodnego rangi  $N$.}

\label{Dirac field}

Zgodnie z powyższymi rozważaniami, współzmiennicza forma (\ref{inf F z psi}) pojemności informacyjnej z  pochodną  kowariantną $D_{\mu}$ ma postać:
\begin{eqnarray}
\label{I every psi N field}
I =  4 \, N\int_{\cal X} d^{4}{\bf x} \sum_{n=1}^{N/2} \sum_{\mu=0}^{3} (D_{\mu}\psi_{n})^{*}D^{\mu}\psi_{n} \; .
\end{eqnarray}
Zatem jedyna TFI (\ref{physical K}) dostępna w metodzie EFI dla równania Kleina-Gordona i każdego pola typu  $\psi$ rangi $N$ (skalarnego czy fermionowego) ma  postać: 
\begin{eqnarray}
\label{TPI every field}
K = K_{KG} = \int_{\cal X} d^{4}{\bf x} \; k_{KG} \equiv 4 \, N\int_{\cal X} d^{4}{\bf x} \sum_{n=1}^{N/2} \sum_{\mu=0}^{3} (D_{\mu}\psi_{n})^{*}D^{\mu}\psi_{n} + Q \; ,
\end{eqnarray} 
gdzie $k_{KG}$ jest gęstością TFI dla równania Kleina-Gordona (por. (\ref{k form dla psi})). \\
\\
Podobnie, wykorzystując (\ref{Fisher_information-kinetic form bez n}) w miejsce (\ref{inf F z psi})  moglibyśmy zapisać $K_{KG}$ dla pola bosonowego, do czego powrócimy jednak później.  \\
Są dwie drogi, którymi analiza oparta o  TFI zadaną  przez (\ref{TPI every field}) może podążać \cite{Frieden,Dziekuje za models building}. Pierwsza związana jest z  $N$-skalarami a druga z polami Diraca. 

\subsubsection*{TFI Kleina-Gordona} 

Wtedy, gdy  rozważamy pole $N$-skalara, to jak wiemy $Q$ jest równe $Q_{S}$ zadanemu  przez (\ref{Q for N scalar}). 
Fakt ten oznacza, że  TFI dla równania ruchu $N$-skalara, sprowadza się do postaci Kleina-Gordona, tzn.:
\begin{eqnarray}
\label{TPI for N-scalar}
K = K_{S} = K_{KG} \;\;\;\;{\rm dla}\;\;\;\; Q = Q_{S} \; , \end{eqnarray}
a wariacyjna zasada informacyjna: 
\begin{eqnarray}
\label{var IP for N-scalar}
\delta_{(\psi^{*})}K_{S} \equiv \delta_{(\psi^{*})}(I + Q_{S}) = 0 \; ,
\end{eqnarray}
zadana przez (\ref{var K}), a w konsekwencji przez (\ref{EL eq dla psi}), prowadzi do równania Kleina-Gordona dla pola skalarnego \cite{Frieden,Mroziakiewicz} (patrz (\ref{row KL dla swobodnego})).

\subsubsection*{TFI  Diraca} 

Jak wiemy, dla pola Diraca, równanie  Kleina-Gordona jest otrzymane drogą kwadratury równania Diraca \cite{Fecko-fizyka matematyczna}.  Zatem, mogłoby się wydawać, że metoda informacyjna nie wybiera sama z siebie właściwej postaci TFI dla pola fermionowego. Sytuacja ma się jednak zgoła inaczej. Przedstawimy ją poniżej \cite{Dziekuje za models building}.\\
\\
{\bf Transformacje izometrii $I$}: Zapiszmy (\ref{I every psi N field}) pól $\psi$, zarówno dla $N$-skalara jak i pola fermionowego rangi $N$, w następującej postaci: 
\begin{eqnarray}
\label{I with g}
I  = 4 N \sum_{n=1}^{N/2} \int_{\cal X} d^{4}{\bf x} \,\sum_{\mu,\,\nu=0}^{3} (D_{\mu}\,\psi_{n}^{*}({\bf x}))\, \eta^{\mu\nu}\,(D_{\nu}\,\psi_{n}({\bf x})) \; ,
\end{eqnarray}
która to postać pozwala zauważyć, że jest ona niezmiennicza ze względu na transformacje izometrii działające w $N/2$ - wymiarowej, zespolonej przestrzeni $\mathcal{C}^{N/2}$ 
pól $\psi\equiv(\psi_{n}({\bf x}))$ 
rangi $N$.   \\
\\
{\bf Dwa typy izometrii $I$}:\\
{\bf Przypadek skalarów}: W przypadku $N$-skalarów $\psi$ tworzących podprzestrzeń w  $\mathcal{C}^{N/2}$ izometrie te są transformacjami identycznościowymi,  pozostawiając pola $\psi$ niezmienionymi. \\
{\bf Przypadek pól Diraca}: Natomiast dla pól fermionowych rangi $N$ określonych na przestrzeni bazowej ${\cal X}$ Minkowskiego, transformacje izometrii tworzą grupę Clifforda $Pin(1,3)$, będącą podzbiorem algebry Clifforda $C(1,3)$). Elementy grupy Clifforda $Pin(1,3)$ działają w podprzetrzeni 
$\mathcal{C}^{N/2}$  spinorów $\psi$. \\
\\
{\bf Macierze Diraca}: Okazuje się, że macierze Diraca $\gamma^{\mu}$, $\mu=0,1,2,3\,$, (por. (\ref{kowariantna postac r Diraca})-(\ref{m Pauliego})),  tworzą spinorową reprezentację ortogonalnej bazy w $C(1,3)$ i 
%
%
spełniają  tożsamości:
\begin{eqnarray}
\label{antykomutator dla m Diraca}
\eta^{\mu\nu} = 1/2 \, \{\gamma^{\mu},\gamma^{\nu}\} \; , \;\;\; \mu, \nu = 0,1,2,3 \; ,
\end{eqnarray}
które są podstawowym związkiem dla iloczynu Clifforda.  W przypadku spinorów rangi $N=8$, baza w $Pin(1,3)$ jest  $2^{N/2}=16$ wymiarowa \cite{Fecko-fizyka matematyczna}.  \\
\\
{\bf Faktoryzacja $K_{KG}$}: Dla trywialnego przypadku $N$-skalarów, postać (\ref{TPI every field}) jest formą podstawową. Jednak w przypadku spinorów rangi $N$, po skorzystaniu z (\ref{antykomutator dla m Diraca}) w (\ref{I with g}), można dokonać  rozkładu fizycznej informacji $K_{KG}$, (\ref{TPI every field}), na składowe. 
Postać jawnego rozkładu $K_{KG}$ na składowe i ich faktoryzację, gdzie każdy z otrzymanych czynników jest elementem  grupy Clifforda $Pin(1,3)$, podał Frieden  \cite{Frieden}. \\
\\
Główny rezultat rozkładu fizycznej informacji $K_{KG}$, z uwzględnieniem pól cechowania w pochodnej kowariantnej $D_{\mu}$,  który czyni zadość ``ograniczeniu'' Kleina-Gordona  (co oznacza, że  wymnożenie czynników i dodanie ich dałoby na powrót (\ref{TPI every field})), ma postać:
%
\begin{eqnarray}
\label{TPI Dirac field}
K = K_{D}\equiv4\, N\int_{\cal X} d^{4}{\bf x} \sum_{n=1}^{N/2}\sum_{\mu=0}^{3}(D_{\mu}\psi_{n})^{*}D^{\mu}\psi_{n} + Q \, , 
\end{eqnarray}
gdzie informacja strukturalna $Q$ jest równa:
\begin{eqnarray}
\label{Q in Dirac} 
\!\!\!\!\!\!\!\!\! Q = Q_{D}  =  \int_{\cal X} d^{4}{\bf x} \, \textit{q}_{D} \equiv  -\,4\, N\int_{\cal X} d^{4}{\bf x} \sum_{n=1}^{N/2}\left[v_{1n}\, v_{2n} + (\frac{m\, c}{\hbar})^{2}\psi_{n}^{*}\,\psi_{n}\right]
+  (pozosta{\ell}e\; cz{\ell}ony) \; , \;\;\;\;
\end{eqnarray}
gdzie $\textit{q}_{D}$ jest zgodnie z (\ref{gestosc q dla niezaleznych Yn}) gęstością informacji strukturalnej, a $\frac{N}{2}=4$ wymiarowe wektory kolumnowe $v_{i} = (v_{i1},v_{i2},v_{i3},v_{i4})^{T}$, $i=1,2$,  
mają składowe:
\begin{eqnarray}
v_{1n} = \sum_{n'=1}^{4} \left(i\, {\mathbf 1} \, D_{0}-\beta\frac{m\, c}{\hbar} + \sum_{l=1}^{3} i\, \alpha^{l}\, D_{l}\right)_{n n'} \!\! \psi_{n'}\;\label{free field eq 1} \; , \;\;\;\;\;\;\;\; n=1,2,3,4 \; 
\end{eqnarray}
oraz  
\begin{eqnarray}
\label{free field eq 2}
v_{2n} = \sum_{n'=1}^{4} \left(-i\, {\mathbf 1}\, D_{0}+\beta^{*}\frac{m\, c}{\hbar}+\sum_{l=1}^{3} i\, \alpha^{l *}\,  D_{l}\right)_{n n'} \!\! \psi_{n'}^{*} \; , \;\;\;\;\; n=1,2,3,4 \; ,
\end{eqnarray}
gdzie macierze $\alpha^{l}$, $l=1,2,3$, oraz $\beta$ są macierzami Diraca (\ref{m Diraca alfa beta})\footnote{Macierze Diraca $\gamma^{\mu}$, $\mu=0,1,2,3\,$, występujące w tzw. kowariantnej formie równania Diraca:
\begin{eqnarray}
\label{kowariantna postac r Diraca}
(i \, \gamma^{\mu} D_{\mu} - m) \psi = 0 \; ,
\end{eqnarray}
wyrażają się poprzez  macierze Diraca $\beta$ oraz $\alpha^{l}$, $l=1,2,3$, w sposób następujący:  $\gamma^{0} = \beta$ oraz $\gamma^{l} = \beta \alpha^{l}\,$, ($\gamma_{\mu} = \sum_{\nu=0}^{3} \eta_{\mu \nu} \gamma^{\nu}$). Macierze Diraca $\vec{\alpha} \equiv \left(\alpha^{1},{\alpha^{2}},{\alpha^{3}}\right)$ oraz $\beta$ mają postać: 
\begin{eqnarray}
\label{m Diraca alfa beta}
{\alpha^{l}}=\left({\begin{array}{cc}
0 & {\sigma^{l}}\\
{\sigma^{l}} & 0\end{array}}\right)\!,\;\; l=1,2,3, \quad\quad \beta = \left({\begin{array}{cc}
\! {\mathbf 1}\! & \!0\!\\
\!0\! & \!{-{\mathbf 1}}\!\end{array}}\right)
\end{eqnarray}
gdzie $\sigma^{l}=\sigma_{l}$, $\left(l=1,2,3\right)$ są macierzami Pauliego,
a ${\mathbf 1}$ jest macierzą jednostkową: 
\begin{eqnarray}
\label{m Pauliego}
\sigma^{1}=\left({\begin{array}{cc}
\!0 & 1\!\\
\!1 & 0\!\end{array}}\right)\!\!,\;\; \sigma^{2}=\left({\begin{array}{cc}
\!0\! & \!{-i}\!\\
\!{i}\! & \!0\!\end{array}}\right)\!\!,\;\;
\sigma^{3}=\left({\begin{array}{cc}
\!1\! & \!0\!\\
\!0\! & \!{-1}\!\end{array}}\right)\!\!,\;\; 
{\mathbf 1} =  \left({\begin{array}{cc}
\!1 & 0\!\\
\!0 & 1\!\end{array}}\right) \;\; .
\end{eqnarray}
},  
a ${\mathbf 1}$ jest $4\times4$ - wymiarową macierzą jednostkową. 
\\
\\
{\bf Zasady informacyjne dla równania Diraca}: W końcu, informacyjna zasada  strukturalna (\ref{zmodyfikowana obserwowana zas strukt z P i z kappa}) dla $\kappa=1$ i gęstości informacji strukturalnej $\textit{q}_{D}$ określonej zgodnie z (\ref{Q in Dirac}) oraz zasada wariacyjna (\ref{var K rozpisana}) mają postać:
\begin{eqnarray}
\label{IPs for Dirac}
\widetilde{\textit{i'}} + \widetilde{\mathbf{C}} + \textit{q}_{D} = 0 \; \;\; {\rm dla} \;\;\; \kappa=1 \;\; \;\;\;\;\; {\rm oraz} \;\;\;\;\;\;\;   \delta_{(\psi^{*})} K_{D} = \delta_{(\psi^{*})} (I + Q_{D}) = 0 \; . \;\;\;
\end{eqnarray}
Powyższe zasady dają na poziomie obserwowanym warunki 
(\ref{strukt row dla psi}) oraz (\ref{EL eq dla psi})  metody EFI, czyli układ dwóch równań różniczkowych, które są  rozwiązywane  samospójnie.  W wyniku otrzymujemy równanie Diraca: 
\begin{eqnarray}
\label{Dirac eq}
v_{1} = \left(i\, D_{0}-\beta\frac{m\, c}{\hbar} + \sum_{l=1}^{3}\alpha^{l}\, i\, D_{l}\right)\,\psi\;
 = 0 \; , \;\;\; {\rm gdzie} \;\;\;  
\psi  = \left({\begin{array}{c}
\psi_{1} \\
\psi_{2} \\
\psi_{3} \\
\psi_{4} 
\end{array}}\right) \; .
\end{eqnarray}
{\bf Generacja masy}: Zwróćmy uwagę, że w trakcie powyższego rozkładu $K_{KG}$ z członu kinetycznego $I$, (\ref{I with g}),  generowane są wszystkie składniki informacji strukturalnej $Q$, (\ref{Q in Dirac}). W trakcie tej procedury człon masowy generowany jest poprzez sprzężenie Fourierowskie (\ref{Fourier transf}) pomiędzy reprezentacją  położeniową i pędową oraz uśrednienie dokonane w reprezentacji energetyczno-pędowej, tak jak to  miało miejsce dla (\ref{m E p}). Zatem również masa pola  Diracowskiego  jest przejawem istnienia Fourierowskiego samosplątania pomiędzy reprezentacją położeniową a pędową, a cała procedura jest odbiciem: \\ 
{\bf i)} założenia analityczności logarytmu funkcji wiarygodności, \\
{\bf ii)}    
podziału na część Fisherowską $I$ oraz (początkowo nieznaną) część strukturalną $Q$ z wysumowaniem po kanałach informacyjnych i uśrednieniem po przestrzeni próby ${\cal B}$ opisanym powyżej  (\ref{expected form of information eq}), 
 \\
{\bf iii)} przejścia z pojemnością informacyjną $I$ do Friedenowskiej postaci kinematycznej z całkowaniem po zakresie przesunięć ${\cal X}$ amplitud rozkładu układu, opisanym w Rozdziale~\ref{The kinematical form of the Fisher information}, \\
{\bf iv)} tautologicznego wygenerowania, zgodnie z zasadą Macha,  informacji strukturalnej $Q$ z informacji Fishera $I$, z uwzględnieniem niezmienniczości $I$ ze względu na transformację amplitud będącą jej izometrią oraz transformację Fouriera pomiędzy reprezentacją położeniową i pędową. 
\\
\\
Przedstawiony schemat generacji masy nie obejmuje wyznaczenia jej wartości. Mówi on raczej o tym czego jej pojawienie się jest wyrazem. 
\\
\\
{\bf Warunek zerowania się $pozosta{\ell}ych \; cz\ell on\acute{o}w$}: W równaniach (\ref{free field eq 1}) oraz (\ref{free field eq 2}) macierz jednostkowa ${\mathbf 1}$, która stoi  przy $D_{0}$ oraz  $\frac{N}{2}\times \frac{N}{2} = 4\times4$ wymiarowe macierze Diraca $\beta$, $\alpha_{l}$, są jednymi z elementów grupy Clifforda $Pin(1,3)$. 
%
%
Jak wspomnieliśmy, Frieden \cite{Frieden} przeprowadził opisaną 
decompozycję  $K_{KG}$, (\ref{TPI every field}), do postaci $K_{D}$ podanej w (\ref{TPI Dirac field}). Równocześnie pokazał, że wyrażenie oznaczone w (\ref{Q in Dirac}) jako  ``$pozosta{\ell}e \; cz\ell ony$''  zeruje się przy założeniu, że {\it macierze $\beta$ i $\alpha_{l}$ spełniają relacje algebry Clifforda}. \\
  \\
{\bf Podsumowanie}: Zauważyliśmy, że pojemność informacyjna $I$ w TFI 
jest zadana przez (\ref{I with g}) zarówno dla pól $N$-skalarów jak i pól fermionowych rangi $N$. 
Kluczową sprawą jest, że 
%
%
dla każdego równania ruchu, określona jest tylko jemu charakterystyczna postać $\textit{q}$. 
Zauważyliśmy, że dla $N$-skalarów, jedynym członem, który tworzy całkowitą fizyczną informacyjnę jest  $K_{KG}$ w postaci (\ref{TPI every field}) bez żadnego ukrytego rozkładu i faktoryzacji, pochodzących z nie-Fourierowskiego splątania. 
W przypadku równania Diraca, informacja fizyczna Kleina-Gordona $K_{KG}$, (\ref{TPI every field}), również wchodzi w całkowitą informację fizyczną Diraca $K_{D}$, 
ale tylko jako jej  szczególna   część  (jak to można również zauważyć z porównania (\ref{TPI Dirac field}), (\ref{Q in Dirac}) oraz  (\ref{TPI every field}) i (\ref{Q for N scalar})),  splątana  Fourierowsko w $K_{F}$ (\ref{free field eq all 2}). Fakt ten jest blisko związany z efektem  EPR-Bohm'a \cite{Khrennikov} opisanym w Rozdziale~\ref{Pojemnosc informacyjna zagadnienia EPR}.
\\
\\
\\
{\bf Postać $Q$ dla równania Diraca}: Tak więc, w zgodzie z (\ref{Q in Dirac}), szczegółowa postać  $Q$ dla równania ruchu Diraca okazała się być inna niż dla 
$N$-skalarów i można ją zapisać następująco: 
\begin{eqnarray}
\label{Q for psi field}
Q = Q_{D} \equiv \mathbb{S}_{q} - \,4\, N\int_{\cal X} d^{4}{\bf x} \sum_{n=1}^{N/2}\left[ (\frac{m\, c}{\hbar})^{2}\psi_{n}^{*}\,\psi_{n}\right]
 \, , 
\end{eqnarray}
gdzie $\mathbb{S}_{q}$ jest charakterystyczną częścią Diracowską działania dla kwadratury równania Diraca, wyznaczoną dla pola Diraca o randze $N=8$ przy wzięciu pod uwagę  wszystkich  
symetrii układu. \\
\\
{\bf Całka działania kwadratury równania Diraca}: Ponieważ jednak całka działania (por. Rozdział~\ref{dzialanie v.s. zasady informacyjne}) dla kwadratury  specyficznej części Diracowskiej zachowuje się następująco: 
\begin{eqnarray}
\label{Sq for psi field}
\!\!\!\!\! \mathbb{S}_{q} = - \,4\, N\int_{\cal X} d^{4}{\bf x} \sum_{n=1}^{N/2}\left(v_{1n}\, v_{2n}\right) + (pozosta{\ell}e\; cz{\ell}ony) = 0 \, ,
\end{eqnarray}
tzn. spełnia 
warunek zerowy, zatem 
$K\equiv K_{D}$ redukuje się i (w kwadraturze) określa  postać $K_{KG}$ dla 
(\ref{TPI every field})\footnote{
Równanie Kleina-Gordona otrzymane z wariacji informacji (\ref{TPI every field jawna postac}) ma postać ($\vec{\nabla}=(\partial/\partial x^{l})\,, \;l=1,2,3$):
\begin{eqnarray}
\label{row KL dla dowolnego N}
- c^{2} \hbar^{2} \; ( \vec{\nabla} - \frac{ie\vec{A}}{c\hbar} ) \cdot ( \vec{\nabla} - \frac{ie\vec{A}}{c\hbar} ) \; \psi_{n} + \hbar^{2} (\frac{\partial}{\partial t} + \frac{ie\phi}{\hbar} )^{2} \; \psi_{n} + m^{2} \, c^{4} \, \psi_{n} = 0 \; .
\end{eqnarray}
Dla pola swobodnego czteropotencjał cechowania $A_{\mu} = (\phi, -\vec{A})$ jest równy zeru i wtedy otrzymujemy:
\begin{eqnarray}
\label{row KL dla swobodnego}
-c^{2}\hbar^{2}\, \nabla^{2} \psi_{n} + \hbar^{2}\frac{{\partial^{2}}}{{\partial t^{2}}} \, \psi_{n} + m^{2}c^{4} \, \psi_{n} = 0 \; .
\end{eqnarray}

}: 
\begin{eqnarray}
\label{TPI every field jawna postac}
K = K_{KG} \equiv 4 \, N \int_{\cal X} d^{4}{\bf x} \sum_{n=1}^{N/2} \left[ \sum_{\mu=0}^{3} (D_{\mu}\psi_{n})^{*}D^{\mu}\psi_{n} -  (\frac{m\, c}{\hbar})^{2}\psi_{n}^{*}\,\psi_{n}\right]
  \; .
\end{eqnarray} 
%
%
{\bf Uwaga o fundamentalnej różnicy pola skalarnego i pola Diraca}: 
Ważną sprawą jest zauważenie, że chociaż postać $I$ dla $N$-skalarów oraz pól fermionowych rangi $N$ wygląda ``powierzchniowo'' tak samo, to jednak układy te 
różnią się istotnie. W samej bowiem rzeczy, podczas gdy {\it $N$-skalar jest rozwiązaniem rówania ruchu, które wynika jedynie z wariacyjnej zasady informacyjnej} (\ref{var IP for N-scalar}), to {\it pole fermionowe jest samospójnym rozwiązaniem zasad informacyjnych, zarówno wariacyjnej jak i strukturalnej}, podanych w (\ref{IPs for Dirac}). 
%
%
%



\subsection{Końcowe uwagi o wkładzie $Q$ w zasadę strukturalną}

\label{few comments on Q}

\vspace{3mm}

Wynik Rozdziału~\ref{structural principle} \cite{Dziekuje informacja_2} związany z wyprowadzeniem strukturalnej zasady  informacyjnej jest ogólny,  
o ile tylko, w celu zagwarantowania słuszności rozwinięcia w szereg Taylora, funkcja log-wiarygodności  $\ln P(\Theta)$ jest analityczną 
funkcją wektorowego parametru położenia  $\Theta$ w zbiorze jego  
wartości $V_{\Theta}$. 
%
%
%
Z kolei, w  Rozdziale~\ref{information transfer} zauważyliżmy, że model rozwiązany przez metodę EFI jest modelem metrycznym z metryką Rao-Fisher'a oraz, że na poziomie całkowym, model metryczny jest  równoważny modelowi analitycznemu. \\
Dla układów, które nie posiadają dodatkowych więzów różniczkowych, wszystkie 
położeniowe stopnie swobody są związane jedynie poprzez analizę modelu metrycznego, wynikającą z równań (\ref{zmodyfikowana obserwowana zas strukt z P i z kappa})-(\ref{var K rozpisana}) oraz wspomnianą  zasadę Macha, generującą człon strukturalny z Fisherowskiego członu kinematycznego, co prowadzi do czynnika efektywności $\kappa = 1$. 
Równania Kleina-Gordona oraz Diraca omawiane w Rozdziałach~\ref{Klein-Gordon scalars} oraz \ref{TPI of the Klein-Gordon equation} są modelami tego typu.  
\\ 
%
Jeśli na układ nałożony jest dodatkowy warunek, który nie wynika ani ze strukturalnej zasady informacyjnej (\ref{zmodyfikowana obserwowana zas strukt z P i z kappa}), $\widetilde{\textit{i'}} + \widetilde{\mathbf{C}} +  \textit{q} = 0$ z $\kappa =1$, ani z zasady wariacyjej (\ref{var K rozpisana}), 
wtedy wzrasta związek strukturalny pomiędzy położeniowymi stopniami swobody co powoduje, że  $\kappa$ wiążąca w równaniu  (\ref{zmodyfikowana obserwowana zas strukt z P i z kappa}) gęstość informacji strukturalnej $\textit{q}$ z gęstością pojemności informacyjnej $\widetilde{\textit{i'}}\,$, musi maleć poniżej wartości~1.  
Własność ta wynika z wklęsłości pojemności informacyjnej $I$ przy mieszaniu układów \cite{Frieden}, które  pojawiaja się np. na skutek wprowadzenia dodatkowego różniczkowego więzu.  
%
%
%
Na przykład, omawiany w Dodatku~\ref{Maxwell field} warunek Lorentza dla pola cechowania Maxwella, który jest równaniem typu równania ciągłości strumienia, pojawia się nie jako konsekwencja równań ruchu badanego pola, ale jako 
ograniczenie szukane na drodze niezależnej statystycznej estymacji. 
Najprawdopodobniej ten dodatkowy warunek\footnote{Jednak to przypuszczenie należy udowodnić.} 
pojawia się jako rezultat rozwinięcia w szereg Taylora ``gołej'' funkcji wiarygodności  $P(\Theta)$ \cite{Dziekuje informacja_2},  podobnie jak wyprowadzone w Rozdziale~\ref{master eq} równanie   master (co sygnalizowałoby spójność całej statystycznej metody estymacyjnej\footnote{Z drugiej 
strony, same równania master leżą w innej części klasycznej statystycznej estymacji, tzn. w obszarze działania teorii procesów stochastycznych \cite{Sobczyk_Luczka}.
}
). 
Zatem warunek Lorentza nałożony na układ redukuje jego symetrię, co  pociąga za sobą pojawienie się zasady strukturalnej  $\widetilde{\textit{i'}} + \widetilde{\mathbf{C}} + \kappa \, \textit{q} = 0$, (\ref{zmodyfikowana obserwowana zas strukt z P i z kappa}), z czynnikiem efektywności $\kappa$ mniejszym niż 1.  
W przypadku równań Maxwella omówionych w Dodatku~\ref{Maxwell field},  zwrócimy uwagę \cite{Dziekuje za models building}, że wartość $\kappa =1/2$ pojawia się  automatycznie w metodzie EFI na skutek samospójnego  rozwiązania równania strukturalnego,  wariacyjnego i  nałożonego  warunku Lorentza. 
%
%

%
%
%
%

\subsection{Zasada najmniejszego działania v.s. zasady informacyjne}

\label{dzialanie v.s. zasady informacyjne}

Powyższe rozważania Rozdziału~\ref{structural inf}  są częściowo poświęcone omówieniu dodatkowego wyniku zastosowania metody EFI, tzn. ustaleniu różnych postaci informacji strukturalnej $Q$. Zauważono, że przy umiarkowanie rozbudowanym aparacie metody EFI,  następuje nie tylko wyprowadzenie, tzn. estymacja, jej wynikowego równania ruchu (lub równania generującego rozkład),  ale jakby przy okazji, pojawienie się postaci obserwowanej informacji strukturalnej $\texttt{q\!F}_{n}(q_{n}({\bf x}))$. \\
\\
{\bf Konstrukcja całki działania}: Z powyższego faktu  wynika  określenie  związku pomiędzy całką działania wraz z zasadą najmniejszego  działania, a całkowitą informacją fizyczną wraz z zasadami informacyjnymi. Okazuje się bowiem, że po wstawieniu $\texttt{q\!F}_{n}(q_{n}({\bf x}))$  z powrotem do $K$ otrzymujemy  całkę  działania $\mathbb{S}$ modelu \cite{Dziekuje za models building}: 
\begin{eqnarray}
\label{S and K connection}
\mathbb{S}(q_{n}({\bf x}),\partial q_{n}({\bf x})) = 
K(q_{n}({\bf x}),\partial q_{n}({\bf x}),\texttt{q\!F}_{n}(q_{n}({\bf x}))) \, .
\end{eqnarray}
\\
%
%
Sens równości (\ref{S and K connection}) określa  poniższa  konstrukcja $\mathbb{S}(q_{n}({\bf x}),\partial q_{n}({\bf x}))$.  Wpierw rozwiązujemy zasady informacyjne, strukturalną i wariacyjną,  znajdując obserwowaną informację strukturalną  $\texttt{q\!F}_{n}(q_{n}({\bf x}))$. 
Następnie możemy otrzymać równanie ruchu na dwa sposoby: \\
\\
{\bf (a) Pierwszy sposób analizy}: 
W metodzie EFI wstawiamy   otrzymane  $\texttt{q\!F}_{n}(q_{n}({\bf x}))$ od razu do równań {\it Eulera-Lagrange'a wynikających z wariacyjnej zasady informacyjnej} z gęstością informacji fizycznej $k(q_{n}({\bf x}),\partial q_{n}({\bf x}),$ $\texttt{q\!F}_{n}(q_{n}({\bf x})))$. \\
\\
{\bf (b) Drugi sposób analizy}: 
W drodze do całki działania teorii pola wstawiamy  otrzymane  $\texttt{q\!F}_{n}(q_{n}({\bf x}))$ do 
$K(q_{n}({\bf x}),\partial q_{n}({\bf x}),$ $\texttt{q\!F}_{n}(q_{n}({\bf x})))$ i stosując zasadę najmniejszego działania dla tak skonstruowanego  $\mathbb{S}(q_{n}({\bf x}),\partial q_{n}({\bf x}))$, {\it otrzymujemy  równania Eulera-Lagrange'a teorii pola}.\\
\\
{\bf Rezultat}: Otrzymane w konsekwencji analizy typu  (a) lub (b) równanie ruchu, bądź równanie generujące 
rozkład\footnote{W przypadku wyprowadzenia równania Boltzmanna w Rozdziale~\ref{rozdz.energia}, 
po rozwiązaniu układu równań strukturalnego (\ref{rownanie strukt E}) i wariacyjnego (\ref{rownanie wariacyjne E}), otrzymuje się postać (\ref{postac mikro Q dla E}) na $\texttt{q\!F}_{n}(q_{n}({\bf x}))$. Jest to postać, która po wstawieniu  do równania wariacyjnego (\ref{rownanie wariacyjne E}) daje równanie generujące (\ref{falowe kappa 1}) amplitudę $q_{n}({\bf x})$. Zasada wariacyjna  (\ref{gggg}) (gdzie  $\texttt{q\!F}_{n}(q_{n}({\bf x}))$ wyznaczone w (\ref{postac mikro Q dla E}) występuje w $k$, (\ref{k dla Boltzmanna})),  dałaby również równanie generujące (\ref{falowe kappa 1}). 

Równość (\ref{S and K connection}) należy więc rozumieć następująco. Znając $\mathbb{S}(q_{n}({\bf x}),\partial q_{n}({\bf x}))$, które występuje po lewej stronie równania (\ref{S and K connection}) jako $K$ ze wstawioną,  samospójnie wyznaczoną postacią $\texttt{q\!F}_{n}(q_{n}({\bf x}))$ i wariując  je dopiero wtedy ze względu na $q_{n}({\bf x})$, otrzymujemy to samo równanie generujące co rozwiązanie obu zasad strukturalnej i wariacyjnej jednocześnie.
}, 
jest w obu podejściach 
takie samo\footnote{Natomiast w ogólności, nie można przyrównać samej zasady wariacyjnej metody EFI dającej równanie  Eulera-Lagrange'a (dla amplitudy $q_{n}({\bf x})$), w którym może być uwikłana obserwowana informacja strukturalna $\texttt{q\!F}_{n}(q_{n}({\bf x}))$, do zasady najmniejszego działania $\mathbb{S}$, która  daje  końcową postać równania Eulera-Lagrange'a dla $q_{n}({\bf x})$ bez żadnego, bezpośredniego śladu postaci  $\texttt{q\!F}_{n}(q_{n}({\bf x}))$.
}. 
\\
%
\\
{\bf Struktura statystyczna teorii}: Jednakże, z powodu definicji $\mathbb{S}$ danej przez {\it lewą stronę} 
równania (\ref{S and K connection}), pojawia się pytanie, czy metoda EFI, która wykorzystuje $K$ dane przez {\it prawą stronę} 
równania (\ref{S and K connection}), daje jakieś dodatkowe informacje w porównaniu do  zasady najmniejszego działania dla $\mathbb{S}$. Potwierdzająca odpowiedź jest następująca. 
Według Rozdziału~\ref{structural inf} metoda EFI  dokonuje (zazwyczaj) rozkładu $K$ na podstawowe bloki podając otwarcie  wszystkie fizyczne i statystyczne pojęcia, i narzędzia analizy. Równość (\ref{S and K connection}), która jest jedynie definicją $\mathbb{S}$, wyraża bowiem jednocześnie fakt, że $K$ ma bardziej złożoną strukturę niż~$\mathbb{S}$. \\
Powyższe stwierdzenie oznacza, że $\mathbb{S}$ oraz zasada najmniejszego działania niosą zarówno  mniejszą {\it fizyczną} informację niż  oryginalne zasady informacyjne (uzupełnione fizycznymi więzami modelu), jak i niższą informację o pojęciach {\it statystycznych}~. \\
{\bf Zmniejszenie informacji statystycznej}: 
Po otrzymaniu rozwiązania równań informacyjnych i ``ściągnięciu'' wyniku do $\mathbb{S}$, 
znikają  z $\mathbb{S}$  nie tylko pojęcia informacji Fishera oraz automatycznie pojemności informacyjnej, lecz i pojęcie całkowitej wewnętrznej dokładności modelu, tzn. informacji Stama. \\
%
%
\\
{\bf Zmniejszenie informacji fizycznej}: W końcu, jeśli rzeczywistość (w znaczeniu matematycznych podstaw modelowych) leżąca poza modelami teorii pola byłaby statystyczna,   wtedy odrzucając ich oryginalny związek z analizą na przestrzeni  statystycznej ${\cal S}$ oraz estymacyjną metodą  EFI, która wybiera fizycznie właściwą (pod)przestrzeń ${\cal S}$, tracimy również oryginalne narzędzia tej analizy służące do konstrukcji działania  $\mathbb{S}$. Zamiast tego zadowalamy się zgadywaniem jego postaci jedynie na podstawie fizycznych warunków wstępnych, zastanawiając się np. jak szczególnym jest pojęcie amplitudy oraz jak wyjątkowa jest zasada nieoznaczoności Heisenberga. \\
\\
{\bf Przykład}: Obok uwagi na  wstępie Rozdziału~\ref{structural inf}, ostatnie eksperymenty ze światłem wydają się mówić, że Fourierowskie częstości nie są widoczne w optycznej lokalizacji najmniejszych jego impulsów. 
W takiej sytuacji zasada Heisenberga $\delta\nu\delta t\geq 1/2$  mogłaby nie reprezentować sobą żadnej granicznej fizycznej rzeczywistości, gdzie $\delta\nu$ oraz $\delta t$ są poprzez transformację Fouriera związanymi z sobą  szerokościami połówkowymi  impulsu w częstości oraz w czasie. Gdyby eksperymenty te potwierdziły się, wtedy zasada Heisenberga straciłaby swoje oparcie w transformacji Fouriera dla zmiennych komplementarnych \cite{Roychoudhuri}. Natomiast jest możliwe, że znalazłaby ona wtedy  swoje oparcie w nierówności Rao-Cramera, która podaje dolne ograniczenie na wariancję estymatora parametru w przypadku pomiaru jednokanałowego \cite{Frieden} (por. Dodatek~\ref{Zasada nieoznaczonosci Heisenberga}).

\chapter[Przykłady z fizyki statystycznej i ekonofizyki oraz efekt EPR-Bohm'a]{Przykłady z fizyki statystycznej i ekonofizyki oraz efekt EPR-Bohm'a}

\label{Przyklady}

Ogólne statystyczne podstawy estymacji MNW zostały przedstawione  w  Rozdziałach~\ref{MNW}-\ref{Entropia wzgledna i IF}. Natomiast w Rozdziałach~\ref{Zasady informacyjne} oraz \ref{Kryteria informacyjne w  teorii pola} skryptu przedstawiono  metodę  EFI  jako szczególny typ estymacyjnej procedury  statystycznej, 
opracowanej w ramach teorii pomiaru.  Metoda EFI pokazuje w jaki
sposób wychodząc z pojęcia funkcji wiarygodności oraz pojemności informacyjnej $I$ (por. Rozdział~\ref{The kinematical form of the Fisher information}) 
%
otrzymać więzy strukturalne wynikające z analitycznej informacyjnej zasady  strukturalnej, która wraz z wariacyjną zasadą informacyjną 
%
%
%
%
%
%
oraz równaniem ciągłości (lub ogólniej, równaniem  typu master por. Rozdział~\ref{Geometryczne sformulowanie teorii estymacji})),  prowadzi do równań różniczkowych teorii \cite{Frieden}. W trakcie procedury otrzymujemy informację strukturalną $Q$ opisywanego układu, która wraz z pojemnością informacyjną $I$ tworzy  informację fizyczną układu $K$, będącą statystycznym poprzednikiem całki działania (por. Rozdział~\ref{dzialanie v.s. zasady informacyjne}). \\ 
Wyprowadzenia równań ruchu lub równań generujących rozkład zasadzają się na potraktowaniu wszystkich warunków nałożonych na układ jako związków na odchylenia (fluktuacje) wartości pomiarowych  od wartości oczekiwanych. Przy tym, analizowane dane pojawiają się jako efekt pomiaru dokonanego przez układ (por. Rozdział~\ref{Podstawowe zalozenie Friedena-Soffera}). \\
Obecny Rozdział dzieli się na dwie części, pierwszą mającą  zastosowanie termodynamiczne  oraz drugą, opisującą zjawisko EPR-Bohm'a. Za wyjątkiem krótkiej analizy rozwiązań metody EFI dla równań transportu Boltzmann'a, obie części łączy wymiar  próby $N=1$.

\section[Wyprowadzenie klasycznej fizyki statystycznej z informacji Fishera]{Wyprowadzenie klasycznej fizyki statystycznej z informacji Fishera}

\label{fizykastatystyczna}

Celem obecnego rozdziału jest wyprowadzenie metodą EFI podstawowych rozkładów klasycznej fizyki sta\-tystycznej. 
Otrzymamy więc równania generujące, z których wyprowadzone zostaną:  rozkład Boltzmanna dla energii, a następnie rozkład Maxwella-Boltzmanna dla pędu. Jako przykład zastosowania  analizy w ekonofizyce, podamy przykład  produkcyjności branż w modelu Aoki-Yoshikawy. Przedstawione rachunki idą śladem analizy Friedena, Soffera, Plastino i Plastino \cite{Frieden}, jednak zostaną one wykonane w oparciu o wprowadzoną w  Rozdziale~\ref{structural principle} strukturalną zasadę informacyjną \cite{Dziekuje informacja_2}. Różnicę interpretacyjną  pomiędzy obu podejściami podano w Rozdziale~\ref{information transfer}. \\
Dodatkowo wyprowadzony zostanie warunek informacyjny na górne ograniczenie  tempa wzrostu entropii Shannona \cite{Frieden}.

\subsection{Fizyczne sformułowanie zagadnienia}

\label{Fizyczne sformulowanie zagadnienia dla predkosci}

Rozważmy gaz składający się z $M$ identycznych cząsteczek o masie
$m$ zam\-knięty w zbiorniku. Temperatura gazu ma stałą wartość
$T$. Ruch cząsteczek jest losowy i oddziałują one ze sobą poprzez
siły potencjalne, zderzając się ze sobą i ściankami naczynia, przy
czym zakładamy, że są to zderzenia sprężyste. Średnia prędkość każdej cząsteczki jest równa zero.\\

Oznaczmy przez $\theta_{\wp} = \left(\theta_{\epsilon}, \vec{\theta}_{\wp}\right)$ czterowektor wartości oczekiwanej energii oraz pędu cząsteczki, gdzie indeks
$\wp$ oznacza pęd.  We współrzędnych kartezjańskich 
$\vec{\theta}_{\wp}=\left(\theta_{\wp_1},\theta_{\wp_2}, \theta_{\wp_3}\right)$. Podobnie jak w (\ref{parameters separation})
wprowadzamy zmienną losową $Y = \left( Y_{\epsilon} \equiv \frac{E}{c}, \;\vec{Y}_{\wp}\right) $, przyjmującą warości ${\bf y}=\left({\bf y}_{\epsilon}\equiv \epsilon/c\,,\;\vec{\bf y}_{\wp}\right)$, której składowe spełniają związki: 
\begin{eqnarray}
\label{E}
{\bf y}_{\epsilon}  \equiv \frac{\epsilon}{c} = \theta_{\epsilon} + {\bf x}_{\epsilon} \; , \;\;\;\; \frac{\epsilon_{0}}{c} \le {\bf y}_{\epsilon} \le \infty 
\end{eqnarray}
\begin{eqnarray}
\label{y_p}
\vec{\bf y}_{\wp}=\vec{\theta}_{\wp} + \vec{\bf x}_{\wp} \; , \;\;\;\; \vec{\bf y}_{\wp} = \left({\bf y}_{\wp_1},{\bf y}_{\wp_2},{\bf y}_{\wp_3}\right)\; , \end{eqnarray}
gdzie 
\begin{eqnarray}
\label{xp to p}
\vec{\bf x}_{\wp} = \vec{\wp} \; 
\end{eqnarray}
oznacza fluktuację pędu, natomiast $c$ oznacza prędkość światła. 
Zmienne i parametry energetyczne $ {\bf y}_{\epsilon}$, $\,{\bf x}_{\epsilon}$ oraz  $\theta_{\epsilon}$  zostały wyrażone w  jednostce współrzędnych pędowych $energia/c$. 
Parametry $\theta_{\epsilon}$ oraz $\vec{\theta}_{\wp}$ są odpowiednimi
wartościami oczekiwanymi energii (z dokładnością do 1/c) oraz pędu, a ${\bf x}_{\epsilon}$ oraz $\vec{\bf x}_{\wp}$ fluktuacjami względem wartości oczekiwanych.\\
\\
Znajdziemy rozkład prawdopodobieństwa dla fluktuacji energii $X_{\epsilon}$ przyjmującej wartości ${\bf x}_{\epsilon}$ oraz fluktualcji pędu $\vec{X}_{\wp}$ przyjmującej wartości $\vec{\wp}$ dla jednej, dowolnej cząsteczki gazu w dowolnej chwili czasu $t$. Ponieważ wartość $t$ nie musi być duża, zatem  rozważamy gaz, który nie koniecznie jest w stanie równowagi.
Będziemy więc szukać postaci nierównowagowego rozkładu  prawdopodobieństwa, odpowiadającego tak postawionemu problemowi.  \\
\\
{\bf Uwaga o czterowektorze energii-pędu}: Jednak po pierwsze, współrzędne czterowektora fluktuacji energii-pędu $\left(X_{\epsilon}, \vec{X}_{\wp} \right)$  nie są statystycznie niezależne, tzn. nie są  niezależnymi stopniami swobody układu. Po drugie, jak się okaże, spełniają one zasadę dyspersyjną typu (\ref{m E p}), 
więc nie tworzą układu zmiennych Fishera (porównaj (\ref{zmienne Fisherowskie})).  Zatem ogólny problem wymagałby  estymacji  odpowiednich równań generujących dla skomplikowanego czasoprzestrzennego zagadnienia, tzn. 
należałoby wyznaczyć łączny rozkład prawdopodobieństwa  współrzędnych $\left(X_{\epsilon}, \vec{X}_{\wp} \right)$, co wykracza poza zakres skryptu. Niemniej np. w przypadku relatywistycznych zjawisk astrofizycznych, taka estymacja może okazać się niezbędna. \\
W skrypcie ograniczymy się jedynie do wyznaczenia brzegowych  rozkładów prawdopodobieństwa dla $X_{\epsilon}$ (i w konsekwencji dla $E$) oraz dla $\vec{X}_{\wp}$, co w nierelatywistycznej granicy jest uzasadnione. \\

\subsection{Informacja kinetyczna i strukturalna oraz sformułowanie zasad informacyjnych}
\label{zasady inf dla energii i predkosci}

Tak więc, statystycznie jedna cząsteczka podlega łącznemu rozkładowi prawdopodobieństwa $p\left({\bf x}_{\epsilon},\vec{\bf x}_{\wp}\right)$,
przy czym współrzędne czterowektora pędu nie są niezależne. Uproszczona analiza skoncentruje się na rozkładach brzegowych, których analiza ze względu na wspomniany brak niezależności nie odtwarza analizy łącznej, chociaż jest słuszna w przybliżeniu   nierelatywistycznym.  \\
Wyznaczymy więc {\it brzegowe amplitudy prawdopodobieństwa} $q\left({\bf x}_{\epsilon}\right)$ oraz $q\left(\vec{\bf x}_{\wp}\right)$: 
\begin{eqnarray}
q_{n}\left({\bf x}_{\epsilon}\right),\quad n=1,...,N_{\epsilon} \;\;\;\; {\rm oraz} \;\;\;\;  q_{n}\left(\vec{\bf x}_{\wp}\right), \;\;\;\;  n=1,...,N_{\wp} \; ,
\end{eqnarray}
a następnie powrócimy do brzegowych rozkładów prawdopodobieństwa 
$p\left({\bf x}_{\epsilon}\right)$ oraz $p\left(\vec{\bf x}_{\wp}\right)$. 
W końcu po wyznaczeniu $p\left({\bf x}_{\epsilon}\right)$ skorzystamy z (\ref{E}), aby otrzymać wymagany rozkład $p\left(\epsilon\right)$. Podobnie, korzystając z (\ref{y_p}) otrzymamy po wyznaczeniu  $p\left(\vec{\bf x}_{\wp}\right)$ rozkład prędkości $p\left(\vec{\bf y}_{\wp}\right)$, przy czym te dwa ostatnie są równe w naszych rozważaniach, ze względu na średnią wartość prędkości cząsteczki $\vec{\theta}_{\wp} = 0$.  \\
\\
{\bf Pojemność informacyjna dla parametrów czterowektor  energii-pędu}: Chociaż $\left({\bf x}_{\epsilon}, \vec{\bf x}_{\wp}\right)$ nie jest czterowektorem we wspomnianym ujęciu Fishera, to jest on czterowektorem w sensie Lorentza. Korzystamy więc z metryki czasoprzestrzeni Minkowskiego w postaci $(\eta_{\nu \mu}) = diag(1,-1,-1,-1)$ zgodnie z  (\ref{metryka M}). 
\\
W ogólnym przypadku rozkładu łącznego oraz  
próbkowania czterowektora energii i pędu, pojemność informacyjna ma  dla parametru wektorowego $\Theta = \left((\theta_{\nu n})_{\nu=0}^{3}\right)_{n=1}^{N}$ postać (\ref{pojemnosc C dla polozenia - powtorka wzoru})\footnote{W zgodzie z ogólną własnością kontrakcji indeksu Minkowskiego dla czterowektora $x~\equiv~(x_{\nu})_{\nu=0}^{3}$ $=~(x_{0},x_{1},x_{2},x_{3})$: 
\[
\sum_{\nu \mu = 0}^{3} \eta_{\nu\mu}\frac{df}{dx_{\nu}}\frac{df}{dx_{\mu}}=\left(\frac{df}{dx_{0}}\right)^{2}-\sum_{k=1}^{3} \left(\frac{df}{dx_{k}}\right)^{2} \; .
\] 
W \cite{Frieden} metryka ma postać Euklidesową, a zmienne mają urojoną współrzędną przestrzenną $\left(x_{0}, i \vec{x} \right)$, por.  Rozdział~\ref{Poj inform zmiennej los poloz}.
}:
\begin{eqnarray}
\label{I dla lacznego E p}
I = \sum\limits_{n=1}^{N} \int_{\cal B} dy \; P\left(y|\Theta\right) \left[ {\left(\frac{\partial\ln P\left(y|\Theta\right)}{\partial\theta_{\epsilon \, n}}  \right)^{2}} - \sum\limits_{k=1}^{3} {\left(\frac{\partial\ln P\left(y|\Theta\right)}{\partial\theta_{\wp_{k} n}} \right)^{2}} \right] \;  ,
\end{eqnarray}
gdzie $y=({\bf y})_{n=1}^{N}$ jest $N$-wymiarową próbą, a ${\cal B}$ przestrzenią próby. 
Zatem pojemności informacyjne $I\left(\Theta_\epsilon\right)$ oraz
$I\left(\Theta_{\vec{\wp}}\right)$ dla rozkładów brzegowych fluktuacji energii oraz pędu mają postać\footnote{Jak w Rozdziale~\ref{The kinematical form of the Fisher information},  indeks $n$ przy współrzędnej pominięto, korzystając z założenia, że rozproszenie zmiennej nie zależy od punktu pomiarowego próby.}: 
\begin{eqnarray}
\label{min}
I\left(\Theta_{\epsilon}\right) = 4 \int_{{\cal X}_{\epsilon}}{d{\bf x}_{{\epsilon}} \sum\limits_{n=1}^{N_{{\epsilon}}}{\left({\frac{{dq_{n}\left({{\bf x}_{{\epsilon}}}\right)}}
{{d{\bf x}_{{\epsilon}}}}}\right)^{2}}}
\end{eqnarray}
oraz 
\begin{eqnarray}
\label{minp}
I\left( \Theta_{\vec{\wp}} \right) = - 4 \int_{{\cal X}_{\wp}} d \vec{\bf x}_{\wp} \sum\limits_{n=1}^{N_{\wp}} \sum\limits_{k=1}^{3}
\left( \frac{ dq_{n} \left( \vec{\bf x}_{\wp} \right) }{ dx_{\wp_{k}}} \right)^{2}
\; ,
\end{eqnarray}
gdzie minus w (\ref{I dla lacznego E p}) i w konsekwencji w (\ref{minp}), wynika zgodnie z Rozdziałem~\ref{Poj inform zmiennej los poloz} z uwzględnienia metryki Minkowskiego  (\ref{metryka M}), natomiast ${\cal X}_{\epsilon}$ oraz ${\cal X}_{\wp}$ są zbiorami wartości odpowiednio zmiennych  $X_{\epsilon}$ oraz $X_{\wp}$. Z poniższych rachunków przekonamy się, że nieuwzględnienie metryki Minkowskiego (gdy $\eta_{00} =1$ to $\eta_{kk} =-1$ dla $k=1,2,3$),  doprowadziłoby w konsekwencji w termodynamicznych rozważaniach do błędnego rozkładu 
prędkości\footnote{Odpowiedź 
na pytanie, czy rozważania termodynamiczne są przyczyną metryki Minkowskiego niezbędnej w relatywistycznej teorii pola, wykracza poza obszar skryptu. Niemniej autor skryptu uważa, że tak się istotnie sprawy mają, tzn. że przestrzeń Euklidesowa z transformacją Galileusza są pierwotne wobec przestrzeni Minkowskiego z transformacją Lorentza. Stąd podejście efektywnej teorii pola Logunova \cite{Denisov-Logunov} do teorii grawitacji jest bliższe teorii pomiaru fizycznego Friedena-Soffera (którą jest EFI). Nieco więcej na ten temat można znaleźć w Dodatku~\ref{general relativity case}.}.\\
%
%
\\
{\bf Zasady informacyjne}: Zasady informacyjne, strukturalna (\ref{zmodyfikowana obserwowana zas strukt z P i z kappa}) oraz  wariacyjna (\ref{var K rozpisana}) mają poniższą postać. Dla energii: 
\begin{eqnarray}
\label{epiE}
\widetilde{\textit{i'}}(\Theta_{\epsilon}) + \widetilde{\mathbf{C}}_{\epsilon} + \kappa \, \textit{q}(\Theta_{\epsilon}) = 0 \; , \;\;\; \delta_{(q_{n})}\left(I(\Theta_{\epsilon}) + Q(\Theta_{\epsilon})\right) = 0 \; , 
\end{eqnarray}
oraz dla pędu: 
\begin{eqnarray}
\label{epip}
\widetilde{\textit{i'}}(\Theta_{\vec{\wp}}) + \widetilde{\mathbf{C}}_{\vec{\wp}} + \kappa \, \textit{q}(\Theta_{\vec{\wp}}) = 0 \; , \;\;\; \delta_{(q_{n})}\left(I(\Theta_{\vec{\wp}}) + Q(\Theta_{\vec{\wp}})\right) = 0 \; . 
\end{eqnarray}
Gęstości  pojemności informacyjnych,  $\widetilde{\textit{i'}}(\Theta_{\epsilon})$ oraz $\widetilde{\textit{i'}}(\Theta_{\vec{\wp}})$, są określone zgodnie z (\ref{gestosc i Amarii}) i (\ref{zmodyfikowana obserwowana zas strukt z P i z kappa}), a gęstości  informacji strukturalnych,  $\textit{q}(\Theta_{\epsilon})$  oraz $\textit{q}(\Theta_{\vec{\wp}})$, są określone zgodnie z (\ref{gestosc q dla niezaleznych Yn}), natomiast  $Q(\Theta_{\epsilon})$ oraz $Q(\Theta_{\vec{\wp}})$ są odpowiednimi informacjami  strukturalnymi. \\
W pierwszej kolejności rozważymy problemem (\ref{epiE}) dla $p({\epsilon})$.\\
\\
{\bf Przypomnienie roli zasad informacyjnych}: W rachunkach metody EFI prowadzących do równania generującego rozkład, obok zasady wariacyjnej (\ref{var K rozpisana}) wykorzystywana jest postać  (\ref{zmodyfikowana obserwowana zas strukt z P i z kappa}) zmodyfikowanej obserwowanej zasady strukturalnej. Warto  pamiętać, że zasada (\ref{zmodyfikowana obserwowana zas strukt z P i z kappa}) wynika z żądania istnienia rozwinięcia Taylora 
logarytmu funkcji wiarygodności wokół prawdziwej wartości parametru (por. Rozdział~\ref{structural principle}) oraz z metryczności przestrzeni statystycznej ${\cal S}$.  Natomiast oczekiwana strukturalna zasada informacyjna (\ref{expected form of information eq}) jest narzędziem pomocniczym w definicji całkowitej fizycznej informacji $K$, (\ref{physical K}), oraz informacyjnej zasady wariacyjnej (\ref{var K rozpisana}).

\subsection{Rozkład Boltzmanna dla energii}

\label{rozdz.energia} 

Poniżej podamy rozwiązanie zasad informacyjnych (\ref{epiE}), strukturalnej oraz wariacyjnej, otrzymując w pierwszym kroku analizy równanie generujące amplitudy dla rozkładu Boltzmanna. \\
\\
Załóżmy wstępnie, że  wartość fluktuacji energii ${\bf x}_{{\epsilon}}$ zmienia się w pewnym  zakresie $\left\langle {\bf x}_{{\epsilon}}^{min}, {\bf x}_{{\epsilon}}^{max}\right\rangle$: 
\begin{eqnarray}
\label{zakres}
{\bf x}_{{\epsilon}}^{min} \le {\bf x}_{{\epsilon}} \le {\bf x}_{{\epsilon}}^{max} \;\; .
\end{eqnarray}
W ten sposób  wpierw  uchwycimy ogólną  zależność amplitudy rozkładu od  ${\bf x}_{{\epsilon}}^{max}$, 
a następnie dokonamy przejścia granicznego, przechodząc z górną granicą fluktuacji energii ${\bf x}_{{\epsilon}}^{max}$ do nieskończoności. \\
\\
{\bf Pojemność informacyjna}  ma postać (\ref{Fisher_information-kinetic form bez n}): 
\begin{eqnarray}
\label{informacjaE}
I\left(\Theta_{{\epsilon}}\right) = 4 \int\limits_{{\bf x}_{{\epsilon}}^{min}}^{{\bf x}_{{\epsilon}}^{max}} {d{\bf x}_{{\epsilon}} \sum\limits _{n=1}^{N}{q_{n}^{'2}\left({\bf x}_{{\epsilon}} \right)}} \; , \;\;\;\; {\rm gdzie} \;\;\;\;  \; q_{n}^{'}\left({\bf x}_{{\epsilon}}\right) \equiv \frac{dq_{n}\left({\bf x}_{{\epsilon}}\right)}{d{\bf x}_{{\epsilon}}} \; ,
\end{eqnarray}
gdzie rozkłady prawdopodobieństwa $p_{n}$ dla fluktuacji energii są powiązane z amplitudami $q_{n}$ zależnością (\ref{amplituda a rozklad}): 
\begin{eqnarray}
p_{n}({\bf x}_{{\epsilon}}) = q_{n}^{2}({\bf x}_{{\epsilon}}) \; .
\end{eqnarray} 
{\bf Informacja strukturalna}  zgodnie z (\ref{Q dla niezaleznych Yn w d4y}) jest następująca: 
\begin{eqnarray}
\label{Q diag dla E}
Q(\Theta_{{\epsilon}}) = \int\limits_{{\bf x}_{{\epsilon}}^{min}}^{{\bf x}_{{\epsilon}}^{max}} d{\bf x}_{{\epsilon}} \, \textit{q}(\Theta_{{\epsilon}}) = \int\limits_{{\bf x}_{{\epsilon}}^{min}}^{{\bf x}_{{\epsilon}}^{max}} d{\bf x}_{{\epsilon}} \, \sum\limits _{n=1}^{N} q_{n}^{2}({\bf x}_{{\epsilon}}) \,\texttt{q\!F}_{n}(q_{n}({\bf x}_{{\epsilon}})) \; .
\end{eqnarray}
\\
{\bf Informacyjna zasada wariacyjna} w (\ref{epiE}) dla pojemności informacyjnej $I$, (\ref{informacjaE}), oraz informacji strukturalnej $Q$, (\ref{Q diag dla E}), ma postać:
\begin{eqnarray}
\label{gggg}
\delta_{(q_{n})} K = \delta_{(q_{n})} (I(\Theta_{{\epsilon}}) + Q(\Theta_{{\epsilon}})) = \delta_{(q_{n})} \left(\;\int\limits_{{\bf x}_{{\epsilon}}^{min}}^{{\bf x}_{{\epsilon}}^{max}} {d{\bf x}_{{\epsilon}} \, k} \right) = 0 \; , 
\end{eqnarray}
gdzie $k$ jest równe: 
\begin{eqnarray}
\label{k dla Boltzmanna}
k = 4 \sum\limits_{n=1}^{N} {\left(q_{n}^{'2} + \frac{1}{4}q_{n}^{2}({\bf x}_{{\epsilon}}) \,\texttt{q\!F}_{n}(q_{n}({\bf x}_{{\epsilon}}))  \right)} \; ,
\end{eqnarray}
zgodnie z ogólną postacią gęstości obserwowanej informacji fizycznej (\ref{k form}) dla amplitud $q_{n}$. \\
\\
{\bf Rozwiązaniem problemu wariacyjnego} (\ref{gggg}) wzgędem $q_{n}({\bf x}_{{\epsilon}})$ jest {\it równanie Eulera-Lagrange'a} (\ref{EL eq}):
\begin{eqnarray}
\label{euler}
\frac{d}{{d{\bf x}_{{\epsilon}}}}\left({\frac{{\partial k}}{{\partial q_{n}^{'}({\bf x}_{{\epsilon}})}}}\right) = \frac{{\partial k}}{{\partial q_{n}({\bf x}_{{\epsilon}})}} \;  \;\;\;\; {\rm dla} \;\; n = 1, 2,..., N \; ,
\end{eqnarray}
gdzie 'prim' oznacza pochodną po ${\bf x}_{{\epsilon}}$,  tzn. $q_{n}^{'}({\bf x}_{{\epsilon}}) \equiv d q_{n}({\bf x}_{{\epsilon}})/d{\bf x}_{{\epsilon}}$. \\
\\
Zatem dla rozważanego problemu, równanie (\ref{euler}) dla $k$ jak w (\ref{k dla Boltzmanna}), przyjmuje postać: 
\begin{eqnarray}
\label{rownanie wariacyjne E}
2 \, q_{n}^{''}({\bf x}_{{\epsilon}}) = \frac{{\partial (\frac{1}{4}q_{n}^{2}({\bf x}_{{\epsilon}}) \,\texttt{q\!F}_{n}(q_{n}({\bf x}_{{\epsilon}})))}}{{\partial q_{n}}} = \frac{{d (\frac{1}{4} q_{n}^{2}({\bf x}_{{\epsilon}}) \,\texttt{q\!F}_{n}(q_{n}({\bf x}_{{\epsilon}})))}}{{dq_{n}}} \; .
\label{rozweulera}
\end{eqnarray}
{\bf Zmodyfikowana obserwowana zasada strukturalna}: 
Po wycałkowaniu (\ref{informacjaE}) przez części,  pojemność $I$ wynosi: 
\begin{eqnarray}
\label{postac I po calk czesci}
I(\Theta_{{\epsilon}}) =  - 4 \int\limits_{{\bf x}_{{\epsilon}}^{min}}^{{\bf x}_{{\epsilon}}^{max}} {d{\bf x}_{{\epsilon}}  \sum\limits _{n=1}^{N}{q_{n}({\bf x}_{{\epsilon}}) \, q_{n}^{''}({\bf x}_{{\epsilon}})}} + C \; , 
\end{eqnarray}
gdzie
\begin{eqnarray}
\label{postac IC z pradem}
C \equiv  4 \int\limits_{{\bf x}_{{\epsilon}}^{min}}^{{\bf x}_{{\epsilon}}^{max}} d{\bf x}_{{\epsilon}}  \, \left( q_{n}\left({\bf x}_{{\epsilon}}\right) q^{'}_{n}\left({\bf x}_{{\epsilon}}\right) \right)^{'}  = 4 \sum \limits_{n=1}^{N}{C_{n}} \; 
\end{eqnarray}
oraz
\begin{eqnarray}
\label{Cn}
C_{n} =  {q_{n}\left({\bf x}_{{\epsilon}}^{max}\right)q_{n}^{'}\left({\bf x}_{{\epsilon}}^{max}\right)-q_{n}\left({\bf x}_{{\epsilon}}^{min}\right)q_{n}^{'}\left({\bf x}_{{\epsilon}}^{min}\right)} \; .
\end{eqnarray}
Zatem widzimy, że {\it zmodyfikowana obserwowana zasada strukturalna} w (\ref{epiE}) jest ze względu na (\ref{postac I po calk czesci})-(\ref{postac IC z pradem}) oraz (\ref{Q diag dla E}), 
następująca: 
\begin{eqnarray}
\label{obserwowana zas strukt dla Boltzmann suma n}
\sum\limits_{n=1}^{N}{\left( - {q_{n}({\bf x}_{{\epsilon}}) q_{n}^{''}({\bf x}_{{\epsilon}}) + \tilde{C}_{n}  + \kappa \frac{1}{4}q_{n}^{2}({\bf x}_{{\epsilon}}) \,\texttt{q\!F}_{n}(q_{n}({\bf x}_{{\epsilon}}))}\right)}  = 0 \; , \;\;\;
\end{eqnarray}
gdzie 
\begin{eqnarray}
\label{Cn tilde}
\tilde{C}_{n} = C_{n}/({\bf x}_{{\epsilon}}^{max} - {\bf x}_{{\epsilon}}^{min}) \; . 
\end{eqnarray}
Zatem na poziomie obserwowanym, dla każdego $n=1,2,...,N$,  otrzymujemy zasadę strukturalną w postaci\footnote{Gdyby nie wycałkować w (\ref{postac I po calk czesci}) $I$ przez części, wtedy z (\ref{obserwowana zas strukt z P i z kappa}) i po skorzystaniu z postaci kinematycznej pojemności, 
zasada strukturalna na poziomie obserwowanym miałaby postać: 
\begin{eqnarray}
\label{mikroskowowa dla E}
q_{n}^{' \,2}({\bf x}_{{\epsilon}}) + \frac{\kappa}{4} q_{n}^{2}({\bf x}_{{\epsilon}}) \,\texttt{q\!F}_{n}(q_{n}({\bf  x}_{{\epsilon}})) = 0 \; .
\end{eqnarray}
}:
\begin{eqnarray}
\label{rownanie strukt E}
- q_{n}({\bf x}_{\epsilon}) q_{n}^{''}({\bf x}_{\epsilon}) + \tilde{C}_{n}  + \kappa \frac{1}{4}q_{n}^{2}({\bf x}_{\epsilon}) \,\texttt{q\!F}_{n}(q_{n}({\bf x}_{\epsilon})) = 0 \; .
\end{eqnarray}
Korzystając z (\ref{rownanie wariacyjne E}) w (\ref{rownanie strukt E}) otrzymujemy:
\begin{eqnarray}
\label{row rozn z q oraz qF}
\frac{1}{2} q_{n}  \frac{{d (q_{n}^{2}({\bf x}_{{\epsilon}}) \,\texttt{q\!F}_{n}(q_{n}({\bf x}_{{\epsilon}})))}}{{dq_{n}}} = \kappa q_{n}^{2}({\bf x}_{{\epsilon}}) \,\texttt{q\!F}_{n}(q_{n}({\bf x}_{{\epsilon}})) + 4 \,\tilde{C}_{n} \; .
\end{eqnarray}
Poniżej, dla uproszczenia zapisu pominiemy przy rozwiązywaniu równania (\ref{row rozn z q oraz qF}) oznaczenie argumentu ${\bf x}_{{\epsilon}}$ w amplitudzie $q_{n}$. Zapiszmy (\ref{row rozn z q oraz qF}) w postaci:
\begin{eqnarray}
\label{row rozn z q oraz qF do wycalkowania}
2\frac{{dq_{n}}}{{q_{n}}}=\frac{{d\left[\frac{1}{4}q_{n}^{2} \,\texttt{q\!F}_{n}(q_{n}) \right]}}{{\kappa \, \left[\frac{1}{4} \, q_{n}^{2} \,\texttt{q\!F}_{n}(q_{n})\right]  +   \tilde{C}_{n}}} \; .
\end{eqnarray}
Rozwiązując powyższe równanie różniczkowe otrzymujemy kolejno: 
\[
2\ln q_{n} + \alpha_{n}^{'} = \frac{1}{\kappa} \ln \left({ \frac{\kappa }{4} \, q_{n}^{2} \, \texttt{q\!F}_{n}(q_{n}) +  \, \tilde{C}_{n}}\right)\]
i po przekształceniu:
\[
2\kappa\ln\alpha_{n}^{''} q_{n} = \ln \left({ \frac{\kappa}{4}q_{n}^{2} \,\texttt{q\!F}_{n}(q_{n}) +  \, \tilde{C}_{n}}\right) \; ,
\]
gdzie $\alpha_{n}^{'}$ jest w ogólności zespoloną stałą całkowania  oraz stała  $ \alpha_{n}^{''} = \exp(\alpha_{n}^{'}/2)$, skąd: 
\begin{eqnarray}
\label{wyn-j}
q_{n}^{2}({\bf x}_{{\epsilon}}) \,\texttt{q\!F}_{n}(q_{n}) =  \frac{4}{\kappa}\left({\alpha_{n}^{2}q_{n}^{2\kappa} -  \, \tilde{C}_{n}}\right) \; ,
\end{eqnarray}
gdzie (w ogólności zespolona) stała $\alpha_{n}^{2}=\alpha_{n}^{'' \, 2\kappa}$. Zatem w rezultacie otrzymaliśmy obserwowaną informację strukturalną,  $\texttt{q\!F}_{n}(q_{n}({\bf x}_{{\epsilon}}))$, jako znaną funkcję amplitudy $q_{n}({\bf x}_{{\epsilon}})$. \\
\\
{\bf Równanie generujące z $\kappa$}: W końcu, korzystając z (\ref{rownanie wariacyjne E}) 
oraz z (\ref{wyn-j}), otrzymujemy równanie różniczkowe dla $q_{n}({\bf x}_{{\epsilon}})$:
\begin{eqnarray}
\label{row generujace ampl dla E}
q_{n}^{''}({\bf x}_{{\epsilon}}) = \alpha_{n}^{2} q_{n}^{2\kappa-1}({\bf x}_{{\epsilon}}) \, .
\end{eqnarray}
{\bf Podsumowanie}: Wariacyjna oraz strukturalna obserwowana zasada informacyjna zostały zapisane w postaci układu równań  (\ref{rownanie wariacyjne E}) oraz  (\ref{rownanie strukt E}), który rozwiązując, dał w rezultacie szukaną postać (\ref{row generujace ampl dla E}) {\it równania generującego} amplitudę $q_{n}({\bf x}_{{\epsilon}})$. Znalezienie postaci tego równania było pośrednim  celem metody EFI.   \\
\\
{\bf Warunek analityczności i metryczności}: Z postaci (\ref{row generujace ampl dla E}) widać, że skoro amplituda $q_{n}({\bf x}_{{\epsilon}})$ jest funkcją rzeczywistą, zatem stała $\alpha_{n}^{2}$ musi być rzeczywista. 
Rozwiązanie równania (\ref{row generujace ampl dla E}) znajdziemy wtedy, gdy współczynnik efektywności wynosi:
\begin{eqnarray}
\label{kappa1}
\kappa = 1 \;\; ,
\end{eqnarray}
czyli dla przypadku, gdy zasada strukturalna dla układu jest konsekwencją analityczności logarytmu funkcji wiarygodności  
(\ref{rozwiniecie w szereg Taylora}) oraz metryczności przestrzeni statystycznej ${\cal S}$. \\
W przypadku $\kappa = 1 $ obserwowana informacja strukturalna (\ref{wyn-j}) przyjmuje prostą postać: 
\begin{eqnarray}
\label{postac mikro Q dla E}
q_{n}^{2}({\bf x}_{{\epsilon}}) \,\texttt{q\!F}_{n}(q_{n}({\bf x}_{{\epsilon}})) = 4 ( {\alpha_{n}^{2}q_{n}^{2}({\bf x}_{{\epsilon}}) -  \tilde{C}_{n}} ) \; ,
\end{eqnarray}
a całkowa postać $SI$ jest następująca:
\begin{eqnarray}
\label{postac calkowa Q dla E}
Q(\Theta_{{\epsilon}}) = 4 \int{d{\bf x}_{{\epsilon}} \sum\limits_{n=1}^{N}{\alpha_{n}^{2}q_{n}^{2}\left({\bf x}_{{\epsilon}}\right)} - 4  \sum\limits_{n=1}^{N}{C_{n}}} \; ,
\end{eqnarray}
co po skorzystaniu z unormowania kwadratu amplitudy: 
\begin{eqnarray}
\label{unormowanie q2}
\int{d{\bf x}_{{\epsilon}} \, q_{n}^{2}\left({\bf x}_{{\epsilon}}\right)} = \int{d{\bf x}_{{\epsilon}} \, p_{n}\left({\bf x}_{{\epsilon}}\right)} = 1 \; ,
\end{eqnarray}
daje:
\begin{eqnarray}
\label{postac Q dla E po war normalizacji qn}
Q(\Theta_{{\epsilon}}) = 4  \sum\limits_{n=1}^{N} ( \alpha_{n}^{2} - C_{n} )  = 4  \sum\limits_{n=1}^{N} \alpha_{n}^{2} - C \; .
\end{eqnarray}
W ostatnim przejściu w  
(\ref{postac Q dla E po war normalizacji qn}) 
skorzystano z postaci stałej $C \equiv 4 \sum \limits_{n=1}^{N}{C_{n}}$
wprowadzonej w (\ref{postac I po calk czesci}). \\
 \\
{\bf Równanie generujące}: Dla rozważanego przypadku $\kappa =1$, równanie generujące  (\ref{row generujace ampl dla E}) ma postać: 
\begin{eqnarray}
\label{falowe kappa 1}
q_{n}^{''}({\bf x}_{{\epsilon}}) = \alpha_{n}^{2}q_{n}({\bf x}_{{\epsilon}}) \; .
\end{eqnarray}
\\
{\bf Sprawdzenie rachunków}: Wstawiając (\ref{falowe kappa 1}) do  (\ref{postac I po calk czesci}),  otrzymujemy następującą postać pojemności informacyjnej: 
\begin{eqnarray}
\label{I  wraz z Ic dla E}
I(\Theta_{{\epsilon}}) = - 4 \sum\limits_{n=1}^{N} {\alpha_{n}^{2}}  + C \; ,
\end{eqnarray}
co wraz z (\ref{postac Q dla E po war normalizacji qn}) oznacza sprawdzenie poprawności rachunku, poprzez spełnienie przez otrzymane rozwiązanie 
oczekiwanego strukturalnego warunku  $I+Q =0$, zgodnie z (\ref{rownowaznosc strukt i zmodyfikowanego strukt}).\\
  \\
{\bf Rozwiązanie równania generującego}: Kolejnym etapem analizy jest rozwiązanie równania generującego (\ref{falowe kappa 1}). \\
Najogólniejsza postać amplitudy $q_{n}({\bf x}_{{\epsilon}})$ będącej rozwiązaniem 
równania (\ref{falowe kappa 1}) jest przy warunku ${\bf x}_{{\epsilon}}^{min} \le {\bf x}_{{\epsilon}} \le {\bf x}_{{\epsilon}}^{max} $, jak w (\ref{zakres}),  następująca:
\begin{eqnarray}
\label{rozw2q}
q_{n}({\bf x}_{{\epsilon}}) = B_{n} \exp \left(\alpha_{n} {\bf x}_{{\epsilon}} \right) + D_{n}\exp\left(-\alpha_{n} {\bf x}_{{\epsilon}} \right) \, ,  \;\;\;\;  {\bf x}_{{\epsilon}}^{min} \le x\le {\bf x}_{{\epsilon}}^{max} \, , \;\;\; B_{n},D_{n} = const. 
\end{eqnarray}
Ponieważ stała $\alpha_{n}^{2}$ jest rzeczywista, zatem wprowadzając nową rzeczywistą stałą $\beta_{n}$, można $\alpha_{n}$ przedstawić jako $\alpha_{n}=\beta_{n}$ i wtedy rozwiązanie (\ref{rozw2q}) ma charakter  czysto eksponencjalny: 
\begin{eqnarray}
\label{rozw ekspo dla E}
q_{n}({\bf x}_{{\epsilon}})=B_{n}\exp\left(\beta_{n}{\bf x}_{{\epsilon}}\right)+D_{n}\exp\left(-\beta_{n}{\bf x}_{{\epsilon}}\right) \; ,
\end{eqnarray}
bądź jako $\alpha_{n}=i\beta_{n}$, i wtedy rozwiązanie (\ref{rozw2q}) ma charakter czysto trygonometryczny: 
\begin{eqnarray}
\label{rozw trygonometryczne dla E}
q_{n}({\bf x}_{{\epsilon}})=B_{n}\exp\left(i\beta_{n}{\bf x}_{{\epsilon}}\right)+D_{n}\exp\left(-i\beta_{n}{\bf x}_{{\epsilon}}\right) \; .
\end{eqnarray}
{\bf Warunek normalizacji dla amplitud}: Tak określone funkcje muszą być dopuszczalne jako amplitudy
prawdopodobieństwa, zatem muszą spełniać {\it warunek normalizacji} dla  gęstości prawdopodobieństwa (\ref{unormowanie q2}). \\
Załóżmy, że wartość fluktuacji energii ${\bf x}_{{\epsilon}}$ nie jest ograniczona od góry, co zrealizujemy jako dążenie ${\bf x}_{{\epsilon}}^{max}$ do nieskończoności.
Jednak kwadrat funkcji trygonometrycznej 
nie może być unormowany do jedności dla ${\bf x}_{{\epsilon}}^{max} \rightarrow\infty$, zatem funkcja trygonometryczna (\ref{rozw trygonometryczne dla E}) nie jest dopuszczalnym rozwiązaniem. \\
{\bf Pozostaje więc rozwiązanie eksponencjalne} (\ref{rozw ekspo dla E}). Ponieważ jednak warunek unormowania (\ref{unormowanie q2}) ma być zadany na  przedziale otwartym ${\bf x}_{{\epsilon}}^{min} \le {\bf x}_{\epsilon} < \infty$, zatem część rozwiązania z dodatnią eksponentą musi być odrzucona ze względu na jej rozbieżność do nieskończoności. 
Skąd otrzymujemy żądanie, że dla $\beta_{n} \ge 0$ stała $B_{n}=0$.\\
\\
Podsumowując, szukana postać amplitudy jest więc następująca: \begin{eqnarray}
\label{qn rozwiazanie dla E}
q_{n}\left({\bf x}_{{\epsilon}}\right) = {D_{n}\exp\left({-\beta_{n} \, {\bf x}_{{\epsilon}}}\right)}\; , \quad \beta_{n}\in\mathbf{R}_{+} \; , \quad {\bf x}_{{\epsilon}}^{min} \le {\bf x}_{{\epsilon}} < \infty \; .
\end{eqnarray}
Z powyższego i z warunku normalizacji (\ref{unormowanie q2}) $\int_{{\bf x}_{{\epsilon}}^{min}}^{\infty}{d{\bf x}_{{\epsilon}} \, q_{n}^{2}\left({\bf x}_{{\epsilon}}\right)}  = 1$,  
wyznaczamy stałą $D_{n}$, otrzymując: 
\begin{eqnarray}
D_{n} = \sqrt{2\beta_{n}}\exp\left({\beta_{n} \, {\bf x}_{{\epsilon}}^{min}}\right)\; .
\end{eqnarray}
\\
{\bf Ostateczna postać amplitudy}: Rozwiązanie to zostało otrzymane dla przypadku  $\beta_{n} = \alpha_{n} \in \mathbf{R}_{+}$, zatem 
ostateczną postacią (\ref{rozw2q}) jest: 
\begin{eqnarray}
\label{rozw cosinus dla ampl dla E}
q_{n}\left({\bf x}_{{\epsilon}}\right)=\sqrt{2\alpha_{n}}\exp\left[{\alpha_{n}\left({{\bf x}_{{\epsilon}}^{min}-{\bf x}_{{\epsilon}}}\right)}\right] \; ,\quad \alpha_{n} \in \mathbf{R}_{+} \; .
\end{eqnarray}
Zauważmy, że 
$\alpha_{n}$ jest w jednostkach $\left[c/energia\right]$. \\
  \\
{\bf Końcowa postać pojemności informacyjnej}:  W końcu możemy wyznaczyć pojemność informacyjną. Wstawiając (\ref{rozw cosinus dla ampl dla E}) do  (\ref{informacjaE}),  otrzymujemy: 
\begin{eqnarray}
\label{I dla E z war > 0}
I(\Theta_{{\epsilon}}) = 4 \sum\limits_{n=1}^{N} {\alpha_{n}^{2}} > 0 \; , \quad \alpha_{n} \in \mathbf{R}_{+} \; ,
\end{eqnarray}
co po porównaniu z (\ref{I  wraz z Ic dla E})  daje wartość stałej $C$ równą:
\begin{eqnarray}
\label{wartosc stalej I_C}
C = 8 \sum\limits_{n=1}^{N} {\alpha_{n}^{2}} \, .
\end{eqnarray}
{\bf Uwaga o stabilności rozwiązania}: Warunek dodatniości pojemności informacyjnej otrzymany w  (\ref{I dla E z war > 0}), dla pojemności informacyjnej związanej z dodatnią częścią sygnatury metryki Minkowskiego, jest istotnym wynikiem z punktu widzenia teorii pomiaru. W rozważanym przykładzie analizy estymacyjnej wartości oczekiwanej energii cząstki gazu, został on otrzymany na gruncie samospójnego rozwiązania równań różniczkowych \cite{Arnold}  informacyjnej obserwowanej zasady  strukturalnej oraz zasady  wariacyjnej. 
Niespełnienie tego warunku 
oznacza niestabilność badanego układu.  \\
\\
Z kolei, wstawiając otrzymaną wartość stałej $C$ do (\ref{postac Q dla E po war normalizacji qn}) otrzymujemy: 
\begin{eqnarray}
\label{Q dla E z war < 0}
Q(\Theta_{{\epsilon}}) = - 4 \sum\limits_{n=1}^{N} {\alpha_{n}^{2}} <0 \; .
\end{eqnarray}
Natomiast sam problem wariacyjny (\ref{gggg}), który  można wyrazić po skorzystaniu z postaci $Q(\Theta_{{\epsilon}})$ w  (\ref{postac Q dla E po war normalizacji qn}) w postaci:
\begin{eqnarray}
\label{sam problem wariacyjny}
\delta_{(q_{n})}\left(I(\Theta_{{\epsilon}}) + 4  \sum\limits_{n=1}^{N} \alpha_{n}^{2} - C \right) = 0 \; ,
\end{eqnarray}
jest ze swojej natury nieczuły na wartość stałej $C$. \\
  \\
{\bf Uwaga o randze amplitudy $N$}: Istotnym zagadnieniem metody EFI jest wielkość próby $N$ pobranej przez układ, tzn. ranga $N$ amplitudy. Do  sprawy  liczby amplitud $q_{n}$ wchodzących w opis układu podejdziemy w najprostszy z możliwych sposobów, sugerując stosowanie dwóch prostych i niewykluczających się  kryteriów:  \\
{\bf (1)} {\bf Kryterium minimalizacji $I$ ze względu na rangę amplitudy $N$} \cite{Frieden}.  Tzn. przy zachowaniu warunku $I>0$, (\ref{informacja Stama vs pojemnosc informacyjna Minkowskiego}),  ranga $N$ może na tyle spaść,  żeby istniało jeszcze samospójne rozwiązanie równań cząstkowych strukturalnej i wariacyjnej zasady informacyjnej. \\
Kryterium to nie oznacza nie realizowania rozwiązań z większą niż minimalna liczbą $N$.  \\
{\bf (2) Kryterium obserwacyjne wyboru rangi amplitudy}  wiąże się z wyborem takiej wartości $N$, dla której otrzymane rozwiązanie ma realizację obserwowaną w eksperymencie. 
We współczesnych teoriach fizyki statystycznej oraz teoriach pola,  realizowane są rozwiązania z niskimi wartościami rangi. Fakt ten zauważyliśmy już w Rozdziale~\ref{Kryteria informacyjne w teorii pola} dla modeli teorii pola. \\
\\
Rozważany w bieżącym rozdziale przykład rozkładu energii 
pozwala na ilustrację kryterium (1).  W kolejnym z rozdziałów znajdziemy  rozwiązanie EFI dla rozkładu prędkości cząsteczki gazu,  stosując również kryterium (1).  Obok stacjonarnego rozwiązania  Maxwella-Boltzmanna z $N=1$, wskażemy na fizyczną interpretację rozwiązań z $N>1$. \\
\\
{\bf Zastosowanie kryterium (1) dla rozkładu energii}: Zauważmy, że ponieważ wszystkie $\alpha_{n}$ w (\ref{I dla E z war > 0}) są rzeczywiste, zatem, przy {\it założeniu} braku wpływu nowych stopni swobody na poprzednie, pojemność informacyjna $I$ wzrasta wraz ze wzrostem $N$. 
Zgodnie z kryterium (1), przyjmijmy dla rozważanego przypadku rozkładu fluktuacji energii, że:
\begin{eqnarray}
\label{minimalne N 1 dla rozkladu E}
N=1 \; .
\end{eqnarray}
Jedyny współczynnik $\alpha_{1}$ oznaczmy teraz $\alpha$, natomiast parametr $\Theta_{{\epsilon}} = (\theta_{{\epsilon}1}) \equiv \theta_{{\epsilon}}$. \\
Zatem z (\ref{rozw cosinus dla ampl dla E}) mamy amplitudę:
\begin{eqnarray}
\label{qn dla N=1}
q\left({\bf x}_{{\epsilon}}\right) = \sqrt{2\alpha}\exp\left[{\alpha\left({{\bf x}_{{\epsilon}}^{min} - {\bf x}_{{\epsilon}}}\right)}\right] \; ,
\end{eqnarray}
której odpowiada rozkład gęstości prawdopodobieństwa fluktuacji energii  ${\bf x}_{{\epsilon}}$: 
\begin{eqnarray}
\label{pn dla N=1}
p\left({\bf x}_{{\epsilon}}\right) = q^{2}\left({\bf x}_{{\epsilon}}\right) = 2\alpha\exp\left[{2\alpha\left({{\bf x}_{{\epsilon}}^{min} - {\bf x}_{{\epsilon}}}\right)}\right] \; , \quad \alpha \in \mathbf{R}_{+} \; .
\end{eqnarray} 
{\bf Rozkład gęstości prawdopodobieństwa energii ${\epsilon}$  cząsteczki} jest końcowym punktem analizy obecnego rozdziału. Ponieważ zgodnie z (\ref{E}) mamy:
\begin{eqnarray}
\label{zakres E}
{\bf y}_{{\epsilon}}  \equiv \frac{{\epsilon}}{c} = \theta_{{\epsilon}} + {\bf x}_{{\epsilon}} \; , \;\;\;\; \frac{{\epsilon}_{0}}{c} \le {\bf y}_{{\epsilon}} < \infty \; , \;\;\;\; {\rm dla} \;\;\;\; {\bf x}_{{\epsilon}}^{min} \leq {\bf x}_{{\epsilon}} < \infty \; , 
\end{eqnarray}
gdzie ${\epsilon}_{0}/c = \theta_{{\epsilon}} + {\bf x}_{{\epsilon}}^{min}$. Zatem $d{\epsilon}/d{\bf x}_{{\epsilon}}=c$, skąd rozkład dla zmiennej ${\epsilon}$ ma postać: 
\begin{eqnarray}
\label{rozklad p od E}
p\left({\epsilon}\right)  = p\left({\bf x}_{{\epsilon}}\right) \frac{1}{|d{\epsilon}/d{\bf x}_{{\epsilon}}|} = 2 \frac{\alpha}{c} \exp \left[{-2\alpha\left({{\epsilon}-{\epsilon}_{0}}\right)/c}\right] \; , \quad {\epsilon}_{0} \le {\epsilon} <\infty \; .
\end{eqnarray}
\\
Pozostało jeszcze określenie stałej $\alpha$. Ponieważ wartość oczekiwana energii wynosi: 
\begin{eqnarray}
\label{srednia E}
\langle {E} \rangle \equiv c\, \theta_{\epsilon} = \int\limits_{{\epsilon}_{0}}^{+\infty}{d{\epsilon} \, p\left({\epsilon}\right) {\epsilon}} \; ,
\end{eqnarray}
więc wstawiając (\ref{rozklad p od E}) do (\ref{srednia E}) otrzymuje się: 
\begin{eqnarray}
\label{stala alfa}
2 \alpha = c \left(\langle E \rangle - {\epsilon}_{0}\right)^{-1} \; .
\end{eqnarray}
Zwróćmy uwagę, że z (\ref{srednia E}) wynika po pierwsze, że $E$ jest estymatotem wartości oczekiwanej $\langle {E} \rangle$ energii cząstki:
\begin{eqnarray}
\label{estymator sredniej E}
\widehat{\langle E \rangle} = E \; ,
\end{eqnarray}
a po drugie, że jest on nieobciążony.\\
\\
{\bf Szukany rozkład gęstości prawdopodobieństwa energii cząstki} ma zatem postać:
\begin{eqnarray}
\label{rozklad koncowy E}
p\left( {\epsilon} \right) = \left\{ \begin{array}{l}
(\left\langle E \right\rangle -{\epsilon}_{0})^{-1} \; \exp\left[-  \left({{\epsilon}-{\epsilon}_{0}}\right)/(\left\langle E \right\rangle - {\epsilon}_{0})\right] \;  \;\;\;\;  {\rm dla} \quad \;\;\; {\epsilon} \ge {\epsilon}_{0} \\
\quad \quad 0  \;\; \quad \quad \quad \quad\quad \quad\quad \quad \;\;  \quad \quad \quad \quad \quad \quad \quad \;\, {\rm dla} \quad \;\;\; {\epsilon} < {\epsilon}_{0}\end{array} \right. \; ,
\end{eqnarray}
gdzie w drugiej linii po prawej stronie zaznaczono fakt nie występowania cząstek z energią mniejszą niż  ${\epsilon}_{0}$. Rozkład  (\ref{rozklad koncowy E}) jest końcowym rezultatem metody EFI. Jego postać daje zasadniczo {rozkład Boltzmanna} dla energii cząsteczki w gazie.   \\
\\
{\bf Rozkład Boltzmanna dla energii cząsteczki}: Aby domknąć temat od strony fizycznej zauważmy, że energia całkowita cząsteczki wynosi ${\epsilon} = {\epsilon}_{kin} + V$, gdzie ${\epsilon}_{kin}$ jest energią kinetyczną cząsteczki a $V$ jej energią potencjalną.  Do potencjału $V$ możemy dodać pewną stałą np. ${\epsilon}_{0}$,  nie zmieniając przy tym fizycznego opisu zjawiska, zatem po przesunięciu ${\epsilon}$ o ${\epsilon}_{0}$  otrzymujemy:
\begin{eqnarray}
\label{E bez E0}
{\epsilon}_{0} = 0 \;  \;\;\;\; {\rm oraz} \;\;\;\; {\epsilon} \ge 0 \; .
\end{eqnarray}
Z kolei, dla cząstki gazu poruszającej się bez obrotu,  {\it twierdzenie o ekwipartycji energii} mówi, że: 
\begin{eqnarray}
\label{ekwipart}
\left\langle E \right\rangle  = \frac{3kT}{2} \; ,
\end{eqnarray}
gdzie $T$ jest tempetaturą bezwzględną gazu. Wstawiając (\ref{E bez E0}) wraz z  (\ref{ekwipart}) do (\ref{rozklad koncowy E}) otrzymujemy:
\begin{eqnarray}
\label{rozklad Boltzmanna z kT}
p\left({\epsilon}\right) = (3kT/2)^{-1} e^{-2{\epsilon}/3kT} \; ,\quad\quad {\epsilon} \ge 0 \; ,
\end{eqnarray} 
czyli  właściwą postać  rozkładu Boltzmnna  dla energii cząsteczki w gazie o temperaturze $T$.\\
  \\
{\bf Informacja Fishera dla $\theta_{\epsilon}$ i DORC dla estymatora $\left\langle  E \right\rangle$}:  W przypadku $N=1$ oraz skalarnego parametru $\theta_{\epsilon}$, pojemność informacyjna (\ref{I dla E z war > 0})  jest równa informacji Fishera $I_{F}$ dla tego parametru. Gdy dolne ograniczenie na energię cząstki wynosi $\epsilon_{0} = 0$, otrzymujemy po skorzystaniu z (\ref{stala alfa}) oraz (\ref{Q dla E z war < 0}) następujący związek: 
\begin{eqnarray}
\label{IF I oraz Q dla E oraz N=1}
I_{F} (\theta_{\epsilon}) = I(\theta_{\epsilon}) = - Q(\theta_{\epsilon})= 4  \alpha^{2}  = \frac{c^{2}}{\left\langle  E \right\rangle^{2}} > 0 \;  .
\end{eqnarray}
Z Rozdziału~\ref{Estymacja w modelach fizycznych na DORC} wiemy, że estymacja powyższego parametru oczekiwanego  $\theta_{\epsilon}$,  dualnego do $2 \alpha$, dla standardowego rozkładu eksponentialnego, a takim jest rozkład Boltzmanna, spełnia DORC w Twierdzeniu Rao-Cramera. Zatem wariancja estymatora tego parametru wynosi:
\begin{eqnarray}
\label{wariancja estymatora theta dla E}
\sigma^{2}(\hat{\theta}_{\epsilon}) = \frac{1}{I_{F} (\theta_{{\epsilon}})} = \frac{\left\langle  E \right\rangle^{2}}{c^{2}} \; ,
\end{eqnarray}
skąd otrzymujemy wariancję estymatora (\ref{estymator sredniej E}) średniej energii cząstki, równą:
\begin{eqnarray}
\label{wariancja estymatora sredniej E}
\sigma^{2}\left(\widehat{\left\langle  E \right\rangle}\right) =  \left\langle  E \right\rangle^{2} \; .
\end{eqnarray}
Na tym kończymy analizę metody EFI dla rozkładu Boltzmanna. \\
\\
{\bf Uwaga o równowadze statystycznej w metodzie} EFI:  Biorąc pod uwagę ograniczenia narzucone na normalizację (\ref{unormowanie q2}) i skończoność wartości oczekiwanej (\ref{srednia E}) rozkładu,  
metoda EFI wyznacza rozkład przy narzuceniu zasad informacyjnych. Poprzez obserwowaną zasadę strukturalną dokonuje ona, dla gładkiej funkcji wiarygodności,  separacji członu estymacyjnego dla parametru $\theta_{\epsilon}$ związanego z gęstością pojemności $\textit{i}$, której całka $I$ jest śladem po metryce Rao-Fishera 
przestrzeni statystycznej ${\cal S}$,  
od członu strukturalnego $\textit{q}$. Jest to wyrazem zasady Macha. Natomiast  poprzez zasadę wariacyjną, metoda EFI dokonuje stabilizacji rozwiązania, wybierając najmniejszą odległość (liczoną wzdłuż geodezyjnej w przestrzeni statystycznej ${\cal S}$)  
wyestymowanego stanu układu od stanu zaobserwowanego. Geometria ${\cal S}$ zależy od pojemności kanału informacyjnego $I$ wyliczonej z uwzględnieniem jej zależności od informacji strukturalnej $Q$.
 \\
\\
{\bf Uwaga}: W związku z powyższym oraz poprzednio podanym kryterium (1) doboru rangi amplitudy $N$ poprzez minimalizację $I$, pojawia się następujące, ogólne spojrzenie na estymacyjny charakter metody EFI. \\
{\bf Estymacja metodą EFI} 
oznacza wybór 
równań generujących rozkład (lub równań ruchu) na skutek działania dwóch czynników. 
Pierwszy z nich wymaga, po pierwsze   
wzrostu informacji o układzie zawartej w gęstości pojemności informacyjnej $\widetilde{\textit{i'}}\,$ określonej w (\ref{gestosc i Amarii}) oraz (\ref{zmodyfikowana obserwowana zas strukt z P i z kappa}), ale tylko na tyle na ile wymaga tego, zawarta w zasadzie strukturalnej  $\widetilde{\textit{i'}} + \widetilde{\mathbf{C}} + \kappa \, \textit{q} = 0$, (\ref{zmodyfikowana obserwowana zas strukt z P i z kappa}), struktura układu opisana gęstością  informacji strukturalnej $\textit{q}$, (\ref{gestosc q dla niezaleznych Yn}), przy, po drugie 
jednoczesnej minimalizacji całkowitej fizycznej informacji $K = I + Q$,  (\ref{physical K}). Oszacowanie poszukiwanego równania następuje więc na skutek żądania samospójności rozwiązania dwóch różniczkowych zasad informacyjnych, (\ref{zmodyfikowana obserwowana zas strukt z P i z kappa}) oraz  (\ref{var K rozpisana}). \\
Natomiast oczekiwana zasada strukturalna $I + \kappa Q =0$, (\ref{condition from K}), jest w myśl związku (\ref{rownowaznosc strukt i zmodyfikowanego strukt}) drugim, metrycznym czynnikiem, domykającym powyższą różniczkową  
analizę statystyczną doboru modelu. Dzieli ona modele statystyczne na dwie grupy: modele metryczne, 
równoważne zgodnie z  (\ref{rownowaznosc strukt i zmodyfikowanego strukt}) modelowi analitycznemu z metryką Rao-Fishera 
 oraz modele, które nie spełniają zasady (\ref{rownowaznosc strukt i zmodyfikowanego strukt}), czyli nierównoważne modelowi z metryką Rao-Fishera. \\
\\
{\bf Związek EFI oraz ME dla standardowego rozkładu  eksponentialnego}: Poprzednio 
otrzymaliśmy 
rozkład Boltzmanna (\ref{rozklad Boltzmanna z max entropii}) jako  szczególny przypadek rozwiązania  zasady maksymalnej entropii (ME), realizowany w sytuacji warunku normalizacji oraz istnienia jedego parametru obserwowanego, którym jest średnia wartość energii cząstki. Szczególna postać standardowego rozkładu eksponentialnego oraz te same warunki brzegowe są przyczyną otrzymania tego samego rozwiązania w metodzie EFI oraz ME. W ogólności metoda EFI wykracza poza całą klasę modeli eksponentialnych metody ME.

\newpage

\subsection{Model Aoki-Yoshikawy dla produkcyjności branż}

\label{Model Aoki-Yoshikawy - ekonofizyka}

Model Aoki i Yoshikawy (AYM) został opracowany w celu opisu produkcyjności branż kraju \cite{Aoki-Yoshikawy,Garibaldi-Scalas}. Rozważmy $g$ ekonomicznych sektorów. Sektor $i$-ty jest scharakteryzowany przez {\it czynnik wielkości produkcji} $n_{i}$, tzn. liczebność, oraz {\it zmienną poziomu produkcyjności} $A$, tzn. wydajność jednostkową, przyjmującą wartości $a_{i}$.\\
Niech liczebność $n_{i}$ będzie liczbą {\it aktywnych} pracowników w $i$-tym sektorze, co oznacza, że praca (robocizna) jest jedynym czynnikiem produkcyjnym. Zmienną losową w AYM jest poziom produkcyjności $A$, której rozkład jest określony poprzez parę $(a_{i}, n_{i})$, $i=1,2...,g$. \\
\\
{\bf Unormowanie jako pierwszy warunek modelu}: Załóżmy, że {\it całkowity zasób czynnika wielkości produkcji}, tzn. liczba dostępnych pracowników, jest w ekonomii zadany jako  wielkość egzogeniczna, czyli nie kontrolowana od wewnątrz lecz zadana z zewnatrz. Niech jego wielkość jest  równa $n$, co traktujemy jako {\it pierwsze ograniczenie} w modelu, tak, że  zachodzi:
\begin{eqnarray}
\label{calkowita produkcji}
\sum_{i=1}^{g} n_{i} = n \; .
\end{eqnarray} 
\\
{\bf Sektory}. Uporządkujmy wielkość produkcyjności $a_{i}$ 
w sektorach od najmniejszej do największej:
\begin{eqnarray}
\label{ciag produkcyjnosci}
a_{1} < a_{2} < ... < a_{g} \; .
\end{eqnarray} 
Ponieważ $a_{i}$ jest poziomem produkcyjności $i$-tego sektora, 
zatem uzysk (wartość produkcji) w $i$-tym sektorze wynosi:
\begin{eqnarray}
\label{produkcji w i tym sektorze} 
z_{i} = a_{i} \, n_{i} \; .
\end{eqnarray} 
Zatem całkowity uzysk $z$ w ekonomii kraju wynosi:
\begin{eqnarray}
\label{calkowity uzysk z}
z = \sum_{i=1}^{g} z_{i} = \sum_{i=1}^{g} a_{i} \, n_{i} \; .
\end{eqnarray} 
{\bf Drugi warunek modelu}: Wiekość $z$ jest interpretowana jako {\it produkt krajowy brutto} (PKB). \\
Założeniem AYM dla wartości PKB jest ustalenie wartości $z$ poprzez egzogenicznie zadany agregatowy popyt $D$ (demand), tzn.:
\begin{eqnarray}
\label{popyt}
z = D \; .
\end{eqnarray} 
Zatem {\it drugie ograniczenie} w modelu ma postać:
\begin{eqnarray}
\label{calkowity uzysk}
\sum_{i=1}^{g} a_{i} \, n_{i}  = D \; .
\end{eqnarray} 
{\bf Celem metody EFI dla modelu AYM} jest, po pierwsze wyznaczenie  równania generującego, a po drugie, teoretycznego rozkładu liczebności dla zmiennej poziomu produkcyjności $A$, tzn. określenie  wektora obsadzeń:
\begin{eqnarray}
\label{rozklad n}
{\bf n} = (n_{1} , n_{2} , ... , n_{g}) \; .
\end{eqnarray} 
{\bf Porównanie analizy dla AYM oraz rozkładu Boltzmanna}: Rozważane zagadnienie jest odpowiednikiem poprzedniego zagadnienia związanego z określeniem rozkładu prawdopodobieństwa energii cząstki gazu. Można je określić jako zagadnienie rozmieszczenia $n$ cząstek gazu na $g$ poziomach energetycznych ${\epsilon}_{i}$ w warunkach równowagi statystycznej, w taki sposób, że zachowane są, liczba cząstek: 
\begin{eqnarray}
\label{calkowita liczba czastek}
\sum_{i=1}^{g} n_{i} = n \; 
\end{eqnarray} 
oraz całkowita energia gazu ${\cal E}$:
\begin{eqnarray}
\label{calkowita energia E}
\sum_{i=1}^{g} {\epsilon}_{i} \, n_{i}  = {\cal E} \;  \;\;\; {\rm lub} \;\;\;\; \left\langle E \right\rangle = \sum_{i=1}^{g} \frac{n_{i}}{n} \,{\epsilon}_{i}  = \frac{{\cal E}}{n} \; .
\end{eqnarray} 
Zatem poziom produkcyjności $a_{i}$ jest analogiem poziomu energetycznego ${\epsilon}_{i}$, natomiast ograniczenie, które dał w AYM  popyt $D$ jest analogiem ograniczenia pochodzącego od wartości całkowitej energii ${\cal E}$ gazu. \\
\\
Dokonajmy następującego przyporządkowania pomiędzy wielkościami opisującymi rozkład energii cząstki gazu oraz rozkład poziomu produkcyjności. Po lewej stronie przyporządkowania  ``$\leftrightarrow$'' jest wielkość dla energii cząstki, po prawej dla produkcyjności pracownika. 
Strzałki $\rightarrow$ w nawiasach oznaczają {\it przejście od ciągłego do dyskretnego} rozkładu zmiennej (lub na odwrót). \\
\\
Zatem, zmiennej energii cząstki $E$ odpowiada poziom produkcyjności pracownika $A$: 
\begin{eqnarray}
\label{przyporzadkowanie A E}
E = ({\epsilon} \rightarrow {\epsilon}_{i}) \leftrightarrow A = (a_{i} \rightarrow a) \; .
\end{eqnarray} 
Rozkładowi prawdopodobieństwa zmiennej energii cząstki $p(\epsilon)$ odpowiada rozkład poziomów produkcyjności pracownika $p(a)$: 
\begin{eqnarray}
\label{przyporzadkowanie Pa Pe}
\left( p (\epsilon) \rightarrow p_{\epsilon_{i}}=\frac{n_{i}}{n} \right) \leftrightarrow \left( p_{i}=\frac{n_{i}}{n}  \rightarrow p (a)\right) \; .
\end{eqnarray} 
Normalizacje rozkładów są sobie przyporządkowane następująco:
\begin{eqnarray}
\label{przyporzadkowanie Na Ne} 
\left(\int_{{\cal Y}_{{\epsilon}}} d\epsilon \, p(\epsilon) = 1 \rightarrow
\sum_{i=1}^{g} p_{\epsilon_{i}} = 1
\right) \leftrightarrow \left( \sum_{i=1}^{g} p_{i} = 1 \rightarrow \int_{{\cal Y}_{a}} da \, p(a) =1 \right) \; .
\end{eqnarray} 
Wartości oczekiwanej 
energii cząstki odpowiada wartość oczekiwana wartość produkcyjności pracownika: 
\begin{eqnarray}
\label{przyporzadkowanie SRa SRe}
\theta_{\epsilon} \equiv \left\langle E \right\rangle = \left(  \int_{{\cal Y}_{{\epsilon}}} d{\epsilon} \, p({\epsilon}) \,\epsilon \rightarrow
\sum_{i=1}^{g} \frac{n_{i}}{n} \,{\epsilon}_{i}    \right) \leftrightarrow \left( \sum_{i=1}^{g} p_{i} \, a_{i}   \rightarrow \int_{{\cal Y}_{a}} da \; p(a) \,a \right) = \left\langle A  \right\rangle \equiv \theta_{A} \,  ,
\end{eqnarray} 
przy czym w AYM wartość oczekiwana produkcyjności jest zadana jako:
\begin{eqnarray}
\label{wartosc oczekiwana produkcyjnosci}
\left\langle A  \right\rangle =  D/n \; .
\end{eqnarray} 
\\
{\bf Określenie zmiennej addytywnych fluktuacji}: Aby analiza w AYM mogła przebiegać dokładnie tak samo jak dla rozkładu Boltzmanna, musimy dokonać jeszcze jednego przejścia po stronie produkcyjności, a mianowicie przejść od poziomu produkcyjności $A$ do jej fluktuacji $X_{a}$ od wartości oczekiwanej $\left\langle A  \right\rangle$, tzn. dokonać addytywnego rozkładu: $Y_{a} \equiv A = \left\langle A  \right\rangle + X_{a}$. \\
\\
Odpowiedni analog pomiędzy fluktuacjami energii i produkcyjności ma więc następującą postać:  Dla energii cząstki zachodzi (\ref{E}):
\begin{eqnarray}
\label{przyporzadkowanie xe}
{\bf y}_{\epsilon} = \frac{\epsilon}{c} =  \theta_{\epsilon} + {\bf x}_{\epsilon} \; ,  \;\; \frac{\epsilon_{0}}{c} \le {\bf y}_{\epsilon} \le \infty \; , \;\;\; {\bf x}_{{\epsilon}}^{min} \le {\bf x}_{{\epsilon}} < \infty \; ,
\end{eqnarray} 
gdzie skorzystano z założenia o nieograniczoności od góry fluktuacji energii (por. (\ref{qn rozwiazanie dla E}) wraz z dyskusją zawartą powyżej), 
natomiast dla produkcyjności pracownika zachodzi:
\begin{eqnarray}
\label{przyporzadkowanie xa}
{\bf y}_{a} \equiv a =  \theta_{A}   + {\bf x}_{a} \; , \;\;\;\; a_{0} \le {\bf y}_{a} \le \infty \; , \;\;\;  {\bf x}_{a}^{min} = a_{0} -\theta_{A} \le {\bf x}_{a} < \infty\; .
\end{eqnarray} 
\\
{\bf Równanie generujące amplitudę produkcyjności}: Możemy teraz przenieść powyższą analizę EFI dla rozkładu Boltzmanna na grunt modelu AYM. Zatem wychodząc z zasady  wariacyjnej (\ref{gggg}) oraz strukturalnej (\ref{obserwowana zas strukt dla Boltzmann suma n}) i odpowiednich analogów pojemności informacyjnej $(\ref{informacjaE})$ oraz informacji strukturalnej (\ref{Q diag dla E}), otrzymujemy zgodnie z analizą Rozdziału~\ref{rozdz.energia}  równanie generujące rozkład produkcyjności (\ref{falowe kappa 1}), które dla wielkości próby  $N=1$ ma postać:
\begin{eqnarray}
\label{produkcyjnosc row generujace}
\frac{d^{2}q({\bf x}_{a})}{d\,{\bf x}_{a}^2} = \alpha^{2} q({\bf x}_{a}) \; ,
\end{eqnarray}
gdzie $q({\bf x}_{a})$ jest amplitudą rozkładu fluktuacji  produkcyjości, a $\alpha$ rzeczywistą stałą. \\
\\
Rachunki analogiczne do przeprowadzonych 
pomiędzy (\ref{falowe kappa 1}) a (\ref{rozklad koncowy E}), z  warunkiem normalizacji  (\ref{przyporzadkowanie Na Ne}),  
które poprzednio doprowadziły do amplitudy (\ref{qn dla N=1}) rangi $N=1$, dają nastepujące rozwiązanie równania (\ref{produkcyjnosc row generujace}) na  amplitudę fluktuacji produkcyjności $X_{a}$ w zakresie (\ref{przyporzadkowanie xa}): 
\begin{eqnarray}
\label{qn dla N=1 dla produkcyjnosci}
q\left({\bf x}_{a}\right) = \frac{1}{\sqrt{ D/n }}\exp \left[ - \,{\frac{{(D/n) - a_{0} + {\bf x}_{a}  }}{2 \, D/n } }\right] \; \;\;\;{\rm dla} \;\;\;\; {\bf x}_{a}^{min} = a_{0} -\theta_{A} \le {\bf x}_{a} < \infty\; 
\end{eqnarray}
oraz w analogii do (\ref{rozklad koncowy E}), rozkład gęstości  produkcyjności $A$: 
\begin{eqnarray}
\label{rozklad koncowy A}
p\left( a \right) = \left\{ \begin{array}{l}
\frac{1}{(D/n) - a_{0}} \; \exp\left( - \, \frac{a - a_{0}}{(D/n) - a_{0}} \right) \;\;\;  \;\;\;\;\;\;  {\rm dla} \quad \;\;\; a \ge a_{0} \\
\quad \quad 0  \;\,    \quad \quad \quad \quad \quad  \quad \quad \quad \quad \quad \quad \;\; {\rm dla} \quad  \;\;\; a < a_{0} \end{array} \right. \; ,
\end{eqnarray}
gdzie w drugiej linii po prawej stronie zaznaczono fakt nie występowania produkcyjności mniejszej niż  $a_{0}$. 
Rozkład  (\ref{rozklad koncowy A}) jest końcowym rezultatem metody EFI dla modelu AYM produkcyjności. \\
\\
{\bf Porównanie wyników metody AYM oraz EFI}:  
Aby postać rozkładu (\ref{rozklad koncowy A}) oddała w pełni  wynik  Aoki i Yoshikawy\footnote{Omówienie  metody Aoki i Yoshikawy zamieszczono na końcu obecnego rozdziału.}, {\bf należy powrócić do dyskretyzacji wartości} $a \rightarrow a_{i}$ zmiennej $A$. \\
Rozwiązali oni powyższy problem  stosując metodę mnożników Lagrange'a z warunkami ograniczającymi (\ref{calkowita produkcji}) oraz (\ref{calkowity uzysk}). Następnie założyli \cite{Aoki-Yoshikawy}, że wartości produkcyjności są dyskretne,  tworząc ciąg arytmetyczny: 
\begin{eqnarray}
\label{ciag a i}
a_{i} = i \; a_{0} \; \;\;\; {\rm gdzie} \;\;\; i=1,2,...,g \;,
\end{eqnarray}
gdzie $a_{0}$ jest najmniejszą produkcyjnością. \\
\\
{\bf Względny  popyt agregatowy $r$}: W końcu, przy założeniach, po pierwsze,  że liczba dostępnych sektorów produkcyjności jest bardzo duża, tzn. $g >> 1$ oraz po drugie, że: 
\begin{eqnarray}
\label{defin r}
r \equiv \frac{D/n}{a_{0}} \; ,
\end{eqnarray}
tzn. {\it agregatowy popyt przypadający na jednego pracownika $D/n$ odniesiony do najmniejszej produkcyjność} $a_{0}$, jest bardzo duży, Aoki i Yoshikawy otrzymali wynik \cite{Aoki-Yoshikawy,Garibaldi-Scalas}:
\begin{eqnarray}
\label{wynik AY na p}
P(i|{\bf n}^{*}) = \frac{n_{i}^{*}}{n} \approx \frac{1}{r-1} \left( \frac{r-1}{r} \right)^{i}\approx (\frac{1}{r} + \frac{1}{r^2}) \; e^{ -  \frac{i}{r}  } \; , \;\;\; i=1,2,... \; , \;\;\; r >> 1 \; ,
\end{eqnarray}
gdzie $n_{i}^{*}$, $i=1,2,...,g$, są współrzędnymi $n_{i}$ wektora obsadzeń  (\ref{rozklad n}), przy których prawdopodobieństwo pojawienia się tego wektora obsadzeń jest maksymalne. \\
\\
{\bf Wynik analizy  metodą AYM}: Rezultat (\ref{wynik AY na p}) podaje {\it prawdopodobieństwo, że  losowo wybrany pracownik jest w $i$-tym sektorze produkcyjności, o ile gospodarka znajduje się w stanie określonym wektorem obsadzeń} ${\bf n}^{*} = (n_{1}^{*},n_{2}^{*},...,n_{g}^{*})$. \\
\\
{\bf Przejście do rozkładu dyskretnego dla wyniku EFI}: Aby porównać wynik (\ref{wynik AY na p}) otrzymany w AYM z wynikiem (\ref{rozklad koncowy A}) otrzymanym w metodzie EFI, przejdźmy w (\ref{rozklad koncowy A}) do rozkładu dyskretnego. W tym celu musimy wycałkować wynik EFI w przedziale $(a_{i},a_{i+1})$, przy czym od razu założymy, że zachodzi $a_{i} = i \; a_{0}$, (\ref{ciag a i}). W rezultacie 
otrzymujemy\footnote{
Przyjmując, że $r>>1$, otrzymujemy,  w podanych granicach,  następujący rozkład prawdopodobieństwa produkcyjności pracownika:
\begin{eqnarray}
\label{rozklad koncowy A dysktretny przybl}
P\left( i \right) \approx  (\frac{1}{r} + \frac{1}{2 r^2}) \, \left(e^{- \frac{i}{r}} \; + \; \frac{1}{r} \right)
 \;\;\;\;\;  {\rm dla} \quad \;\;\; i= 1,2,...\;  \;\;\;{ \rm oraz} \;\;\; a_{0} > 0  \; , \;\;\; r >> 1 \, .
\end{eqnarray}
}:
\begin{eqnarray}
\label{rozklad koncowy A dysktretny}
P\left( i \right) = \int_{i a_{0}}^{(i+1) a_{0}} \! da \; p\left( a \right) = \left( 1 - e^{-1/(r - 1 )} \right) e^{- (i - 1)/(r-1) }
 \;\;\;\;\;  {\rm dla} \quad \;\;\; i=1,2,... \; .
\end{eqnarray} \\
{\bf Porównanie modeli dla $a_{0}=0$}. Niech  $\delta a$ jest stałą szerokością sektorów produkcyjności.  Wtedy, w przypadku gdy $a_{0}=0$, wzór (\ref{rozklad koncowy E}) metody EFI prowadzi w miejsce (\ref{rozklad koncowy A dysktretny}) do rozkładu:
\begin{eqnarray}
\label{rozklad koncowy A dysktretny a0 = 0}
P\left( i \right) = \int_{(i-1) \delta a}^{i \delta a} \! da \; p\left( a \right) = \left( -1 + e^{1/\,\tilde{r}} \, \right) e^{- i/\,\tilde{r} }
\; , \;\;\;\; i=1,2,... \;  \;\;\;  {\rm dla} \quad \; a_{0}=0 \; ,
\end{eqnarray}
gdzie  zamiast (\ref{defin r}) wprowadziliśmy:
\begin{eqnarray}
\label{defin r tilda}
\tilde{r} \equiv \frac{D/n}{\delta a} \; ,
\end{eqnarray}
jako {\it agregatowy popyt przypadający na jednego pracownika $D/n$ odniesiony do szerokości sektora produkcyjność} $\delta a$. W granicy  $\tilde{r} >> 1$ z (\ref{rozklad koncowy A dysktretny a0 = 0}) otrzymujemy: 
\begin{eqnarray}
\label{rozklad koncowy A dysktretny a0 = 0 oraz r duze}
P\left( i \right) \approx  \left( \frac{1}{\tilde{r}} + \frac{1}{2 \;\tilde{r}^2} \right) \, e^{- i/\,\tilde{r} } \; , \;\;\;\; i=1,2,... \; , \;\;\;  {\rm dla} \quad \; a_{0}=0 \; , \;\;\; \tilde{r} >> 1 \; .
\end{eqnarray}\\
\\
\\
{\bf Wniosek z porównania wyników  metody AYM oraz EFI}: W granicy dużych $\tilde{r}$ oba wyniki się schodzą. 
Jednakże granicę najmniejszej produkcyjności $a_{0}$ metoda EFI ujmuje inaczej niż AYM. To znaczy, formuła EFI (\ref{rozklad koncowy A dysktretny a0 = 0 oraz r duze}) daje dla $a_{0}=0$ inną kwadratową poprawkę w $\tilde{r}$ niż rezultat (\ref{wynik AY na p}) modelu AYM dla $r=\tilde{r}$ oraz $a_{0} \rightarrow 0$. \\
\\
{\bf Uwaga}: Ponadto  wynik (\ref{rozklad koncowy A dysktretny a0 = 0}) jest dokładny, natomiast 
w AYM dyskretyzacja poziomu produktywności jest tylko wybiegiem technicznym, gdyż zmienna ta jest z natury ciągła, do czego i tak w końcu odwołuje się metoda AYM  przy  wyznaczaniu mnożników Lagrange'a,  przechodząc z powodów rachunkowych w (\ref{calkowita produkcji}) oraz (\ref{calkowity uzysk}) z  $g$ do nieskończoności.\\
 \\
{\small {\bf Uzupełnienie}: {\it Analiza Aoki i Yoshikawy dla produkcyjności.}  
Podajmy sposób wyprowadzenia rozkładu wektora obsadzeń w AYM metodą czynników Lagrange'a \cite{Garibaldi-Scalas}. 
Wprowadźmy wektor ${\bf H}^{(n)}$ {\it indywidualnych} przypisań, posiadający tyle składowych ilu jest pracowników w całej gospodarce kraju:
\begin{eqnarray}
\label{wektor idywidualnych przypisan}
{\bf H}^{(n)} \equiv ( H_{1}, H_{2},..., H_{n} ) = {\bf h} \equiv ( h_{1}, h_{2},..., h_{n} ) \; .
\end{eqnarray} 
Każda ze współrzędnych $H_{i}$, $i=1,2,...,n$, może przyjmować wartości $h_{i} = s$, gdzie $s \in \left\{ 1,2,...,g \right\}$, co oznacza, że $i$-ty pracownik jest aktywny w $s$-tym sektorze gospodarki. Zatem wektor ${\bf h}$ podaje  jedną konfigurację indywidualnych przypisań. 
Jeśli ustalimy wektor obsadzeń  ${\bf n}$, (\ref{rozklad n}), to liczba  $W$ różnych ${\bf h}$ (indywidualnych konfiguracji przypisań),  realizujących ten sam ustalony wektor obsadzeń  ${\bf n}$,   wynosi \cite{podrecznik z kombinatoryki}:
\begin{eqnarray}
\label{liczba konfiguracji W dla wekt obsadzen}
W({\bf H}|{\bf n}) = \frac{n!}{\prod_{i=1}^{g} n_{i}!} \; .
\end{eqnarray} 
Boltzman zauważył, że gdy układ znajduje się w równowadze statystycznej to, prawdopodobieństwo $\pi({\bf n})$, że znajduje się on w stanie  o określonym wektorze obsadzeń ${\bf n}$ (tzn. że pojawił się taki właśnie wektor obsadzeń), jest proporcjonalne do liczby jego możliwych realizacji, tzn. do $W({\bf H}|{\bf n})$. Zatem: 
\begin{eqnarray}
\label{prawdopodobienstwo pojawieniem sie wektora n}
\pi({\bf n}) =  W({\bf H}|{\bf n}) \;P({\bf H}|{\bf n}) =  \frac{n!}{\prod_{i=1}^{g} n_{i}!} \;\prod_{i=1}^{g} p^{n_{i}} =
\frac{n!}{\prod_{i=1}^{g} n_{i}!} \;p^{n} = {\cal K} \, \frac{n!}{\prod_{i=1}^{g} n_{i}!}  
\; ,
\end{eqnarray} 
gdzie ${\cal K}$ jest właściwą normalizacyjną stałą, a $p$ jest {\it prawdopodobieństwem} zajęcia przez $i$-tego pracownika określonego $s$-tego sektora obsadzeń, {\it które zostało przyjęte jako takie samo dla wszystkich indywidualnych konfiguracji tych obsadzeń}.   
W celu rozwiązania  postawionego problemu maksymalizacji  prawdopodobieństwa $\pi({\bf n})$ z warunkami (\ref{calkowita produkcji}) oraz (\ref{calkowity uzysk}), rozwiązujemy poniższy układ $g$ równań:
\begin{eqnarray}
\label{MNW z warunkami}
\frac{\partial  }{\partial n_{i}} \left[ \ln \pi({\bf n})  + \nu \left( \sum_{i=1}^{g}n_{i} -n \right) - \beta \left( \sum_{i=1}^{g} a_{i}n_{i} - D \right) \right] = 0\; ,
\end{eqnarray}
otrzymując jako rozwiązanie wektor obsadzeń ${\bf n}$:
\begin{eqnarray}
\label{n*}
n_{i} = n_{i}^{*} = e^{\nu} \, e^{-\beta a_{i}} \; , \;\;\; i=1,2,...,g \; .
\end{eqnarray}
Stałe $\nu$ oraz $\beta$ otrzymujemy wykorzystując (\ref{n*}) w (\ref{calkowita produkcji}) oraz (\ref{calkowity uzysk}). \\
Uwaga: Aby zapisać $\prod_{i=1}^{g} n_{i}!$ w formie nadającej się do minimalizacji, korzystamy z przybliżenia Stirling'a:
\begin{eqnarray}
\label{wzor Stirlinga}
\ln \left[ \prod_{i=1}^{g} n_{i}! \right] \approx  \sum_{i=1}^{g} n_{i} (\ln n_{i} -1) \; ,
\end{eqnarray}
 słusznego dla dużego układu z $n>>1$ oraz $n_{i}>>1$, $i=1,2,...,g$.\\
\\
{\bf Układ w równowadze statystycznej}: Przedstawiona w tym przypisie  metoda znajdowania {\it rozkładu w równowadze statystycznej} zakłada spełnienie hipotezy wyrażonej równaniem (\ref{prawdopodobienstwo pojawieniem sie wektora n})  i mówiącej, że {\it wszystkie stany opisane wektorem indywidualnych przypisań ${\bf H}^{(n)}\,$, a spełniające warunki (\ref{calkowita produkcji}) oraz (\ref{calkowity uzysk}) są równie prawdopodobne}. 
Rozkład opisany wektorem ${\bf n}^{*}$ jest w tym ujęciu sednem definicji rozkładu, będącego w równowadze statystycznej.
}

\vspace{3mm}

\subsection{Rozkład Maxwella-Boltzmanna dla prędkości}

Poniżej wyznaczymy rozkładu prędkości cząsteczki w gazie. 
Wychodząc z informacji Fishera (\ref{minp}) dla pędu oraz posługując się wariacyjną i strukturalną zasadą informacyjną (\ref{epip}) otrzymamy równanie generujące i znajdziemy jego rozwiązania, tzn. postać amplitud dla rozkładu prędkości.  \\
\\
{\bf Wartość współczynnika efektywności}: Ze względu na spójność rozważań dla czterowektora pędu, przyjęcie współczynnika $\kappa=1$ w rozważaniach dla energii skutkuje przyjęciem $\kappa=1$ w analizie dla rozkładu pędu. \\
\\
{\bf Pojemność informacyjna parametru} $\Theta_{\vec{\wp}}$ została podana w (\ref{minp}): 
\begin{eqnarray}
I\left(\Theta_{\vec{\wp}}\right) =  -
4 \int_{{\cal X}_{\wp}} {\! d\vec{\bf x}_{\wp}
\sum\limits _{n=1}^{N}{  {\sum\limits_{k=1}^{3}{\left(\! {\frac{{\partial q_{n}(\vec{\bf x}_{\wp})}}{{\partial x_{\wp_{k}}}}}\right)^{2}  \! }}}}  \, , \;\;\; \nonumber
\end{eqnarray} 
gdzie znaczenie znaku ``minus'' w definicji pojemności informacyjnej dla części pędowej zostało omówione w Rozdziale~\ref{zasady inf dla energii i predkosci}.\\
\\
{\bf Informacja strukturalna i zależność $\texttt{q\!F}_{n}$ od amplitudy oraz prędkości}: $Q$ dla parametru  $\Theta_{\vec{\wp}}$ ma postać:
\begin{eqnarray}
\label{inform strukt dla B-M}
Q\left(\Theta_{\vec{\wp}}\right) =  
\int_{{\cal X}_{\wp}} {\! d\vec{\bf x}_{\wp}
\sum\limits _{n=1}^{N} 
 q_{n}^{2}(\vec{\bf x}_{\wp}) \,\texttt{q\!F}_{n}(q_{n}(\vec{\bf x}_{\wp}), \vec{\bf x}_{\wp}) }  \, . \;\;\;
\end{eqnarray} 
Wprowadzenie do obserwowanej 
informacji strukturalnej $\texttt{q\!F}_{n}(q_{n}(\vec{\bf x}_{\wp}),\vec{\bf x}_{\wp})$ jej jawnej zależności nie tylko od amplitud $q_{n}$, ale również od pędu $\vec{\bf x}_{\wp}$, pozwala na 
rozważenie szerszego zakresu zagadnień niż to miało miejsce dla przypadku rozkładu energii, mianowicie pozwala na rozważenie  rozwiązań {\it nierównowagowych}.\\
\\
{\bf Obserwowana, strukturalna zasada informacyjna}: 
Podobnie jak to uczyniliśmy w (\ref{postac I po calk czesci}) dla energii, tak i teraz 
dla pojemności informacyjnej $I\left(\Theta_{\vec{\wp}}\right)$,  (\ref{minp}), dokonamy całkowania przez części. Zakładając dodatkowo, że amplitudy dla prędkości $q_{n}(\vec{\bf x}_{\wp})$ znikają w $\pm$ nieskończoności, {\bf oczekiwana  strukturalna zasada informacyjna}  
$\,\widetilde{\textit{i'}}(\Theta_{\vec{\wp}}) + \widetilde{\mathbf{C}}_{\vec{\wp}} + \kappa \, \textit{q}(\Theta_{\vec{\wp}}) = 0$ w  (\ref{epip}) przyjmuje  postać: 
\begin{eqnarray}
\label{strukt zas dla pedu calkow czesci}
\!\!\!\!\! \widetilde{\textit{i'}}\left(\Theta_{\vec{\wp}}\right) + \textit{q}\left(\Theta_{\vec{\wp}}\right) =  4 
\sum\limits _{n=1}^{N} \left[  q_{n}(\vec{\bf x}_{\wp})\sum_{k=1}^{3}{\frac{{\partial^{2} q_{n}(\vec{\bf x}_{\wp})}}{{\partial x_{\wp_{k}}^{2}}}}  \! +  \frac{1}{4}\;
 q_{n}^{2}(\vec{\bf x}_{\wp}) \,\texttt{q\!F}_{n}(q_{n}(\vec{\bf x}_{\wp}), \vec{\bf x}_{\wp})  \right] = 0 \, , \;\;\,
\end{eqnarray} 
gdzie fakt, że 
$\widetilde{\mathbf{C}}_{\vec{\wp}} = 0$, nie wnosząc tym samym  wkładu w powyższą zasadę strukturalną, 
wynika ze znikania amplitud w nieskończoności. \\
\\
{\bf Wariacyjna zasada informacyjna} w (\ref{epip}) przyjmuje postać:
\begin{eqnarray}
\label{wariacyjna zas dla pedu}
\!\!\!\!\! \delta_{(q_{n})} \! \left(I\left(\Theta_{\vec{\wp}}\right) + Q\left(\Theta_{\vec{\wp}}\right)\right) =  4 \!\! \int_{{\cal X}_{\wp}}{\!\! d\vec{\bf x}_{\wp}\sum\limits _{n=1}^{N}{\left[ - {\sum\limits_{k=1}^{3}{ \!\left( \!{\frac{{\partial q_{n}(\vec{\bf x}_{\wp})}}{{\partial x_{\wp_{k}}}}}\right)^{2} + \! \frac{1}{4} 
 q_{n}^{2}(\vec{\bf x}_{\wp}) \,\texttt{q\!F}_{n}(q_{n}(\vec{\bf x}_{\wp}), \vec{\bf x}_{\wp}) }}\right]}} = 0 \, . \;\;\, 
\end{eqnarray} 
\\
{\bf Równania Eulera-Lagrange'a}: Rozwiązaniem $N$-funkcyjnego problemu wariacyjnego (\ref{wariacyjna zas dla pedu}) jest układ równań Eulera-Lagrange'a:
\begin{eqnarray}
\label{row E-L dla pedu}
\sum\limits _{m=1}^{3}{\frac{\partial}{{\partial x_{\wp_{m}}}}\left({\frac{{\partial \, k \,(q_{n}(\vec{\bf x}_{\wp}), \vec{\bf x}_{\wp}) }}{{\partial \left(\frac{{\partial\, q_{n}(\vec{\bf x}_{\wp})}}{{\partial x_{\wp_{m}}}} \right) }}}\right)}=\frac{{\partial k(q_{n}(\vec{\bf x}_{\wp}), \vec{\bf x}_{\wp})}}{{\partial q_{n}(\vec{\bf x}_{\wp})}},\quad\quad n=1, 2,..., N \; ,
\end{eqnarray}
gdzie 
\begin{eqnarray}
\label{gestosc k dla pedu}
k \,(q_{n}(\vec{\bf x}_{\wp}), \vec{\bf x}_{\wp}) = 4 {\left[ - {\sum\limits_{k=1}^{3}{ \!\left( \!{\frac{{\partial q_{n}(\vec{\bf x}_{\wp})}}{{\partial x_{\wp_{k}}}}}\right)^{2} + \! \frac{1}{4} 
 q_{n}^{2}(\vec{\bf x}_{\wp}) \,\texttt{q\!F}_{n}(q_{n}(\vec{\bf x}_{\wp}), \vec{\bf x}_{\wp}) }}\right]} \;  
\end{eqnarray}
jest gęstością informacji fizycznej (\ref{k form}) zdefiniowaną w Rozdziale~\ref{equations of motion}. \\
\\
{\bf Nieujemność $k$}: Zauważmy, że z (\ref{gestosc k dla pedu}) i z żądania nieujemności informacji fizycznej $k$ na  poziomie obserwowanym, wynika:
\begin{eqnarray}
\label{qF dodatnia dla rozkl pedu}
\texttt{q\!F}_{n}(q_{n}(\vec{\bf x}_{\wp}), \vec{\bf x}_{\wp}) \geq 0 \; , 
\end{eqnarray}
tzn.  {\it nieujemność obserwowanej  informacji strukturalnej w analizie estymacyjnej wartości oczekiwanej pędu cząsteczki gazu.}  \\
\\
{\bf Uwaga}: Poniżej, ze względu na uproszczenie zapisu, pominiemy zaznaczenie fluktuacji pędu $\vec{\bf x}_{\wp}$ w argumencie amplitudy  $q_{n}(\vec{\bf x}_{\wp})$.\\
  \\
{\bf Układ równań różniczkowych}: 
Dla każdego $n=1,2,...,N$, zasada strukturalna (\ref{strukt zas dla pedu calkow czesci}) jest na poziomie obserwowanym  następująca: 
\begin{eqnarray}
\label{uppro}
q_{n}\sum_{m=1}^{3}{\frac{{\partial^{2}q_{n}}}{{\partial x_{\wp_{m}}^{2}}}} + \frac{1}{4} q_{n}^{2} \,\texttt{q\!F}_{n}(q_{n}, \vec{\bf x}_{\wp}) 
 = 0 \; .
\end{eqnarray}
Natomiast każde z $N$ równań Eulera-Lagrange'a (\ref{row E-L dla pedu}) ma postać następującego równania różniczkowego:
\begin{eqnarray}
\label{roznp}
\sum\limits_{m=1}^{3}{\frac{{\partial^{2}q_{n}}}{{\partial x_{\wp_{m}}^{2}}}} = - \frac{1}{2}\frac{{\partial (
\frac{1}{4} q_{n}^{2} \,\texttt{q\!F}_{n}(q_{n}, \vec{\bf x}_{\wp}) 
) }}{{\partial q_{n}}} \; .
\end{eqnarray}
Równanie strukturalne (\ref{uppro}) wraz z równaniem Eulera-Lagrange'a (\ref{roznp}) posłuży do wyprowadzenia równania generującego rozkład Maxwella-Boltzamnna.\\
\\
{\bf Wyprowadzenie równania generującego}: Równania (\ref{uppro}) oraz (\ref{roznp}) pozwalają wyeliminować występującą w nich sumę,  dając równanie:
\begin{eqnarray}
\label{row rozniczkowe na qF dla p}
\frac{1}{2}\frac{{\partial (\frac{1}{4} q_{n}^{2} \,\texttt{q\!F}_{n}(q_{n}, \vec{\bf x}_{\wp}))}}{{\partial q_{n}}} = \frac{{\frac{1}{4} q_{n}^{2} \,\texttt{q\!F}_{n}(q_{n}, \vec{\bf x}_{\wp}) }}{{q_{n}}} \; ,
\end{eqnarray}
które po obustronnym scałkowaniu prowadzi do rozwiązania: 
\begin{eqnarray}
\label{rozw dla qF}
\frac{1}{4} \, q_{n}^{2} \,\texttt{q\!F}_{n}(q_{n}, \vec{\bf x}_{\wp})  = q_{n}^{2} \, f_{n}\left({\vec{\bf x}_{\wp}}\right) \; ,
\end{eqnarray}
lub
\begin{eqnarray}
\label{rozw dla samego qF}
\texttt{q\!F}_{n}(q_{n}, \vec{\bf x}_{\wp})  = 4 \, f_{n}\left({\vec{\bf x}_{\wp}}\right) \geq 0\; .
\end{eqnarray}
Funkcja $f_{n}(\vec{\bf x}_{\wp})$ nie zależy od amplitudy $q_{n}$ i pojawiła się w wyniku całkowania równania (\ref{row rozniczkowe na qF dla p}) jako pewna stała całkowania w znaczeniu jej niezależności od amplitudy $q_{n}$. Zaznaczona nieujemność funkcji  $f_{n}\left({\vec{\bf x}_{\wp}}\right)$ w jej dziedzinie  wynika z nieujemności  $\texttt{q\!F}_{n}(q_{n}, \vec{\bf x}_{\wp})$ otrzymanej w (\ref{qF dodatnia dla rozkl pedu}).\\
\\
{\bf Równanie generujące}: Wykorzystując (\ref{rozw dla qF}) w  (\ref{uppro}) eliminujemy obserwowaną informację strukturalną, otrzymując {\it równanie generujące} dla amplitudy rozkładu (fluktuacji) prędkości $\vec{\bf x}_{\wp}$ cząsteczki w gazie: 
\begin{eqnarray}
\label{rpw generujace dla q  pedu}
\nabla^{2} q_{n}(\vec{\bf x}_{\wp}) = - q_{n}(\vec{\bf x}_{\wp}) \, f_{n}\left({\vec{\bf x}_{\wp}}\right) \; ,
\end{eqnarray}
gdzie $\nabla^{2} \equiv \sum_{m=1}^{3} {\partial^{2}}/{\partial x_{\wp_{m}}^{2}}$ jest operatorem Laplace'a ze względu na współrzędne pędowe $x_{\wp_{m}}$.\\
 \\
Równanie (\ref{rpw generujace dla q  pedu}), podobnie jak
(\ref{falowe kappa 1}), pojawiło się jako rozwiązanie strukturalnej i wariacyjnej zasady informacyjnej. \\
Rozwiązanie $q_{n}$  równania (\ref{rpw generujace dla q  pedu}) jest {\it samospójnym} rozwiązaniem sprzężonego układu równań różniczkowych (\ref{uppro}) i (\ref{roznp}),  utworzonych przez  parę zasad informacyjnych. \\
  \\
{\bf Rozwiązanie równania generującego}: Poniżej rozwiążemy równanie generujące (\ref{rpw generujace dla q  pedu}), czyniąc kilka fizycznych założeń odnośnie postaci funkcji $f_{n}(\vec{\bf x}_{\wp})$. Jej postać wpływa na postać otrzymanych amplitud $q_{n}(\vec{\bf x}_{\wp})$, a zatem również na rozkład $p(\vec{\bf x}_{\wp})$. \\
\\
{\bf Fizyczne założenia o postaci $f_{n}(\vec{\bf x}_{\wp})$}: Załóżmy, że każda gęstość rozkładu prawdopodobieństwa $p_{n}(\vec{x})$ jest taką samą funkcją parzystą  każdej współrzędnej  $x_{\wp_{i}}$ (fluktuacji) pędu  $\vec{\bf x}_{\wp}$. W  konsekwencji  {\it układ jest  izotropowy}, tzn. rozkład prawdopodobieństwa dla pędu nie zależy od kierunku w bazowej przestrzeni położeń. \\
Następnie  zakładamy {\it nierelatywistyczne przybliżenie}, co oznacza, że prędkości cząsteczek są dużo mniejsze od prędkości światła $c$,  czyli fluktuacja pędu cząsteczki $x_{\wp_{i}}$ jest również mała w porównaniu z $mc$. \\
Z powyższych założeń wynika ogólna postać funkcji  $f_{n}(\vec{\bf x}_{\wp})$. Otóż jej rozwinięcie w szereg potęgowy ma tylko składowe parzyste modułu fluktuacji pędu $|\vec{\bf x}_{\wp}|$, a ponieważ wartość $|\vec{\bf x}_{\wp}|$ jest mała, zatem szereg ten obetniemy na drugim wyrazie:
\begin{eqnarray}
\label{postac funkcji ABp}
f_{n}(\vec{\bf x}_{\wp}) = A_{n} + B \, |\vec{\bf x}_{\wp}|^{2} \; ,\quad \quad A_{n}, B = const. 
\end{eqnarray}
{\bf Równanie generujące z $f_{n}$}: Podstawiając (\ref{postac funkcji ABp}) do (\ref{rpw generujace dla q  pedu}), otrzymujemy następującą postać równania generującego:
\begin{eqnarray}
\label{rownanie gener z f kwadrat}
\nabla^{2}q_{n}\left({\vec{\bf x}_{\wp}}\right)+\left(A_{n} + B |\vec{\bf x}_{\wp}|^{2}\right)q_{n}\left({\vec{\bf x}_{\wp}}\right) = 0 \; \;\; {\rm dla} \;\;\; n=1,2,...,N \; .
\end{eqnarray}
Przejdźmy do nowego indeksu: 
\begin{eqnarray}
\label{index n'}
n^{'}:= n-1 \; = \; 0, 1,..., N-1  \; .
\end{eqnarray}
Wtedy w miejsce (\ref{rownanie gener z f kwadrat}) otrzymujemy   równanie:
\begin{eqnarray}
\label{rownanie gener z f kwadrat z n'}
\nabla^{2} q_{n'} \left( \vec{\bf x}_{\wp} \right) + \left(A_{n'} + B |\vec{\bf x}_{\wp}|^{2} \right) q_{n'} \left( \vec{\bf x}_{\wp} \right) = 0 \; \;\; {\rm dla} \;\;\; n' = 1 ,  2,..., N-1 \; , 
\end{eqnarray}
które po dokonaniu separacji zmiennych kartezjańskich i faktoryzacji amplitudy:  
\begin{eqnarray}
\label{separacja}
q_{n'}\left({\vec{\bf x}_{\wp}}\right) = q_{n'_{1}}\left(x_{{\wp}_{1}}\right) q_{n'_{2}}\left(x_{{\wp}_{2}}\right) q_{n'_{3}}\left(x_{{\wp}_{3}}\right) \; ,
\end{eqnarray}
przechodzi w równoważny mu układ trzech równań różniczkowych:  
\begin{eqnarray}
\label{row gen po faktoryzacji dla p}
q_{n'_{i}}^{''}\left({x_{\wp_i}}\right)+\left({A_{n'_{i}} + B \, x_{\wp_i}^{2}}\right)q_{n'_{i}}\left({x_{\wp_i}}\right) = 0 \; , \quad i=1,2,3,   \;\;\;
\sum\limits _{i=1}^{3}{A_{n'_{i}}} \equiv A_{n'} \; .
\end{eqnarray}
Gdy stałe równania (\ref{row gen po faktoryzacji dla p}) mają postać: 
\begin{eqnarray}
\label{stale row gener dla ni'}
A_{n_{i}^{'}}=\frac{n_{i}^{'} + 1/2}{a_{0}^{2}} \, , \;\;\; B = -\frac{1}{{4a_{0}^{4}}} \, , \;\;\;  a_{0}=const. \, ,  \;\;\; n_{i}^{'}=0,1,... \; ,
\end{eqnarray}
wtedy ma ono rozwiązanie. \\
\\
{\bf Postać rozwiązania}: Ponieważ każde z równań (\ref{row gen po faktoryzacji dla p}) jest {\bf równaniem Helmholtz'a}, zatem jego rozwiązaniami są paraboliczno-cylindryczne funkcje   \cite{Ab,kamke}:
\begin{eqnarray}
\label{rozwiazanie dla q z ni' dla pedu}
q_{n_{i}^{'}}\left({x_{\wp_{i}}}\right) =  e^{-x_{\wp_{i}}^{2}/\left(4 a_{0}^{2} \right)} \; 2^{- n_{i}^{'}/2} \; H_{n_{i}^{'}} \left( \frac{x_{\wp_{i}}}{a_{0} \sqrt{2}} \right) \, , \;\;\; {\rm gdzie} \;\;\; n_{i}^{'}=0,1,...  \; , \;\;\;  i=1,2,3 \; ,
\end{eqnarray}
gdzie $H_{n_{i}^{'}}$ są wielomianami Hermite'a:
%
%
\begin{eqnarray}
\label{wiel Hermite}
H_{j}\left( t \right) = j!  \sum\limits_{m=0}^{\left[j/2\right]} \left({-1}\right)^{m} \, \frac{\left( 2 \,t \right)^{j-2m}}{m!\left( j-2m \right)! } \; , \;\;\; j = 0,1,... \; ,
\end{eqnarray}
gdzie $\left[j/2\right]$ oznacza liczbę całkowitą nie  przekraczającą $j/2$. \\
\\
{\bf Amplitudy fluktuacji  pędu}: Wstawiając (\ref{rozwiazanie dla q z ni' dla pedu}) do (\ref{separacja}) otrzymujemy szukaną postać amplitudy fluktuacji  pędu: 
\begin{eqnarray}
\label{rozwiazanie dla q z n' dla pedu}
\!\!\!\!\!\!\!\!\!\! q_{n^{'}}\left({\vec{\bf x}_{\wp}}\right) & = &  e^{ - |\vec{\bf x}_{\wp}|^{2}/\left(4 a_{0}^{2} \right)} \; 2^{- n^{'}/2}
\; \cdot \nonumber \\
& & \nonumber \\
& \cdot & \!\!\!\!\!\!\!\!\!\! \sum\limits _{{\scriptstyle {\quad ijk\hfill\atop {\scriptstyle i+j+k=n^{'}\hfill}}}}{\!\!\!\!\!\! a_{n_{ijk}^{'}}} \; H_{i}\left(\frac{x_{\wp_{1}}}{a_{0}\sqrt{2}}\right)H_{j}\left(\frac{x_{\wp_{2}}}{a_{0}\sqrt{2}}\right)H_{k}\left(\frac{x_{\wp_{3}}}{a_{0}\sqrt{2}}\right) \, , \;\;\; {\rm gdzie} \;\;\; a_{n_{ijk}^{'}} =  const. \; ,  \;\;
\end{eqnarray}
przy czym:
\begin{eqnarray}
\label{zwiazek n z ni'}
\sum\limits _{i=1}^{3}{n_{i}^{'} = n^{'}}  = 0, 1,..., N-1 \; .
\end{eqnarray}
\\
{\bf Rozkład fluktuacji  pędu}: Funkcja rozkładu prawdopodobieństwa ma więc zgodnie z (\ref{p jako suma po qn2 przez N}) postać:
\begin{eqnarray}
\label{rozwiazanie dla prawdop z n' dla pedu}
\!\!\!\!\!\!\!\!\!\!\!\!\!\!\!\!\! 
p\left({\vec{\bf x}_{\wp}}\right) &=&   \frac{1}{N}\sum_{n'=0}^{N-1}{q_{n'}^{2}}\left({\vec{\bf x}_{\wp}}\right) = p_{0} \, e^{- \left| \vec{\bf x}_{\wp} \right|^{2}/ \left({2a_{0}^{2}}\right)} \;\; \cdot \nonumber \\
&\cdot& \left\{ 1+\sum_{n^{'}=1}^{N-1}{2^{-n^{'}}\left[\sum\limits _{{\scriptstyle {\quad ijk\hfill\atop {\scriptstyle i+j+k=n^{'}\hfill}}}}{b_{n_{ijk}^{'}}
H_{i}\left(\frac{x_{\wp_{1}}}{a_{0} \sqrt{2}}\right)
H_{j}\left(\frac{x_{\wp_{2}}}{a_{0} \sqrt{2}}\right)
H_{k}\left(\frac{x_{\wp_{3}}}{a_{0} \sqrt{2}}\right)}\right]^{2}}\right\} \; ,
\end{eqnarray}
gdzie $p_{0}=a_{0_{000}}^{2}/N$, a liczba 1 w nawiasie klamrowym pochodzi z $n^{'}=0$ w sumie w pierwszej równości, natomiast 
nowe stałe $b_{n_{ijk}^{'}}$ są proporcjonalne do stałych $a_{n_{ijk}^{'}}$. \\
Po raz pierwszy takie wyprowadzenie postaci funkcji rozkładu gęstości prawdopodobieństwa dla fluktuacji pędu zostało podane  w \cite{Frieden}. \\
%
\\
{\bf Rozwiązania równowagowe i nierównowagowe}: Wartość $N=1$, dla której (\ref{rozwiazanie dla prawdop z n' dla pedu}) jest rozkładem Gaussa, daje równowagowy {\it rozkład Maxwella-Boltzmanna dla (fluktuacji) pędu cząsteczki gazu}.
%
%
Pozostałe rozwiązania dla $N\ge2$ dają rozwiązania nierównowagowe \cite{Frieden}. Jednak i one są stacjonarne ze względu na fakt bycia samospójnymi rozwiązaniami układu sprzężonych równań różniczkowych zasady strukturalnej (\ref{uppro}) i wariacyjnej (\ref{roznp}). \\
Wcześniej, rozkłady (\ref{rozwiazanie dla prawdop z n' dla pedu}) z $N\ge2$  zostały odkryte przez Rumer'a i Ryskin'a \cite{Rumer and Ryskin-1980} jako  rozwiązania równania transportu Boltzmann'a. \\
\\
{\bf Interferencja rozwiązań}: Zwróćmy uwagę, że rozwiązanie (\ref{rozwiazanie dla prawdop z n' dla pedu}) implikuje  interferencję pomiędzy wyrażeniami  iloczynowymi występującymi  w amplitudach (\ref{rozwiazanie dla q z n' dla pedu}). Interferencja pojawiła się więc jako cecha charakterystyczna dla rozwiązywanego równania różniczkowego oraz  wprowadzenia amplitud do opisu układu, a nie jako cecha charakterystyczna wyłącznie mechaniki kwantowej.\\
\\ 
{\bf Rozkład Maxwella-Boltzmanna dla prędkości z $N=1$}: 
%
Założyliśmy na wstępie, że  zbiornik zawierający  cząsteczki gazu jest w spoczynku, w pewnym inercjalnym układzie współrzędnych. Zatem średnia prędkość cząsteczki wynosi zero, tzn. $\vec{\theta}_{\wp} = 0$ i z (\ref{y_p}) otrzymujemy $\vec{\bf y}_{\wp} = \vec{\bf x}_{\wp} = \vec{\wp}$. \\
\\
Tak wiec, dla $N=1$ rozkład gęstości prawdopodobieństwa pędu  (\ref{rozwiazanie dla prawdop z n' dla pedu}) przyjmuje postać: 
\begin{eqnarray}
\label{row}
p \left( \vec{\wp} \right) = p_{0} \; e^{- |\vec{\wp}\,|^{2}/(2a_{0}^{2})} = p_{0}\; e^{- \wp_{1}^{2}/(2a_{0}^{2})} e^{- \wp_{2}^{2}/(2a_{0}^{2})} e^{- \wp_{3}^{2}/(2a_{0}^{2})} \; ,
\end{eqnarray}
gdzie $\vec{\wp} = (\wp_{1}, \wp_{2}, \wp_{3})$ oraz $-\infty<\wp_{i}<\infty$, $i=1,2,3$. 
Wyznaczmy wartość stałej $a_{0}$ dla przypadku zerowej 
energii potencjalnej. Z warunku normalizacji $\int d \vec{\wp} \, p\left(\vec{\wp} \right)=1\,$ wyznaczamy, że stała $p_{0}$  w (\ref{row}) wynosi:
\begin{eqnarray}
p_{0}=\frac{1}{{\left({2\pi}\right)^{{3\mathord{\left/{\vphantom{32}}\right.\kern -\nulldelimiterspace}2}}a_{0}^{3}}}\label{p0} \; .
\end{eqnarray}
Ponieważ wtedy $\left\langle E\right\rangle = \left\langle \vec{\wp}^{\,2}\right\rangle/(2m)\,$, a z twierdzenia o ekwipartycji energii wiemy, że $\left\langle E\right\rangle = 3kT/2$,  zatem 
\begin{eqnarray}
\label{wartoczekp}
\left\langle {\vec{\wp}^{\,2}}\right\rangle = 3 \, m \, k T .
\end{eqnarray}
Mając wartość $p_{0}$, wyznaczmy wartość oczekiwaną: 
\begin{eqnarray}
\left\langle {\vec{\wp}^{2}}\right\rangle =\int{d\vec{\wp} \, p\left({\vec{\wp}}\right)\vec{\wp}^{2}} \, 
\end{eqnarray}
i przyrównajmy ją do (\ref{wartoczekp}). W rezultacie otrzymujemy: \begin{eqnarray}
\label{a0}
{a_{0}^{2}=mc^{2}kT} \; .  
\end{eqnarray}
{\bf Rozkładu  wartości pędu}: Kolejnym krokiem jest wyznaczenie rozkładu prawdopodobieństwa wartości pędu $\wp = |\vec{\wp}\,|$. Przechodząc od współrzędnych  kartezjańskich pędu $\left(\wp_{1},\wp_{2},\wp_{3}\right)$ do współrzędnych sferycznych  $\left(\wp,\theta,\phi\right)$, otrzymujemy:
\begin{eqnarray}
\label{pj}
p\left({\wp,\theta,\phi}\right)=\left|J \,\right|p\left({\wp_{1},\wp_{2},\wp_{3}}\right) \; , \;\;\; 
J = \wp^{2}\sin\theta \; ,
\end{eqnarray}
gdzie $J$ jest jakobianem przejścia. Wstawiając teraz (\ref{p0}) wraz z (\ref{a0}) do (\ref{row}) otrzymujemy rozkład 
$p\left({\wp_{1},\wp_{2},\wp_{3}}\right)$ ze znanymi już stałymi $a_{0}$ oraz $p_{0}$. Wynik ten  wstawiając do (\ref{pj}) i   wycałkowując po zmiennych $\theta$ oraz $\phi$, otrzymujemy szukany rozkład gęstości prawdopodobieństwa dla wartości pędu: 
\begin{eqnarray}
\label{maxbol}
p\left(\wp\right)=\sqrt{{2\mathord{\left/{\vphantom{2\pi}}\right.\kern -\nulldelimiterspace}\pi}}\left({mkT}\right)^{{{-3}\mathord{\left/{\vphantom{{-3}2}}\right.\kern -\nulldelimiterspace}2}}\wp^{2}e^{{{-\wp^{2}}\mathord{\left/{\vphantom{{-\wp^{2}}{2mkT}}}\right.\kern -\nulldelimiterspace}{2mkT}}} \, .
\end{eqnarray}
{\bf Prawo Maxwella-Boltzmanna}: Rozkład (\ref{maxbol})  wyraża  tzw. prawo Maxwella-Boltzmanna, a podstawiając   $\wp =m v$ w  (\ref{maxbol}), w miejsce wartości pędu, otrzymujemy  rozkład Maxwella-Boltzmanna dla wartości prędkości $v$ cząsteczki gazu.


\subsection{Informacja Fishera jako ograniczenie dla wzrostu entropii}

Rozważmy układ zawierający \textit{jedną} lub \textit{więcej} cząstek poruszających
się w sposób losowy, w pewnym zamkniętym obszarze. Niech rozważany obszar będzie izolowany, tzn. żadna cząstka ani z obszaru nie ucieka ani do niego nie przechodzi.  Brzeg ($\vec{B}$) obszaru jest zadany  wektorem wodzącym $\vec{b} \in \vec{B}$. 
W pewnej chwili układ dokonuje pomiaru jego czterowektora 
położenia $t, \vec{y}$, zgodnie z 
rozkładem gęstości prawdopodobieństwa 
$p(\vec{y}, t)$\footnote{
Jeśli układ opisany jest funkcją falową $\psi(\vec{y}, t)$, wtedy 
$p(\vec{y}, t) =  |\psi(\vec{y}, t)|^{2}$.
}.  
\\
\\
Niech $(y^{1},y^{2},y^{3})$ są współrzędnymi kartezjańskimi wektora $\vec{y}$ {\bf położenia przestrzennego}, a $r=|\vec{y}|$ jego długością. 
Ponieważ założyliśmy, że cząstka znajduje się gdzieś wewnątrz obszaru więc:
\begin{eqnarray}
\label{pt brzegowe = 1}
p\left(t \right) = \int d \vec{y} \, p\,(\vec{y}, t)   = 1 \; .
\end{eqnarray}
Oznaczmy przez $S_{H}\left(t\right)$ entropię Shannona układu (\ref{eboltmana})
w chwili czasu $t$: 
\begin{eqnarray}
S_{H}\left(t\right) = - \int d\vec{y}\; p(\vec{y}, t) 
\ln p(\vec{y}, t) \; .
\label{st}
\end{eqnarray}
{\bf Drugą zasada termodynamiki}: Można pokazać  (por. Dodatek~\ref{Wyprowadzenie drugiej zasady termodynamiki}), że entropia Shannona spełnia drugą zasadę termodynamiki:
\begin{eqnarray}
\frac{{dS_{H}\left(t\right)}}{{dt}} \ge 0 \; ,
\label{2zasterm}
\end{eqnarray}
która określa {\it dolną granicę tempa zmiany entropii}. Poniższe
rozważania poświęcone są znalezieniu ograniczenia na jego górną granicę. \\
\\
{\bf Równanie ciągłości strumienia}: Ponieważ żadna cząstka nie opuszcza ani nie wpływa do obszaru, zatem
spełnione jest równanie ciągłości strumienia prawdopodobieństwa: 
\begin{eqnarray}
\label{ciaglosc}
\frac{\partial p(\vec{y}, t)}{\partial t} + \vec{\nabla} \cdot \vec{P}\left({\vec{y},t}\right)=0 \; , 
\end{eqnarray}
gdzie $\vec{P}\left({\vec{y},t}\right)$ jest {\it prądem prawdopodobieństwa}, którego konkretna postać zależy od rozważanego układu. \\
 \\
{\bf Warunki brzegowe}: Z kolei określmy warunki brzegowe spełniane przez gęstość prawdopodobieństwa i jego prąd. 
Będą nam one przydatne dla rozwiązania postawionego sobie zadania. \\
Ponieważ żadne cząstki nie przechodzą przez granicę obszaru, zatem $\vec{P}$ spełnia warunek brzegowy Dirichleta:  
\begin{eqnarray}
\label{warbrzegP}
\vec{P} \left( {\vec{y},t} \right) \left|\begin{array}{l}
\!\!\!_{{\vec{y} \in {\vec{B}}}}\end{array} \right. = 0 \; .
\end{eqnarray}
Załóżmy dodatkowo, że jeśli brzeg obszaru znajduje się w nieskończoności, to: 
\begin{eqnarray}
\label{Pr0}
\mathop{\lim}\limits _{\vec{y}\to\infty}\vec{P}\left({\vec{y},t}\right) \to 0 \; , \;\;\;\; {\rm szybciej \;\; ni\dot{z}} \;\;\;\; 1/r^{2} \; .
\end{eqnarray}
Ponieważ obszar jest odizolowany, zatem prawdopodobieństwo, że cząstka znajduje się na jego granicy znika, co oznacza, że $p(\vec{y}, t)$ również spełnia warunek brzegowy Dirichleta: 
\begin{eqnarray}
\label{warbrzegp}
p(\vec{y}, t)\left|\begin{array}{l}
\!\!\!_{{\vec{y} \in \vec{B}}}\end{array}\right. = 0 \; .
\end{eqnarray}
Ze względu na normalizację: 
\begin{eqnarray}
\int{d\vec{y}\; p(\vec{y}, t)} = 1 \; ,
\end{eqnarray}
zakładamy, że w przypadku brzegu nieskończenie oddalonego,  
$p(\vec{y}, t)$ spełnia warunek: 
\begin{eqnarray}
\label{pr0}
\mathop{\lim}\limits _{\vec{y}\to\infty}p\left({\vec{y}\left|\; t\right.}\right)\to 0 \; , 
\;\;\;\; {\rm szybciej \;\; ni\dot{z}} \;\;\;\; 1/r^{3} \;\; .
\end{eqnarray}
W końcu dla domknięcia potrzebnych warunków brzegowych przyjmujemy:
\begin{eqnarray}
\label{zeroPp}
\vec{P}\ln p\left|\begin{array}{l}
\!\!\!_{{\vec{y} \in \vec{B}}}\end{array}\right. = 0 \; .
\end{eqnarray}
  \\
{\bf Wyprowadzenie ograniczenia na tempo wzrostu entropii}: Przystąpmy teraz do sedna rachunków. Różniczkowanie (\ref{st}) po $\frac{\partial }{\partial t}$
daje: 
\begin{eqnarray}
\frac{\partial S_{H}}{\partial t} = - \frac{\partial}{{\partial t}}\int{d\vec{y}\; p\ln p} = - \int{d\vec{y}\; \frac{\partial p}{\partial t}\ln p} - \int{ d\vec{y}\; p \frac{1}{p} \frac{\partial p}{\partial t}} \; .
\label{dr}
\end{eqnarray}
Druga całka po prawej stronie daje po skorzystaniu z warunku unormowania:  
\begin{eqnarray}
\frac{\partial}{{\partial t}}\int{d\vec{y}\; p} = 0 \; .
\end{eqnarray}
Oznaczmy przez $\left(P^{1},P^{2},P^{3}\right)$ kartezjańskie składowe prądu $\vec{P}$. 
Podstawiając (\ref{ciaglosc}) do pierwszej całki po prawej stronie
w (\ref{dr}) otrzymujemy: 
\begin{eqnarray}
\label{pierwsza}
\frac{\partial S_{H}}{\partial t} = \int{d\vec{y}\; \vec{\nabla} \cdot\vec{P}\ln p}=\int{\int{\int{dy^{3} \,dy^{2}\,dy^{1} \left[{\frac{\partial}{{\partial y^{1}}}P^{1} + \frac{\partial}{{\partial y^{2}}}P^{2} + \frac{\partial}{{\partial y^{3}}}P^{3}}\right]\ln p}}} \; ,
\end{eqnarray}
gdzie $\nabla_{i} = \frac{\partial}{\partial y^{i}} \,$. \\
\\
{\bf Całkując przez części} wewnętrzne całki $\int{dy^{i} \frac{\partial}{\partial y^{i}} P^{i} }$ w (\ref{pierwsza}) dla trzech składników $i=1,2,3$,  otrzymujemy dla każdego z nich:
\begin{eqnarray}
\int{dy^{i} (\frac{\partial}{{\partial y^{i}}} P^{i}) \ln p} = P^{i} \ln p \left|\begin{array}{l}
\!\!\!_{{\vec{y} \in \vec{B}}} \end{array}\right. - \int{dy^{i}\frac{{P^{i}}}{p} \frac{\partial p}{\partial y^{i}} } = - \int{dy^{i} P^{i} \frac{\partial p}{\partial y^{i}} \frac{1}{p} } \; , \;\;\; i = 1,2,3 \; ,
\end{eqnarray}
gdzie skorzystano z (\ref{zeroPp}). Zatem:
\begin{eqnarray}
\label{dS po dt iloczyn}
\frac{\partial S_{H}}{\partial t} = - \int{dy^{3}dy^{2}dy^{1} \left( P^{1} \frac{\partial p}{\partial y^{1}} + P^{2} \frac{\partial p}{\partial y^{2}} + P^{3} \frac{\partial p}{\partial y^{3}}\right) \frac{1}{p} } = - \int{d\vec{y} \left( \vec{P} \cdot \vec{\nabla} p \right) \frac{1}{p} } \; .
\end{eqnarray}
Z powyższego, po prostych przekształceniach mamy: 
\begin{eqnarray}
\label{kwadrat dS po dt}
\left(\frac{\partial S_{H}}{\partial t}\right)^{2} = \left[ \int d\vec{y} \left(\frac{\vec{P}}{\sqrt{p}}\right) \cdot \left(\frac{\sqrt{p}\,  \vec{\nabla} p}{p} \right) \right]^{2} =  \left[ \, \sum_{i=1}^{3} \int d\vec{y} \left(\frac{P^{i}}{\sqrt{p}}\right) \; \left(\frac{\sqrt{p}\;  \nabla_{i} p}{p} \right) \right]^{2} \; .
\end{eqnarray}
\\
{\bf Z nierówności Schwartza} otrzymujemy:
\begin{eqnarray}
\label{nierownosc Schwartza}
\left[ \, \sum_{i=1}^{3} \int d\vec{y} \left(\frac{P^{i}}{\sqrt{p}}\right) \; \left(\frac{\sqrt{p}\;  \partial_{i} p}{p} \right) \right]^{2} \leq
\left[ \sum_{i=1}^{3} \int d\vec{y} \left(\frac{P^{i}}{\sqrt{p}}\right)^{2} \right] \left[ \sum_{i=1}^{3} \int d\vec{y}  \left(\frac{\sqrt{p}\;  \nabla_{i} p}{p} \right)^{2} \right]
\; .
\end{eqnarray}
{\bf Ogólna postać ograniczenia na tempo wzrostu entropii}:  Ostatecznie z (\ref{kwadrat dS po dt}) i (\ref{nierownosc Schwartza}) oraz po zastępieniu sumowań po $i$ znakiem iloczynów skalarnych, dostajemy: 
\begin{eqnarray}
\label{st2}
\left(\frac{\partial S_{H}}{\partial t}\right)^{2} \le \int{d\vec{y}\;\frac{{\vec{P}\cdot\vec{P}}}{p}}\int{d\vec{y}\;\frac{{\vec{\nabla} p\cdot\vec{\nabla} p}}{p}} \; .
\end{eqnarray}
\\
{\bf Przejście do postaci z pojemnością informacyjną}: Ponieważ gradiant $\vec{\nabla} \equiv (\frac{\partial }{\partial y^{1}}, \frac{\partial }{\partial y^{2}}, \frac{\partial }{\partial y^{3}})$ zawiera różniczkowanie po wartościach pomiarowych $y^{i}$,  a nie fluktuacjach $x^{i}$, jak to jest w kinematycznej postaci  pojemności informacyjnej Friedena-Soffera, zatem musimy uczynić 
dodatkowe założenie.\\
{\bf Założenie niezmienniczości rozkładu ze względu na przesunięcie}:   Przejdźmy do addytywnych przesunięć, $\vec{x} = \vec{y} - \vec{\theta}$. Ponieważ  $\vec{x} = (x^{i})_{i=1}^{3}$ oraz $\vec{y} = (y^{i})_{i=1}^{3}$ różnią się o stałą wartość oczekiwaną położenia $\vec{\theta}=(\theta^{i})_{i=1}^{3}$, zatem 
{\it zakładając dodatkowo , że  rozkład $p$ jest niezmienniczy  ze względu na przesunięcie} $\vec{\theta}$, otrzymujemy:
\begin{eqnarray}
\label{niezm na przes}
\frac{\partial p }{\partial y^{i}} = \frac{\partial p}{\partial x^{i}} \; , \;\;\; i=1,2,3 \; .
\end{eqnarray}
Przy powyższym założeniu, zachodzi:
\begin{eqnarray}
\label{poj info od t}
\int{d\vec{y}\;\frac{{\vec{\nabla} p(\vec{y}, t)\cdot\vec{\nabla} p(\vec{y}, t)}}{p(\vec{y}, t)}} = \int{d\vec{x}\;\frac{{\vec{\nabla} p(\vec{x}, t)\cdot\vec{\nabla} p(\vec{x}, t)}}{p(\vec{x}, t)}} \; 
\end{eqnarray}
i okazuje się, że  druga całka po prawej stronie nierówności w (\ref{st2}) jest pojemnością informacyjną trzech kanałów przestrzennych dla $N=1$ w chwili czasu $t$:  
\begin{eqnarray}
\label{pojI}
I \equiv I \left(t\right)=\int{d\vec{x}\;\frac{\nabla p\left(\vec{x},t \right) \cdot \nabla p\left(\vec{x},t \right)}{p\left(\vec{x},t \right)}} \; .
\end{eqnarray}
\\
{\bf Ograniczenie na tempo wzrostu $S_{H}$ układu z niezmienniczością przesunięcia}: Korzystając z powyższej postaci pojemności informacyjnej, możemy zapisać (\ref{st2}) następująco:
\begin{eqnarray}
\label{ograniczenieS}
\left(\frac{\partial S_{H}}{\partial t}\right)^{2} \le I(t) \int{d\vec{x}\;\frac{{\vec{P}\cdot\vec{P}}}{p}} \; , \;\;\; {\rm lub} \;\;\; \left(\frac{\partial S_{H}}{\partial t}\right) \le \sqrt{I(t)} \;\; \sqrt{\int{d\vec{x}\;\frac{{\vec{P}\cdot\vec{P}}}{p}}}  \; ,
\end{eqnarray}
\\
Pokazaliśmy zatem, że {\it przy założeniu niezmienniczości rozkładu za względu na przesunięcie}, tempo wzrostu entropii jest obustronnie ograniczone. 
Ograniczenie dolne (\ref{2zasterm}) wyraża zasadę niemalenia entropii Shannona w czasie. Jej  termodynamicznym odpowiednikiem jest twierdzenie H Boltzmanna.  \\
\\
{\bf Wniosek}: Nierówność (\ref{ograniczenieS}) oznacza, że  {\it  ograniczenie górne  tempa wzrostu entropii jest proporcjonalne do pierwiastka z pojemności informacyjnej} (\ref{pojI}) {\it dla pomiaru położenia $\vec{y}$}. Jest to jednen z nowych wyników teorii pomiaru otrzymany przez Friedena, Soffera, Plastino i Plastino \cite{Frieden}. Jego termodynamiczne konsekwencje czekają na weryfikację. \\
\\
W \cite{Frieden} podano przykłady zastosowania tego twierdzenia dla strumienia  cząstek klasycznych, strumienia w elektrodynamice klasycznej i strumienia cząstek ze spinem 1/2.  Poniżej zostanie podany wynik analizy dla tego ostatniego przypadku. 

\subsubsection{Wynik dla strumienia cząstek ze spinem 1/2}

W Rozdziale~\ref{Dirac field} pokazaliśmy, że metoda EFI daje dla  relatywistycznej cząstki o spinie połówkowym równanie ruchu  Diraca (\ref{Dirac eq}). 
Pod nieobecność pola elektromagnetycznego wynikające z niego równanie ciągłości  ma postać:
\begin{eqnarray}
\label{rownciaglpsi}
\frac{\partial}{{\partial t}}p\left({\vec{y}, t}\right) + \vec{\nabla}\cdot\vec{P}\left({\vec{y}, t}\right) = 0 \; ,
\end{eqnarray}
gdzie gęstość prawdopodobieństwa $p$ oraz gęstość prądu prawdopodobieństwa są równe odpowiednio:
\begin{eqnarray}
\label{psi}
p\left({\vec{y}, t}\right)=\psi^{\dagger}\psi,\quad\psi\equiv\psi\left({\vec{y}, t}\right) \; ,
\end{eqnarray}
oraz 
\begin{eqnarray}
\label{psi+}
\vec{P}\left({\vec{y}, t}\right) =  c \, \psi^{\dagger} \, \vec{\alpha} \,\psi \; ,
\end{eqnarray}
gdzie $\vec{\alpha} \equiv \left(\alpha^{1},{\alpha^{2}},{\alpha^{3}}\right)$ są  macierzami  Diraca (\ref{m Diraca alfa beta}), natomiast $\psi^{\dagger} = (\psi_{1},\psi_{2},\psi_{3},\psi_{4})^{*}$ jest polem sprzężonym hermitowsko do bispinora Diraca $\psi$ (Rozdział~\ref{Dirac field}). \\
\\
{\bf Ograniczenie na tempo wzrostu entropii}: W  \cite{Frieden} pokazano, że tempo wzrostu entropii  (\ref{ograniczenieS}) ma w tym przypadku postać: 
\begin{eqnarray}
\label{nierownosc S I c}
{\frac{{\partial S_{H}}}{{\partial t}}}\le c\sqrt{I(t)} \; .
\end{eqnarray}
\\
{\bf Wniosek}: Nierówność ta oznacza, że dla układu, który posiada niezmienniczość przesunięcia, 
wzrost entropii Shannona rozkładu w jedno\-stce czasu (czyli tempo spadku informacji) jest ograniczony przez skończoną prędkość światła jak również przez pojemność informacyjną, jaką posiada układ (np. swobodny elektron) o swoich  współrzędnych czasoprzestrzennych. \\
Inny sposób interpretacji 
(\ref{nierownosc S I c}) polega na zauważeniu, że dostarcza ona definicji prędkości światła $c$ jako górnego ograniczenia stosunku tempa zmiany entropii do pierwiastka informacji Fishera, która jest przecież na wskroś statystycznym pojęciem informacyjnym. 

\newpage

\section[Zastosowanie wprowadzonego formalizmu do analizy paradoksu EPR]{Zastosowanie wprowadzonego formalizmu do analizy paradoksu EPR}

\label{paradoks EPR}

\begin{figure}[ht]
\begin{center}
\includegraphics[width=70mm,height=35mm]{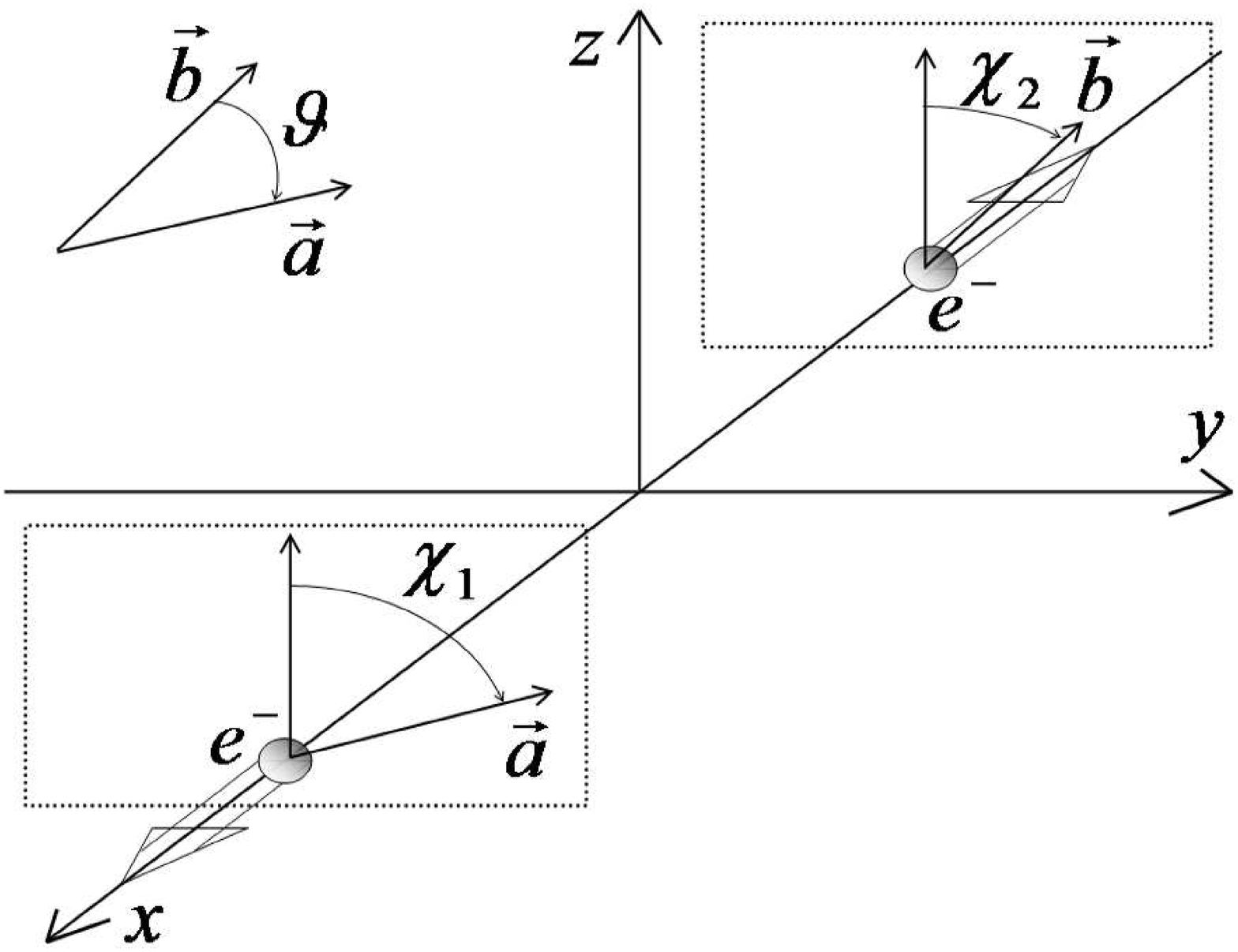}
\end{center}
\caption{Eksperyment EPR-Bohma. Zaznaczono elektrony rejestrowane w urządzeniach  Sterna-Gerlacha; 
opis w tekście Rozdziału. W lewym górnym rogu rysunku zaznaczono kąt $\vartheta$ pomiędzy kierunkami $\vec{a}$ oraz $\vec{b}$  urządzeń Sterna-Gerlacha. }
\label{fig:5} 
\end{figure}

Przedstawiona w tym rozdziale analiza paradoksu EPR pochodzi od Friedena \cite{Frieden}. Zamiesz\-czamy ją  ze względu na to, że  efekt ten jest dość powszechnie utożsamiany z własnością teorii kwantowych, ale również z powodu uporządkowania warunków brzegowych zagadnienia zawartego w pracy orginalnej \cite{Frieden}, a przedstawionego w pracy  \cite{Mroziakiewicz} oraz wnioskowania uzgodnionego z przedstawioną w skrypcie fizyczną interpretacją informacji fizycznej $K$ \cite{Dziekuje informacja_1, Dziekuje informacja_2}. \\
  \\
{\bf Opis eksperymentu EPR-Bohm'a}: Rozpocznijmy od omówienia eksperymentu EPR-Bohm'a. Rozważmy 
źródło molekuł o spinie zero, które rozpadają się na parę identycznych cząstek o spinie $1/2$ lecących w przeciwnych kierunkach. Taka, początkowa dla rozważań eksperymentu EPR-Bohm'a, konfiguracja dwucząstkowej molekuły może być efektywnie przygotowana jako stan końcowy w rozpraszaniu $e^{-}e^{-} \to e^{-}e^{-}$, gdzie
spiny początkowych elektronów  procesu są ustawione przeciwnie (równolegle i antyrównolegle względem osi $z$), a ich początkowe pędy wzdłuż osi $y$ wynoszą $\vec{p}\,$ oraz $-\vec{p}$ \cite{Manoukian}. Istnieje niezerowa amplituda, że dwie rozproszone cząstki (tu wyprodukowane elektrony), poruszają się z pędami wzdłuż osi $x$, jak na Rysunku~\ref{fig:5}. 
\\
W eksperymencie EPR-Bohm'a dokonywany jest pomiar spinu  rozproszonych cząstek, spinu cząstki 1 wzdłuż wektora jednostkowego $\vec{a}$, tworzącego z osią $z$ kąt
$\chi_{1}$ oraz spinu cząstki 2 wzdłuż wektora jednostkowego $\vec{b}$,  
tworzącego z osią $z$ kąt $\chi_{2}$. 
Niech analizator ,,$a$'', będący urządzeniem typu Sterna-Gerlacha,    
mierzy rzut $S_{a}$
spinu $\vec{S}_{1}$ cząstki 1 na kierunek 
$\vec{a}$ i podobnie analizator ,,$b$'' mierzy rzut $S_{b}$
spinu $\vec{S}_{2}$ 
cząstki 2 na kierunek $\vec{b}$. Kąt pomiędzy płaszczyznami wektorów $\vec{a}$ oraz $\vec{b}$, zawierającymi oś $x$, 
wynosi $\vartheta = \chi_{1}-\chi_{2}$, $0\le  \vartheta < 2\pi$. \\
Poniżej wyprowadzimy warunki brzegowe dla metody EFI, uważając aby nie odwoływać się do widzenia rzeczywistości przez pryzmat mechaniki kwantowej
\footnote{Lecz biorąc pod uwagę ogólne stwierdzenie, że jeśli $P_{12}$  jest  łącznym rozkładem prawdopodobieństwa 
pewnych dwóch zmiennych losowych ``1''  oraz ``2'', a ${P_{1}}$ oraz ${P_{2}}$ ich rozkładami brzegowymi, to  jeśli w ogólności nie są one względem siebie niezależne (tzn.  łączne prawdopodobieństwo nie jest iloczynem 
brzegowych), wtedy:
\begin{eqnarray}
\label{subaddytywnosc informacji}
I_{F}\left({P_{12}}\right) \ge I_{F} \left({P_{1}}\right) + I_{F} \left({P_{2}}\right)\equiv \tilde{C} \, 
\end{eqnarray} 
gdzie $\tilde{C}$ jest pojemnością informacyjną złożonego układu. 
Relacja (\ref{subaddytywnosc informacji}) oznacza, że jeśli występują jakiekolwiek korelacje między zmiennymi to, jeśli znamy wynik doświadczenia dla pierwszej zmiennej to maleje ilość infor\-macji koniecznej do określenia wyniku doświadczenia dla drugiej z nich, tzn.  istnienie korelacji w układzie zwiększa  informację Fishera $I_{F}$ o parametrach charakteryzujących rozkład układu. 
}. 

\subsection{Warunki brzegowe}

\label{Warunki brzegowe}

W celu rozwiązania równań różniczkowych EFI konieczne jest ustalenie warunków brzegowych na prawdopodobieństwa. Wynikają one z przesłanek fenomenologicnych,  zasad zachowania i symetrii przestrzennej  badanego układu. 
\\
\\
{\bf Założenie} {\it o istnieniu dokładnie dwóch możliwych rzutów spinu cząstki ze spinem $\hbar/2$ na dowolny wybrany kierunek w przestrzeni}: Niech ``$+$'' oznacza obserwowaną wartość rzutu spinu $S_{a}=+\hbar/2$ natomiast ``$-$'' oznacza $S_{a} = - \hbar/2$. \\
\\
{\bf Uwaga o fenomenologii rejestracji spinu cząstki}: Jest to jedyne miejsce, w którym uciekamy się do opisu fenomenologicznego, odwołując się do nieklasycznej fizyki zjawiska, mianowicie zakładamy, że rzutowanie spinu cząstki na określony kierunek przestrzenny jest skwantowane zgodnie z wymiarem reprezentacji grupy obrotów (fakt istnienia takich reprezentacji wynika z rozważań entro\-pijnych \cite{Frieden}). \\
Jednak nie oznacza to, że nie istnieje model teoriopolowy, który by takie kwantowanie rzutowania opisywał. Musiałby on jedynie zakładać, że po pierwsze spinowe stopnie swobody (z ciągłym rozkładem jego kierunku przed pomiarem) nie mają (prostego) charakteru czasoprzestrzennego \cite{dziekuje za neutron}, a po drugie, że  bezwładność cząstki związana ze spinowymi stopniami swobody jest bardzo mała w porównaniu z siłą sprzężenia tych spinowych stopni swobody z otoczeniem (np. aparaturą traktowaną jako rezerwuar), którego moment pędu ignorujemy.
Wtedy przejście cząstki ze spinem przez jakąkolwiek aparaturę pomiarową typu Sterna-Gerlacha porządkowałoby jej spin w sposób dyskretny. Oczywiście zmiana momentu pędu aparatury  już nas ``nie interesuje''. \\
 \\
%
{\bf Łączną  przestrzeń zdarzeń} $\Omega_{ab}$ stanów spinowych pary cząstek 1 i 2 przyjmujemy   
jako następującą:
\begin{eqnarray}
\label{4 zdarzenia EPR}
S_{a}S_{b} \equiv S_{ab} \in \Omega_{ab} = \left\{ S_{++},S_{--},S_{+-},S_{-+} \right\} \equiv \left\{(++),(--),(+-),(-+)\right\} \; .
\end{eqnarray}
Załóżmy, że możemy zdefiniować cztery łączne warunkowe 
prawdopodobieństwa\footnote{W przypadku braku wspólnej przestrzeni zdarzeń $\Omega_{AB}$ dla zmiennych losowych powiedzmy $A$ oraz $B$, nie można by zdefiniować łącznego rozkładu prawdopodobieństwa $P(A,B)$ dla tych dwóch zmiennych, pomimo istnienia ich rozkładów brzegowych $P(A)$ oraz $P(B)$, co oznacza, że nie dało by się  dokonać ich łącznego  pomiaru.  
Również na ogół, pomimo istnienia łącznego rozkładu brzegowego $P(A,B)$ dla zmiennych $A$ oraz $B$, oraz rozkładu brzegowego $P(B,C)$ dla zmiennych $B$ oraz $C$, nie istnieje rozkład łączny $P(A,B,C)$ dla zmiennych $A$, $B$ i $C$.  Zwróćmy uwagę na fakt, że w dowodzie nierówności Bella \cite{Bell,Khrennikov} przyjmuje się za oczywisty fakt istnienia łącznego rozkładu $P(A,B,C)$. Możliwość taka istnieje zawsze, gdy wspólna przestrzeń zdarzeń $\Omega_{ABC}$ tych trzech zmiennych losowych istnieje i jest iloczynem kartezjańskim  $\Omega_A \times \Omega_B \times \Omega_C$. Z drugiej strony, nierówności Bell'a  są znane w klasycznej teorii prawdopodobieństwa od czasów Boole'a jako test, który w przypadku ich niespełnienia świadczy o niemożliwości konstrukcji łącznego rozkładu prawdopodobieństwa.  Rozumowanie to można rozszerzyć na dowolną liczbę zmiennych losowych \cite{Khrennikov,Accardi}.  \\
W pełnym opisie rzeczywistego eksperymentu EPR-Bohm'a, a nie tylko w eksperymencie typu "`gedanken"', powinny występować obok dwóch   zmiennych losowych rzutów spinów mierzonych w analizatorach "`a"' oraz "`b"', również dwie zmienne losowe kątowe mierzone dla tych cząstek w chwili ich produkcji. 
}
$P\left(S_{ab}|\vartheta \right)$: 
\begin{eqnarray}
\label{spinprawdop}
P\left(++|\vartheta\right) \; , \;\; P\left(--|\vartheta\right) \; , \;\; 
P\left(+-|\vartheta\right) \; , \;\; P\left(-+|\vartheta\right) \; .
\end{eqnarray}
{\bf Warunek normalizacji  prawdopodobieństwa} $P\left(S_{a}S_{b}|\vartheta\right)$  w eksperymencie EPR-Bohm'a można, ze względu na wykluczanie się różnych zdarzeń $S_{ab}$, (\ref{4 zdarzenia EPR}), zapisać następująco: 
\begin{eqnarray}
\label{normalizacja P daje wsp w qab}
P\left(\bigcup\limits _{ab}{S_{a}S_{b}|\vartheta}\right)} = \sum\limits_{ab}{P\left(S_{a}S_{b}|\vartheta\right) = 1 \;  \;\;\;\;\; {\rm dla \;\; ka\dot{z}dego} \;\; \vartheta \in \langle 0, 2\pi)\; .
\end{eqnarray}
W analizie estymacyjnej postulujemy, że prawdopodobieństwo $P\left(S_{ab}\left|\vartheta \right.\right)\left|\begin{array}{l}
_{\!\!\! \vartheta = \hat{\vartheta}}
\end{array}\right.$ jest funkcją estymatora $\hat{\vartheta}$ parametru $\vartheta$.  \\
\\
{\bf Metoda estymacji}: Korzystając z proporcjonalności  prawdopodobieństw(\ref{spinprawdop}) do obserwowanej liczebności zdarzeń w detektorze, oszacowuje się wartość kąta $\vartheta$. Jednak, aby to uczynić, trzeba mieć model analitycznych  formuł na prawdopodobieństwa  (\ref{spinprawdop}). Poniżej wyprowadzimy je metodą EFI. \\
  \\
{\bf Sformułowanie warunków brzegowych}. \\
\\
(1) Ponieważ zdarzenia (\ref{4 zdarzenia EPR}) wykluczają się wzajemnie i rozpinają całą przestrzeń zdarzeń, więc pierwszym warunkiem brzegowym jest {\bf warunek normalizacji}: 
\begin{eqnarray}
\label{normalizacja prawd dla EPR}
\sum_{ab}{P\left(S_{ab}\right)} = P\left(S_{++}\right)+P\left(S_{+-}\right)+P\left(S_{-+}\right)+P\left(S_{--}\right) = 1  \; ,
\end{eqnarray}
spełniony niezależnie od wartości kąta $\vartheta$.\\
\\
Dla każdego zdarzenia $ab$ prawdopodobieństwa $P\left(S_{ab}\right)$ oraz $P\left(S_{ab}|\vartheta\right)$ są ze sobą związane następująco:
\begin{eqnarray}
\label{srednia z  PSab theta}
P\left(S_{ab}\right) = \int\limits_{0}^{2\pi} {P\left(S_{ab}|\vartheta\right) r\left(\vartheta\right) d\vartheta} \; ,
\end{eqnarray}
gdzie uśrednienie nastąpiło z tzw. {\it funkcją  niewiedzy} $r(\vartheta)$. 
Ze względu na warunki unormowania (\ref{normalizacja P daje wsp w qab}) oraz (\ref{normalizacja prawd dla EPR}) 
otrzymujemy jej możliwą postać:
\begin{eqnarray}
\label{prawdkat}
r(\vartheta)=\frac{1}{2\pi} \; ,\quad\quad0\le \vartheta < 2 \pi \; ,
\end{eqnarray}
która oznacza, że w zakresie aparaturowej zmienności ustawienia wartości kąta $\vartheta \in \left\langle 0, 2 \pi \right)$, z powodu naszej niewiedzy,  jest możliwa w równym stopniu każda jego wartość. \\
\\
{\bf Uwaga o wartości kąta $\vartheta$}: Wartość $\vartheta$ jest związana z ustawieniem aparatury pomiarowej Sterna-Gerlacha ,,$a$'' i ,,$b$'', i trudno ją (na serio) traktować jako wielkość posiadającą rozproszenie. Zasadnicza $\vartheta$ jest parametrem charakterystycznym dla przeprowadzonego eksperymentu. Niemniej przez niektórych $r(\vartheta)$ jest widziane jako  ,,prawdopodobieństwo'' znane a priori, co oznacza, że wyprowadzoną w ten sposób mechanikę kwantową należałoby  traktować jako statystyczną teorię Bayesowską \cite{stany koherentne}. \\
 \\
(2) Kolejne warunki wynikają z {\bf symetrii układu  i zasady zachowania cał\-kowitego spinu} przy czym w  eksperymencie {\it względny orbitalny moment pędu wynosi zero}. \\
\\
Rozważmy prosty przypadek $\vartheta=0$, gdy obie płaszczyzny, w których ustawione są urządzenia Sterna-Gerlacha
są tak samo zorientowane. Z warunku zachowania całkowitego spinu wynika:
\begin{eqnarray}
\label{vartheta zero}
P\left(++|0\right)=P\left(--|0\right)=0 \; ,\quad \vartheta=0
\end{eqnarray}
co oznacza, że w tym ustawieniu aparatury nigdy nie zobaczymy obu spinów jednocześnie skierowanych w górę czy w dół. Analogicznie,  jeśli kąt $\vartheta=\pi$, to warunek:
\begin{eqnarray}
\label{vartheta pi}
P\left(+-|\pi\right)=P\left(-+|\pi\right)=0,\quad \vartheta = \pi \; 
\end{eqnarray}
oznacza, że w tym przypadku nigdy nie zobaczymy spinów ustawionych
jeden w górę drugi w dół. W konsekwencji z zasady zachowania całkowitego spinu otrzymujemy, że jeśli $\vartheta=0$ lub $\vartheta=\pi$, to zawsze obserwacja jednego spinu daje nam całkowitą wiedzę o drugim. 
W tym przypadku stany spinów wyraźnie nie są niezależne, są stanami 
skorelowanymi. 
Wniosek ten jest intuicyjnie zawarty w sposobie ich przygotowania.\\
  \\
Następnie, ponieważ $P\left(S_{b}|\vartheta\right)$ jest prawdopodobieństwem brzegowym wystąpienia określonej wartości rzutu spinu cząstki 2, zatem nie zależy ono od $S_{a}$, czyli od  orientacji rzutu spinu cząstki 1, więc  nie zależy również od kąta $\vartheta$ pomiędzy wektorami $\vec{a}$ oraz $\vec{b}$. Mamy więc: 
\begin{eqnarray}
\label{Sb}
P\left(S_{b}|\vartheta\right) = C = const. 
\end{eqnarray}
Stąd 
\begin{eqnarray}
\label{warunek poczatkowy dla Sb}
\!\!\!\!\!\!\! P\left(S_{b}\right) = \int\limits_{0}^{2\pi} {P\left(S_{b}|\vartheta\right) r\left(\vartheta\right) d\vartheta} = C \int\limits_{0}^{2\pi} {r\left(\vartheta\right) d\vartheta} = C \quad \;\; {\rm dla} \;\;\; S_{b} = +,- \;\, , 
\end{eqnarray}
gdzie $r(\vartheta)$ jest określone w (\ref{prawdkat}).
 \\
  \\
Z warunku unormowania prawdopodobieństwa zdarzenia pewnego, mamy
\begin{eqnarray}
\label{cpol}
P\left(S_{b}=+\right)+P\left(S_{b} = - \right) = C + C = 2C = 1 \;\;\; {\rm  czyli} \;\;\; C = \frac{1}{2} \; .
\end{eqnarray}
Zatem z ({\ref{Sb}) otrzymujemy, 
że\footnote{Podobnie dla $S_{a}$, wychodząc z $P\left(S_{a}|\vartheta\right) = C' = const. $ i postępując analogicznie jak przy przejściu od (\ref{Sb}) do (\ref{cpol}), otrzymujemy:
\begin{eqnarray}
\label{pol dla Sa}
P\left(S_{a}|\vartheta\right) = \frac{1}{2} \; .
\end{eqnarray}
}
: 
\begin{eqnarray}
\label{pol}
P\left(S_{b}|\vartheta\right) = \frac{1}{2} \; ,
\end{eqnarray}
natomiast z (\ref{warunek poczatkowy dla Sb}):
\begin{eqnarray}
\label{Sb wycalkowane}
P\left(S_{b}\right) = \frac{1}{2} \; .
\end{eqnarray}
\\
Inną ważną własnością symetrii przestrzennej eksperymentu jest  brak preferencji występowania spinu skierowanego w górę czy w dół, tzn: \begin{eqnarray}
\label{symetriaEPR}
P\left(S_{+-}|\vartheta\right)=P\left(S_{- +}|\vartheta\right)\quad {\rm oraz} \quad P\left(S_{++}|\vartheta\right)=P\left(S_{--}|\vartheta\right) \; ,
\end{eqnarray}
co oznacza, że jeśli obserwowalibyśmy eksperyment dla układu odwróconego względem osi $x$ o kąt $\pi$, to statystyczny wynik byłby dokładnie taki sam. Z (\ref{symetriaEPR}) otrzymujemy: \begin{eqnarray}
\label{++ = - -}
P\left(S_{++}\right)=\int{P\left(S_{++}|\vartheta\right)r\left(\vartheta\right)d\vartheta}=\int{P\left(S_{--}|\vartheta\right)r\left(\vartheta\right)d\vartheta}=P\left(S_{--}\right)
\end{eqnarray}
oraz w sposób analogiczny:
\begin{eqnarray}
P\left(S_{+-}\right)=P\left(S_{-+}\right)\label{22} \; .
\end{eqnarray} 
Dodatkowo otrzymujemy: 
\begin{eqnarray}
\label{doda}
P\left(S_{+-}\right) &=& \int{P\left(S_{+-}|\vartheta\right) r\left(\vartheta\right)d\vartheta} = \int{P\left(S_{+-}|\vartheta+\pi\right) r\left(\vartheta+\pi\right)d\vartheta}  \nonumber \\
& = & \int{P\left(S_{++}|\vartheta\right) r\left(\vartheta+\pi\right)
d\vartheta} = P\left(S_{++}\right) \; , 
\end{eqnarray}
gdzie w drugiej równości skorzystano ze zwykłej zamiany zmiennych
$\vartheta \to \vartheta + \pi$, w trzeciej równości z tego, że $P\left(S_{+-}|\vartheta+\pi\right) = P\left(S_{++}|\vartheta\right)$, 
a w ostatniej z $r(\vartheta)=1/{(2\pi)} = r(\vartheta+\pi)$.\\ 
\\
Jako konsekwencja (\ref{++ = - -})-(\ref{doda}) otrzymujemy: 
\begin{eqnarray}
P\left(S_{-+}\right) = P\left(S_{--}\right) \; .
\label{doda2}
\end{eqnarray}
Ze względu na (\ref{++ = - -})-(\ref{doda2}) oraz (\ref{normalizacja prawd dla EPR}) otrzymujemy: 
\begin{eqnarray}
P\left(S_{ab}\right) = \frac{1}{4} \; .
\label{laczneEPR}
\end{eqnarray}
W końcu wzór Bayes'a na {\it prawdopodobieństwo warunkowe} daje: 
\begin{eqnarray}
\label{warunkSab}
P\left(S_{a}|S_{b}\right)=\frac{P\left(S_{ab}\right)}{P\left(S_{b}\right)} = \frac{1/4}{1/2} = \frac{1}{2}\quad {\rm dla\; ka\dot{z}dego }\;\; S_{a}, S_{b} \; .
\end{eqnarray}
\\
{\bf Końcowe uwagi o warunkach brzegowych}: Zauważmy, że jak dotąd w wyprowadzeniu zależności (\ref{vartheta zero})
do (\ref{warunkSab}) nie skorzystano z EFI. Są to bowiem  warunki brzegowe dla równań metody EFI. Zależności te wynikały
z początkowej obserwacji istnienia dla cząstki o spinie połówkowym dokładnie dwóch możliwych rzutów spinu na dowolny kierunek (por. Uwaga na początku rozdziału), zasady zachowania całkowitego momentu pędu oraz z symetrii układu. Przy wyborze rozwiązania równania generującego rozkład, skorzystamy jeszcze z warunku geometrycznej symetrii układu przy obrocie o kąt  $\vartheta = 2 \pi$.\\

\subsection{Pojemność informacyjna dla zagadnienia EPR-Bohm'a}

\label{Pojemnosc informacyjna zagadnienia EPR}

\vspace{3mm}

Poniżej wyprowadzimy wyrażenia dla prawdopodobieństw (\ref{spinprawdop}) jako wynik  estymacji rozkładów metodą  EFI. \\
\\
{\bf Wielkość mierzona przez obserwatora}: Obserwator zewnętrzny mierzy jedynie wartość rzutu spinu  jednej cząstki \textit{powiedzmy,
że cząstki} 1, to znaczy $S_{a}$,  natomiast nie mierzy wartości rzutu spinu drugiej cząstki
$S_{b}$, jest ona traktowana jako wielkość nieznana, ale ustalona.
Również wartość $S_{b}$ nie jest estymowana z obserwacji $S_{a}$.\\
\\
{\bf Określenie przestrzeni położeń dla EFI}: Podobnie jak  poprzednio układ sam próbkuje swoimi Fisherowskimi  kinetycznymi stopniami swobody dostępną mu przestrzeń położeń. Tym razem jest to jednowymiarowa przestrzeń kąta $\vartheta$. \\
%
%
Na podstawie danych pomiarowych 
wartości rzutu spinu $S_{a}$, esty\-mujemy metodą EFI kąt $\vartheta$
pomiędzy dwoma płaszczyznami zaznaczonymi na rysunku \ref{fig:5}.
Kąt ten jest więc traktowany jako nieznany parametr i jest estymowany z obserwacji $S_{a}$. \\
 \\
{\bf Określenie funkcji wiarygodności i przestrzeni próby}:  Ponieważ dokonujemy pomiaru $S_{a}$, a $S_{b}$ oraz $\vartheta$ są 
nieznanymi, ale ustalonymi wielkościami, 
zatem funkcję wiarygodności dla powyżej postawionego problemu zapiszemy następująco: 
\begin{eqnarray}
\label{finkcja wiaryg dla vartheta}
P\left(S_{a}|S_{b}, \vartheta\right) \; \rightarrow \;\;\;\; {\rm funkcja \;\; wiarygodno\acute{s}ci \;\; pr\acute{o}by \;\; dla} \;\;  \vartheta \, .
\end{eqnarray}
Jej postaci szukamy, wykorzystując metodę EFI. Jednak postać informacji  Fishera, która byłaby miarą precyzji estymacji kąta $\vartheta$ opartej o pomiar $S_{a}$, wynikać będzie z jej pierwotnej postaci (\ref{postac I koncowa w pn}) dla wymiaru próby $N=1$.  \\
\\
Skoro więc $\vartheta$ jest estymowanym parametrem, zatem  pełni on teraz taką rolę jak parametrem $\theta$ w wzorze (\ref{postac I koncowa w pn}), w którym,  obok podstawienia  $\theta\to \vartheta$, należy dokonać następujących podstawień: 
\begin{eqnarray}
\label{N oraz wartosci y w funkcji wiaryg dla EPR}
n = N = 1 \;  \;\;\; {\rm oraz} \;\;\; \int{dy^{1}} \to  \sum\limits_{a} \; ,
\end{eqnarray}
gdzie przestrzenią (jednowymiarowej) próby jest zbiór 
$\left\{-1,+1\right\}$ wartości zmiennej losowej $S_{a}$.\\
\\
{\bf Amplitudy prawdopodobieństwa} $q_{ab}$ są zdefiniowane jak zwykle następująco:                                    
\begin{eqnarray}
\label{inffishEPR2}
P\left(S_{a}\left|{S_{b},\vartheta}\right.\right) = q_{ab}^{2}\left(\vartheta\right) \; .
\end{eqnarray}
\\
{\bf Oczekiwana IF parametru $\vartheta$. Pojemność informacyjna kanału $(\vartheta, S_{b})$}  jest równa:
\begin{eqnarray}
\!\!\!\!\!\!\!\! I_{b} &= & \sum\limits_{a=-}^{+}{\frac{1}{P\left(S_{a}\left|{S_{b}, \vartheta} \right.\right)}\left(\frac{\partial P \left(S_{a}\left|{S_{b}, \vartheta}\right.\right)}{\partial \vartheta}\right)^{2}} = \sum\limits_{a=-}^{+}{\frac{1}{q_{ab}^{2}} \left(\frac{\partial q_{ab}^{2}}{\partial \vartheta}\right)^{2}} = \sum\limits_{a=-}^{+}{\frac{1}{q_{ab}^{2}}\left(2q_{ab}\frac{\partial q_{ab}}{\partial \vartheta}\right)^{2}}  \;\;\;\; \nonumber \\
\!\!\!\!\!\!\!\!  &=& 4\sum\limits_{a=-}^{+}{\left(\frac{\partial q_{ab}}{\partial \vartheta}\right)^{2}} \;  \;\;\;\;
\end{eqnarray}
i podsumowując:
\begin{eqnarray}
\label{inffishEPR}
I_{b} \equiv I\left({S_{b},\vartheta}\right) = 4 \sum\limits_{a=-}^{+}{q_{ab}^{'2}\left(\vartheta\right)} \; ,\quad S_{b} = \left({+,-}\right) \; , \;\;\; {\rm gdzie} \;\;\; q_{ab}^{'} \equiv \frac{dq_{ab}}{d\vartheta} \, ,
\end{eqnarray}
gdzie sumowanie po $a$ odpowiada całkowaniu w informacji Fishera po przestrzeni bazowej. \\
Zwrócmy uwagę, że $\vartheta$ w powyższej formule  na $I_{b} \equiv I\left({S_{b},\vartheta}\right)$ jest wyznaczone dla konkretnej wartości $\vartheta$ i konkretnej wartości $S_{b}$ rzutu spinu cząstki 2.  Zatem jest to pojemność informacyjna jednego kanału $(\vartheta, S_{b})$ czyli {\it informacja Fishera parametru} $\vartheta$ w tym kanale:
\begin{eqnarray}
\label{IF dla vartheta w EPR}
I_{F}(\vartheta) = I_{b} \, . 
\end{eqnarray}
Powróćmy do warunków (\ref{N oraz wartosci y w funkcji wiaryg dla EPR}). Pierwszy z nich  oznacza, że ranga amplitudy pola jest równa $N=1$, drugi oznacza, że {\it przestrzeń bazowa to teraz dwupunktowy zbiór wartości rzutu spinu cząstki} 1 {\it na kierunek '$a$'.} W poniższych uwagach odniesiemy się do tych elementów analizy. \\
  \\
{\bf Uwaga o sumowaniu po $a$ w przestrzeni próby}: Jako, że $\vartheta$ jest parametrem, a pomiary w próbie są związane z obserwacjami
$S_{a}$, zatem zapis w (\ref{inffishEPR}) wymaga pewnego wyjaśnienia.
Wyrażenie (\ref{inffishEPR}) reprezentuje dwa równania dla dwóch
możliwych wartości $S_{b}$. Ponieważ zmienną w pomiarze jest rzut
spinu cząstki 1, zatem sumowanie przebiega po dwóch możliwych wartościach
spinu $S_{a}=+,-.$ 
Tak więc, zapis $I\left({S_{b},\vartheta}\right)$
wskazuje na {\it informację zawartą w przestrzeni próby zmiennej}  $S_{a}$ dla cząstki $1$ na temat nieznanego kąta $\vartheta$ w obecności pewnej nieznanej, lecz ustalonej
wartości rzutu spinu $S_{b}$ cząstki $2$. \\
\\
{\bf Całkowita pojemność informacyjna $I_{1_{a}}$ dla parametru $\vartheta$}: Ponieważ kąt $\vartheta$ jest
parametrem, którego wartość może się zmieniać w sposób ciągły w przedziale
$\langle0,2\pi)$, stąd zgodnie z (\ref{inffishEPR}) mamy nieskończoną 
liczbę kanałów 
informacji Fishera związanych z wartościami $\vartheta$. Dla każdego z tej nieskończonej liczby  kanałów są jeszcze dwa kanały informacyjne związane z możliwymi wartościami dla $S_{b}$. Aby poradzić sobie z taką sytuacją metoda EFI wykorzystuje 
pojedynczą, skalarną informację (oznaczmy ją $I_{1_a}$), nazywaną pojemnością informacyjną, wprowadzoną w Rozdziale~\ref{Pojecie kanalu informacyjnego}. Wielkość tą konstruujemy dokonując
sumowania informacji po wszystkich możliwych kanałach. 
Zatem,  suma przebiegać będzie po wszystkich wartościach kąta $\vartheta_{k}$ przy nieznanym rzucie spinu $S_{b}$ cząstki 2.\\ 
Tak więc, po pierwsze,  pojemność informacyjna $I_{n k}$ dla jednego kanału, wprowadzona po raz pierwszy w (\ref{krzy4}), przyjmuje postać: 
\begin{eqnarray}
\label{pojemnosc jednego kanalu w EPR}
I_{b k} = I \left(S_{b},\vartheta_{k}\right) \; .
\end{eqnarray}
Po drugie, w wyniku sumowania po wszystkich możliwych  wartościach kąta $\vartheta_{k}$ oraz rzutach spinu $S_{b}$ cząstki 2,  otrzymujemy {\it całkowitą pojemność informacyjną dla parametru} $\vartheta$: 
\begin{eqnarray}
\label{pojemnoscEPR}
& & I_{1_{a}} \equiv \sum\limits_{b=-}^{+} \sum\limits_{k}{I\left({S_{b},\vartheta_{k}}\right)} \rightarrow \nonumber \\ 
& & \rightarrow I_{1_{a}} = \sum\limits_{b=-}^{+}{\int\limits_{0}^{2\pi} d\vartheta\;I\left({S_{b},\vartheta}\right)} = 4 \sum\limits_{b=-}^{+}\sum\limits_{a=-}^{+} {\int\limits _{0}^{2\pi}{d\vartheta}\; q_{ab}^{'2}\left(\vartheta\right)} \equiv 4\sum\limits _{ab}{\int\limits _{0}^{2\pi}{d\vartheta}\; q_{ab}^{'2}\left(\vartheta\right)} \; ,
\end{eqnarray}
gdzie całkowanie pojawia się z powodu zastąpienia sumy dla dyskretnego indeksu $k$ całkowaniem po ciągłym zbiorze wartości parametru $\vartheta$. 
Oznacza to, że po zrozumieniu czym jest pojedynczy $k$ - ty kanał związany z $\vartheta$, wykonujemy w celu otrzymania całkowitej pojemności informacyjnej  całkowanie, 
które przebiega zgodnie z (\ref{prawdkat}) od $0$ do $2\pi$. W
ostatniej linii wykorzystano (\ref{inffishEPR}). \\
Sumowanie w (\ref{pojemnoscEPR})
po $ab$ przebiega po wszystkich możliwych kombinacjach łącznej przestrzeni stanów $S_{ab}$ określonej w (\ref{4 zdarzenia EPR}). Indeks '$1_a$' w $I_{1_a}$ oznacza, że jest to pojemność informacyjna jednej cząstki, przy czym wyróżniliśmy cząstkę 1, dla której pomiar dokonywany jest w analizatorze $a$. {\bf Pojemność informacyjna $I_{1_a}$ wchodzi do estymacyjnej procedury} EFI.\\
\\
{\bf Uwaga o randze $N=1$}. Zachodzi pytanie: Jeśli okaże się, że tradycyjne formuły mechaniki kwantowej dla eksperymentu EPR-Bohm'a pojawią się dla powyżej określonej  estymacji z $N=1$ (a tak się istotnie stanie), to czyżby pomiar odziaływania cząstki z aparaturą Sterna-Gerlach'a miał mieć termodynamiczny charakter  oddziaływania małego układu  z termostatem, w zrozumieniu podanym na samym początku Rozdziału~\ref{Warunki brzegowe}? W końcu, jak wiemy z poprzednich rozdziałów, rozwiązania z $N=1$ odnoszą się do zjawisk termodynamicznych. 



\newpage

\subsection{Informacja strukturalna. Amplituda prawdopodobieństwa}

\label{Informacja strukturalna EPR}

Biorąc pod uwagę ogólną postać informacji strukturalnej (\ref{Q dla niezaleznych Yn w d4y}) oraz uwagi Rozdziału~\ref{Pojemnosc informacyjna zagadnienia EPR} odnośnie konstrukcji postaci $I_{1_a}$, (\ref{pojemnoscEPR}), zauważamy, że  {\it informacja strukturalna}  $Q_{1_a}$ dla obserwowanej 
cząstki ma postać: 
\begin{eqnarray}
\label{strukturalnaEPR}
Q_{1_a} \equiv \sum\limits_{ab}{\int\limits_{0}^{2\pi}{d\vartheta}\;  q_{ab}^{2}(\vartheta) \, \texttt{q\!F}_{ab}(q_{ab})} \; .
\end{eqnarray}
Poniższa analiza związana z wyprowadzeniem amplitud $q_{ab}$ jest 
podobna do analizy dla rozkładu Boltzmanna w Rozdziale \ref{fizykastatystyczna}. \\
\\
{\bf Wariacyjna zasada informacyjna}: Dla  pojemności informacyjnej $I$, (\ref{pojemnoscEPR}), oraz informacji strukturalnej $Q$, (\ref{strukturalnaEPR}) wariacyjna zasada informacyjna ma postać: 
\begin{eqnarray}
\label{zskalarnaEPR}
\delta_{(q_{ab})} K \equiv \delta_{(q_{ab})}\left( I_{1_a} + Q_{1_a}\right) = \delta_{(q_{ab})} \left(\;\int\limits _{0}^{2\pi} {d\vartheta \, k} \right)  = 0 \; ,
\end{eqnarray}
gdzie zgodnie z ogólną postacią (\ref{k form}), gęstość całkowitej informacji fizycznej $k$  dla amplitud $q_{ab}$ jest równa: 
\begin{eqnarray}
\label{k EPR}
k = 4 \sum\limits_{ab}{\left(q_{ab}^{'2} + \frac{1}{4} q_{ab}^{2}(\vartheta)  \texttt{q\!F}_{ab}(q_{ab})\right)} \; .
\end{eqnarray}
Rozwiązaniem problemu {\it wariacyjnego} (\ref{zskalarnaEPR}) względem $q_{ab}$ jest {\it równanie Eulera-Lagrange'a} (\ref{EL eq})  (por. (\ref{euler})): 
\begin{eqnarray}
\label{row E-L dla EPR}
\frac{d}{d\vartheta}\left(\frac{\partial k}{\partial q_{ab}^{'}(\vartheta)}\right)=\frac{\partial k}{\partial q_{ab}} \; .
\end{eqnarray}
Z równania (\ref{row E-L dla EPR}) dla $k$ jak w (\ref{k EPR}) otrzymujemy, dla każdego dwu\-cząstkowego łącznego stanu spinowego $S_{ab}$: 
\begin{eqnarray}
\label{rozweularaEPR}
q_{ab}^{''}=\frac{1}{2}\frac{{d (\frac{1}{4} q_{ab}^{2}  \texttt{q\!F}_{ab}(q_{ab}))}}{{dq_{ab}}} \; .
\end{eqnarray}
Ponieważ $q_{ab}^{2}(\vartheta) \texttt{q\!F}_{ab}(q_{ab})$ jest jawnie jedynie funkcją $q_{ab}$, więc różniczka zupełna zastąpiła pochodną cząstkową po $q_{ab}$ występującą w (\ref{row E-L dla EPR}).\\
\\
{\bf Zmodyfikowana obserwowana zasada strukturalna}: 
Po wycałkowaniu (\ref{pojemnoscEPR}) przez części,  pojemność $I$ wynosi (por. (\ref{postac I po calk czesci}), (\ref{Cn}), (\ref{Cn tilde})): 
\begin{eqnarray}
\label{IEPR}
I_{1_a} = 4\sum\limits _{ab}{\int\limits_{0}^{2\pi} {d\vartheta\left({\tilde{C}_{ab}-q_{ab}q_{ab}^{''}}\right)}} \; ,
\end{eqnarray}
gdzie 
\begin{eqnarray}
\label{stalaCEPR}
\tilde{C}_{ab} = \frac{1}{{2\pi}}\left({q_{ab}\left({2\pi}\right)q_{ab}^{'} \left({2\pi}\right)-q_{ab}\left(0\right)q_{ab}^{'}\left(0\right)}\right) \; .
\end{eqnarray}
Z powodu braku dodatkowych więzów $\kappa =1$, zatem {\it zmodyfikowana obserwowana zasada strukturalna}  $\widetilde{\textit{i'}} + \widetilde{\mathbf{C}}_{EPR} +  \textit{q} = 0$, (\ref{zmodyfikowana obserwowana zas strukt z P i z kappa}), 
jest ze względu na (\ref{IEPR}) oraz (\ref{strukturalnaEPR}), następująca: 
\begin{eqnarray}
\label{u}
4\sum\limits_{ab} \left(- q_{ab}q_{ab}^{''} + \tilde{C}_{ab} + \frac{1}{4} q_{ab}^{2} \texttt{q\!F}_{ab}(q_{ab}) \right) = 0  \; .
\end{eqnarray}
Poniżej przekonamy się, że $\widetilde{\mathbf{C}}_{EPR} = 4 \sum \tilde{C}_{ab} = 0$.  \\
\\
Dla każdego dwu\-cząstkowego łącznego stanu spinowego $S_{ab}$, obserwowana zasada strukturalna (\ref{u}) ma więc postać (por. \ref{rownanie strukt E}): 
\begin{eqnarray}
\label{mikroEPR}
 - q_{ab}q_{ab}^{''} + \tilde{C}_{ab} + \frac{1}{4} q_{ab}^{2} \texttt{q\!F}_{ab}(q_{ab}) = 0 \; .
\end{eqnarray}
Wraz z równaniem Eulera-Lagrange'a (\ref{rozweularaEPR}), 
równanie (\ref{mikroEPR}) posłuży do wyprowadzenia równania generującego rozkład.  \\
\\
{\bf Wyprowadzenia równania generującego}: Wykorzystując w  (\ref{mikroEPR}) związek (\ref{rozweularaEPR}), otrzymujemy (por.  (\ref{row rozn z q oraz qF})) dla każdego dwu\-cząstkowego łącznego stanu spinowego $S_{ab}$: 
\begin{eqnarray}
\frac{1}{2}q_{ab}\frac{{d(\frac{1}{4} q_{ab}^{2} \texttt{q\!F}_{ab}(q_{ab}))}}{{dq_{ab}}} = \frac{1}{4} q_{ab}^{2} \texttt{q\!F}_{ab}(q_{ab}) + \tilde{C}_{ab} \; .
\end{eqnarray}
Zapiszmy powyższe równanie w wygodniejszej formie: 
\begin{eqnarray}
\frac{{2dq_{ab}}}{{q_{ab}}}=\frac{{d \left(\frac{1}{4} q_{ab}^{2}  \texttt{q\!F}_{ab}(q_{ab})\right)}}{{\frac{1}{4} q_{ab}^{2} \texttt{q\!F}_{ab}(q_{ab}) + \tilde{C}_{ab}}} \; ,
\end{eqnarray}
z której po obustronnym wycałkowaniu otrzymujemy: 
\begin{eqnarray}
\label{jEPR}
\frac{1}{4} q_{ab}^{2}(\vartheta) \texttt{q\!F}_{ab}(q_{ab}) = \frac{{q_{ab}^{2}(\vartheta)}}{{A_{ab}^{2}}} - \tilde{C}_{ab} \; ,
\end{eqnarray}
gdzie $A_{ab}^{2}$ jest  w ogólności zespoloną stałą. \\
\\
{\bf Równanie generujące}: Podstawiając (\ref{jEPR}) do (\ref{rozweularaEPR}) otrzymujemy szukane  różniczkowe równanie generujące dla amplitud $q_{ab}$: 
\begin{eqnarray}
\label{row generujace dla amplitud w EPR}
q_{ab}^{''}(\vartheta) = \frac{{q_{ab}(\vartheta)}}{{A_{ab}^{2}}} \; ,
\end{eqnarray}
będące konsekwencją obu zasad informacyjnych, 
strukturalnej i wariacyjnej. \\
\\
{\bf Rozwiązanie równania generującego}: Ponieważ amplituda $q_{ab}$ jest rzeczywista,  
więc $A_{ab}^{2}$ też musi być rzeczywiste. 
Podobnie jak w rozdziale \ref{rozdz.energia},  ponieważ stała $A_{ab}^{2}$ jest rzeczywista, więc można ją przedstawić za pomocą innej rzeczywistej stałej $a_{ab}$ 
jako $A_{ab}=a_{ab}$ lub $A_{ab} = i \, a_{ab}$. 
Zatem istnieją dwie klasy
rozwiązań równania (\ref{row generujace dla amplitud w EPR}). \\
\\
Dla $A_{ab} = a_{ab}$, rozwiązanie (\ref{row generujace dla amplitud w EPR}) ma charakter czysto {\it eksponencjalny}: 
\begin{eqnarray}
q_{ab}(\vartheta) = B_{ab}^{''}\exp\left(-\frac{\vartheta}{a_{ab}}\right) + C_{ab}^{''}\exp\left(\frac{\vartheta}{a_{ab}}\right) \; ,\quad A_{ab}=a_{ab} \; ,
\end{eqnarray}
gdzie stałe $B_{ab}^{''}$ oraz $C_{ab}^{''}$ są rzeczywiste. \\
\\
Natomiast dla
$A_{ab} = i\, a_{ab}$, rozwiązanie (\ref{row generujace dla amplitud w EPR}) ma charakter {\it trygonometryczny}:
\begin{eqnarray}
\label{qabEPR}
q_{ab}(\vartheta) = B_{ab}^{'}\sin\left(\frac{\vartheta}{a_{ab}}\right) + C_{ab}^{'}\cos\left(\frac{\vartheta}{a_{ab}}\right)\; ,\quad A_{ab} = i \, a_{ab} \; ,
\end{eqnarray}
gdzie $a_{ab}$, $B_{ab}^{'}$, $C_{ab}^{'}$ są rzeczywistymi stałymi.\\
 \\
{\bf Warunek niezmienniczości przy obrocie o $2 \pi$}: W rozważanym obecnie 
przypadku\footnote{
Przypomnijmy sobie analizę przeprowadzoną w rozdziale (\ref{rozdz.energia}) dla rozkładu energii cząsteczki gazu. 
Wtedy ze względu na nieograniczony zakres argumentu amplitudy,   wybraliśmy rozwiązanie o charakterze eksponencjalnym.
} 
wartości $\vartheta$ są kątem z ograniczonego zbioru $\langle0, 2\pi)$. Zatem z geometrycznej symetrii układu przy obrocie o kąt $2\pi$ wynika, że rozkład $P\left(S_{a}|S_{b},\vartheta\right)$
jest również funkcją okresową zmiennej $\vartheta$. Ze względu na (\ref{inffishEPR2}) 
warunek ten oznacza, że funkcja $q_{ab}(\vartheta)$ powinna być okresowa, \textit{zatem wybieramy rozwiązanie o charakterze} {\bf trygonometrycznym}, przy czym 
Funkcje sin oraz cos w (\ref{qabEPR}) są funkcjami bazowymi tworzącymi amplitudę prawdopodobieństwa $q_{ab}(\vartheta)$.\\
  \\
{\bf Funkcje bazowe} na przestrzeni amplitud powinny być ortogonalne. Z  postaci $q_{a}$ w (\ref{qabEPR}) wynika, że funkcjami bazowymi są funkcje  'sin' oraz 'cos'. Ponieważ $q_{a}$ jest  określone na przestrzeni parametru $\vartheta \in \left\langle 0, 2 \pi\right)$, zatem  {\it warunek ortogonalności funkcji bazowych} na tej przestrzeni: 
\begin{eqnarray}
\int\limits _{0}^{2\pi}{d\vartheta\sin\left({\vartheta\mathord{\left/{\vphantom{\vartheta{a_{ab}}}}\right.\kern -\nulldelimiterspace}{a_{ab}}}\right)}\cos\left({\vartheta\mathord{\left/{\vphantom{\vartheta{a_{ab}}}}\right.\kern -\nulldelimiterspace}{a_{ab}}}\right)=0\label{ortogonalEPR}\end{eqnarray}
daje po wycałkowaniu postać stałych $a_{ab}$: 
\begin{eqnarray}
\label{warAEPR}
a_{ab}=\frac{2}{{n_{ab}}} \; ,\quad\quad {\rm gdzie} \;\;\; n_{ab} = 1,2,... \;\; .
\end{eqnarray}
\\
{\bf Warunek minimalizacji pojemności $I$}: Warunek (\ref{warAEPR}) jest warunkiem na dopuszczalne wartości
$a_{ab} = -i A_{ab}$ (por. (\ref{qabEPR})), ale jak widać nie wskazuje on na żadną z nich jednoznacznie. 
Może to nastąpić, gdy ustalimy wartość $n_{ab}$, co uczynimy ograniczając rozważania do  zapostulowanego w Rozdziale~\ref{Poj inform zmiennej los poloz}  warunku minimalizacji informacji kinetycznej $I \rightarrow min$. \\
\\
Zacznijmy od wyznaczenia stałej $\tilde{C}_{ab}$, (\ref{stalaCEPR}). Po  wstawieniu do  (\ref{stalaCEPR}) amplitudy (\ref{qabEPR}) z (\ref{warAEPR}),  otrzymujemy:
\begin{eqnarray}
\label{postac stalej Cab}
\!\!\!\!\!\!\!\!\!\!\!\! \tilde{C}_{ab} \!\! &=& \!\! \frac{n_{ab}}{4 \pi}  \, \sin (\pi  \, n_{ab})  \\
&\cdot& \!\! \left((B_{ab}^{'}-C_{ab}^{'})
(B_{ab}^{'}+C_{ab}^{'}) \cos (\pi \, n_{ab}) - 2 B_{ab}^{'} C_{ab}^{'} \sin (\pi \, n_{ab}) \right) = 0 \, 
, \;\;\; {\rm gdzie} \;\;\; 
n_{ab} = 1,2,... \;\; , \;\; \nonumber
\end{eqnarray} 
skąd po skorzystaniu z $I_{1_a} = 4\sum\limits{\int {d\vartheta({\tilde{C}_{ab}-q_{ab}q_{ab}^{''}})}}$, (\ref{IEPR}), oraz równania generującego  (\ref{row generujace dla amplitud w EPR}) pozwalającego wyeliminować  $q_{ab}^{''}$, otrzymujemy użyteczną postać pojemności informacyjnej: 
\begin{eqnarray}
\label{postac Iab w Aab i qab}
I_{1_{a}} = - 4\sum\limits _{ab}{\frac{1}{{A_{ab}^{2}}}\int\limits _{0}^{2\pi}{d\vartheta\; q_{ab}^{2}(\vartheta)}} \; .
\end{eqnarray}
Należy jeszcze wyznaczyć całkę w (\ref{postac Iab w Aab i qab}):
\begin{eqnarray}
\label{piEPR}
\int\limits _{0}^{2\pi}{d\vartheta\; q_{ab}^{2}\left(\vartheta\right)} \!\! &\equiv& \!\!
\int\limits_{0}^{2\pi} {d\vartheta\; P \left({S_{a}\left|{S_{b}, \vartheta} \right.} \right)} 
= \int\limits_{0}^{2\pi}{d\vartheta\frac{P\left(S_{a},S_{b},\vartheta\right)}{p\left(\vartheta,S_{b}\right)}} = \int\limits_{0}^{2\pi} {d\vartheta \frac{{p\left({\vartheta,S_{a}\left|{S_{b}}\right.}\right)P\left({S_{b}}\right)}}{{p\left({\vartheta,S_{b}}\right)}}} \nonumber \\
&=& \int{d\vartheta\frac{{p\left({\vartheta,S_{a}\left|{S_{b}}\right.}\right)\frac{1}{2}}}{{p\left({S_{b}\left|\vartheta\right.}\right)r\left(\vartheta\right)}}}=\int\limits _{0}^{2\pi}{d\vartheta\frac{{p\left({\vartheta,S_{a}\left|{S_{b}}\right.}\right)\frac{1}{2}}}{{\frac{1}{2}\frac{1}{{2\pi}}}}} \nonumber \\ &=& 2\pi\int\limits _{0}^{2\pi}{d\vartheta\; p\left({\vartheta,S_{a}\left|{S_{b}}\right.}\right)} 
= 2\pi P \left({S_{a}\left|{S_{b}}\right.}\right)=2\pi\frac{1}{2} = \pi \; ,
\end{eqnarray}
gdzie w pierwszej linii i) wpierw skorzystano z definicji amplitudy,  $q_{ab}^{2}\left(\vartheta\right) = P\left(S_{a}\left|{S_{b},\vartheta}\right.\right)$,  (\ref{inffishEPR2}), ii) następnie
z definicji prawdopodobieństwa warunkowego, iii) znowu z definicji prawdopodobieństwa warunkowego:
\begin{eqnarray}
\label{p od vartheta Sa warunek Sb}
p\left({\vartheta,S_{a}\left|{S_{b}}\right.}\right) = \frac{P\left(S_{a},S_{b},\vartheta\right)}{ P\left({S_{b}}\right)} \; ,
\end{eqnarray}
następnie w drugiej linii z $p\left(\vartheta, S_{b} \right) = p\left(S_{b}\left|\vartheta \right. \right) r(\vartheta)\,$ (por. Uwaga poniżej (\ref{prawdkat})) oraz z wyrażeń (\ref{prawdkat}), (\ref{pol}) oraz w trzeciej linii z:
\begin{eqnarray}
\label{calka z p od vartheta Sa warunek Sb}
P(S_{a}|S_{b}) = \int_{0}^{2 \pi} d \vartheta \, p\left({\vartheta,S_{a}\left|{S_{b}}\right.}\right)  \; ,
\end{eqnarray}
a na koniec z (\ref{warunkSab}). \\
Wykorzystując (\ref{postac Iab w Aab i qab}), (\ref{piEPR}) oraz  $A_{ab} = i \, a_{ab}$ i  (\ref{warAEPR}), możemy  wyrazić informację $I_{1_a}$ poprzez stałe $n_{ab}$, 
otrzymując: 
\begin{eqnarray}
\label{I1piEPR}
\!\!\! I_{1_a} = -Q_{1_a} = - 4\sum\limits _{ab}{\frac{1}{{A_{ab}^{2}}}\int\limits _{0}^{2\pi}{d\vartheta\; q_{ab}^{2}}} = 4\pi\sum\limits_{ab}{\frac{1}{{a_{ab}^{2}}}} = \pi\sum\limits_{ab}{n_{ab}^{2}} \; , \;\; {\rm gdzie} \;\; n_{ab} = 1,2,... \;\, , \;\;
\end{eqnarray}
przy czym drugą z powyższych równości, a zatem i pierwszą, otrzymano korzystając z postaci informacji strukturalnej  (\ref{strukturalnaEPR}), (\ref{jEPR})  oraz z (\ref{postac stalej Cab}).  Związek $Q_{1_a} = - I_{1_a}$ jest wyrazem ogólnego warunku (\ref{rownowaznosc strukt i zmodyfikowanego strukt}) spełnienia przez metodę EFI oczekiwanej zasady strukturalnej (\ref{ideal condition from K}).\\
\\
Z powyższego związku wynika, że warunek minimalizacji $I_{1_a}$  będzie spełniony dla:
\begin{eqnarray}
\label{n_ab minimal I}
n_{ab}=1 \; , \;\;\;\; I_{1_a} \rightarrow min \;\; , 
\end{eqnarray}
dla dowolnych $S_{a}$ oraz $S_{b}$. 
Warunek $n_{ab}=1$ zgodnie z (\ref{warAEPR}) odpowiada następującej wartości $a_{ab}$: 
\begin{eqnarray}
\label{ss}
a_{ab} = 2 \quad\quad {\rm dla\;\; dowolnych} \;\;\; S_{a}\;\; {\rm oraz} \;\; S_{b} \; .
\end{eqnarray}
Sumowanie w (\ref{I1piEPR}) przebiega po wszystkich $S_{a}$ i $S_{b}$, zatem  dla $n_{ab} = 1$, otrzymujemy wartość minimalną $I_{1_a}$: 
\begin{eqnarray}
\label{I EPR minimalna}
I_{1_a \,(min)} = \pi \sum\limits_{a=-}^{+} \sum\limits_{b=-}^{+} {n_{ab}^{2}} = 4 \pi \; , \;\; {\rm gdzie} \;\; n_{ab} = 1 \; . 
\end{eqnarray}
Tak więc otrzymaliśmy minimalną wartość pojemności informacyjnej dla parametru $\vartheta$.

{\bf Wyznaczenie stałych w amplitudzie}: W wyrażeniu (\ref{qabEPR}), które dla $a_{ab} = 2$ przyjmuje postać: 
\begin{eqnarray}
\label{qab dla a=2}
q_{ab}(\vartheta) = B_{ab}^{'}\sin\left(\frac{\vartheta}{2}\right) + C_{ab}^{'}\cos\left(\frac{\vartheta}{2}\right) \; , 
\end{eqnarray}
występują jeszcze stałe $B_{ab}^{'}$ oraz $C_{ab}^{'}$, które również musimy wyznaczyć. 
Wyznaczymy je korzystając z wcześniej ustalonych w (\ref{vartheta zero}), wartości łącznego prawdopodobieństwa warunkowego  $P\left(S_{ab}|\vartheta\right)$ 
dla  $\vartheta=0$. \\ 
Wyraźmy wpierw $P\left(S_{ab}|\vartheta\right)$ poprzez amplitudę $q_{ab}$:
\begin{eqnarray}
\label{PqEPR}
P\left(S_{ab}|\vartheta\right) \equiv P\left(S_{a}S_{b}|\vartheta\right) = \frac{P\left(S_{a}S_{b},\vartheta\right)}{r(\vartheta)} = P\left(S_{a}|S_{b},\vartheta\right)\frac{P\left(S_{b} \, , \vartheta\right)}{r(\vartheta)}\nonumber \\
=P\left(S_{a}|S_{b},\vartheta\right)P\left(S_{b}|\vartheta\right)=q_{ab}^{2}(\vartheta)P\left(S_{b}|\vartheta\right)=\frac{1}{2}\, q_{ab}^{2}(\vartheta) \; ,
\end{eqnarray}
gdzie skorzystano z (\ref{pol}). 
Z  równania (\ref{PqEPR}) wynika, że: 
\begin{eqnarray}
\label{q Sab zero}
q_{ab}(\vartheta) = 0 \; \;\; {\rm je\acute{s}li} \;\;\;  P\left(S_{ab}|\vartheta\right) = 0 \; .
\end{eqnarray}
Korzystając z (\ref{q Sab zero}) oraz   $P\left(++|0\right)=P\left(--|0\right)=0$, (\ref{vartheta zero}), i wstawiając (\ref{qab dla a=2}) dla $\vartheta=0$ do (\ref{PqEPR}), otrzymujemy:
\begin{eqnarray}
\label{ce}
C_{++}^{'}=C_{--}^{'}=0 \; .
\end{eqnarray}
Ponadto,  korzystając z symetrii geometrycznej eksperymentu, $P\left(S_{+-}|\vartheta\right)=P\left(S_{- +}|\vartheta\right)$ oraz $P\left(S_{++}|\vartheta\right)$ $=P\left(S_{--}|\vartheta\right)$,  zapisanej w (\ref{symetriaEPR}), otrzymujemy: 
\begin{eqnarray}
\label{wspol B prim dla q w EPR}
B_{++}^{'} = B_{--}^{'}  \;\;\;  {\rm oraz} \;\;\; B_{+-}^{'} = B_{-+}^{'} \; ,
\end{eqnarray}
oraz
\begin{eqnarray}
\label{C+- rowne C+- EPR}
C_{+-}^{'} = C_{-+}^{'}  \; .
\end{eqnarray}
{\bf Analiza z wykorzystaniem metryki Rao-Fishera}: 
Poniżej przekonamy się, że wyznaczenie pozostałych stałych $B_{ab}^{'}$ oraz $C_{ab}^{'}$ wymaga  dodatkowego założenia, odnoszącego się do postaci metryki Rao-Fishera na przestrzeni statystycznej ${\cal S}$, szukanego statystycznego modelu. \\
%
%
Szukanym rozkładem prawdopodobieństwa w eksperymencie EPR-Bohm'a jest  dyskretny rozkład  (\ref{spinprawdop})  określony  na  przestrzeni zdarzeń $S_{a}S_{b} \equiv S_{ab} \in \Omega_{ab}$, (\ref{4 zdarzenia EPR}), i unormowany zgodnie z (\ref{normalizacja P daje wsp w qab}) do jedności. Zatem zbiór możliwych wyników to: 
\begin{eqnarray}
\label{wyniki lacznego pomiaru Sa i Sb}
i \equiv ab = (++),(--),(+-),(-+) \;\; .
\end{eqnarray} 
Aplitudy prawdopodobieństwa związane z rozkładem (\ref{spinprawdop}) mają zgodnie z (\ref{PqEPR}) postać:
\begin{eqnarray}
\label{rozklad na przestrzeni statystycznej EPR}
\tilde{q}_{i} \equiv \frac{1}{\sqrt{2}}\, q_{ab}(\vartheta) =  \sqrt{P\left(S_{ab}|\vartheta\right)} \; .
\end{eqnarray}
Zgodnie z (\ref{qab dla a=2}) amplitudy $q_{ab}(\vartheta)$ mają następujące pochodne: 
\begin{eqnarray}
\label{pochodna qab dla EPR}
\frac{\partial q_{ab}}{\partial \vartheta} =  \frac{1}{2}\left(B_{ab}^{'}\cos\left(\frac{\vartheta}{2}\right) - C_{ab}^{'}\sin\left(\frac{\vartheta}{2}\right) \right)\; .
\end{eqnarray}
Dla $\aleph=4$ wyników (\ref{wyniki lacznego pomiaru Sa i Sb}),  z ogólnego związku (\ref{metryka Rao-Fishera w ukladzie wsp - amplitudy}) określającego metrykę Rao-Fishera $g_{ab} =  4 \sum_{i=1}^{\aleph} \frac{\partial q^{i}}{\partial \theta^{a}} \frac{\partial q^{i}}{\partial \theta^{b}} \, $, otrzymujemy po skorzystaniu z 
(\ref{rozklad na przestrzeni statystycznej EPR}) oraz (\ref{pochodna qab dla EPR})  
następującą postać metryki $g_{\vartheta \, \vartheta}$ indukowanej z rozkładu  (\ref{spinprawdop}): 
\begin{eqnarray}
\label{metryka Rao-Fishera dla EPR}
g_{\vartheta \, \vartheta}(\vartheta) &=&  4 \sum_{i=1}^{\aleph=4} \frac{\partial \tilde{q}_{i}}{\partial \vartheta} \frac{\partial \tilde{q}_{i}}{\partial \vartheta} = 4 \sum\limits_{ab} \frac{\partial (\frac{1}{\sqrt{2}} q_{ab})}{\partial \vartheta} \frac{\partial (\frac{1}{\sqrt{2}} q_{ab})}{\partial \vartheta}  \nonumber \\ &=&  \frac{1}{2}  \sum_{ab} \left( (B_{ab}^{'})^{2}  + \left((C_{ab}^{'})^{2} - (B_{ab}^{'})^{2}\right) \sin^{2} (\frac{\vartheta}{2})  -   B_{ab}^{'} \, C_{ab}^{'}\sin(\vartheta)   \right) \; 
\end{eqnarray}
na jednowymiarowej przestrzeni statystycznej ${\cal S}$ parametryzowanej w  bazie $\vartheta$. \\
\\
Po skorzystaniu z (\ref{wspol B prim dla q w EPR}), postać metryki $g_{\vartheta \, \vartheta}(\vartheta)$, (\ref{metryka Rao-Fishera dla EPR}), prowadzi  dla $\vartheta=0$ do warunku:
\begin{eqnarray}
\label{g dla vartheta = 0}
g_{\vartheta \, \vartheta}(\vartheta=0) =  \frac{1}{2}  \sum_{ab} (B_{ab}^{'})^{2}  =   (B_{++}^{'})^{2} +  (B_{+-}^{'})^{2}  \; \;\;\; {\rm dla } \;\; \vartheta=0 \; ,
\end{eqnarray}
natomiast dla $\vartheta = \pi$ do warunku:
\begin{eqnarray}
\label{g dla vartheta = pi}
g_{\vartheta \, \vartheta}(\vartheta=\pi) =  \frac{1}{2}  \sum_{ab}   (C_{ab}^{'})^{2}   =     (C_{+-}^{'})^{2}     
\;\;\;\; {\rm  dla } \;\; \vartheta = \pi \; ,
\end{eqnarray}
gdzie skorzystano również z (\ref{ce}).\\
\\
{\bf Centralne założenie statystyczne dla eksperymentu EPR-Bohm'a}: 
Załóżmy, że {\it w eksperymencie } EPR-Bohm'a {\it metryka Rao-Fishera $g_{\vartheta \, \vartheta}$ na ${\cal S}$ jest niezależna od  wartości   parametru}  $\vartheta$, tzn.:
\begin{eqnarray}
\label{Centralne zal stat dla EPR}
g_{\vartheta \, \vartheta}(\vartheta) = g_{\vartheta \, \vartheta} = const. \; . 
\end{eqnarray} 
W szczególnym przypadku warunek   (\ref{Centralne zal stat dla EPR}) oznacza, że $g_{\vartheta \, \vartheta}(\vartheta=0) = g_{\vartheta \, \vartheta}(\vartheta=\pi)$, co uwzględniając w zależnościach  (\ref{g dla vartheta = 0}) oraz (\ref{g dla vartheta = pi}) daje:
\begin{eqnarray}
\label{rownanie dla B i C}
(B_{++}^{'})^{2} +  \, (B_{+-}^{'})^{2} =   (C_{+-}^{'})^{2}  \geq 0 
\; .
\end{eqnarray}
Warunek ten oznacza, że $C_{+-}^{'} \neq 0 \, $,  
gdyż w przeciwnym wypadku, tzn. dla $C_{+-}^{'}=0$, z warunku (\ref{rownanie dla B i C}) oraz z (\ref{ce}) otrzymalibyśmy zerowanie się wszystkich współczynników  $B_{ab}^{'}$ oraz  $C_{ab}^{'}$, co odpowiadałoby trywialnemu przypadkowi braku rozwiązania dla zagadnienia EPR. 
Zatem otrzymujemy:
\begin{eqnarray}
\label{C+- niezerowe EPR}
C \equiv C_{+-}^{'} = C_{-+}^{'} \neq 0 \; ,
\end{eqnarray}
gdzie równości współczynników wynika z (\ref{C+- rowne C+- EPR}).\\
\\
W końcu niezależność $g_{\vartheta \, \vartheta}$ od  wartości $\vartheta$, (\ref{Centralne zal stat dla EPR}), daje po skorzystaniu z postaci $g_{\vartheta \, \vartheta}$, (\ref{metryka Rao-Fishera dla EPR}), oraz z warunków (\ref{ce}), (\ref{wspol B prim dla q w EPR}) i (\ref{C+- rowne C+- EPR}), warunek: 
\begin{eqnarray}
\label{warunek z g dla dowolnego vartheta 1}
\sum_{ab} \left( (C_{ab}^{'})^{2} - (B_{ab}^{'})^{2}   \right) 
= 2 \,(  \,(C_{+-}^{'})^{2}  - (B_{++}^{'})^{2} - \,(B_{+-}^{'})^{2} )  = 0 \; ,
\end{eqnarray}
czyli warunek pokrywający się z (\ref{rownanie dla B i C}) oraz:
\begin{eqnarray}
\label{warunek z g dla dowolnego vartheta 2}
\sum_{ab} B_{ab}^{'} \,C_{ab}^{'} =  2 (B_{+-}^{'} \,C_{+-}^{'}) = 0 \; .
\end{eqnarray}
Warunek (\ref{warunek z g dla dowolnego vartheta 2}) wraz z (\ref{C+- niezerowe EPR}) oznacza:
\begin{eqnarray}
\label{zerowanie B+-}
B_{+-}^{'} =  B_{-+}^{'} = 0 \; ,
\end{eqnarray}
gdzie skorzystano również z (\ref{wspol B prim dla q w EPR}).
Po uwzględnieniu (\ref{zerowanie B+-}) w (\ref{warunek z g dla dowolnego vartheta 1}), otrzymujemy: 
\begin{eqnarray}
\label{rownosc B2  i C2}
(B_{++}^{'})^{2} = (C_{+-}^{'})^{2} \; .
\end{eqnarray}
W końcu zauważmy, że ze względu na (\ref{zerowanie B+-}) oraz  (\ref{C+- niezerowe EPR}), warunek istnienia nietrywialnego  rozwiązania oznacza, że: 
\begin{eqnarray}
\label{nie zerowanie B++}
B \equiv B_{++}^{'} = B_{--}^{'} \neq 0,
\end{eqnarray}
gdzie ponownie w równości skorzystano z (\ref{wspol B prim dla q w EPR}).  
  \\
Podstawiając otrzymane wyniki dla współczynników $B_{ab}^{'}$ oraz $C_{ab}^{'}$ do (\ref{qab dla a=2}), otrzymujemy:
\begin{eqnarray}
\label{wynikqEPR}
q_{++}(\vartheta) = q_{--}(\vartheta)=B\sin\left(\vartheta/2\right) \; , \;\;
q_{-+}(\vartheta) = q_{+-}(\vartheta)=C\cos\left(\vartheta/2\right) \; .
\end{eqnarray}
Jak widać, równość współczynników w związkach (\ref{C+- niezerowe EPR}) oraz (\ref{nie zerowanie B++}) jest odbiciem równości  odpowiednich amplitud, wynikającej z symetrii odbicia przestrzennego (\ref{symetriaEPR}) oraz  wzoru (\ref{PqEPR}). \\
\\
Musimy jeszcze wyznaczyć stałe $B$ oraz $C$. Z powodu 
warunku normalizacji  prawdopodobieństwa  $P\left(S_{a}S_{b}|\vartheta\right)$, (\ref{normalizacja P daje wsp w qab}), otrzymujemy ze względu na (\ref{PqEPR}) równanie:
\begin{eqnarray}
\label{qkwa}
\frac{1}{2}\left(q_{++}^{2}(\vartheta)+q_{--}^{2}(\vartheta)+q_{-+}^{2}(\vartheta)+q_{+-}^{2}(\vartheta)\right) = 1 \; .
\end{eqnarray}
Korzystając z (\ref{wynikqEPR}) i (\ref{qkwa}) mamy: \begin{eqnarray}
B^{2}\sin^{2}\left(\vartheta/2\right)+C^{2}\cos^{2}\left(\vartheta/2\right)=\left(B^{2}-C^{2}\right)\sin^{2}\left(\vartheta/2\right)+C^{2} = 1 \; .
\end{eqnarray}
{\bf Końcowa postać amplitud}: Porównując współczynniki stojące przy odpowiednich funkcjach zmiennej
$\vartheta$ po lewej i prawej stronie drugiej równości powyższego wyrażenia, otrzymujemy: 
\begin{eqnarray}
\label{postac B oraz C}
B^{2} = C^{2} = 1 \; ,
\end{eqnarray}
co po wstawieniu do (\ref{wynikqEPR}) daje ostatecznie rozwiązanie równania generującego (\ref{row generujace dla amplitud w EPR}):
\begin{eqnarray}
\label{qEPR}
q_{++}(\vartheta) = q_{--}(\vartheta) = \pm \sin\left(\vartheta/2\right) \; , \quad \;\; q_{-+}(\vartheta) = q_{+-}(\vartheta) = \pm \cos\left(\vartheta/2\right) \; .
\end{eqnarray}
{\bf Wynik na prawdopodobieństwo w eksperymencie EPR-Bohm'a}: 
Podstawienie amplitudy (\ref{qEPR}) do $P\left(S_{ab}|\vartheta\right) = \frac{1}{2} \, q_{ab}^{2}(\vartheta) \,$, (\ref{PqEPR}), 
daje wynik na łączne prawdopodobieństwo otrzymania określonej
kombinacji rzutów spinów przy zadanej wartości kąta $\vartheta$: \begin{eqnarray}
\label{wynikEPR}
P\left(++|\vartheta\right) = P\left(--|\vartheta\right)=\frac{1}{2}\sin^{2}\left(\vartheta/2\right)  , \;\; P\left(+-|\vartheta\right) = P\left(-+|\vartheta\right)=\frac{1}{2}\cos^{2}\left(\vartheta/2\right)  ,
\end{eqnarray}
{\it który jest przewidywaniem mechaniki kwantowej} \cite{Manoukian}. \\
\\
Na koniec, podstawiając wartości otrzymanych współczynników $B_{ab}^{'}$ oraz $C_{ab}^{'}$ do (\ref{metryka Rao-Fishera dla EPR}), otrzymujemy wartość stałej $g_{\vartheta \vartheta}$ dla jedynej składowej metryki Rao-Fishera w  (\ref{Centralne zal stat dla EPR}):
\begin{eqnarray}
\label{metryka w EPR}
g_{\vartheta \, \vartheta}(\vartheta) = g_{\vartheta \, \vartheta} = 1 \; ,
\end{eqnarray}
która zgodnie (\ref{stalosc il wewn dla Euklidesowego ukl wsp}) jest metryką samo-dualnego (Euklidesowego) układu współrzędnych $\vartheta$ na  jednowymiarowej przestrzeni statystycznej ${\cal S}$.\\
\\
{\bf Różnica pomiędzy przedstawionym wyprowadzeniem a analizą w  \cite{Frieden}}: Wynik (\ref{wynikEPR}) w ramach medody EFI, został oryginalnie wyprowadzony w \cite{Frieden}, jednakże powyższy sposób wyprowadzenia  różni się w dwóch miejscach. Po pierwsze, warunki  brzegowe zostały ujętne w sposób bardziej przejrzysty  \cite{Mroziakiewicz}, a po drugie w \cite{Frieden} odwołano się do własności ortogonalności wprowadzonych tam kwantowych amplitud, o czym wspomnimy na końcu rozdziału, a czego w powyższym wyprowadzeniu uniknięto, wprowadzając w to miejsce {\it warunek niezależności metryki Rao-Fishera od wartości parametru $\vartheta$}.   \\
\\
{\bf Wnioski}: W (\ref{laczneEPR}) wyznaczono łączne prawdopodobieństwo
$P\left({S_{ab}}\right)=\frac{1}{4}$. Z drugiej strony w (\ref{Sb wycalkowane}) 
otrzymano $P(S_{b})=1/2$ (i analogicznie $P(S_{a})=1/2$), z czego wynika, że: 
\begin{eqnarray}
\label{niezalezne}
P\left({S_{ab}}\right)=P\left({S_{a}}\right)P\left({S_{b}}\right) \; ,
\end{eqnarray}
który to warunek oznacza niezależność spinowych zmiennych  $S_{a}$ 
oraz $S_{b}$. \\
\\
Natomiast z (\ref{wynikEPR}) widać, że efekt korelacji spinu zmienia się bardzo mocno wraz z wartością kąta $\vartheta$   
i w samej rzeczy porównanie (\ref{wynikEPR}) z (\ref{pol dla Sa}) i (\ref{pol}) daje warunek: 
\begin{eqnarray}
\label{nierownosc P z brzegowymi dla zaleznosci od kata w EPR}
P\left(S_{ab}|\vartheta\right)\neq P\left(S_{a}|\vartheta\right)P\left(S_{b}|\vartheta\right) \; .
\end{eqnarray}
Relacje  (\ref{niezalezne}) oraz   (\ref{nierownosc P z brzegowymi dla zaleznosci od kata w EPR}) 
nie są jednak w sprzeczności.  
Istotnie, 
ponieważ prawdopodobieństwa $P\left({S_{ab}}\right)$, $P\left({S_{a}}\right)$ oraz $P\left({S_{b}}\right)$ są wyznaczone na skutek uśrednień po wszystkich wartościach kąta  $\vartheta$, zatem można było oczekiwać, że po dokonaniu tych uśrednień korelacja zmniejszy się.  Zauważmy, że w (\ref{niezalezne}) 
nie ma zależności od $\vartheta$, co wynika z tego, że  prawdopodobieństwo $P\left({S_{ab}}\right)$  z definicji określa równoczesne pojawienie się kombinacji rzutów spinów 
$S_{ab}$ niezależnie od informacji 
o zmiennej kątowej $\vartheta$. 
Podobnie, $P\left({S_{a}}\right)$ oraz $P\left({S_{b}}\right)$ określają prawdopodobieństwa odpowiadających im zdarzeń w sytuacji pozbycia się informacji o kącie $\vartheta$.\\ 
Równanie  (\ref{niezalezne}) mówi więc, że  w sytuacji uśrednienia po kącie $\vartheta$, czyli wtedy gdy zmienna ta jest pod kontrolą, 
zmienne $S_{a}$ oraz $S_{b}$ rzutów spinów są niezależne, więc są  również całkowicie nieskorelowane. \\
Natomiast wynik (\ref{nierownosc P z brzegowymi dla zaleznosci od kata w EPR}) dla eksperymentu EPR-Bohm'a zachodzi wtedy, gdy dokonujemy estymacji kąta $\vartheta$ metodą EFI, czyli w sytuacji oczywistego braku uśrednienia po $\vartheta$. 
 \\ 
 \\
{\bf Porównanie stwierdzeń o uśrednieniu}: Powyższy wynik  ma interesującą 
fizyczną interpretację. Mianowicie twierdzenie Ehrenfesta mówi, że
średnie wartości kwantowych operatorów są równe ich klasycznym odpowiednikom. 
W jego świetle wyrażenie (\ref{niezalezne}) mówi, że w przypadku
uśrednienia po kącie $\vartheta$ stany splątane EPR-Bohm'a, na powrót ulegają klasycznej separacji. W analizie statystycznej mówimy, że po wyeliminowaniu wpływu zmiennej trzeciej, którą jest kąt $\vartheta$, okazało się, że zmienne rzutów spinów $S_{a}$ i $S_{b}$ są nieskorelowane. \\
  \\
{\bf Uwaga o wyprowadzeniu Friedena \cite{Frieden}}: Warunek niezależności wyprowadzenia formuł (\ref{wynikqEPR}) od mechaniki kwantowej można by nieco osłabić, tzn. na tyle, aby statystyczność teorii była dalej widoczna. Frieden  uczynił to, jak następuje \cite{Frieden}: \\
Pokażmy, że $q_{ab}(\vartheta)$ jest proporcjonalna do  ``{\it kwantowej}'' {\it amplitudy}
$\psi_{ab}(\vartheta)$,  parametryzowanej parametrem  $\vartheta$ dla stanów  (\ref{4 zdarzenia EPR}). 
Niech $P\left(S_{ab},\vartheta\right)$ jest prawdopodobieństwem pojawienia się konfiguracji~$(a,b)$ rzutu spinów, 
{\it podczas gdy} parametr wynosi $\vartheta$. Nie jest to  prawdopodobieństwo łączne zajścia zdarzenia: ``pojawiła się konfiguracji $(a,b)$ rzutu spinów oraz kąt $\vartheta$'', gdyż ten ostatni nie jest zmienną losową. \\
\\
Funkcję $\psi_{ab}(\vartheta)$ określimy tak, że kwadrat jej modułu spełnia związek: 
\begin{eqnarray}
\label{k}
|\psi_{ab}(\vartheta)|^{2} \equiv |\psi_{ab}\left({\vartheta\left|{S_{ab}} \right.}\right)|^{2} 
\equiv p\left({\vartheta\left|{S_{ab}}\right.}\right) = \frac{P\left(S_{ab},\vartheta\right)}{P\left(S_{ab}\right)} \; ,
\end{eqnarray}
tzn. {\bf jest prawdopodobieństwem, że skoro pojawiłaby się łączna konfiguracja spinów $S_{ab}$, to wartość kąta wynosi} $\vartheta$.\\ Problem polega na tym, że $q_{ab}(\vartheta)$ opisuje losowe zachowanie zmiennej spinowej $S_{a}$, a nie $\vartheta$, co oznacza, że $q_{ab}(\vartheta)$ jest  amplitudą typu $\psi_{\vartheta}({ab})$,  a nie $\psi_{ab}({\vartheta})$. \\
Aby pokazać, że $q_{ab}(\vartheta)$ jest proporcjonalna do $\psi_{ab}(\vartheta)$, skorzystajmy z (\ref{k}), a następnie z twierdzenia Bayesa oraz definicji prawdopodobieństwa warunkowego:
\begin{eqnarray}
\label{d}
\!\!\!\!\!\!\!\! |\psi_{ab}(\vartheta)|^{2} & = & \frac{P\left(S_{ab},\vartheta\right)}{P\left(S_{ab}\right)} = \frac{P(S_{ab}\left|{\vartheta}\right.) r(\vartheta)}{P(S_{ab}\left|{S_{b}}\right.)P\left(S_{b}\right)} = \frac{P(S_{ab}\left|{\vartheta}\right.) r(\vartheta)}{P(S_{a}\left|{S_{b}}\right.)P\left(S_{b}\right)} =  \frac{{P(S_{a}\left|{S_{b},\vartheta}\right.)P\left({S_{b}}\right)r(\vartheta)}}{{P(S_{a}\left|{S_{b}}\right.)P\left({S_{b}}\right)}} \;\;\; \nonumber \\
\!\!\!\!\!\!\!\!  &=&  \frac{{q_{ab}^{2}\left(\vartheta\right) r(\vartheta)}}{{P(S_{a}\left|{S_{b}}\right.) }} \; . \;\;\; 
\end{eqnarray}
Korzystając z (\ref{prawdkat}), (\ref{warunkSab}) oraz (\ref{inffishEPR2}), 
otrzymujemy więc z (\ref{d}), że \begin{eqnarray}
|\psi_{ab}(\vartheta)|^{2} = \frac{1}{\pi}\, q_{ab}^{2}(\vartheta)  \; ,
\label{proporcEPR}
\end{eqnarray}
skąd 
\begin{eqnarray}
\psi_{ab}(\vartheta) = \frac{e^{i\alpha}}{\pi}\, q_{ab}(\vartheta) \; , \quad \alpha \in \mathbf{R} \; .
\end{eqnarray}
Zatem, $\psi_{ab}(\vartheta) \propto q_{ab}(\vartheta)$, co oznacza, że\footnote{W mechanice kwantowej powiedzielibyśmy, że $\psi_{ab}(\vartheta)$ oznacza amplitudę prawdopodobieństwa  zdarzenia, że wartość kąta wynosi $\vartheta$, o 
ile pojawiłaby się łączna konfiguracja spinów $S_{ab}$.
}, 
skoro 
pojawiła się łączna konfiguracja spinów $S_{ab}$, to amplituda prawdopodobieństwa $\psi_{ab}(\vartheta)$, że wartość kąta wynosi $\vartheta$, jest 
proporcjonalna do amplitudy prawdopodobieństwa $q_{ab}(\vartheta)$ zaobserwowania 
rzutu spinu $S_{a}$ cząstki 1 pod warunkiem, że rzut spinu cząstki
2 wynosiłby $S_{b}$, a kąt $\vartheta$. Zdanie to jest wyrazem splątania, które pojawiło się w kwanto-mechaniczno opisie eksperymentu EPR-Bohm'a.  \\
\\
W koncu, ze względu na (\ref{proporcEPR}),  amplitudy  $q_{ab}(\vartheta)$ są proporcjonalne do amplitud $\psi_{ab}(\vartheta)$. Zatem Frieden zarządał {\bf ortogonalności amplitud kwantowych} $\psi_{++}$ i $\psi_{+-}$, skąd automatycznie wyniknęła ortogonalność amplitud $q_{++}$ oraz $q_{+-}$:
\begin{eqnarray}
\int\limits _{0}^{2\pi}{d\vartheta\; q_{++}q_{+-}}=\int\limits _{0}^{2\pi}{d\vartheta\; B_{++}^{'}\sin\left({\vartheta\mathord{\left/{\vphantom{\vartheta2}}\right.\kern -\nulldelimiterspace}2}\right)\left[{B_{+-}^{'}\sin\left({\vartheta\mathord{\left/{\vphantom{\vartheta2}}\right.\kern -\nulldelimiterspace}2}\right)+C_{+-}^{'}\cos\left({\vartheta\mathord{\left/{\vphantom{\vartheta2}}\right.\kern -\nulldelimiterspace}2}\right)}\right]} = 0 \; ,
\label{ortogEPR}\end{eqnarray}
co pozwoliło na wyprowadzenie zerowania się  $B_{+-}^{'}$ oraz  $B_{-+}^{'}$, (\ref{zerowanie B+-}), a dalej (postępując już jak powyżej) otrzymanie formuł  (\ref{wynikEPR}).

\subsection{Niepewność wyznaczenia kąta}

\label{Niepewnosc wyznaczenia kata}

Istnieje jeszcze jedna sprawa dotycząca analizy estymacyjnej,  opisanej powyższymi  rachunkami, która  wymaga podkreślenia. \\
\\
{\bf Rozkład eksperymentalny}: Otóż w rzeczywistym pomiarze  badacz otrzymuje wartości rzutu spinu $S_{a}$ z określonymi częstościami, które są oszacowaniami rozkładu prawdopodobieństwa  (\ref{wynikEPR}), co z kolei pozwala na punktowe oszacowanie kąta $\vartheta$.   \\
Analiza statystyczna metody EFI, która doprowadziła do  (\ref{wynikEPR}) jest, zgodnie z postulatem EFI, 
sprawą układu dokonującego próbkowania przestrzeni pomiarowej położeń i estymacji oczekiwanych parametrów\footnote{Jest tak w całej analizie EFI nie uwzględniającej wpływu urządzenia pomiarowego.}. 
{\it Jednak jako metoda statystyczna i ona podlega pod ograniczenie Rao-Cramera  dokładności oszacowania estymowanego parametru, którym w tym przypadku jest $\vartheta$}.  Sytuacja ta ma  następujące konsekwencje. \\
\\
{\bf Wewnętrzny błąd  estymacji metody EFI parametru $\vartheta$}: 
Zauważmy, że nierówność Rao-Cramera ma postać, $\sigma^{2}_{\theta} \, F \ge  1/I_{F}\left(\vartheta\right) \,$, (\ref{tw R-C dla par skalarnego}), gdzie tym razem  estymator $F \equiv \hat{\vartheta}$. Z nierówności tej otrzymujemy warunek na wariancję estymatora $\hat{\vartheta}$ kąta $\vartheta$: 
\begin{eqnarray}
\label{R-C dla IF dla EPR}
\mathop{\sigma^{2}}\left(\hat{\vartheta}\right)\ge\frac{1}{I_{F}(\vartheta)} \; ,
\end{eqnarray}
gdzie $I_{F}(\vartheta) = I_{b}$ jest informacją Fishera (\ref{IF dla vartheta w EPR}), a nie pojemnością informacyjną $I_{1_a}$, (\ref{pojemnoscEPR}), czy w konsekwencji jej wartością minimalną  (\ref{I EPR minimalna}). 
Bez analizowania postaci estymatora $\hat{\vartheta}$, nierówność Rao-Cramera (\ref{R-C dla IF dla EPR}) określa DORC dla jego wariancji, o ile tylko estymator ten jest nieobciążony. \\
\\
Wiemy też, że ponieważ pojemność informacyjna jest sumą po kanałach z odpowiadających im informacji Fishera, więc: 
\begin{eqnarray}
\label{relacja pojemnosci i IF dla EPR}
I_{1_a} \ge I_{F}(\vartheta) \; .
\label{okok}
\end{eqnarray}
Z (\ref{relacja pojemnosci i IF dla EPR}) oraz z (\ref{R-C dla IF dla EPR}) otrzymujemy: 
\begin{eqnarray}
\label{I oraz IF oraz var dla EPR}
\frac{1}{I_{1_a}}\le\frac{1}{I_{F}(\vartheta)} \le \mathop{\sigma^{2}}\left(\hat{\vartheta}\right) \; .
\end{eqnarray}
Ponieważ pojemność kanału $I_{1_a}$ według  (\ref{I EPR minimalna}) wynosi: 
\begin{eqnarray}
I_{1_a} = 4 \pi \; ,
\end{eqnarray}
więc podstawiając tą wartość do nierówności (\ref{I oraz IF oraz var dla EPR}) otrzymujemy:
\begin{eqnarray}
\label{Rao-Cramer w EPR}
\mathop{\sigma^{2}}\left(\hat{\vartheta}\right)\ge\frac{1}{4\pi} \approx 0.08 \; {\rm rad}^{\,2} \; .
\end{eqnarray}
{\bf Wniosek}: Nierówność (\ref{Rao-Cramer w EPR}) stwierdza, że obserwacja jednej wartości
rzutu spinu \textit{przy kompletnej nieznajomości kąta} $\vartheta$ (stąd
(\ref{prawdkat})) daje o nim małą, lecz skończoną informację. Błąd
jego estymacji $\sqrt{0,08}\; rad=0,28\; rad$ jest dość duży, co jest związane z płaską funkcją ``niewiedzy''  
$r(\vartheta)$ określoną w  (\ref{prawdkat}).

\subsubsection{Wpływ zaszumienia pomiaru} 

\label{zaszumienie pomiaru}

Analiza EFI w obliczu  pomiaru i płynącego z tego faktu zaszumienia danych wewnętrznych EFI nie jest przedmiotem tego skryptu.  Zainteresowany czytelnik znajdzie omówienie tego tematu w \cite{Frieden,Mroziakiewicz}. Poniżej zamieszczam wniosek z analizy  Mroziakiewicz \cite{Mroziakiewicz} dotyczący tego problemu dla  powyższej analizy eksperymentu EPR-Bohm'a. \\
\\
W  pomiarze stanu układu przez zewnętrznego obserwatora, otrzymujemy dane zaszumione przez układ pomiarowy. Tzn. dane pomiarowe rzutu spinu (oznaczmy je $\bar{S}_{a}$) są generowane przez prawdziwe wartości wielkości $S_{a},S_{b},\vartheta$ w obecności szumu aparatury pomiarowej.  Szum ten powstaje co prawda w urządzeniach Sterna-Gerlacha $a$ oraz $b$, ale ze względu na założenie podane na samym początku Rozdziału~\ref{Warunki brzegowe}, z innych przyczyn niż (przyjęte jako równe zeru) fluktuacje rzutu spinu.  \\
\\
{\bf IF uwzględniająca zaszumienie}: Uzyskana przy tych danych informacja Fishera $I_{zasz}$ o parametrze $\vartheta$, uwzględniająca {\it zaszumienie} pomiarowe, jest generowana przez informację Fishera $I_{F}(\vartheta) = I_{b}$  (\ref{IF dla vartheta w EPR}). W tym sensie  {\it informacja Fishera $I_{F}(\vartheta)$, (\ref{IF dla vartheta w EPR}), procedury estymacyjnej EFI jest częścią informacji, która przejawia się w pomiarze}. Zatem obok nierówności (\ref{okok}) zachodzi również: 
\begin{eqnarray}
I_{F}(\vartheta) \ge I_{zasz}(\vartheta) \; .
\end{eqnarray}
Ze względu na to, że $I_{zasz}(\vartheta)$ wchodzi w (\ref{I oraz IF oraz var dla EPR}) w miejsce $I_{F}(\vartheta)$, warunek ten oznacza pogorszenie jakości estymacji w porównaniu 
z (\ref{Rao-Cramer w EPR}). 


\subsection{Informacja $Q$ jako miara splątania}

Tak jak we wszystkich problemach estymacyjnych rozwiązanych metodą EFI, tak i w eksperymencie EPR-Bohm'a wykorzystano przy estymacji kąta $\vartheta$, 
obserwowaną zmodyfikowaną informacyjną zasadę strukturalną $\widetilde{\textit{i'}} + \widetilde{\mathbf{C}} + \kappa \, \textit{q} = 0$,  (\ref{zmodyfikowana obserwowana zas strukt z P i z kappa}), która ze względu na $\kappa=1$  w eksperymencie 
EPR-Bohm'a\footnote{Dodatkowo dla eksperymentu EPR-Bohm'a $\widetilde{\mathbf{C}} \equiv \widetilde{\mathbf{C}}_{EPR}=0$, zgodnie z (\ref{postac stalej Cab}). 
}, 
daje  na poziomie oczekiwanym związek $Q_{1_a} = - I_{1_a}$,  (\ref{I1piEPR}), zgodnie z ogólnym warunkiem (\ref{rownowaznosc strukt i zmodyfikowanego strukt}) metody EFI. \\
\\
{\bf Dostępność pomiarowa $I_{1_a}$}: Ponieważ nie mierzony był stan cząstki $2$, tzn. rzut spinu  $S_{b}$ w czasie naszej analizy mógł przyjąć wartość $+$ lub $-$, zatem pojemność informacyjna $I_{1_a}$ jest informacją o kącie  $\vartheta$, zawartą w obserwacji  rzutu spinu cząstki $1$ dokonanej przez układ i zgodnej z dokładnością do szumu z obserwacją  dokonaną przez zewnętrznego 
obserwatora\footnote{Zaszumienie z  aparatury Sterna-Gerlacha, o którym wspomniano powyżej, dla uproszczenia pomijamy.
}.  \\
\\
{\bf Nieobserwowalna wewnętrza struktura układu odbita w $Q_{1_a}$}: Natomiast informacja strukturalna $Q_{1_a}$ (\ref{strukturalnaEPR}) dla cząstki $1$, jako pozostała część informacji fizycznej $K$, jest informacją zawartą w nieobserwowanej  wewnętrznej strukturze informacyjnej całego układu. Sugeruje to istnienie łącznej, {\it nieseparowalnej} informacji strukturalnej dla cząstek $1$ i $2$. Zatem  informacja o łącznej strukturze układu, to nie to samo co suma informacji o jego składowych. \\
\\
{\bf Splątanie stanów cząstek}: Ze względu na $Q_{1_a} = - I_{1_a}$, przestrzeń danych odbita w $I_{1_a}$, tzn. obserwacje częstości rzutu spinu cząstki $1$, jest 
na poziomie informacji, związana z 
wewnętrzą przestrzenią  konfiguracyjną układu odbitą w $Q_{1_a}$. Co prawda, ze względu na brak bezpośredniego wglądu w relacje wewnątrz układu, możemy jedynie ze skończoną dokładnością (Rozdział~\ref{Niepewnosc wyznaczenia kata}) wnioskować o kącie $\vartheta$, jednak sam fakt możliwości takiego wnioskowania określa sytuację nazywaną {\it splątaniem stanów} obu cząstek. \\
\\
{\bf Wniosek ogólny}: Wynik ten prowadzi do stwierdzenia, że strukturalna (wewnętrzna) zasada informacyjna $I = - \kappa Q$, stosowana dla każdego rozważanego problemu EFI, opisuje\footnote{W przypadku braku separowalności I oraz Q na sumy odpowiednio pojemności informacyjnych oraz informacji strukturalnych, właściwych dla podukładów.
} 
splątanie przestrzeni danych  obserwowanych\footnote{Na przykład,  częstości  rejestracji określonych rzutów spinu $S_{a}$ w eksperymencie EPR-Bohm'a.} z nieobserwowaną konfiguracją układu cząstek\footnote{Na przykład, kątem $\vartheta$ w eksperymencie EPR-Bohm'a.}. Postuluje to  wykorzystywanie EFI  jako formalizmu do predykcji wystąpienia stanów splątanych  i to nie tylko w przypadku problemu EPR-Bohm'a. Tak więc, informacja strukturala $Q$ jest informacją o ,,splątaniu'' widocznym w korelacji danych przestrzeni pomiarowej
z przestrzenią konfiguracyjną układu. \\
\\
Na koniec wspomnijmy, że inne traktowanie informacji fizycznej $K$ w \cite{Frieden} niż w obecnym skrypcie, sprawia, że wnioski tam otrzymane są inne\footnote{Wnioski w pracy \cite{Frieden} są następujące: Przekaz [związanej] informacji J do {\it cząstki obserwowanej} wzbudza wartość spinu $S_{a}$  do poziomu danej w próbce, zgodnie z końcowym prawem (\ref{wynikEPR}). Oznacza to, że przekaz informacji sprawia, że cząstka {\it spełnia} końcowe prawo (\ref{wynikEPR}) [opisujące zachowanie się] spinu. Przyczyną jest jakiś nieznany mechanizm oddziaływania, być może pewna ``bogata i złożona struktura, która może odpowiadać na informację i kierować zgodnie z tym, swoim własnym ruchem'' . W ten sposób, przekaz [związanej] informacji J wiąże z sobą stany spinowe nieobserwowanej i obserwowanej cząstki. A zatem J (zgodnie z wymogiem) ``nadaje postać'' własnościom spinu {\it nie}obserwowanej cząstki. }.


\chapter[Zakończenie]{Zakończenie}

Celem skryptu była prezentacja uogólnienia MNW  oraz 
zastosowania IF w jej części wnioskowania statystycznego dotyczącego estymacji niepartametrycznej metodą EFI, zaproponowaną przez Friedena i Soffera do  opisu zjawisk fizycznych i ekonofizycznych. Podstawy takiej analizy statystycznego opisu zjawisk zostały wprowadzonych w latach 20 ubiegłego wieku przez
Fishera. \\
Początkowo Fisher wprowadził
MNW poszerzoną o pojęcie IF w celu 
rozwiązania problemu estymacji punktowej i przedziałowej parametru rozkładu zmiennej losowej w sytuacji małej próby. Jego statystystyczna metoda doboru modeli, konstruowana niezależnie od ówcześnie rozwijanych teorii fizycznych, okazała się jednak sięgać o wiele dalej, w obszar estymacji równań fizycznych teorii pola, przy czym okazało się, że wielkość (małej) próby jest  w tej estymacji cechą charakterystyczną modeli. Celowi temu poświęcona jest główna część niniejszego skryptu. \\
\\
Jesteśmy w punkcie, w którym można już dać pewne wstępne podsumowanie metody EFI jako procedury statystycznej budowania modeli.
Jest więc EFI metodą statystycznej  estymacji równań ruchu teorii pola lub równań generujących rozkłady fizyki statystycznej. 
Gdy  szukane  równanie ruchu, wyestymowane metodą EFI poprzez rozwiązanie jej równań (\ref{zmodyfikowana obserwowana zas strukt z P i z kappa}) oraz (\ref{var K rozpisana}), zostało otrzymane, wtedy może być dalej prowadzone poszukiwanie rozkładu prawdopodobieństwa 
na analitycznych warstwach przestrzeni statystycznej ${\cal S}$, tzn. na (pod)rozmaitości w ${\cal S}$ z metryką Rao-Fishera. \\
Może się zdarzyć, że łączny  rozkład prawdopodobieństwa 
jest analizowany przez EFI na całej położe\-niowo-pędowej 
przestrzeni fazowej układu\footnote{Gdyby 
na przykład chcieć wyznaczyć metodą EFI łączny rozkład energii i pędu cząstki gazu.
}
i to dla więcej niż jednego pola. 
%
%
Ma to np. miejsce, gdy obok pola 
Diraca, obecne jest również pole cechowania (Rozdział~\ref{foliation of S}). 
Wtedy  na warstwie rozkładu fermionowego, w jego pochodnych kowariantnych, pole 
cechowania\footnote{Dla brzegowego rozkładu pola cechowania wyznaczonego z łącznego rozkładu wszystkich 
pól.
} 
musi być samospójnie wzięte pod uwagę.  \\
%
Sedno metody EFI jest zawarte w ogólnej postaci informacji fizycznej $K$ zadanej przez równania (\ref{TPI diag}) oraz (\ref{k form}) (lub (\ref{k form dla psi})). Jest ona  funkcją zarówno obserwowanej informacji strukturalnej $\texttt{q\!F}_{n}(q_{n}({\bf x}))$ jak i amplitud $q_{n}({\bf x})$ (lub $\psi_{n}({\bf x})$) rozkładów, wraz z ich pochodnymi. Przy tak ogólnym  zrozumieniu $K$, różnorodność równań EFI jest konsekwencją różnych  warunków wstępnych dyktowanych przez fizykę. Mogą się one wyrażać np. poprzez równania ciągłości,  które same są wynikiem estymacji statystycznej \cite{Dziekuje informacja_2} (Rozdział~\ref{master eq}), pewne symetrie  charakterystyczne dla zjawiska (Rozdział~\ref{Poj inform zmiennej los poloz}) oraz warunki normalizacyjne. 
Wstępne założenia fizyczne mogą, poza ewentualnością  nałożenia dodatkowych równań ciągłości,  wskazywać również na zastosowanie jedynie wariacyjnej zasady informacyjnej,  
jak to ma miejsce w przypadku równania Kleina-Gordona dla pola skalarnego  (Rozdział~\ref{Klein-Gordon scalars}), bądź na obie zasady i ich samospójne rozwiązanie, jak to miało miejsce w pozostałych przypadkach. 
\\
\\
Podsumowując, przeszliśmy w skrypcie przez trzy etapy statystycznej rozbudowy zastosowania MNW oraz IF. Na pierwszym, wstępnym etapie,  przedstawiono zastosowanie MNW oraz  IF w analizie doboru modeli  (Rozdział~\ref{MNW}) oraz omówiono podstawy podejścia geometrii różniczkowej do konstrukcji przestrzeni statystycznej, obrazując poznany aparat statystyczny przykładami estymacji modeli  eksponentialnych.  Na drugim etapie, wprowadzono pojęcie entropii względnej i pojemności informacyjnej oraz wprowadzono  strukturalną i wariacyjną zasadę informacyjną  (Rozdział~\ref{Entropia wzgledna i IF}-\ref{Zasady informacyjne}) metody EFI, a w trzecim  przedstawiono wykorzystanie tych pojęć do wyprowadzenia podstawowych równań modeli fizycznych (Rozdziały~\ref{Kryteria informacyjne w  teorii pola}-\ref{Przyklady}, Dodatek~\ref{Maxwell field}-\ref{general relativity case}), włączając w to przykład ekonofizycznego opisu zjawiska  (Rozdział~\ref{Model Aoki-Yoshikawy - ekonofizyka}).

\chapter[Dodatki]{Dodatki}

\section[Dodatek: Zasada nieoznaczoności Heisenberga]{Dodatek: Zasada nieoznaczoności Heisenberga}

\label{Zasada nieoznaczonosci Heisenberga}

Poniższe wyprowadzenie zasady nieoznaczoności Heisenberga zostało przedstawione w \cite{Frieden,Mania}. 
Zasada Heisenberga stwierdza, że w określonej chwili czasu \textit{t} nie można  z dowolną dokładnością wyznaczyć jednocześnie położenia ${\bf y}$ i pędu ${\bf y}_{p}$ cząstki, tzn. że położenie i pęd cząstki są {\it rozmyte}, co możemy zapisać następująco:
\begin{eqnarray}
\label{U1.1}
\sigma_{\theta}^{2} \cdot \sigma _{p}^{2} \ge \left(\frac{\hbar}{2} \right)^{2} \; ,
\end{eqnarray}
gdzie  $\sigma_{\theta}^{2} $ i $\sigma_{p}^{2} $ są wariancjami położenia i pędu cząstki względem ich wartości oczekiwanych  $\theta$ oraz $\theta_{p}$. 

\subsubsection*{Wyprowadzenie nierówności (\ref{U1.1}) dla pola rangi $N=2$}

Załóżmy, że  dokonujemy   estymacji tylko w jednym przestrzennym  kanale informacyjnym przy założeniu, że pozostałe czasoprzestrzenne parametry rozkładu są znane.  Powyższa relacja może być wyprowadzona przy odwołaniu się do  własności informacji  Fishera. 
Wartości położenia ${\bf y}$ zmiennej $Y$ są {\it rozmyte} (bądź fluktuują) wokół jej wartości oczekiwanej $\theta$, skąd wartości ${\bf x} \in {\cal X}$ zmiennej odchyleń $X$ spełniają (jak zwykle w skrypcie), związek:
\begin{eqnarray}
\label{U1.0}
{\bf x} = {\bf y} - \theta \; .
\end{eqnarray}
Załóżmy, że jest  spełniona nierówność Rao-Cramera (\ref{tw R-C dla par skalarnego}):
\begin{eqnarray}
\label{U1.2}
\sigma_{\theta}^{2} \cdot I_{F}(\theta) \ge 1 \; , \;\;\; {\rm gdzie} \;\;\;\;  \sigma_{\theta}^{2} \equiv E\left((\hat{\theta }(Y)-\theta )^{2}\right)  \; ,
\end{eqnarray}
gdzie $I_{F}(\theta) $ jest informacją Fishera parametru 
$\theta$ dla {\it pojedynczego pomiaru} (z powodu odwołania się do skalarnej wersji twierdzenia Rao-Cramera), która po skorzystaniu z postaci kinematycznej (\ref{Fisher_information-kinetic form bez n}) w Rozdziale~\ref{The kinematical form of the Fisher information}, ma postać:
\begin{eqnarray}
\label{U1.3}
I_{F} = 4 \int_{\cal X} d{\bf x}\left(\frac{\partial q({\bf x})}{\partial {\bf x}} \right)^{2}  \, .
\end{eqnarray}
Rozważmy zespoloną amplitudę $\psi$, (\ref{amplitudapsi}), dla pola rangi $N=2$: 
\begin{eqnarray}
\label{pole psi rangi 2}
\psi ({\bf x}) = \psi_{1}({\bf x}) = \frac{1}{{\sqrt{2}}} \left( q_{1}({\bf x}) + i \, q_{2}({\bf x}) \right) \; , 
\end{eqnarray}
która ma jako część rzeczywistą i urojoną dwie rzeczywiste amplitudy $q_{i}$, $i=1,2$. Założymy, że  $N=2$ - wymiarowa próba dla zmiennej odchyleń  $X$ jest prosta, skąd amplitudy $q_{i}({\bf x})$, $i=1,2$, są takie same i równe $q({\bf x})$:
\begin{eqnarray}
\label{prosta proba}
q_{1}({\bf x}) = q_{2}({\bf x}) = q({\bf x}) \; .
\end{eqnarray}
Zespoloną funkcję $\psi \left({\bf x}\right)$ można zapisać poprzez jej transformatę Fouriera, przechodząc z reprezentacji położeniowej ${\bf x}$ do pędowej ${\bf p}$:
\begin{eqnarray}
\label{U1.4}
\psi \left({\bf x}\right) = \frac{1}{\sqrt{2\pi \hbar} } \int d{\bf p} \; \phi({\bf p}) \, \exp \left(i \,{\bf p}\, {\bf x} \right) \, ,
\end{eqnarray}
gdzie ${\bf p}$ jest rozmyciem (fluktuacją) pędu ${\bf y}_{p}$ wokół jego wartości oczekiwanej $\theta_{p}\,$, tzn.:
\begin{eqnarray}
\label{rozmycie pedu}
\;\;\;\; {\bf p} = {\bf y}_{p} - \theta_{p} \; ,
\end{eqnarray}
a pęd cząstki jest mierzony w tej samej chwili co jego   położenie.\\
Dla amplitudy $N=2$ pojemność informacyjna $I$, (\ref{inf F z psi}), ma postać:
\begin{eqnarray}
\label{U1.5}
I =  8 \int d{\bf x} \frac{d\psi^{\,*}({\bf x})}{d{\bf x}} \frac{d\psi ({\bf x})}{d{\bf x}}  = 8 \int d{\bf x}\left|\frac{d\psi \left({\bf x}\right)}{d{\bf x}} \right|^{2}   \; .
\end{eqnarray}
Przedstawiając $\psi$ w postaci: 
\begin{eqnarray}
\label{U1.6}
\psi({\bf x}) =\left|\psi({\bf x}) \right|\exp \left(i S({\bf x})\right) \; 
\end{eqnarray}
i wykonując różniczkowanie:
\begin{eqnarray}
\label{U1.7}
\frac{d\psi \left({\bf x}\right)}{d{\bf x}} =\frac{d\left|\psi \left({\bf x}\right)\right|}{d{\bf x}} \;e^{iS\left({\bf x}\right)} +i\left|\psi \left({\bf x}\right)\right|\; e^{i S\left({\bf x}\right)} \; \frac{d S\left({\bf x}\right)}{d{\bf x}} \; ,
\end{eqnarray}
możemy (\ref{U1.5}) przekształcić następująco:
\begin{eqnarray}
\label{U1.8}
\begin{array}{l} {I = 8\int d{\bf x}\left|\frac{d\psi \left({\bf x}\right)}{d{\bf x}} \right|^{2}  =8\int d{\bf x}\frac{d\psi ^{*} \left({\bf x}\right)}{d{\bf x}} \frac{d\psi \left({\bf x}\right)}{d{\bf x}}  =}  \\ 
{\quad = 8 \int d{\bf x}\left(\frac{d\left|\psi \left({\bf x}\right)\right|}{d{\bf x}} e^{-iS\left({\bf x}\right)} -i\left|\psi \left({\bf x}\right)\right|e^{-iS\left({\bf x}\right)} \frac{dS\left({\bf x}\right)}{d{\bf x}} \right)\left(\frac{d\left|\psi \left({\bf x}\right)\right|}{d{\bf x}} e^{iS\left({\bf x}\right)} +i\left|\psi \left({\bf x}\right)\right|e^{iS\left({\bf x}\right)} \frac{dS\left({\bf x}\right)}{d{\bf x}} \right) =}  \\ 
{\quad =8\int d{\bf x}\left[\left(\frac{d\left|\psi \left({\bf x}\right)\right|}{d{\bf x}} \right)^{2} +i\left|\psi \left({\bf x}\right)\right|e^{iS\left({\bf x}\right)} \frac{d\left|\psi \left({\bf x}\right)\right|}{d{\bf x}} \frac{dS\left({\bf x}\right)}{d{\bf x}} -i\left|\psi \left({\bf x}\right)\right|e^{iS\left({\bf x}\right)} \frac{dS\left({\bf x}\right)}{d{\bf x}} \frac{d\left|\psi \left({\bf x}\right)\right|}{d{\bf x}} +\right.  }  \\ 
{\quad \left. +\left|\psi \left({\bf x}\right)\right|^{2} \left(\frac{dS\left({\bf x}\right)}{d{\bf x}} \right)^{2} \right]=8\int d{\bf x}\left[\left(\frac{d\left|\psi \left({\bf x}\right)\right|}{d{\bf x}} \right)^{2} +\left|\psi \left({\bf x}\right)\right|^{2} \left(\frac{dS\left({\bf x}\right)}{d{\bf x}} \right)^{2} \right] } \; . \end{array}
\end{eqnarray}
Zajmijmy się teraz pierwszym składnikiem pod ostatnia całką w (\ref{U1.8}). Ponieważ norma amplitudy $\psi$ jest równa:
\begin{eqnarray}
\label{norma psi}
|\psi| = \sqrt{\frac{1}{2} \,( q_{1}^{2} + q_{2}^{2} )} \; ,
\end{eqnarray}
zatem:
\begin{eqnarray}
\label{U1.9}
\left(\frac{d\left|\psi \left({\bf x}\right)\right|}{d{\bf x}} \right)^{2} = \frac{1}{2} \left(\frac{d}{d{\bf x}} \sqrt{q_{1}^{2} + q_{2}^{2} }  \,\right)^{2} =\left|q_{1} = q_{2} = q\right|=\left(\frac{dq}{d{\bf x}} \right)^{2} \; .
\end{eqnarray}
Podstawiając powyższy wynik do całki (\ref{U1.8}) i korzystając z (\ref{U1.3}),  otrzymamy następującą równość:
%
\begin{eqnarray}
\label{U1.10}
I &=& 8\int d{\bf x} \left|\frac{d\psi \left({\bf x}\right)}{d{\bf x}} \right|^{2}  = 8\int d{\bf x}\left[\left(\frac{d\left|\psi \left({\bf x}\right)\right|}{d{\bf x}} \right)^{2} +\left|\psi \left({\bf x}\right)\right|^{2} \left(\frac{dS\left({\bf x}\right)}{d{\bf x}} \right)^{2} \right] 
 \nonumber \\ 
& =& 8\int d{\bf x}\left(\frac{dq}{d{\bf x}} \right)^{2}  +8\int d{\bf x}\left|\psi \left({\bf x}\right)\right|^{2} \left(\frac{dS\left({\bf x}\right)}{d{\bf x}} \right)^{2}  = 2I_{F} + 8 \, E\left[ \left(\frac{dS\left({\bf x}\right)}{d{\bf x}} \right)^{2} \right] 
\; .
\end{eqnarray}
Z (\ref{U1.10}) oraz (\ref{U1.5}) wynika więc następująca nierówność:
\begin{eqnarray}
\label{U1.11}
2 I_{F} \le I \;\;\; {\rm lub} \;\;\; I_{F} \le 4\int d{\bf x}\left|\frac{d\psi \left({\bf x}\right)}{d{\bf x}} \right|^{2}  \;
\; .
\end{eqnarray}
Wykorzystując transformatę Fouriera (\ref{U1.4}),  
otrzymujemy z (\ref{U1.11}) po przejściu   do reprezentacji pędowej:
\begin{eqnarray}
\label{U1.12}
I_{F} \le \frac{4}{\hbar^{2} } \int d{\bf p}\,  \left|\phi \left({\bf p}\right)\right|^{2} \, {\bf p}^{2}
\; .
\end{eqnarray}
gdzie $\left|\phi \left({\bf p}\right)\right|^{2} $ jest brzegową gęstością prawdopodobieństwa $P({\bf p})$ odchyleń pędu. \\
Zatem całka  po prawej stronie (\ref{U1.12}) jest wartością oczekiwaną $E({\bf p}^{2})$ dla ${\bf p}^{2}$:
\begin{eqnarray}
\label{U1.13}
I_{F} \le \frac{4}{\hbar^{2} } \, E\left( {\bf p}^{2} \right) = \left(\frac{2}{\hbar} \right)^{2} \,E\left[ ({\bf y}_{p} - \theta_{p})^{2} \right] \equiv \left(\frac{2}{\hbar} \right)^{2} \sigma_{p}^{2} 
\; ,
\end{eqnarray}
gdzie $\sigma_{p}^{2} $ jest wariancją pędu cząstki, a pierwsza równość  wynika z tego, że ${\bf p}$ jest odchyleniem  (fluktuacją) pędu ${\bf y}_{p}$ od jego wartości oczekiwanej $\theta_{p}\,$, (\ref{rozmycie pedu}).
Podstawiając powyższy wynik do nierówności Rao-Cramera, 
$\sigma_{\theta}^{2} \cdot I_{F}(\theta) \ge 1$,  (\ref{U1.2}), otrzymujemy (\ref{U1.1}):
\begin{eqnarray}
\label{U1.14}
\sigma_{\theta}^{2} \cdot \sigma _{p}^{2} \ge \left({\raise0.7ex\hbox{$ \hbar $}\!\mathord{\left/{\vphantom{\hbar 2}}\right.\kern-\nulldelimiterspace}\!\lower0.7ex\hbox{$ 2 $}} \right)^{2} \; .  
\end{eqnarray}
c.n.d.

\newpage

\section[Dodatek: Równanie Schrödingera]{Dodatek: Równanie Schrödingera}

\label{Rownanie Schrodingera} 

Równania Schrödingera wyprowadzimy jako nierelatywistyczną
granicę równania Kleina-Gordo\-na. Wyprowadzenie go  jako nierelatywistycznej granicy 
równania Diraca byłoby merytorycznie bardziej uzasadnione \cite{Sakurai 2}, jednak celem poniższego wyprowadzenia jest zwrócenie uwagi na relatywistyczne pochodzenie spinu ele\-ktronu. \\
  \\
Rozpoczniemy od separacji zmiennych czasowych i przestrzennych występujących w równaniu (\ref{row KL dla swobodnego}) dla pola rangi $N=2$, zapisując po lewej stronie wszystkie wyrazy zawierające pochodną czasową, a po prawej pochodną po współrzędnych przestrzennych:
\begin{eqnarray}
\label{klseparacja}
& &\left({\hbar^{2}\frac{{\partial^{2}}}{{\partial t^{2}}} + 2\,i\,\hbar \,e\,\phi\frac{\partial}{{\partial t}} + i \,\hbar \,e\,\frac{{\partial\phi}}{{\partial t}} - e^{2}\phi^{2}}\right) \psi \nonumber \\
& & = \left({ c^{2}\hbar^{2} \vec{\nabla}^{2} - 2\,i\,e\,c\, \hbar\left({\vec{A} \cdot \vec{\nabla}}\right) - i\,e\,c \,\hbar \left({\vec{\nabla} \cdot \vec{A}}\right) - e^{2}{\vec{A}}^{2} - m^{2}c^{4}} \right) \psi \, ,
\end{eqnarray}
gdzie skorzystano z wyrażenia 
\begin{eqnarray}
{\vec{\nabla}\cdot\left({\vec{A}\psi}\right)=\psi\vec{\nabla}\cdot\vec{A}+\vec{A}\cdot\vec{\nabla}\psi} \; .
\end{eqnarray}
Następnie skorzystajmy z nierelatywistycznej reprezentacji funkcji
falowej \cite{Sakurai 2}:  
\begin{eqnarray}
\label{nonrelat}
\psi\left({{\bf x},t}\right) = \tilde{\psi}\left({{\bf x},t}\right)e^{-imc^{2}t/\hbar} \; ,
\end{eqnarray}
gdzie wydzieliliśmy z $\psi$ wyraz zawierający energię spoczynkową
$mc^{2}$, otrzymując nową funkcję falową $\tilde{\psi}$. Kolejnym krokiem jest zastosowanie (\ref{nonrelat}) w (\ref{klseparacja}). Różniczkując (\ref{nonrelat}) po czasie otrzymujemy: 
\begin{eqnarray}
\label{1roz}
\frac{\partial}{{\partial t}} \psi = \left({\frac{\partial}{{\partial t}}\tilde{\psi}-\frac{{imc^{2}}}{\hbar}\tilde{\psi}}\right) e^{{{-imc^{2}t}/\hbar}} \;\; ,
\end{eqnarray}
a po kolejnym różniczkowaniu:
\begin{eqnarray}
\label{2roz}
\frac{{\partial^{2}}}{{\partial t^{2}}}\psi = \left({\frac{{\partial^{2}}}{{\partial t^{2}}}\tilde{\psi} - 2\frac{{imc^{2}}}{\hbar}\frac{\partial}{{\partial t}}\tilde{\psi} - \frac{{m^{2}c^{4}}}{{\hbar^{2}}}\tilde{\psi}}\right) e^{{{-imc^{2}t}/\hbar}} \;\; .
\end{eqnarray}
Stosując nierelatywistyczne ($n.r$) przybliżenie: 
\begin{eqnarray}
\label{enr}
\frac{E_{n.r}}{mc^{2}} << 1 \; ,
\end{eqnarray}
gdzie $E_{n.r}$ jest określone poprzez poniższe zagadnienie własne:
\begin{eqnarray}
- \hbar^{2} \frac{\partial^{2}}{\partial t^{2}}\, \tilde{\psi} = E_{n.r}^{2} \, \tilde{\psi} \; ,
\end{eqnarray}
możemy pominąć pierwszy wyraz po prawej stronie (\ref{2roz}).\\  Odwołajmy się do przybliżenia stałego, słabego potencjału, dla którego zachodzi:
\begin{eqnarray}
\label{ephi}
e \phi << mc^{2} \; ,
\end{eqnarray}
oraz
\begin{eqnarray}
\label{stacphi}
\frac{\partial\phi}{\partial t} = 0 \; .
\end{eqnarray}
Następnie, wstawiając (\ref{1roz}) oraz (\ref{2roz}) do (\ref{klseparacja}), i wykorzystując (\ref{enr}), (\ref{ephi}) oraz  (\ref{stacphi}), otrzymujemy po 
przemnożeniu przez $-1/{2mc^{2}}$ i {\it oznaczeniu} $\tilde{\psi}$ jako $\psi$,  równanie: 
\begin{eqnarray}
\label{juz}
{i\hbar\frac{\partial}{\partial t}\psi-e\phi\psi = -\frac{\hbar^{2}}{2m}\vec{\nabla}^{2}\psi+\frac{ie\hbar}{mc}\vec{A}\cdot\vec{\nabla}\psi+\frac{ie\hbar}{2mc}\left(\vec{\nabla}\cdot\vec{A}\right)\psi + \frac{e^{2}{\vec{A}}^{2}}{2mc^{2}}\,\psi} \; .
\end{eqnarray}
Jest to równanie Schödingera dla nierelatywistycznej funkcji falowej $\psi$ z potencjałem elektromagnetycznym $\vec{A}$ 
oraz $\phi$. \\
W przypadku zerowania się części wektorowej potencjału $\vec{A}$ i dla niezerowego potencjału skalarnego $\phi$, równanie to 
przyjmuje znaną postać: 
\begin{eqnarray}
\label{schrodinger}
i\hbar\frac{\partial}{{\partial t}}\psi = -\frac{{\hbar^{2}}}{{2m}}\vec{\nabla}^{2}\psi + e \,\phi \, \psi \; .
\end{eqnarray}
\\
\\
{\bf Pytanie}: {\it Powiedz dlaczego powyższe wyprowadzenie równania Schödingera z równania Kleina-Gordona, świadczy o relatywistycznym charakterze spinu elektronu?}

\newpage

\section[Dodatek: Rezultaty EFI dla elektrodynamiki Maxwella oraz teorii grawitacji]{Dodatek: Rezultaty EFI dla elektrodynamiki Maxwella oraz teorii grawitacji}

\subsection{Dodatek: Pole cechowania Maxwella}

\label{Maxwell field}

Poniżej zaprezentujemy rezultat metody EFI otrzymany w zapisie Friedena-Soffera dla wyprowadzenia równań Maxwella ($M$).  
Punktem wyjścia jest pojemność informacyjna (\ref{postac I dla p po x bez n}):
\begin{eqnarray}
I = \sum_{n=1}^N {\int_{\cal X} d^{4}{\bf x} \frac{1}{{p_{n} \left(  {\bf x}  \right)}} \sum\limits_{\nu=0}^{3}  {\left( {\frac{{\partial p_{n} \left(  {\bf x}  \right)}}{{\partial {\bf x}_{\nu} }}}  {\frac{{\partial p_{n} \left(  {\bf x}  \right)}}{{\partial {\bf x}^{ \nu} }}} \right) } } \; .\nonumber
\end{eqnarray}  
Rozważmy równania ruchu Maxwella dla pola rangi $N=4$ z amplitudą rzeczywistą $q_{n}$, $n=1,2,3,4$. Zakładamy, że pola cechowania są proporcjonalne do tych rzeczywistych amplitud  \cite{Frieden}:
\begin{eqnarray}
\label{amplitudy dla Maxwella}
q_{\nu}({\bf x})=a\, A_{\nu}({\bf x}) \; , \;\;\; {\rm gdzie} \;\;\; \nu \equiv n-1 = 0,1,2,3 \; , 
\end{eqnarray}
gdzie $a$ jest pewną stałą. \\
Wykorzystując  metrykę Minkowskiego $(\eta^{\nu\mu})$,  definujemy amplitudy $q^{\nu}({\bf x})$ {\it dualne} do $q_{\nu}({\bf x})$:
\begin{eqnarray}
\label{amplitudy dla Maxwella dual}
q^{\nu}({\bf x}) \equiv \sum_{\mu=0}^{3} \eta^{\nu \mu} q_{\mu}({\bf x}) = a\, \sum_{\mu=0}^{3}  \eta^{\nu \mu} A_{\mu}({\bf x}) \equiv a\,A^{\nu}({\bf x}) \; , \;\;\; {\rm gdzie} \;\;\; \nu \equiv n-1 = 0,1,2,3 \; ,
\end{eqnarray}
gdzie wprowadzono dualne pola cechowania $A^{\mu}({\bf x})$:
\begin{eqnarray}
\label{pola cech dualne dla Maxwella}
A^{\nu}({\bf x}) \equiv \sum_{\mu=0}^{3}  \eta^{\nu \mu} A_{\mu}({\bf x})  \; , \;\;\; {\rm gdzie} \;\;\; \nu  = 0,1,2,3 \; .
\end{eqnarray}
Amplitudy $q_{\nu}({\bf x})$  są związane z punktowymi rozkładami prawdopodobieństwa $p_{n} \left({\bf x}  \right)$ następująco: 
\begin{eqnarray}
\label{rozklad p n dla Maxwella}
p_{n} \left({\bf x}  \right) \equiv p_{q_{\nu}} \left(  {\bf x}  \right)  =  q_{\nu}({\bf x}) q_{\nu}({\bf x}) = a^{2} A_{\nu}({\bf x}) A_{\nu}({\bf x}) \; , \;\;\; {\rm gdzie} \;\;\; \nu \equiv n-1 = 0,1,2,3 \; .
\end{eqnarray}
%
%
Widzimy więc, że w metodzie EFI dla elektrodynamiki Maxwella, indeks próby $n$ staje się indeksem czasoprzestrzennym. Zatem postać pojemności informacyjnej musi uwzględniać fakt dodatkowej  estymacji w kanałach czasoprzestrzennych, przyjmując zgodnie z ogólnymi zaleceniami  Rozdziału~\ref{Poj inform zmiennej los poloz},  postać współzmienniczą:
\begin{eqnarray}
\label{postac I dla p po x  suma po Mink Maxwell}
I &=& \sum_{\mu=0}^{3} {\int_{\cal X} d^{4}{\bf x}\, \eta^{\mu\mu}\frac{1}{{p_{q_{\mu}} \left(  {\bf x}  \right)}} \sum\limits_{\nu=0}^{3}  {\left( {\frac{{\partial p_{q_{\mu}} \left(  {\bf x}  \right)}}{{\partial {\bf x}_{\nu} }}}  {\frac{{\partial p_{q_{\mu}} \left(  {\bf x}  \right)}}{{\partial {\bf x}^{ \nu} }}} \right) } } \nonumber \\ 
&=&  4 \sum_{\mu=0}^{3} {\int_{\cal X} d^{4}{\bf x}\,  \sum\limits_{\nu=0}^{3}  {\left( {\frac{{\partial q_{\mu} \left(  {\bf x}  \right)}}{{\partial {\bf x}_{\nu} }}}  {\frac{{\partial q^{\mu} \left(  {\bf x}  \right)}}{{\partial {\bf x}^{ \nu} }}} \right) } }  \; .
\end{eqnarray}  
Dalej postępujemy jak w \cite{Frieden}. Stosujemy obie zasady informacyjne, strukturalną (\ref{zmodyfikowana obserwowana zas strukt z P i z kappa}) z $\kappa=1/2$ oraz wariacyjną (\ref{var K rozpisana}): 
\begin{eqnarray}
\label{IPs for Maxwell}
\widetilde{\textit{i'}} + \widetilde{\mathbf{C}} + \kappa \, \textit{q} = 0 \;, \;\;\; {\rm gdzie} \;\;\; \kappa=1/2 \;  \;\;\;\; {\rm oraz } \;\;\;\; \delta_{(q)}(I + Q) = 0 \; ,
\end{eqnarray}
które rozwiązujemy samospójnie, wraz z nałożonym dodatkowo warunkiem Lorentza:
\begin{eqnarray}
\label{warunek Lorentza}
\partial_{\mu} A^{\mu}=0 \; .
\end{eqnarray}
Zgodnie z (\ref{postac I dla p po x  suma po Mink Maxwell}) oraz (\ref{amplitudy dla Maxwella dual})  pojemność informacyjna $I$ jest 
następująca: 
%
\begin{eqnarray}
\label{IF for Maxwell}
I =  4\, a^{2} \sum_{\mu=0}^{3} \int_{\cal X} \! d^{4}{\bf x}\,  \sum\limits_{\nu=0}^{3}  {\left( {\frac{{\partial A_{\mu} \left(  {\bf x}  \right)}}{{\partial {\bf x}_{\nu} }}}  {\frac{{\partial A^{\mu} \left(  {\bf x}  \right)}}{{\partial {\bf x}^{ \nu} }}} \right) }  \, .
\end{eqnarray}
Właściwa postać informacji strukturalnej $Q\equiv Q_{M}$ dla równań ruchu Maxwella, została wyprowadzona w \cite{Frieden}. \\
  \\
{\it Warunki fizyczne}: W celu otrzymania zarówno $Q_{M}$, jak i wartości $\kappa$, muszą być przyjęte pewne fizyczne założenia \cite{Frieden}. {\it Po pierwsze} jest to warunek Lorentza, {\it po drugie}, pewna wstępna postać  $Q_{M}$. {\it Po trzecie}, wymagane jest również założenie o braku dodatkowych źródeł pola elektromagnetycznego w przestrzeni wolnej od czterowektora prądu $j^{\,\mu}=(c\rho,\vec{j}\,)$. 
 \\
 \\
Drugi z tych warunków, zapisany zgodnie z notacją zawartą  w (\ref{k form}), wyraża się żądaniem następującej faktoryzacji obserwowanej 
informacji strukturalnej:
\begin{eqnarray}
\label{qF M}
 \,a^{2}\, A_{\nu}\,A^{\nu}\,\texttt{q\!F}_{\nu}(A_{\nu}(x)) =  \frac{a}{\kappa}\, A^{\nu}\, F_{\nu}(j_{\nu}) \; ,
\end{eqnarray}
która jest następnie wykorzystana w (\ref{IPs for Maxwell}). Jeżeli  funkcja $F_{\nu}(j_{\nu})$ spełnia założenie o zależności jedynie od czterowektora prądu $j_{\,\nu}$ \cite{Frieden}, wtedy z zasad informacyjnych metody EFI, (\ref{IPs for Maxwell}), wynika zarówno wartość współczynnika efektywności $\kappa=1/2$, jak i równanie ciągłości strumienia  $\partial^{\nu}j_{\nu}=0$ \cite{Dziekuje za models building}. W efekcie informacja strukturalna dla równania Maxwella jest równa \cite{Frieden}:
\begin{eqnarray}
\label{Q in electrodyn}
Q = Q_{M} \equiv\,- \frac{64\pi}{c}\, a\,\sum_{\mu=0}^{3}\int_{\cal X} d^{4}{\bf x}\, j_{\mu}\, A^{\mu} \; . 
\end{eqnarray}
W końcu, dla pojemności informacyjnej $I$ jak w (\ref{IF for Maxwell}) i informacji strukturalnej $Q$ jak w  (\ref{Q in electrodyn}) oraz dla  $a=2$,  metoda EFI daje  wektorowe równanie falowe w cechowaniu Lorentza:
\begin{eqnarray}
\label{wave eq for A}
\Box A^{\mu} = (4\pi/c)j^{\mu} \; ,
\end{eqnarray}
co w konsekwencji prowadzi do znanej postaci równań Maxwella dla pola magnetycznego i elektrycznego \cite{Jackson,Frieden}. 
 \\
 \\ 
Podkreślmy, że proporcjonalność $q_{\nu}({\bf x}) = a A_{\nu}({\bf x})$ oraz  warunek normalizacji:
\begin{eqnarray}
\label{A normalization}
(1/4)\sum_{\nu=0}^{3}\int_{\cal X} d^{4} {\bf x} \, q_{\nu}^{2}({\bf x})=\sum_{\nu=0}^{3}\int_{\cal X} d^{4} {\bf x} \, A_{\nu}^{2}({\bf x}) = 1 \; ,
\end{eqnarray}
stawiają pytanie o znaczenie lokalizacji fotonu oraz istnienie jego funkcji falowej, które są ostatnio mocno dyskutowane w literaturze dotyczącej optyki. Dyskusja zawarta w  \cite{Roychoudhuri} wspiera głównie pogląd, który stawia równania Maxwella na tych samych podstawach co równanie Diraca w sformułowaniu pierwszej kwantyzacji. Fakt ten był wcześniej zauważony w pracy Sakurai \cite{Sakurai 2}. Przyjmując interpretację funkcji falowej fotonu jako mającą te same podstawy co występująca w  (\ref{free field eq all 2}) funkcja falowa cząstek materialnych z masą $m=0\,$,  z trzeciego z powyższych {\it fizycznych warunków} można by zrezygnować \cite{Frieden}. \\
\\
Zauważmy, że 
normalizacja\footnote{Normalizacja 
czteropotencjału $A_{\nu}$ zadana przez (\ref{A normalization}) do jedności mogłaby nie zachodzić  \cite{Leonhardt}. Warunkiem koniecznym dla $q_{\nu}({\bf x})$ jest, aby niezbędne średnie mogły być wyliczone. Porównaj tekst poniżej (\ref{observed IF}).
}
(\ref{A normalization}) czteropotencjału $A_{\nu}$ uzgadnia wartość stałej proporcjonalności\footnote{Otrzymanej wcześniej jako wynik uzgodnienia rezultatu metody EFI z równaniami 
Maxwella.
}
$a=2$ z wartością $N=4$ dla pola świetlnego. \\
  \\
Normalizacja (\ref{A normalization}), nałożona jako warunek na rozwiązanie równania (\ref{wave eq for A}), jest spójna z narzuceniem warunku początkowego Cauchy'ego we współrzędnej czasowej (por. dyskusja w \cite{Frieden}). 
Gdy jednocześnie  we współrzędnych  przestrzennych narzucony jest  warunek Dirichlet'a (lub Neumann'a), wtedy te mieszane czasowo-przestrzenne warunki brzegowe nie są współzmiennicze, co skutkuje tym, że rozwiązanie równania (\ref{wave eq for A}) nie jest współzmiennicze. \\
Jednakże tylko z mieszanymi warunkami brzegowymi rozwiązanie to jest jednoznaczne  \cite{Morse-Feshbach,Frieden}, co dla metody EFI jest warunkiem  koniecznym, gdyż  jest ona metodą estymacji statystycznej. Fakt ten stoi  w opozycji do przypadku, gdy warunek Dirichlet'a (lub Neumann'a) jest nałożony współzmienniczo zarówno we współrzędnych  przestrzennych jak i współrzędnej czasowej, gdyż co prawda otrzymane  rozwiązanie jest wtedy współzmiennicze, jednak  nie jest ono jednoznaczne.  \\
\\
W Rozdziale~\ref{Klein-Gordon scalars} powiedzieliśmy, że transformacja Fouriera tworzy rodzaj samosplątania pomiędzy dwoma reprezentacjami,  położeniową i pędową, dla realizowanych wartości zmiennych układu  występujących w  $I$ \cite{Frieden}. Strukturalna zasada informacyjna 
wyjaśnia to splątanie, zachodzące pomiędzy pędowymi stopniami swobody, jako spowodowane masą układu zgodną z (\ref{m E p}) lub (\ref{m E p dla q}).\\
%
%
Rozważmy bezmasową cząstkę, np. foton. Jeśli w zgodzie z powyższymi rozważaniami dla pola Max\-wella, relacja (\ref{free field eq all 2 real amplitudes}) określająca informację Fouriera miałaby być spełniona dla cząstki o masie   $m=0\,$  oraz dla amplitud 
interpretowanych zgodzie z  (\ref{amplitudy dla Maxwella}) i (\ref{A normalization}) jako charakteryzujących foton, 
wtedy również ze strony eksperymentu należałoby się spodziewać zarówno weryfikacji kwestii związanej z naturą funkcji falowej fotonu \cite{Roychoudhuri} jak i sygnaturą metryki czasoprzestrzeni.  W samej rzeczy, w metryce czasoprzestrzeni Minkowskiego (\ref{metryka M}), zgodnie z (\ref{m E p dla q}) jedyną  możliwością, aby cząstka była bezmasowa, jest zachodzenie warunku    $E^{2}/c^{2}-\vec{\wp}^{\,2} = 0$ 
dla wszystkich jej monochromatycznych, Fourierowskich mod, o ile  tylko mody te miałyby posiadać fizyczną interpretację dla cząstki bezmasowej. Warunek ten  
oznaczałby jednak, że mody Fourierowskie nie byłyby ze sobą splątane  \cite{Dziekuje za channel} (w przeciwieństwie do tego co zachodzi dla cząstki masowej) i w zasadzie powinna istnieć możliwość dokonania detekcji każdego  indywidualnego moda rozkładu Fouriera. Zatem, jeśli częstość  indywidualnego moda Fouriera impulsu świetlnego nie zostałaby zarejestrowana,  to mogłoby to oznaczać, że nie jest on  obiektem  fizycznym. Fakt ten  mógłby doprowadzić do problemów dla kwantowej interpretacji fotonu, określonego jako fizyczna realizacja konkretnego Fourierowskiego moda. Problem ten został ostatnio zauważony w związku z przeprowadzonymi eksperymentami optycznymi  \cite{Roychoudhuri}, z których wynika, że Fourierowski rozkład częstości impulsu świetlnego nie reprezentuje rzeczywistych optycznych częstości, 
co sugeruje, że być może rzeczywisty foton jest ``ziarnem  elektromagnetycznej substancji'' nie posiadającym Fourierowskiego przedstawienia.  \cite{Roychoudhuri,Dziekuje_Jacek_nova_2}. } 

\vspace{5mm}

\subsection{Dodatek: Metoda EFI dla teorii grawitacji}

\label{general relativity case}

Poniżej przedstawimy jedynie główne wyniki związane z konstrukcją EFI dla teorii grawitacji. Wychodząc z ogólnej postaci pojemności (\ref{postac I dla p po x bez n}) i postępując analogicznie jak powyżej dla pola Maxwella przy definicji amplitud dualnych,  otrzymujemy  amplitudową postać dla pojemności informacyjnej metody EFI. 
%
Następnie postępujemy już jak w \cite{Frieden}, gdzie zostało podane wyprowadzenie słabej (tzn. falowej) granicy równań ruchu Einsteina, dla przypadku pól z rangą $N=10$, a  amplitudy  $q_{n}({\bf x}) \equiv q_{\nu\mu}({\bf x})$ w liczbie dziesięć, są  rzeczywiste i  symetryczne w indeksach $\nu,\mu=0,1,2,3$. 
Zatem, pojemności informacyjna jest następująca:
\begin{eqnarray}
\label{I in gen rel}
I =  4 \int_{\cal X} d^{4}{\bf x} \sum_{\nu,\mu=0}^{3} \sum_{\gamma=0}^{3} \frac{\partial q_{\nu\mu}({\bf x})}{\partial {\bf x}_{\gamma}}\frac{\partial q^{\nu\mu}({\bf x})}{\partial {\bf x}^{\gamma}} \; ,
\end{eqnarray}
gdzie amplitudy dualne  mają postać $q^{\delta \tau}({\bf x}) =\sum_{\nu,\mu=0}^{3} \eta^{\delta \nu} \eta^{\tau \mu} q_{\nu\mu}({\bf x})$. 
Rozwiązując samospójne równania różniczkowe obu zasad informacyjnych, strukturalnej i wariacyjnej, wraz z nałożonym warunkiem Lorentza, $\sum_{\nu=0}^{3}\partial_{\nu} q^{\nu\mu}({\bf x})=0$, który redukuje współczynnik efektywności do wartości $\kappa=1/2$, otrzymujemy następującą postać informacji strukturalnej:
\begin{eqnarray}
\label{Q in gen rel}
Q = - \,8 \int_{\cal X} d^{4}{\bf x}\,\frac{1}{L^{4}}\sum_{\nu\,\mu=0}^{3} \bar{h}_{\nu\mu}\left[\frac{16\,\pi\, G}{c^{4}} \, T^{\nu\mu} - 2\,\Lambda\,\eta^{\nu\mu}\right] \; ,
\end{eqnarray}
gdzie
\begin{eqnarray}
\label{h and q in gen rel}
\bar{h}_{\nu\mu}({\bf x})\equiv L^{2}q_{\nu\mu}({\bf x}) \; ,
\end{eqnarray}
natomiast $\eta_{\nu\mu}$ jest metryką Minkowskiego, $G$ jest stałą grawitacyjną, 
$T_{\nu\mu}$ jest tensorem energii-pędu, 
$\Lambda$ jest tzw. stałą kosmologiczną, a stała $L$ jest charakterystyczną skalą, na której  amplitudy $\bar{h}_{\nu\mu}$ są 
uśrednione.\\
Rozwiązanie metody EFI pojawia się w postaci równania falowego dla amplitud $\bar{h}_{\nu\mu}$: 
\begin{eqnarray}
\label{h equation in gravit}
\Box \bar{h}_{\nu\mu}=\frac{16\,\pi\, G}{c^{4}}T_{\nu\mu}-2\,\Lambda\,\eta_{\nu\mu} \; ,
\end{eqnarray}
gdzie $\Box \equiv \sum_{\nu=0}^{3} \partial^{\nu}\partial_{\nu} = \sum_{\mu, \,\nu=0}^{3} \eta^{\mu \nu} \partial_{\mu}\partial_{\nu}$ jest operatorem d'Alemberta.  Równania (\ref{h equation in gravit}) mają postać właściwą dla równań ruchu w granicy słabego pola, w  ogólnej teorii względności. 
Rezultatem EFI jest również równanie ciągłości strumienia, $\sum_{\nu=0}^{3}\partial^{\nu} T_{\nu\mu}({\bf x})=0$, dla tensora 
$T_{\nu\mu}$. \\
 
Pozostaje pytanie o rozwiązanie dla dowolnie silnego pola. Jedna z odpowiedzi jest następująca. Ponieważ $\Box\bar{h}_{\nu\mu}$ jest jednoznaczną liniową aproksymacją tensora rangi drugiej  $2 R_{\nu\mu} - g_{\mu\nu}R$, gdzie $R_{\nu\mu}$ jest tensorem Ricci'ego \cite{Misner-Thorne-Wheeler},  $g_{\nu\mu}$ jest tensorem metrycznym, a $R = \sum_{\nu=0}^{3} R^{\nu}_{\,\nu}$, zatem można by uznać, że równanie Einsteina wyłania się jako jedyne możliwe, w tym sensie, że jego linearyzacją jest równanie falowe słabego pola (\ref{h equation in gravit}) otrzymane w EFI \cite{Frieden}. \\
Jednakże postać rozwiązania (\ref{h equation in gravit}) metody EFI mogłaby również sugerować inną selekcję przyszłego modelu grawitacji. Otóż w całym formalizmie EFI amplitudy są podstawą do definicji pola, a nie metryki czasoprzestrzeni. Zatem bardziej naturalnym wydaje się  zinterpretowanie  $\bar{h}_{\nu\mu}$ jako   pola 
grawitacyjnego\footnote{A nie 
tzw. metryki słabego pola 
pochodzącej od liniowej części  zaburzenia 
metryki. Metryka słabego pola ma postać: $\bar{h}_{\nu\mu} = h_{\nu\mu} - \frac{1}{2} \eta_{\mu \nu} h$, gdzie $h=\sum_{\mu, \nu =0}^{3} \eta^{\mu \nu} h_{\mu \nu}$ oraz $h_{\mu\nu} = g_{\mu\nu} - \eta_{\mu \nu}$, dla $|h_{\mu\nu}| << 1$.
}. 
Tak więc równanie (\ref{h equation in gravit}) metody EFI dla grawitacji leży bliżej innej teorii  grawitacji, nazywanej  ``szczególną teorią względności grawitacji'', która prowadzi do efektywnej teorii grawitacji typu Logunov'a \cite{Denisov-Logunov}, co poprzez widoczny związek z teoriami cechowania 
czyni ją ``bogatszą'' niż samą ogólną teorię względności. Problem statystycznego porównania obu teorii grawitacji jest kwestią przyszłych prac.

\newpage

\section[Dodatek: Informacyjny odpowiednik  drugiej zasady termodynamiki: 
Twierdzenie~$I$]{Dodatek: Informacyjny odpowiednik  drugiej zasady termodynamiki: \\Twier\-dzenie~$I$}

\label{Wyprowadzenie drugiej zasady termodynamiki}

Niech próba będzie $N=1$-wymiarowa. Podobnie jak przechodzi się z (\ref{I dla pn jeden parametr}) do (\ref{dys}) w celu otrzymania dyskretnej postaci 
informacji 
Fishera\footnote{Pojemność 
informacyjna  (\ref{postac I dla p po x bez n}) dla $N=1$ i skalarnego parametru $\theta$ wynosi:
\begin{eqnarray}
\label{I kinetyczny_dodatek}
I_{F} = \int dx \frac{1}{p_{\theta}(x)} \left(\frac{\partial p_{\theta}(x)}{\partial x}\right)^{2} \; ,
\end{eqnarray} 
i jest to informacja Fishera parametru $\theta$, 
gdzie informację o parametrze $\theta$ pozostawiono  w indeksie  rozkładu. 
Interesujący  związek informacji Fishera z entropią Kullbacka-Leiblera pojawia się na skutek zmiany  rozkładu zmiennej losowej  spowodowanego  nie  zmianą parametru rozkładu, ale  
zmianą   wartości $x$ na $x + \Delta x$. 
Zastąpmy więc (\ref{I kinetyczny_dodatek}) sumą Riemanna (\ref{zwiazek I oraz S}) i wprowadźmy wielkość:  
\begin{eqnarray}
\label{delta_dodatek}
\delta_{\Delta x} \equiv \frac{p_{\theta}\left({x_{k} + \Delta x}\right)}{p_{\theta}(x_{k})} - 1 \; .
\end{eqnarray}
Postępując dalej podobnie jak  przy przejściu od (\ref{delta}) do (\ref{I porownanie z Sn}) (tyle, że teraz rozkłady różnią się z powodu zmiany wartości fluktuacji $x$), otrzymujemy (\ref{iewf}).
}, 
tak można też pokazać, że można  przejść do jej następującej dyskretnej postaci:
\begin{eqnarray}
\label{zwiazek I oraz S}
I_{F} = \Delta x\sum_{k}{\frac{1}{p\left(x_{k}\right)}\left[\frac{p\left(x_{k}+ \Delta x \right) - p\left(x_{k}\right)}{\Delta x_{k}}\right]^{2}} \; .
\end{eqnarray}
Postępując w sposób analogiczny jak w przypadku wyprowadzenia  (\ref{I porownanie z Sn}) z (\ref{dys}), otrzymujemy:
\begin{eqnarray}
\label{iewf}
I_{F} = -\frac{2}{{\left({\Delta x}\right)^{2}}} \; S_{H}\left({p\left(x\right)| p\left({x+\Delta x}\right)}\right) \ .
\end{eqnarray}
%
%
Niech zmiana rozkładu prawdopodobieństwa z $p \equiv p(x)$ w chwili $t$ do $p_{\Delta x} \equiv p\left({x+\Delta x}\right)$ następuje na skutek infinitezymalnej zmiany czasu o $dt$. Jeśli więc dla rozkładów $p$ oraz $p_{\Delta x}$ entropia względna spełnia warunek: 
\begin{eqnarray}
\label{wzrost S wzglednego z czasem}
\frac{ d S_{H} \left({p\,| p_{\Delta x}}\right)(t) }{dt} \ge 0 \; ,
\end{eqnarray}
wtedy ze względu na (\ref{iewf}) informacja Fishera spełnia {\bf warunek}:
\begin{eqnarray}
\label{twierdzenie I}
\frac{{dI_{F} \left(t\right)}}{{dt}} \le 0
\end{eqnarray}
{\bf nazywny $I$-twierdzeniem}. Oznacza on, że informacja Fishera $I_{F}$ dla parametru $\theta$ maleje monotonicznie z czasem. \\
\\
{\bf Uwaga}: Dowód (\ref{wzrost S wzglednego z czasem}) oraz faktu, że relacja ta  istotnie prowadzi do drugiej zasady termodynamiki w rozumienia twierdzenia $H$, czyli dowód, który trzeba przeprowadzić dla entropii Shannona, można znaleźć w pracach na temat dynamiki  układów otwartych \cite{uklady otwarte} . \\
%
%
%
%
\\
Wprowadzenie twierdzenia $I$ jako mającego związek z twierdzeniem $H$ jest jak widać nieprzypadkowe. Ale również nieprzypadkowe jest  podobieństwo działające w drugą stronę. A mianowicie, pojemność informacja jest związana z rangą pola  $N$, co było widoczne w całej treści skryptu. W istniejącej literaturze można znaleźć wyprowadzenia pokazujące,  że istnieje entropijny odpowiednik zasady nieoznaczoności Heisenberga, oraz związek informacji Shannona z wymiarem reprezentacji grupy obrotów dla pól fermionowych rangi $N$. 

\subsection[Dodatek: Temperatura Fishera]{Dodatek:  Temperatura Fishera}
\label{Temperatura Fishera}

Jako kolejną ilustrację nieprzypadkowego podobieństwa rachunków informacyjnych i entropijnych rozważmy definicję temperatury Fishera dla parametru $\theta$. Otóż, informacja Fishera parametru $\theta$ pozwala na zdefiniowanie {\it temperatury Fishera} $T_{\theta}$ związanej z estymacją tego parametru:
\begin{eqnarray}
\label{temperatura I}
\frac{1}{T_{\theta}} \equiv - k_{\theta} \frac{\partial I_{F}}{\partial \theta} \; .
\end{eqnarray}
Tak określona temperatura $T_{\theta}$ jest miarą czułości informacji Fishera na zmianę parametru $\theta$, podobnie jak temperatura Boltzmanna $T$ (dla której zachodzi $\frac{1}{T} \equiv  \frac{\partial H_{B}}{\partial E} $) jest miarą czułości entropii Boltzmanna $H_{B}$ na zmianę energii $E$ układu. Temperatura Fishera znalazła swoje zastosowanie między innymi w badaniu rynków finansowych \cite{temperature 1,temperature ostatnia}.

{}

\printindex

\end{document}